%% file: Thesis.tex
\documentclass[
11pt,
a4paper,
twoside
]{report}
\input{Thesis_Preamble}

\begin{document}

\pagestyle{plain}

\pagenumbering{roman}


{\pagestyle{empty}
	
\input{Thesis_FrontCover} 
\cleardoublepage

\input{Thesis_FrontCoverFinal}

\cleardoublepage
}

\justifying
\setlength\parskip{6pt}
\setlength\parindent{18pt}

\phantomsection
\addcontentsline{toc}{chapter}{Abstract}
\input{Thesis_Abstract}

\cleardoublepage

\phantomsection
\addcontentsline{toc}{chapter}{Resumo}
\input{Thesis_Abstract_PT}

\cleardoublepage

\phantomsection
\addcontentsline{toc}{chapter}{Acknowledgments}
\input{Thesis_Acknowledgements}
\cleardoublepage

{
\hypersetup{linkcolor=black}
\setlength\parskip{0pt}

\tableofcontents
\cleardoublepage 

\phantomsection
\addcontentsline{toc}{chapter}{List of Publications}
\input{Thesis_Publications}

\cleardoublepage

%

}

\setcounter{page}{1}
\pagenumbering{arabic}


\input{Thesis_Introduction}
\cleardoublepage

\part[Random matrix theory of Markovian dissipation]{Random matrix theory of Markovian dissipation
\\\vspace{3cm}\normalsize\normalfont \begin{chapquote}[1.5cm]{Craig et al. (2022)\footnote{N. Craig et al., \textit{Snowmass Theory Frontier Report}, arXiv:2211.05772 (2022).}}
	Theory addresses our essential human curiosity to understand Nature at the deepest level, encompassing not only the actual laws of nature but also all the conceivable ones.
\end{chapquote}
}
\cleardoublepage

\input{Thesis_RandomLindblad}
\cleardoublepage

\input{Thesis_SYKLindblad}
\cleardoublepage

\input{Thesis_Kraus}
\cleardoublepage

\part[Symmetry classes of dissipative quantum matter]{Symmetry classes of dissipative quantum matter
\\\vspace{3cm}\normalsize\normalfont \begin{chapquote}[1.5cm]{Arruda Furtado (1885)}
	Knowing is classifying. Nothing can be known unless it can be classified, that is, compared and studied in its differences and similarities with other things.
\end{chapquote}
}
\cleardoublepage

\input{Thesis_Correlations}

\cleardoublepage

\input{Thesis_ClassificationSYK}

\cleardoublepage

\input{Thesis_ClassificationLindblad}

\cleardoublepage


\part[Miscellaneous topics]{Miscellaneous topics
	\\\vspace{3cm}\normalsize\normalfont \begin{chapquote}[1.5cm]{Marthe Troly-Curtin (1912)}
		Time you enjoy wasting is not wasted time.
	\end{chapquote}
}
\cleardoublepage

\input{Thesis_Circuits}

\cleardoublepage

\input{Thesis_QLaguerre}
\cleardoublepage

\input{Thesis_Conclusions}
\cleardoublepage


\phantomsection
\addcontentsline{toc}{part}{Appendices}
\part*{Appendices}
\cleardoublepage

\appendix



\input{Thesis_App_KeldyshLindblad}

\cleardoublepage

\input{Thesis_App_Kraus_GinUECUE} 
\cleardoublepage

\input{Thesis_App_KrausTransposition} 
\cleardoublepage

\input{Thesis_App_SYKCombinatorics} 
\cleardoublepage

\input{Thesis_App_Asymptotics_WSYK}

\cleardoublepage


\phantomsection
\addcontentsline{toc}{chapter}{References}
\setlength{\bibsep}{0pt plus 0.3ex}

\vfill
\cleardoublepage

\end{document}

%% file: Thesis_Preamble.tex
\usepackage[utf8]{inputenc}
\usepackage[T1]{fontenc}
\usepackage[english]{babel}

\usepackage{amsmath}  
\usepackage{amsthm}   
\usepackage{amsfonts}
\usepackage{amssymb}

\usepackage{graphicx}
\usepackage[dvipsnames]{xcolor}
\usepackage{tabularx}
\usepackage{multirow}
\usepackage{booktabs}

\usepackage{bm}
\usepackage{bbm}

\usepackage{tikz}
\usepackage[compat=1.1.0]{tikz-feynman} 

\usepackage{physics}

\usepackage{enumerate}
\usepackage{mathrsfs}
\usepackage[export]{adjustbox}
\usepackage[math]{cellspace}
\usepackage{url}

\usepackage[numbers,sort&compress]{natbib} 

\usepackage{ragged2e} 
\usepackage{setspace}

\usepackage{parskip}
\usepackage{hyperref} 
\usepackage[figure,table]{hypcap}
\usepackage{notoccite}
\hypersetup{colorlinks,
			linkcolor=BrickRed,
            citecolor=MidnightBlue,
            urlcolor=MidnightBlue,
            bookmarks=true,
            bookmarksopen=true, 
            bookmarksopenlevel=0,   
            bookmarksnumbered=true,
	        pdftitle={Thesis Lucas Sa},
            pdfauthor={Lucas Sa}
        }
    
\usepackage{geometry}	
\makeatletter
\newenvironment{chapquote}[2][2cm]
{\setlength{\@tempdima}{#1}%
	\def\chapquote@author{#2}%
	\noindent\parshape 1 \@tempdima \dimexpr\textwidth-2\@tempdima\relax%
	\itshape}
{\par\normalfont\hfill---\ \chapquote@author\hspace*{\@tempdima}\par\bigskip}
\makeatother

\graphicspath{{./}}

\newcounter{ls}

\renewcommand*\d{\mathop{}\!\mathrm{d}}
\newcommand{\dirac}[1]{\,\delta\!\left(#1\right)}
\newcommand{\heav}[1]{\,\Theta\!\left(#1\right)}

\newcommand{\av}[1]{\left\langle#1\right\rangle}

\newcommand{\conj}[1]{{#1}^*}
\newcommand{\pd}{\partial}
\renewcommand{\vec}[1]{\underline{#1}}
\renewcommand{\i}{\mathrm{i}}
\renewcommand{\(}{\left(}
\renewcommand{\)}{\right)}
\newcommand{\id}{\mathbbm{1}}

\newcommand{\eff}{\mathrm{eff}}
\newcommand{\scT}{\mathcal{T}}
\newcommand{\scC}{\mathcal{C}}
\newcommand{\scP}{\mathcal{P}}
\newcommand{\scQ}{\mathcal{Q}}
\newcommand{\scU}{\mathcal{U}}

\newcommand{\scI}{\mathcal{I}}
\newcommand{\scK}{\mathcal{K}}
\newcommand{\scL}{\mathcal{L}}
\newcommand{\scA}{\mathcal{A}}
\newcommand{\scB}{\mathcal{B}}
\newcommand{\scLH}{\mathcal{L}_{\mathrm{H}}}
\newcommand{\scLD}{\mathcal{L}_{\mathrm{D}}}
\newcommand{\scLJ}{\mathcal{L}_{\mathrm{J}}}
\newcommand{\swap}{\mathcal{S}}
\newcommand{\tphi}{\tilde{\phi}}
\newcommand{\balpha}{{\overline{\alpha}}}

\newcommand{\sT}{\mathcal{T}}
\newcommand{\sP}{\mathcal{P}}
\newcommand{\sQ}{\mathcal{Q}}
\renewcommand{\tt}{\mathbf{t}}

\newcommand{\RH}{\check{\mathcal R}}
\newcommand{\RHp}{\mathcal R}

\newcommand{\sL}{\mathcal{L}}

\newcommand{\sN}{\mathcal{N}}

\newcommand{\sD}{\mathcal{D}}
\newcommand{\sC}{\mathcal{C}}
\newcommand{\rmT}{\mathrm{T}}
\newcommand{\rmTb}{{\bar{\mathrm{T}}}}
\newcommand{\rmK}{\mathrm{K}}
\newcommand{\rmR}{\mathrm{R}}
\newcommand{\rmA}{\mathrm{A}}
\newcommand{\rmH}{\mathrm{H}}

\newcommand{\Omegatilde}{\widetilde{\Omega}}
\newcommand{\be}{\begin{eqnarray}}
\newcommand{\ee}{\end{eqnarray}}

\newcommand{\edoc}{\end{document}}

\newcommand{\eref}[1]{(\ref{#1})}

\newcommand{\sign}{{\rm sign}}

\newcommand{\diag}{\hbox{diag}}
\renewcommand{\(}{\left(}
\renewcommand{\)}{\right)}
\newcommand{\bmat}{\left ( \begin{array}{cc} }
\newcommand{\emat}{\end{array} \right ) }

\newcommand{\beq}{\begin{equation}}
\newcommand{\beqs}{\begin{equation*}}
\newcommand{\eeq}{\end{equation}}
\newcommand{\eeqs}{\end{equation*}}

\newcommand{\sR}{\mathcal{R}}
\newcommand{\sS}{\mathcal{S}}

\newcommand{\cH}{\mathcal{H}}
\newcommand{\adiag}{\mathrm{antidiag}}
\renewcommand{\mod}{\mathrm{\,mod\,}}

\renewcommand{\va}{{\vec{a}}}
\renewcommand{\vb}{{\vec{b}}}
\newcommand{\vc}{{\vec{c}}}
\newcommand{\vd}{{\vec{d}}}
\newcommand{\vast}{{\vec{a}^{\raisebox{-2.5pt}{$\scriptstyle\prime$}}}}
\newcommand{\vbst}{{\vec{b}^{\raisebox{-2.5pt}{$\scriptstyle\prime$}}}}
\newcommand{\must}{{\mu^{\raisebox{-2.5pt}{$\scriptstyle\prime$}}}}
\newcommand{\qb}{{\hat{q}}}
\newcommand{\sM}{\mathcal{M}}
\newcommand{\sigmaH}{\sigma_\mathrm{H}}
\newcommand{\sigmaL}{\sigma_\mathrm{L}}
\newcommand{\EH}{E_{0\mathrm{H}}}
\newcommand{\EL}{E_{0\mathrm{L}}}

\newcommand{\matAIIId}{
	\begin{pmatrix}  & A \\ B &  \end{pmatrix}
}
\newcommand{\matAIId}{
	\begin{pmatrix} A & B \\ C & A^\top \end{pmatrix},\ \begin{cases}B=-B^\top\\C=-C^\top\end{cases}
}
\newcommand{\matC}{
	\begin{pmatrix} A & B \\ C & -A^\top \end{pmatrix},\ \begin{cases}B=B^\top\\C=C^\top\end{cases}
}
\newcommand{\matAIdp}{
	\begin{pmatrix} & A \\ A^\top & \end{pmatrix}
}
\newcommand{\matAIIdp}{
	\begin{pmatrix}
		&         & A & B \\
		&         & C & D \\
		D^\top & -B^\top &   &   \\
		-C^\top &  A^\top &   &
	\end{pmatrix}
}
\newcommand{\matAIdm}{
	\begin{pmatrix} & A \\ B & \end{pmatrix},\ \begin{cases}A=A^\top\\B=B^\top\end{cases}
}
\newcommand{\matAIIdm}{
	\begin{pmatrix} & A \\ B & \end{pmatrix},\ \begin{cases}A=-A^\top\\B=-B^\top\end{cases}
}

\newcommand{\matAIm}{
	\begin{pmatrix} & A \\ A^* & \end{pmatrix}
} 

\newcommand{\matAIp}{
	\begin{pmatrix} & A \\ B & \end{pmatrix},\ \begin{cases}A=A^*\\B=B^*\end{cases}
}

\newcommand{\matBDId}{
	\begin{pmatrix}
		A & B \\
		B^\top &  C
	\end{pmatrix},\
	\begin{cases}
		A=A^\dagger=A^*=A^\top\\
		B=-B^*\\
		C=C^\dagger=C^*=C^\top\\
	\end{cases}
}

\newcommand{\matBDIalt}{
	\begin{pmatrix}
		A & B \\
		-B^\top &  C
	\end{pmatrix},\
	\begin{cases}
		A=-A^\dagger=-A^\top=A^*\\
		B=-B^*\\
		C=-C^\dagger=-C^\top=C^*
	\end{cases}
}

\newcommand{\matCIalt}{
	\begin{pmatrix}
		A & B \\
		-B^* &  -A^*
	\end{pmatrix},\
	\begin{cases}
		A=A^\dagger\\
		B=B^\top
	\end{cases}
}

\newcommand{\matBDIdpp}{
	\begin{pmatrix}
		& & A & B \\
		& & C &  D \\
		A^\top & C^\top & & \\
		B^\top & D^\top & &
	\end{pmatrix},\
	\begin{cases}
		A=A^*\\
		B=-B^*\\
		C=-C^*\\
		D=D^*
	\end{cases}
}

\newcommand{\matBDIdmm}{
	\begin{pmatrix}
		& & A & B \\
		& & B^\top &  C \\
		A^* & -B^* & & \\
		-B^\dagger & C^* & &
	\end{pmatrix},\
	\begin{cases}
		A=A^\top\\
		C=C^\top
	\end{cases}
}

\newcommand{\matBDIdpm}{
	\begin{pmatrix}
		& A \\
		A^\top &  
	\end{pmatrix},\
	A=A^\dagger
}

\newcommand{\matBDIdmp}{
	\begin{pmatrix}
		& A \\
		B &  
	\end{pmatrix},\
	\begin{cases}
		A=A^\dagger=A^\top=A^*\\
		B=B^\dagger=B^\top=B^*
	\end{cases}
}

%% file: Thesis_FrontCover.tex

\thispagestyle{empty}

\includegraphics[width=0.25\textwidth]{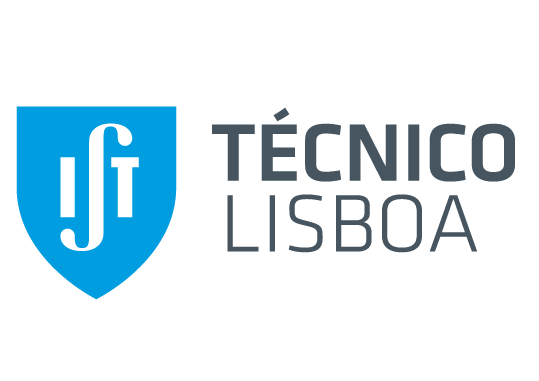}

\begin{center}

{\LARGE
\textbf{UNIVERSIDADE DE LISBOA}\\
\vspace{0.2cm}
\textbf{INSTITUTO SUPERIOR TÉCNICO}
}\\

\vspace{3.5cm}
{\Huge \textbf{Signatures of dissipative quantum chaos}}\\
\vspace{2.5cm}
{\LARGE \textbf{Lucas de Barros Pacheco Seara de Sá}} \\
\vspace{2.0cm}

{%
\Large
\begin{tabular}{ll}
	Supervisor:    & Doctor Pedro José Gonçalves Ribeiro \\
	Co-supervisor: & Doctor Toma\v{z} Prosen
\end{tabular} } \\

\vspace{2.0cm}

{
\Large	
Thesis approved in public session to obtain the PhD Degree in
\\
\textbf{Physics}
\\
\vspace{0.5cm}
Jury final classification
\\
\vspace{0.2cm}
\textbf{Pass with Distinction and Honour}
\vspace{1.5cm}
}

\vfill
{\Large \textbf{2023}}
\end{center}

%% file: Thesis_FrontCoverFinal.tex

\thispagestyle{empty}

\includegraphics[width=0.25\textwidth]{IST_A_RGB_POS}

\begin{center}
	
{\LARGE
\textbf{UNIVERSIDADE DE LISBOA}\\
\vspace{0.2cm}
\textbf{INSTITUTO SUPERIOR TÉCNICO}
}\\
	
\vspace{1.0cm}
{\Huge \textbf{Signatures of dissipative quantum chaos}}\\
\vspace{1.0cm}
{\LARGE \textbf{Lucas de Barros Pacheco Seara de Sá}} \\
\vspace{1cm}
	
{%
\Large
\begin{tabular}{ll}
Supervisor:    & Doctor Pedro José Gonçalves Ribeiro \\
\vspace{-0.3cm}
Co-supervisor: & Doctor Toma\v{z} Prosen
\end{tabular} } \\
	
\vspace{0.8cm}

{
\Large	
Thesis approved in public session to obtain the PhD Degree in
\\
\textbf{Physics}
\\
\vspace{+0.5cm}
Jury final classification
\\
\vspace{0.1cm}
\textbf{Pass with Distinction and Honour}
}

\vspace{0.8cm}
{\Large \textbf{Jury}}\\

\end{center}

{
\large
\vspace{0.1cm}
	\textbf{Chairperson:}\\
	\-\hspace{0.6cm}Doctor José Luís Rodrigues Júlio Martins, Instituto Superior Técnico, Universidade de Lisboa \\
	\textbf{Members of the committee:}\\
	\-\hspace{0.6cm}Doctor Shinsei Ryu, Princeton University, EUA \\ 
	\-\hspace{0.6cm}Doctor Sebastian Diehl, Institute for Theoretical Physics, University of Cologne, Alemanha\\
	\-\hspace{0.6cm}Doctor Pedro Domingos Santos Sacramento, Instituto Superior Técnico, Universidade de Lisboa \\
	\-\hspace{0.6cm}Doctor Pedro José Gonçalves Ribeiro, Instituto Superior Técnico, Universidade de Lisboa\\
}

\begin{center}
\vspace{0.2cm}
{\Large
\textbf{Funding Institutions}\\
\vspace{0.1cm}
Fundação para a Ciência e a Tecnologia
}
	
\vfill
{\Large \textbf{2023}}

\end{center}

%% file: Thesis_Abstract.tex

\section*{Abstract}

\justifying{
Understanding the far-from-equilibrium dynamics of dissipative quantum systems, where dissipation and decoherence coexist with unitary dynamics, is an enormous challenge with immense rewards. Often, the only realistic approach is to forgo a detailed microscopic description and search for signatures of universal behavior shared by collections of many distinct, yet sufficiently similar, complex systems. Quantum chaos provides a powerful statistical framework for addressing this question, relying on symmetries to obtain information not accessible otherwise. This thesis examines how to reconcile chaos with dissipation, proceeding along two complementary lines.
In Part I, we apply non-Hermitian random matrix theory to open quantum systems with Markovian dissipation and discuss the relaxation timescales and steady states of three representative examples of increasing physical relevance: single-particle Lindbladians and Kraus maps, open free fermions, and dissipative Sachdev-Ye-Kitaev (SYK) models. For systems with single-body quantum chaos, we establish the universality of their steady states. For the strongly-interacting SYK model, we find the relaxation to be dissipation-driven at strong dissipation (and compute it analytically) and chaos-driven, with an anomalously large gap, at weak dissipation.
In Part II, we investigate the symmetries, correlations, and universality of many-body open quantum systems, classifying several models of dissipative quantum matter. We use the non-Hermitian SYK Hamiltonian as a prototypical toy model with both generic features of the universal quantum ergodic state reached around the Heisenberg time and nonuniversal, but still generic, properties of many-body quantum dissipative systems in their approach to ergodicity. Moreover, we go beyond effective non-Hermitian Hamiltonians, and consider the symmetry classification of many-body Lindbladians. We obtain a tenfold classification in the absence of unitary symmetries, realized in realistic spin-chain models, which can be enriched by the presence of unitary symmetries. In all examples, we investigate the spectral correlations and confirm that they are described by universal random matrix theory---the hallmark of quantum chaotic behavior.
From a theoretical viewpoint, this thesis lays out a generic framework for the study of the universal properties of realistic, chaotic, and dissipative quantum systems. From a practical viewpoint, it provides the concrete building blocks of dynamical dissipative evolution constrained by symmetry, with potential technological impact on the fabrication of complex quantum structures.
}

\vfill

\noindent\textbf{\large Keywords:} open quantum systems, quantum chaos, random matrix theory, symmetry classifications, Sachdev-Ye-Kitaev model.

%% file: Thesis_Abstract_PT.tex

\section*{Resumo}

\justifying{
Compreender a dinâmica fora do equilíbrio de sistemas quânticos dissipativos, onde a dissipação e a decoerência coexistem com a dinâmica unitária, é um enorme desafio. Frequentemente, a única abordagem realista é renunciar a uma descrição microscópica detalhada e, ao invés, procurar por assinaturas de comportamento universal partilhadas por conjuntos de sistemas complexos distintos mas suficientemente semelhantes. O caos quântico fornece o enquadramento estatístico para abordar esta questão, recorrendo a simetrias para obter informação não acessível de outra forma. Esta tese examina como conciliar caos com dissipação, procedendo ao longo de duas linhas complementares.
Na Parte I, aplicamos a teoria das matrizes aleatórias não-Hermíticas a sistemas quânticos abertos com dissipação Markoviana e discutimos as escalas de tempo de relaxação e estados estacionários de três exemplos representativos com crescente relevância física: Lindbladianos e mapas de Kraus de uma partícula, fermiões livres abertos e modelos de Sachdev-Ye-Kitaev (SYK) dissipativos. Para sistemas com caos quântico a um corpo, estabelecemos a universalidade dos seus estados estacionários. Para o modelo SYK com interações fortes, descobrimos que o relaxamento é dominado pela dissipação no regime de dissipação forte e dominado pelo caos, com um hiato anormalmente grande, no regime de dissipação fraca.
Na Parte II, investigamos as simetrias, correlações e universalidade de sistemas quânticos abertos de muitos corpos, classificando vários modelos de matéria quântica dissipativa. Usamos o Hamiltoniano SYK não-Hermítico como um modelo simplificado prototípico com não só um estado estacionário ergódico universal mas também propriedades não universais, mas ainda genéricas, de sistemas dissipativos quânticos de muitos corpos na sua aproximação à ergodicidade. Além disso, vamos além dos Hamiltonianos não-Hermíticos efetivos e consideramos a classificação de simetria dos Lindbladianos de muitos corpos. Encontramos uma classificação com dez classes na ausência de simetrias unitárias, realizada em modelos realistas de cadeias de spin, a qual pode ser enriquecida por simetrias unitárias. Em todos os exemplos, investigamos as correlações espectrais e confirmamos que elas são descritas pela teoria das matrizes aleatórias---o traço distintivo do comportamento caótico quântico.
Do ponto de vista teórico, esta tese desenvolve um enquadramento genérico para o estudo das propriedades universais de sistemas quânticos realistas, caóticos e dissipativos. Do ponto de vista prático, fornece as componentes para a construção de evoluções dinâmicas dissipativas restringidas por simetrias, com potencial impacto tecnológico na fabricação de estruturas quânticas complexas.
}

\vfill

\noindent\textbf{\large Palavras-chave:} sistemas quânticos abertos, caos quântico, teoria das matrizes aleatórias, classificações de simetria, modelo de Sachdev-Ye-Kitaev

%% file: Thesis_Acknowledgements.tex

\section*{Acknowledgments}

{\setlength{\parindent}{0pt}
	
First and foremost, I have to thank my thesis supervisors, Pedro Ribeiro and Toma\v{z} Prosen, for guiding me throughout the last five years, for their enthusiastic encouragement on my progress, sharp critique, and general advice on all types of matters, for encouraging me to set my own research agenda, and for always finding some time for me in their (quite) busy schedules. This work could not exist without them.

I extend my warmest gratitude and appreciation to Antonio Garc\'ia-Garc\'ia and Jac Verbaarschot, who, in many ways, took on a role of third and fourth supervisors, not only through scientific collaboration but also regarding my academic development. I have benefited immensely from Antonio's endless impetus and source of fresh ideas and Jac's insight and patience in working out every last detail.

I would also like to thank my other collaborators in the work described in this thesis: Tankut Can, Jo{\~a}o Costa, Andrea De Luca, Can Yin, and Jie Ping Zheng. I have thoroughly enjoyed the many discussions we had. 

Finally, thank you to the many people whom I have discussed my work with, and learned from, and, in particular, to Shinsei Ryu for hosting me in Princeton during my exchange there. A thank you note also to my fellow members in the q.m@t group at T\'ecnico, who made the last five years the most pleasant experience.

My PhD studies were supported by Funda\c{c}\~ao para a Ci\^encia e a Tecnologia (FCT-Portugal) through Grant No.\ SFRH/BD/147477/2019. I acknowledge the computer resources of FMF in Ljubljana. The numerical computations of this work were mainly performed on FMF's cluster \textit{Olimp} and supported by the H2020 ERC Advanced Grant 694544-OMNES. I also acknowledge partial support from the QuantERA II Programme that has received funding from the European Union’s Horizon 2020 research and innovation programme under Grant Agreement No.\ 101017733.

}

%% file: Thesis_Publications.tex

\chapter*{List of Publications}

{
\setlength\parskip{6pt}
\setstretch{1.0}


\begin{itemize}
 \item
	A.~M.~Garc\'ia-Garc\'ia, \textbf{L.~Sá}, J.~J.~M.~Verbaarschot, and C.~Yin\\
	\textit{Sixfold way of traversable wormholes in the Sachdev-Ye-Kitaev model}\\
	\href{https://arxiv.org/abs/2305.09663}{[arXiv:2305.09663]}
\item
	J.~Costa, P.~Ribeiro, A.~de Luca, T.~Prosen, and \textbf{L.~Sá}\\
	\textit{Spectral and steady-state properties of fermionic random quadratic Liouvillians}\\
	\href{https://scipost.org/SciPostPhys.15.4.145}{SciPost Phys.\ \textbf{15}, 145 (2023)}
	\quad
	\href{https://arxiv.org/abs/2210.07959}{[arXiv:2210.07959]}
\item
    \textbf{L.~Sá}, P.~Ribeiro, and T.~Prosen\\
    \textit{Symmetry Classification of Many-Body Lindbladians: Tenfold Way and Beyond}\\
    \href{https://journals.aps.org/prx/abstract/10.1103/PhysRevX.13.031019}{Phys.\ Rev.\ X \textbf{13}, 031019 (2023)}
    \quad \href{https://arxiv.org/abs/2212.00474}{[arXiv:2212.00474]}
\item
	A.~M.~Garc\'ia-Garc\'ia, \textbf{L.~Sá}, J.~J.~M.~Verbaarschot, and J.~P.~Zheng\\
	\textit{Keldysh wormholes and anomalous relaxation in the dissipative Sachdev-Ye-Kitaev model}\\
	\href{https://journals.aps.org/prd/abstract/10.1103/PhysRevD.107.106006}{Phys.\ Rev.\ D \textbf{107}, 106006 (2023)}
	\quad
	\href{https://arxiv.org/abs/2210.01695}{[arXiv:2210.01695]}
\item
	A.~M.~Garc\'ia-Garc\'ia, \textbf{L.~Sá}, and J.~J.~M.~Verbaarschot\\
	\textit{Universality and its limits in non-Hermitian many-body quantum chaos using the Sachdev-Ye-Kitaev model}\\
	\href{https://doi.org/10.1103/PhysRevD.107.066007}{Phys.\ Rev.\ D \textbf{107}, 066007 (2023)}
	\quad
	\href{https://arxiv.org/abs/2211.01650}{[arXiv:2211.01650]}
\item
    \textbf{L.~Sá}, P.~Ribeiro, and T.~Prosen\\
    \textit{Lindbladian dissipation of strongly-correlated quantum matter}\\
    \href{https://journals.aps.org/prresearch/abstract/10.1103/PhysRevResearch.4.L022068}{Phys.\ Rev.\ Res.\ \textbf{4}, L022068 (2022)}
    \quad
    \href{https://arxiv.org/abs/2112.12109}{[arXiv:2112.12109]}
\item
    A.~M.~Garc\'ia-Garc\'ia, \textbf{L.~Sá}, and J.~J.~M.~Verbaarschot\\
    \textit{Symmetry Classification and Universality in Non-Hermitian Many-Body Quantum Chaos by the Sachdev-Ye-Kitaev Model}\\
    \href{https://journals.aps.org/prx/abstract/10.1103/PhysRevX.12.021040}{Phys.\ Rev.\ X \textbf{12}, 021040 (2022)}
    \quad
    \href{https://arxiv.org/abs/2110.03444}{[arXiv:2110.03444]}
\item
	\textbf{L.~Sá} and A.~M.~Garc\'ia-Garc\'ia\\
	\textit{Q-Laguerre spectral density and quantum chaos in the Wishart-Sachdev-Ye-Kitaev model}\\
	\href{https://doi.org/10.1103/PhysRevD.105.026005}{Phys.\ Rev.\ D \textbf{105}, 026005 (2022)}
	\quad
	\href{https://arxiv.org/abs/2104.07647}{[arXiv:2104.07647]}
\item
    \textbf{L.~Sá}, P.~Ribeiro, and T.~Prosen\\
    \textit{Integrable nonunitary open quantum circuits}\\
    \href{https://journals.aps.org/prb/abstract/10.1103/PhysRevB.103.115132}{Phys.\ Rev.\ B \textbf{103}, 115132 (2021)}
    \quad
    \href{https://arxiv.org/abs/2011.06565}{[arXiv:2011.06565]}
\item 
    \textbf{L.~Sá}, P.~Ribeiro, T.~Can, and T.~Prosen\\
    \textit{Spectral transitions and universal steady states in random Kraus maps and circuits}\\
    \href{https://doi.org/10.1103/PhysRevB.102.134310}{Phys. Rev. B \textbf{102}, 134310 (2020)}
    \quad
    \href{https://arxiv.org/abs/2007.04326}{[arXiv:2007.04326]}
\end{itemize}
\clearpage
\begin{itemize}
\item 
	\textbf{L.~Sá}, P.~Ribeiro, and T.~Prosen\\
	\textit{Complex Spacing Ratios: A Signature of Dissipative Quantum Chaos}\\
	\href{https://link.aps.org/doi/10.1103/PhysRevX.10.021019}{Phys.\ Rev.\ X \textbf{10}, 021019 (2020)}
	\quad
	\href{https://arxiv.org/abs/1910.12784}{[arXiv:1910.12784]}
\item 
	\textbf{L.~Sá}, P.~Ribeiro, and T.~Prosen\\
	\textit{Spectral and steady-state properties of random Liouvillians}\\
	\href{https://iopscience.iop.org/article/10.1088/1751-8121/ab9337}{J.\ Phys.\ A: Math.\ Theor.\ \textbf{53}, 305303 (2020)}
	\quad
	\href{https://arxiv.org/abs/1905.02155}{[arXiv:1905.02155]}
\end{itemize}

%% file: Thesis_Introduction.tex

\chapter{Introduction}
\label{chapter:introduction}

The butterfly effect is a well-known manifestation of chaos: a butterfly flapping its wings in Brazil could set off a tornado in Texas~\cite{lorenz1963}. In the quantum world of microscopic particles, the uncertainty and complexity of outcomes are embodied by a statistical framework known as quantum chaos~\cite{guhr1998,stockmann1999,haake2013}: complicated quantum systems behave as if they were random. Broadly speaking, quantum chaos aims to answer two questions:
\begin{enumerate}[(i)]
	\item How to model a generic interacting quantum system?
	\item What do generic quantum systems have in common?
\end{enumerate}
It applies to systems as diverse as heavy nuclei ~\cite{bohigas1988,mitchell2010}, quantum computers~\cite{landsman2019Nature,mi2021Science,block2021PRX}, and black holes~\cite{maldacena2016JHEP,cotler2017JHEP}, and has been one of the driving forces in the development of nuclear and condensed matter physics and, more recently, quantum gravity. 

Until recently, the theory of quantum chaos still missed a key ingredient: openness. No physical system is truly isolated from its environment (be it an experimental apparatus or the rest of the universe), and the interaction of a system with its environment has dramatic consequences, such as dissipation (loss of energy by the system) or decoherence (loss of its very quantum nature)~\cite{breuerpetruccione,weiss2012,braun2001}. 
Openness can be a hindrance, e.g., for quantum computation, with relaxation and decoherence limiting the accuracy of near-term devices~\cite{shor1995PRA,preskill2018quantum}. Or it can be a resource, as in atomic and mesoscopic physics, where experimental advances now allow us to engineer special states of matter stabilized by dissipation~\cite{diehl2008NatPhys,verstraete2009NatPhys,barreiro2011Nat,diehl2011NatPhys}.

In this thesis, we examine how to reconcile chaos and dissipation in strongly-interacting quantum systems. It reports my contributions to the development of the field of \textit{dissipative quantum chaos}, along two complementary lines (corresponding to the first two parts of the thesis, with some additional topics covered in Part III): random matrix theories of Markovian dissipation and symmetry classes of dissipative quantum matter. To formulate them, we first need to review their two constituent ingredients: quantum chaos (through a brief historical overview in Sec.~\ref{sec:intro_quantum_chaos}) and open quantum systems (Sec.~\ref{sec:intro_OQS}). These two strands of research are beautifully united by non-Hermitian random matrix theory (RMT), which saw a rapid development in the 1990s, laying what we can now understand as the foundations of dissipative quantum chaos. We review its main results in Sec.~\ref{sec:intro_nonHermitian}. After a dormant period, dissipative quantum chaos saw a renaissance in 2018--2023, which coincided with the writing of this thesis. In Sec.~\ref{sec:intro_developments}, we review these developments. Finally, an outline and the main results of the thesis are given in Sec.~\ref{sec:intro_outline}.

\section{Quantum chaos}
\label{sec:intro_quantum_chaos}

A (nonlinear) classical dynamical system is chaotic if it shows exponential sensitivity to initial conditions (i.e., phase-space trajectories diverge exponentially in time for a small deviation in initial conditions). A natural question is how to extend the notion of chaos to a quantum setting. At first sight, this seems impossible, as quantum mechanics is a linear theory, and therefore, small deviations in initial states cannot grow exponentially. On the other hand, for a particular limit of quantum mechanics---i.e., classical mechanics---to exhibit chaos, there must exist some underlying quantum mechanism responsible for it. This state of affairs led to the development of quantum chaos as the search for---and the quantitative understanding of---so-called \emph{signatures of quantum chaos}: quantities which distinguish quantum systems with an underlying notion of chaos from those without. Quite remarkably, to this day, there is still no general and rigorous definition of what a quantum chaotic system is, and the presence of signatures of quantum chaos is used as the \emph{de facto} definition of quantum chaotic behavior.

Many different signatures of quantum chaos have been proposed over the years; a non-exhaustive list includes typicality and the eigenstate thermalization hypothesis (ETH)~\cite{d'alessio2014}, scrambling of quantum information and quantum complexity~\cite{sekino2008JHEP,swingle2016PRA,maldacena2016JHEP,roberts2017JHEP}, operator growth~\cite{nahum2018PRX,keyserlingk2018PRX}, and sensitivity to changes in the dynamics~\cite{gorin2006,bin2020PRL}. The most popular signature is, however, spectral statistics. Indeed, 
in the search for an understanding of quantum chaos, no other tool has been more useful than random matrix theory (RMT)~\cite{guhr1998,mehta2004,haake2013,forrester2010}. 

\subsection{Statistical theories of spectra}
\label{sec:intro_spectra}

Describing strongly-interacting many-body quantum systems is a daunting task. Not only do we not possess the tools to solve most interacting models, the vast majority of actual physical systems is so complicated that we do not even know how to write down an accurate Hamiltonian describing it. The best we can often do, then, is to propose a simplified toy model that captures the main features we are interested in. In the 1950s, Wigner took this reasoning to its logical extreme and, for cases where our ignorance is maximal, proposed to describe the Hamiltonian of a complicated system---in his case, the compound nucleus---by a random matrix~\cite{wigner1951b,wigner1955}, retaining only a small set of symmetries (e.g., particle number and angular momentum conservation, time-reversal symmetry), but otherwise randomizing all degrees of freedom. This opened the possibility of developing the statistical mechanics of such complicated systems. 
However, while in ordinary statistical mechanics, in the absence of detailed knowledge of a \textit{state} one resorts to an ensemble of states compatible with the dynamics of the system, this new statistical approach renounces the exact description of the system itself, replacing it with an ensemble of systems compatible with some set of symmetries~\cite{dyson1962a}.
Wigner's proposal essentially started the mathematical field of RMT,\footnote{While Wishart introduced correlated random matrices in the context of multivariate analysis~\cite{wishart1928}, it was the work done in the context of nuclear physics that really led to the accelerated expansion of RMT, see the collection of reprints of Ref.~\cite{porter1965}.} whose applications now pervade all of mathematics and physics---for a survey see Ref.~\cite{akemann2011oxford}. In hindsight, it also marks the birth of the physical theory of \textit{quantum chaos}.

Obviously, by using a description in terms of an \textit{ensemble} of Hamiltonians, it is not possible to capture all the microscopic details of a system. The simplest questions we can ask about the spectrum of such a system pertain to the distribution of its eigenvalues $E_k$, $k=1,\dots,D$, as measured by the mean level density \begin{equation}
\varrho(E)=\frac{1}{D}\left\langle\sum_{k=1}^D \delta(E-E_k)\right\rangle,
\end{equation}
where $\langle\cdot\rangle$ denotes an appropriate average,\footnote{
	For a random matrix or a disordered physical system, $\langle\cdot\rangle$ denotes averaging over different Hamiltonians of the ensemble. For a physical system without disorder, there is a single Hamiltonian and so no ensemble averaging is possible. In this case, $\langle\cdot\rangle$ denotes an energy average over the spectrum (also called a running average). That is, we consider an interval $\Delta E$ with $d$ levels ($1\ll d\ll D$) inside which the mean level density is approximately constant and define the average of some function $g(E)$ to be $\langle g(E)\rangle=(1/\Delta E)\int_E^{E+\Delta E}\d E' g(E')$. This is \emph{not}, \textit{a priori}, the same as ensemble averaging, however, for us to be able to compare the predictions of RMT with actual data, we must assume that both averages are actually equal. This assumption goes by the name of \emph{ergodic hypothesis}~\cite{guhr1998}.
} which specifies the probability of finding a level in the interval $[E,E+\d E]$. Unfortunately, this quantity is highly system-dependent and does not describe accurately the level density of actual physical systems. Indeed, if the random matrix modeling the Hamiltonian has Gaussian entries, then its spectral density is given by the Wigner semicircle distribution~\cite{wigner1955,wigner1958},
\begin{equation}
\varrho_\mathrm{SC}(E)=\frac{2}{\pi E_0^2}\sqrt{E_0^2-E^2},
\end{equation}
supported on $-E_0<E<E_0$, where $E_0$ is the spectral edge of the random matrix (it sets the energy scale). Realistic physical systems such as heavy nuclei and black holes have, instead, an exponential growth of eigenstates above the ground state, given by the Bethe formula~\cite{bethe1936}
\begin{equation}
\varrho_\mathrm{Bethe}\propto \exp\{\sqrt{\gamma(E-E_0)}\},
\end{equation}
for $E\gtrsim E_0$, with $\gamma$ a constant and $E_0$ again the (many-body) ground-state energy. For finite systems, the level density in the bulk is also not semicircular and a density closer to Gaussian is expected~\cite{garcia-garcia2017PRD}.

At first sight, the disagreement between the level densities of random matrices and heavy nuclei---the systems RMT was introduced to describe---invalidates the applicability of the theory. A more refined random matrix model, nowadays known as the Sachdev-Ye-Kitaev model~\cite{sachdev1993,kitaev2015TALK1,kitaev2015TALK2,kitaev2015TALK3,sachdev2015PRX}, was shown to overcome this difficulty and also gives the correct density of states (see Sec.~\ref{sec:intro_SYK}). 

More importantly, RMT enjoyed great success at the time because it correctly predicted the distribution of the \emph{spacings} $s_j=E_{j}-E_{j-1}$ of the ordered eigenvalues $E_j$. Indeed, after removing the dependence on the level density\footnote{
In regions with a high spectral density, the levels are naturally closer (small spacings) than in regions with low spectral density (large spacings). For us to be able to compare these two regions and measure the actual correlations, the spectrum has to be made flat, i.e., the dependence on the spectral density removed, see Sec.~\ref{sec:intro_RMT} for a more detailed description of the procedure.
} (a procedure known as unfolding, described in Sec.~\ref{sec:intro_RMT}), the level spacing distribution of both large symmetric random matrices considered by Wigner~\cite{wigner1955} and heavy nuclei~\cite{bohigas1983} is well-described by Wigner's surmise~\cite{wigner1957b}
\begin{equation}
\label{eq:intro_Wigner_surmise}
P(s)=\frac{\pi}{2}\,s\,e^{-\pi s^2/4}.
\end{equation}
The defining feature of Eq.~(\ref{eq:intro_Wigner_surmise}) is the vanishing of $P(s)$ as $s\to 0$, a feature known as level repulsion. On the contrary, uncorrelated random numbers can be arbitrarily close to each other and one finds
\begin{equation}
P(s)=e^{-s},
\end{equation}
a result known as Poisson statistics,\footnote{
	The set of uncorrelated variables forms a Poisson process and the probability of finding $n$ levels in a given interval of length $s$ is given by a Poisson distribution $P_n(s)=(s^n/n!)e^{-s}$. The special case of $n=0$ (i.e., there being an empty spacing) gives the exponential distribution.}
which has a finite density as $s\to 0$, a feature known as level clustering. This distinction will be important when formulating the quantum chaos conjecture (see Sec.~\ref{sec:intro_BGS}).

Linear level repulsion, as in Eq.~(\ref{eq:Wigner_surmise}), is not the only possibility in random Hamiltonians, as three degrees of level repulsion exist, depending on its symmetry properties. The relevant symmetry is an antiunitary symmetry $T$, $T\i T^{-1}=-\i$, that commutes with the Hamiltonian, $THT^{-1}=H$, and can square to either $\pm1$. Conventionally, it is called time-reversal symmetry (TRS). If no such operator exists (a situation we denote with $T^2=0$), then the Hamiltonian has arbitrary complex entries (up to the requirement of Hermiticity) and is diagonalized by a unitary transformation. If there exists TRS and $T^2=+1$ then there exists a basis in which the Hamiltonian is real symmetric and is diagonalized by an orthogonal transformation. Finally, for a TRS-invariant system with $T^2=-1$, there exists a basis in which the Hamiltonian has real quaternionic entries, it is diagonalized by a unitary symplectic transformation, and all its eigenvalues are doubly degenerate (Kramers degeneracy).\footnote{An example of such systems are systems with spin-orbit coupling, which have TRS but broken rotational invariance.}
If we further require that the entries of $H$ are all independent random variables (up to the requirements of Hermiticity), the resulting random-matrix ensembles are known as the Gaussian unitary, orthogonal, and symplectic ensembles (GUE, GOE, and GSE), respectively. This is the classical Wigner-Dyson classification or Dyson's threefold way~\cite{dyson1962i,dyson1962ii,dyson1962iii,dyson1962a}. 
The Wigner-Dyson ensembles are characterized by the so-called Dyson index $\beta$, which counts the number of independent degrees of freedom in each entry of $H$. It is thus given by $\beta=1$, $2$, and $4$, for the GOE, GUE, and GSE respectively. The behavior under TRS (labeled by the Dyson index) completely determines the short-range (local) level statistics deep in the bulk of the spectrum. For instance, the consecutive level spacing distributions are given by~\cite{guhr1998}
\begin{equation}\label{eq:Wigner_surmise}
P_\beta(s)=a_\beta s^\beta e^{-b_\beta s^2},
\qquad
a_\beta=2\,\frac{\Gamma^{\beta+1}\left(\frac{\beta+2}{2}\right)}{\Gamma^{\beta+2}\left(\frac{\beta+1}{2}\right)},
\qquad
b_\beta=\frac{\Gamma^{2}\left(\frac{\beta+2}{2}\right)}{\Gamma^{2}\left(\frac{\beta+1}{2}\right)},
\end{equation}
with $\Gamma$ denoting the Gamma function.\footnote{
	To be precise, this result is derived for $2\times2$ matrices; the formulas for arbitrary $N$ are known in terms of infinite products~\cite{mehta2004}. Numerical results show, however, that the $N\to\infty$ level statistics differ little from Wigner's surmise, at most $1$--$2\%$.
}

It is remarkable that the statistical properties of eigenvalue correlations (here measured by the spacing distribution) are independent of any details of the Hamiltonian (either the microscopic details of a physical theory or the distribution of the entries of a random Hamiltonian) and depend only on a single symmetry property. The natural question we address next is under which conditions this universal RMT behavior is observed or, more bluntly, what is `chaotic' in quantum chaos.

\subsection{The quantum chaos conjecture}
\label{sec:intro_BGS}

As discussed above, the original motivation for the introduction of random matrices was the \emph{complexity} of the nuclear system under consideration. Perhaps the most striking feature shared by random matrices and complex systems is level repulsion. The presence or lack of level repulsion is not hard to understand from simple perturbation theory: if, upon the variation of a parameter of the model, two distinct levels come close, they will generically cross if independent, while they avoid each other (repel) if the two corresponding eigenstates are coupled---the interaction lifts the degeneracy.

The quantization of integrable (which possess a complete set of
conserved quantities) or chaotic (which do not) classical systems provides another example where a similar reasoning holds~\cite{percival1973}. Since the motion in phase space of an integrable classical system lives on invariant tori that do not overlap, one could assume that, upon quantization, the corresponding eigenstates are independent and level crossings are not avoided. On the other hand, the trajectory of a chaotic system eventually covers the phase space uniformly and, upon quantization, all states should thus be mixed, leading to level repulsion. Numerical experiments on quantum billiards supported these observations~\cite{mcdonald1979,casati1980}, although the precise degree of level repulsion could not be established.

While these observations led to the hypothesis that the distinction between ``regular'' (uncorrelated levels) and ``irregular'' (correlated and repelling levels) spectra could be origin for the distinction between chaotic and integrable classical motion, they did not address what the exact form of the level spacing distribution (or any other statistical property) should be.\footnote{Although an early attempt was made in Ref.~\cite{casati1980}.} Going beyond the mere existence of level clustering, Berry and Tabor conjectured~\cite{berry1977} that integrable systems have, in the semiclassical limit, \emph{exactly} the same statistical properties as uncorrelated random variables (Poisson statistics). The final step was then taken by Bohigas, Giannoni, and Schmit, who conjectured~\cite{bohigas1984} that, in the semiclassical limit, the statistical properties of a chaotic system match \emph{exactly} those of a random matrix with the appropriate symmetries (GOE if time-reversal symmetric or GUE if not).

The Bohigas-Giannoni-Schmit (BGS) substantially enlarged the set of physical systems described by RMT and has driven the field of quantum chaos for several decades.
It is by now universally accepted, given the enormous body of numerical and experimental evidence supporting it---for surveys see, e.g., Refs.~\cite{haake2013,guhr1998,ullmo2016}.\footnote{Some nongeneric exceptions, e.g., arithmetic billiards~\cite{bogomolny1992}, are known, though.} 
Great effort has also been employed to analytically derive the BGS conjecture, with the most successful approach based on semiclassical periodic orbit theory. Using a diagonal approximation, Berry~\cite{berry1985} was able to find the leading-order RMT behavior of the spectral form factor. More refined calculations extended this result to second order in Refs.~\cite{sieber2001,sieber2002} and, finally, to arbitrary order in Refs.~\cite{heusler2004,muller2004,muller2005,muller2009}.

After a decades-long focus on single-particle systems with few degrees of freedom, the focus of quantum chaos has shifted back to many-body systems in recent years. The semiclassical reasoning can be naturally extended for bosonic many-body systems~\cite{engl2014,engl2015,dubertrand2016,akila2017,waltner2017,rammensee2018,richter2022}, but many systems of interest (e.g., fermionic and spin-$1/2$) lack a well-defined semiclassical limit, and understanding the emergence of RMT in such systems is an ongoing effort~\cite{kos2018PRX,chan2018PRL,chan2018PRX,garratt2021PRX}.
Given the success of RMT and (single-particle) quantum chaos, the agreement with random matrix statistics is nowadays often used as the definition of a quantum chaotic many-body system.

\subsection{Random matrix theory and signatures of quantum chaos}
\label{sec:intro_RMT}

Once chaos in a quantum system is identified with the emergence of RMT, the quest for signatures of quantum chaos can then be framed as the search for the spectral observables where the RMT behavior (or the lack thereof) is most easily manifested. In Sec.~\ref{sec:intro_spectra}, we have discussed the role of level spacings as such a signature.

We start by specifying the probability distribution function $P(H)$ of the matrices $H$ of our RMT, typically of the form $P(H)=\exp\{-\Tr V(H)\}$, for some potential $V$. By an orthogonal/unitary/symplectic transformation, one can rotate $H$ into its diagonal basis, with eigenvalues $E_1,\dots,E_D$ and eigenvectors organized into an $D\times D$ matrix $U$:
\begin{equation}
P(H)\d H=e^{-\Tr V(H)}|\Delta(\Lambda)|^\beta \d \Lambda \d U.
\end{equation}
The change of basis introduces the ubiquitous Vandermonde interaction $\abs{\Delta}^\beta\equiv\prod_{j<k}\abs{E_j-E_k}^\beta$, independently of the specific choice of weight $P(H)$. In particular, if the spectral support of $H$ is compact (such as in the circular ensembles~\cite{dyson1962i}), the weight may be chosen flat. For convenience, the most common choice is a Gaussian potential, defining the Gaussian ensembles discussed in Sec.~\ref{sec:intro_spectra}.
Assuming a rotationally invariant matrix ensemble, $P(W H W^{-1})=P(H)$ for all $W$, the eigenvectors can be trivially integrated out. What remains is the joint eigenvalue distribution, given by
\begin{equation}\label{eq:joint_prob}
P(E_1,\dots,E_D)\propto \prod_{i<j}\abs{E_i-E_j}^\beta\prod_k\exp{-V(E_k)},
\end{equation}
which obviously contains all the information about the correlations.\footnote{The correlations arise because the Vandermonde interaction renders the probability nonfactorizable, $P(E_1,\dots,E_D)\neq P_1(E_1)\cdots P_N(E_D)$.} 
This is still a too complicated object to work with.
By integrating out $(D-k)$ eigenvalues\footnote{Any $(D-k)$ eigenvalues can be integrated out because the joint eigenvalue distribution is invariant under the exchange of its arguments.} we obtain the $k$-point distribution function 
\begin{equation}
R_k(E_1,\dots,E_k)=\int_{-\infty}^{+\infty}\d E_{k+1}\cdots\d E_DP(E_1,\dots,E_D),
\end{equation}
which gives the probability of finding any $k$ eigenvalues with values $E_1,\dots,E_k$ independently of the values of the remaining $(D-k)$ eigenvalues. Several approaches exist to compute $R_k$, e.g., recursively using the orthogonal polynomial method~\cite{mehta2004}, mapping the problem to a Coulomb gas in 2d and using electrostatics and statistical mechanics~\cite{dyson1962ii,forrester2010}, or expressing $R_k$ in terms of $k$-point Green's functions (also known as resolvents), which can then be computed using diagrammatic expansions~\cite{thooft1974,brezin1978}, supersymmetry~\cite{efetov1983,efetov1999,verbaarschot1985}, replicas~\cite{kamenev1999}, or Keldysh field theory~\cite{altland2000,kamenevbook}.

The 1-point correlation function---the average density of states, $\varrho(E)=R_1(E)$---is highly dependent on the choice of the weight of the ensemble, but higher correlation functions display universality~\cite{guhr1998}.
The simplest universal correlation function is the two-point function,
\begin{equation}
\varrho_2(E_1,E_2)=\frac{1}{D}\av{\sum_{j,k=1}^D \delta(E_1-E_j)\delta(E_2-E_k)}
= \varrho(E_1)\delta(E_1-E_2)+R_2(E_1,E_2).
\end{equation}
It is often more convenient to work with the Fourier transform of the 2-point function, known as the spectral form factor (SFF)~\cite{mehta2004,leviandier1986,brezin1997,prange1997}, which can also be directly related to the quantum properties of the system through the time evolution operator $U_t=\exp\{-\i H t\}$,
\begin{equation}
K_c(t)=\av{\left|\frac{1}{D}\Tr U_t\right|^2}_c=
\av{\sum_{j,k}e^{-\i\left(E_j-E_k\right)t}}_c
=1+\frac{1}{D}\int\d E_1 \d E_2 \, R_{2c}(E_1-E_2)\, e^{-\i(E_1-E_2) t},
\end{equation}
where the subscript $c$ denotes the connected component of the SFF and of the two-point function, $R_{2c}(E_1,E_2)=R_2(E_1,E_2)-R_1(E_1)R_1(E_2)$ and we have assumed translational invariance of the spectrum. In the large-$N$ limit, the SFF can be computed for the three Gaussian or circular ensembles~\cite{mehta2004,guhr1998}, yielding
\begin{align}
&K_c^{(\beta=2)}=
\begin{cases}
\tau,\qquad &\tau\leq 1\\
1,\qquad &\tau\geq1
\end{cases},
\\
&K_c^{(\beta=1)}=
\begin{cases}
2\tau-\tau\log\(2\tau+1\),\qquad &\tau\leq1\\
2-\tau\log\(\frac{2\tau+1}{2\tau-1}\),\qquad &\tau \geq1
\end{cases},
\\
&K_c^{(\beta=4)}=
\begin{cases}
\tau/2-(\tau/4)\log\(|\tau-1|\),\qquad &\tau\leq2\\
1,\qquad &\tau \geq2 
\end{cases},
\end{align}
where $\tau=t/t_\mathrm{H}$ and $t_\mathrm{H}$, which is inversely proportional to the mean level spacing, is the Heisenberg time. For very long times $t\gg t_\mathrm{H}$, the SFF probes the discreteness of the spectrum and becomes flat (it asymptotes to $1$). For times $t\sim t_\mathrm{H}$, the SFF probes energies separated by up to a few level spacings and, therefore, measures level repulsion. As was the case of the level spacing distribution, the SFF is capable of distinguishing the behavior under TRS, while being insensitive to any microscopic details. 
For many systems, the agreement with RMT extends to substantially shorter times, $t<t_\mathrm{H}$, due to so-called spectral rigidity---directly related to the power-law tails of the two-level correlation function---, which is responsible for the ramp of the SFF, $K_c^{(\beta)}\sim2t/\beta t_\mathrm{H}$, whose slope depends solely on the Dyson index. 
For sufficiently short times, level statistics of realistic quantum chaotic Hamiltonians deviate from the RMT prediction.
The timescale that marks the onset of these deviations and
delimits the region of universal quantum chaotic dynamics---related to the so-called dip or correlation hole~\cite{leviandier1986,wilkie1991,alhassid1992,schiulaz2019PRB} of the
connected SFF---depends on details of the dynamics. For disordered systems, where it is called the Thouless time, it is related to the typical diffusion time needed for a single particle to cross the sample and only after it do physical systems exhibit ergodic behavior. However, still in the context of disordered systems, this timescale is sometimes determined by ensemble fluctuations and is not directly related to the type of motion~\cite{brody1981RMP,jia2020JHEP}. The determination of Thouless time in many-body systems is a topic of great current interest~\cite{garcia-garcia2016PRD,garcia-garcia2018PRD,kos2018PRX,gharibyan2018JHEP,chan2018PRL}.

Despite the correlation functions being the natural object to work with theoretically, when performing experiments or numerical simulations, other spectral observables (which are built out of one or more correlation functions) are easier to access. The most popular spectral observable is, arguably, the spacing distribution $P(s)$, already discussed in Sec.~\ref{sec:intro_spectra}, which is a function of all $R_k$ with $k\geq2$. We have already seen that, for the Gaussian ensembles, $P(s)$ is very well described (both qualitatively and quantitatively) by the distribution for $2\times2$ matrices, the Wigner surmise, Eq.~(\ref{eq:Wigner_surmise}).
Because of universality, this result holds for any ensemble with the same symmetries of the GOE/GUE/GSE. In obtaining the Wigner surmise, not only the distribution itself but also its first moment were normalized to one, $\int_0^\infty\d s P(s)=\int_0^\infty \d s\, sP(s)=1$. Setting $\langle s\rangle=1$ is achieved by changing the scale to the so-called unfolded scale~\cite{guhr1998,haake2013}: we change variables to a new level sequence, $e_1,\dots,e_D$, with
\begin{equation}
e_j=\int_{-\infty}^{E_j} \d E \varrho(E),
\end{equation}
for which the mean level density is flat and, thus, fluctuations can be uniformly compared across the spectrum and with numerical and experimental data.

Unfolding the spectrum is a nontrivial procedure. It requires an analytic expression (or accurate estimate) of the level density $\varrho$, which one might not have in general. Unfolding can be easily achieved in theoretical calculations for the classical random matrix ensembles and simple systems like quantum billiards~\cite{bohigas1984,bohigas1984b,stockmann1999}, but it is not always possible for more complex systems, like many-body systems. Although the spectrum can be numerically unfolded, an alternative simpler way to overcome the local dependence on the level density is to consider ratios of spectral observables. It is expected that the ratios are automatically independent of the local density of states (the dependence of numerator and denominator should ``cancel out''), rendering the unfolding unnecessary. In this spirit, ratios of consecutive spacings,
\begin{equation}
\label{eq:rstat}
r_i=\min\left(\frac{E_{i+1}-E_i}{E_i-E_{i-1}},\frac{E_i-E_{i-1}}{E_{i+1}-E_i}\right),
\end{equation}
were first considered in Ref.~\cite{oganesyan2007PRB} and extensively applied in comparing real data to the theoretical predictions. In Refs.~\cite{atas2013,atas2013long}, analytic expressions for the ratio distributions were obtained, in the form of a Wigner-like surmise,
\begin{equation}
\label{eq:surmise}
P(r)=\frac{2}{Z_\beta}\frac{(r+r^2)^\beta}{(1+r+r^2)^{3\beta/2}},
\end{equation}
with $Z_1=8/27$, $Z_2=4\pi/81\sqrt{3}$, and $Z_4=4\pi/729\sqrt{3}$ (with level repulsion $P(r)\propto r^\beta$ as $r\to0$). For Poisson statistics, $P(r)=2/(1+r^2)$ (exhibiting level clustering as $r\to0$).

While the spacing statistics measure short-range correlations, long-range fluctuations are captured, for instance, by the number variance $\Sigma^2(L)$~\cite{dyson1962iv}, which is defined as the variance of the number of levels in an interval of length $L$. For random matrices one finds $\Sigma^2(L)\propto \log L$ at large $L$, while for Poisson statistics $\Sigma^2(L)\propto L$. One can, therefore, also distinguish a chaotic system from an integrable one by the large-$L$ behavior of the number variance. $\Sigma^2(L)$ can be reduced to an integral of the 2-point function and hence measures only 2-level correlations. 
A related spectral observable is the $\Delta_3$ statistic, also introduced in Ref.~\cite{dyson1962iv}, which measures the deviation of the cumulative level distribution from a straight line. Finally, as discussed, the SFF also captures long range correlations, and in recent years, although not self-averaging~\cite{prange1997}, has become the preferred tool for this task.

\subsection{Symmetries and the tenfold way}
\label{sec:intro_symmetries_AZ}

As seen above, the bulk local-level statistics (such as the spacing ratio distribution), are solely determined by the behavior of the Hamiltonian under TRS.
By Wigner's theorem, symmetries in quantum mechanics are implemented by either unitary or antiunitary operators that commute with the Hamiltonian and it would thus seem we have exhausted all available symmetries: unitary operators commuting with the Hamiltonian give conserved quantities and antiunitary operators commuting with the Hamiltonian determine its bulk level statistics. However, we can relax our requirements and allow for involutive transformations, i.e., a unitary operator $P$ and an antiunitary operator $C$ that \emph{anticommute} with the Hamiltonian, $PHP^{-1}=-H$ and $CHC^{-1}=-H$, with $P\i P=\i$ and $C\i C^{-1}=-\i$~\cite{mehta1968,mehta1983,verbaarschot1993PRL,verbaarschot1994PRL,altland1997}. These operators implement spectral mirror symmetries, since eigenvalues always come in pairs $(E,-E)$ (they are mirrored across the origin). $C$ is known as a particle-hole symmetry (PHS) and, like TRS, can square to either $\pm1$.\footnote{We denote the absence of PHS as $C^2=0$.} It can also always be taken to commute with TRS. $P$ is known as a chiral symmetry (CS) and can always be taken to square to $+1$ (its absence is denoted as $P^2=0$). Note that the composition $TC$ is unitary and anticommutes with $H$, thus implementing a CS; for this reason, CS is only an independent symmetry in the absence of both TRS and PHS.

\begin{table}[t]
	\centering
	\caption{Altland-Zirnbauer classes. For each class, we list its name (Cartan label), the defining symmetries (TRS, PHS, and CS), the corresponding Gaussian ensemble, and the level repulsion exponents $\alpha$ and $\beta$. For the Wigner-Dyson ensembles (GUE, GOE, GSE), $\alpha$ is not defined.}
	\label{tab:intro_AZ}
	\begin{tabular}{@{}Sl Sc Sc Sc Sl Sc Sc@{}}
		\toprule
		Class & $T^2$ & $C^2$ & $P^2$ & RMT Ens. & $\beta$ & $\alpha$ \\ \midrule
		A     & $0$   & $0$   & $0$   & GUE      & $2$     & ---      \\
		AI    & $+1$  & $0$   & $0$   & GOE      & $1$     & ---      \\
		AII   & $-1$  & $0$   & $0$   & GSE      & $4$     & ---      \\
		AIII  & $0$   & $0$   & $+1$  & chGUE    & $2$     & $1+2\nu$ \\
		BDI   & $+1$  & $+1$  & $+1$  & chGOE    & $1$     & $\nu$    \\
		CII   & $-1$  & $-1$  & $+1$  & chGSE    & $4$     & $3+4\nu$ \\
		D     & $0$   & $+1$  & $0$   & BdG      & $2$     & $0$      \\
		C     & $0$   & $-1$  & $0$   & BdG      & $2$     & $2$      \\
		CI    & $+1$  & $-1$  & $+1$  & BdG      & $1$     & $1$      \\
		DIII  & $-1$  & $+1$  & $+1$  & BdG      & $4$     & $1$      \\ \bottomrule
	\end{tabular}
\end{table}

We say the symmetry class of the Hamiltonian is labeled by the squares of TRS and PHS and the presence or absence of CS. Because each of TRS and PHS can square to $0$ or $\pm1$, and we have an additional case with both absent but with CS present, we have $3\times3+1=10$ classes, the celebrated Altland-Zirnbauer tenfold way~\cite{altland1997}, which are tabulated in Table~\ref{tab:intro_AZ}. It is possible to establish a one-to-one correspondence between an AZ class and a symmetric space~\cite{zirnbauer1996,heinzner2005}, and since there are ten (nonexceptional) families of the latter, this shows that we have exhausted all possible symmetry classes.\footnote{The correspondence with symmetric spaces also gives the classes their names in terms of Cartan labels A, AI, etc.} Just as the GUE, GOE, and GSE give the simplest ensembles of random matrices in classes A, AI, and AII (complex, real, and quaternionic Gaussian random matrices respectively), we can assign a simple Gaussian random matrix ensemble to each of the seven classes with spectral mirror symmetry, see the fourth column of Table~\ref{tab:intro_AZ}. The five chiral classes (with $P^2=1$) have, in some basis, an off-diagonal structure
$
H=\begin{pmatrix}
0 & h
\\
h^\dagger & 0
\end{pmatrix}
$.
If the matrix $h$ is a real, complex, or quaternionic matrix, it belongs to class BDI, AIII, and CII, respectively. If $h$ is taken to be Gaussian, this defines the chiral Gaussian orthogonal, unitary, and symplectic ensembles (chGOE, chGUE, and chGSE), respectively~\cite{verbaarschot1993PRL,verbaarschot1994PRL}. 
A distinctive feature of these ensembles is the existence of exact zero modes. If $h$ is taken to be a $n\times m$ rectangular matrix [$H$ is then $(n+m)\times (n+m)$], then $\nu=|n-m|$ eigenvalues are identically zero for the chGOE and chGUE, while the GSE has $2\nu$ zero modes owing to Kramers degeneracy.
Similarly, the four remaining classes (D, C, CI, and DIII) also have a representative Gaussian ensemble each; however, these have no established name, being instead collectively known as the Bogoliubov-de Gennes (BdG) ensembles~\cite{altland1997}.

Because PHS and CS lead to a reflection of the spectrum around the origin, we can expect that correlations near this point (also known as the hard edge) are distinct from the Wigner-Dyson statistics found deep in the bulk. This expectation turns out to be correct, and the seven classes with spectral mirror symmetry are characterized by a second level-repulsion exponent $\alpha$, in addition to the Dyson index $\beta$, which measures the level repulsion from the origin, $P_{0}(s)\propto s^\alpha$. The values of $\alpha$ and $\beta$ are tabulated in Table~\ref{tab:intro_AZ}---for a derivation see Refs.~\cite{altland1997,haake2013}. 
The three Wigner-Dyson classes do not have a well-defined exponent $\alpha$, as the origin is not a special point. Remarkably, for the seven classes with spectral mirror symmetry, not only the correlations but also the spectral density are universal~\cite{verbaarschot1993PRL} in a microscopic region around the origin (up to a few tens of level spacings), with known analytic expressions~\cite{gnutzmann2006}.

An alternative signature of chaos that distinguishes these seven classes is the distribution of the eigenvalue closest to the origin, denoted $E_1>0$, in units of its average value, $x_1=E_1/\av{E_1}$, which again does not require unfolding. From now on, we always consider $\nu=0$. The distribution $P(x_1)$ is known exactly for the chiral RMT ensembles~\cite{forrester1993NPB,wilke1998PRD,nishigaki1998PRD,damgaard2001PRD,akemann2009PRE,sun2020PRL}:
\begin{align}
\label{eq:emin_CI}
&P_\mathrm{AIII}(x_1)=a\ b\ x_1 \exp\(-b^2x_1^2/2\),
\qquad
a=b=\sqrt{\pi/2},
\\
\label{eq:emin_BDI}
&P_\mathrm{BDI}(x_1)=a(2+bx_1)\exp\(-\frac{b^2x_1^2}{8}-\frac{bx_1}{2}\),
\qquad a=b/4=\sqrt{\frac{\pi e}{8}}\ \mathrm{erfc}(1/\sqrt{2}),
\\
\label{eq:emin_CII}
&P_\mathrm{CII}(x_1)=a\ b^{3/2}\ x_1^{3/2} \exp\(-b^2x_1^2/2\)I_{3/2}(b x),
\qquad
a=\sqrt{\pi/2}b=\pi/2\sqrt{e},
\end{align}
where $I_n$ is a modified Bessel function of the first kind.
The constant $a$ fixes the normalization of $P(x_1)$, while $b$ fixes the mean of $x_1$ to be one. Note that $P(x_1)\propto x_1^\alpha$ as expected. 
For $\alpha=1$, $P(x_1)$ is independent of $\beta$~\cite{sun2020PRL} and thus $P_\mathrm{CI}(x_1)=P_{\mathrm{DIII}}(x_1)=P_\mathrm{AIII}(x_1)$.
The exact expressions for classes C and D are obtained from the derivative of a Fredholm determinant~\cite{sun2020PRL}, but simple closed-form expressions can be derived from a Wigner-like surmise~\cite{akemann2009PRE} for $2\times2$ matrices. These are given by~\cite{sun2020PRL}
\begin{align}
\label{eq:emin_C}
\begin{split}
&P_\mathrm{C}(x_1)=a b^2 x_1^2 \exp{-2 b^2 x_1^2}
\left[30 b x_1 - 4 b^3 x_1^3
+ \sqrt{\pi} \exp{b ^2 x_1^2} \mathrm{erfc}(b x_1)
\left(15 - 12 b^2 x_1^2 + 4 b^4 x_1^4\right)\right],
\\
&a=\frac{10-5\sqrt{2}}{3\pi^{3/2}},
\qquad
b=\frac{10-5\sqrt{2}}{2\sqrt{\pi}},
\end{split}
\\
\label{eq:emin_D}
\begin{split}
&P_\mathrm{D}(x_1)=a b^2\exp{-2 b^2 x_1^2}
\left[6 b x_1 - 4 b^3 x_1^3
+ \sqrt{\pi} \exp{b ^2 x_1^2} \mathrm{erfc}(b x_1)
\left(3 - 4 b^2 x_1^2 + 4 b^4 x_1^4\right)\right],
\\
&a=\frac{7-4\sqrt{2}}{2\pi^{3/2}},
\qquad
b=\frac{7-4\sqrt{2}}{2\sqrt{\pi}}.
\end{split}
\end{align}

\subsection{The Sachdev-Ye-Kitaev model}
\label{sec:intro_SYK}

We are by now convinced that structureless random matrices, as considered by Wigner, give the right level correlations and provide a tool to study the universal correlations of quantum chaotic physical systems. Next, we return to the question of which, if any, random model could lead also to the right spectral density (while preserving RMT correlations). 

In the 1970s, it was realized that the undesirable feature of Wigner-Dyson random matrices was that, for a $N$-body system, it contained up to $N$-body interactions, while realistic systems have few-body (typically one- or two-body) interactions. Following this reasoning, Bohigas, Flores, French, Mon, and Wong, among others, proposed the so-called (two-body) embedded ensemble~\cite{french1970PhysLettB,french1971PhysLettB,bohigas1971PhysLettB,bohigas1971PhysLettB2,mon1975AnnPhys,brody1981RMP} of Hamiltonians of the form
\begin{equation}
H=\sum_{i<j,k<\ell}^N J_{ij;k\ell}c_i^\dagger c_j^\dagger c_k c_\ell,
\end{equation}
where $c_i^\dagger$ ($c_i$) are fermionic creation (annihilation) operators and $J_{ij;k\ell}$ are real Gaussian random variables. As expected, the spectral correlations matched those of RMT (in particular those of the GOE), while the spectral density was found numerically to be Gaussian in the bulk. Whether the Bethe formula is satisfied above the ground state remained unanswered at the time.

Twenty years later, a variant of the embedded ensemble, namely a random exchange spin model~\cite{sachdev1993}, made a return in the study of quantum spin liquids. Many of the features that would make the model popular in the future, such as the structure of the large-$N$ limit and the near-conformal character of its solutions, were worked out in Refs.~\cite{sachdev1993,parcollet1999PRB,georges2000PRL,georges2001PRB,chowdhury2022RMP} and the existence of a nonzero zero-temperature entropy and the potential connection to holography were put forward in Ref.~\cite{sachdev2010PRL}.

The modern reincarnation of the model, dubbed the Sachdev-Ye-Kitaev (SYK) model~\cite{kitaev2015TALK1,kitaev2015TALK2,kitaev2015TALK3,sachdev2015PRX}, consists of $N$ Majorana fermions with random all-to-all $q$-body interactions in zero dimensions:
\begin{equation}
H=-\i^{q/2} \sum_{i_1<\cdots <i_q}^N J_{i_1\cdots i_q} \psi_{i_1}\cdots \psi_{i_q}.
\end{equation}
Here, $J_{i_1\cdots i_q}$ is a totally antisymmetric tensor of real random Gaussian variables and $\psi_i$ are Majorana operators satisfying the Clifford algebra $\{\psi_i,\psi_j\}=\delta_{ij}$ and $\psi_i^\dagger=\psi_i$.
The SYK model saw a surge of interest because of its connection to two-dimensional quantum gravity, after it was shown to be maximally chaotic, exactly-solvable at strong coupling, and near conformal~\cite{polchinski2016JHEP,maldacena2016JHEP,maldacena2016PRD,jevicki2016JHEP,jevicki2016JHEP2,bagrets2016NuclPhysB,bagrets2017NuclPhysB,altland2018NuclPhysB,gross2017JHEP,kourkoulou2017ARXIV,kitaev2018JHEP}.

Despite having strong interactions, the SYK model is still analytically tractable because the interactions are random and all-to-all. In this case, we can reduce a complicated lattice problem with $N$ fermionic degrees of freedom to a impurity problem with a single mean-field (collective) degree of freedom $G(t,t')=(1/N)\sum_i \psi_i(t)\psi_i(t')$. Working either diagrammatically or with the saddle-point approximation of a path integral, $G$ and its conjugate field $\Sigma$ satisfy a pair of relatively simple nonlinear differential equations. In the large-$N$ limit, the mean-field description becomes exact and, although an analytic solution is only possible in some special cases, one can always solve them numerically in a self-consistent fashion.

As desired, the SYK model was found to display an exponential growth of low-energy excitations typical of black holes and heavy nuclei~\cite{erdos2014MPAG,garcia-garcia2017PRD,cotler2017JHEP,stanford2017JHEP}, that is, its low-energy spectral density reproduces the Bethe formula. 
Indeed, using combinatorial techniques, it is possible to show~\cite{cotler2017JHEP,garcia-garcia2017PRD,erdos2014MPAG} that for large $N$, the spectral density of the SYK model is given by the weight function of the $Q$-Hermite polynomials~\cite{ismail1987EJC},
\begin{equation}\label{eq:intro_spectral_density_SYK}
\varrho_\mathrm{SYK}(E)
=(Q;Q)_\infty(-Q;Q)_\infty^2\frac{2}{\pi E_0}
\sqrt{1-E^2/E_0^2}
\prod_{k=1}^\infty
\(1-\frac{4E^2/E_0^2}{2+Q^{k}+Q^{-k}}\),
\end{equation}
supported on $-E_0\leq E\leq E_0$, with (dimensionless) ground-state energy $E_0=2/(\sqrt{1-Q})$ and $Q$-deformation parameter ($-1\leq Q \leq 1$)
\begin{equation}
Q=\binom{N}{q}^{-1}\sum_{s=0}^q 
(-1)^{q+s}\binom{q}{s}\binom{N-q}{q-s}.
\end{equation}
$(a;Q)_\infty=\prod_{k=0}^\infty \(1-aQ^k\)$ is the $Q$-Pochhammer symbol. This result is exact in the double-scaling limit $q$, $N\to\infty$ with $q^2/N$ fixed and expected to hold also for fixed $q$ up to $1/N$ corrections. For $q=4$, it is indistinguishable from the numerically-exact spectral density for values of $N$ as small as $N=8$~\cite{garcia-garcia2017PRD}. 
Note that the spectral density is of the form of the Wigner semicircle distribution times a $Q$-dependent multiplicative correction. It interpolates between the Wigner semicircle when $Q\to0$ (random-matrix limit) and a Gaussian density as $Q\to1$. For fixed $0<Q<1$, we can take the large-$N$ limit explicitly in Eq.~(\ref{eq:intro_spectral_density_SYK}). Deep in the bulk, $|E|\ll E_0$, one finds the asymptotic Gaussian density~\cite{garcia-garcia2017PRD}
\begin{equation}
\varrho_\mathrm{SYK}^{(\mathrm{bulk})}(E)\approx 
c_N \exp\left[
\frac{-2(E/E_0)^2}{|\log Q|}
\right],
\end{equation}
with $c_N$ some constant that can be determined exactly.
On the other hand, close to the ground state $E\approx -E_0$, the density is given by~\cite{cotler2017JHEP,garcia-garcia2017PRD}
\begin{equation}
\varrho_\mathrm{SYK}^{(\mathrm{Bethe})}(E)\approx 
2c_N\exp\left[-\frac{\pi^2}{2|\log Q|}\right]
\sinh\left[\frac{2\pi\sqrt{2(1-E/E_0)}}{|\log Q|}\right],
\end{equation}
which gives the desired exponential growth of states (Bethe formula). By closing the circle of Wigner's original quest for a model of the nucleus, the SYK model perfectly illustrates the power of theories of random Hamiltonians.

\begin{table}[t!]
	\centering
	\caption{Symmetry classification of the SYK Hamiltonian into Altland-Zirnbauer classes for all $q$ and even $N$. Only class AIII is missing.}
	\label{tab:intro_SYK_classes}
	\begin{tabular}{@{}l cccc@{}}
		\toprule
		$N\,\mathrm{mod}\,8$   & 0   & 2    & 4   & 6    \\ \midrule
		$q\,\mathrm{mod}\,4=0$ & AI  & A    & AII & A    \\
		$q\,\mathrm{mod}\,4=1$ & BDI & CI   & CII & DIII \\
		$q\,\mathrm{mod}\,4=2$ & D   & A    & C   & A    \\
		$q\,\mathrm{mod}\,4=3$ & BDI & DIII & CII & CI   \\ \bottomrule
	\end{tabular}
\end{table}

What was also unexpected, but absolutely remarkable, is that the SYK model realizes nine out of the ten Altland-Zirnbauer classes by tuning its parameters ($q\mod4$ and $N\mod8$)~\cite{you2017PRB,garcia-garcia2016PRD,cotler2017JHEP,li2017JHEP,kanazawa2017JHEP,garcia-garcia2018PRD,behrends2019PRB,behrends2020PRL,sun2020PRL}, see Table~\ref{tab:intro_SYK_classes}.\footnote{
	To be precise, for odd $q$, $H$ corresponds to a supersymmetric charge. The corresponding Hamiltonian is $H^2$.
} Only class AIII is missing, although it can be realized in a chiral extension of the model, to be discussed in Ch.~\ref{chapter:QLaguerre}.
This makes the SYK model one of the most paradigmatic models of many-body quantum chaos, as it allows us to explore the properties of all symmetry classes in a single model. For $q>2$, the local level correlations agree spectacularly with RMT~\cite{garcia-garcia2016PRD,garcia-garcia2016PRD}, both in the bulk of the spectrum and near the edges, signaling the quantum chaotic nature of the SYK model. The SFF shows agreement with RMT for long times~\cite{cotler2017JHEP}, while the deviations at short times were understood analytically as arising mostly from collective modes of the spectrum~\cite{jia2020JHEP}. For $q=2$, the SYK Hamiltonian is integrable, as it describes a free theory, and (many-body) level correlations are Poissonian. The interplay of single-body chaos and many-body integrability in random quadratic fermionic Hamiltonians was also studied~\cite{cotler2017JHEP,magan2016PRL,lydzba2020PRL,liao2020PRL,winer2020PRL}.

Finally, the SYK model also captures many features of non-Fermi liquids~\cite{song2017PRL,zhang2017PRB,gnezdilov2018PRB,can2019PRB,chowdhury2022RMP} and wormholes~\cite{maldacena2018ARXIV,garcia-garcia2019PRD,kim2019PRX,qi2020JHEP,plugge2020PRL,lantagne-hurtubise2020PRR,sahoo2020PRR,klebanov2020JHEP,lensky2021JHEP,garcia-garcia2021PRD2,haenel2021PRB}.
These developments placed it in a prominent position at the intersection of high-energy physics and condensed matter as one of the few analytically tractable models of both holography and strongly interacting quantum matter. It will play a central role in this thesis.
Several experimental implementations have been proposed~\cite{danshita2017PTEP,chew2017prb,pikulin2017PRX,chen2018PRL,franz2018NatRM,wei2021PRA}, and its practical and technological relevance highlighted~\cite{garcia-alvarez2017PRL,luo2019npjQI,babbush2019PRA,rossini2020PRL,rosa2020JHEP,behrends2022PRL}. A (debated) experimental implementation was reported in Ref.~\cite{jafferis2022Nature}.

\section{Open quantum systems}
\label{sec:intro_OQS}

The controlled manipulation of a large number of quantum degrees of freedom is evermore turning into an experimental reality. Nevertheless, statistical deviations in controlling protocols and interactions with the environment are, to some degree, unavoidable. The ensuing relaxation and decoherence effects are responsible for computation errors and limit the accuracy of sensing devices. In addition to dissipation and decoherence, contact with different environments may induce currents of otherwise conserved quantities, such as energy and charge, and the observables of the system typically attain a steady state.

In Sec.~\ref{sec:intro_quantum_chaos}, we considered isolated and perfect quantum systems, which are described by a vector $\ket{\psi}$ in Hilbert space and whose unitary dynamics are generated by an Hermitian Hamiltonian $H$. The time evolution is given by the application of a unitary map $U_t=\exp\{-\i H t\}$ that propagates an initial state $\ket{\psi(0)}$ to a final state $\ket{\psi(t)}=U_t \ket{\psi(0)}$, or alternatively, by solving a time-local linear differential equation (the Schr\"odinger equation), $\partial_t\ket{\psi}=-\i H\ket{\psi}$.

The focus of this thesis is, however, on those systems that are not only extremely complex but also coupled to some external environment.
In principle, the joint system-plus-bath evolution still follows the Schr\"odinger equation for the total system. However, in practice, a microscopic description in terms of the whole set of degrees of freedom of both the system and its environment is often impossible. Even if we could proceed with such a complete description, it would contain much more information than we are interested in. One thus seeks a description in terms of the open system's degrees of freedom only, by ``integrating out'' the environment degrees of freedom. Consequently, the state of the system is described not by a pure state $\ket{\psi}$, but by a density matrix $\rho$, an Hermitian positive-semidefinite matrix of unit trace, such that its eigenvalues---which give the probabilities of finding the system in the different states at time $t$---are real, positive, and add up to one, as probabilities should. For concreteness, we will consider finite-dimensional systems, whose dimension is denoted as $N$.

\subsection{Quantum channels and Kraus operators}
\label{sec:intro_Kraus}

The time evolution of the system's reduced density matrix is given by a nonunitary superoperator\footnote{
	If $\mathcal{H}$ is the Hilbert space of pure states of dimension $N$ with (orthonormal) basis $\ket{n}$, then the Liouville space $\mathcal{K}$ of linear operators on $\mathcal{H}$ (that is, the space of density matrices or mixed states) is also a Hilbert space of dimension $N^2$ with Hilbert-Schmidt inner product $\langle A,B\rangle=\mathrm{Tr}(A^\dagger B)=\langle B,A\rangle^*$, for some operators $A$ and $B$ in $\mathcal{K}$. A superoperator~$\Phi$ is a linear map $\Phi:\mathcal{K}\to\mathcal{K}$.
} $\Phi_t$,
\begin{equation}
\label{eq:intro_def_map}
\rho(t)=\Phi_t[\rho(0)].
\end{equation} 
While in general nonunitary, $\Phi_t$ should send physical density matrices (Hermitian, positive, and unit trace) to physical density matrices, i.e., it should be an Hermiticity-preserving, trace-preserving, and completely positive (CPTP) map,\footnote{
	A positive map sends positive density operators to positive density operators. Complete positivity is a stronger condition meaning that the map is positive even when acting only on part of a larger system. That is, the map $\Phi$ is completely positive if $\Phi\otimes \mathbbm{1}_k$ is positive for all $k$, where $\mathbbm{1}_k$ is the $k$-dimensional identity. This property ensures the positivity of the density matrix even when entangled with an external environment.
} also known as quantum channel~\cite{nielsen2002,bengtsson2017}.
Any dynamical map $\Phi_t$ with the above properties can be written in the so-called Kraus representation (or operator-sum representation)~\cite{kraus1983} as follows. Any linear superoperator $\Phi_t$ can be written in the form~\cite{gorini1976}
\begin{equation}
\Phi_t(\rho)=\sum_{\mu=1}^{N^2}A_\mu(t) \rho B_\mu^\dagger(t),
\end{equation}
for some (non-Hermitian) linear operators $A_\mu(t)$ and $B_\mu(t)$. Now, if $\Phi_t$ is Hermiticity-preserving, we have $[\Phi_t(\rho)]^\dagger=\Phi_t(\rho)=\Phi_t(\rho^\dagger)$ and we can change basis to write\footnote{
	To see this, choose a complete orthonormal basis of operators $G_\mu$ (i.e., $\Tr[G_\mu^\dagger G_\nu]=\delta_{\mu\nu}$, $\mu,\nu=1,\dots,N^2$), decompose $A_\mu(t)=\sum_{\nu=1}^{N^2} a_{\mu\nu}(t)G_\nu$ and $B_\mu=\sum_{\nu=1}^{N^2} b_{\mu\nu}(t)G_\nu$, and define $c_{\mu\nu}(t)=\sum_{\kappa=1}^{N^2} a_{\mu\kappa}(t)b_{\nu\kappa}^*(t)$. We then have $[\Phi_t(\rho)]^\dagger=\sum_{\mu,\nu=1}^{N^2} c_{\mu\nu}^*(t) G_\nu \rho^\dagger G_\mu^\dagger=\sum_{\mu,\nu=1}^{N^2} c_{\mu\nu}(t) G_\mu \rho^\dagger G_\nu^\dagger=\Phi_t(\rho^\dagger)$, whence it follows that $c_{\mu\nu}(t)=c_{\nu\mu}^*(t)$. Since $c(t)$ is an Hermitian matrix, it can be diagonalized $c(t)=U(t) \varepsilon(t) U^\dagger(t)$, with a unitary matrix $U(t)$ and a real diagonal matrix $\varepsilon(t)$. Finally, we define the Kraus operators $K_\mu(t)=\sum_{\nu=1}^{N^2} U_{\mu \nu}(t)G_\nu$ and the stated result follows. The orthonormality of the Kraus operators follows from the unitarity of the change of basis.
}
\begin{equation}
\Phi_t(\rho)=\sum_{\mu=1}^{N^2}\varepsilon(t) K_\mu(t) \rho K_\mu^\dagger(t), 
\end{equation}
where $\varepsilon(t)$ is real and the Kraus operators $K_\mu$ are orthonormal, $\Tr[K_\mu^\dagger(t) K_\nu(t)]=\delta_{\mu\nu}$. 
For the map to be trace-preserving, we have that $\Tr[\rho]=\Tr[\Phi_t(\rho)]$, whence it follows that the Kraus operators satisfy\footnote{
	To see this, we write explicitly $\Tr[\rho]=\Tr[\Phi_t(\rho)]=\sum_{\mu=1}^{N^2}\varepsilon(t) \Tr[\rho K_\mu^\dagger(t)K_\mu(t)]$. Since this equation must hold for all $\rho$, the stated result follows.
}
\begin{equation}
\sum_{\mu=1}^{N^2} \varepsilon(t)K_\mu^\dagger(t) K_\mu(t)=\id.
\end{equation}
Finally, for the map to be completely positive, all $\varepsilon(t)$ must be non-negative.\footnote{
	To see this, consider and arbitrary state $\ket{\phi}$ and define $\ket{\chi_\mu}=K_\mu^\dagger\ket{\phi}$. Then $\bra{\phi}\Phi_t(\rho)\ket{\phi}=\sum_{\mu=1}^{N^2}\varepsilon_\mu(t)\bra{\chi_\mu}\rho\ket{\chi_\mu}$. Since $\rho$ is positive $\bra{\chi_\mu}\rho\ket{\chi_\mu}\geq0$ and, hence, $\bra{\phi}\Phi_t(\rho)\ket{\phi}\geq0$ for an arbitrary $\ket{\phi}$ only if all $\varepsilon_\mu(t)\geq0$. This shows the positivity of $\Phi_t$. By the same argument, $\Phi_t\otimes \id$ is also positive and hence $\Phi_t$ is completely positive.
} In summary, any CPTP map admits the Kraus form,
\begin{equation}
\Phi_t(\rho)=\sum_{\mu=1}^{N^2} K_\mu(t) \rho K_\mu^\dagger(t),
\qquad 
\sum_{\mu=1}^{N^2} K_\mu^\dagger(t) K_\mu(t)=\id.
\end{equation}
In the closed limit, $\Phi_t$ is completely determined by the unitary operator $U_t$, $\Phi_t(\cdot)=U_t \cdot U_t^\dagger$, $U_t^\dagger U_t=\id$. Finally, the Stinespring dilation theorem~\cite{stinespring1955} establishes that tracing out the environment from the joint system-plus-environment unitary evolution gives a CPTP map and so, our procedure to reduce the description to only the system's degrees of freedom indeed provides a physically meaningful prescription.

\subsection{The Lindblad equation}

In analogy with the Schr\"odinger equation, we can also seek a description in terms of a quantum master equation, 
\begin{equation}
\label{eq:intro_master_eq}
\partial_t \rho(t)=\mathcal{L}_t[\rho(t)],
\end{equation}
where the superoperator $\mathcal{L}$ is known as the Liouvillian. Remarkably, if the quantum evolution satisfies the mild assumptions that (i) the system and the bath are initially separable (i.e., the initial total density matrix factorizes as $\rho(0)\otimes \rho_\mathrm{B}$) and (ii) the quantum channel $\Phi_t$ is invertible for all times (i.e., initially distinct states remain distinct for arbitrarily long times), then it can be cast in the time-local form of Eq.~(\ref{eq:intro_master_eq})~\cite{andersson2007,hall2014PRA}.\footnote{In some cases, even if $\Phi_t$ is not invertible it is possible to write down a time-local master equation, but the procedure is not as straightforward~\cite{andersson2007}.} Indeed, if assumption (i) holds, then the system's reduced density matrix satisfies the Nakajima-Zwanzig equation~\cite{nakajima1958,zwanzig1960,breuerpetruccione}
\begin{equation}
\label{eq:intro_nakajima_zwanzig}
\partial_t \rho(t)=\int_0^t \d s\, \mathcal{K}(t,s)[\rho(s)],
\end{equation}
where the explicit form of the memory kernel $\mathcal{K}(t,s)$ is typically very hard to obtain. If, moreover, assumption (ii) holds, then $\rho(s)=\Phi_s[\rho(0)]=\Phi_s\{\Phi_t^{-1}[\rho(t)]\}$, and we can write Eq.~(\ref{eq:intro_nakajima_zwanzig}) as
\begin{equation}
\partial_t \rho(t)=\mathcal{L}_t [\rho(t)], \qquad
\mathcal{L}_t=\int_0^t \d s\, \mathcal{K}(t,s)\circ\Phi_s\circ\Phi_t^{-1},
\end{equation}
i.e., in the desired time-local form. Moreover, we can differentiate Eq.~(\ref{eq:intro_def_map}) and, again with assumption (ii), use $\rho(0)=\Phi_t^{-1}[\rho(t)]$ to write $\partial_t \rho(t)=(\partial_t\Phi_t)\{\Phi_t^{-1}[\rho(t)]\}$. Comparing with Eq.~(\ref{eq:intro_master_eq}), we can reduce a quantum channel to a Liouvillian through
\begin{equation}
\mathcal{L}_t=(\partial_t \Phi_t)\circ\Phi_t^{-1}.
\end{equation}

A particular case of interest, which is the main focus of this thesis, is that of Markovian master equations, which assume the Gorini-Kossakowski-Sudarshan-Lindblad (or just Lindblad, for short) form~\cite{belavin1969,lindblad1976,gorini1976}.
In Markovian systems~\cite{breuerpetruccione,rivas2014RepProgPhys,breuer2016RMP}, the bath is memoryless, which simplifies the description considerably. Any information that flows from the system to the bath is instantaneously scrambled among all the bath degrees of freedom and, consequently, the bath exerts no backreaction on the system. At each point in time, the future evolution of the system depends only on its current state and not the entire previous history.\footnote{
	The notion of Markovianity can be formulated in terms of the divisibility of the dynamical map $\Phi_t$~\cite{wolf2008,rivas2014RepProgPhys,breuer2016RMP}.
	Let us restrict our attention to the cases where the inverse of the dynamical map $\Phi_t$ exists for all times $t\geq0$. In that case we can define a map $\Phi_{t,s}=\Phi_t \circ \Phi_s^{-1}$, for $0\leq s\leq t$, that propagates a state $\rho(s)$ to $\rho(t)$. Obviously $\Phi_{t}=\Phi_{t,0}=\Phi_{t,s}\Phi_{s,0}$. The existence of the inverse $\Phi_t$ for all times thus implies the divisibility of the quantum evolution. While $\Phi_{t,0}$ and $\Phi_{s,0}$ are completely positive by assumption, $\Phi_{t,s}$ need not be because the inverse $\Phi_s^{-1}$ of a completely positive map is not always completely positive. If, for all $s$, $\Phi_{t,s}$ is positive, then $\Phi_t$ is said to be P-positive. If, more strongly, $\Phi_{t,s}$ is completely positive, $\Phi_t$ is said CP-divisible. Markovian evolution corresponds to P-divisible dynamical maps~\cite{wissmann2015PRA,breuer2016RMP}.	
}

One possibility to arrive at the Lindblad equation is to start from the joint system-plus-bath Schr\"odinger equation and trace our the bath degrees of freedom~\cite{breuerpetruccione}.
One then assumes that the bath is much larger than the system and that the coupling between them is weak (Born approximation), leading to a factorization of the total density matrix into system and bath parts at all times. One further assumes that there is a clear separation of the system and environment timescales (Markov approximation), with the bath much faster than the system, leading the future state to depend only on the present and not the past states, effectively rendering the bath memoryless. A standard procedure~\cite{breuerpetruccione} is then to apply second-order perturbation theory and take the trace over the bath degrees of freedom. The resulting Liouvillian is of Lindblad form~\cite{belavin1969,lindblad1976,gorini1976},
\begin{equation}
\label{eq:intro_Lindblad}
\partial_t\rho=\scL (\rho)= -\i \comm{H}{\rho}+\sum_{m=1}^{M}\(2L_m\rho L_m^\dagger-\acomm{L_m^\dagger L_m}{\rho}\),
\end{equation}
with a (renormalized) system Hamiltonian $H$ and $M$ traceless jump operators $L_m$, $m=1,\dots,M$, that couple the system to its environment. Following a procedure analogous to that outlined in Sec.~\ref{sec:intro_Kraus}, it is possible to show~\cite{hall2014PRA} that Eq.~(\ref{eq:intro_Lindblad}) is the most general first-order linear equation that generates an Hermiticity-preserving CPTP map~\cite{lindblad1976,gorini1976,alicki2007}.
Lifting the requirement of complete positivity, the Liouvillian generator of any Hermiticity- and trace-preserving quantum master equation, including non-Markovian ones, can be written in generalized Lindblad form,
\begin{equation}
\scL (\rho)= -\i \comm{H(t)}{\rho}+\sum_{m=1}^{M}\gamma_m(t)\(2L_m(t)\rho L_m^\dagger(t)-\acomm{L_m^\dagger(t) L_m(t)}{\rho}\)
\end{equation}
with time-dependent Hamiltonian and jump operators and, possibly, negative decoherence rates $\gamma_m(t)$ (also time dependent). These negative rates are associated with non-Markovianity, since it is possible for the system to recohere, i.e., for information to flow back from the system into the bath.
While systems with Markovian dissipation generated by a Lindbladian\footnote{Since all our Liouvillians will be of Lindblad form, we will use the terms Liouvillian and Lindbladian interchangeably.} of the form~(\ref{eq:intro_Lindblad}) will be the main focus of this thesis, this result shows that the implications of our work are far-reaching.

\subsection{Other sources of non-Hermiticity}

Although it is the focus of this thesis, non-Hermiticity does not arise only from coupling a system to a bath and integrating out the latter. Non-Hermitian effective descriptions of quantum Hamiltonians appear in a multitude of other problems, e.g.: the Euclidean QCD Dirac operator at nonzero chemical potential~\cite{stephanov1996PRL,halasz1997PRD,kanazawa2021PRD}, the scattering matrix of open quantum systems from quantum dots
\cite{alhassid2000} to compound nuclei~\cite{verbaarschot1985}, flux lines depinned from columnar defects by a transverse magnetic field in superconductors~\cite{hatano1996PRL,hatano1997PRB}, and continuously-monitored systems with postselection of the null measurement outcome~\cite{plenio1998,daley2014}---see Ref.~\cite{ashida2020} for a broad review of non-Hermitian physics.

Finally, we mention the possibility of non-Hermiticity arising in nonconventional quantum mechanics. In 1998, Bender and Boettcher proposed~\cite{bender1998PRL} that some non-Hermitian operators that are not Hermitian but invariant under the joint action of parity and time-reversal---so-called PT symmetry---could still have real eigenvalues and lead to consistent unitary dynamics~\cite{bender2007RPP}. The simplest example is the single-particle Schr\"odinger Hamiltonian $H=\hat{p}^2+\i \hat{x}^3$. A generalized notion of PT symmetry~\cite{bender2002JPhysA} involves coupling two copies of a system, with one the time-reversed of the other, in such a way that the losses in one are compensated by the gains in the other~\cite{benderbook}. One can then see such systems as intermediate between closed equilibrium systems and open systems. PT-symmetric systems have been experimentally realized, e.g., in optics, acoustics, and metamaterials---see Ref.~\cite{elganainy2018NatPhys} for a review.

\subsection{The spectrum of the Lindbladian}

Obtaining the dynamics or the steady state of a strongly-interacting dissipative quantum system is an enormous challenge. Nevertheless, over the last two decades, many instances of integrable open systems were obtained and solved. Exactly solvable Lindbladian dynamics were found for quadratic systems~\cite{prosen2008,prosen2008prl,prosen2010jstat,prosen2010njp,zunkovivc2010,prosen2010jphysa,znidaric2010jphysa,prosen2011ilievski,zunkovic2014}, free systems subject to dephasing~\cite{znidaric2010,znidaric2011,temme2012,eisler2011}, stationary states of boundary-driven interacting spin chains~\cite{prosen2011a,prosen2011b,prosen2012,buca2012,prosen2013prl,prosen2013njp,karevski2013,popkov2013,prosen2014,ilievski2014,ilievski2014JSTAT,buca2014,lenarcic2015,popkov2015,prosen2015REVIEW,buca2016,ilievski2017,matsui2017,buca2018,popkov2020prl,popkov2020pre}, interacting Liouvillians that can be mapped to Bethe-ansatz integrable non-Hermitian Hamiltonians~\cite{medvedyeva2016,rowlands2018,shibata2019a,shibata2019b,ziolkowska2020SciPost}, collective spin models~\cite{ribeiro2019}, triangular Lindbladians~\cite{torres2014PRA,nakagawa2020PRL,buca2020NJP}, and systems with a special Hilbert space structure~\cite{essler2020PRE}. Using an integrable Trotterization~\cite{vanicat2018,vanicat2018b}, the first example of an integrable quantum channel was found by us in Ref.~\cite{sa2021PRB}, which will be discussed in Ch.~\ref{chapter:circuits}. Afterward, this method was shown to be a powerful and reliable new avenue for constructing exact solutions in open systems~\cite{deleeuw2021PRL,deleeuw2022ARXIV,su2022PRB}.

An alternative approach is to consider those systems that are complex enough for a statistical approach to be again applicable. Such systems and the methods used to study them form the subject now known as \emph{dissipative quantum chaos}. Most of the remainder of this thesis will be concerned with the spectral properties of such systems. In the following we set up the eigenvalue problem and list some general properties of the spectrum. For concreteness, we focus on the Lindbladian spectrum. The spectrum of the quantum channels can be then obtained by exponentiation.

As the generator of open quantum dynamics, the Lindbladian is a non-Hermitian operator and, consequently, has a complex spectrum. Indeed, the adjoint Lindbladian is defined through $\Tr[\mathcal{L}^\dagger(A)B]=\Tr[A \mathcal{L}(B)]$, for some operators $A$ and $B$. Using the cyclic property of the trace, we have
\begin{equation}
\label{eq:intro_adjoint}
\scL^\dagger (A)= +\i \comm{H}{A}+\sum_{m=1}^{M}\(2L_m^\dagger A L_m-\acomm{L_m^\dagger L_m}{A}\).
\end{equation}
Comparing with Eq.~(\ref{eq:intro_Lindblad}), we immediately conclude $\mathcal{L}\neq\mathcal{L}^\dagger$.
If we formally integrate the time-independent Lindblad equation~(\ref{eq:intro_Lindblad}), the state of an $N$-dimensional system at time $t$ is
\begin{equation}
\label{eq:intro_formal_solution}
\rho(t)=e^{t\mathcal{L}}\rho(0)=\sum_{\alpha=0}^{N^2-1} \langle \tilde{\rho}_\alpha,\rho(0)\rangle e^{\Lambda_\alpha t}\rho_\alpha\,,
\end{equation}
where we have expanded the density matrix in the eigenbasis of the Lindbladian. Here, $\langle \tilde{\rho}_\alpha,\rho(0)\rangle$ is a constant that depends on the initial conditions and is written in terms of the Hilbert-Schmidt inner product, $\langle A,B\rangle=\Tr[A^\dagger B]$ for some operators $A$ and $B$. The right eigenoperators
of $\mathcal{L}$, respecting $\mathcal{L}(\rho_{\alpha})=\Lambda_{\alpha}\rho_{\alpha}$, with $\alpha=0,\dots,N^{2}-1$, are denoted by $\rho_{\alpha}$, with
$\Lambda_{\alpha}$ the respective complex eigenvalues, ordered by decreasing real part, $\Re(\Lambda_0)>\Re(\Lambda_1)>\cdots\Re(\Lambda_N^2-1)$. $\tilde{\rho}_\alpha$ are left eigenvectors, satisfying $\mathcal{L}^{\dagger}(\tilde{\rho}_\alpha^{\dagger})=\Lambda_\alpha\tilde{\rho}_\alpha^{\dagger}$. Because the Liouvillian is non-Hermitian, left and right eigenoperators do not coincide. Left and right eigenoperators are also not orthogonal, $\langle\rho_\alpha,\rho_\beta\rangle\neq\delta_{\alpha,\beta}$ and $\langle\tilde{\rho}_\alpha,\tilde{\rho}_\beta\rangle\neq\delta_{\alpha,\beta}$, but instead biorthogonal, $\langle\tilde{\rho}_\alpha,\rho_\beta\rangle=\delta_{\alpha\beta}$.

Now, $\mathcal{L}$ is not an arbitrary non-Hermitian operator. First, because the dynamics are Hermiticity-preserving, i.e., $(\partial_t\rho)^\dagger=\partial_t\rho$, we have $[\mathcal{L}(\rho)]^\dagger=\mathcal{L}(\rho^\dagger)$ and, hence, using the right eigenvalue equation, that the eigenvalues of $\mathcal{L}$ are real or come in complex conjugated pairs.
Second, all eigenvalues satisfy $\Re\left(\Lambda_{\alpha}\right)\le\Lambda_{0}=0$. This is intuitively clear, as otherwise $\rho(t)$ would have unbounded moments as $t\to\infty$.
Third, there is always at least one vanishing eigenvalue, $\Lambda_0=0$. To see this, we note that, using Eq.~(\ref{eq:intro_adjoint}), $\mathcal{L}^\dagger(\id)=0$, i.e., the identity operator is a left eigenvector of $\mathcal{L}$ with eigenvalue zero. The corresponding right eigenvector $\rho_0$ is a steady state, left invariant
by the evolution. In the absence of additional symmetries (so-called Liouvillian strong symmetries~\cite{buca2012}), the steady state is unique and, since all other states decay in time with rate $|\Re\left(\Lambda_\alpha\right)|$, the system relaxes to the steady state, $\lim_{t\to\infty}\rho(t)=\rho_0$. For finite systems, the relaxation is exponential, with a rate set by the spectral gap $\Delta=|\Re(\Lambda_1)|$.

\section{Non-Hermitian RMT and the foundations of dissipative quantum chaos}
\label{sec:intro_nonHermitian}

Because open quantum systems are described by non-Hermitian operators, applying to them the ideas and methodology of quantum chaos required developing non-Hermitian RMT.
This idea was first pursued by Ginibre in 1965~\cite{ginibre1965}, who, without any specific physical motivation in mind at the time, proposed to study the statistical properties of three ensembles of $N\times N$ matrices with independent and identically distributed Gaussian random variables (that correspond to lifting the Hermiticity constraints in the Wigner-Dyson ensembles): complex asymmetric matrices form the Ginibre unitary ensemble (GinUE); real asymmetric matrices, the Ginibre orthogonal ensemble (GinOE); and real quaternionic matrices, the Ginibre symplectic ensemble (GinSE).

As is the case for Hermitian random matrices, from the joint eigenvalue distribution~\cite{ginibre1965,lehmann1991PRL}, one can compute $k$-point correlation functions, in particular, the spectral density~\cite{ginibre1965,mehta2004,sommers1988}, which, in the large-$N$ limit is (with the appropriate normalization of the matrix elements),
\begin{equation}
\varrho_{\mathrm{Gin}}(z,z^*)=\frac{1}{\pi}\Theta\(1-|z|\),
\end{equation}
for all three ensembles. Here $\Theta$ is the Heaviside step function. This is the first example of Girko's circular law~\cite{girko1985}, which specifies that the limiting eigenvalue density for a wide class of non-Hermitian random matrices is constant inside the unit disk.
Contrary to Hermitian Gaussian matrices, non-Hermitian matrices have a sharp edge, where the level density jumps discontinuously from a finite value to zero.
Taking the real and imaginary parts of the entries to have different variances defines the so-called elliptic Ginibre ensembles~\cite{lehmann1991PRL}, whose eigenvalues are supported inside an ellipse. By varying the variance of the imaginary parts, this ensemble interpolates between the GinUE and the GUE~\cite{fyodorov1997,fyodorov1998,fyodorov2003}.

While non-Hermitian matrices have in general complex eigenvalues, the eigenvalues of the GinOE and GinSE come in complex-conjugated pairs (because of the symmetry $TH^*T^{-1}=H$, with $T$ the identity for the GinOE and $T$ the symplectic unit for the GinSE). At finite $N$, the GinOE and GinSE have, therefore, special features on and near the real axis, where the level density deviates from the flat one found deep in the bulk of the spectrum. More precisely, there is a depletion of complex eigenvalues $z+\i y$ near the real axis, with the level density showing a repulsion $\propto |y|$ for the GinOE~\cite{edelman1997} and $\propto |y|^2$ for the GinSE~\cite{kanzieper2002}. Additionally, the GinOE can have a finite number $0\leq k\leq N$ of purely real eigenvalues. The average number of real eigenvalues of an $N\times N$ GinOE matrix is $\propto \sqrt{N}$ in the large-$N$ limit~\cite{sommers1988,edelman1994}, while its full distribution is also known~\cite{edelman1997,kanzieper2005PRL,forrester2007PRL}.

Inspired by the success of RMT in modeling chaotic quantum Hamiltonians, Grobe, Haake, and Sommers~\cite{grobe1988} investigated the spectral properties of dissipative quantum maps (specifically, dissipative kicked tops) with regular and chaotic classical limits. They found that an integrable quantum map has the same statistics as a Poisson process in the plane, in particular, linear level repulsion, $P(s)\propto s$ for $s\to0$;\footnote{
	While this seems rather distinct from the absence of level repulsion for integrable Hermitian Hamiltonians, the level repulsion is a consequence of the area element $s$ in the two-dimensional complex plane. The actual levels are still independent, as in the Hermitian case.
}
while a chaotic quantum map shows the same spacing distribution as a random matrix from the GinUE, in particular, \emph{cubic level repulsion}, $P(s)\propto s^3$ for $s\to0$.
Quite remarkably, the three Ginibre ensembles show the same cubic level repulsion~\cite{grobe1988,grobe1989}, irrespective of the real, complex, or quaternionic nature of their entries. It was conjectured~\cite{grobe1989} that cubic level repulsion is universal (up to logarithmic corrections) for any non-Hermitian random matrix.
Based on these observations, Ref.~\cite{grobe1988} proposed a dissipative version of the quantum chaos conjecture, thus laying the foundations of dissipative quantum chaos.

Despite the pioneering contributions of Grobe, Haake, and Sommers, a systematic application of RMT to the non-Hermitian \emph{generators} and \emph{maps} of dissipative quantum systems and the investigation of the many non-Hermitian universality classes were left mostly unexplored and had to wait for another 30 years (see the next section). Nevertheless, great progress was made in the following years in non-Hermitian RMT, driven, once again, mostly by applications in nuclear physics, in particular, scattering systems and QCD. Indeed, in the latter case, the continuum Euclidean Dirac operator at nonzero chemical potential is modeled by non-Hermitian chiral matrices, the so-called chiral Ginibre ensembles (chGinUE, chGinOE, and chGinSE)~\cite{halasz1997PRD}, while the lattice Wilson-Dirac operator can be described by random matrices from what are today called non-Hermitian classes AIII~\cite{kieburg2012PRL,kieburg2013PRD} and BDI$^\dagger$~\cite{kieburg2015PRD}.

In the context of chaotic scattering, particular attention was devoted to the study of the non-Hermitian Hamiltonian $H_\mathrm{eff}=H-\i g\Gamma$~\cite{sokolov1988,sokolov1989}, where $H$ and $\Gamma=A A^T$ are $N\times N$ real symmetric matrices and  $\Gamma$ is, in addition, positive definite (and parametrized in terms of the $N\times M$ real matrix $A$), and $g$ measures the degree of non-Hermiticity. The eigenvalues of $H_\mathrm{eff}$ give the poles of the $S$-matrix and, therefore, the location and width of scattering resonances. To tackle the spectral problem of $H_\mathrm{eff}$, new techniques had to be developed, as the powerful Green's function (or resolvent) methods mentioned in Sec.~\ref{sec:intro_RMT} fail for non-Hermitian matrices, whose eigenvalue support is not a line but a dense domain in the complex plane. This problem can be overcome by considering matrix-valued resolvents, using either the method of Hermitization~\cite{feinberg1997a,feinberg1997b,feinberg2006} or quaternionic methods~\cite{janik1997a,janik1997b,jarosz2006} (quaternionic free probability is reviewed in App.~\ref{app:kraus_GinUECUE}). Then, using replicas~\cite{haake1992}, supersymmetry~\cite{lehmann1995}, diagrammatics~\cite{janik1997a}, or free probability~\cite{janik1997b}, the boundary of the spectral support of $H_{\mathrm{eff}}$ could be computed analytically in the limit $N \to \infty$ as the solution to the following quartic equation:
\begin{equation}
\label{eq:intro_boundary}
x^2=\frac{4m}{gy}-\(\frac{m}{y}-\frac{g}{1+gy}+\frac{1}{g}\)^2,
\end{equation}
where $m=M/N$ and $x+\i y$ represents a point in the complex plane.
In particular, a phase transition, where the spectrum splits into two disconnected components at a critical value of $g$ was found.

Besides eigenvalue correlations, also eigenvector statistics attracted considerable attention, as they are relevant for, e.g., the stability of nonequilibrium systems~\cite{bourgade2020} and non-Hermitian diffusion~\cite{grela2014PRL,grela2018JPhysA}.
As seen above, the eigenvectors of non-Hermitian matrices are not orthogonal but biorthogonal. As such, the overlaps of different eigenvectors are nontrivial. Chalker and Mehlig~\cite{chalker1998PRL,mehlig2000JMP} introduced the overlap matrix
\begin{equation}
O_{\alpha \beta}=\langle \tphi_\alpha | \tphi_\beta \rangle
\langle \phi_\alpha | \phi_\beta \rangle,
\end{equation}
where $\tphi_\alpha$ and $\phi_\alpha$ are the left and right eigenvectors of a non-Hermitian operator, respectively, and computed the asymptotic behavior of its lowest moments. Most of the work has focused on the heavy-tail distribution of the diagonal elements $O_{\alpha \alpha}$ for the GinUE~\cite{bourgade2020}. It was also found soon after that the matrix-valued resolvent mentioned above contains, in addition to the one-point eigenvalue function, the one-point eigenvector correlation function~\cite{janik1999PRE,jarosz2006,nowak2018JHEP}.

Just as in the Hermitian case, there is a very rich mathematical structure behind the non-Hermitian random matrix ensembles, with many analytical results obtained since the 1990s, in particular for the Ginibre and chiral Ginibre ensembles---for a recent mathematical review, see Refs.~\cite{byun2022,byun2023}. Notwithstanding, a more sustained development of dissipative quantum chaos had to wait until around 2018.

\section{Recent developments in dissipative quantum chaos}
\label{sec:intro_developments}

The project of this thesis started five years ago. In 2018, several groups kickstarted the field of dissipative quantum chaos, which has flourished since. Many exciting developments took place in parallel to the results reported in this thesis, which we review in this section.

\subsection{Random matrix theory of Markovian dissipation}

Given the tremendous success of RMT in the description of complex closed quantum systems, along the lines of Wigner's work six decades ago, it is perhaps surprising that the same ideas in the context of dissipative systems had remained mostly unexplored. This changed in late 2018, with the first proposal of a random Lindblad operator~\cite{denisov2019PRL}. By randomly sampling both the Hamiltonian of the system and the jump operators that connect it to the bath from featureless RMT ensembles, it was found that random Liouvillians have a universal lemon-shaped spectral support~\cite{denisov2019PRL,sa2019JPhysA,lange2021,tarnowski2021PRE}, whose boundary~\cite{denisov2019PRL} and spectral distribution~\cite{tarnowski2021PRE} could be computed analytically using quaternionic free probability. The spectral gap for systems with jump operators sampled from different distributions was extensively studied in Ref.~\cite{can2019JPhysA,can2019PRL}, while the dependence of spectral and steady-state properties of the random Liouvillian with system size, dissipation strength, and the number of jump operators was addressed in Ref.~\cite{sa2019JPhysA}. Moreover, the lemon-like shape of the spectrum of random Lindbladians is different from the spindle-like shape of the spectrum of random classical Markov (or Kolmogorov) generators~\cite{timm2009,denisov2019PRL,nakerst2023PRE} and the transition between the two (super-decoherence) was studied in Ref.~\cite{tarnowski2021PRE}. Finally, Ref.~\cite{tarnowski2023OSID} studied random classical and quantum Markov generators that satisfy detailed balance and, hence, have real spectra.

Along similar lines, complex Markovian systems with a stroboscopic time evolution, modeled by random Kraus maps, were also considered. Different sampling schemes for structureless quantum maps were proposed in Refs.~\cite{bruzda2009,bruzda2010,kukulski2021JMP} (see also Ref.~\cite{horvat2009JSTAT} for classical random stochastic matrices), but these maps lacked a notion of dissipation strength, difficulting the comparison with random Lindbladians. One-parameter~\cite{sa2020PRB} and two-parameter~\cite{matsoukas-roubeas2023ARXIV} random Kraus maps unveiled spectral transitions and a rich spectral phase diagram. Remarkably, the steady-state properties of random Lindbladians and random Kraus maps coincide~\cite{sa2020PRB}, hinting towards the universality of steady states of random dissipative systems.

While the results for fully random dynamics are encouraging, physically motivated models have few-body interactions, rendering them very different. Indeed, considering jump operators with few-body interactions has clarified the role of locality in the separation of dissipative timescales~\cite{wang2020PRL,sommer2021PRR,hartmann2023ARXIV} and metastability~\cite{li2022PRB}, and allowed an accurate description of the noise of the IBM quantum computing platform~\cite{sommer2021PRR}.
Moreover, a model of random open free fermions~\cite{costa2022}, was shown to host two regimes: an ergodic one with the same universal properties as structureless random Lindbladians, and a nonergodic one with suppressed dissipation.
Random Lindbladians with different notions of spatial locality lead to a finer hierarchy of relaxation timescales~\cite{wang2020PRL} and to different phases of relaxation dynamics~\cite{orgad2022ARXIV}.

Finally, with the goal of describing dissipation in strongly-correlated quantum matter, Ref.~\cite{sa2022PRR} initiated the study of a Lindbladian version of the SYK model, the first exactly-solvable example of a non-integrable Liouvillian. We found that the dissipative SYK model dissipates exponentially fast and computed the rate analytically~\cite{sa2022PRR}. The rich physics of this model was subsequently unveiled in Refs.~\cite{kulkarni2022PRB,garcia2023PRD2,kawabata2023PRB}, namely anomalous relaxation~\cite{garcia2023PRD2}, a connection to novel wormhole configurations in gravity~\cite{garcia2023PRD2}, and dynamical phase transitions~\cite{kawabata2023PRB}.

\subsection{Symmetries classes of dissipative quantum matter}

In parallel to these developments on the RMT modeling of the non-Hermitian generators of open quantum systems, the study of the symmetries, correlations, and universality of dissipative quantum chaos began in earnest around the same time.

The classification of non-Hermitian matrices in terms of their global symmetries was first attempted in Refs.~\cite{bernard2002a,bernard2002,magnea2008}, after it was realized that non-Hermitian matrices admit more possible antiunitary symmetries than Hermitian ones (essentially, complex conjugation and transposition become distinct transformations). Years later, the subject was revisited~\cite{sato2012PTP,lieu2018PRB,gong2018PRX,kawabata2019NatComm} and the role of different antiunitary symmetries was clarified, finally culminating in a 38-fold classification for matrices with a point-gap spectrum~\cite{kawabata2019PRX,zhou2019PRB} and a 54-fold classification for matrices with a line-gap spectrum~\cite{liu2019PRB,ashida2020}. As was already the case in closed quantum systems, open versions of the SYK model provide concrete realizations of many of these classes~\cite{garcia2022PRX,garcia2023PRD,kawabata2023PRXQ,garcia2023ARXIVb}.

Non-Hermitian Hamiltonians provide only an effective description of open quantum dynamics and a fundamental question was how the classification is constrained by the conservation of trace, Hermiticity, and complete positivity in Lindbladian dynamics. Using causality arguments, Ref.~\cite{lieu2020PRL} argued that there are ten classes of single-particle Lindbladians. However, they did not consider shifting the spectral origin~\cite{prosen2012PRL,prosen2012PRA}, which avoids the causality restrictions, as pointed out in Ref.~\cite{kawasaki2022PRB}. Once this possibility is accounted for, all 54 classes of non-Hermitian Hamiltonians can be implemented at the level of single-particle spectra. Ref.~\cite{altland2021PRX} considered the invariance of the dynamics under linear or antilinear, canonical or anticanonical transformations of fermionic creation and annihilation operators, leading to a tenfold classification of the single-particle tensors that appear in the Hamiltonian
and jump operators, including in the presence of interactions. A full many-body classification of Lindbladians, based on its matrix representations was performed in Refs.~\cite{sa2023PRX,kawabata2023PRXQ}. In particular, Ref.~\cite{sa2023PRX} showed that there is a tenfold classification without unitary symmetries, which can be enriched by the presence of the latter.

\subsection{Correlations and universality in dissipative quantum chaos}

With this rich symmetry classification established, it was immediately realized that the RMT correlations were only understood for a very small subset of classes. So, while  the universal cubic level repulsion of Ginibre ensembles was extended to the full spacing distribution in Ref.~\cite{akemann2019}, it was also established that noncubic level repulsion can exist in non-Hermitian ensembles with different symmetries~\cite{hamazaki2020PRR} and that it is the behaviour of the matrix under transposition, not complex conjugation, that determines the local level correlations. Two other universality classes, called AI$^\dagger$ and AII$^\dagger$, exist beyond Ginibre (class A) and Wigner-like surmises were computed for their spacing distributions~\cite{hamazaki2020PRR}. In classes with complex-conjugation symmetry, also the behavior of eigenvalues on or close to the real axis can differ from the Ginibre ensemble~\cite{xiao2022PRR}.

As discussed in Sec.~\ref{sec:intro_RMT}, computing the spacing distribution requires unfolding the spectrum, which, furthermore, can be ambiguous for non-Hermitian matrices~\cite{akemann2019}. This issue was circumvented with the introduction of complex spacing ratios (CSRs)~\cite{sa2020PRX}, which, beside level repulsion also measure local angular correlations. The CSR distribution was obtained numerically for the three universality classes in Refs.~\cite{sa2020PRX,kanazawa2021PRD,garcia2022PRX} and computed analytically for class A from a Wigner-like surmise in Ref.~\cite{sa2020PRX} and in the limit $N\to\infty$ in Ref.~\cite{dusa2022PRE}. CSRs are now the standard tool for analyzing short-range correlations of non-Hermitian spectra.

Nonlocal correlations over different energy scales have also been considered. To this end, several different non-Hermitian generalization of the SFF were proposed (with unfortunately similar names). For Hermitian systems, the SFF can be defined, equivalently, as (i) the trace of the system's unitary quantum channel $U_t(\cdot) U_t^\dagger$, (ii) the Fourier transform of the two-point correlation function of the unitary evolution operator $U_t$, or (iii) the survival probability of a coherent Gibbs state under the unitary evolution $U_t$. For non-Hermitian systems, these three definitions do not coincide and which definition is preferable depends on the specific problem under consideration. 
The dissipative form factor (DFF), introduced in Ref.~\cite{can2019JPhysA}, generalizes property (i) and is given by the trace of the quantum channel $\Phi_t$. The late-time behavior of the DFF is controlled by the spectral gap and was used to compute the latter for random Liouvillians~\cite{can2019JPhysA}, while the finite-time behavior captures dynamical phase transitions~\cite{kawabata2023PRB}. It does not, however, measure the correlations of eigenvalues in the complex plane.
The dissipative spectral form factor (DSFF) was defined~\cite{li2021PRL} (see also Refs.~\cite{fyodorov1997,braun2001}) as the complex Fourier transform of the two-point correlator of $\Phi_t$, thus generalizing (ii), and, as such, measures spectral correlations of the dissipative map. It was computed analytically for class A in Ref.~\cite{li2021PRL} and numerically for classes AI$^\dagger$ and AII$^\dagger$ in Ref.~\cite{garcia2023PRD}. It accurately captures the quantum chaos properties of several single- and many-body physical systems~\cite{li2021PRL,chan2022NatComm,shivam2023PRL,gosh2022PRB,garcia2023PRD}, while its limitations (e.g., nonstationarity and the potential absence of a correlation hole) were highlighted in Ref.~\cite{garcia2023PRD}.
The non-Hermitian analogue of the SFF as employed in Refs.~\cite{xu2021PRB,cornelius2022PRL,matsoukas-roubeas2023ARXIV,matsoukas-roubeas2023ARXIVb} (see also Refs.~\cite{xu2019PRL,delcampo2020JHEP,matsoukas-roubeas2023JHEP}) follows route (iii) and measures the survival probability of a Gibbs state under the dissipative map $\Phi_t$. It is a convenient tool to study the effect of decoherence on Hamiltonian chaos, which, for the particular case of energy dephasing, can be either suppressed~\cite{xu2021PRB} or enhanced~\cite{cornelius2022PRL}.

Finally, the Chalker-Mehlig eigenvector overlaps have further elucidated some aspects of dissipative quantum chaos, as they were found to be empirical detectors of antiunitary symmetries~\cite{sa2023PRX} and were used in establishing the violation of the standard eigenstate thermalization hypothesis by non-Hermitian Hamiltonians~\cite{cipolloni2023ARXIV}.

\section{Outline and main results of the thesis}
\label{sec:intro_outline}

This thesis is divided into three parts and presents the details of some of the developments discussed in the previous section, namely, those in Refs.~\cite{sa2019JPhysA,sa2020PRX,sa2020PRB,sa2021PRB,sa2022PRD,garcia2022PRX,sa2022PRR,garcia2023PRD,garcia2023PRD2,sa2023PRX,costa2022}.

Part I, \textit{Random matrix theory of Markovian dissipation}, addresses the question (i) \textit{How to model a generic dissipative quantum system?}
Embracing the quantum chaos perspective, we apply RMT to model the generators of open quantum systems, both in continuous and discrete time, with increasingly realistic features.

We start, in Ch.~\ref{chapter:randomLindblad}, by reviewing some known results about random Lindblad dynamics with single-body quantum chaos. We will focus on results previously obtained by us for unstructured (first-quantized) quantum systems and quadratic many-body systems that will be important for the following chapters. Of particular importance are the scaling of the spectral gap with dissipation strength and the nature of the nontrivial steady states. 

In Ch.~\ref{chapter:SYKLindblad}, we then move to random Lindblad dynamics with many-body quantum chaos, modeled by the strongly-interacting dissipative SYK model. We compute the gap of this system and find dissipation-driven relaxation in the strong-dissipation regime and anomalous relaxation in the weakly dissipative regime.

Afterward, in Ch.~\ref{chapter:kraus}, we move to the complementary setting of discrete-time dissipative Floquet dynamics, and we investigate random quantum channels. We study both the spectral properties, finding a phase transition in the spectrum that we can characterize analytically, and the steady-state properties, which we find to be identical to those of random Lindblad dynamics with single-body quantum chaos, indicating universality.

Part II, \textit{Symmetry classes of dissipative quantum matter}, is concerned with the question (ii) \textit{What do generic quantum systems have in common?} If a system is chaotic, it is expected to display the statistical correlations of a random matrix with the same symmetries. For this reason, symmetry classifications offer key universal information. In this part, we provide the classification of several models of dissipative quantum matter.

We open this part with a study of symmetries, correlations, and universality in dissipative quantum chaos in Ch.~\ref{chapter:correlations}. We first present a detailed derivation of the 54-fold classification of non-Hermitian Hamiltonians. Next, we review the increasingly popular complex spacing ratios (previously discussed in detail in Ref.~\cite{saMScThesis}). Then we study bulk correlations over longer energy scales using the dissipative spectral form factor. Finally, we propose eigenvector overlaps as an empirical detector of antiunitary symmetry.

To test universality and its limits in many-body dissipative quantum chaos, one needs a system that is both complex enough to test the results of RMT and simple enough for analytical and numerical treatment. We propose that this model is the non-Hermitian SYK model, which is the subject of Ch.~\ref{chapter:classificationSYK}. We show that it belongs to one of nine possible symmetry classes without reality conditions. We then investigate level statistics on several timescales, finding exact random matrix universality at long times and deviations at short times.

In Ch.~\ref{chapter:classificationLindblad}, we present the tenfold classification of many-body Lindbladians without unitary symmetries, which can be enriched by the presence of the latter, and build simple experimentally-relevant examples in all ten classes.


Part III, \textit{Miscellaneous topics}, covers topics that, while closely related to it, are somewhat outside the main narrative of this thesis. We study systems that, strictly speaking, do not pertain to the realm of dissipative quantum chaos, as they are either not dissipative or not chaotic.

In Ch.~\ref{chapter:circuits}, we study \textit{nonchaotic} dissipative quantum systems, namely, integrable nonunitary open quantum circuits. We propose the first integrable nonunitary circuit, rigorously prove its integrability, and investigate integrability-breaking deformations. 

In Ch.~\ref{chapter:QLaguerre}, we consider a \textit{nondissipative} quantum chaotic system, the Wishart-Sachdev-Ye-Kitaev model. We derive an analytic approximation for its spectral density in terms of $Q$-Laguerre polynomials and investigate its quantum chaotic properties, finding in particular chGUE statistics, which are not found in the standard SYK model.

In Ch.~\ref{chapter:conclusions} we present our concluding remarks and point to unexplored directions in dissipative quantum chaos. 

Some technical details are presented in five appendices.	

%% file: Thesis_RandomLindblad.tex

\chapter{Random Lindblad dynamics I. Single-body quantum chaos}
\label{chapter:randomLindblad}

In this chapter, we review the spectral and steady-state properties of random Liouvillians with single-particle quantum chaos. 

First, in Sec.~\ref{sec:RL_RMT_Liouvillian}, we study the most generic systems with Markovian dissipation, focusing on a random matrix theory (RMT) ensemble of completely unconstrained stochastic Liouvillians. The global spectral features, the spectral gap, and the steady-state properties follow three different regimes as a function of the dissipation strength and, within each regime, we determine the scaling with the dissipation strength and system size. We find that, for two or more dissipation channels, the spectral gap vanishes linearly in the weak dissipation limit and then increases monotonically with dissipation.
The spectral distribution of the steady state is Poissonian at low dissipation strength and conforms to that of a random matrix once the dissipation is sufficiently strong, with a perturbative crossover between the two limits.

Then, in Sec.~\ref{sec:RL_quadratic_Lindblad}, we consider random open quantum systems with restricted one-body interactions, modeled by quadratic fermionic Liouvillian operators. The Hamiltonian dynamics is modeled by a generic random quadratic operator, i.e., as a  featureless superconductor of class D, whereas the Markovian dissipation is described by $M$ random linear jump operators. By varying the dissipation strength and the ratio of dissipative channels per fermion, $m=M/(2N_F)$, where $N_F$ is the number of fermions, we find two distinct phases where the support of the single-particle spectrum has one or two connected components. 
In the strongly dissipative regime, this transition occurs for $m=1/2$ and is concomitant with a qualitative change in both the steady state and the spectral gap that rules the long-time dynamics. 
Above this threshold, the spectral gap and the steady-state purity qualitatively agree with the RMT (i.e., non-quadratic) case studied previously. Below $m=1/2$, the spectral gap closes in the thermodynamic limit and the steady state decouples into an ergodic and a nonergodic sector, yielding a non-monotonic steady-state purity as a function of the dissipation strength.

This chapter is based on Refs.~\cite{sa2019JPhysA,costa2022}. Because these results were already discussed at length elsewhere~\cite{saMScThesis,costaMScThesis}, we will only review the main results that are important for Lindbladians with many-body chaos, to be addressed in Ch.~\ref{chapter:SYKLindblad}}, and for the universality of steady states, which is discussed in Ch.~\ref{chapter:kraus}. We refer the reader to Refs.~\cite{saMScThesis,costaMScThesis} for further details.

\section{RMT Lindbladians}
\label{sec:RL_RMT_Liouvillian}

We start our investigation of random open quantum systems with Markovian dissipation by studying, in the spirit of Wigner's approach, fully unstructured random Liouvillians. They are characterized by a fully random Hamiltonian and a set of fully random jump operators. More precisely, for the Liouvillian 
\begin{equation}
\label{eq:lindblad-1}
\mathcal{L}\left(\rho\right)=-i\left[H,\rho\right]
+\sum_{\mu=1}^{M}\(2L_\mu \rho L_\mu^\dagger-L_\mu^\dagger L_\mu \rho-\rho L_\mu^\dagger L_\mu \),
\end{equation}
we sample the $N\times N$ Hamiltonian $H$ from the GUE with unit variance, i.e.,
\begin{equation}
\label{eq:RL_ensemble_H}
P\left(H\right)\propto \exp\left\{-\frac{1}{2}\Tr\left(H^{2}\right)\right\}.
\end{equation}
To sample the $M$ independent jump operators, coupled to the environment with dissipation strength $g^2/2$, we define a complete orthogonal basis, $\left\{ G_{i}\right\} $ with $i=0,\dots,N^{2}-1$, for the space of operators acting on an Hilbert space of dimension $N$, respecting $\Tr\left[G_{i}^{\dagger}G_{j}\right]=\delta_{ij}$, with $G_{0}=\mathbbm{1}/\sqrt{N}$ proportional to the identity. Each jump operator can then be decomposed as
\begin{equation}
L_{\mu}=\frac{g}{\sqrt{2}}\sum_{j=1}^{N^{2}-1}G_{j}w_{j\ell},
\end{equation}
where $w$ is an $(N^2-1)\times M$ complex matrix drawn from the GinUE,
\begin{equation}
\label{eq:RL_ensemble_L}
P\left(w\right)\propto \exp\left\{-\frac{1}{2}\Tr\left(w^{\dagger}w\right)\right\}. 
\end{equation}
Note that $L_{\mu}$ is taken to be traceless, i.e., orthogonal to $G_{0}$, to ensure that the dissipative term in Eq.~(\ref{eq:lindblad-1}) does not contribute to the Hamiltonian dynamics.\footnote{For each realization of the system, our operators are time-independent. The statistical approach then corresponds to ensemble-averaging over random matrices. This is to be contrasted with the different approach of time-averaging perturbations that evolve randomly in time, but have a fixed direction in matrix space.} Effectively, the random Liouvillian~(\ref{eq:lindblad-1}) with Hamiltonian~(\ref{eq:RL_ensemble_H}) and jump operators~(\ref{eq:RL_ensemble_L}) models (large) single-particle chaotic quantum systems and, thus, our model probes single-particle quantum chaos.

In Ref.~\cite{sa2019JPhysA}, we studied the statistical properties of the spectrum of $\mathcal{L}$ drawn from an ensemble of random Lindblad operators parametrized by system size dimension $N$, number of decay channels $M$, and effective dissipation strength
\begin{equation}
g_{\mathrm{eff}}^2=\sqrt{\left(2M\beta N\right)}g^2.
\end{equation}
(Here and throughout, $\beta=2$, see Ref.~\cite{sa2019JPhysA} for $\beta=1$, which shows identical results.)
The right eigenvectors of $\mathcal{L}$, respecting $\mathcal{L}\left(\rho_{\alpha}\right)=\Lambda_{\alpha}\rho_{\alpha}$, with $\alpha=0,\dots,N^{2}-1$, are denoted by $\rho_{\alpha}$, with $\Lambda_{\alpha}$ the respective eigenvalue.
By construction, $\Re\left(\Lambda_{\alpha}\right)\le\Lambda_{0}=0$, and $\rho_{0}$, if unique,\footnote{The steady state is unique in the absence of any additional symmetries. Since we are considering the less structured Liouvillian possible the steady states we find are unique by construction.} is the asymptotic steady state, left invariant by the evolution. The spectrum and eigenvectors of $\mathcal{L}$ were obtained by exact diagonalization.

\begin{figure}[t]
	\centering 
	\includegraphics[width=\textwidth]{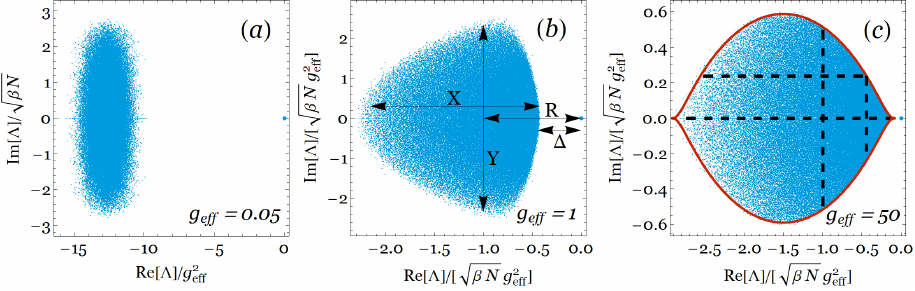}
	\caption{\label{fig:spectrum}Spectrum of a random Lindblad operator for different values of $g_\mathrm{eff}$, computed for $N=80$ and $M=2$. $R$, $X$, $Y$ and $\Delta$ are, respectively, the center for mass of the spectrum, the standard deviation along the real and imaginary axes, and the spectral gap. The horizontal (vertical) dashed lines in $(c)$ correspond to $0$, $Y$ ($R$, $R+X$), and the solid red boundary is explicitly computed in Ref.~\cite{denisov2019PRL}.}
\end{figure}

While in later chapters of this thesis we are mainly interested in the statistical properties of the spectral gap and the steady state, for completeness, we show the spectrum of a random Lindblad operator in the complex plane computed for different values of $g_\mathrm{eff}$ in Fig.~\ref{fig:spectrum}. 
The boundaries of the spectrum evolve from an ellipse, for small $g$, to a lemon-like shape at large $g$. 
In Fig.~\ref{fig:spectrum}~$(c)$ we plot (solid line) the spectral boundary, explicitly computed in Ref.~\cite{denisov2019PRL} for the case of maximal number of jump operators, $M=N^{2}-1$, using quaternionic free probability.
These results, obtained here for $M=2$, indicate that the lemon-shaped spectral boundary is ubiquitous in the strong dissipation regime. The spectral distribution inside the spectral support (again, for the maximal rank case $M=N^{2}-1$) has since also been analytically obtained in Ref.~\cite{tarnowski2021PRE}.

\subsection{Spectral gap}
\label{subsec_RL_spectral_gap}

\begin{figure}[tbp]
	\centering \includegraphics[width=0.7\textwidth]{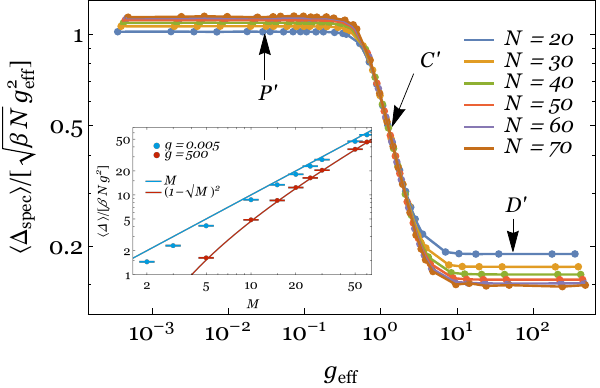}
	\caption{\label{fig:gap}Average spectral gap as a function of $g_{\mathrm{eff}}$ plotted for different values of $N$ for $M=2$. Inset: evolution of the spectral gap with the number of jump operators $M$, for $N=60$ and $g=0.005$ (blue) and $g=500$ (red); the full lines correspond to the analytic predictions.}
\end{figure}

We now turn to the study of the spectral gap $\Delta_{\mathrm{spec}}$, which is a particularly important spectral feature since it determines the long-time relaxation asymptotics. Figure~\ref{fig:gap} shows the average spectral gap, $\langle\Delta_{\mathrm{spec}}\rangle$, for $M=2$ as a function of $g_{\mathrm{eff}}$ for different values of $N$. In Ref.~\cite{sa2019JPhysA}, we found that, in the thermodynamic limit, the distribution of the gap becomes sharply peaked around its mean; hence, the latter accurately describes the long-time dynamics.

We find three qualitatively different regimes (marked as P$'$, C$'$, D$'$ in Fig.~\ref{fig:gap}). For regimes P$'$ and D$'$, the average gap behaves as $\langle\Delta_{\mathrm{spec}}\rangle\propto g_{\mathrm{eff}}^{2}$; for the crossover regime C$'$, the gap varies as $\langle\Delta_{\mathrm{spec}}\rangle\propto g_{\mathrm{eff}}$. Importantly, at small dissipation, the spectral gap closes linearly with dissipation strength $g_\mathrm{eff}^2$. While expected perturbatively for finite systems (i.e., finite $N$), this property survives even in the thermodynamic limit (i.e., if the limit $N\to\infty$ is taken before $g_\mathrm{eff}\to0$). For systems with \emph{single-particle} quantum chaos (here and in Secs.~\ref{sec:RL_quadratic_Lindblad} and \ref{sec:q2}), the spectral gap completely rules the long-time asymptotic dynamics of the system, and hence, relaxation becomes exponentially slow in the limit of a weak coupling to the bath. However, as we will see in Ch.~\ref{chapter:SYKLindblad}, for system with \emph{many-body} quantum chaos, relaxation of many-body observables is ruled by a distinct gap that does not close as dissipation decreases, leading to anomalously fast relaxation.

The spectral gap is described by a single scaling function for all $g_\mathrm{eff}$ that can be computed analytically in the double-scaling limit $N\to\infty$, $M\to\infty$ using holomorphic Green's function methods~\cite{saMScThesis}. It is found that $\Delta_{\mathrm{spec}}=-\sqrt{8N}\tilde{y}$, where $\tilde{y}$ is the smallest positive real solution of the quartic equation [cf., Eq.~(\ref{eq:intro_boundary})]:
\begin{equation}
\label{eq:RL_4eq}
\frac{4M}{\tilde{g}^2\tilde{y}}=\(\frac{M}{\tilde{y}}-\frac{\tilde{g}^2}{1+\tilde{g}^2\tilde{y}}+\frac{1}{\tilde{g}^2}\)^2,
\end{equation}
with $\tilde{g}^2=-\sqrt{N/2}g^2$. The limiting $N\to\infty$ values of the spectral gap for small and large $g_{\mathrm{eff}}$ are found to be
\begin{equation}\label{eq:asymptotics_gap}
\Delta_\mathrm{spec}=
\begin{cases}
2 N g^2 M\,, &\mathrm{when}\ g\to0\,,\\
2 N g^2 (1-\sqrt{M})^2, &\mathrm{when}\ g\to\infty\,.
\end{cases}
\end{equation}
These predictions describe the spectral gap increasingly well for growing $M$; for $M=2$, although not exact, they give a good estimate, see inset in Fig.~\ref{fig:gap}. Notwithstanding that the above two results are derived in the limits $g_{\mathrm{eff}}\to0$ and $g_{\mathrm{eff}}\to\infty$, they provide a remarkable description for the whole P$'$ and D$'$ regimes, respectively.

For the special case $M=1$, although three regimes are also present~\cite{sa2019JPhysA,saMScThesis}, the scaling of the gap changes in the strongly dissipative regime. In particular, the spectral gap \emph{closes} as $\av{\Delta_\mathrm{spec}}\propto g_\mathrm{eff}^{-2/3}$ when $g_\mathrm{eff}\to\infty$, signaling a dissipation-suppressed (not enhanced) relaxation. The special role of the $M=1$ case can be understood as a remnant of the decoupling transition observed for quadratic random Liouvillians, to be discussed in Sec.~\ref{sec:RL_quadratic_Lindblad}.

\subsection{Steady state}
\label{subsec:RL_steady_state}

\begin{figure}[t]
	\centering \includegraphics[width=\textwidth]{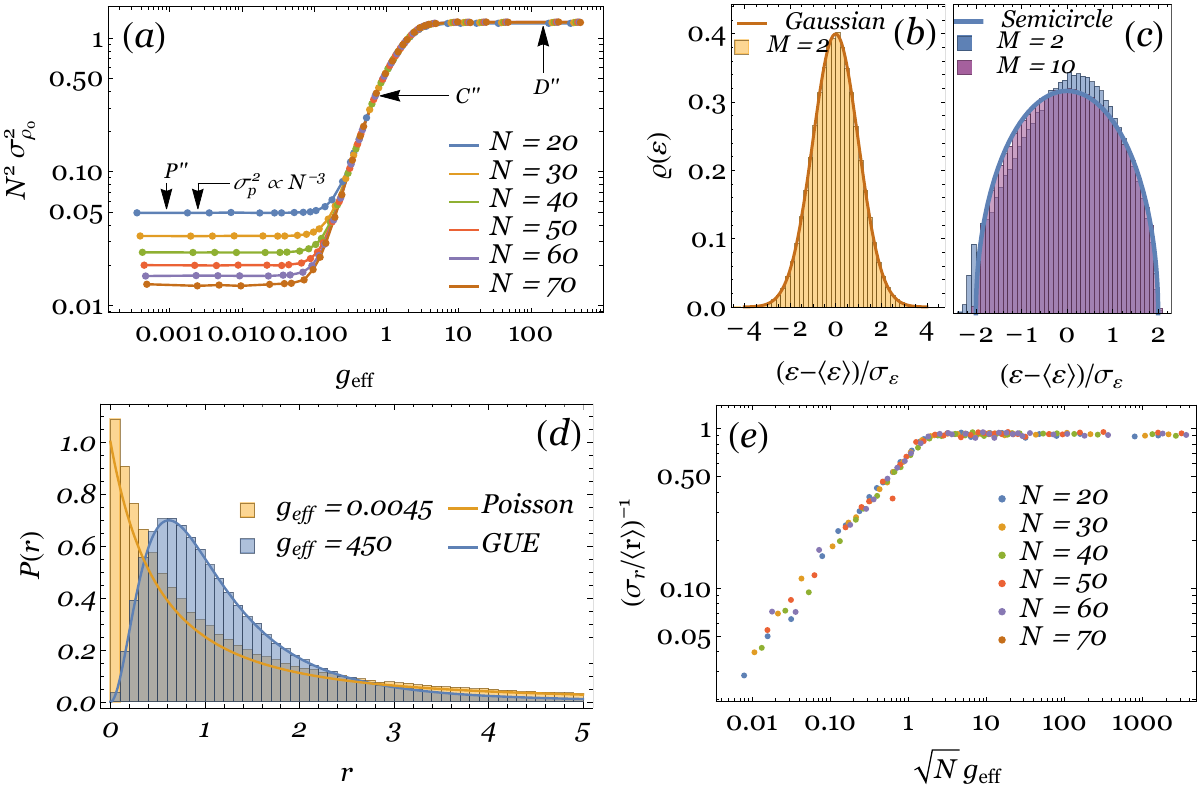}
	\caption{$(a)$: Variance of steady-state probabilities as a function of $g_{\mathrm{eff}}$ plotted for different values of $N$ for $M=2$. $(b)-(c)$: spectral density for the effective Hamiltonian $\mathcal{H}$ for $g_\mathrm{eff}=0.0045$ and $g_\mathrm{eff}=450$, respectively. $(d)$: statistics of level spacing ratios for weak and strong dissipation and comparison with approximate analytic predictions for Poisson and RMT statistics (solid lines). $(e)$: ratio of the first two moments of the distribution of $r$, $(\sigma_{r}/\langle r\rangle)^{-1}$, for $M=2$.}
	\label{fig:steady_state}
\end{figure}

We next characterize the steady state $\rho_{0}$. First, we consider the variance, $\sigma_{\rho_{0}}^{2}$, of the eigenvalues of $\rho_{0}$. This quantity is related to the difference between the purity of the steady state, $\mathcal{P}_{0}=\Tr\left(\rho_{0}^{2}\right)$, which quantifies the degree of mixing of $\rho_{0}$, and that of a fully-mixed state $\mathcal{P}_{\mathrm{FM}}=1/N$, $\mathcal{P}_{0}-\mathcal{P}_{\mathrm{FM}}=N\sigma_{\rho_{0}}^{2}$.
Figure~\ref{fig:steady_state}~$(a)$ shows the variance~$\sigma_{\rho_{0}}^{2}$ as a function of $g_{\mathrm{eff}}$ for $M=2$. Here again, three different regimes (marked P$''$, C$''$, and D$''$) can be observed. In the perturbative weak-dissipation regime P$''$, we observe $\sigma_{\rho_{0}}^{2}\propto N^{-3}$, independently of $g_\mathrm{eff}$. In the crossover regime C$''$, $\sigma_{\rho_{0}}^{2}\propto N^{-2}g_{\mathrm{eff}}^{2}$, while in the strongly dissipative regime D$''$, $\sigma_{\rho_{0}}^{2}\propto N^{-2}$ again independently of $g_\mathrm{eff}$.
These scalings imply that, up to subleading $1/N$ corrections, the steady state is fully mixed in regime P$''$, $\langle\mathcal{P}_{0}\rangle=\mathcal{P}_{\mathrm{FM}}+\mathcal{O}\left(1/N^{2}\right)$, while in regimes C$''$ and D$''$, $\mathcal{P}_{0}$ is only proportional to $\mathcal{P}_{\mathrm{FM}}$. At large dissipation, the steady state can be very well described by a random Wishart matrix, in agreement with general results of the entanglement spectrum of random bipartite systems~\cite{zyczkowski2001JPhysA,sommers2004JPhysA}. Purity can then be computed straightforwardly. This computation will be discussed in detail for the random Kraus channel of Ch.~\ref{chapter:kraus}, which, remarkably, displays exactly the same steady-state properties. These results, together with those of Sec.~\ref{sec:RL_quadratic_Lindblad} point towards the universality of the steady states of chaotic Markovian open quantum systems. The case with $M=1$ is again qualitatively different from $M>1$ at strong dissipation, the purity $\langle\mathcal{P}_0\rangle$ being of order $N^0$~\cite{sa2019JPhysA,saMScThesis}, signaling a steady state closer to a pure state than to a fully-mixed one. 

Next, we investigate the steady state's spectrum. For that, it is useful to introduce the effective Hamiltonian $\mathcal{H}=-\log\rho_{0}$. We denote the eigenvalues of $\mathcal{H}$ by $\varepsilon_i$ and present their spectral density $\varrho(\varepsilon)$ in Figs.~\ref{fig:steady_state}~$(b)$ and $(c)$, corresponding to a point in regime P$''$ and D$''$, respectively. In the weak coupling regime, P$''$, $\varrho(\varepsilon)$ is well described by a Gaussian, while at strong coupling, D$''$, it is described by a fixed-trace Marchenko-Pastur distribution, which, for large $M$ is increasingly well described by a Wigner semicircle distribution. As for the purity, the spectrum of the steady state coincides with that of random Kraus operators and will be discussed in greater detail in Ch.~\ref{chapter:kraus}.

Figure~\ref{fig:steady_state}~$(d)$ presents the probability distribution, $P(r)$, of adjacent spacing ratios, $r_{i}=s_{i}/s_{i-1}$, with $s_{i}=\varepsilon_{i+1}-\varepsilon_{i}$, which automatically unfolds the spectrum of $\mathcal{H}$~\cite{atas2013}. The analytic predictions for the GUE and for the Poisson distribution~\cite{atas2013} (full lines) are given for comparison. The agreement of the numerical data of the points in the P$''$ and D$''$ regimes, respectively, with the Poisson and GUE predictions is remarkable. Within regime C$''$, we observe a crossover between these two regimes.
To illustrate the crossover in the spectral properties of $\mathcal{H}$ with $g_{\mathrm{eff}}$, we provide in Fig.~\ref{fig:steady_state}~$(e)$ the ratio of the first two moments of the distribution of $r$, $(\sigma_{r}/\langle r\rangle)^{-1}$ which can distinguish between Poisson and GUE statistics. Since in the Poissonian case $P(r)=1/(1+r)^{2}$, the $n$-th moment of the distribution diverges faster than the $(n-1)$-th and thus $\sigma_{r}/\langle r\rangle\to\infty$. On the other hand, for the GUE this ratio is given by a finite number of order unity, $\sigma_{r}^2/\langle r\rangle^2=256\pi^{2}/(27\sqrt{3}-4\pi)^{2}-1\simeq1.160$. Figure~\ref{fig:steady_state}~$(e)$ shows that Poisson statistics are only attained in the dissipationless limit $N^{1/2}g_{\mathrm{eff}}\to0$. On the other hand, the GUE values are attained for $g_{\mathrm{eff}}N^{1/2}\simeq1$. Thus, in the thermodynamic limit, the effective Hamiltonian $\mathcal{H}$ is quantum chaotic for all finite values of $g_{\mathrm{eff}}$.

\section{Open free fermions: Random quadratic Lindbladians}
\label{sec:RL_quadratic_Lindblad}

As mentioned above, the RMT Lindbladian discussed in Sec.~\ref{sec:RL_RMT_Liouvillian} makes no assumptions about the open dynamics besides their complexity. They thus pertain to systems with single-particle quantum chaos. While these results are encouraging, actual physical systems are very different, with the dynamics constrained by spatial locality or few-body interactions. Focusing on the latter (see Ref.~\cite{orgad2022ARXIV} for an example of the former), the simplest example is that of open free fermions, i.e., open quantum systems with one-body interactions that render the Lindbladian quadratic, which we discuss in the following. While many-body, all properties of these systems are fully determined by their single-particle interactions. As such, these random Liouvillians still describe systems with single-particle quantum chaos. The case of two-body interactions, which entails genuine many-body quantum chaos, with unique emerging properties such as anomalous relaxation, is deferred to Ch.~\ref{chapter:SYKLindblad}.

We focus our attention on a system with $N_F$ complex fermions satisfying $\{c_i,c_j^\dagger\}=\delta_{ij}$. The Liouvillian 
\begin{equation}
\label{eq:Lindblad_quad}
\mathcal{L} \left[  \rho \right] = - i \left [\hat{H}, \rho \right] + \sum_{\mu=1}^M \left ( 2 \hat{L}_{\mu} \rho \hat{L}_{\mu}^{\dagger} -  \{\hat{L}_{\mu}^{\dagger} \hat{L}_{\mu}, \rho\}\right),
\end{equation}
with Hamiltonian $\hat{H}$ and $M$ jump operators $\hat{L}_\mu$
is said to be quadratic if 
\begin{equation}
\label{eq:RL_H_and_l}
\hat{H} = \frac{1}{2} \sum_{i,j=1}^{2 N_F} C_i^{\dagger} H_{ij} C_j
\qquad\text{and}\qquad
\hat{L}_{\mu} = \sqrt{g} \sum_{j=1}^{2 N_F} l_{\mu j} C_j,
\end{equation}
with $C=\{c_1,\dots,c_{N_F},c_1^{\dagger},\dots,c_{N_F}^{\dagger} \}^T$ a vector of fermionic creation and annihilation operators that satisfies $ \{C_i, C_j^{\dagger}\} =  \delta_{ij}$.
The quadratic Lindblad operator obtained from this construction ensures that the dynamics preserve the Gaussian form of an initial density matrix. Thus, the time evolution of the $2^{N_F}\times2^{N_F}$ density matrix can be encoded by its second moments' matrix---the correlation matrix---of size $2N_F \times 2N_F$. 
Analogously to quadratic Hamiltonian systems, it is possible to construct a single-particle basis whose dimension scales linearly with the number of fermionic modes, $N_F$. Many-body observables, such as the Liouvillian's many-body spectrum and steady-state correlators, can be straightforwardly computed from single-particle quantities.  
Moreover, the single-body spectrum can be identified with that of a non-Hermitian Hamiltonian~\cite{prosen2008}, leaving the determination of the Liouvillian's spectral properties only dependent on the specification of the single-particle Hamiltonian~$H$ and jump operators~$l$. After a change of basis into the Majorana basis, the single-particle Hamiltonian $H$ is drawn from a Gaussian ensemble of antisymmetric matrices, while the single-particle jump operators are drawn from the GinUE. In the following, we review the main properties of random quadratic Liouvillians and compare them with the ones of RMT Liouvillians, discussed in Sec.~\ref{sec:RL_RMT_Liouvillian}. For details, we refer the reader to Ref.~\cite{costa2022}.

\subsection{Phase diagram}

\begin{figure}[t]
	\centering
	\includegraphics[width=\textwidth]{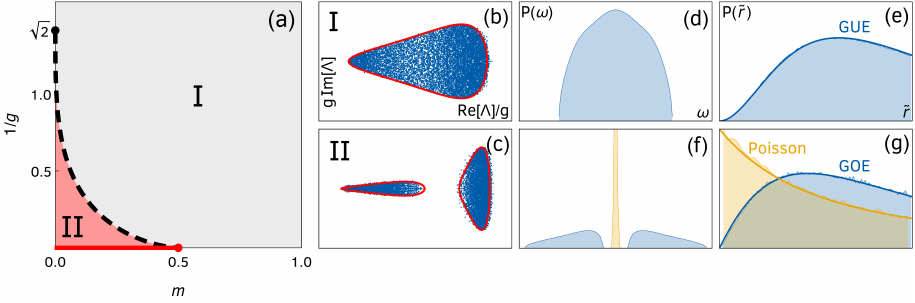}
	\caption{Schematic representation of the main results of this section, the single-particle spectral and steady-state properties of random quadratic Liouvillians. (a) Phase diagram in the $1/g$ versus $m$ plane. Phases I and II are separated by the critical line $g_\mathrm{c}(m)$ (dashed line). In the limit $g\to\infty$, the phase transition occurs at $m=1/2$, with the spectral gap vanishing for $m<1/2$ (red line). (b, c) Single-body spectrum in the complex plane, with a single connected component in phase I (b) and two disconnected components in phase II (c). The spectral boundary (red line) can be obtained analytically. (d, e) Steady-state properties for $g\to\infty$ and $m>1/2$ (phase I). The steady state has a single sector with the single-particle effective Hamiltonian eigenvalues, $\omega$, distributed according to RMT (d) and GUE statistics for the spacing ratio $\tilde{r}$ (e). (f, g) Steady-state properties for $g\to\infty$ and $m<1/2$ (phase II, red line). The steady-state spectrum splits into an ergodic and a nonergodic sectors, in which the effective Hamiltonian's eigenvalues follow, respectively, RMT and a normal distribution (f). The $\tilde{r}$ statistics are Poisson in the nonergodic sector and interpolate from GUE to GOE in the ergodic sector as $m\to0$ (g).}
	\label{fig:main_results}
\end{figure}

In the thermodynamic limit $N_F\to\infty$, the single-particle properties of random quadratic Liouvillians, specified by Eqs.~(\ref{eq:Lindblad_quad}) and (\ref{eq:RL_H_and_l}), are determined by two parameters: the dissipation strength $g$ and the ratio of the number of jump operators to the number of fermions $m=M/(2N_F)$. In Fig.~\ref{fig:main_results}(a), we plot the phase diagram of this system in the $1/g$ versus $m$ plane. For large enough $m$ and small enough $g$, the system is in phase I, characterized by a single-body spectrum supported on a simply-connected region of the complex plane, see Fig.~\ref{fig:main_results}(b). When $g$ is increased or $m$ decreased across a critical value, a phase transition occurs and, in phase II, the single-body spectrum splits into two disconnected components, see Fig.~\ref{fig:main_results}(c). The existence of these two regions signals the existence of an intermediate period of metastability, during which an extensive number of modes coexist without (considerable) decay. The critical line separating the two phases [dashed line in Fig.~\ref{fig:main_results}(a)] and the boundaries of the single-body spectral support can be computed analytically~\cite{haake1992,lehmann1995,janik1997a,janik1997b}. 

For very strong dissipation $g\to\infty$, the transition occurs at $m=1/2$ and corresponds to the decoupling of some fermionic degrees of freedom from the dynamics. Indeed, the Hamiltonian contribution vanishes when $g\to\infty$ and there is an insufficient number of jump operators ($M<2N_F$) to couple all $2N_F$ fermionic creation and annihilation operators to the environment. Below the transition [$m<1/2$, red line in Fig.~\ref{fig:main_results}(a)] the decoupled fermions exhibit nonergodic features, discussed in more detail below. As $g$ is lowered to a finite value, the Hamiltonian starts to couple the dissipatively decoupled fermions and the critical value of $m$ decreases. At weak dissipation $g<1/\sqrt{2}$, the Hamiltonian contribution is strong enough to couple all fermions and there is no transition.

\subsection{Spectral gap}

The spectral gap, which sets the (inverse) timescale of relaxation to the steady state, coincides for the single- and many-body spectra. It can also be obtained analytically and assumes a very simple form in the limits $g\to0$ and $g\to\infty$. For weak dissipation, the spectral gap closes linearly with $g$ and $m$ (for all $m$), as expected from perturbation theory. On the other hand, for large dissipation, the spectral gap acts as an order parameter of the transition at $m=1/2$. For $m<1/2$, in the limit $g \to \infty$ the gap closes like $1/g$ leading to a gapless Liouvillian, signaling a slow approach to the steady-state. For $m>1/2$, the gap has the linear scaling in $g$, typical of a dissipation-driven relaxation, growing with $m$ as $(\sqrt{2m} - 1)^2$, as dictated by the Marchenko-Pastur law. At the critical point $m=1/2$, the gap closes as $(m/g)^{1/3}$. For large but finite $g$, the gap is nonzero for any value of $m$, but still exhibits qualitatively different behaviors above and below the transition. 

\subsection{Steady state}

The steady state, to which the system relaxes in the long-time limit, is Gaussian for quadratic Liouvillians and is thus fully characterized by its single-particle properties. In the limit $g\to0$ (where there is no decoupling transition), the steady-state single-body spectrum is Gaussian, fully mixed, and displays Poisson statistics irrespectively of the value of $m$. As $g$ increases, there is a perturbative crossover well below the threshold $g=1/\sqrt{2}$ for the appearance of the decoupling transition, and the steady state becomes distributed according to RMT [see Fig.~\ref{fig:main_results}(d)], is mixed but not fully mixed, and exhibits GUE statistics [see Fig.~\ref{fig:main_results}(e)]. The purity interpolates monotonously between the two limits $g\to0$ and $g\to\infty$.

When $g$ is further increased, the steady state is also influenced by the decoupling transition. In the limit $g\to\infty$, the results can be obtained analytically through perturbation theory. Above the transition, $m>1/2$, the properties attained after the perturbative crossover do not change. However, for $m<1/2$, the spectrum of the steady state also splits into two independent sectors, see Fig.~\ref{fig:main_results}(f). The sector of the fermions coupled to the environment retains the properties of $m>1/2$, except for the spectral statistics which, remarkably, crossover from GUE to GOE as $m\to0$, see Fig.~\ref{fig:main_results}(g). The spectrum of the decoupled sector, on the other hand, is composed of uncorrelated Gaussian random variables displaying Poisson statistics (nonergodic behavior), see Fig.~\ref{fig:main_results}(g). These two sectors are well separated for small-enough $m$, see Fig.~\ref{fig:main_results}(f), but overlap for larger $m$.

We emphasize that the nonergodic features of the steady state were obtained in the limit $g\to\infty$. However, they leave strong imprints in the dynamics at large but finite $g$ and $N_F$. Indeed, the interplay of the two sectors leads to a decrease of the purity, with a nonmonotonic behavior as a function of $g$, and to an interpolation between RMT and Poisson statistics. At the present point in time, it is not clear whether the nonergodic features of the steady state survive in the thermodynamic limit $N_F\to\infty$ at large but finite $g$ (i.e., whether the red line in Fig.~\ref{fig:main_results}(a) is smoothly connected to the rest of the diagram), which would require a more detailed finite-size scaling analysis and is deferred to future work.

\subsection{Comparison with the RMT Lindbladian}

Finally, we note that the properties of random quadratic Liouvillians above the transition are quantitatively similar to those identified in the RMT Liouvillians discussed in Sec.~\ref{sec:RL_RMT_Liouvillian}. Because the single-body spectral gap coincides with the many-body spectral gap, the results of the two models can be directly compared. In the weak dissipation regime, the same linear growth with the dissipation parameter is found in both cases and it is the expected perturbative result. The strong dissipation regime is, as we have seen, richer. 
In the RMT case, the role of the parameter $m$ is played not by the ratio of dissipation channels to the number of degrees of freedom ($2N_F$ here), but by the number of channels itself $m \to M/2$. The regime $m<1/2$ is, therefore, inaccessible since $M$ is a positive integer. For RMT Liouvillians with more than one decay channel (corresponding to the ergodic phase $m>1/2$) we also observed a linear-in-$g$ growth of the gap, which again is the expected perturbative result for the dissipation-dominated dynamics. 
The case of a single decay channel, $M=1$ (corresponding to $m=1/2$ here), shows the same closing of the gap with $\sim g^{-1/3}$ (once one accounts for different normalization conventions) as for the RMT Liouvillian.
As we now understand the difference between $m>1/2$ and $m<1/2$ as a transition in which some degrees of freedom become decoupled from the environment and, thus, protected from dissipation, with $m=1/2$ corresponding to the critical transition point, our findings shine a new light on the special role played by RMT Liouvillians with a single jump operator, $M=1$.
Remarkably, not only the scaling with $g$ but also that with $m$ coincide in both cases (including quantitative agreement of prefactors) and we conclude that the spectral gap coincides in quadratic and fully random Liouvillians in the mutually accessible regimes ($m\geq1/2$). Going beyond the spectral gap, unconstrained fully-random Liouvillians also exhibit a connected spectrum and the steady state exhibits a nonergodic-to-ergodic crossover and a corresponding change in purity as a function of $g$.

\section{Summary and outlook}

In this chapter, we reviewed the single-body spectral and steady-state properties of structureless random Lindbladians and quadratic fermionic random Lindbladians. In the quadratic case, our analysis focused on the phase transition observed in the single-body spectrum and its repercussions in the steady-state properties. More precisely, in phase I, the spectral and steady-state properties of quadratic and fully-random Liouvillians are qualitatively similar: the spectrum is formed by a single connected component; the gap grows linearly with dissipation strength; the steady-state purity is monotonic with dissipation strength; and there is a nonergodic-to-ergodic crossover in the steady-state level statistics, the full steady state being ergodic for sufficiently strong dissipation. In phase II, there are qualitative differences: the single-body spectrum splits into two disconnected components at a finite system-environment coupling strength; the spectral gap closes for strong dissipation; the purity is non-monotonic with dissipation strength; and the steady state decouples into ergodic and nonergodic sectors. 

In summary, our work identifies a regime of universal random Markovian dissipation but also illustrates the possibility of nonergodic behavior in quadratic open quantum systems and the potential to suppress dissipation even in the presence of strong system-environment coupling.
A natural extension of this work is to ask whether these nonergodic features survive interactions. Their robustness could be addressed, for instance, with the Sachdev-Ye-Kitaev Lindbladian~\cite{sa2022PRR,kulkarni2022PRB,garcia2023PRD2,kawabata2023PRB}. 
Moreover, it is also not clear whether the nonergodic features of the steady state survive in the thermodynamic limit $N_F\to\infty$ at large but finite $g$. This interesting question would require a more detailed finite-size scaling analysis and is deferred to future work.
Other interesting possibilities are to consider bosonic Liouvillians and non-Markovian generators. Finally, further work is needed to determine whether the properties of stochastic Markovian dissipative models that we described are also present in more realistic models of open quantum systems.

%% file: Thesis_SYKLindblad.tex

\chapter{Random Lindblad dynamics II. Many-body quantum chaos}
\label{chapter:SYKLindblad}

After having considered open quantum systems with single-particle quantum chaos in the previous chapter, we now turn to strongly-interacting quantum systems, which display many-body quantum chaos. In this chapter, we propose the Sachdev-Ye-Kitaev (SYK) Lindbladian as a paradigmatic solvable model of many-body dissipative quantum chaos. More concretely, we study the nonequilibrium dynamics of $N$ strongly-coupled fermions with $q$-body random all-to-all interactions, coupled to a Markovian environment through jump operators either linear or quadratic in the Majoranas. Analytical progress is possible by developing a dynamical mean-field theory for the Liouvillian time evolution on the Keldysh contour (Sec.~\ref{sec:SYK_action}). By disorder-averaging the interactions (Sec.~\ref{sec:SYK_effective_action}), we derive an (exact) effective action for two collective fields (Green's function and self-energy). In the large-$N$ limit, we obtain the saddle-point equations satisfied by the collective fields (Sec.~\ref{sec:SYK_SD}), which determine the typical timescales of the dissipative evolution, particularly, the relaxation gap that rules the relaxation of the system to its steady state.
For strong dissipation, the system relaxes exponentially at a rate linear in the coupling, characteristic of a dissipation-driven relaxation, with a prefactor that we compute analytically (Sec.~\ref{sec:SYK_strong_dissipation}). We then solve the saddle-point equations numerically for $q=4$ and weak coupling (Sec.~\ref{sec:SYK_weak_dissipation}). There are oscillatory corrections to the exponential relaxation and we observe an anomalously large decay rate with a finite value even in the absence of an explicit coupling to the environment.
The enhanced relaxation rate is caused by the chaotic internal strong interactions of the SYK model. We confirm this by analytically solving the integrable $q=2$ case (Sec.~\ref{sec:q2}), for which the gap vanishes linearly as the dissipation strength goes to zero.

This chapter is based on Refs.~\cite{sa2022PRR,garcia2023PRD2}.

\section{Keldysh approach to the SYK Lindbladian}
\label{sec:SYK_action}

We study a Hermitian SYK Hamiltonian with a $q$-body interaction of infinite range coupled to a Markovian environment. The SYK model is defined by the Hamiltonian 
\begin{align}\label{eq:sykq}
H &=
-\i^{q/2}
\sum_{i_1<\cdots < i_q}^N
J_{i_1\cdots i_q}
\psi^{i_1} \cdots \psi^{i_q},
\end{align}
where $J_{i_1\cdots i_q}$ are random numbers extracted from a Gaussian distribution of zero average and variance $\langle J_{i_i \cdots i_q}^2 \rangle = (q-1)!J^2/N^{q-1}$, with $i_1,\dots,i_q=1,\dots,N$. $J_{i_1\cdots i_q}$ is totally antisymmetric in its indices. We will mostly focus on the cases $q = 2$ (integrable) and $q=4$ (chaotic). The $\psi^i$ are Majorana operators defined by the commutation relation $\{\psi^i, \psi^j\}=\delta_{ij}$, with ${\psi^i}^\dagger=\psi^i$.

The real-time evolution of the density matrix $\rho$ of this open SYK system is simply $d\rho/dt = \mathcal{L}(\rho)$, where $\mathcal{L}$ is the Liouvillian, which can be conveniently expressed in the Lindblad form: 
\begin{align} \label{eq:liou0}
\sL \(\rho\)= -\i \comm{H}{\rho}+\sum_{m=1}^{M}\(2L_m\rho L_m^\dagger-\acomm{L_m^\dagger L_m}{\rho}\).
\end{align}

In the following, we will take the Markovian environment to be described by jump operators that are either linear in the Majorana operators,
\be
\label{jump_ope}
L_i = \sqrt{\mu/2}\psi^i,
\ee
with $i = 1, \dots, N$ and $\mu$ a positive real number, or quadratic in Majoranas, 
\begin{equation}\label{eq:H_L_SYK_def}
L_m=\i\sum_{i<j}^{N}\ell_{m,ij}\psi^i\psi^j,
\end{equation}
with $m=1,\dots,M$ and $\ell_{m,ij}$ antisymmetric (in $i$ and $j$) independent Gaussian random variables with zero mean and variance
\begin{equation}
\label{eq:J_l_moments}
\av{\abs{\ell_{m,ij}}^2}=\frac{\gamma^2}{N^2}.
\end{equation}
Notice the nontrivial scaling of the quadratic SYK couplings, which is required for a nontrivial theory in the large-$N$ limit (see Sec.~\ref{sec:SYK_effective_action} below for details).
We further define the positive-definite matrix
\begin{equation}\label{eq:Gamma_def}
\Gamma_{ijkl}=\sum_{m=1}^M\ell_{m,ij}\ell_{m,kl}^*,
\end{equation}
which satisfies $\Gamma_{ijkl}=-\Gamma_{jikl}=\Gamma_{jilk}=-\Gamma_{ijlk}=\Gamma_{klij}^*$. If we let $N,M\to\infty$ with $m=M/N$ fixed, $\Gamma_{ijkl}$ also becomes Gaussian distributed. Then Eq.~(\ref{eq:J_l_moments}) implies that the mean and the variance of $\Gamma_{ijkl}$ are, respectively,
\begin{equation}\label{eq:Gamma_moments}
\av{\Gamma_{ijij}}=\frac{m\gamma^2}{N}
\quad\text{and}\quad
\av{\abs{\Gamma_{ijkl}}^2}_\mathrm{con}= \frac{m\gamma^4}{N^3}.
\end{equation}
$J_{i_1\cdots i_q}$ must be real to ensure Hermiticity of the Hamiltonian, while $\ell_{m,ij}$ can generally be complex but, for simplicity, from now on we restrict ourselves to Hermitian jump operators (i.e., real $\ell_{m,ij}$).
The scales $J$ and $\gamma$ or $\mu$ measure the strength of the unitary and dissipative contributions to the Liouvillian, respectively.
The Hamiltonian describes coherent long-range $q$-body interactions, while each $L_m$ gives an independent channel for incoherent two-body interactions.

For quadratic jump operators, the full Lindbladian is then given by
\begin{equation}\label{eq:Liouv_op}
\begin{split}
&\sL(\rho)=
\i^{q/2+1}\sum_{i_1<\cdots<i_q}^N J_{i_1\cdots i_q}\(
\psi^{i_1}\cdots \psi^{i_q}\rho-\rho\psi^{i_1}\cdots \psi^{i_q}
\)\\
&-\sum_{\substack{i<j\\k<l}}^N
\Gamma_{ijkl}\(
2\psi^i\psi^j\rho\psi^k\psi^l
-\psi^k\psi^l\psi^i\psi^j\rho
-\rho\psi^k\psi^l\psi^i\psi^j
\).
\end{split}
\end{equation}
For the linear jump operators, it is given by
\begin{equation}\label{eq:Liouv_op_mu}
\begin{split}
&\sL(\rho)=
\i^{q/2+1}\sum_{i_1<\cdots<i_q}^N J_{i_1\cdots i_q}\(
\psi^{i_1}\cdots \psi^{i_q}\rho-\rho\psi^{i_1}\cdots \psi^{i_q}
\)+\mu\sum_{i=1}^N \psi^i \rho \psi^i-\frac{1}{2}\mu N\rho.
\end{split}
\end{equation}

For sufficiently long times, and taking into account that both our types of jump operators are Hermitian, the density matrix will relax to the infinite-temperature (fully-mixed) state,
\be
\label{eq:steady-state}
\rho_\infty =\frac{1}{2^{N/2}} \sum_k |k\rangle \langle k|,
\ee
characterized by $\mathcal{L}(\rho_\infty)=0$. We note that for jump operators more general than those of Eqs.~(\ref{jump_ope}) or (\ref{eq:H_L_SYK_def}), the system's density matrix may decay to either a mixed state at finite temperature or a nonequilibrium steady state, depending on the details of the Hamiltonian.
Our main interest is in the approach to the steady state~(\ref{eq:steady-state}) and, for that purpose, we study the retarded Green's function 
\be 
\i G^\rmR(t) \delta_{ij} =  \Theta(t)\left\langle \Tr\left[\rho_{\infty}\{\psi^i(t),\psi^j\}\right]\right\rangle,
\label{steady}
\ee
where $\Theta(t)$ is the Heaviside function, $\av{\cdots}$ denotes the average over both $J_{i_1\cdots i_q}$ and $\Gamma_{ijkl}$ and the (Heisenberg-picture) Majorana operator $\psi^i(t)$ satisfies the adjoint Lindblad equation, $\pd_t\psi^i=\sL^\dagger(\psi^i)$.
In general, we expect the system to relax exponentially to its steady state at a rate $\Delta$ [also known as the (relaxation) gap]. In that case, we have 
\begin{equation}
\i G^\mathrm{R}(t)\propto e^{-\Delta t}.
\end{equation}

We now switch to the Keldysh path-integral representation of the Majorana Liouvillian (see App.~\ref{app:KeldyshLindblad} for a derivation and Ref.~\cite{sieberer2016} for the bosonic version).
We introduce real Grassmann fields $a_{i}(z)$ living on the closed-time contour $z\in\sC=\sC^+\cup\sC^-$, where real time runs from $-\infty$ to $+\infty$ (branch $\sC^+$) and then back again to $-\infty$ (branch $\sC^-$).
The Grassmann field $a_i(t^+)$ [$a_i(t^-)$], with $t^+\in\sC^+$ ($t^-\in\sC^-$), propagates forward (backward) in time and is the path-integral representation of a Majorana operator acting on the density matrix from the left (right). Using Eq.~(\ref{eq:Liouv_op}), we can immediately write down the partition function:
\begin{equation}\label{eq:path_integral}
Z=\int \prod_{i=1}^N\sD a_i\ e^{\i S[a_i]},
\end{equation}
where we omitted an initial-state contribution that is irrelevant for the long-time dissipative dynamics. For the quadratic jump operators, the Lindblad-Keldysh action is
\begin{equation}\label{eq:C_Keldysh_action}
\begin{split}
\i S[a_i]&=
\i\int_\sC \d z\,\frac{1}{2} \sum_{i=1}^N
a_{i}(z)\,\i\pd_za_{i}(z)
+\i \int_\sC \d z
\sum_{i_1<\cdots<i_q}^N \i^{q/2}J_{i_1\cdots i_q}a_{i_1}(z)\cdots a_{i_q}(z)
\\
&+\int_\sC \d z \d z'\,K(z,z')\sum_{\substack{i<j\\k<l}}^N
\Gamma_{ijkl} a_i(z)a_j(z)a_k(z')a_l(z'),
\end{split}
\end{equation}
while for the linear jump operators it is
\begin{equation}\label{eq:C_Keldysh_action_mu}
\begin{split}
\i S[a_i]&=\i\int_\sC \d z \,\frac{1}{2}\sum_{i=1}^N a_i(z)\i\partial_z a_i(z)
+\i\int_\sC \d z\,\sum_{i_1<\cdots<i_q}^N  \i^{q/2}  J_{i_1\cdots i_q}a_{i_1}(z)\cdots a_{i_q}(z)
\\
&+\mu\int_\sC d z\, d z'\, K(z,z')\sum_{i=1}^N a_i(z)a_i(z').
\end{split}
\end{equation}
The memory kernel $K(z,z')$ allows for both Markovian and non-Markovian dissipative dynamics. Comparing Eqs.~(\ref{eq:Liouv_op})~and~(\ref{eq:C_Keldysh_action}) [or (\ref{eq:Liouv_op_mu})~and~(\ref{eq:C_Keldysh_action_mu})] we can read off the lesser and greater components of the Markovian kernel:
\begin{align}
\label{eq:K_kernel_<}
K^<(t_1,t_2)&\equiv K(t_1^+,t_2^-)=2\dirac{t_1-t_2},
\\
\label{eq:K_kernel_>}
K^>(t_1,t_2)&\equiv K(t_1^-,t_2^+)=0,
\end{align}
respectively.
Equations~(\ref{eq:C_Keldysh_action})--(\ref{eq:K_kernel_>}) can also be derived microscopically by tracing out the environment in a unitary system-plus-environment theory, see App.~\ref{app:KeldyshLindblad}.

\section{Effective action for collective fields}
\label{sec:SYK_effective_action}

To proceed, we average over the random couplings $J_{ijkl}$ and $\Gamma_{ijkl}$, in the limit $N\to\infty$. In the case of quadratic jump operators, we additionally take the limit $M\to\infty$ with $m=M/N$ fixed. For concreteness, we consider the case of quadratic jump operators in detail and give the result for linear jump operators at the end of the section.

We are interested in the averaged partition function,
\begin{align}
\av{Z}&=\int \prod_{i=1}^N\sD a_i\
\av{e^{\i S[a_i]}}_{J,\Gamma},
\end{align}
where the average is performed over both the unitary and dissipative disorder (i.e., over $J_{i_1\cdots i_q}$ and $\Gamma_{ijkl}$, respectively). The disorder average of the unitary contribution to the action is straightforward because the random variables $J_{i_1\cdots i_q}$ are, as usual, chosen Gaussian with mean and variance
\begin{equation}\label{eq:SM_J_moments}
\av{J_{i_1\cdots i_q}}=0
\quad\text{and}\quad
\av{J_{i_1\cdots i_q}^2}=\frac{(q-1)!J^2}{N^{q-1}},
\end{equation}
respectively.
Then, the averaged unitary contribution to the path integral reads as
\begin{equation}\label{eq:av_Herm_Ham}
\begin{split}
&\av{
	\exp\left\{\i \int_\sC \d z\sum_{i_1<\cdots<i_q}^N
	\i^{q/2} J_{i_1\cdots i_q}a_{i_1}(z)\cdots a_{i_q}(z)\right\}
}_J
\\
&=\exp\left\{
-\frac{\i^q}{2}\int_\sC\d z\d z' \sum_{i_1<\cdots<i_q}^N
\av{J_{i_1\cdots i_q}^2}
a_{i_1}(z)\cdots a_{i_q}(z)a_{i_1}(z')\cdots a_{i_q}(z')
\right\}
\\
&=
\exp\left\{
-\frac{\i^q}{2}\int_\sC \d z \d z'\, \frac{(q-1)!J^2}{N^{q-1}}\frac{1}{q!}\(\sum_{i=1}^N a_i(z)a_i(z')\)^q
\right\}
\\
&=\exp\left\{-N\frac{\i^qJ^2}{2q}\int_\sC \d z \d z' 
\left[G(z,z')\right]^q
\right\},
\end{split}
\end{equation}
where we used the definition of the (mean-field) Green's function:
\begin{equation}\label{eq:SM_G_def}
G(z,z')=-\frac{\i}{N}\sum_{i=1}^N  a_{i}(z) a_{i}(z').
\end{equation}

The disorder average of the dissipative contribution cannot be carried out in full generality since the action is not linear in the random variables $\ell_{m,ij}$. Nonetheless, if the number of decay channels $M$ is large, then, using Eq.~(\ref{eq:Gamma_def}) and the Central Limit Theorem (CLT), the random variables $\Gamma_{ijkl}$ become Gaussian-distributed with nonzero mean. Therefore, we disorder-average over $\Gamma_{ijkl}$ instead of over $\ell_{m,ij}$.
(If working in a perturbative approach, this can be understood as a two-loop computation of the averaged partition function. The one-loop computation, i.e., quadratic order in $\ell_{m,ij}$, takes only the mean of $\Gamma_{ijkl}$ into account, and not its variance, and corresponds to evolution under an average Liouvillian.)
By choosing the mean and variance of the independent Gaussian variables $\ell_{m,ij}$ to be
\begin{equation}
\av{\ell_{m,ij}}=0
\quad\text{and}\quad
\av{\abs{\ell_{m,ij}}^2}=\sigma_\ell^2,
\end{equation}
and assuming $M$ is large (the exact scaling of $M$ and $\sigma_\ell^2$ with $N$ will be determined consistently below), it follows from Eq.~(\ref{eq:Gamma_def}) that the only nonzero mean and variance of $\Gamma_{ijkl}$ are
\begin{equation}\label{eq:SM_Gamma_moments}
\av{\Gamma_{ijij}}=M\sigma_\ell^2
\quad\text{and}\quad
\av{\abs{\Gamma_{ijkl}}^2}_\mathrm{con}
=\av{\abs{\Gamma_{ijkl}}^2}-
\abs{\av{\Gamma_{ijkl}}}^2=
M\sigma_\ell^4,
\end{equation}
respectively.
Under these conditions, the dissipative contribution to the averaged partition function reads as
\begin{equation}\label{eq:av_nonHerm_Ham}
\begin{split}
&\av{\exp\left\{
	\int_\sC \d z \d z'\,K(z,z') \sum_{\substack{i<j\\k<l}}^N
	\Gamma_{ijkl} a_i(z)a_j(z)a_k(z')a_l(z')
	\right\}}_\Gamma\\
&=\exp\left\{
\int_\sC \d z \d z'\,K(z,z')
\sum_{i<j}^N
\av{\Gamma_{ijij}} a_i(z)a_j(z)a_i(z')a_j(z')
\right\}\\
&\times	
\exp\Bigg\{\frac{1}{2}\int_\sC \d z \d z' \d w \d w'\,
K(z,z') K(w,w')
\\
&\left.\times
\sum_{\substack{i<j\\k<l}}^N
\av{\Gamma_{ijkl}\Gamma_{klij}}_\mathrm{con} a_i(z)a_j(z)a_k(z')a_l(z')a_k(w)a_l(w)a_i(w')a_j(w')
\right\}\\
&=\exp\left\{
\frac{M\(N\sigma_\ell\)^2}{2} \int_\sC \d z \d z'\, K(z,z') \left[G(z,z')\right]^2\right\}\\
&\times
\exp\left\{\frac{M\(N\sigma_\ell\)^4}{8}
\int_\sC \d z \d z' \d w \d w'\,
K(z,z') K(w,w')
\left[G(z,w')\right]^2\left[G(z',w)\right]^2
\right\}.
\end{split}
\end{equation}

We can now choose the scalings of $M$ and $\sigma_\ell^2$ such that (i) all three contributions to the action (unitary, one-loop dissipative, and two-loop dissipative) have the same (linear) scaling with $N$ and (ii) the CLT is applicable. These conditions are uniquely satisfied for the choice of Eq.~(\ref{eq:J_l_moments}),
\begin{equation}\label{eq:dissipative_scalings}
M=mN
\quad\text{and}\quad
\sigma_\ell^2=\frac{\gamma^2}{N^2}.
\end{equation}

Next, we make the Green's function dynamical by enforcing the definition~(\ref{eq:SM_G_def}) in the path integral. We introduce the self-energy, $\Sigma(z,z')=-\Sigma(z',z)$ conjugate to $G(z,z')$. Then, using the integral representation of the matrix Dirac delta and the associated resolution of the identity,
\begin{equation}
\begin{split}
1&=\int \sD G \dirac{G(z,z')+\frac{\i}{N}\sum_{i}a_i(z)a_i(z')}
\\
&=\int \sD G \sD \Sigma \exp\left\{
-\frac{N}{2}\int_\sC \d z \d z'\, \Sigma(z,z')\(
G(z,z')+\frac{\i}{N}\sum_{i}a_i(z)a_i(z')
\)\right\},
\end{split}
\end{equation}
the averaged partition function reads as
\begin{equation}
\av{Z}=\int \sD G \sD \Sigma e^{\i S_1[G,\Sigma]}
\(
\int \sD a\, e^{\i S_2[a,\Sigma]}
\)^N,
\end{equation}
with actions
\begin{equation}\label{eq:S_1}
\begin{split}
\i S_1[G,\Sigma]=
\frac{N}{2}&\left\{
- \int_\sC \d z \d z'\,\Sigma(z,z')G(z,z')
-\frac{\i^q J^2}{q}\int_\sC \d z \d z' \left[G(z,z')\right]^q
\right.\\
&\ +\frac{m\gamma^4}{4}
\int_\sC \d z \d z' \d w \d w'\,
K(z,z') K(w,w')
\left[G(z,w')\right]^2\left[G(z',w)\right]^2
\\
&\left.
\ +\,m\gamma^2 \int_\sC \d z\, K(z,z') \left[G(z,z')\right]^2
\right\}
\end{split}
\end{equation}
and
\begin{equation}\label{eq:S_2}
\i S_2[a,\Sigma]=
-\frac{1}{2} \int_\sC \d z \d z'
a(z)\left[\dirac{z-z'}\pd_z+\i \Sigma(z,z')\right]a(z').
\end{equation}
The action~(\ref{eq:S_2}) is quadratic in the Grassmann fields, which can, therefore, be integrated out. This integration yields $\(\det[\pd+\i\Sigma]\)^{N/2}\propto \(\det[\i\pd-\Sigma]\)^{N/2}$. We finally arrive at the effective action for $(G,\Sigma)$ on the time-contour $\sC$:
\begin{equation}
\av{Z}=\int \sD G\, \sD\Sigma \ e^{\i S_\eff[G,\Sigma]},
\end{equation}
\begin{equation}\label{eq:C_S_eff}
\begin{split}
\i S_\eff[G,\Sigma]=
\,\frac{N}{2}&\Bigg\{
\Tr\log\(\i\pd-\Sigma\)
-\int_\sC \d z \d z'\, \Sigma(z,z')\,G(z,z')
-\frac{\i^qJ^2}{q}\int_\sC \d z\d z'\, \left[G(z,z')\right]^q
\\
&+\frac{m\gamma^4}{4}\int_\sC \d z \d z' \d w \d w' \,
K(z,z')\,K(w,w')
\left[G(z,w')\right]^2 \left[G(z',w)\right]^2
\\
&+m\gamma^2\int_\sC \d z \d z' \,
K(z,z')\left[G(z,z')\right]^2
\Bigg\}.
\end{split}
\end{equation}

For the case of linear jump operators, since the jump operators are not disordered, no averaging over the dissipative contribution is required. Because the Hamiltonian is the same as before, the averaging over the unitary contribution is identical. The resulting effective action for the collective fields is thus
\begin{equation}
\label{srt}
\begin{split}
\i S_\eff[G,\Sigma]=\frac{N}{2}\Bigg\{
&\Tr\log(i\partial-\Sigma)-
\int_\sC d z\, d z'\, \Sigma(z,z')G(z,z')-\frac{\i^q J^2}{q}\int_\sC d z\, d z'\, [G(z,z')]^q
\\
&+\i\mu\int_\sC d z\, dz'\, K(z,z')G(z,z')
\Bigg\}.
\end{split}
\end{equation}

\section{Schwinger-Dyson equations}
\label{sec:SYK_SD}

\subsection{Contour Schwinger-Dyson equations}

Variation of Eq.~(\ref{eq:C_S_eff}) with respect to $\Sigma(z,z')$ and $G(z,z')$ (recall that both are antisymmetric in their contour indices) leads to the Schwinger-Dyson (SD) equations on $\sC$. Variation with respect to $\Sigma(z,z')$ gives the Dyson equation
\begin{align}
\label{eq:C_SD_Sigma}
\(\i\pd-\Sigma\)\cdot &\,G=\id_{\sC},
\end{align}
where $\id_{\sC}$ is the identity on the Keldysh contour.
Variation with respect to $G(z,z')$ gives the equation for the self-energy, which, for the case of quadratic jump operators reads as
\begin{align}
\begin{split}\label{eq:C_SD_G}
	\Sigma(z,z')=&-\i^qJ^2\left[G(z,z')\right]^{q-1}
	+m\gamma^2\left[K(z,z')+K(z',z)\right]G(z,z')
	\\
	&+\frac{m \gamma^4}{2} G(z,z') \int_{\sC} \d w \d w' \,
	\left[K(z,w)K(w',z')+K(w,z)K(z',w')
	\right]\left[G(w,w')\right]^2.
	\end{split}
	\end{align}
Eq.~(\ref{eq:C_SD_Sigma}) is to be understood as a matrix equation, while Eq.~(\ref{eq:C_SD_G}) acts on each matrix element individually. The two equations above are exact for the SYK Lindbladian in the large-$N$, large-$M$ limit. For linear jump operators, the equation for the self-energy is replaced by
\begin{align}
\begin{split}\label{eq:C_SD_G_linear}
\Sigma(z,z')=&-\i^qJ^2\left[G(z,z')\right]^{q-1}+\i\mu\left[K(z,z')-K(z',z)\right].
\end{split}
\end{align}

\subsection{Real-time Majorana Green's functions}

We now move back from contour times $(z,z')$ to real times $(t_1,t_2)$. 
The Keldysh-contour Green's function $\mathcal{G}$ is defined as the contour-ordered two-point correlation function
\begin{equation}
\mathcal{G}_{ij}(z,z')=-\i \av{T_z\, \psi^i(z) \psi^j(z')}.
\end{equation}
At the saddle point of the SYK model (i.e., at the mean-field level) it coincides with the collective field $G$ defined in Eq.~(\ref{eq:SM_G_def}).
From now on, we assume that we are at the saddle-point, hence $G\,\delta_{ij}=\mathcal{G}_{ij}$ and we refer to $G$ as the Green's function. We obtain the different components of the Green's function in real time by restricting $(z,z')$ to the two branches of the contour, $\sC^+$ and $\sC^-$. We begin by defining the greater and lesser Majorana Green's functions,
\begin{align}\label{eq:SM_G^>}
&G^>(t_1,t_2)=
G(t_1^-,t_2^+)=
-\frac{\i}{N}\sum_{i=1}^N a_i(t_1^-) a_i(t_2^+)
\intertext{and}
\label{eq:SM_G^<}
&G^<(t_1,t_2)=
G(t_1^+,t_2^-)=
-\frac{\i}{N}\sum_{i=1}^N a_i(t_1^+) a_i(t_2^-)
=-G^>(t_2,t_1)
\end{align}
where, as before, $a_i(t^+)$ is a Majorana fermion propagating forward in time (along the contour $\sC^+$) and $a_i(t^-)$ propagates backward (along $\sC^-$). Unlike for complex fermions, only one of $G^>$ and $G^<$ is independent~\cite{eberlein2017PRB,babadi2015PRX}, say, the greater component, while the lesser component satisfies $G^<(t_1,t_2)=-G^>(t_2,t_1)$. Next, we have the time-ordered Green's function,
\begin{equation}\label{eq:SM_G^T}
\begin{split}
G^\rmT(t_1,t_2)&=
G(t_1^+,t_2^+)=
-\frac{\i}{N}\sum_{i=1}^N a_i(t_1^+) a_i(t_2^+)
\\
&=\heav{t_1-t_2}G^>(t_1,t_2)+\heav{t_2-t_1}G^<(t_1,t_2)
\\
&=\heav{t_1-t_2}G^>(t_1,t_2)-\heav{t_2-t_1}G^>(t_2,t_1),
\end{split}
\end{equation}
and the anti-time-ordered Green's function,
\begin{equation}\label{eq:SM_G^Tb}
\begin{split}
G^\rmTb(t_1,t_2)&=
G(t_1^-,t_2^-)=
-\frac{\i}{N}\sum_{i=1}^N a_i(t_1^-) a_i(t_2^-)
\\
&=\heav{t_2-t_1}G^>(t_1,t_2)+\heav{t_1-t_2}G^<(t_1,t_2)
\\
&=\heav{t_1-t_2}G^<(t_1,t_2)-\heav{t_2-t_1}G^<(t_2,t_1).
\end{split}
\end{equation}
The time-ordered and anti-time-ordered components are also related to the greater and lesser components through the identity
\begin{equation}
[G^\rmT(t_1,t_2)]^2+[G^\rmTb(t_1,t_2)]^2=[G^>(t_1,t_2)]^2+[G^<(t_1,t_2)]^2,
\end{equation}
which we will use below.

Another useful set of Green's functions (obtained after the so-called Keldysh rotation) are the retarded Green's function,
\begin{equation}\label{eq:SM_G^R}
\begin{split}
G^\rmR(t_1,t_2)
&=\heav{t_1-t_2}\(G^>(t_1,t_2)-G^<(t_1,t_2)\)
\\
&=\heav{t_1-t_2}\(G^>(t_1,t_2)+G^>(t_2,t_1)\),
\end{split}
\end{equation}
the advanced Green's function,
\begin{equation}\label{eq:SM_G^A}
\begin{split}
G^\rmA(t_1,t_2)
&=\heav{t_2-t_1}\(G^<(t_1,t_2)-G^>(t_1,t_2)\)
\\
&=-\heav{t_2-t_1}\(G^>(t_1,t_2)+G^>(t_2,t_1)\)
\\
&=-G^\rmR(t_2,t_1),
\end{split}
\end{equation}
and the Keldysh Green's function,
\begin{equation}\label{eq:SM_G^K}
\begin{split}
G^\rmK(t_1,t_2)
&=G^>(t_1,t_2)+G^<(t_1,t_2)
\\
&=G^>(t_1,t_2)-G^>(t_2,t_1)
\\
&=-G^\rmK(t_2,t_1).
\end{split}
\end{equation}
The components of the contour self-energy $\Sigma(z,z')$ satisfy the same relations as in Eqs.~(\ref{eq:SM_G^<})--(\ref{eq:SM_G^K}).

Next, we change variables to $t=t_1-t_2$ and $\sT=(t_1+t_2)/2$. For long times, $\sT\to\infty$, the system loses any information about its initial state and relaxes to the steady state. The Green's function $G^>$ depends now only on $t$, and we move to Fourier space with continuous frequencies $\omega$, using the convention
\begin{equation}
\begin{split}
A(t)&=\int \frac{\d\omega}{2\pi}\, \hat{A}(\omega)\,e^{-\i \omega t},
\\
\hat{A}(\omega)&=\int \d t\, A(t)\,e^{\i \omega t},
\end{split}
\end{equation}
for a function $A(t)$ and its Fourier transform $\hat{A}(\omega)$ (we omit the hat on the Fourier transform whenever no confusion arises). For reference, in this convention, the Fourier transform of the step function is 
\begin{equation}
\hat{\Theta}(\omega)=-\sP \frac{1}{\i \omega}+\pi\dirac{\omega}.
\end{equation}

\subsection{Keldysh rotation}

We further perform a Keldysh rotation by defining the real quantities~\cite{ribeiro2015PRL}:
\begin{align}
\label{eq:SM_rho^+-_def}
\rho^{\pm}(\omega)
&=-\frac{1}{2\pi \i}\(G^>(\omega)\pm G^<(\omega)\)
=-\frac{1}{2\pi \i}\(G^>(\omega)\mp G^>(-\omega)\),
\\
\label{eq:SM_sigma^+-_def}
\sigma^{\pm}(\omega)
&=-\frac{1}{2\pi \i}\(\Sigma^>(\omega)\pm \Sigma^<(\omega)\)
=-\frac{1}{2\pi \i}\(\Sigma^>(\omega)\mp \Sigma^>(-\omega)\),
\end{align}
and their Hilbert transforms
\begin{align}
\label{eq:SM_rhoH_def}
\rho^\rmH(\omega)&=-\frac{1}{\pi}\sP\!\int \d \nu\, \frac{\rho^-(\nu)}{\omega-\nu},
\\
\label{eq:SM_sigmaH_def}
\sigma^\rmH(\omega)&=-\frac{1}{\pi}\sP\!\int \d \nu\, \frac{\sigma^-(\nu)}{\omega-\nu}.
\end{align}
The spectral function $\rho^-(\omega)$ is symmetric and normalized as $\int d \omega\, \rho^{-}(\omega)=1$.
Close to the steady state, $\rho^+(\omega)$ and $\rho^-(\omega)$ are related by a fluctuation-dissipation-like relation, $\rho^+(\omega)=\tanh(\beta\omega/2)\, \rho^-(\omega)$, where $\beta=0$ is the temperature of the infinite-temperature steady state. Hence, $\rho^+(\omega)$ vanishes identically. Note that for more general jump operators that lead to finite-temperature equilibrium or nonequilibrium steady states we have, in general, $\rho^+(\omega)\neq0$.
In terms of $\rho^\pm$, the components of the Green's function read as [in the last equality of each line we use the special property $\rho^+(\omega)=0$ of our model]
\begin{align}
\label{eq:SM_G>_rho}
G^{>}(\omega)
&=-\pi \i\(\rho^+(\omega)+\rho^-(\omega)\)
=-\pi \i \rho^-(\omega),
\\
\label{eq:SM_G<_rho}
G^{<}(\omega)
&=-\pi \i\(\rho^+(\omega)-\rho^-(\omega)\)
=\pi \i \rho^-(\omega),
\\
\label{eq:SM_GT_rho}
G^{\rm T}(\omega) &=-\pi\(\rho^{\rmH}(\omega) + \i \rho^+(\omega)\)
=-\pi \rho^\rmH(\omega), \\
G^{\rm \bar T}(\omega) &=\pi\(\rho^{\rmH}(\omega) - \i \rho^+(\omega)\)=\pi \rho^\rmH(\omega), \\
\label{eq:SM_GR_rho}
G^\rmR(\omega)&=-\pi\(\rho^\rmH(\omega)+\i \rho^-(\omega)\),
\\
G^\rmA(\omega)&=-\pi\(\rho^\rmH(\omega)-\i \rho^-(\omega)\),
\\
G^\rmK(\omega)&=-2\pi \i\rho^+(\omega)=0.
\end{align}
Exactly the same relations hold for the self-energies $\sigma^\pm$. In real time $t$, our main quantity of interest is $G^\rmR(t)$ which can be obtained from the spectral function by
\begin{equation}\label{eq:GR_rho-}
\i G^\rmR(t)=\heav{t}\int_{-\infty}^{+\infty} d \omega\,  \rho^-(\omega)\,e^{-\i\omega t}.
\end{equation}

\subsection{Real-time Schwinger-Dyson equations}
 
We can now rewrite the saddle-points in terms of $\rho^\pm(\omega)$ and $\sigma^\pm(\omega)$. The Dyson equation on the Keldysh contour, Eq.~(\ref{eq:C_SD_Sigma}), can be written as
\begin{equation}\label{eq:SM_Dyson}
\int_\sC \d z'' \(
G_0^{-1}(z,z'')-\Sigma(z,z'')
\)G(z'',z')=
\delta(z-z'),
\end{equation}
where $G_0^{-1}(z,z')=\i\delta(z-z')\pd_z$.
Restricting $(z,z')$ to $\sC^+\times\sC^-$ and $\sC^-\times\sC^+$ (i.e., applying Langreth's rules) the right-hand side of Eq.~(\ref{eq:SM_Dyson}) vanishes and, noting also that $G_0^{-1>}=G_0^{-1<}=0$, we obtain, in Fourier space,
\begin{align}
\Sigma^>(\omega)G^\rmA(\omega)=\(\omega-\Sigma^\rmR\)G^>(\omega),\\
\Sigma^<(\omega)G^\rmA(\omega)=\(\omega-\Sigma^\rmR\)G^<(\omega).
\end{align}
Taking the sum and the difference of these equations, we can relate $\rho^\pm$ with $\sigma^\pm$:
\begin{equation}\label{eq:SM_SDE_rho+-}
\begin{split}
\rho^\pm(\omega)
&=\sigma^\pm(\omega)\,
\frac{G^\rmA(\omega)}{\omega-\Sigma^\rmR(\omega)}
=\sigma^\pm(\omega)\,
\frac{G^\rmA(\omega)\(\omega-\Sigma^\rmA(\omega)\)}{\abs{\omega-\Sigma^\rmR(\omega)}^2}
\\
&=\frac{\sigma^\pm(\omega)}{
	\(\omega+\pi \sigma^\rmH(\omega)\)^2
	+\(\pi \sigma^-(\omega)\)^2},
\end{split}
\end{equation}
where we used the equivalent of Eqs.~(\ref{eq:SM_G^R}) and (\ref{eq:SM_G^A}) for the self-energy and the fact that the Dyson equation is diagonal for the retarded and advanced components, $(\omega-\Sigma^\rmA)G^\rmA=(\omega-\Sigma^\rmR)G^\rmR=1$.

Restricting the SD equation for the self-energy, Eq.~(\ref{eq:C_SD_G}), to $(z,z')=(t_1^-,t_2^+)$ and $(z,z')=(t_1^+,t_2^-)$ and using Eqs.~(\ref{eq:K_kernel_<}) and (\ref{eq:K_kernel_>}), it reads as:
\begin{align}
\label{eq:Sigma>}
\Sigma^>(t_1,t_2)=
G^>(t_1,t_2)\left\{
-\i^q J^2\left[G^>(t_1,t_2)\right]^{q-2}
-m\gamma^4\left[G^<(t_1,t_2)\right]^2
+2m\gamma^2\dirac{t_1-t_2}
\right\},
\\
\label{eq:Sigma<}
\Sigma^<(t_1,t_2)=
G^<(t_1,t_2)\left\{
-\i^q J^2\left[G^<(t_1,t_2)\right]^{q-2}
-m\gamma^4\left[G^>(t_1,t_2)\right]^2
+2m\gamma^2\dirac{t_1-t_2}
\right\}.
\end{align}
The preceding equations are given by multiplication in time and hence a convolution in Fourier space. Straightforward algebraic manipulation using Eqs.~(\ref{eq:SM_G>_rho}) and (\ref{eq:SM_G<_rho}) and setting $\rho^+(\omega)=\sigma^+(\omega)=0$ gives the final SD equations for quadratic jump operators:
\begin{align}
\rho^-(\omega)&=\frac{\sigma^-(\omega)}{
	\left[\omega+\pi \sigma^\rmH(\omega)\right]^2
	+\left[\pi \sigma^-(\omega)\right]^2},
\label{sdreal_rho}
\\
\sigma^-(\omega)&=\frac{J^2}{2^{q-2}}\left(\rho^-\right)^{*(q-1)}(\omega)+\frac{m\gamma^4}{4}\left(\rho^-\right)^{*3}(\omega)+\frac{m\gamma^2}{\pi}.
\label{sdreal_sigma}
\intertext{The case of linear jump operators proceeds similarly and we find:}
\rho^-(\omega)&=\frac{\sigma^-(\omega)}{
	\left[\omega+\pi \sigma^\rmH(\omega)\right]^2
	+\left[\pi \sigma^-(\omega)\right]^2}.
\label{sdreal_rho_linear}
\\
\sigma^-(\omega)&=\frac{J^2}{2^{q-2}}\left(\rho^-\right)^{*(q-1)}(\omega)+\frac{\mu}{\pi}.
\label{sdreal_sigma_linear}
\end{align}
Above, $(\rho^-)^{*n}(\omega)$ denotes the $n$-fold convolution of the spectral function with itself,
\begin{equation}
(\rho^-)^{*n}(\omega)=
\int \prod_{j=1}^{n-1}\d \nu_j\, 
\rho^-(\omega-\sum_{j=1}^{n-1} \nu_j)
\prod_{j=1}^{n-1} \rho^-(\nu_j).
\end{equation}

In the remainder of the chapter we solve the SD equations (\ref{sdreal_rho})--(\ref{sdreal_sigma_linear}). First, in Sec.~\ref{sec:SYK_strong_dissipation}, we find an approximate analytic solution in the strong-dissipation regime, where the relaxation is dissipation-driven. In Sec.~\ref{sec:SYK_weak_dissipation}, we numerically solve the SD equations for $q=4$ and find an anomalously large decay rate (or gap) for weak dissipation, a regime where the strongly-coupled internal SYK dynamics dominate the relaxation. For $q=2$ and linear jump operators, the Green's functions can be evaluated analytically for all dissipation strengths and, because the model is not quantum chaotic, no anomalous relaxation is found, which is discussed in Sec.~\ref{sec:q2}.

\section{Strong dissipation: Dissipation-driven relaxation}
\label{sec:SYK_strong_dissipation}

\subsection{Analytic relaxation gap from a Lorentzian ansatz}

In the strong dissipation regime ($J/\mu\ll1$ for linear jump operators and $J/m\gamma^2\ll1$ for quadratic ones), the relaxation gap can be determined analytically from a Lorentzian ansatz for the spectral function (alternatively, an exponential ansatz for the retarded Green's function):
\begin{equation}\label{eq:rho-_ansatz_largegamma}
\rho^-(\omega)=\frac{1}{\pi}\frac{\Delta}{\omega^2+\Delta^2}.
\end{equation}
The gap $\Delta$ is to be computed self-consistently.
Because the Lorentzian is stable under convolution, Eq.~(\ref{sdreal_sigma}) leads to the following self-energy for quadratic jump operators,
\begin{equation}\label{eq:sigma-_ansatz_largegamma}
\sigma^-(\omega)=
\frac{J^2}{2^{q-2}\pi}\frac{(q-1)\Delta}{\omega^2+\left[(q-1)\Delta\right]^2}
+\frac{m\gamma^4}{4\pi}\frac{3\Delta}{\omega^2+(3\Delta)^2}
+\frac{m\gamma^2}{\pi},
\end{equation}
while Eq.~(\ref{sdreal_sigma_linear}) gives the self-energy for linear jump operators,
\begin{equation}\label{eq:sigma-_ansatz_largegamma_linear}
\sigma^-(\omega)=
\frac{J^2}{2^{q-2}\pi}\frac{(q-1)\Delta}{\omega^2+\left[(q-1)\Delta\right]^2}
+\frac{\mu}{\pi}.
\end{equation}
Since we are interested in the low-frequency response, we set $\omega=0$. The regime of validity of this approximation can be determined self-consistently and is discussed in the next subsection. Plugging Eqs.~(\ref{eq:rho-_ansatz_largegamma}) and (\ref{eq:sigma-_ansatz_largegamma}) back into the Dyson equation, Eq.~(\ref{sdreal_rho_linear}), we find an analytic expression for the relaxation gap of the SYK Lindbladian with quadratic jump operators:
\begin{equation}\label{eq:analytical_gap}
\Delta=\frac{m\gamma^2}{2}\(
1+\sqrt{\frac{3m+1}{3m}+\frac{1}{2^{q-4}(q-1)}\(\frac{J}{m\gamma^2}\)^2}
\).
\end{equation}
Proceeding identically, the gap for linear jump operators is found to be
\begin{equation}\label{eq:analytical_gap_linear}
\Delta=\frac{\mu}{2}\(
1+\sqrt{1+\frac{1}{2^{q-4}(q-1)}\(\frac{J}{\mu}\)^2}
\).
\end{equation}
From Eqs.~(\ref{eq:analytical_gap}) and (\ref{eq:analytical_gap_linear}), we see that in the strong dissipation limit ($J/m\gamma^2\ll1$ or $J/\mu\ll1$), which is their the regime of validity (see the next subsection), the relaxation is dissipation-driven since, in the strict limit of infinite dissipation, $J/m\gamma^2\to0$ or $J/\mu\to0$, the gap is trivially linear in the dissipation strength ($m\gamma^2$ or $\mu$). Yet, it is remarkable that we can compute both the prefactor of the linear scaling and the corrections for large but finite dissipation analytically, given the strongly-correlated nature of the dissipative SYK model. In Sec.~\ref{sec:SYK_weak_dissipation}, we will solve the SD equations numerically for all dissipation strength. We will find that, at large $m\gamma^2$ or $\mu$, the analytic approximation of this section agrees excellently with the numerical solution, whereas for low enough $m\gamma^2$ or $\mu$, there are oscillatory corrections to the exponential relaxation. We can also establish the breakdown of the purely exponential relaxation at intermediate $m\gamma^2$ or $\mu$ by examining the accuracy of the Lorentzian approximation, which we do next.

\subsection{Accuracy of the Lorentzian approximation}

We check the accuracy of the Lorentzian approximation for the special case $q=4$ and quadratic jump operators. The more general case proceeds similarly. We again assume a Lorentzian form for the spectral function $\rho^-(\omega)$,
\begin{equation}\label{eq:SM_rho-_ansatz_largegamma}
\rho^-(\omega)=\frac{1}{\pi}\frac{\Gamma(\omega)}{\omega^2+\Gamma(\omega)^2},
\end{equation}
where the width $\Gamma$ is now frequency-dependent. The Lorentzian ansatz holds if $\Gamma$ is frequency-independent. Proceeding as in the previous section, we find that $\Gamma$ has to satisfy the self-consistency equation
\begin{equation}\label{eq:self_consistency_Gamma}
\Gamma=\frac{
	\(\omega^2+\Gamma^2\)
	\(m\gamma^2+\frac{J^2+m\gamma^4}{4}\frac{3\Gamma}{\omega^2+9\Gamma^2}\)
}{
	\omega^2\(
	1+\frac{J^2+m\gamma^4}{4}\frac{1}{\omega^2+9\Gamma^2}
	\)^2+\(
	m\gamma^2+\frac{J^2+m\gamma^4}{4}\frac{3\Gamma}{\omega^2+9\Gamma^2}
	\)^2
}.
\end{equation}
Note that, contrarily to before, we have not set $\omega=0$. 

\begin{figure}[t]
	\centering
	\includegraphics[width=0.7\textwidth]{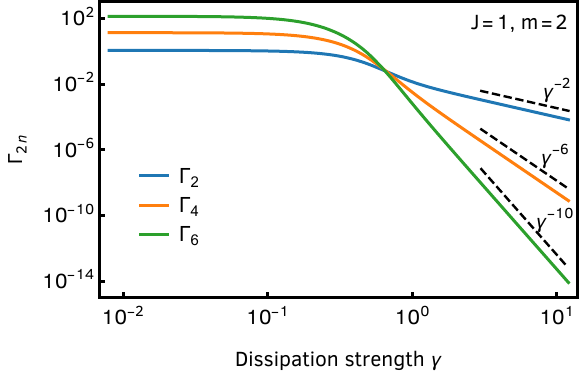}
	\caption{Three lowest-order corrections to the Lorentzian ansatz, $\Gamma_{2n}$ with $n=1,2,3$, as a function of dissipation strength $\gamma$, for $q=4$, $J=1$, and $m=2$. For large dissipation, the corrections have a power-law decay, $\Gamma_{2n}\propto \gamma^{2-4n}$. This supports the claim that the Lorentzian ansatz is exact at $\gamma=\infty$ and very accurate for $m\gamma^2\gg J$. At low $\gamma$, the corrections to the ansatz become dominant and the approximation breaks down.}
	\label{fig:SM_Lorentzian_coeffs}
\end{figure}

Next, we expand $\Gamma$ in powers of $\omega$, $\Gamma=\Delta+\Gamma_2\omega^2+\Gamma_4\omega^4+\cdots$. At zeroth order in $\omega$, Eq.~(\ref{eq:self_consistency_Gamma})
is solved by Eq.~(\ref{eq:analytical_gap}). The higher-order coefficients, $\Gamma_2$, $\Gamma_4,\dots$, have to vanish for our ansatz to be consistent, since in deriving the self-consistency equation we implicitly assumed $\Gamma$ to be constant. At each order, by using the solution for lower-order coefficients, we obtain an algebraic equation for the coefficient $\Gamma_{2n}$. For instance, at order $\omega^2$, we use the result at order $\omega^0$ [Eq.~(\ref{eq:analytical_gap})], to find that Eq.~(\ref{eq:self_consistency_Gamma}) yields a linear equation for $\Gamma_2$ solved by
\begin{equation}
\begin{split}
\Gamma_2=\,\frac{2}{3m\gamma ^2} &\left[\left(\frac{12m^2+9m+1}{9m^2}-\frac{4m+7/3}{3m}\sqrt{\frac{3 m+1}{3m}+\frac{J^2}{3m^2\gamma^4}}\right)\right.
\\
+&\left.\,\frac{J^2}{3m^2\gamma^4} \left( \frac{9m+2}{3m}-\frac{7}{3}\sqrt{\frac{3 m+1}{3m}+\frac{J^2}{3m^2\gamma ^4}}\right)
+ \frac{J^4}{9m^4\gamma^8}\right]
\\\times&
\left[\left(\frac{1}{3m}+\frac{J^2}{3m^2\gamma^4}\right)^2 \sqrt{\frac{3 m+1}{3m}+\frac{J^2}{3m^2\gamma^4}}\right]^{-1},
\end{split}
\end{equation}
which decays as $1/\gamma^2$ as $m\gamma^2/J\to\infty$. We can proceed analogously for higher-order coefficients. Although it may in principle be possible to show exactly that $\lim_{\gamma\to\infty}\Gamma_{2n}=0$ for all $n$, here we limit ourselves to the first three corrections, $\Gamma_2$, $\Gamma_4$, and $\Gamma_6$, plotted in Fig.~\ref{fig:SM_Lorentzian_coeffs}, which decay as $\gamma^{2-4n}$. This procedure indicates the exactness of the Lorentzian approximation as $\gamma\to\infty$ and also provides a quantitative measure for its accuracy at finite $\gamma$.

\section{Weak dissipation: Anomalous relaxation in many-body dissipative quantum chaos}
\label{sec:SYK_weak_dissipation}

For $q=4$, we solved the SD equations numerically in a self-consistent manner by proposing an ansatz for $\rho^-(\omega)$ and $\sigma^-(\omega)$ and then iterating the SD equations until convergence is achieved~\cite{ribeiro2015PRL}. We give the details on the numerical method in Sec.~\ref{sec:SYK_method}. We give a brief overview of the solutions for the case of quadratic jump operators in Sec.~\ref{sec:q4_quadratic} and then analyze them in more detail for the case of linear jump operators in Sec.~\ref{sec:q4_linear}.

\subsection{Numerical method}
\label{sec:SYK_method}

We solved Eqs.~(\ref{sdreal_rho}) and (\ref{sdreal_sigma}) iteratively on a linearly-discretized frequency grid $\Lambda=\{-\omega_\mathrm{\max},$ $-\omega_\mathrm{\max}+\Delta\omega,\dots,\omega_\mathrm{\max}\}$, with $\omega_\mathrm{max}=1000$ and $\Delta\omega=0.05$. We used the following procedure:
\begin{enumerate}
	\item Given a spectral function $\rho^-_i(\omega)$, we compute a new self-energy $\sigma_{i+1}^-(\omega)$ from Eq.~(\ref{sdreal_sigma}) by interpolating $\rho^-(\omega)$ [with $\rho^-(\omega\notin\Lambda)=0$] and numerically evaluating the triple convolution.
	\item We evaluate the Hilbert transform $\sigma_{i+1}^\rmH(\omega)$, Eq.~(\ref{eq:SM_sigmaH_def}), using the trapezoid rule.
	\item We compute the new spectral function using Eq.~(\ref{sdreal_rho}). To ensure there is monotone convergence, we do a partial update,
	\begin{equation}
	\rho^-_{i+1}(\omega)=(1-\eta_\mathrm{mix})\,\rho^-_i(\omega)+
	\eta_\mathrm{mix}\,\frac{\sigma_{i+1}^-(\omega)}{
		\(\omega+\pi \sigma_{i+1}^\rmH(\omega)\)^2
		+\(\pi \sigma_{i+1}^-(\omega)\)^2},
	\end{equation}
	with $\eta_\mathrm{mix}=0.1$.
	\item We repeat steps 1.--3.\ until the solution converges, in the sense that the total difference between two successive iterations is less than some prescribed accuracy, 
	\begin{equation}
	\sum_{\omega\in\Lambda}\abs{\rho^-_{i+1}(\omega)-\rho^-_{i}(\omega)}<\epsilon=10^{-4}.
	\end{equation}
	At each step, we also checked the normalization of the spectral function, $\int \d\omega\, \rho^-(\omega)=1$ to within the prescribed accuracy.
\end{enumerate}
For a given $J$ and $m$, we started from moderately high value of $\gamma$, for which we expect the Lorentzian ansatz to be accurate. We used this ansatz as the initial seed for the algorithm outlined above. We then successively lowered $\gamma$ until $\gamma=0$, at each step using a previously converged solution as the new seed. The case of linear jump operators, Eqs.~(\ref{sdreal_rho_linear}) and (\ref{sdreal_sigma_linear}), proceeds in the same way.

As a check on our method, we confirmed that the system indeed relaxes to the infinite-temperature steady state. To that end, we solved Eqs.~(\ref{eq:SM_SDE_rho+-})--(\ref{eq:Sigma<}) for $q=4$ and quadratic jump operators with a nonzero initial seed for $\rho^+(\omega)$ and found that the equations converged to $\rho^+(\omega)=0$, while $\rho^-(\omega)$ coincides with the solution of Eqs.~(\ref{sdreal_rho}) and (\ref{sdreal_sigma}).

To go back to the frequency domain, we computed Eq.~(\ref{eq:GR_rho-}) using the trapezoid rule. With the frequency grid $\Lambda$ described above, we were able to study the decay of the retarded Green's function down to $\i G^\rmR(t)\sim10^{-6}$.

\subsection{Quadratic jump operators}
\label{sec:q4_quadratic}

\begin{figure}[t]
	\centering
	\includegraphics[width=0.8\columnwidth]{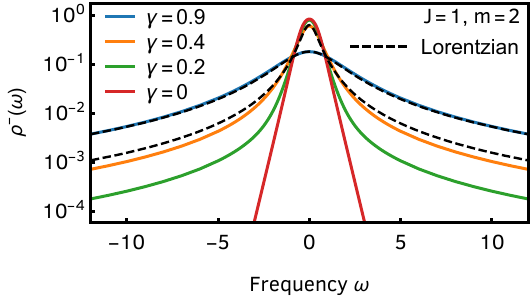}
	\caption{Spectral function~$\rho^-(\omega)$ obtained from the numerical solution of the SD equations, Eqs.~(\ref{sdreal_rho}) and (\ref{sdreal_sigma}), for $J=1$, $m=2$, and different $\gamma$. For large $\gamma$, the solution is well described by a Lorentzian (dashed lines) with the width computed analytically, Eq.~(\ref{eq:analytical_gap}). For intermediate $\gamma$ (e.g., $\gamma=0.4$), the Lorentzian ansatz still gives a reasonable description of the result (especially for low frequencies), but it fails for low dissipation.}
	\label{fig:rhominus}
\end{figure}

\begin{figure}[t]
	\centering
	\includegraphics[width=0.95\columnwidth]{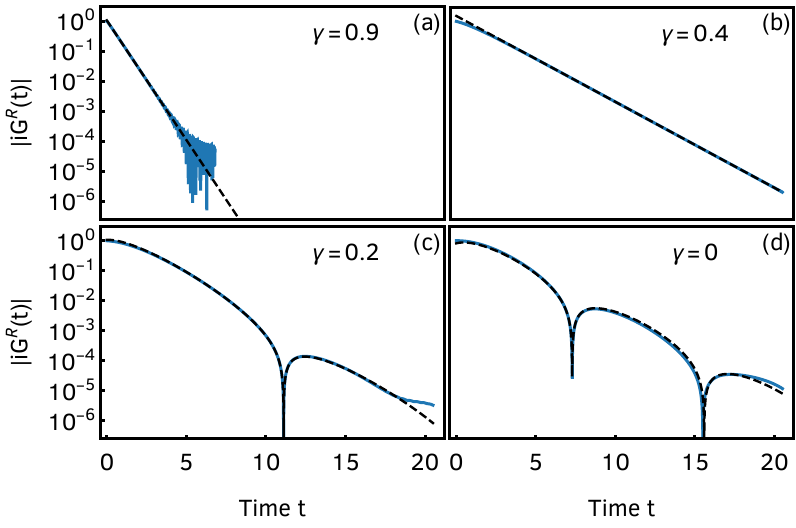}
	\caption{Retarded Green's function as a function of time for the Lindbladian with quadratic jump operators, $J=1$, $m=2$, and different $\gamma$. The full blue curves are obtained by Fourier transforming the numerical solution for $\rho^-(\omega)$, as prescribed in Eq.~(\ref{eq:GR_rho-}). The dashed black lines give the best asymptotic fit to Eq.~(\ref{eq:GR_fit}).}
	\label{fig:Gret}
\end{figure}

The results for $J=1$, $m=2$, and different values of $\gamma$ are plotted in Fig.~\ref{fig:rhominus}. For large-enough $\gamma$, the spectral function is well approximated by a Lorentzian. Fourier transforming back to the time domain [Eq.~(\ref{eq:GR_rho-})], see Figs.~\ref{fig:Gret}(a) and \ref{fig:Gret}(b), this implies a well-defined gap $\Delta$ (i.e., relaxation rate), as the retarded Green's function decays exponentially, $\i G^\rmR(t)=\heav{t}\exp\{-\Delta t\}$.

\begin{figure}[t]
	\centering
	\includegraphics[width=0.8\columnwidth]{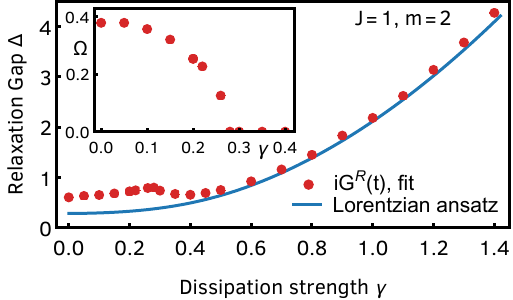}
	\caption{Relaxation gap~$\Delta$ as a function of dissipation strength $\gamma$ for the Lindbladian with quadratic jump operators, $J=1$ and $m=2$. The red dots are obtained from the fit to Eq.~(\ref{eq:GR_fit}), while the blue line is the analytical result from the Lorentzian ansatz, Eq.~(\ref{eq:analytical_gap}). The two agree for large $\gamma$ but saturate to different values as $\gamma\to0$. Inset: Frequency~$\Omega$ of the oscillatory correction as a function of $\gamma$. For $\gamma\gtrsim0.28$, the period of oscillations either diverges ($\Omega=0$) or becomes longer than the numerically-accessible time window.}
	\label{fig:gapPlot}
\end{figure}

The comparison of Eqs.~(\ref{eq:rho-_ansatz_largegamma}) and (\ref{eq:analytical_gap}) with the numerical solution is given in Fig.~\ref{fig:rhominus}. For $J=1$ and $m=2$, already for $\gamma=0.9$, there is excellent agreement. Accordingly, we see a clear exponential decay of $\i G^\rmR(t)$ in Fig.~\ref{fig:Gret}~(a). For intermediate $\gamma$, say, $\gamma=0.4$, there are noticeable deviations in the tails, but the low-frequency part of $\rho^-$ is still perfectly described by Eqs.~(\ref{eq:rho-_ansatz_largegamma}) and (\ref{eq:analytical_gap}) and $\i G^\rmR(t)$ still decays exponentially, see Fig.~\ref{fig:Gret}~(b). For small $\gamma$, the tails of the spectral function are very far from Lorentzian.
This signals possible power-law or oscillatory corrections to the asymptotic decay of $G^\rmR(t)$ (depending on the precise form of $\rho^-(\omega)$, which cannot be determined analytically), see Figs.~\ref{fig:Gret}(c) and \ref{fig:Gret}(d).
We can extract the relaxation gap from $\i G^\rmR(t)$ by fitting the numerical results to an exponential function with power-law and oscillatory corrections. We found the former to be negligible in general, but the latter to be relevant for small $\gamma$, i.e., 
\begin{equation}\label{eq:GR_fit}
\i G^\rmR(t)=A\, e^{-\Delta t}\cos(\Omega t+\phi)
\end{equation}
gives an excellent fit for $t\gg 1$
with fitting parameters $A$, $\Delta$, $\Omega$, and $\phi$. The resulting gap is plotted in Fig.~\ref{fig:gapPlot} as a function of $\gamma$. We conclude that, for large $\gamma$, $\Delta$ grows quadratically, in agreement with Eq.~(\ref{eq:analytical_gap}), while it starts to deviate from the Lorentzian ansatz at intermediate values $\gamma\approx0.5$. As $\gamma$ further decreases, our results are consistent (within the numerically accessible time window) with a bifurcation of the real gap $\Delta$ to a pair of complex-conjugated gaps $\Delta\pm\i\Omega$ at $\gamma\approx0.28$, see inset of Fig.~\ref{fig:gapPlot}. Remarkably, as $\gamma\to0$, $\Delta$ saturates to a finite value, indicating that even an infinitesimally small amount of dissipation leads to relaxation at a finite rate---anomalous relaxation. This is admissible given that we took the thermodynamic limit first. Although the Lorentzian ansatz and the numerical solution saturate to different values when $\gamma\to0$, the former still gives a qualitatively correct picture for the relaxation rate of the SYK model across all dissipation scales.

\subsection{Linear jump operators}
\label{sec:q4_linear}

A natural question to ask is to what extent the anomalous behavior we have found in the weak-coupling limit is universal or whether it is a particularity of model considered above. To answer this question, we turn to the case of linear and nonrandom jump operators. As we will see below, the phenomenology we find here is similar, but more accentuated, than in the quadratic case above, and we analyze it in detail.

\begin{figure}[t]
	\centering
	\includegraphics[width=0.95\columnwidth]{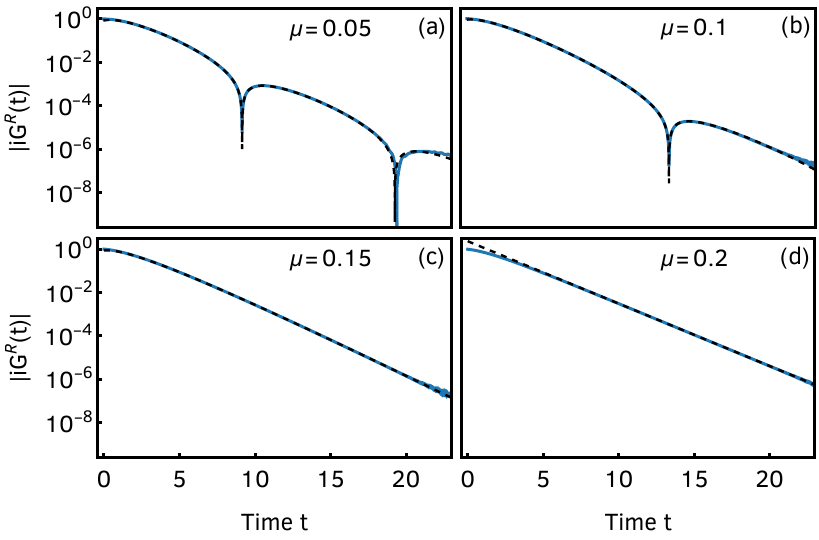}
	\caption{Retarded Green's function as a function of time for the Lindbladian with linear jump operators, $J=1$, and different $\mu$. The full blue curves are obtained by Fourier transforming the numerical solution for $\rho^-(\omega)$, as prescribed in Eq.~(\ref{eq:GR_rho-}). The dashed black lines give the best asymptotic fit to Eq.~(\ref{eq:GR_fit}).}
	\label{fig:GretLinear}
\end{figure}

\begin{figure}[t]
	\centering
	\includegraphics[width=\columnwidth]{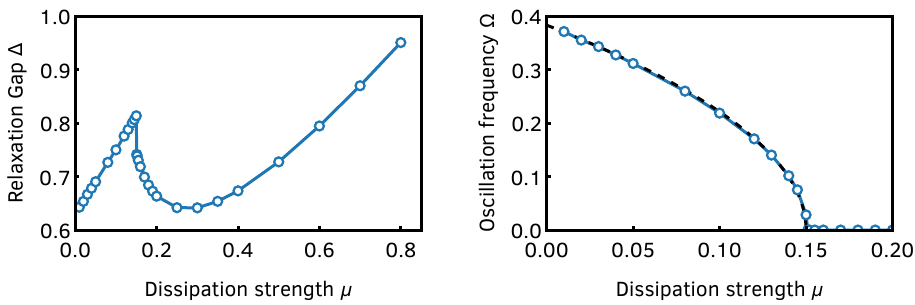}
	\caption{Relaxation gap~$\Delta$ (left) and oscillation frequency $\Omega$ (right) as a function of dissipation strength $\mu$ for the Lindbladian with linear jump operators, $J=1$, obtained from the fit to Eq.~(\ref{eq:GR_fit}). The $\mu$ dependence of the frequency (right), is very well fitted (dashed curve) by $\Omega(\mu)\approx \sqrt{\mu_c-\mu}$, with $\mu_c \approx 0.15$. The oscillations in the Green's function terminate with a second-order phase transition at $\mu = \mu_c$.}
	\label{fig:comEuLor}
\end{figure}

In Fig.~\ref{fig:GretLinear}, we depict the retarded Green's functions for different values of the coupling to the environment $\mu$. As before, we find an excellent agreement with a fit to $A e^{-\Delta t}\cos{(\Omega t+\phi)}$ [Eq.~(\ref{eq:GR_fit})] with $A$, $\Delta$, $\Omega$, and $\phi$ fitting parameters. The frequency of the oscillations, $\Omega$, vanishes for sufficiently large $\mu > 0.15$. The exponential decay, controlled by $\Delta$, persists for any coupling strength $\mu$. However, we stress that, especially for zero or small $\mu$, $\Delta$ does not coincide with the spectral gap $\Delta_\mathrm{spec}$ (i.e., the gap in the spectrum between the steady state and the first excited state). It is rather an ``order parameter'' characterizing the relaxation that results from ensemble averaging. As $\mu$ increases, $\Delta$ approaches the spectral gap $\Delta_\mathrm{spec}$. A more complete understanding of the relation of the gap $\Delta$ to the spectral gap $\Delta_\mathrm{spec}$ can be obtained from the finite-$N$ Green's functions, for which we refer the reader to App.~E of Ref.~\cite{garcia2023PRD2}.
In Fig.~\ref{fig:comEuLor}, we show the gap $\Delta$ (left) and the frequency of the oscillations $\Omega$ (right) of the Green's function as a function of the coupling $\mu$. Both $\Delta$ and $\Omega$ have an intriguing $\mu$ dependence that we now analyze in detail.

For $\mu = 0$, the gap is finite and the frequency of the oscillations in the Green's function is largest. Physically, this corresponds to a complex order parameter whose real part controls the exponential decay and the imaginary part the frequency of oscillation.
A nonzero gap in the limit $\mu\to0$ corresponds to a finite decay rate even in the absence of a coupling to the bath, an unexpected phenomenon that we previously referred to as anomalous relaxation.

As $\mu$ increases, the imaginary part of the excited eigenvalues becomes smaller. Therefore, a larger $\mu$ is expected to reduce the value of $\Omega(\mu)$. Indeed, we find that the frequency decreases monotonically with $\mu$. 
Interestingly, $\Omega(\mu)$ shows excellent agreement
with a simple ansatz $\Omega(\mu) \approx \sqrt{\mu_c- \mu}$ with $\mu_c \approx 0.15$. For $\mu >\mu_c$, the decay of Green's function is purely exponential, which implies the vanishing of the frequency of oscillations. This is consistent with the existence of a second-order phase transition at $\mu = \mu_c$.

For a small but finite $\mu/\mu_c < 1$, $\Delta(\mu) - \Delta(0) \sim \mu$ increases linearly. This makes sense physically since a stronger coupling to the bath results in a larger decay rate. 
We think this linear behavior is related to the fact that the energy of the first excited state is always real and equal to $4\mu$. However, the overall effect on $\Delta$ is relatively small compared with the $\mu = 0$ contribution, namely, the spectral gap $\Delta_\mathrm{spec} \equiv \min|\Re\Lambda_i|$, where $\Lambda_i$ are the eigenvalues of $\scL$, is only a small contribution to the relaxation gap $\Delta$. 
The linearity suggests that the slope can be obtained by a perturbative treatment in $\mu$.

At a critical value $\mu_c\approx 0.15$, corresponding to the vanishing of the oscillation frequency, the linear increase of $\Delta$ stops abruptly.
A further increase in the coupling $\mu$ leads to a sharp, but likely continuous, drop in $\Delta$. This is rather unexpected as an increase in the coupling to the bath should always lead to an increase in the decay rate as the approach to equilibrium should occur faster. However, we clearly observe a drop as $\mu \gtrsim 0.15$ increases. A possible explanation of this nonmonotonic behavior is that the oscillations become imaginary, $\Omega \to i \Omega$, for $\mu > \mu_c$, so that they effectively become a negative contribution to the gap. 
For a sufficiently large $\mu$, in agreement also with Ref.~\cite{kulkarni2022PRB}, the decay rate $\Delta$ approaches a linear dependence $\Delta\approx\mu$ [see Eq.~(\ref{eq:analytical_gap_linear})], which is the expected result \cite{sieberer2016} for a strong coupling to the environment. In this region, the spectrum is mostly real and $\Delta$ is close to the spectral gap $\Delta_\mathrm{spec}$.

\section{Absence of anomalous relaxation for \texorpdfstring{$q=2$}{q=2}}
\label{sec:q2}

Finally, we ask whether the anomalous relaxation is related to the fact that the system is strongly interacting and quantum chaotic. In order to answer this question we turn to the study of the model with $q=2$ and linear jump operators, which is quadratic in Majorana operators and, therefore, not quantum chaotic. Because of its quadratic nature, its SD equations admit an exact solution for all dissipation strength.

The SD equations for the retarded component of the Green's function are:
\begin{equation}
(\omega-\Sigma^\mathrm{R})G^\mathrm{R}=1
\qquad \text{and} \qquad
\Sigma^\mathrm{R}=-\pi(\sigma^\mathrm{H}+\i\sigma^-).
\end{equation}
Using the SD equation for the self-energy, Eq.~(\ref{sdreal_sigma_linear}), which for $q=2$ reads as 
\begin{equation}
\sigma^-=J^2 \rho^- +\frac\mu\pi,
\end{equation}
with Hilbert transform $\sigma^\mathrm{H}=J^2\rho^\mathrm{H}$, 
the retarded self-energy is
\begin{equation}
	\Sigma^\mathrm{R}=-\pi\(J^2 \rho^\mathrm{H}+\i J^2 \rho^-+\frac{\i\mu}{\pi}\)=J^2 G^\mathrm{R} -\i \mu,
\end{equation} 
where, to arrive at  the last equality, we employed Eq.~(\ref{eq:SM_GR_rho}).
Inserting the retarded self-energy into the Dyson equation, we find that the retarded Green's function satisfies the quadratic equation
\begin{equation}
J^2 (G^\mathrm{R})^2-(\omega+i\mu)G^\mathrm{R}+1=0.
\end{equation}
The solution that vanishes at $\omega \to \pm \infty$ is given by
\be
\label{eq:SYK_q2GR}
G^\rmR(\omega) = -\frac \omega {2 J^2} +\frac{\sign(\omega)}{2J^2} \sqrt{(\omega+\i\mu)^2-4J^2}
\ee
and Fourier transforming back to the time domain, we find
\begin{equation}\label{eq:SYK_q2GR_time}
\i G^\mathrm{R}(t)=\Theta(t)\, e^{-\mu t}\,\frac{J_1(2Jt)}{Jt},
\end{equation}
where $J_1$ is a Bessel function of the first kind.
For completeness, we also write down the single-particle spectral function of the model:
\begin{equation}
\rho^-(\omega)=-\frac{1}{\pi}\mathrm{Im}G^\mathrm{R}(\omega)
=\frac{-\mu+\frac{1}{\sqrt{2}}\sqrt{
		4J^2+\mu^2-\omega^2+\sqrt{
			(4J^2+\mu^2-\omega^2)^2-4\mu^2\omega^2
}}}{2\pi J^2}.
\label{eq:SYK_q2rhom}
\end{equation}

\begin{figure}[t]
	\centering
	\includegraphics[width=\textwidth]{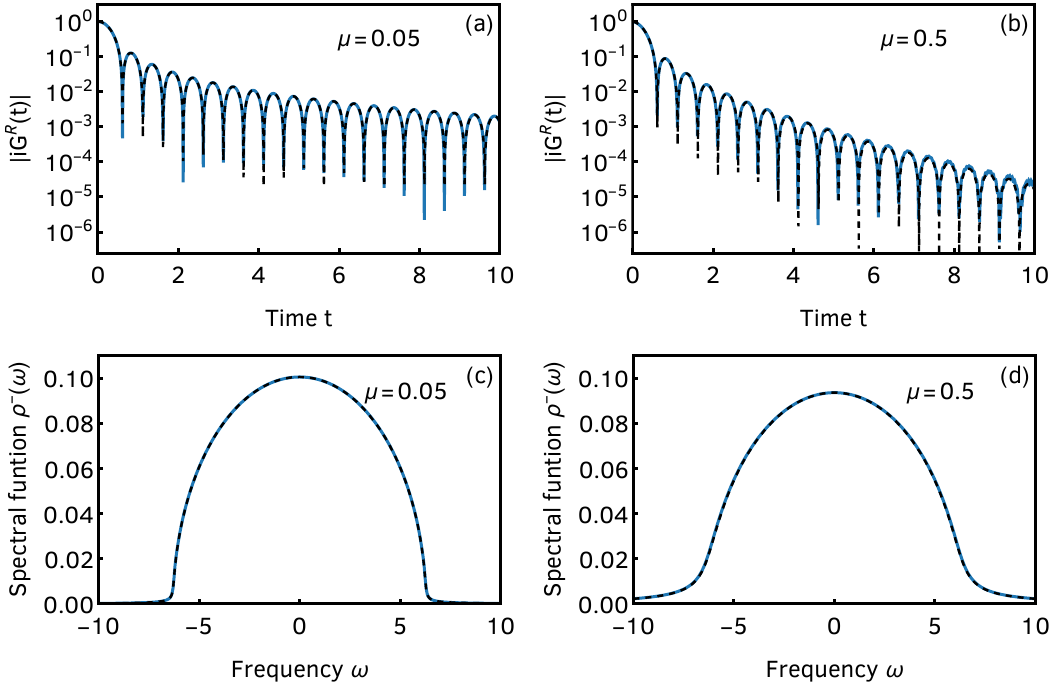} 
	\caption{Comparison of the numerical and analytical solutions of the $q=2$ SD equations with linear jump operators and $J=\pi$. Top row: Retarded Green's function as a function of time, obtained numerically (blue full curve) using the procedure of Sec.~\ref{sec:SYK_method} and analytically (dashed black curve) in Eq.~(\ref{eq:SYK_q2GR_time}). Bottom row: spectral function as a function of frequency. The analytic prediction is that of Eq.~(\ref{eq:SYK_q2rhom}). We find excellent agreement in all cases. The oscillations persist for large $\mu$ and the gap closes for small $\mu$, which is an indication that the relaxation is always dissipation-driven.
	}\label{fig:gllq2com}
\end{figure}

We conclude that an exponential decay with oscillatory corrections persists as in the $q = 4$ case, but, in stark contrast to it, the oscillations do not vanish at a critical $\mu_c$, there is a power-law correction to the exponential decay and, crucially, the rate of the exponential decay is simply $\Delta=\mu$. Therefore, the gap vanishes for $\mu = 0$. Since we do not observe an anomalously large value for the gap, the approach to equilibrium is dissipation-driven for all values of $\mu$.
The reason is the integrable nature of the $q = 2$ SYK, which requires a coupling to the environment to reach a steady state. Therefore, in the absence of any coupling to the environment, (many-body) quantum chaos and strong interactions seem necessary conditions to reach a steady state.
Finally, we note that, for $q = 2$, the first excited state in the finite-$N$ spectrum is equal to $-\mu$.
So, unlike $q = 4$, the resulting spectral gap
is equal to the decay rate of the Green's function, $\Delta=\Delta_\mathrm{spec}=\mu$.

In Fig.~\ref{fig:gllq2com}, we compare the analytical solution discussed above with the numerical solution of the SD equations using the procedure of Sec.~\ref{sec:SYK_method}. They are indistinguishable without using any fitting parameters.

\section{Summary and outlook}

In summary, we studied the dissipative dynamics of the SYK model in the framework of the Lindbladian quantum master equation, for both linear and quadratic jump operators.
We found exponential relaxation to the infinite-temperature steady state (with possible oscillatory corrections) and  analytically computed the relaxation gap in the limit of strong dissipation.
We have also shown that the decay rate is finite even if the coupling to the environment is zero provided that the dynamics is quantum chaotic (for $q=4$), a result that signals an anomalously fast relaxation. Only in the limit of strong coupling to the bath, is the approach to the steady state controlled by the environment.

In contrast, for a dissipative integrable SYK model ($q=2$), the decay rate, which we have obtained analytically, is always dominated by the coupling to the environment. Indeed, contrarily to the quantum chaotic case, it vanishes if there is no coupling to the bath, showing no anomalous relaxation. Moreover, also unlike the quantum chaotic case, there are oscillatory corrections for all values of dissipation strength.

Our work paves the way for further analytical investigations of dissipative strongly correlated quantum matter, as many interesting questions remain unanswered. 
Our work can be used to study more general setups with non-Markovian dissipation by tuning the kernel $K(z,z')$. 
Moreover, going away from the scalings of Eq.~(\ref{eq:J_l_moments}) and considering $1/N$ corrections and non-Hermitian jump operators allows for nontrivial steady states. An analysis of the spectral and steady-state properties of general SYK Lindbladians, based on exact diagonalization along the lines of Refs.~\cite{sa2019JPhysA,sa2020PRB}, is a natural next step. 
Another interesting avenue of research is to investigate more general jump operators in order to characterize the conditions for the observation of a phase at weak or zero coupling to the environment, in which the decay rate is not controlled by the environment.

%% file: Thesis_Kraus.tex

\chapter{Random quantum channels: Spectral transitions and universal steady states}
\label{chapter:kraus}

As seen in previous chapters, the study of dissipation and decoherence in generic open quantum systems led to the investigation of spectral and steady-state properties of random Lindbladian dynamics. A natural question is, then, how realistic and universal those properties are. In this chapter, we address these issues by considering a different description of dissipative quantum systems, namely, the discrete-time Kraus map representation of completely positive quantum dynamics.
Through random matrix theory (RMT) techniques and numerical exact diagonalization, we study random Kraus maps, allowing for a varying dissipation strength. We find the spectrum of the random Kraus map to be either an annulus or a disk inside the unit circle in the complex plane, with a transition between the two cases taking place at a critical value of dissipation strength.
The eigenvalue distribution and the spectral transition are well described by an effective RMT model that we can solve exactly in the thermodynamic limit, by means of non-Hermitian RMT and quaternionic free probability. The steady state, on the contrary, is not affected by the spectral transition. It has, however, a perturbative crossover regime at small dissipation, inside which the steady state is characterized by uncorrelated eigenvalues. At large dissipation, the steady state is well described by a random Wishart matrix. The steady-state properties thus coincide with those already observed for random Lindbladian dynamics, indicating their universality.
Quite remarkably, the statistical properties of the local Kraus circuit are qualitatively the same as those of the nonlocal Kraus map, indicating that the latter, which is more tractable, already captures the realistic and universal physical properties of generic open quantum systems.

The rest of the chapter is organized as follows. In Sec.~\ref{sec:kraus_model} we define the model studied. Spectral and steady-state properties are separately analyzed in detail (both numerically and analytically) in Secs.~\ref{sec:kraus_spectral} and \ref{sec:kraus_steady}, respectively. We then study the same quantities for the extended 1D Kraus circuit in Sec.~\ref{sec:1d}.
Appendix~\ref{app:kraus_GinUECUE} presents a derivation of the spectral support and eigenvalue distribution of the RMT model used to analytically describe the 0D Kraus map.

This chapter is based on Ref.~\cite{sa2020PRB}.

\section{Parametric random Kraus operators}
\label{sec:kraus_model}

Consider an $N$-dimensional quantum system in some initial state $\rho_0$. The action of a quantum dynamical map $\Phi$ (with $k$ decay channels) leads to a new state $\rho_1=\Phi\left(\rho_0\right)=\sum_{\mu=1}^kK_\mu\rho_0 K_\mu^\dagger$, where the Kraus operators $K_\mu$ are subjected to the trace-preservation constraint $\sum_{\mu=1}^kK_\mu ^\dagger K_\mu=\mathbbm{1}$~\cite{nielsen2002,bengtsson2017}. The successive action of the quantum map, $t$ times, leads to the final state $\rho_t=\Phi\left(\rho_{t-1}\right)=\Phi^t\left(\rho_0\right)$. The superoperator $\Phi$ admits the matrix representation
\begin{equation}\label{eq:def_map}
\Phi=\sum_{\mu=1}^k K_\mu\otimes K_\mu^*.
\end{equation}

We parametrize the deviation of the Kraus map from unitarity through a parameter $p\in[0,1]$, such that $p=0$ corresponds to unitary evolution, while $p=1$ corresponds to the case of a structureless quantum map studied in Refs.~\cite{bruzda2009,bruzda2010}. To this end, we consider two types of Kraus operators $K_\mu$ (in total $k=1+d$):
\begin{equation}\label{eq:Kraus_def}
\begin{split}
&K_0=\sqrt{1-p}\,U,\qquad \text{with}\quad  U^\dagger U=\mathbbm{1},\\
&K_j=\sqrt{p}\,M_j,\qquad \text{with}\quad \sum_{j=1}^d M_j^\dagger M_j=\mathbbm{1}.
\end{split}
\end{equation}
Here, $U$ is an $N\times N$ unitary matrix, while the $d$ Kraus operators $M_j$ are constructed as truncations of enlarged $Nd\times Nd$ unitary matrices~\cite{zyczkowski2000JPhysA}, following Ref.~\cite{bruzda2009}: generate an $Nd\times Nd$ random unitary matrix $V$, formed by $d^2$ blocks $V_{ij}$, $i,j=1,\dots,d$, of dimension $N\times N$,
\begin{equation}\label{eq:blocksU}
V=
\begin{pmatrix}
V_{11} & V_{12} & \cdots & V_{1d} \\
V_{21} & V_{22} & \cdots & V_{2d} \\
\vdots & \vdots & \ddots & \vdots \\
V_{d1} & V_{d2} & \cdots & V_{dd}
\end{pmatrix},
\end{equation}
and take the $d$ Kraus operators to be the blocks of the first (block-) column, i.e., $M_j=V_{j1}$. The constraint $\sum_{j=1}^dM_j ^\dagger M_j=\mathbbm{1}$ is automatically satisfied because of the orthonormality of the columns of $V$.\footnote{
	More generally, we could take the $j$th Kraus operator as any (normalized) linear combination of (block-) columns of $V$, i.e., $M_j=\sum_{\alpha=1}^d\psi_\alpha V_{j\alpha}$, with $\sum_{\alpha=1}^d\abs{\psi_\alpha}^2=1$. Without loss of generality, we set $\psi_1=1$ and $\psi_\alpha=0$ for $\alpha\neq1$ from now on.}
By construction, the Kraus operators $K_\mu$, $\mu=0,\dots,d$, satisfy $\sum_\mu K_\mu^\dagger K_\mu=\mathbbm{1}$. In what follows, to construct a random Kraus map, we draw both $U$ and $V$ from the circular unitary ensemble (CUE)~\cite{dyson1962i,haake2013}.

In the next two sections, we consider the most general random quantum map, without imposing any spatial structure. The quantum system can be understood as $N$ sites on a fully connected graph (and hence interpreted either as a 0D or $\infty$D system). Quantum maps defined on a 1D lattice are addressed in Sec.~\ref{sec:1d}. The matrix representation of the 0D quantum dynamical map is thus given by
\begin{equation}\label{eq:0d_map}
\Phi=(1-p)\,U\otimes U^*+p\sum_{j=1}^dM_j\otimes M_j^*.
\end{equation}

\section{Spectral properties}
\label{sec:kraus_spectral}

\begin{figure}[t]
	\centering
	\includegraphics[width=\textwidth]{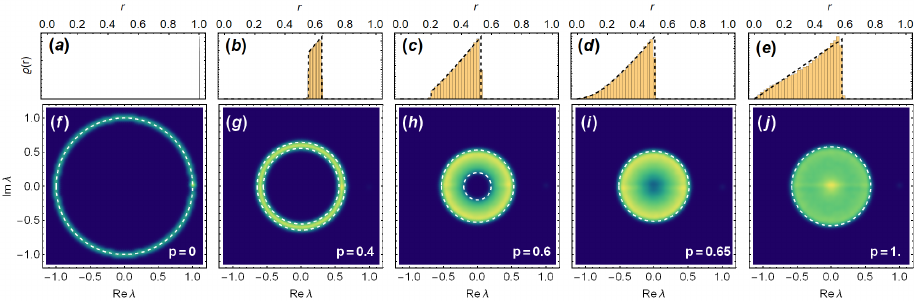}
	\caption{Global spectrum of the fully-connected quantum map, for different values of $p\in[0,1]$, $N=50$, and $d=3$. The eigenvalues are obtained from exact diagonalization of the map (\ref{eq:0d_map}). Ensemble averaging is performed such as to always obtain at least $10^5$ eigenvalues. (a)--(e): radial eigenvalue distribution (density); the dashed black line is given by the theoretical prediction of Eq.~(\ref{eq:pKraus_rhoR_RMT}). (f)--(j): eigenvalue density in the complex plane; the outer (inner) dashed line depicts the outer (inner) circular boundary of radius $R_+$ ($R_-$), given by Eq.~(\ref{eq:pKraus_rpm_analytic}).}
	\label{fig:pKraus_global_spectrum}
\end{figure}

\begin{figure}[t]
	\centering
	\includegraphics[width=0.6\columnwidth]{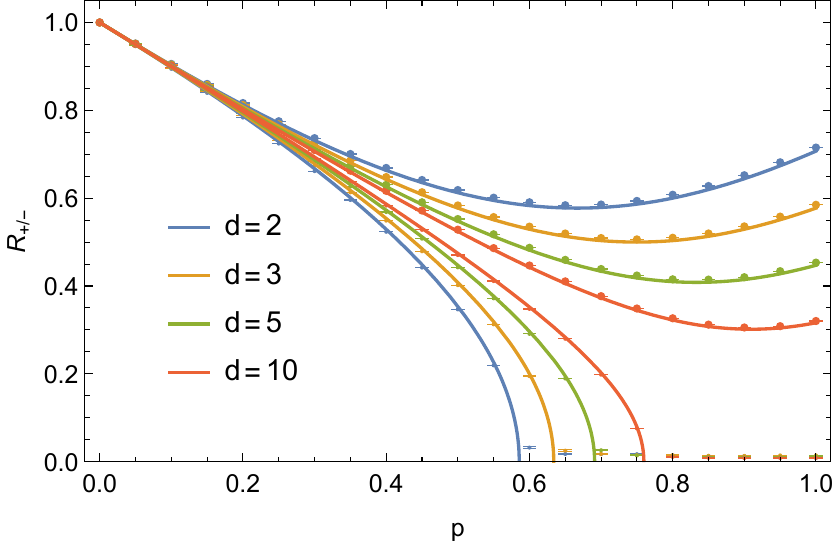}
	\caption{Inner and outer radius of the eigenvalue support of the fully-connected quantum map, as a function of $p$, for different $d$ and $N=60$. The points are the numerical results from exact diagonalization of $\Phi$ from Eq.~(\ref{eq:0d_map}), the solid lines give the analytical expressions from Eq.~(\ref{eq:pKraus_rpm_analytic}).}
	\label{fig:pKraus_rINOUT}
\end{figure}

\subsection{Numerical results}
The spectrum of $\Phi$ in the complex plane, obtained by exact diagonalization, is plotted in Fig.~\ref{fig:pKraus_global_spectrum}, for different values of $p$, $N=50$, and $d=3$. We note that the spectral distribution of the operator $\Phi$ is expected to be self-averaging when $N\to\infty$. We indeed observed that, for the system sizes considered, single realizations are already very close to the ensemble average that we plot in Fig.~\ref{fig:pKraus_global_spectrum}.

For $p=0$ all the spectral weight lies on the unit circle and for $p=1$ it uniformly covers the disk of radius $1/\sqrt{d}$. For intermediate values, it forms an annulus with well-defined inner and outer radius, $R_+$ and $R_-$, respectively. The annulus closes to a disk, i.e., $R_-=0$, at a finite value of $p_c<1$ depending only on $d$ and not on $N$, in the limit of large $N$. 
Using the ansatz explained in Sec.~\ref{subsec:RMT_model}, 
the value of $p_c$, and more generally the functions $R_\pm(p)$ can be computed analytically in the large-$N$ limit. 
Figure~\ref{fig:pKraus_rINOUT} shows the inner and outer radius of the eigenvalue distribution for several values of $d$ ($d=2$, $3$, $5$ and $10$).
With increasing $p$, the spectrum first contracts into a circle of radius smaller than $1/\sqrt{d}$, before spreading again and attaining a radius $1/\sqrt{d}$ at exactly $p=1$. 
The value of $p$ for which the minimum value is attained goes to $p=1$ as $d$ increases. 

Because of the rotational invariance of the spectrum, the spectral gap of $\Phi$, corresponding to the slowest decaying mode, is given by $\Delta = - \log\abs{R_+}$. 
The isolated eigenvalue at $\lambda = 1$ corresponds to the fixed point of $\Phi$, i.e., the steady state $\rho_{\text{SS}}$. This eigenvalue is not visible in the plots due to its vanishing spectral weight. 
For asymptotic long times, $\Delta$ is the rate at which a density matrix approaches the steady state $\rho_{\text{SS}}$ under repeated application of $\Phi$, i.e., $\rho_t - \rho_{\text{SS}} \propto e^{-\Delta t} $. 
Interestingly, $\Delta$ is a non-monotonous function of $p$, contradicting the naive expectation that a larger non-Hermitian component yields faster relaxation. 
The eigenmodes of $\Phi$ with the fastest decaying rates correspond to the smallest $\abs{\lambda}$. For $p>p_c$, a number of such modes instantaneously vanish after the first application of $\Phi$. On the contrary, for $p<p_c$ all modes have a finite lifetime.     

\subsection{RMT model for the quantum map}
\label{subsec:RMT_model}

The form of Eq.~(\ref{eq:0d_map}) makes it difficult to directly determine  the spectral density of $\Phi$. 
Here, in order to analytically characterize the eigenvalue distribution of this quantum map, we analyze instead a tractable effective RMT model given by
\begin{equation}\label{eq:Phieff}
\tilde \Phi =(1-p)\,\mathbb{U}+\frac{p}{\sqrt{d}}\mathbb{G},
\end{equation}
where $\mathbb{U}$ is an $N^2\times N^2$ Haar-random unitary matrix and $\mathbb{G}$ is an $N^2\times N^2$ random matrix drawn from the Ginibre Unitary Ensemble (GinUE)~\cite{ginibre1965,haake2013} with unit variance.\footnote{
	The effective model~(\ref{eq:Phieff}) does not account for the eigenvalue $1$ (steady state), thus describing only the annular/disk-shaped bulk of the spectrum.
} 
The effective model~(\ref{eq:Phieff}) becomes exact in the double limit $N,d\to\infty$ and can be justified as follows. 

We start with the unitary contribution to Eqs.~(\ref{eq:0d_map}) and (\ref{eq:Phieff}). To analytically compute the spectral properties of $\tilde \Phi$, only two properties of $\mathbb{U}$ are used: it is unitary and it has a flat spectrum (see the Appendix for details of the computation). Therefore, for the purpose of our calculations, we can approximate $U\otimes U^*$ by $\mathbb{U}$ if the former also possesses these two properties. Since $U\otimes U^*$ is trivially unitary, it only remains to show that it has a flat spectrum, at least in the large-$N$ limit.
Now, $U\otimes U^*$ has eigenvalues $\exp{i\varphi_{\alpha\beta}}\equiv\exp{i(\theta_\alpha-\theta_\beta)}$, $\alpha,\beta=1,\dots,N$, where $\exp{i\theta_\alpha}$ are the eigenvalues of $U$, with flat spectral density $\varrho_U(\theta)=1/(2\pi)$ on the unit circle. In the large-$N$ limit, the spectral density of $U\otimes U^*$ is (we denote the spectral density of $U\otimes U^*$ by $\varrho_\otimes(\varphi)$ and the two-point function of $U$ by $R_2(\theta_1,\theta_2)$):
\begin{equation}
\begin{split}\label{eq:rho_otimes}
\varrho_{\otimes}(\varphi)
&=\int\d\theta_1\d\theta_2\, R_2(\theta_1,\theta_2) \dirac{\varphi-(\theta_1-\theta_2)}
=2\pi R_2(\varphi)\\
&=\frac{1}{2\pi}\frac{1}{1-1/N}\left(
1-\frac{1}{N^2}\left(\frac{\sin(N\varphi/2)}{\sin(\varphi/2)}\right)^2
\right),
\end{split}
\end{equation}
where we have used the translational invariance of the CUE two-point function, 
\begin{equation}
\begin{split}
&R_2(\theta_1,\theta_2)
=R_2(\theta_1-\theta_2)\\
=&\frac{1}{N(N-1)}\left[
\left(\frac{N}{2\pi}\right)^2-
\left(\frac{1}{2\pi}\frac{\sin(N(\theta_1-\theta_2)/2)}{\sin((\theta_1-\theta_2)/2)}\right)^2
\right].
\end{split}
\end{equation}
In the large-$N$ limit, the second term inside the brackets in Eq.~(\ref{eq:rho_otimes}) converges to $(2\pi/N)\dirac{\varphi}$, and hence, in this limit, $\varrho_\otimes(\varphi)\to1/(2\pi)$, as claimed.

Regarding the dissipative term in Eq.~(\ref{eq:Phieff}), it was conjectured in Ref.~\cite{bruzda2009} that it can be approximated, for large $N$ and large $d$, by $(1/\sqrt{d})\,\mathbb{G}$.\footnote{
	Our results show that $d=3$ can already be considered as the large-$d$ limit if $N$ is the order of a few tens.}
Alternatively, one can note that each $M_j$ is a truncation of a Haar-random unitary and hence its entries are independent and identically distributed Gaussian random variables with zero mean and variance $\sigma_M^2/N$, where $\sigma_M^2=1/(2d)$~\cite{zyczkowski2000JPhysA}. ($M_j$ is supported on a disk of radius $1/\sqrt{d}$, but its eigenvalue density is not flat inside this disk; it is instead flat on the hyperbolic plane, whence there is an increase of the spectral density near the boundary of the disk.) Now, let $K_j=M_j\otimes M_j^*$. Then, the first moment of $K_j$ is 
\begin{equation}
\mu^K_1=
\frac{1}{N^2}\left\langle \Tr K_j \right\rangle =
\frac{1}{N^2}\left\langle \abs{\Tr M_j}^2 \right\rangle =
\frac{2\sigma_M^2}{N}=
\frac{1}{N d},
\end{equation}
which is zero in the large-$N$ limit. The second moment (equivalently the second cumulant) is 
\begin{equation}
\mu^K_2=
\frac{1}{N^{2}}\left\langle \Tr K_j^2-\langle\Tr K_j\rangle^2\right\rangle =
4\sigma_M^4=
\frac{1}{d^2}
\end{equation}
plus corrections which vanish when $N\to\infty$. So, we arrive at the claim that $K_j$ is represented by a random matrix whose entries are random variables with zero mean and variance $4\sigma_M^4/N^2$. By the central limit theorem of non-Hermitian matrices~\cite{nica2006,mingo2017}, taking the sum of $d$ such matrices (which are almost independent since the unitary constraints become less relevant as the dimensions of the matrices grow) results in a $N^2\times N^2$ (real) Ginibre matrix, whose matrix elements have variance $4d\sigma_M^4/N^2=(dN^2)^{-1}$, supported in a disk of radius $1/\sqrt{d}$. If the matrix elements of $K_j$ were normally distributed, then this result would also follow directly from free probability at arbitrary and finite $d$. Since the entries of $K_j$ are \emph{not} normally distributed, we have to resort to the $d\to\infty$ limit and then propose to extend the result to small $d$. This last step is justified \emph{a posteriori} by the remarkable agreement between numerical small-$d$ results and analytical large-$d$ predictions.

In App.~\ref{app:kraus_GinUECUE}, we study the general GinUE-CUE crossover ensemble of matrices of the form $\Phi=a\mathbb{G}+b\mathbb{U}$, $a,b\in\mathbb{R}$, and compute its spectral support and eigenvalue distribution. The results derived there can be readily used to model the quantum map by setting $a=p/\sqrt{d}$ and $b=(1-p)$. 

Since the spectrum of both GinUE and CUE matrices is isotropic in the large-$N$ limit, so is that of the quantum map. In perfect agreement with the numerical results of the previous section, see Fig.~\ref{fig:pKraus_rINOUT}, we find the spectrum to be supported on an annulus whose inner and outer radii are given by 
\begin{equation}\label{eq:pKraus_rpm_analytic}
R_\pm=\frac{1}{\sqrt{d}}\sqrt{(1-p)^2d\pm p^2}.
\end{equation}
The annulus-disk transition in the spectrum occurs at $R_-(p_c)=0$, i.e., $p_c=1/(1+1/\sqrt{d})$. For $p<p_c$, $R_-$ is no longer defined and the spectrum remains a disk (in which case, we conventionalize $R_-\equiv0$). The function $R_+(p)$ is not a monotonic function of $p$, its minimum, $(d+1)^{-1/2}$, being at $p_m=1/(1+1/d)\neq p_c$. The radial eigenvalue distribution, $\varrho(r)=2\pi r\varrho(r,\theta)$, $r\in[R_-,R_+]$, is given by,
\begin{equation}\label{eq:pKraus_rhoR_RMT}
\varrho(r)=2r\frac{d}{p^2}\left(1-\frac{(1-p)^2d}{\sqrt{p^4+4(1-p)^2d^2 r^2}}\right).
\end{equation}
The near-perfect agreement of the analytical radial distribution with the exact-diagonalization results can be seen in Fig.~\ref{fig:pKraus_global_spectrum}~(a)--(e). The small residual deviations are due to finite-$d$ effects (Fig.~\ref{fig:pKraus_global_spectrum} shows data with $d=3$).

\section{Steady-state properties}
\label{sec:kraus_steady}

The steady state $\rho_\mathrm{SS}$ is the (in general unique) fixed point of the quantum map $\Phi$, $\Phi(\rho_\mathrm{SS})=\rho_\mathrm{SS}$. 
The model described above supports nontrivial (i.e., non-fully mixed) steady states. 
We find that the steady-state properties are similar to those of a random Lindbladian with non-Hermitian jump operators, discussed in Ch.~\ref{chapter:randomLindblad}. 
This is an important result as it corroborates that the properties of $\rho_\mathrm{SS}$ of non-trivial generic quantum dynamical processes are solely determined by universality arguments.

In order to characterize the steady state, we consider the following measures:
\begin{enumerate}
	\item Steady-state spectrum (steady-state probability distribution, $P_\mathrm{SS}(\lambda) =(1/N)\Tr [ \delta(\lambda - \rho_\mathrm{SS} ) ] $).
	\item Rényi entropies. We consider the $n$th moment of the eigenvalue distribution through the $n$th Rényi entropy $S_n=S_n(\rho_{\rm SS})$:
	\begin{equation}
	S_n(\rho)=-\frac{1}{n-1}\log\left(\Tr\rho^n\right).
	\end{equation}
	In particular, the first Rényi entropy gives the von Neumann entropy $S_1=-\Tr\left(\rho_\mathrm{SS}\log\rho_\mathrm{SS}\right)$, while the second Rényi entropy is related to the purity of the steady state, $\mathcal{P}_\mathrm{SS}=\Tr\rho_\mathrm{SS}^2=e^{-S_2}$.
	\item Entanglement spectrum. We define an effective Hamiltonian $\mathcal{H}_\mathrm{SS}=-\log\rho_\mathrm{SS}$ and study its spectrum, instead of the spectrum of the steady state itself. Of particular interest are its spectral statistics (e.g., level spacing ratios), that can distinguish an ergodic steady state from a regular one.
\end{enumerate}

In the following, we first analyze the limiting cases of very large ($p=1$) and very small ($p\to0^+$) dissipation separately, exactly determining their steady-state spectral distributions. We then examine the crossover regime interpolating between these two limits, paying special attention to the purity and the correlations in the entanglement spectrum.

\begin{figure}[tp]
	\centering
	\includegraphics[width=\columnwidth]{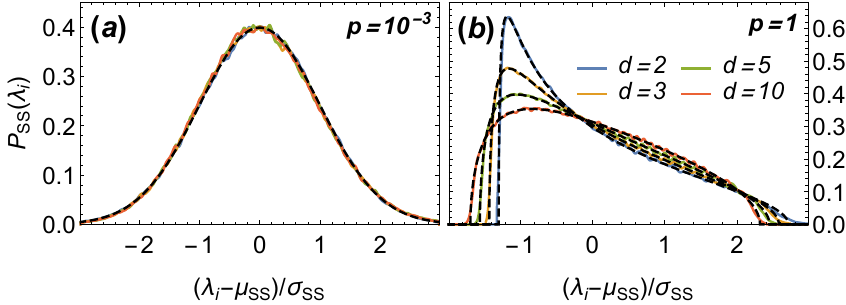}
	\caption{Steady-state eigenvalue distribution (eigenvalues centered at their mean, $\mu_\mathrm{SS}$, and rescaled by their standard-deviation, $\sigma_\mathrm{SS}$, at (a) small and (b) large dissipation, for $N=40$ and $d=2,3,5,10$. Coloured full lines correspond to a smoothed histogram of eigenvalues obtained numerically by exact diagonalization of 5000 steady-state density matrices. The (rescaled and recentered) distributions conform to a normal distribution $\mathcal{N}(0,1)$ at small dissipation and to the Marchenko-Pastur law~(\ref{eq:SS_MP}) at large dissipation (black dashed lines).}
	\label{fig:pKraus_SS_dist}
\end{figure}

\subsection{Large dissipation}

At large dissipation, $p\to1$, the steady-state distribution conforms to a Marchenko-Pastur distribution~\cite{marchenko1967}, see Fig.~\ref{fig:pKraus_SS_dist}~(b), in agreement with general results of the entanglement spectrum of random bipartite systems~\cite{zyczkowski2001JPhysA,sommers2004JPhysA,znidaric2006JPhysA,zyczkowski2011JPhysA,yang2015PRL} (where one takes the partial trace over all environment degrees of freedom, after an infinite-time evolution under joint system-environment unitary dynamics). More precisely, we have to consider a fixed-trace Wishart ensemble to account for probability conservation, $\Tr\rho_\mathrm{SS}=1$. Fixing $\Tr\rho_\mathrm{SS}=1$ leads to a rescaled Marchecko-Pastur probability distribution~\cite{forrester2010,nadal2011JStatPhys}, given by
\begin{equation}\label{eq:SS_MP}
P_\mathrm{SS}(\lambda)\big\rvert_{p=1}=P_\mathrm{MP}(\lambda;\lambda_\pm)=\frac{1}{2\pi\kappa}\frac{1}{\lambda}\sqrt{(\lambda_+-\lambda)(\lambda-\lambda_-)},
\end{equation}
where $\lambda_\pm=\kappa\left(\sqrt{d}\pm1\right)^2$. The only free parameter, $\kappa$, is fixed by the trace normalization ($\mu_\mathrm{SS}=1/N$), yielding $\kappa=1/(Nd)$, whence $\lambda_\pm=(1/N)(1\pm1/\sqrt{d})^2$. Higher (non-central) moments of the distribution are readily found to be~\cite{mingo2017} 
\begin{equation}\label{eq:moments_Wishart}
\Tilde{\mu}_n=\frac{\kappa^n}{n}\sum_{\ell=1}^n \binom{n}{\ell-1}\binom{n}{\ell}d^\ell.
\end{equation}
Of particular importance for what follows is the variance $\sigma^2_\mathrm{MP}\equiv\Tilde{\mu}_2-1/N^2=1/(N^2d)$.

\subsection{Small dissipation}

At small dissipation, $p\to0^+$, the steady-state eigenvalues are \emph{not} correlated, having instead independent Gaussian distributions, see Fig.~\ref{fig:pKraus_SS_dist}~(a). To show this, we study the steady state perturbatively. At exactly $p=0$, there is an $N$-fold degeneracy of the unit eigenvalue, which gets lifted by any amount of non-unitarity. Nonetheless, we expect the sector of degenerate eigenstates to completely determine the steady-state properties as long as the shift in the eigenvalues due to dissipation is smaller than the typical eigenvalue spacing at unitarity, i.e., for $N p\lesssim 2\pi$. We, therefore, expect a perturbative crossover regime (on the scale $1/N$) towards the Marchenko-Pastur regime. The crossover regime is thus highly suppressed in the thermodynamic limit $N\to\infty$. 

Let $U$ be diagonal in the basis $\{\ket{\alpha}\}$, $\alpha=1,\dots,N$, such that $U\ket{\alpha}=\exp{i\theta_\alpha}\ket{\alpha}$. We evaluate the steady-state-defining equality $\rho_\mathrm{SS}=\Phi(\rho_\mathrm{SS})$ in this basis:
\begin{equation}
\begin{split}
\rho_{\alpha\beta}=
&(1-p)\exp{i(\theta_\alpha-\theta_\beta)}\rho_{\alpha\beta}\\
&+p\sum_{j=1}^d\sum_{\gamma,\delta=1}^N\bra{\alpha}M_j\ket{\beta}\rho_{\gamma\delta}\bra{\delta}M_j^\dagger\ket{\beta}.
\end{split}
\end{equation}
From the preceding discussion, at very small $p$ we can restrict ourselves to the degenerate subspace (with zero phase), i.e., to diagonal elements of $\rho_\mathrm{SS}$. The constant-in-$p$ terms cancel and we obtain
\begin{equation}\label{eq:stochastic_eq}
\rho_{\alpha\alpha}=
\sum_{\gamma=1}^N\left(\sum_{j=1}^d\abs{(M_j)_{\alpha\gamma}}^2\right)\rho_{\gamma\gamma}
\equiv \sum_{\gamma=1}^N T_{\alpha\gamma}\rho_{\gamma\gamma}.
\end{equation}
This equation has the immediate interpretation of a classical probability equation: the diagonal elements of $\rho_\mathrm{SS}$ form the invariant probability measure of the random stochastic matrix $T$~\cite{zyczkowski2003JPhysA,horvat2009JSTAT,chafai2010}. That $T$ is a stochastic matrix, i.e.
\begin{subequations}\label{eq:stochastic_eq_conditions}
	\begin{align}
	\mathbb{R}\ni T_{\alpha\gamma}\geq0,\\
	\sum_{\alpha=1}^N T_{\alpha\gamma}=1,
	\end{align}
\end{subequations}
follows immediately from its definition in Eq.~(\ref{eq:stochastic_eq}) and from the orthonormality of rows and columns of $V$ (recall from Sec.~\ref{sec:kraus_model} that the $M_j$ are truncations of the unitary $V$). The distribution of the entries of $T$ can also be immediately inferred. Given that $(M_j)_{\alpha\gamma}=(V_{j1})_{\alpha\gamma}$ are entries of a $(Nd\times Nd)$ Haar-random unitary, which are known to be complex-normal distributed~\cite{pereyra1983}, the entries of $T$ are the sum of the squares of $2d$ real normal-distributed random variables with zero mean and variance $1/(2Nd)$. Therefore, $(2Nd)T_{\alpha\gamma}$ follows a $\chi^2$-distribution with $2d$ degrees of freedom. Note that the matrix elements $T_{\alpha\gamma}$ thus have mean $1/N$, as required from Eq.~(\ref{eq:stochastic_eq_conditions}).

By the Perron-Frobenius theorem~\cite{bengtsson2017}, the maximal eigenvalue of $T$ is $1$ and the corresponding eigenvector (the invariant probability measure, or the diagonal entries of $\rho_\mathrm{SS}$ in our case) has real non-negative entries.\footnote{
	The conditions for the Perron-Frobenius theorem to hold, namely that $M$ is irreducible and aperiodic are met almost surely because the entries of $M$ are, with probability one, nonzero (since they are the squares of the entries of a Haar-random unitary).
}

We now show that the steady-state probabilities of a random stochastic matrix are normally-distributed, following Ref.~\cite{horvat2009JSTAT}. 
We assume that, in the $N\to\infty$ limit, $\rho_{\alpha\alpha}$ and $T_{\alpha\gamma}$ become independent random variables. Then, for fixed $\alpha$ and $\gamma$, $T_{\alpha\gamma}\rho_{\gamma\gamma}$ (no sum over $\gamma$) has a product distribution. We denote the distributions of $\rho_{\alpha\alpha}$, $T_{\alpha\gamma}$, and $T_{\alpha\gamma}\rho_{\gamma\gamma}$ by $P_\rho$, $P_T$, and $P_{T\rho}$, respectively. The mean and variance of these distributions are $\mu_\rho$, $\sigma_\rho^2$, etc. The product-distribution moments satisfy $\mu_{T\rho}=\mu_T\mu_\rho$ and
\begin{equation}\label{eq:sigma_product}
\sigma_{T\rho}^2=\left(\sigma_T^2+\mu_T^2\right)\left(\sigma_\rho^2+\mu_\rho^2\right)-\mu_T^2\mu_\rho^2.
\end{equation}
Now, $\rho_{\alpha\alpha}$ is the sum of $N$ such independently distributed matrices [recall Eq.~(\ref{eq:stochastic_eq})] and, when $N\to\infty$, by the central limit theorem, it is normally-distributed, $\rho_{\alpha\alpha}\sim\mathcal{N}(\mu_\rho,\sigma_\rho)$, with $\mu_{\rho}=N\mu_{T\rho}$ and $\sigma^2_\rho=N\sigma^2_{T\rho}$. This procedure turned the stochastic steady-state equation into a self-consistent condition fixing $\sigma_\rho^2$: the Gaussian distribution of $\rho$ is completely determined by the first two moments of $P_{T\rho}$, which in turn depend only on the two lowest moments of $P_T$ and on $P_\rho$ itself. Substituting $\sigma^2_\rho=N\sigma^2_{T\rho}$ into Eq.~(\ref{eq:sigma_product}), we find
\begin{equation}\label{eq:sigma_rho}
\sigma^2_\mathrm{P}\equiv\sigma^2_\rho
=\frac{\sigma^2_T}{N}\frac{1}{1-N\sigma^2_T-\frac{1}{N}}
=\frac{1}{N^3d}\frac{1}{1-\frac{1}{N}\left(1+\frac{1}{d}\right)},
\end{equation}
where we used $\sigma^2_T=1/(N^2d)$ for the $\chi^2$-distributed random variable $T_{\alpha\gamma}$ and the subscript P distinguishes this perturbative variance from the Marchenko-Pastur variance, $\sigma^2_\mathrm{MP}$, obtained above.

The classical-probability-equation structure of the quantum dynamical equation resulting from perturbation theory in the degenerate subspace at small dissipation was already identified in Refs.~\cite{sa2019JPhysA,saMScThesis} (at the level of the continuous-time classical Markov generator) and used to study the spectral gap of a random Liouvillian; however, its steady-state properties were not investigated. From these results (see also the related Refs.~\cite{timm2009,horvat2009JSTAT}), we see that, at small deviations from unitarity, both spectral and steady-state properties are found to depend solely on the first two moments of a random matrix of small size (i.e., of order $N$ instead of $N^2$).

\begin{figure}[tp]
	\centering
	\includegraphics[width=\columnwidth]{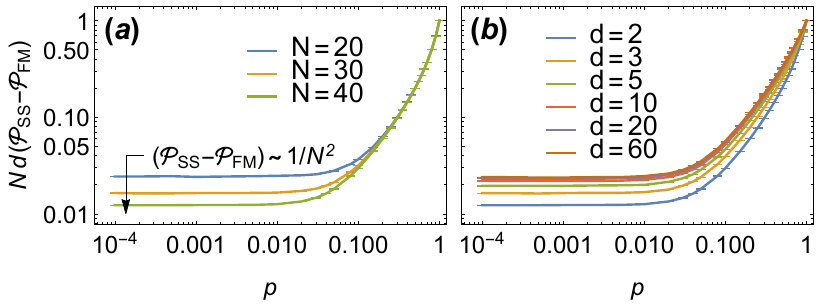}
	\caption{Difference between the steady-state purity and the purity of the fully-mixed state, as a function of $p$. Numerical results obtained by exact diagonalization of 5000 steady-state density matrices. (a): fixed $d=2$ and varying $N$. The purity difference scales as $1/N$ for large-enough $p$ and with $1/N^2$ in the small-$p$ perturbative regime. (b): fixed $N=40$ and varying $d$. For large $d$, the purity scales as $1/d$ across all dissipation regimes (the curves collapse to a single universal curve for all $p$).}
	\label{fig:pKraus_SS_purity}
\end{figure}

\subsection{Purity}

Figure~\ref{fig:pKraus_SS_purity} shows the difference $\mathcal{P}_\mathrm{SS}-\mathcal{P}_\mathrm{FM}$ between the steady-state purity, $\mathcal{P}_\mathrm{SS}$, and that of the fully mixed state, $\mathcal{P}_\mathrm{FM}=1/N$, as a function of $p$ (also note that $\mathcal{P}_\mathrm{SS}-\mathcal{P}_\mathrm{FM}=N\sigma_\mathrm{SS}^2$). By rescaling the purity difference by $1/N$, curves of different $N$ collapse to a universal curve in the large-$p$ limit. At exactly $p=1$, Eq.~(\ref{eq:moments_Wishart}), gives $\av{\mathcal{P}_\mathrm{SS}-\mathcal{P}_\mathrm{FM}}=N\sigma^2_\mathrm{MP}=\mathcal{P}_\mathrm{FM}/d$. This scaling holds for a finite range of $p$, but as $p$ is decreased, the individual curves depart from the universal curve and enter the perturbative crossover regime, characterized by a purity difference proportional to $1/N^2$, as follows from Eq.~(\ref{eq:sigma_rho}) in the large-$N$ limit. Thus, in the small-$p$ regime, the steady state can be considered fully mixed, since $\av{\mathcal{P}_\mathrm{SS}-\mathcal{P}_\mathrm{FM}}\ll \mathcal{P}_\mathrm{FM}$. 

In the thermodynamic limit, the universal Marchenko-Pastur curve covers the entire range of $p$, the fully-mixed state, $\mathcal{P}_\mathrm{SS}-\mathcal{P}_\mathrm{FM}=0$, being achieved only at exactly $p=0$. Also in the thermodynamic limit, the purity scales as $1/d$ for all dissipation strengths, at least for large enough $d$ (see Fig.~\ref{fig:pKraus_SS_purity}~(b)).

\begin{figure}[tp]
	\centering
	\includegraphics[width=\columnwidth]{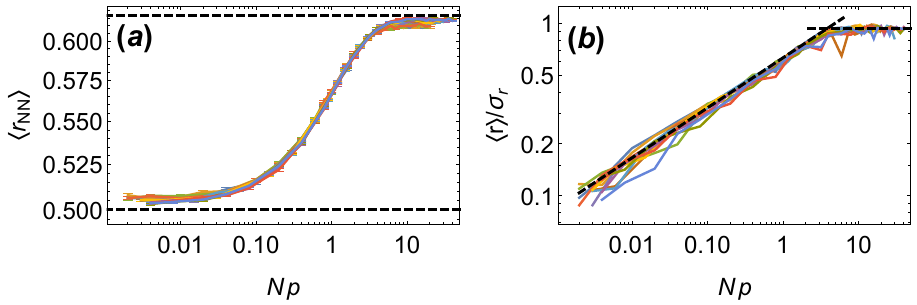}
	\caption{Level spacing statistics of the entanglement Hamiltonian as a function of $Np$. (a): average of the NN/NNN ratio. The dashed lines correspond to the theoretical values for Poisson statistics ($\av{\abs{r_\mathrm{NN/NNN}}}=1/2$, exact) and RMT statistics ($\av{\abs{r_\mathrm{NN/NNN}}}\approx 0.617$, approximate). (b): ratio of the first two moments of the distribution of the consecutive level spacing ratio, $\av{r}/\sigma_r$. The horizontal dashed line corresponds to the theoretical RMT value for the unitary symmetry class ($\av{r}/\sigma_r\approx0.928$), while the other dashed line gives the power-law decrease towards the Poisson limit ($\av{r}/\sigma_r=0$). The curves of different colors correspond to all the combinations of $N=30,40,50$ and $d=2,3,5,10$, collapsed to a single universal curve, clearly showing the crossover regime to scale as $p\sim1/N$.}
	\label{fig:pKraus_SS_cross}
\end{figure}

\subsection{Perturbative crossover}

The perturbative crossover is best seen in the spectral statistics of the entanglement Hamiltonian $\mathcal{H}_\mathrm{SS}$, which are captured by level-spacing statistics~\cite{guhr1998}. Furthermore, to automatically unfold the spectrum, we consider ratios of spacings. We denote the eigenvalues of $\mathcal{H}_\mathrm{SS}$ by $\varepsilon_i$. The nearest-neighbour to next-to-nearest-neighbour (NN/NNN) spacing ratio~\cite{sa2020PRX,srivastava2018} is defined by $r_\mathrm{NN/NNN}=(\varepsilon^\mathrm{NN}_i-\varepsilon_i)/(\varepsilon^\mathrm{NNN}_i-\varepsilon_i)$, where $\varepsilon^\mathrm{NN(NNN)}_i$ denotes the level nearest (next-to-nearest) to $\varepsilon_i$. In Fig.~\ref{fig:pKraus_SS_cross}~(a), we show the average value of $\abs{r_\mathrm{NN/NNN}}$ as a function of dissipation strength, which clearly distinguishes the Poisson statistics of uncorrelated levels ($\av{\abs{r_\mathrm{NN/NNN}}}=1/2$) from the random matrix statistics in the unitary class (a Wigner-like surmise gives an approximate value of $\av{\abs{r_\mathrm{NN/NNN}}}\approx0.617$~\cite{sa2020PRX}). As the size of the system increases, there is no sharp transition between both limits. Alternatively, we can consider the ratios of consecutive spacings~\cite{oganesyan2007PRB,atas2013,atas2013long}, $r_i=(\varepsilon_{i+1}-\varepsilon_i)/(\varepsilon_i-\varepsilon_{i-1})$. For uncorrelated levels, the distribution of $r$, which can be computed exactly, $P_r(r)=1/(1+r)^2$, has all moments but the first undefined (diverging), whence the ratio of the first two moments $\av{r}/\sigma_r$ is zero. On the other hand, for random matrix statistics, the latter is given by a finite value of order one (a Wigner-like surmise calculation gives the result $\av{r}/\sigma_r\approx0.928$~\cite{atas2013}). This is illustrated in Fig.~\ref{fig:pKraus_SS_cross}~(b): outside the perturbative regime, $Np\gtrsim2\pi$, $\av{r}/\sigma_r$ has the RMT value, and there is a power-law crossover to $\av{r}/\sigma_r=0$, which is strictly attained only for $Np=0$. The exponent of the power-law decay, $\av{r}/\sigma_r\sim (Np)^\alpha$, is numerically found to be $\alpha\approx0.3$.

From the discussion above, we conclude that no signature of the spectral transition, at finite $p$, is imprinted on the steady state. On the contrary, the steady-state properties are highly universal, and strongly resembling those of random Lindbladians once the different parametrizations of non-unitarity are taken into account.

\section{1D random Kraus circuits}
\label{sec:1d}

\begin{figure}[tp]
	\centering
	\includegraphics[width=\columnwidth]{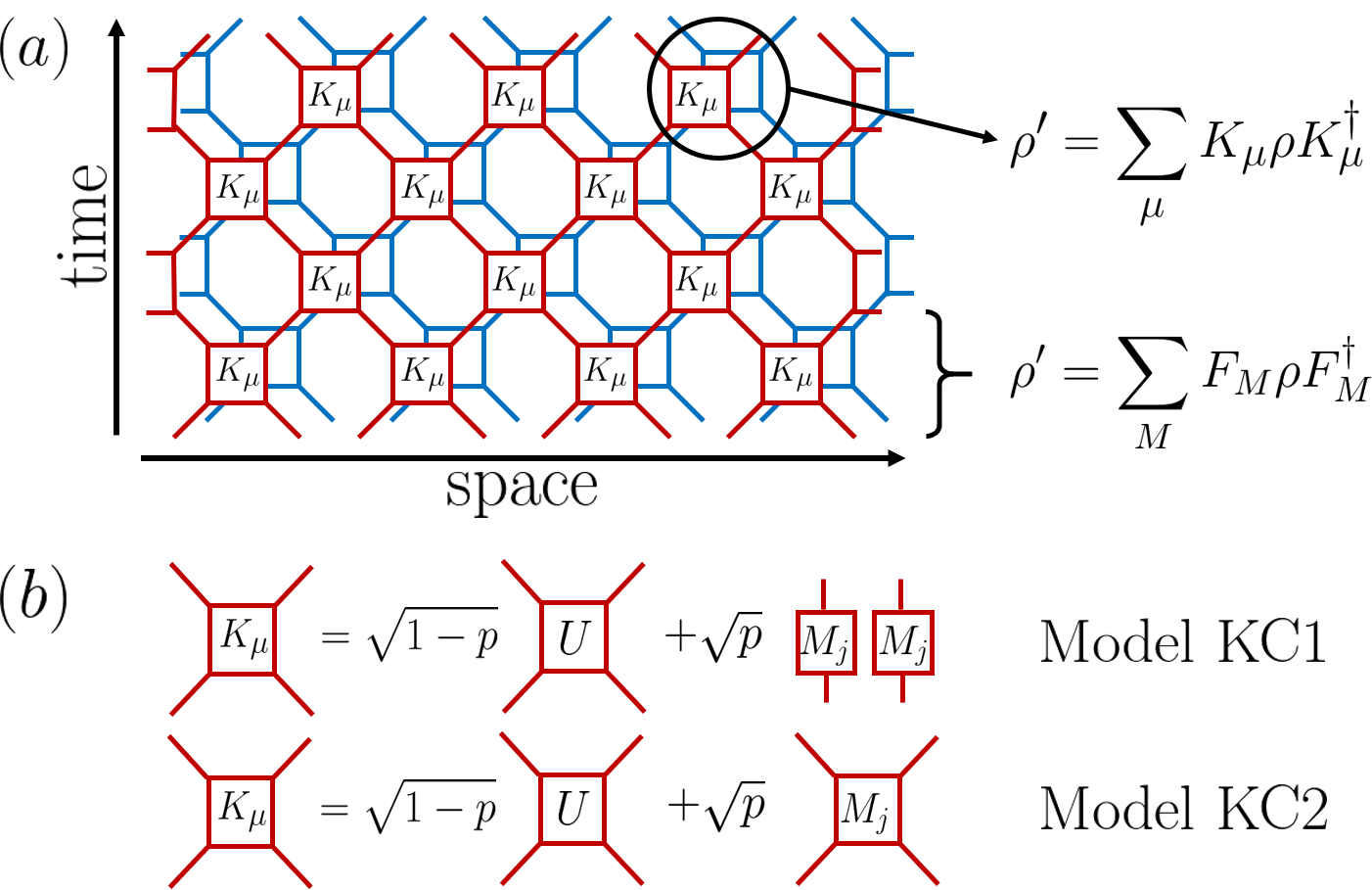}
	\caption{1D local Kraus circuit. (a): ``space-time'' structure of the quantum circuit. The red and blue layers represent the two copies of the system (i.e., one tensor-product factor each in the matrix representation). A local Hilbert space $\mathcal{H}_j$, of dimension $q$, lives on each wire connecting two Kraus operators $K_\mu$, represented by a four-legged square. A full row of the circuit is described by Kraus operators $F_M$, while one time-step of this Floquet Kraus circuit is given by two rows. Periodic boundary conditions are imposed. (b): each ``brick'' represents a two-body Kraus map with Kraus operators of the form of Eq.~(\ref{eq:Kraus_def}). We consider two models (both with two-body unitary dynamics) differing in whether the dissipative contribution is a genuine two-body Kraus operator (model KC2) or factors into one-body operators (KC1).}
	\label{fig:kraus_circuit}
\end{figure}

Next, we consider a one-dimensional system of $L$ qudits of dimension $q$, with local Hilbert space $\mathcal{H}_j=\mathbb{C}^{q}$. For convenience $L$, is taken to be even. 
The time-evolution superoperator corresponds to two rows of the Kraus \emph{circuit} schematically depicted in Fig.~\ref{fig:kraus_circuit}~(a). Note that the same local two-site Kraus map is applied everywhere along the space and time directions.
The composition of two subsequent rows (even and odd) yields one time-step of the Floquet Kraus dynamics. 
One row of the quantum circuit (half a time-step), $\Phi$, can be written in terms of global Kraus operators $F_M$, $\Phi(\rho)=\sum_M F_M\rho F_M^\dagger$, where $M=(\mu_1,\mu_2,\dots,\mu_{L/2})$ is a multi-index, which, in our model, factorize into local two-body Kraus operators $K_{\mu_j}$, of the form of Eq.~(\ref{eq:Kraus_def}), $F_M=K_{\mu_1}\otimes\cdots\otimes K_{\mu_{L/2}}$. The full one-time-step quantum map (two rows of the circuit) is correspondingly given by $\mathbb{T}\Phi\mathbb{T}^\dagger \Phi$, where we have introduced the one-site translation operator $\mathbb{T}$, defined by its action on the computational-basis states:
\begin{equation}
\mathbb{T}\ket{s_1,s_2,\dots,s_L}=\ket{s_L,s_1,\dots,s_{L-1}}, 
\end{equation}
with the quantum number $s_j$ being the spin-$q$ of each site.
We consider two models, both with two-site unitary dynamics, but differing in the number of sites on which the dissipative Kraus operators act: in model KC1, the dissipative two-body Kraus operator $M_j$ factorizes into two one-body Kraus operators $M_{j_1}\otimes M_{j_2}$, while in model KC2 $M_j$ is a genuine two-body operator, see Fig.~\ref{fig:kraus_circuit}~(b).

\begin{figure*}[tp]
	\centering
	\includegraphics[width=\textwidth]{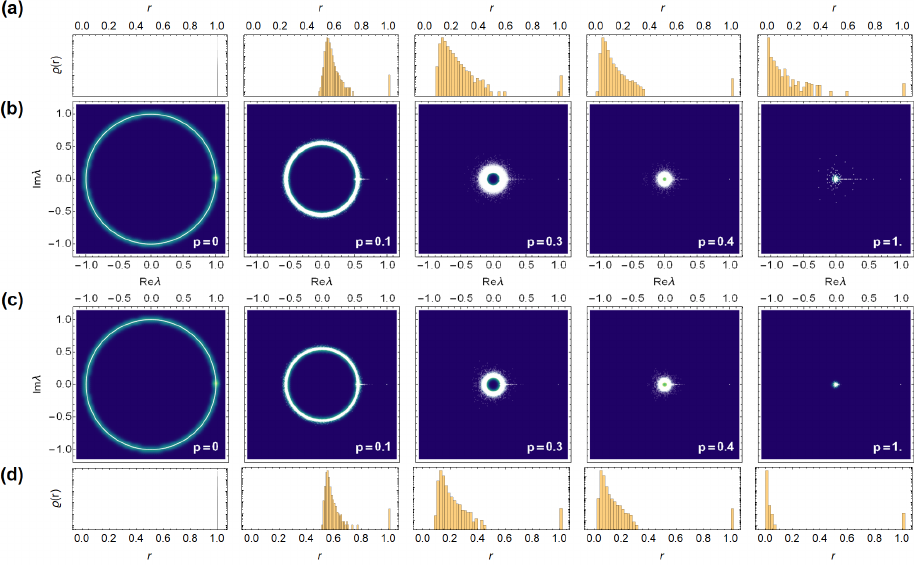}
	\caption{Spectral density of the one-dimensional Kraus circuit ($L=6$ and $q=2$), for different values of $p\in[0,1]$. (a) and (b): model KC1 with $d=3$; (c) and (d): model KC2 with $d=15$. (The radial histograms in (a) and (d) have a logarithmic scale.) The qualitative features (including annulus-disk spectral transition) of the spectra of Fig.~\ref{fig:pKraus_global_spectrum} are also present here. Note that while for the global map the spectrum has sharp and well-defined boundaries (inner and outer radii), this is no longer the case here, with several isolated eigenvalues lying outside a more diffusive boundary (a scatter plot of the individual eigenvalues is superimposed on the spectral density). Note also that the circuit is more contractive here because we introduce interactions by applying \emph{two} layers of quantum maps.}
	\label{fig:pKC_global_spectrum}
\end{figure*}

For the local Kraus circuit, we study the same quantities analyzed for the fully-connected quantum map. Quite remarkably, some of the properties of the local version are qualitatively similar to those of the global map. Furthermore, both models KC1 and KC2 also display essentially the same features. This indicates that the statistical properties of chaotic quantum maps are quite generic.

Figure~\ref{fig:pKC_global_spectrum} shows the spectrum of the quantum maps $\mathbb{T}\Phi\mathbb{T}^\dagger \Phi$ for $q=2$ and $L=6$, increasing $p$, and both KC1 ($d=3$) and KC2 ($d=15$). As before, the spectrum evolves from being supported on the unit circle, at $p=0$ to an annulus for $0<p<p_c$, and then undergoing a transition at $p=p_c$ to a disk spectral support. While this qualitative behavior is exactly that of the fully-connected model, the effective RMT model discussed in Sec.~\ref{subsec:RMT_model} does not quantitatively describe the spectral density, because the map is built from tensoring several maps of \emph{small} dimension. In particular, while the boundaries of the support were very sharp before, there are now several isolated eigenvalues lying outside a diffuse boundary (this can be seen in Fig.~\ref{fig:pKC_global_spectrum}, where we have superimposed a scatter plot of the individual eigenvalues on the spectral density). Besides, the spectrum is not even approximately flat inside its support.

Another qualitative difference from the structureless random Kraus maps is the appearance of a (zero-measure) set of eigenvalues along the real positive axes. 
This feature can be understood as follows. General hermiticity-preserving operators (such as $\Phi$) can be brought to a real representation by a trivial similarity transformation. Therefore, the dissipative part of the map should be modelled by a real Ginibre matrix (drawn from the GinOE), instead of by a complex Ginibre matrix (drawn form the GinUE).\footnote{
	However, because the bulk distribution and the correlation statistics are the same for the GinUE and the GinOE and the former is considerably easier to work with, we conveniently modelled $\mathbb{G}$ by a GinUE matrix without much loss of accuracy.}
Now, the matrices from the GinOE have nonzero spectral weight on the real axis, but it is suppressed in the large-$N$ limit~\cite{kanzieper2005PRL,forrester2007PRL}. Therefore, for the 0D map, where the convergence to the large-$N$ limit is faster, the real spectral weight is strongly surpressed. In contrast, for the 1D circuit, spectral weight along the real line is still visible within the system sizes available to us. This is so because $\Phi$ is constructed by tensoring together many maps of small dimension---each with a finite real spectral weight---which attain the large-$N$ limit in a slower fashion. Finally, the spectral gap is determined by the largest of these real eigenvalues and remains finite in the thermodynamic limit for $p<0$.

\begin{figure}[t]
	\centering
	\includegraphics[width=\columnwidth]{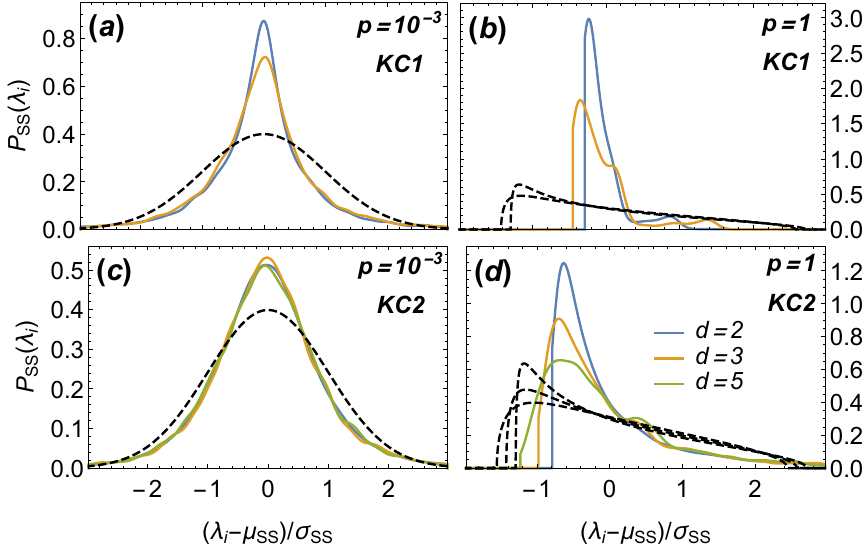}
	\caption{Steady-state eigenvalue distribution (eigenvalues centered at their mean, $\mu_\mathrm{SS}$, and rescaled by their standard-deviation, $\sigma_\mathrm{SS}$), for $L=6$ and $d=2,3,5$. We show results for model KC1 [KC2] at (a) [(c)] small and (b) [(d)] large dissipation. The qualitative features are the same as in Fig.~\ref{fig:pKraus_SS_dist} are still present, although an exact description in terms of Gaussian (a,c) or Marchenko-Pastur (b,d) distributions is no longer possible (see dashed lines).}
	\label{fig:pKC_SS_dist}
\end{figure}

We now turn to the steady state of the random Kraus circuit. In Fig.~\ref{fig:pKC_SS_dist} we show the spectrum of the steady state for small ($p=10^{-3}$) and large $(p=1)$ dissipation. As before, at small $p$, the steady-state spectrum has a symmetric peak around $1/N$, with a variance that is again found to scale as $1/N^3$. For large $p$, the distribution is no longer symmetric, there being a longer tail on the right, and the variance scales with $1/N^2$. 
Even though the main qualitative features of the 0D fully-connected model, it is important to remark that the steady-state spectrum is \emph{not} accurately described by a normal distribution (Marchenko-Pastur law), at small (large) dissipation, see Fig.~\ref{fig:pKC_SS_dist}. These deviations, most notably the bumps in the distribution at large $p$, should be attributed to the small dimension of the local Hilbert space we could access, hindering an exact description in terms of large-$N$ RMT. 

\section{Summary and outlook}
\label{sec:kraus_conclusion}

We have analyzed the spectral and steady-state properties of two kinds of stochastic Floquet non-unitary dynamics: a 0D global (or spatially structureless) Kraus map and a 1D circuit composed of local Kraus operators. 
By changing a single parameter $p\in[0,1]$, loosely regarded as the dissipation strength, the family of Kraus maps we consider interpolates between a random unitary operator, at $p=0$, and a generic quantum stochastic map with $d$ channels, at $p=1$.    

Although qualitatively similar, the 0D and 1D cases show some important differences.
For 0D, all the spectral weight (except for the eigenvalue one corresponding to the steady state) is supported either on an annulus, for $p<p_c$, or on a disk, for $p>p_c$. 
We determined the exact eigenvalue density using an ansatz that perfectly describes the numerical data.
The spectral gap, which determines the asymptotic decay rate to the steady state, coincides with the outer radius of the spectral support and shows a curious non-monotonic behavior as a function of $p$. 
For 1D, the spectrum depicts the same overall features at low (annulus) and large (disk) values of $p$.  
However, the support of the spectral weight is not sharply defined, rendering the annulus-ring transition difficult to determine numerically with the available system sizes. The spectral gap is determined by a set of eigenvalues of vanishing relative spectral weight laying along the real positive axis. 

Regarding the properties of the steady state, the 0D and 1D cases are again qualitatively very similar. For the 0D case, at low dissipation, the steady-state eigenvalue distribution is Gaussian, whereas it conforms to a Marchenko-Pastur law at large dissipation. These features are in correspondence with a Poissonian (small $p$) or GUE (large $p$) level-spacing distributions. 
The crossover between these two regimes is captured by a scaling function of $N p$. Therefore, in the thermodynamic limit, an infinitesimal amount of dissipation renders the spectrum of the steady-state Marchenko-Pastur distributed with GUE level-spacing statistics. 
The fact that the same behavior has previously been found for random Lindblad operators, see Ch.~\ref{chapter:randomLindblad} indicates that such steady-state features are to be generically expected for stochastic open quantum systems. 

Our analysis shows that, although some global features are similar, the unstructured or local nature of the evolution operators imprints characteristic signatures in the spectrum.
Whether these are present in more realistic physical models is an important question for further studies.  It would also be interesting to search for signatures of the annulus-ring transition on the dynamics. 

Finally, our results point to the existence of rather universal steady-state properties of stochastic Markovian dissipative models. Here, again, further work is needed to determine whether these are also present in realistic models of open quantum systems.

%% file: Thesis_Correlations.tex

\chapter{Symmetries, correlations, and universality in dissipative quantum chaos}
\label{chapter:correlations}

The interplay of symmetries, correlations, and dynamics lies at the heart of our understanding of complex interacting quantum many-body systems. It provides a compact and powerful framework for obtaining universal information not otherwise available for generic quantum systems.
Hamiltonians are classified by a reduced number of global antiunitary symmetries and unitary involutions. The behavior under time-reversal, particle-hole, and chiral symmetries places them in one of a reduced number of symmetry classes. In turn, the Bohigas-Giannoni-Schmit conjecture~\cite{bohigas1984} states that, if the system is chaotic, the Hamiltonian displays the statistical behavior of a random matrix from the same symmetry class. Finally, the correlations of random matrices are universal and solely determined by their symmetry class. Certain spectral observables can thus be inferred solely from the knowledge of invariance under simple symmetry transformations.

In this chapter, we discuss how these ideas are applied in the context of dissipative quantum chaos. We start, in Sec.~\ref{sec:correlations_symmclasses}, by giving a detailed derivation of the 54- and 38-fold classifications of non-Hermitian operators. Contrary to the Hermitian Altland-Zirnbauer (AZ), there are several distinct (but equivalent) ways to label symmetry classes. While our final classification result agrees with the seminal works~\cite{bernard2002,kawabata2019PRX}, our derivation follows a slightly different route and the labeling of classes is closer to the original AZ classification. After establishing the general symmetry classification, the sections after are devoted to obtaining signatures of the symmetries on random matrix correlations. We first discuss the consequences of antiunitary symmetries on the global spectrum of a non-Hermitian operator, see Sec.~\ref{sec:correlations_consequences}. We then consider short- and long-range, bulk and hard-edge, and eigenvalue- and eigenvector-based correlation functions, leading to a rich characterization of universal correlations in dissipative quantum chaos. In Sec.~\ref{sec:correlations_CSR} we address short-range bulk correlations using the complex spacing ratio distribution, which captures the three distinct degrees of level repulsion in a complex spectrum~\cite{hamazaki2020PRR}: A, AI$^\dagger$, and AII$^\dagger$. We then move to correlations at the hard edge of the spectrum induced by chiral symmetry, see Sec.~\ref{sec:correlations_hard_edge}, namely the distribution of the (normalized) eigenvalue closest to the origin. We provide numerical results for all classes but also find the exact analytic result for class AIII$^\dagger$. We note that the correlations on and near the real (or imaginary axis) have been conjectured to also display universality for random matrices and have been studied recently in Ref.~\cite{xiao2022PRR}. However, we will not consider them in this thesis. To study correlations over longer energy scales, we consider the dissipative spectral form factor (DSFF)~\cite{li2021PRL}, which is also sensitive to the three classes of level repulsion. In particular, we present the first (numerical) results for the DSFF in the non-Ginibre classes AI$^\dagger$ and AII$^\dagger$. Finally, in Sec.~\ref{sec:correlations_overlaps}, we show that the overlaps of eigenvector pairs related by symmetry serve as a detector of certain antiunitary symmetries. 
In subsequent chapters, we will obtain the symmetry classifications of several concrete models and use the spectral observables introduced in this chapter to study random matrix theory (RMT) universality in them.

This chapter is based on parts of Refs.~\cite{sa2020PRX,garcia2022PRX,garcia2023PRD,sa2023PRX}.

\section{Non-Hermitian symmetry classification: Bernard-LeClair classes}
\label{sec:correlations_symmclasses}

We saw in Ch.~\ref{chapter:introduction} that Hermitian Hamiltonians are classified by their antiunitary symmetries, either commuting time-reversal symmetry (TRS) or anticommuting particle-hole symmetry (PHS), both of which can square to either $\pm1$. In the absence of both, their unitary composition, anticommuting chiral symmetry (CS), can still act nontrivially, leading to a total of 10 Hermitian symmetry classes, the Altland-Zirnbauer classes~\cite{altland1997}.

Similarly, the non-Hermitian symmetry classification follows from the behavior of the irreducible blocks of a non-Hermitian operator under involutive antiunitary symmetries. More precisely, if there is a unitary $\scU$ that commutes with the Hamiltonian $H$,
\begin{equation}
\scU H\scU^{-1}=H,
\end{equation}
we can block diagonalize (reduce) $H$ into blocks (sectors) of fixed eigenvalue of $\scU$. Since different sectors are independent, we consider a single one. Inside this block, the unitary symmetry $\scU$ has a fixed eigenvalue and, therefore, acts trivially as the identity (up to a phase). If no further unitary symmetries exist, the block is irreducible. 

We look for the existence of antiunitary operators $\scT_\pm$, such that $H$ satisfies
\begin{alignat}{99}
	\label{eq:nHsym_Tp_new}
	&\scT_+ H \scT_+^{-1} = +H,\qquad 
	&&\scT_+^2=\pm 1,
	\\
	\label{eq:nHsym_Tm_new}
	&\scT_- H \scT_-^{-1} = -H,\qquad 
	&&\scT_-^2=\pm 1.
\end{alignat}
Since $H$ is non-Hermitian, it can also be related to its adjoint $H^\dagger\neq H$ through antiunitary operators. To this end, we look for the existence of antiunitaries $\scC_\pm$ implementing:
\begin{alignat}{99}
	\label{eq:nHsym_Cp_new}
	&\scC_+ H^\dagger \scC_+^{-1} = +H,\qquad 
	&&\scC_+^2=\pm 1,
	\\
	\label{eq:nHsym_Cm_new}
	&\scC_- H^\dagger\scC_-^{-1} = -H,\qquad 
	&&\scC_-^2=\pm 1.
\end{alignat}
We do not consider the existence of more than one antiunitary of a given kind, since their product is unitary and commutes with $H$ and we assumed $H$ to be irreducible. 
The combined action of antiunitaries of different types gives rise to unitary involutions: the composition of $\scT_+$ and $\scT_-$ (or $\scC_+$ and $\scC_-$) is a unitary transformation that anticommutes with $H$ (chiral symmetry); the composition of $\scT_+$ and $\scC_+$ (or $\scT_- $ and $\scC_-$) is a unitary similarity transformation between $H$ and $H^\dagger$ (pseudo-Hermiticity); and the composition of $\scT_+$ and $\scC_-$ (or $\scT_-$ and $\scC_+$) unitarily maps $H$ to $-H^\dagger$ (anti-pseudo-Hermiticity).
In the absence of antiunitary symmetries, these unitary involutions can still act on their own and we look for unitary operators $\scP$ and $\scQ_\pm$, such that $H$ transforms as
\begin{alignat}{99}
\label{eq:nHsym_P_new}
&\scP H \scP^{-1} = -H,\qquad 
&&\scP^2=1,
\\
\label{eq:nHsym_Qp_new}
&\scQ_+ H^\dagger \scQ_+^{-1} =  +H,\qquad 
&&\scQ_+^2=1,
\\
\label{eq:nHsym_Qm_new}
&\scQ_- H^\dagger \scQ_-^{-1} = - H,\qquad 
&&\scQ_-^2=1.
\end{alignat}
The symmetries of Eqs.~(\ref{eq:nHsym_Tp_new})--(\ref{eq:nHsym_Qm_new}) describe two independent flavors of time-reversal ($\scT_+$ and $\scC_+$) and particle-hole ($\scT_-$ and $\scC_-$) symmetries, chiral or sublattice symmetry ($\scP$) and pseudo- and anti-pseudo-Hermiticity ($\scQ_+$ and $\scQ_-$). In the Bernard-LeClair (BL) classification scheme~\cite{bernard2002}, $\scT_\pm$ are referred to as K symmetries, $\scC_\pm$ as C symmetries, $\scP$ as P symmetry, and $\scQ_\pm$ as Q symmetries.
In the Hermitian case, the classification simplifies to the symmetries of Eqs.~(\ref{eq:nHsym_Tp_new}), (\ref{eq:nHsym_Cm_new}), and (\ref{eq:nHsym_P_new}), and leads to the tenfold classification of Altland and Zirnbauer~\cite{altland1997}.

To see why the antiunitary operators can only square to $\pm1$, we act on $H$ twice with $\scT_\pm$:
\begin{equation}
\label{eq:correl_squareT}
\scT_\pm^2 H (\scT_\pm^{2})^{-1} = H.
\end{equation}
Since $H$ is irreducible, the commuting unitary symmetry $\scT_\pm^2$ must be trivial, i.e., $\scT_\pm^2=e^{\i \theta}\id$. We have $\scT_\pm=\scT_\pm^{-1}\scT_\pm^2=\scT_\pm^2\scT_\pm^{-1}$, or, equivalently, $\scT_\pm^{-1} e^{\i\theta}=e^{\i\theta}\scT_\pm^{-1}$, which implies $e^{\i\theta}=e^{-\i\theta}=\pm1$. The same argument holds for $\scC_\pm$. 
The unitary involutions, on the other hand, can always be chosen to square to $+1$. To see this, say for $\scP^2$, note that the argument of Eq.~(\ref{eq:correl_squareT}) implies that $\scP^2=e^{\i\varphi}$. Because $\scP$ is unitary, we can simply redefine $\scP\to e^{-\i\varphi/2}\scP$ to absorb the phase and set $\scP^2=+1$.

In principle, besides the square of its symmetries, the labeling of a symmetry class can depend on their commutation relations. In the case of antiunitary symmetries, we can see that this information is already contained in their squares. To see this, we first note that we can have either none, one, two, or four antiunitary symmetries. There cannot be classes with three antiunitary symmetries, because their threefold composition gives another antiunitary symmetry of the missing type. We need, therefore, consider only the commutation relations of two or four symmetries. Starting with the case of two, we consider for concreteness the pair of antiunitaries to be $\scT_+$ and $\scC_+$. Now, since $H$ is irreducible, the two unitary involutions $\scC_+\scT_+$ and $\scT_+\scC_+$ cannot the independent and must, at most, differ by a phase, $\scC_+\scT_+=e^{\i\alpha}\scT_+\scC_+$. We can redefine $\scT_+\to e^{-\i\alpha/2}\scT_+$, which changes neither the action on $H$ nor the value of $\scT_+^2$, to fix  $\scC_+\scT_+=\scT_+\scC_+$. We conclude that any pair of antiunitary symmetries can be made to commute. Next, we consider the case of four antiunitary symmetries. Proceeding as before, without loss of generality, we can redefine $\scT_+$ and $\scC_-$ such that $\scT_+$ commutes with both $\scC_+$ and $\scC_-$, but still have $\scC_-\scC_+=e^{\i\gamma}\scC_+\scC_-$.\footnote{
	We have $\scC_+\scT_+=e^{\i\alpha}\scT_+\scC_+$, $\scT_+\scC_-=e^{\i\beta} \scC_-\scT_+$, and $\scC_-\scC_+=e^{\i\gamma} \scC_+\scC_-$. With the redefinitions $\scT_+\to e^{-\i\alpha/2}\scT_+$ and $\scC_-\to e^{-\i\beta/2}\scC_-$ we can fix $\scC_+\scT_+=\scT_+\scC_+$ and $\scT_+\scC_-=\scC_-\scT_+$. However, we cannot redefine $\scC_+$ to fix the commutation relation with $\scC_-$ without spoiling the one with $\scT_+$.
} The composition of the three gives $\scT_-=e^{\i\delta}\scT_+\scC_+\scC_-$ and the fourfold composition $\scT_-\scT_+\scC_+\scC_-$ is a phase, in agreement with the irreducibility of $H$. Taking the square, we find $\scT_-^2=e^{\i\gamma}\scT_+^2\scC_+^2\scC_-^2$, whence it follows that $e^{\i\gamma}=\pm1$ and the square $\scT_-$ contains the same information as the commutation relation of $\scC_+$ and $\scC_-$. Similarly, the commutation relations of $\scT_-$ with $\scC_\pm$ can be expressed in terms of $e^{\i \gamma}$. With a redefinition $\scT_-\to e^{\i\delta/2}\scT_-$, we can also fix $\scT_-\scT_+=\scT_+\scT_-$. In summary, we can always choose one of the four antiunitary symmetries to commute with the other three, while the remaining commutation relations are fixed by the squares of the symmetries.
Finally, in the absence of antiunitary symmetries, we consider unitary involutions. For the same reason as before, there are either one or three unitary involutions and we just need to check the commutation relations in the latter case. The threefold composition $\scP\scQ_+\scQ_-=e^{\i \chi}$ is again a phase. Consequently, we can express the commutation relations of $\scQ_-$ with $\scP$ and $\scQ_+$ in terms of the commutation relation of $\scQ_+$ and $\scP$, which is, therefore, the only information we must determine. Since the phases of $\scQ_+$ and $\scP$ are fixed upon setting $\scP^2=\scQ_+^2=+1$, the phase $e^{\i\chi}$ in the commutation relation $\scP\scQ_+=e^{\i\chi}\scQ_+\scP$ also labels a symmetry class. Writing $\scP=e^{\i\chi}\scQ_+\scP\scQ_+$, we have $1=\scP^2=e^{2\i\chi}$, whence $e^{\i\chi}=\pm1$ and $\scP$ and $\scQ$ either commute or anticommute.

\begin{table}[t!]
	\centering
	\caption{Non-Hermitian symmetry classes with antiunitary symmetries. For each class, we list the square of its antiunitary symmetries and its name under the Kawabata-Shiozaki-Ueda-Sato classification~\cite{kawabata2019PRX}. We have adopted a shorthand notation, where the commutation relations of $\scP$ symmetry are indicated with a subscript in the class name (e.g., class AI$_+$ is denoted AI + $\mathcal{S}_+$ in Ref.~\cite{kawabata2019PRX}). Some classes have no standard name and are instead named after their equivalent class under $H\sim \i H$ (see Table~\ref{tab:correlations_classes_equiv}). Moreover, many classes have two equivalent names (by regarding them to be of type AZ or AZ$^\dagger$ in the notation of Ref.~\cite{kawabata2019PRX}).}
	\label{tab:correlations_classes_anti}
	\begingroup
	\renewcommand*{\arraystretch}{1.0375}
	\begin{tabular}[t]{@{}Sl Sc Sc Sc Sc Sl@{}}
		\toprule
		\#  & $\scT_+^2$ & $\scC_-^2$ & $\scC_+^2$ & $\scT_-^2$ & Class                            \\ \midrule
		1  & ---        & ---        & ---        & ---         & A                                  \\
		2  & $+1$       & ---        & ---        & ---         & AI                                \\
		3  & $-1$       & ---        & ---        & ---         & AII                               \\
		4  & ---        & $+1$       & ---        & ---         & D                                 \\
		5  & ---        & $-1$       & ---        & ---         & C                                 \\
		6  & ---        & ---        & $+1$       & ---         & AI$^\dagger$                      \\
		7  & ---        & ---        & $-1$       & ---         & AII$^\dagger$                     \\
		8  & ---        & ---        & ---        & $+1$        & D$^\dagger$                       \\
		9  & ---        & ---        & ---        & $-1$        & C$^\dagger$                       \\
		10 & $+1$       & $+1$       & ---        & ---         & BDI                               \\
		11 & $+1$       & $-1$       & ---        & ---         & CI                                \\
		12 & $-1$       & $+1$       & ---        & ---         & DIII                              \\
		13 & $-1$       & $-1$       & ---        & ---         & CII                               \\
		14 & $+1$       & ---        & $+1$       & ---         & no name ($\equiv$ BDI$^\dagger$)  \\
		15 & $+1$       & ---        & $-1$       & ---         & no name ($\equiv$ DIII$^\dagger$) \\
		16 & $-1$       & ---        & $+1$       & ---         & no name ($\equiv$ CI$^\dagger$)   \\
		17 & $-1$       & ---        & $-1$       & ---         & no name ($\equiv$ CII$^\dagger$)  \\
		18 & $+1$       & ---        & ---        & $+1$        & AI$_+$ / D$^\dagger_+$            \\
		19 & $+1$       & ---        & ---        & $-1$        & AI$_-$ / C$^\dagger_-$            \\
		20 & $-1$       & ---        & ---        & $+1$        & AII$_-$ / D$^\dagger_-$           \\
		21 & $-1$       & ---        & ---        & $-1$        & AII$_+$ / C$^\dagger_+$           \\
		22 & ---        & $+1$       & $+1$       & ---         & D$_+$ / AI$^\dagger_+$            \\
		23 & ---        & $+1$       & $-1$       & ---         & D$_-$ / AII$^\dagger_-$           \\
		24 & ---        & $-1$       & $+1$       & ---         & C$_-$ / AI$^\dagger_-$            \\
		25 & ---        & $-1$       & $-1$       & ---         & C$_+$ / AII$^\dagger_+$           \\ \bottomrule
	\end{tabular}
	\endgroup
	\hspace{+2ex}
	\begin{tabular}[t]{@{}Sl Sc Sc Sc Sc Sl@{}}
		\toprule
		\#  & $\scT_+^2$ & $\scC_-^2$ & $\scC_+^2$ & $\scT_-^2$ & Class                             \\ \midrule
		26 & ---        & $+1$       & ---        & $+1$        & no name ($\equiv$ BDI)            \\
		27 & ---        & $+1$       & ---        & $-1$        & no name ($\equiv$ DIII)           \\
		28 & ---        & $-1$       & ---        & $+1$        & no name ($\equiv$ CI)             \\
		29 & ---        & $-1$       & ---        & $-1$        & no name ($\equiv$ CII)            \\
		30 & ---        & ---        & $+1$       & $+1$        & BDI$^\dagger$                     \\
		31 & ---        & ---        & $+1$       & $-1$        & CI$^\dagger$                      \\
		32 & ---        & ---        & $-1$       & $+1$        & DIII$^\dagger$                    \\
		33 & ---        & ---        & $-1$       & $-1$        & CII$^\dagger$                      \\
		34 & $+1$       & $+1$       & $+1$       & $+1$        & BDI$_{++}$ / BDI$^\dagger_{++}$   \\
		35 & $+1$       & $+1$       & $+1$       & $-1$        & BDI$_{-+}$ / CI$^\dagger_{+-}$   \\
		36 & $+1$       & $+1$       & $-1$       & $+1$        & BDI$_{+-}$ / DIII$^\dagger_{-+}$  \\
		37 & $+1$       & $+1$       & $-1$       & $-1$        & BDI$_{--}$ / CII$^\dagger_{--}$   \\
		38 & $+1$       & $-1$       & $+1$       & $+1$        & CI$_{+-}$ / BDI$^\dagger_{-+}$    \\
		39 & $+1$       & $-1$       & $+1$       & $-1$        & CI$_{--}$ / CI$^\dagger_{--}$     \\
		40 & $+1$       & $-1$       & $-1$       & $+1$        & CI$_{++}$ / DIII$^\dagger_{++}$   \\
		41 & $+1$       & $-1$       & $-1$       & $-1$        & CI$_{-+}$ / CII$^\dagger_{+-}$    \\
		42 & $-1$       & $+1$       & $+1$       & $+1$        & DIII$_{-+}$ / BDI$^\dagger_{+-}$  \\
		43 & $-1$       & $+1$       & $+1$       & $-1$        & DIII$_{++}$ / CI$^\dagger_{++}$   \\
		44 & $-1$       & $+1$       & $-1$       & $+1$        & DIII$_{--}$ / DIII$^\dagger_{--}$ \\
		45 & $-1$       & $+1$       & $-1$       & $-1$        & DIII$_{+-}$ / CII$^\dagger_{-+}$  \\
		46 & $-1$       & $-1$       & $+1$       & $+1$        & CII$_{--}$ / BDI$^\dagger_{--}$   \\
		47 & $-1$       & $-1$       & $+1$       & $-1$        & CII$_{+-}$ / CI$^\dagger_{-+}$    \\
		48 & $-1$       & $-1$       & $-1$       & $+1$        & CII$_{-+}$ / DIII$^\dagger_{+-}$  \\
		49 & $-1$       & $-1$       & $-1$       & $-1$        & CII$_{++}$ / CII$^\dagger_{++}$   \\ \bottomrule
	\end{tabular}
	\vspace{+10pt}
	\caption{Non-Hermitian symmetry classes without antiunitary symmetries but with unitary involutions. For each class, we list the square of its unitary involutions, the commutation relation of $\scP$ and $\scQ_+$ (denoted as $\scP\scQ_+=\epsilon_{\scP\!\scQ_+}\scQ_+\scP$), and its name under the Kawabata-Shiozaki-Ueda-Sato classification~\cite{kawabata2019PRX}. One class has no standard name and is instead named after its equivalent class under $H\sim \i H$ (see Table~\ref{tab:correlations_classes_equiv}).}
	\label{tab:correlations_classes_noanti}
	\vspace{+8pt}
	\begin{tabular}{@{}Sl Sc Sc Sc Sc Sl@{}}
		\toprule
		\#  & $\scP^2$ & $\scQ_+^2$ & $\scQ_-^2$ & $\epsilon_{\scP\!\scQ_+}$ & Class                   \\ \midrule
		50  & $+1$     & ---        & ---        & ---                       & AIII$^\dagger$          \\
		51  & ---      & $+1$       & ---        & ---                       & AIII                    \\
		52  & ---      & ---        & $+1$       & ---                       & no name ($\equiv$ AIII) \\
		53  & $+1$     & $+1$       & $+1$       & $+1$                      & AIII$_+$                \\
		54  & $+1$     & $+1$       & $+1$       & $-1$                      & AIII$_-$                \\ \bottomrule
	\end{tabular}
\end{table}

Having determined all the information that labels a non-Hermitian symmetry class, we are now in a position to tabulate all possible classes:
\begin{itemize}
	\item \textit{No symmetries.} One class. 
	\item \textit{One antiunitary symmetry.} We have four possible choices of symmetries, and each can square to $\pm1$. We have $4\times2=8$ classes.
	\item \textit{Two antiunitary symmetries.} We have $6$ different pairs of symmetries, and each squares to $\pm1$. This gives $6\times2^2=24$ classes.
	\item \textit{Four antiunitary symmetries.} We must take all symmetries, and since each squares to $\pm1$, we get $2^4=16$ classes.
	\item \textit{One unitary involution.} We have three possibilities to choose from, each with a fixed square, hence 3 classes.
	\item \textit{Three unitary involutions.} All three involutions have their square fixed, and we must only specify the commutation relation of $\scP$ and $\scQ_+$, yielding 2 classes.
\end{itemize}
In total, we get $1+8+24+16+3+2=54$ classes, in agreement with Refs.~\cite{liu2019PRB,ashida2020}. They are tabulated in Tables~\ref{tab:correlations_classes_anti} and \ref{tab:correlations_classes_noanti} (we follow the Kawabata-Shiozaki-Ueda-Sato nomenclature~\cite{kawabata2019PRX}). We note that different (equivalent) labelings of the classes are possible, see Refs.~\cite{bernard2002,magnea2008,kawabata2019PRX,zhou2019PRB,ashida2020}. For example, we could specify two antiunitary symmetries, their squares, and the commutation relations with chiral symmetry~\cite{kawabata2019PRX}, or one antiunitary symmetry, its square, and the commutation relations with chiral symmetry and pseudo-Hermiticity~\cite{zhou2019PRB}.

In certain contexts, it might be desirable to identify the matrices $H$ and $\i H$ as belonging to the same class, since they share the same eigenvectors and have only the spectrum rotated by $\pi/2$.\footnote{
	When a certain axis of symmetry has a special role, we might prefer to work with the 54 classes and not make this identification. This is the case of the real axis in the Lindbladian spectrum, see Ch.~\ref{chapter:classificationLindblad}. For a discussion of the difference between point-gap and line-gap spectra, see, e.g., Refs.~\cite{kawabata2019PRX,ashida2020}.
} The multiplication by $\i$ does not change the action of $\scC_\pm$ or $\scP$ on $H$, but interchanges the role of $\scT_+$ and $\scT_-$ and of $\scQ_+$ and $\scQ_-$. For instance, classes AI (whose only symmetry is $\scT_+^2=+1$) becomes the same as D$^\dagger$ (whose only symmetry is $\scT_-^2=+1$). In total, there are 16 such identifications, listed in Table~\ref{tab:correlations_classes_equiv}, which reduces the total number of classes from 54 to 38, in agreement with Ref.~\cite{kawabata2019PRX}.

\clearpage

\begin{table}[t]
	\centering
	\caption{Equivalence of symmetry classes under the identification $H\sim \i H$. The numbering of the classes follows Tables~\ref{tab:correlations_classes_anti} and \ref{tab:correlations_classes_noanti}. Since there are 16 equivalences, the number of classes is reduced from 54 to 38.}
	\vspace{+8pt}
	\label{tab:correlations_classes_equiv}
	\begin{tabular}{@{}ll@{}}
		\toprule
		Equivalence  & Classes                                                                \\ \midrule
		2 $\equiv$ 8   & AI $\equiv$ D$^\dagger$                                                  \\
		3 $\equiv$ 9   & AII $\equiv$ C$^\dagger$                                                 \\
		10 $\equiv$ 26 & BDI                                                                    \\
		11 $\equiv$ 28 & CI                                                                     \\
		12 $\equiv$ 26 & DIII                                                                   \\
		13 $\equiv$ 27 & CII                                                                    \\
		14 $\equiv$ 30 & BDI$^\dagger$                                                          \\
		15 $\equiv$ 32 & DIII$^\dagger$                                                         \\
		16 $\equiv$ 31 & CI$^\dagger$                                                           \\
		17 $\equiv$ 33 & CII$^\dagger$                                                           \\
		19 $\equiv$ 20 & AI $_-$ / C$_-^\dagger$ $\equiv$ AII$_-$ / D$^\dagger_-$                    \\
		35 $\equiv$ 42 & BDI $_{-+}$ / CI$_{+-}^\dagger\equiv$ DIII$_{-+}$ / BDI$^\dagger_{+-}$   \\
		37 $\equiv$ 44 & BDI $_{--}$ / CII$_{--}^\dagger$ $\equiv$ DIII$_{--}$ / DIII$^\dagger_{--}$ \\
		39 $\equiv$ 46 & CI $_{--}$ / CI$_{--}^\dagger$ $\equiv$ CII$_{--}$ / BDI$^\dagger_{--}$     \\
		41 $\equiv$ 48 & CI $_{-+}$ / CII$_{+-}^\dagger$ $\equiv$ CII$_{-+}$ / DIII$^\dagger_{+-}$   \\
		51 $\equiv$ 52 & AIII                                                                   \\ \bottomrule
	\end{tabular}
\end{table}

\section{Global consequences of antiunitary symmetries}
\label{sec:correlations_consequences}

\begin{figure}[tbp]
	\centering
	\includegraphics[width=0.7\columnwidth]{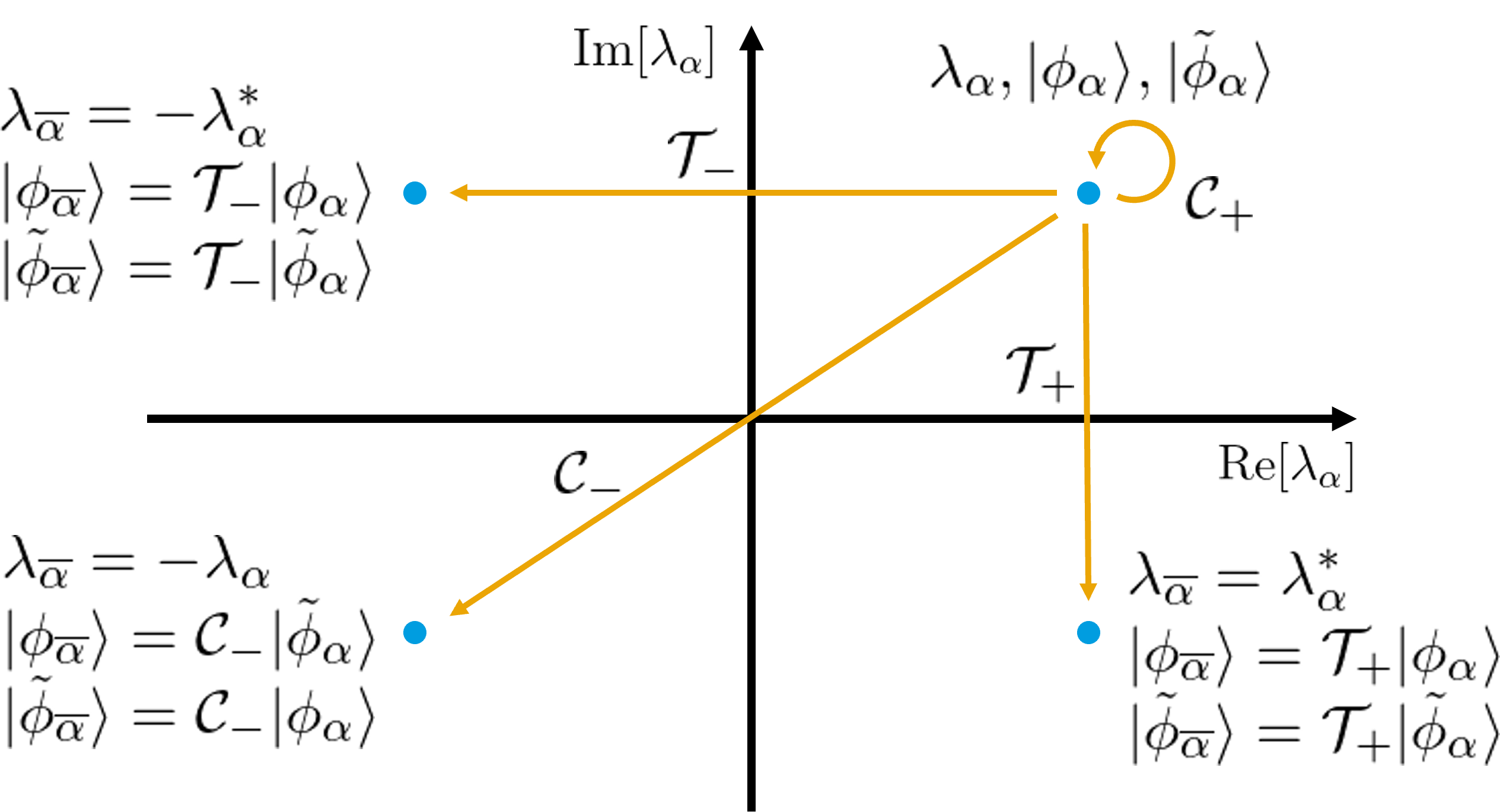}
	\caption{Schematic action of the antiunitary symmetries $\scT_\pm$ and $\scC_\pm$ on the eigenvalues and eigenvectors of $H$. The eigenvalue problem is defined in Eqs.~(\ref{eq:evL_new}) and (\ref{eq:evLd_new}) and we depict a representative eigenvalue $\lambda_\alpha$ in the complex plane, together with its image $\lambda_\balpha$ under the symmetries $\scT_\pm$ and $\scC_\pm$. $\scT_+$, $\scT_-$, and $\scC_-$ reflect the spectrum across the real axis, imaginary axis, and origin, respectively, while $\scC_+$ maps an eigenvalue to itself. $\scT_\pm$ map right eigenvectors to right eigenvectors (and left eigenvectors to left eigenvectors), while $\scC_\pm$ map left eigenvectors to right eigenvectors (and vice versa).}
	\label{fig:spectral_consequences}
\end{figure}

With the allowed classes of $H$ determined, we now consider the properties of each class, which are dictated by the antiunitary symmetries that label them. In this section, we discuss the global constraints that antiunitary symmetries $\scT_\pm$ and $\scC_\pm$ impose on the eigenvalues and eigenvectors of $H$~\cite{hamazaki2020PRR}, which are schematically summarized in Fig.~\ref{fig:spectral_consequences}. In particular, we show that $\scT_+$, $\scT_-$, and $\scC_-$ symmetries reflect the spectrum of $H$ across the real axis, imaginary axis, and origin, respectively. We also set the notation for the remainder of the chapter.

We denote the eigenvalues of $H$ by $\lambda_\alpha$ and the, in general distinct, right and left eigenvectors by $\ket{\phi_\alpha}$ and $|\tphi_\alpha\rangle$, respectively, i.e., 
\begin{align}
	\label{eq:evL_new}
	&H \ket{\phi_\alpha}=\lambda_\alpha \ket{\phi_\alpha},
	\\
	\label{eq:evLd_new}
	&H^\dagger |\tphi_\alpha\rangle=\lambda_\alpha^* |\tphi_\alpha\rangle.
\end{align}

Let us first consider the presence of a $\scT_+$ symmetry. Applying $\scT_+$ to Eq.~(\ref{eq:evL_new}) and using Eq.~(\ref{eq:nHsym_Tp_new}), we obtain
\begin{equation}
H \left(\scT_+\ket{\phi_\alpha}\right)=
\lambda_\alpha^* \left(\scT_+\ket{\phi_\alpha}\right),
\end{equation}
that is, $\scT_+\ket{\phi_\alpha}$ is also a right eigenvector of $H$ with complex-conjugated eigenvalue. Consequently, the spectrum of $\scL'$ is symmetric about the real axis. 

Proceeding similarly, we find that if there is a $\scT_-$ symmetry, $\scT_-\ket{\phi_\alpha}$ is a right eigenvector with eigenvalue $-\lambda_\alpha^*$; i.e., the spectrum is symmetric about the imaginary axis.

If there is an antiunitary symmetry $\scC_\pm$, then the operator implementing it connects left and right eigenvectors. Indeed, for $\scC_+$ we have
\begin{equation}
H^\dagger \left(\scC_+\ket{\phi_\alpha}\right)=
\lambda_\alpha^* \left(\scC_+\ket{\phi_\alpha}\right);
\end{equation}
that is, $\scC_+\ket{\phi_\alpha}$ is a left eigenvector of $H$ with the same eigenvalue as $\ket{\phi_\alpha}$. Accordingly, this symmetry does not affect the global shape of the spectrum and affects only local bulk correlations~\cite{hamazaki2020PRR}. Furthermore, if $\scC_+^2=-1$, then each eigenvalue is doubly degenerate, a phenomenon dubbed non-Hermitian Kramers degeneracy. 

Finally, if there is a $\scC_-$ symmetry, $\scC_-\ket{\phi_\alpha}$ is a left eigenvector of $H$ with eigenvalue $-\lambda_\alpha$; i.e., $\scC_-$ reflects the spectrum across the origin.

In the following sections, we discuss spectral observables that display RMT universality of the different classes. Therefore, it is desirable to have representative random matrices for each class, such that we can compute (numerically) the RMT predictions. For example, matrices from class A have no symmetry and, therefore, are arbitrary complex matrices, while matrices from class AI are asymmetric and real if we choose $\scT_+=\id$. The canonical choice is then to take all independent entries from a Gaussian distribution. We give one possible matrix realization for each RMT ensemble we use in the respective sections.

\section{Short-range bulk correlations: Complex spacing ratios}
\label{sec:correlations_CSR}

We now move to the random matrix signatures of the different antiunitary symmetries. First, we consider bulk local level statistics, which are sensitive to the value of $\scC_+^2$~\cite{hamazaki2020PRR} (we denote the absence of the symmetry as $\scC_+^2=0$). Local level statistics are most conveniently captured by the distribution of complex spacing ratios (CSRs)~\cite{sa2020PRX}, which, contrary to the bare complex spacing distribution~\cite{akemann2019} and the DSFF~\cite{li2021PRL}, do not require unfolding. CSRs have become a popular measure of dissipative quantum chaos, ranging from studies of random Lindbladians~\cite{wang2020PRL,tarnowski2021PRE,sa2022PRR} to nonunitary quantum circuits~\cite{sa2021PRB}, non-Hermitian Anderson transitions~\cite{huang2020PRB,luo2021PRL}, and two-color QCD~\cite{kanazawa2021PRD}, among others.
We define the CSR as~\cite{sa2020PRX}
\begin{equation}
\label{eq:corr_def_CSR}
z_\alpha=\frac{
	\lambda_\alpha^{\mathrm{NN}}-\lambda_\alpha}{
	\lambda_\alpha^{\mathrm{NNN}}-\lambda_\alpha
},
\end{equation}
where $\lambda_\alpha^{\mathrm{NN}}$ and $\lambda_\alpha^{\mathrm{NNN}}$ are the nearest and next-to-nearest neighbors of $\lambda_\alpha$ in the complex plane. By definition, $z_\alpha$ is constrained to the unit disk. Since they are defined in terms of the two nearest eigenvalues, CSRs only measure correlations up to a few level spacings. As a consequence, they can only be sensitive to the symmetry $\scC_+$. Indeed, the other three antiunitary symmetries correlate eigenvalues that are, in a many-body system, exponentially many level spacings apart (reflected across the real or imaginary axis, or the origin).

\begin{figure}[tbp]
	\centering
	\includegraphics[width=\columnwidth]{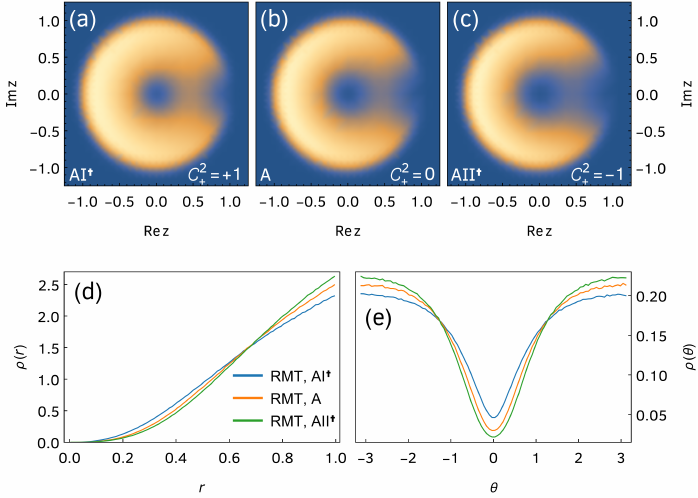}
	\caption{CSR distribution for random matrices in the three bulk classes A, AI$^\dagger$, and AII$^\dagger$. The distribution of $z$ in the complex plane (a)--(c) has a characteristic donut-like shape. The hole at the origin and the low probability at small angles $\theta=0$ are a sign of level repulsion and increase from class AI$^\dagger$ to A to AII$^\dagger$. We obtain these distributions numerically from exact diagonalization of an ensemble of $2^{15}\times2^{15}$ random matrices with $2^8$ realizations. We note that good analytical approximations exist for class A~\cite{sa2020PRX,dusa2022PRE}, but not for classes AI$^\dagger$ and AII$^\dagger$. A more quantitative comparison can be done by studying the marginal (d) radial and (e) angular distributions.}
	\label{fig:CSR_RMT}
\end{figure}

For random matrices, the CSR distribution acquires a characteristic donut-like shape, with the details of the distribution only depending on the value of $\scC_+^2$; see Figs.~\ref{fig:CSR_RMT}(a)--(c). The three types of level repulsion are usually denoted as A ($\scC_+^2=0$), AI$^\dagger$ ($\scC_+^2=+1$), and AII$^\dagger$ ($\scC_+^2=-1$)~\cite{hamazaki2020PRR}, as a reference to the simplest class that displays it.
Both the hole at $\abs{z}=0$ and the suppression of probability for small angles is a signature of level repulsion: not only do levels not like to be close to each other, they do not want to align and tend to spread out evenly in the complex plane. 
The CSR distribution was computed analytically for class A from a Wigner-like surmise in Ref.~\cite{sa2020PRX} and in the limit $N\to\infty$ in Ref.~\cite{dusa2022PRE}. In both cases, the expressions are very involved and we will resort to the numerical distributions instead.
On the other hand, for a set of uncorrelated random points in the complex plane (the equivalent of Poisson statistics) the numerator and denominator in Eq.~(\ref{eq:corr_def_CSR}) are independent and, hence, $z$ has a flat distribution in the unit disk. 

To make a more quantitative comparison with the RMT results, it is convenient to consider the marginal radial, $\rho(r)$, and angular, $\rho(\theta)$, distributions of the CSR expressed as $z_\alpha=r_\alpha \exp\{\i \theta_\alpha\}$. They are shown in Figs.~\ref{fig:CSR_RMT}(d) and (e) for the three bulk RMT ensembles. For a Poisson process on the plane, they are given by $\rho_\mathrm{Poi}(r)=2r$ and $\rho_\mathrm{Poi}(\theta)=1/(2\pi)$. For even a simpler measure, we compute the first moment of the marginal radial distribution $\rho(r)$, $\av{r}=\int \d r\, r \rho(r)$. The values of $\av{r}$ for the three universal bulk statistics (A, AI$^\dagger$, and AII$^\dagger$) are given in Table~\ref{tab:corr_numerics} (left). For a Poisson process, we have instead $\av{r}=2/3$.

\begin{table}[tbp]
	\caption{Universal single-number signatures of the non-Hermitian universality classes without reality conditions. The first radial moment~$\av{r}$ of the CSR distribution measures the bulk level repulsion of the three universal bulk classes A, AI$^\dagger$, and AII$^\dagger$, while the ratio $R_1$ in Eq.~(\ref{eq:r1}) gives the repulsion between the hard edge and the eigenvalue with smallest absolute value for the seven classes with spectral inversion symmetry. The values of $\av{r}$ are obtained by numerical exact diagonalization of $2^{15}\times2^{15}$ random matrices of the corresponding class, averaging over an ensemble of $2^8$ realizations. Meanwhile, the value of $\av{r}$ has been obtained analytically for class A~\cite{dusa2022PRE} in the thermodynamic limit. Its value is equal to $\av{r}=0.73866$, which is in agreement with our numerical value. In order to compute the ratio $R_1$ in Eq.~(\ref{eq:r1}), we numerically diagonalize $10^7$ $100\times100$ matrices of the corresponding universality class. Note that for the CSR distribution (and its moments), it was shown in Ref.~\cite{sa2020PRX} that they have large finite-size corrections for Gaussian-distributed random matrices; hence, very large matrices have to be considered to converge to the universal result of the thermodynamic limit. In contrast, we have verified numerically that the smallest eigenvalue distribution (and thus $R_1$) converges to a universal distribution very rapidly with the system size; therefore, using relatively small matrices is justified.}
	\label{tab:corr_numerics}
	\begingroup
	\renewcommand*{\arraystretch}{1.05}
	\begin{tabular}[t]{@{}l ccc@{}}
		\toprule
		Class    & A      & AI$^\dagger$  & AII$^\dagger$ \\ \midrule
		$\av{r}$ & 0.7384  & 0.7222         & 0.7486         \\ \bottomrule
	\end{tabular}
	\endgroup
	\quad
	\begin{tabular}[t]{@{}l ccccccc@{}}
		\toprule
		Class & AIII$^\dagger$ & D      & C      & AI$^\dagger_+$ & AII$^\dagger_+$ & AI$^\dagger_-$ & AII$^\dagger_-$ \\ \midrule
		$R_1$ & 1.129          & 1.228  & 1.102  & 1.222          & 1.096          & 1.123          & 1.138            \\ \bottomrule
	\end{tabular}
\end{table}

\section{Hard-edge correlations: Distribution of the eigenvalue closest to the origin}
\label{sec:correlations_hard_edge}

Through the use of CSRs, we can only distinguish three universality classes of bulk correlations. In addition, the classes with spectral inversion
symmetry (chiral or particle-hole) show universal repulsion from the spectral origin, the so-called hard edge for real spectra (cf.\ the discussion of Sec.~\ref{sec:intro_symmetries_AZ}). This universal behavior can be captured by zooming in on the eigenvalues closest to the origin, on a scale of up to a few level spacings, the so-called microscopic limit. 
For these additional correlations, we will restrict ourselves to the classes without reality conditions, i.e., without $\scT_\pm$ or $\scQ_\pm$ symmetries, for which the analysis is more complicated because of the existence of additional exactly real or imaginary eigenvalues. There are ten such classes (A, AI$^\dagger$, AII$^\dagger$, AIII$^\dagger$, AI$_+$, AII$_+$, C, D, AI$_-$, and AII$_-$, see Tables~\ref{tab:correlations_classes_anti}, \ref{tab:correlations_classes_noanti}, and \ref{tab:correlations_sym_class_noreal}).\footnote{This counting holds for both the 38- and 54-classifications, since there are no symmetries of type $\scT_\pm$ or $\scQ_\pm$ to identify.} This particular non-Hermitian tenfold way will appear in the classification of the non-Hermitian Sachdev-Ye-Kitaev model in Ch.~\ref{chapter:classificationSYK}.

{
	\setlength\cellspacetoplimit{0.5ex}
	\setlength\cellspacebottomlimit{0.5ex}
	\begin{table}[t!]
		\centering
		\caption{Non-Hermitian symmetry classes without reality conditions. For each class, we list its antiunitary and chiral symmetries, an explicit matrix realization~\cite{magnea2008}, its name under the Kawabata-Shiozaki-Ueda-Sato classification~\cite{kawabata2019PRX}, and the corresponding Hermitian ensemble. In the fourth column, we have adopted a shorthand notation, where the commutation relations of the $\scP$ symmetry are indicated with a subscript in the class name (class AI$_+$, say, is denoted AI + $\mathcal{S}_+$ in Ref.~\cite{kawabata2019PRX}). In the matrix realizations, $A$, $B$, $C$, and $D$ are arbitrary non-Hermitian matrices unless specified otherwise, and empty entries correspond to zeros. In the last column, we list the AZ classes~\cite{altland1997} that are in one-to-one correspondence with the non-Hermitian classes without reality conditions.
		}
		\label{tab:correlations_sym_class_noreal}
		\begin{tabular}{ Sc Sc Sc Sc Sl Sl@{}}
			\toprule
			$\sC_+^2$     & $\sC_-^2$  & $\Pi^2$ & Matrix realization & Class   & Hermitian corresp.     \\ \midrule
			$0$           & $0$        & $0$     & $A$                & A               & GUE (A)         \\
			$0$           & $0$        & $1$     & $\matAIIId$        & AIII$^\dagger$  & chGUE (AIII)    \\
			$+1$          & $0$        & $0$     & $A=A^\top$         & AI$^\dagger$    & GOE (AI)        \\
			$-1$          & $0$        & $0$     & $\matAIId$         & AII$^\dagger$   & GSE (AII)       \\
			$0$           & $+1$       & $0$     & $A=-A^\top$        & D               & BdG (D)       \\
			$0$           & $-1$       & $0$     & $\matC$            & C               & BdG (C)       \\
			$+1$          & $+1$       & $1$     & $\matAIdp$         & AI$^\dagger_+$  & chGOE (BDI)     \\
			$-1$          & $-1$       & $1$     & $\matAIIdp$        & AII$^\dagger_+$ & chGSE (CII)     \\
			$+1$          & $-1$       & $1$     & $\matAIdm$         & AI$^\dagger_-$  & BdG (CI)    \\
			$-1$          & $+1$       & $1$     & $\matAIIdm$        & AII$^\dagger_-$ & BdG (DIII)  \\ \bottomrule
		\end{tabular}
	\end{table}
}

The distribution of the eigenvalue with the smallest modulus, $P_1(\abs{E_1})$, gives, when combined with the bulk CSR distribution, a measure to uniquely distinguish the ten non-Hermitian symmetry classes without reality conditions. To avoid unfolding and carry out a parameter-free comparison with actual numerical data, we normalize the distribution to unity and rescale $|E_1|$ by its average. This quantity is the non-Hermitian counterpart of Eqs.~(\ref{eq:emin_CI})--(\ref{eq:emin_D}). Of the ten classes without reality conditions, three do not have a reflection symmetry across the origin and, hence, no well-defined universal $P_1(\abs{E_1})$. For the remaining seven classes, $P_1(\abs{E_1})$ is universal and can be computed numerically.
For class AIII$^\dagger$, the joint eigenvalue distribution is known~\cite{osborn2004PRL} and we can compute the distribution $P_1(|E_1|)$ analytically, see below.


A convenient way to capture the hard-edge universality by a single number is to consider the ratio (normalized variance)
\begin{equation}\label{eq:r1}
R_1=\frac{\av{\abs{E_1}^2}}{\av{\abs{E_1}}^2}=\frac{\int \d |E| \, |E|^2 P_1(|E|)}{\(\int \d |E| \, |E|\, P_1(|E|)\)^2},
\end{equation}
following the proposal by Sun and Ye for Hermitian random matrices~\cite{sun2020PRL}. The values of $R_1$ for the seven non-Hermitian classes with inversion symmetry listed in Table~\ref{tab:correlations_sym_class_noreal} are tabulated in Table~\ref{tab:corr_numerics} (right).

\subsection{Analytical results for class \texorpdfstring{AIII$^\dagger$}{AIIIdg}}

\begin{figure}[tbp]
	\centering
	\includegraphics[width=\textwidth]{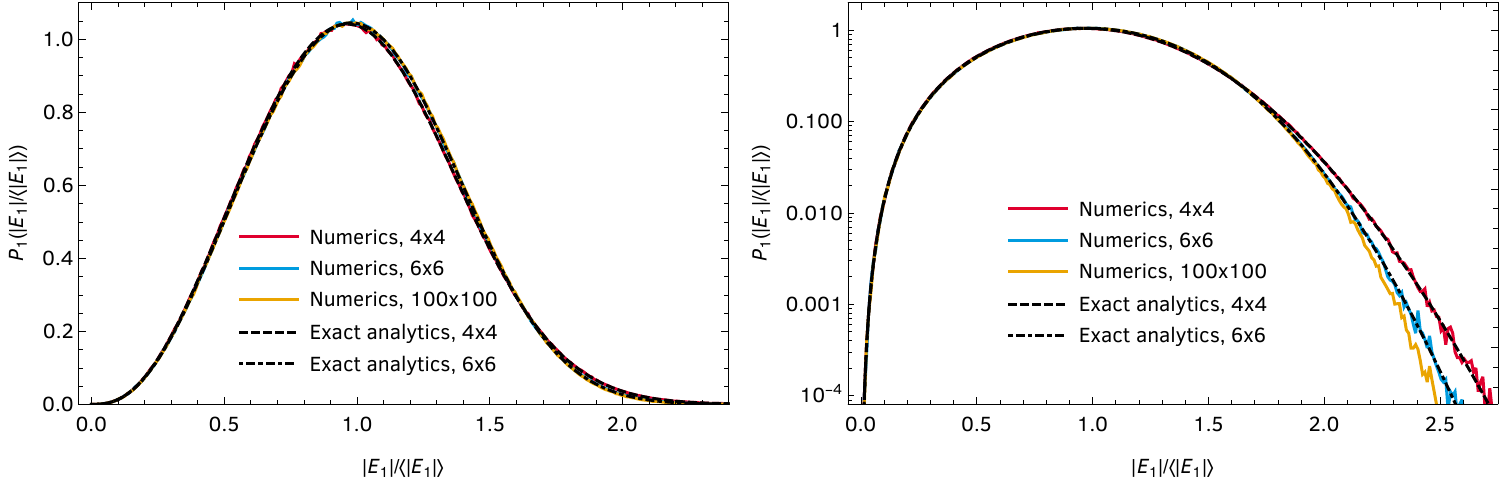}
	\caption{Distribution of the eigenvalues with smallest absolute value, $P_1(|E_1|/\av{|E_1|})$, for random matrices from class AIII$^\dagger$, on linear (left panel) and logarithmic (right panel) scales. The solid, colored curves are obtained from the exact diagonalization of $10^7$ random matrices of different sizes, while the dashed and dot-dashed black lines are the exact analytic results for $D=2$ [$4\times 4$ matrices, Eq.~(\ref{eq:AIIId_P1_lambda_1})] and $D=3$ ($6\times6$ matrices).}
	\label{fig:Emin_surmise_AIIId}
\end{figure}

We now derive the analytical prediction for the distribution of $\abs{E_1}$ in class AIII$^\dagger$. We consider a random matrix~$W$ from class AIII$^\dagger$ with square $D\times D$ off-diagonal blocks (recall Table~\ref{tab:correlations_sym_class_noreal}). The $2D$ eigenvalues of $W$ come in symmetric pairs and are denoted $E_1$, $-E_1$, $E_2$, $-E_2,\dots,E_D$, $-E_D$. We introduce $D$ new variables $z_j=E_j^2$ whose joint distribution is given by~\cite{osborn2004PRL}
\begin{equation}\label{eq:AIIId_jpdf}
P_\mathrm{joint}(z_1,\dots, z_D)\propto
\prod_{j=1}^D K_0(2D |z_j|) \prod_{1\leq j<k\leq D} \abs{z_j-z_k}^2.
\end{equation}
Because the distribution~(\ref{eq:AIIId_jpdf}) is invariant under permutations of eigenvalues $z_j$, we can choose $z_1$ as the eigenvalue with the smallest absolute value. Then, by construction, the distribution of $z_1$ is obtained by integrating out all the remaining eigenvalues outside the disk centered at the origin and with radius $\abs{z_1}$:
\begin{equation}\label{eq:AIIId_P1_z1}
P_1(|z_1|)\propto |z_1| K_0(2D|z_1|)
\int_0^{2\pi}\prod_{j=1}^D\d \phi_j \int_{\abs{z_1}}^{+\infty} 
\prod_{j=2}^D\d |z_j| |z_j|  K_0(2D|z_j|)
\prod_{1\leq j<k\leq D}
\abs{|z_j|e^{\i \phi_j}-|z_k|e^{\i \phi_k}}^2,
\end{equation}
where we denote $z_j=|z_j|e^{\i\phi_j}$ in polar coordinates. We proceed by evaluating the integrals in Eq.~(\ref{eq:AIIId_P1_z1}) for the smallest possible values of~$D$, i.e., $D=2,3,4,\dots$. In the spirit of the Wigner surmise for the spacing distribution, we expect the results to converge to the universal large-$D$ limit very quickly with $D$. Indeed, the $D=2$ result is already almost indistinguishable from large-$D$ numerical calculations on a linear scale, as we now show. Setting $D=2$ in Eq.~(\ref{eq:AIIId_P1_z1}), performing the angular integration of $\abs{|z_1|e^{\i\phi_1}-|z_2|e^{\i\phi_2}}^2$, and changing variables back to the chiral eigenvalue $|E_1|=\sqrt{|z_1|}$, we obtain
\begin{equation}\label{eq:AIIId_P1_lambda_1}
P_1(|E_1|)=
\sN \(c|E_1|\)^3 K_0\(\(c|E_1|\)^2\)
\(
\(c|E_1|\)^4 
\int_{\(c|E_1|\)^2}^{+\infty} \d x\, x K_0(2 x) +
\int_{\(c|E_1|\)^2}^{+\infty} \d x\, x^3 K_0(2 x)
\),
\end{equation}
where the normalization constant is equal to $\sN=32 c$. The arbitrary energy scale can be chosen as $c=7/9$ such that $\av{|E_1|}=\int \d |E_1| |E_1| P_1(|E_1|)=1$. The remaining integrals over Bessel functions could be expressed in closed form in terms of Bessel and Lommel functions, but their precise form is very complicated. The two integrals can easily be evaluated numerically to high accuracy. For small $|E_1|$, $P_1(|E_1|)$ behaves as $\sim |E_1|^3\log |E_1|$.
In Fig.~\ref{fig:Emin_surmise_AIIId}, we show how the $D = 2$ surmise (corresponding to $4\times4$ random matrices) compares with exact diagonalization results ($10^7$ disorder realizations), both in a linear (left panel) and logarithmic scale (right panel). We see perfect agreement with the $4\times4$ numerical result, as expected since we are performing an exact calculation. On a linear scale, it is also hard to distinguish it from the numerical results for large $D=50$ ($100\times100$ matrices), while deviations in the right tail can be noted on a logarithmic scale.

The procedure can be easily improved by considering $D = 3$ and $D = 4$. The resulting expressions are similar to Eq.~(\ref{eq:AIIId_P1_lambda_1}) but involve additional  integrals over Bessel functions and quickly become very cumbersome. In Fig.~\ref{fig:Emin_surmise_AIIId}, we compare the distribution computed for $D = 3$ (not written out) with the corresponding numerics, again seeing perfect agreement. Moreover, we see a fast convergence towards the large-$D$ result (e.g., $D=50$). For most practical purposes, the $D = 2$ result suffices.

\section{Long-range correlations: Dissipative spectral form factor}

So far, we have only considered local level correlations. For real spectra, some widely used long-range correlators are the number variance, the $\Delta_3$ statistic, and the spectral form factor (SFF)~\cite{mehta2004}. In this section, we study the DSFF, which was recently proposed~\cite{li2021PRL} as a measure of quantum chaos in dissipative systems, see also Refs.~\cite{shivam2023PRL,gosh2022PRB}. 

The DSFF is an observable that involves the entire spectrum so, in this case, unfolding (i.e., reparametrizing the eigenvalues such that the spectral density is constant in the new variables) is essential for making quantitative comparisons between different systems, especially when the spectral density is far from flat. For complex spectra, unfolding is ambiguous and we choose to unfold to constant density in the unit disk.\footnote{
	For example, we could have unfolded the eigenvalues $\lambda_k$ so that the density of ${\rm Re}(\lambda_k e^{\i\theta})$ (i.e., the projection of the eigenvalues along some axis with angle $\theta$) becomes constant. However, since analytical results are only available for constant density inside the unit disk we unfold the eigenvalues this way.
} For a radially symmetric (i.e., isotropic) spectrum, we only have to reparametrize the absolute value of the eigenvalues. 
When the eigenvalue density is not strictly isotropic, generically the spectrum is still locally isotropic. In that case, unfolding ambiguities can be resolved by requiring that local spectral isotropy is preserved. For the Ginibre ensembles, there is no need to unfold the spectrum because, well away from the spectral edge, the spectral density is constant and a rescaling to the unit disk is enough. 

Since our goal is to identify universal features of the quantum dynamics, we only consider connected two-point correlators, denoted by a subscript $c$. A SFF for complex eigenvalues was first
introduced in Ref.~\cite{fyodorov1997}:
\begin{equation}
\begin{split}
K_c(t,s_1,s_2)&=\frac 1D\left \langle \sum_{k,l=1}^D e^{\frac \i2 \lambda_k (t+s_1)+\frac \i2 \lambda_k^*(t-s_1)}
e^{-\frac \i 2 \lambda_l (t+s_2)-\frac \i 2 \lambda_l^*(t-s_2)} \right \rangle_c
\\
&=
\int \d x_1 \d x_2 \d y_1 \d y_2\, e^{\i(x_1-x_2)t + \i y_1 s_1 -\i y_2 s_2}
\varrho_{2c}(x_1+\i y_1,x_2+\i y_2).
\label{kfyod}
\end{split}
\end{equation}
where $\varrho_{2c}$ is the connected two-point correlation function, $\lambda_1 = x_1+\i y_1$, $\lambda_2 = x_2+\i y_2$, and $D$ is the number of eigenvalues in the spectrum.
For $s_1=s_2=0$, this becomes the SFF of the real parts of the eigenvalues.
More generally, setting $t=\tau \cos\theta$, $s_1=-\i \tau \sin\theta$, and $s_2=\i\tau\sin\theta$ in Eq.~\eref{kfyod}, we can define the SFF of the eigenvalues projected onto the direction defined by the angle $\theta$, dubbed the DSFF, as proposed in Ref.~\cite{li2021PRL}:
\begin{equation}
\begin{split}
K_c(\tau, \theta) 
&= \frac1D
\left \langle
\left| \sum_{k=1}^D e^{\i\tau {\rm Re}(e^{\i\theta} \lambda_k)} \right|^{2}
\right \rangle_{c}
\\
&=
\int \d^2 \lambda_1 \d^2 \lambda_2\ e^{\i\tau {\rm Re}[e^{\i\theta} (\lambda_1-\lambda_2)]}
\varrho_{2c}(\lambda_1,\lambda_2).
\label{form}
\end{split}
\end{equation}
If the spectral properties are axially symmetric, the DSFF does not depend on the angle $\theta$ onto which the eigenvalues are projected. By averaging over $\theta$ it is possible to increase the statistics of the
DSFF~\cite{li2021PRL}. Analytically, it is convenient to set $\theta=0$.

\begin{figure}[tbp]
	\centerline{\includegraphics[width=\textwidth]{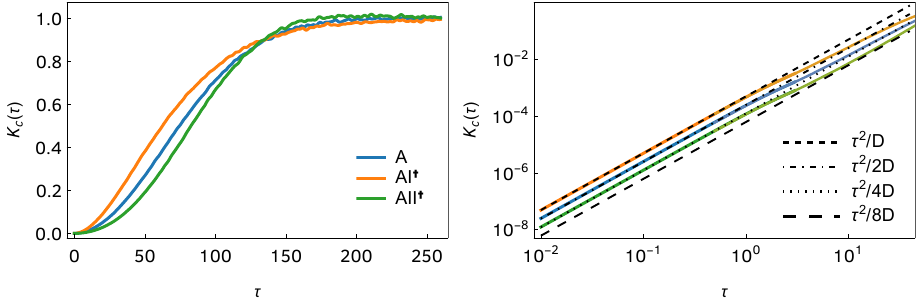}}
	\caption{The DSFF of non-Hermitian matrices from the three bulk universality classes, A, AI$^\dagger$, and AII$^\dagger$. The solid curves correspond to the numerically obtained results for $2048\times2048$ matrices drawn from the respective RMT ensemble, averaging over $10^4$ realizations. The right panel shows a magnification of the region close to the origin, where the spectral DSFF shows quadratic growth in time (dashed lines) with a coefficient that decreases by a factor of 2 going from the perturbative to the nonperturbative domain.}
	\label{fig:A_AId_AIId}
\end{figure}
 
The DSFF is a bulk observable, as it is defined as a sum over all eigenvalues, so it can identify only three different classes A, AI$^\dagger$, and AII$^\dagger$. As in the Hermitian case, where the SFF of the GOE, GUE, and GSE distinguishes the increasing degrees of level repulsion, so does the DSFF of classes AI$^\dagger$, A, and AII$^\dagger$. For long times (small energy differences), the DSFF thus contains the same information as the CSR distribution. However, by going to shorter times (larger energy differences) we can probe nonuniversal features of a complex quantum system.

In Ref.~\cite{fyodorov1997}, the DSFF was calculated for the weak non-Hermiticity limit of the complex elliptic Ginibre ensemble, corresponding to random matrices in class A, but, as stated by the authors, their calculations are also valid in the case of interest here, termed the strong non-Hermiticity limit. They found that, in the large-$D$ limit, and setting $\theta=0$ (since the spectrum is rotationally invariant), see also Ref.~\cite{li2021PRL}:
\be
K_c(\tau) =1 -e^{-\tau ^2/4D}.
\label{form-an}
\ee
The DSFF saturates to its asymptotic value $1$ at a time of the order of $\sqrt{D}$, i.e., the Heisenberg time is $t_\mathrm{H}\sim\sqrt{D}$, in contrast with the Hermitian case, where $t_\mathrm{H}\sim D$. For small $\tau$, the DSFF shows a quadratic growth. However, we note that the limits $D	\to\infty$ and $\tau\to0$ do not commute and hence there are \emph{two} different quadratic regimes, a perturbative and a nonperturbative one. While not appreciated in Ref.~\cite{li2021PRL}, this effect is visible in finite-size numerical experiments.

Before giving a more detailed discussion of these analytical results, we discuss the cases of the two other universality classes, AI$^\dagger$ and AII$^\dagger$, which are much less understood and for which no analytical results are available. Some numerical results were presented in Ref.~\cite{gosh2022PRB} for class AI$^\dagger$, while the DSFF for class AII$^\dagger$ has not been investigated before.
We obtained them numerically and plot them in Fig.~\ref{fig:A_AId_AIId}. As for class A, for classes AI$^\dagger$ and AII$^\dagger$, there is also an early quadratic growth, albeit with a different prefactor---for class AI$^\dagger$ it is twice the prefactor of A, while for class AII$^\dagger$ it is half. 
As can be seen from Fig.~\ref{fig:A_AId_AIId} (right), in all three cases, the prefactor of the $\tau^2$ dependence decreases by a factor of 2 going from the perturbative to the nonperturbative domain---see Sec.~\ref{sec:correlations_DSFF_analytics} for an explanation of this phenomenon for class A.
The approach to the late-time plateau is slower for AI$^\dagger$ than for A, while it is faster for AII$^\dagger$.
Contrary to the Hermitian case, there is no nonanalyticity in the DSFF around the Heisenberg time.

\begin{figure}[tbp]
	\centerline{\includegraphics[width=0.6\textwidth]{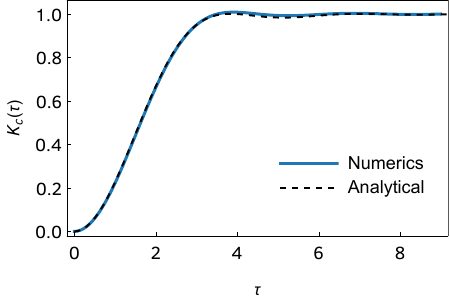}}
	\caption{The DSFF of uncorrelated random variables uniformly distributed on the unit disk. The full blue line corresponds to the numerically obtained DSFF for $10^4$ sets of $2048$ independent random complex numbers, while the dashed black line gives the analytical prediction of Eq.~(\ref{eq:DSFF_Poisson}). There is a perfect agreement between the two.}
	\label{fig:dsff_Poisson}
\end{figure}

Finally, let us consider the DSFF for spectra with
Poisson statistics (i.e., 2d uncorrelated points) unfolded to constant density inside the complex unit disk.
The connected two-point correlation function is given by
\be
\varrho_{2c}(\lambda_1,\lambda_2) = \bar \varrho (\lambda_1) \delta^2(\lambda_1-\lambda_2)
- \frac 1D \bar \varrho(\lambda_1)\bar \varrho(\lambda_2),
\ee
with $\bar \varrho(\lambda_1)=D/\pi$ inside the unit disk.
The connected DSFF is given by ($\theta=0$):
\begin{equation}
\begin{split}\label{eq:DSFF_Poisson}
K_c(\tau)
&= \frac 1{D} \int \d^2\lambda_1 \d^2\lambda_2\, \varrho_{2c}(\lambda_1,\lambda_2) e^{\i\tau\Re(\lambda_1-\lambda_2)}
\\
&=1- \left |\frac 1{D}\int \d^2\lambda\,\bar \varrho(\lambda)\, e^{\i\tau x} \right |^2
\\
&=1-  \left |\frac 1{\pi}\int_{-1}^1  \d x \int_{-\sqrt{1-x^2}}^{+\sqrt{1-x^2}} \d y\,  e^{\i\tau x}\right |^2
\\
&=1-  \left |\frac{2}{\pi}\int_{-1}^1  \d x \sqrt{1-x^2}\, e^{\i\tau x} \right |^2
\\
&= 1 -  4\left(\frac{J_1(\tau)}{\tau}\right )^2.
\end{split}
\end{equation}
In Fig.~\ref{fig:dsff_Poisson}, we show a numerical verification of the result of Eq.~(\ref{eq:DSFF_Poisson}), finding perfect agreement. 

\subsection{Analytical results for class A}
\label{sec:correlations_DSFF_analytics}

We now discuss the analytical results for class A. The simplest RMT ensemble in this class is the GinUE, whose connected two-point correlation function is given by~\cite{ginibre1965,mehta2004}
\be
\varrho_{2c}(\lambda_1,\lambda_2) = \bar \varrho(\lambda_1)\delta^2(\lambda_1-\lambda_2)- |K(\lambda_1,\lambda_2)|^2,
\ee
with kernel
\be
K(\lambda_1,\lambda_2) =\frac D\pi e^{-\frac D2 (|\lambda_1|^2 +|\lambda_2|^2)}\sum_{k=0}^{D-1}\frac{(D \lambda_1 \lambda_2^*)^k}{k!}.
\ee

The spectrum of the GinUE is isotropic and, hence, we set $\theta=0$. The connected DSFF is given by
\begin{equation}
\begin{split}
K_c(\tau) 
&= \frac 1D\int \d^2\lambda_1 \d^2\lambda_2\, e^{\i\tau {\rm Re} (\lambda_1-\lambda_2)} \varrho_{2c}(\lambda_1,\lambda_2)
\\
&=1 - \frac{D}{\pi^2}\sum_{p,s=0}^{D-1}\frac{D^{p+s}}{p! s!} \left|I(\tau,p,s)\right|^2,
\end{split}
\end{equation}
where
\begin{equation}
\begin{split}
I(\tau,p,s) &= \int \d^2\lambda\, e^{\i\tau (\lambda+\lambda^*)/2} \lambda^p \lambda^{* s} e^{-D |\lambda|^2}
\\
&=\i^{|p-s|}\frac{\pi}{D} 
\frac{ {\rm max}(p,s)!}{|p-s|!}
\frac{\tau^{|p-s|}}{ 2^{|p-s|}D^{ {\rm max}(p,s)}}
~_1F_1\({\rm max}(p,s)+1,|p-s|+1,\frac {-\tau^2}{4D}\)
\end{split}
\end{equation}
and ${}_1F_1$ is a hypergeometric function.
Our final expression for the DSFF is given by
\begin{equation}
\label{kcgin}
K_c(\tau)=1- \frac 1D\sum_{p,s=0}^{D-1}\frac{D^{p+s}}{p! s!} \left[\frac{ {\rm max}(p,s)!}{|p-s|!} \frac {\tau^{|p-s|}}{ 2^{|p-s|}D^{ {\rm max}(p,s) }} ~_1F_1\left ({\rm max}(p,s)+1,|p-s|+1,\frac {-\tau^2}{4D} \right ) \right    ]^2.
\end{equation}
This result differs from the expression quoted in Ref.~\cite{li2021PRL} by the factor $2^{-|p-s|}$.
Asymptotically, for large $D$, it simplifies to Eq.~(\ref{form-an})
\be
K_c(\tau) = 1-e^{-\tau^2/4D},
\label{klarge}
\ee
which is in agreement with Ref.~\cite{li2021PRL}.

We note that the large-$D$ limit of the DSFF~\eref{kcgin} does not commute with the $\tau\to 0 $ limit. The
Taylor expansion of Eq.~\eref{klarge} gives
\be
\label{eq:Kc_nonpert}
K_c(\tau) = \frac {\tau^2}{4D}-\frac {\tau^4}{32 D^2} + O(\tau^6),
\ee
while from the large-$D$ result \eref{kcgin} we obtain
\be
\label{eq:Kc_pert}
K_c(\tau) = \frac {\tau^2}{2D}- \frac {D+3}{32 D^2}\tau^4 + O(\tau^6).
\ee
The coefficient of the first term in the latter expansion is equal to the perturbative coefficient
\be
\frac 1D \left\langle \sum_{k=1}^D {\rm Re} (\lambda_k)  \sum_{l=1}^D {\rm Re} (\lambda_l)\right\rangle_c.
\ee
For a real spectrum belonging to the GUE, we also find that for sufficiently small $\tau$, the SFF $K(\tau) \sim \tau^2$ with the perturbative prefactor given by the analogous expression. 
The transition from the perturbative regime~(\ref{eq:Kc_pert}) to the nonperturbative regime~(\ref{eq:Kc_nonpert}) is visible in finite-size numerical results, see Fig.~\ref{fig:A_AId_AIId}.

\section{Eigenvector correlations: Overlaps of symmetry-connected states}
\label{sec:correlations_overlaps}

The final spectral observable we consider is based not on eigenvalue correlations but on eigenvector correlations. We consider the Chalker-Mehlig eigenvector overlap matrix~\cite{chalker1998PRL,mehlig2000JMP}:
\begin{equation}
O_{\alpha\beta}=\langle \tphi_\alpha|\tphi_\beta\rangle \braket{\phi_\beta}{\phi_\alpha},
\end{equation}
where $\ket{\phi_\alpha}$ and $|\tphi_\alpha\rangle$ are the right and left eigenvectors defined in Eqs.~(\ref{eq:evL_new}) and (\ref{eq:evLd_new}). Note that this definition applies only in the case of classes without non-Hermitian Kramers degeneracy. If each eigenvalue is doubly degenerate and the eigenspace dimension is two, each element $O_{\alpha\beta}$ becomes itself a $4\times4$ matrix, and additional care has to be exerted. We defer such considerations to future work and do not consider this case further in this thesis.

The overlap matrix, in particular the distribution of its entries and the first moments, have been intensely investigated for the Ginibre ensembles~\cite{chalker1998PRL,mehlig2000JMP,janik1999PRE,fyodorov2018CMP,bourgade2020,akemann2020Acta}. The diagonal overlaps $O_{\alpha\alpha}$ are sensitive to $\scC_+^2$, a claim we have confirmed numerically. However, since local eigenvalue statistics---measured, for instance, by CSR---are already sensitive to this symmetry, we do not employ it here.

Instead, we propose that the off-diagonal overlaps $O_{\alpha\balpha}$, where $\{|\phi_\alpha\rangle,|\tphi_\alpha\rangle\}$ and $\{\ket{\phi_\balpha},|\tphi_\balpha\rangle\}$ are connected by an antiunitary symmetry $\scT_\pm$ or $\scC_-$, are sensitive to the value of the square of that symmetry. More concretely, we have the following results.
\begin{enumerate}
	\item If $|\tphi_\balpha\rangle\propto \scC_-\ket{\phi_\alpha}$, the overlaps $O_{\alpha\balpha}$ (denoted $O_{\alpha\balpha}^{(\scC_-)}$ for clarity) are all non-negative if $\scC_-^2=+1$, and all nonpositive if $\scC_-^2=-1$. If $\scC_-^2=0$ (i.e., if the symmetry is absent and the eigenvectors are independent), the overlaps are still real for spectra with dihedral symmetry, and the fraction of positive and negative matrix elements is $1/2$ each. For classes with no $\scC_-$ symmetry and no dihedral symmetry, the overlaps are complex.
	\item If $\ket{\phi_\balpha}\propto \scT_-\ket{\phi_\alpha}$ and $\scT_-^2=-1$, the overlaps $O_{\alpha\balpha}$ (denoted $O_{\alpha\balpha}^{(\scT_-)}$) all vanish identically. If $\scT_-^2=+1$ or $0$ they assume arbitrary complex values.
\end{enumerate}

These two statements can be proven in general by a variation of the proof of Kramers degeneracy. 
Let us denote by $\scA$ one of the four antiunitary operators $\scT_\pm$ or $\scC_\pm$. Then, for any two vectors $\psi$ and $\phi$, we have
\begin{equation}
\label{eq:antiunitary_braket}
\begin{split}
\braket{\psi}{\scA\phi}
=\braket{\scA \psi}{\scA^2 \phi}^*
=\scA^2 \braket{\scA \psi}{\phi}^*
=\scA^2 \braket{\phi}{\scA \psi},
\end{split}
\end{equation}
where we use the antiunitarity of $\scA$ and the fact that $\scA^2$ is either $\pm 1$.
In order to prove assertion 1, following Sec.~\ref{sec:correlations_consequences}, we note that 
\begin{equation}
|\tphi_\balpha\rangle =\scC_-|\phi_\alpha\rangle,\quad
|\phi_\balpha\rangle =\scC_-|\tphi_\alpha\rangle,
\end{equation}
where, without loss of generality, we set a possible proportionality constant to one. 
Then, using Eq.~(\ref{eq:antiunitary_braket}), the overlap matrix reads
\begin{equation}
\begin{split}
O^{(\scC_-)}_{\alpha\balpha}
&=\bra{\tphi_\alpha}\scC_-\ket{\phi_\alpha}
\bra{\tphi_\alpha}\scC_-^\dagger\ket{\phi_\alpha}
\\
&=\scC_-^2 \bra{\phi_\alpha}\scC_-\ket{\tphi_\alpha}
\bra{\tphi_\alpha}\scC_-^\dagger\ket{\phi_\alpha}
\\
&=\scC_-^2 \abs{\bra{\phi_\alpha}\scC_-\ket{\tphi_\alpha}}^2,
\end{split}
\end{equation}
and we conclude that the overlap matrix element $O^{(\scC_-)}_{\alpha\balpha}$ has the same sign as $\scC_-^2$, proving assertion~1. In order to prove assertion~2, we note instead the relation between the two right eigenvectors:
\begin{equation}
\ket{\phi_\balpha}=\scT_-\ket{\phi_\alpha}.
\end{equation}
Using Eq.~(\ref{eq:antiunitary_braket}), it immediately follows that
\begin{equation}
\bra{\phi_\alpha}\scT_- \ket{\phi_\alpha}
=\scT_-^2\bra{\phi_\alpha}\scT_- \ket{\phi_\alpha}.
\end{equation}
If $\scT_-^2=-1$, this matrix element and, consequently, the overlap 
\begin{equation}
O^{(\scT_-)}_{\alpha\balpha}=\bra{\tphi_\alpha}\scT_-\ket{\tphi_\alpha}\bra{\phi_\alpha}\scT_-^\dagger\ket{\phi_\alpha}
\end{equation}
vanish identically, proving assertion 2.

Very importantly, explicit knowledge of the operator $\scC_\pm$ or $\scT_\pm$ is not required to compute the respective eigenvector overlaps. To construct the overlap matrix, the eigenvalues are ordered by increasing real part and, for each pair of complex conjugated eigenvalues, by increasing imaginary part. With this ordering, the overlaps $O^{(\scC_+)}_{\alpha\balpha}$ lie on the main diagonal $O_{\alpha\alpha}$ of the matrix $O_{\alpha\beta}$; the overlaps $O^{(\scC_-)}_{\alpha\balpha}$ are the antidiagonal elements $O_{\alpha,D-\alpha+1}$, where $D$ is the sector dimension; the overlaps $O^{(\scT_+)}_{\alpha\balpha}$ are the elements $O_{2\alpha-1,2\alpha}$; and, finally, the overlaps $O^{(\scT_-)}_{\alpha\balpha}$ are the elements $O_{2\alpha-1,D-2\alpha+1}$ and $O_{2\alpha,D-2\alpha+2}$.

\section{Summary and outlook}

In this chapter, we reviewed the role of CSR as a tool to analyze universal spectral features of non-Hermitian systems (integrable and chaotic). We found that angular correlations between levels in dissipative systems provide a clean signature of quantum chaos: Uncorrelated random variables, which describe integrable systems, have a flat, and hence isotropic, ratio distribution in the complex plane, while for RMT ensembles, there is a suppression of small angles and spacings in the large-$D$ limit. Besides distinguishing ergodic from nonergodic correlations, they provide a signature of the symmetry $\scC_+^2$.

We then looked at correlations near the hard edge (origin) of the spectrum. To resolve the value of $\scC_-^2$ and the presence or not of $\scP$ symmetry in classes without reality conditions, we considered the distribution of the eigenvalue with the smallest absolute value, which is universal in these classes. We computed the exact analytical distribution for class AIII$^\dagger$, and gave a simple surmise for the smallest matrix dimension that is extremely accurate in the large-$D$ limit.

We also studied long-range correlations through the use of the DSFF of unfolded eigenvalues. For random matrices, it provides the same information as CSRs but, since it involves correlations over all length scales in the spectrum, it will allow us to better characterize deviations from universality in physical systems. We obtained the first results for the DSFF of bulk universality classes AI$^\dagger$ and AII$^\dagger$ numerically, the DSFF of unfolded Poisson level analytically, and uncovered two distinct quadratic regimes in the small-time limit of the DSFF.  

Finally, we proposed the eigenvector overlaps between states connected by the antiunitary symmetry of interest as a useful new signature of non-Hermitian antiunitary symmetries. Importantly, they can be computed even when the explicit form of the symmetry transformation is not known.
The role of these off-diagonal eigenvector overlaps as a measure of dissipative quantum chaos deserves further study. We used the sign of different overlaps as a proxy for the existence or absence of a given non-Hermitian antiunitary symmetry, but did not study in any detail their distributions. While a numerical study is the natural first step, an analytical investigation following Chalker and Mehlig~\cite{chalker1998PRL} might be possible.

The spectral observables we introduced here will be used in the following chapters to study the correlations of physical models of dissipative quantum matter. While they allow us to uniquely distinguish all the examples discussed there, they are not yet enough to uniquely characterize all 38 classes of non-Hermitian matrices. An interesting future goal is to obtain such a complete characterization.

%% file: Thesis_ClassificationSYK.tex

\chapter{Symmetry classification of the non-Hermitian Sachdev-Ye-Kitaev model: Universality and its limits}
\label{chapter:classificationSYK}

In this chapter, we initiate our study of universality and its limits in many-body dissipative quantum chaos. As a paradigmatic example, we consider a non-Hermitian $q$-body Sachdev-Ye-Kitev (nHSYK) model with $N$ Majorana and provide its complete symmetry classification (Sec.~\ref{sec:1SYK_symm_class}). Depending on $q$ and $N$, we identify 9 out of the 10 non-Hermitian Bernard-LeClair (BL) classes without reality conditions~\cite{bernard2002,kawabata2019PRX}.

We also investigate level statistics on several timescales, combining numerical exact diagonalization results and analytical calculations. 
First, in Sec.~\ref{sec:1SYK_universality}, we study the local level correlations of the nHSYK model, which probe the dynamics around the Heisenberg time, both in the bulk of the spectrum (Sec.~\ref{sec:1SYK_CSR}) and, for classes with one, near the hard edge (Sec.~\ref{sec:1SYK_hard-edge}). We find that these correlations are perfectly described by random matrix theory for $q > 2$, including for classes that involve bulk correlations of type ${\rm AI}^\dagger$ and ${\rm AII}^\dagger$, beyond the Ginibre ensembles. For $q = 2$, local correlations are given by Poisson statistics instead.
These results provide strong evidence of striking universal features in many-body dissipative quantum chaos.

Then, we investigate whether universality extends to timescales much shorter than the Heisenberg time. To this end, in Sec.~\ref{sec:1SYK_long-range}, we study the behavior of the dissipative spectral form factor (DSFF) and find good agreement between nHSYK model and random matrix theory (RMT) for $q > 2$ starting at a timescale that decreases sharply with $q$. The source of deviation from universality, identified analytically, is ensemble fluctuations and is not related to the quantum dynamics. For fixed $q$ and large enough $N$, these fluctuations become dominant up until after the Heisenberg time, so that the DSFF of unfolded eigenvalues is no longer useful for the study of quantum chaos. For $q = 2$, we find saturation to Poisson statistics at a timescale of $\log D$, compared to a scale of $\sqrt D$ for $ q>2$, where $D$ is the dimension of the Hilbert space.

This chapter is based on Refs.~\cite{garcia-garcia2022PRX,garcia2023PRD}.

\section{Symmetry classification of the nHSYK model}
\label{sec:1SYK_symm_class}

We study a single non-Hermitian SYK (nHSYK) model of $N$ Majorana fermions in $(0 + 1)$ dimensions with $q$-body interactions in Fock space, with complex instead of real couplings:
\begin{align}\label{eq:1SYK_hami}
H \, &= \, \sum_{i_1<i_2<\cdots<i_q}^{N} (J_{i_1i_2\cdots i_q}+\i M_{i_1i_2\cdots i_q}) \, \psi_{i_1} \, \psi_{i_2} \, \cdots \, \psi_{i_q},
\end{align}
where $J_{i_1\cdots i_q}$ and $ M_{i_1\cdots i_q}$ are Gaussian-distributed random variables with zero average and variance $\sigma^2 =1/6(2N)^{q-1}$.
The Majorana fermions $\psi_i$ can be represented by $2^{N/2}$-dimensional Hermitian Dirac $\gamma$ matrices and satisfy $\{ \psi_{i}, \psi_{j} \} = 2\delta_{ij}$.
We work in a basis where the odd- and even-numbered Majoranas are represented by real symmetric and purely imaginary antisymmetric matrices, respectively.
We shall consider the cases $q=2,\;3,\;4$, and $6$, and the number of fermions, $N$, is always even. To be precise, we note that for odd $q$, $H$ corresponds to a supercharge (not Hamiltonian) operator.

As described in Ch.~\ref{chapter:correlations}, the Hamiltonian of Eq.~(\ref{eq:1SYK_hami}) is classified by the unitary and antiunitary involutions of its irreducible blocks, which we reproduce here for convenience:
\begin{alignat}{99}
\label{eq:nHsym_TRS}
&\sT_+ H \sT_+^{-1} = H,\qquad 
&&\sT_+^2=\pm 1, \qquad
&& \sT_+ \;\text{antiunitary},
\\
\label{eq:nHsym_PHSd}
&\sT_- H \sT_-^{-1} = -H,\qquad 
&&\sT_-^2=\pm 1, \qquad
&& \sT_- \;\text{antiunitary},
\\
\label{eq:nHsym_TRSd}
&\sC_+ H^\dagger \sC_+^{-1} = H,\qquad 
&&\sC_+^2=\pm 1, \qquad 
&& \sC_+ \;\text{antiunitary},
\\
\label{eq:nHsym_PHS}
&\sC_- H^\dagger \sC_-^{-1} = -H,\qquad 
&&\sC_-^2=\pm 1, \qquad
&& \sC_-\; \text{antiunitary},
\\
\label{eq:nHsym_CSd}
&\Pi H \Pi^{-1} = -H,\qquad 
&&\Pi^2=1, \qquad 
&&\Pi\; \text{unitary},
\\
\label{eq:nHsym_pH}
&\scQ_+ H^\dagger \scQ_+^{-1} = H,\qquad 
&&\scQ_+^2=1, \qquad
&&\scQ_+\;\text{unitary},
\\
\label{eq:nHsym_apH}
&\scQ_- H^\dagger \scQ_-^{-1} = -H,\qquad 
&&\scQ_-^2=1, \qquad
&&\scQ_-\;\text{unitary}.
\end{alignat}
Note that in this chapter, we denote the unitary chiral operator with $\Pi$, since $\scP$ is the widespread notation for one of the charge-conjugation operators, to be introduced below.

For the nHSYK model, we consider the charge-conjugation operators
\cite{you2017PRB,cotler2017JHEP,kanazawa2017JHEP,sun2020PRL}
\begin{align}\label{eq:antiunitary_conj}
\sP=K\prod_{i=1}^{N/2}\psi_{2i-1}
\quad \text{and} \quad
\sR=K\prod_{i=1}^{N/2}\i\psi_{2i},
\end{align}
where $K$ denotes the complex-conjugation operator, which square to
\begin{align}\label{eq:PR_square}
\sP^2=(-1)^{\frac 12 N/2(N/2-1)}
\quad \text{and} \quad 
\sR^2=(-1)^{\frac 12 N/2(N/2+1)}.
\end{align}
The combination of these two operators yields the Hermitian operator,
\begin{equation}\label{eq:chiral_op}
\sS=\sP\sR=\i^{N^2/4}\prod_{i=1}^{N}\psi_{i},
\end{equation}
which squares to the identity. 

The operators $\sP$ and $\sR$ act on Majorana fermions as $\sP \psi_i\sP^{-1}=-(-1)^{N/2}\psi_i$ and $\sR \psi_i \sR^{-1}=(-1)^{N/2}\psi_i$. The complex couplings $J_{i_1\cdots i_q}+\i M_{i_1\cdots i_q}$ are invariant under transposition, and, hence, the nHSYK Hamiltonian has the involutive symmetries
\begin{align}
\label{eq:P_transform_H}
\sP H^\dagger \sP^{-1}&=(-1)^{q(q+1)/2}(-1)^{qN/2}H,
\\
\label{eq:R_transform_H}
\sR H^\dagger \sR^{-1}&=(-1)^{q(q-1)/2}(-1)^{qN/2}H,
\\
\label{eq:S_transform_H}
\sS H \sS^{-1}&=(-1)^q H.
\end{align}
Comparing with Eqs.~(\ref{eq:nHsym_TRS})--(\ref{eq:nHsym_apH}), we see that $\sP$ and $\sR$ play the role of either $\sC_+$ or $\sC_-$, while $\sS$ either commutes or anticommutes with the Hamiltonian.
For the nHSYK model, the many-body matrix elements are manifestly complex, and no antiunitary symmetries that map $H$ back to itself exist. Then, only the involutive symmetries~(\ref{eq:nHsym_TRSd})--(\ref{eq:nHsym_CSd}) can occur.\footnote{In the Hermitian SYK model, the couplings are real and thus invariant under complex conjugation. However, in that case, $\sT_\pm$ and $\sC_\pm$ are equivalent to each other.}
The nHSYK model thus belongs to one of the ten BL symmetry classes without reality conditions~\cite{bernard2002}, which are in one-to-one correspondence with the Hermitian Altland-Zirnbauer (AZ) classes~\cite{altland1997} and were tabulated in Table~\ref{tab:correlations_sym_class_noreal}.
Following the Kawabata-Shiozaki-Ueda-Sato nomenclature~\cite{kawabata2019PRX}, they are dubbed A, AIII$^\dagger$, AI$^\dagger$, AII$^\dagger$, D, C, AI$^\dagger_+$, AII$^\dagger_+$, AI$^\dagger_-$, and AII$^\dagger_-$.
The difference between the AZ classes and the BL classes without reality conditions is that complex conjugation is replaced by transposition and the Hermiticity constraint is lifted. For example, a Hermitian real symmetric Hamiltonian is to be replaced by a non-Hermitian complex symmetric Hamiltonian.

The nHSYK symmetry classification can be performed systematically by evaluating Eqs.~(\ref{eq:PR_square}) and (\ref{eq:P_transform_H})--(\ref{eq:S_transform_H}) for different values of $q\mod4$ and $N\mod8$. Note that while the physical interpretation of the operators $\sP$, $\sR$, and $\sS$ is different in the SYK and nHSYK models, the defining relations in Eqs.~(\ref{eq:PR_square}) and (\ref{eq:P_transform_H})--(\ref{eq:S_transform_H}) are formally the same. It follows that the symmetry classification of the former~\cite{you2017PRB,garcia-garcia2016PRD,cotler2017JHEP,li2017JHEP,kanazawa2017JHEP,garcia-garcia2018PRD,behrends2019PRB,behrends2020PRL,sun2020PRL} also holds for the latter, provided that one replaces any reality condition by a transposition one. We now investigate in more detail the dependence of these symmetries of the odd or even nature of $q$ in the nHSYK Hamiltonian~(\ref{eq:1SYK_hami}).

\paragraph*{Even $q$.}
According to Eq.~(\ref{eq:S_transform_H}), $H$ commutes with $\sS$ (which is proportional to the fermion parity operator), the Hilbert space is split into sectors of conserved even and odd parity, and the Hamiltonian is block-diagonal. There is no chiral symmetry. From Eqs.~(\ref{eq:P_transform_H}) and (\ref{eq:R_transform_H}), we see that $H$ transforms similarly under both $\sP$ and $\sR$ (when they act within the same block) and it suffices to consider the action of one, say $\sP$. We have the
commutation relation
\be
\sS\sP =(-1)^{\frac N2} \sP\sS.
\ee
\begin{itemize}
	\item When $N\mod8=2,6$, $\sP$ is a fermionic operator that anticommutes with $\sS$. Note that $\sP$ is not an involutive symmetry of the Hamiltonian in a diagonal block representation, as it maps blocks of different parity into each other. The two blocks are the transpose of each other and have no further constraints (class A or complex Ginibre).
	
	\item When $N\mod8=0,4$ and $q\mod4=0$, $\sP$ is a bosonic operator that commutes with $\sS$. Each block of the Hamiltonian has the involutive symmetry, $\sP H^\dagger \sP^{-1}=+H$. If $N\mod8=0$, $\sP^2=1$, and we
	can find a basis in which the Hamiltonian is symmetric. This is the universality class of complex symmetric matrices, also known as AI$^\dagger$. 
	If $N\mod8=4$, $\sP^2=-1$, we can find a basis in which $H^\top = I H I^{-1} $, with $I$ the symplectic unit matrix. This class is AII$^\dagger$.
	
	\item When $N\mod8=0,4$ and $q\mod4=2$, $\sP$ is again a bosonic
	operator that commutes with $\sS$. Within each block, we have the involutive symmetry, $\sP H^\dagger \sP^{-1}=-H$. If $N\mod8=0$, $\sP^2=1$, we can find a basis in which the Hamiltonian becomes antisymmetric and the universality class is given by that of complex antisymmetric matrices (non-Hermitian class D); if $N\mod8=4$, $\sP^2=-1$, and we can find a basis where $H^\top =-I H I^{-1}$. Complex matrices satisfying this constraint belong to non-Hermitian class C.
\end{itemize}

\paragraph*{Odd $q$.}

In this case, $\sS$ is a chiral symmetry operator that anticommutes with $H$,
so that $H$
acquires an off-diagonal block structure in a chiral basis.
The operators $\sP$ and $\sR$ now act differently on $H$ (one
satisfies $X H^\dagger X^{-1} = H $ and the other $X H^\dagger X^{-1} = -H$ with
$X$ either $\sP$ or $\sR$). 
Hence both must be considered if we use
the square of $\sP$, $\sR$, and $\sS$ to classify the matrices. However, for the derivation of the block structure, as given in Table~\ref{tab:correlations_sym_class_noreal}, we of course only need either
$\sP$ or $\sR$, and $\sS$.
\begin{itemize}
	\item When $N\mod8=0$, both $\sP$ and $\sR$ are bosonic operators squaring to $+1$. Since they commute with $\sS$, and one of them satisfies $X H^\dagger X^{-1} = H$ ($\sP$ if $q\mod4=3$, $\sR$ if $q\mod4=1$), the Hamiltonian in a suitable basis is a complex matrix with vanishing diagonal blocks and off-diagonal blocks that are the transpose of each other (class AI$^\dagger_+$), irrespective of whether $q\mod4=1$ or $3$.
	
	\item When $N\mod8=4$, both $\sP$ and $\sR$ are bosonic operators commuting with $\sS$ and squaring to $-1$. Hence the off-diagonal blocks $A$ and $B$ of $H$ are related by
	$B^\top = I A I^{-1}$. This is the universality class AII$^\dagger_+$, irrespective of whether $q\mod4=1$ or $3$.
	
	\item When $N\mod8=2,6$ we have that $\{\sS,\sP\}=0$ and  $\{\sS,\sR\}=0$. So $\sP$ and $\sR$ have vanishing diagonal blocks, and depending on $N$, $\sP^2=1$ and $\sR^2=-1$ ($N\mod8=2$), or $\sP^2=-1$ and $\sR^2=1$ ($N\mod8=6$). We choose the operator that squares to $1$. Since it anticommutes with $\sS$, it has the block structure
	\be
	X=\bmat 0 & x^{-1} \\ x & 0 \emat.
	\ee
	If $X H^\dagger X^{-1} =H$ the blocks of $H$ satisfy $ x A^\top x^{-1} = A$ and
	$ x B^\top x^{-1} = B$, so we can find a basis in which the blocks are symmetric.
	This is the case for $ N \mod 8=2$ and $q\mod4 =1$ or $ N \mod 8=6$ and $q\mod4 =3$
	(class AI$^\dagger_-$).
	If $X H^\dagger X^{-1} = -H$ the blocks of $H$ satisfy $ x A^\top x^{-1} = -A$ and
	$ x B^\top x^{-1} =- B$, so we can find a basis in which the blocks are skew-symmetric. This is the case for $ N \mod 8=2$ and $q\mod4 =3$ or $N \mod 8=6$ and $q\mod4 =1$ (class AII$^\dagger_-$).
\end{itemize}

The complete symmetry classification of the nHSYK Hamiltonian
for all $q$ and even $N$ in terms of
BL classes
is summarized in Table~\ref{tab:nHSYK_class}. Note that the tenth symmetry class AIII$^\dagger$ is not realized by the nHSYK Hamiltonian~(\ref{eq:1SYK_hami}).\footnote{It is, however, realized in the Wishart nHSYK model; see Ch.~\ref{chapter:QLaguerre}.}

Since for $q>2$ the nHSYK is quantum chaotic, we expect that spectral correlations follow RMT. In the next section, we show that this is indeed the case. By studying local bulk and hard-edge spectral correlations for different values of $q$ and $N$, we find excellent agreement with the predictions of non-Hermitian RMT for the respective universality class. However, we will also see in Sec.~\ref{sec:1SYK_long-range}, that there are limits to universality and that, for sufficiently long distances in the spectrum, deviations from RMT occur.

\begin{table}[tbp]
	\centering
	\caption{Complete symmetry classification of the nHSYK Hamiltonian into BL classes without reality conditions for all $q$ and even $N$.}
	\label{tab:nHSYK_class}
	\begin{tabular}{@{}l cccc@{}}
		\toprule
		$N\,\mathrm{mod}\,8$   & 0              & 2               & 4               & 6               \\ \midrule
		$q\,\mathrm{mod}\,4=0$ & AI$^\dagger$   & A               & AII$^\dagger$   & A               \\
		$q\,\mathrm{mod}\,4=1$ & AI$^\dagger_+$ & AI$^\dagger_-$  & AII$^\dagger_+$ & AII$^\dagger_-$ \\
		$q\,\mathrm{mod}\,4=2$ & D              & A               & C               & A               \\
		$q\,\mathrm{mod}\,4=3$ & AI$^\dagger_+$ & AII$^\dagger_-$ & AII$^\dagger_+$ & AI$^\dagger_-$  \\ \bottomrule
	\end{tabular}
\end{table}

\section{Universality in many-body quantum chaos}
\label{sec:1SYK_universality}

\subsection{Bulk correlations}
\label{sec:1SYK_CSR}

We initiate our analysis of spectral correlations by studying complex spacing ratios (CSRs). Because of their short-range nature, these ratios probe the quantum dynamics for late timescales of the order of the Heisenberg time.

\begin{figure}[tbp]
	\centering
	\includegraphics[width=\textwidth]{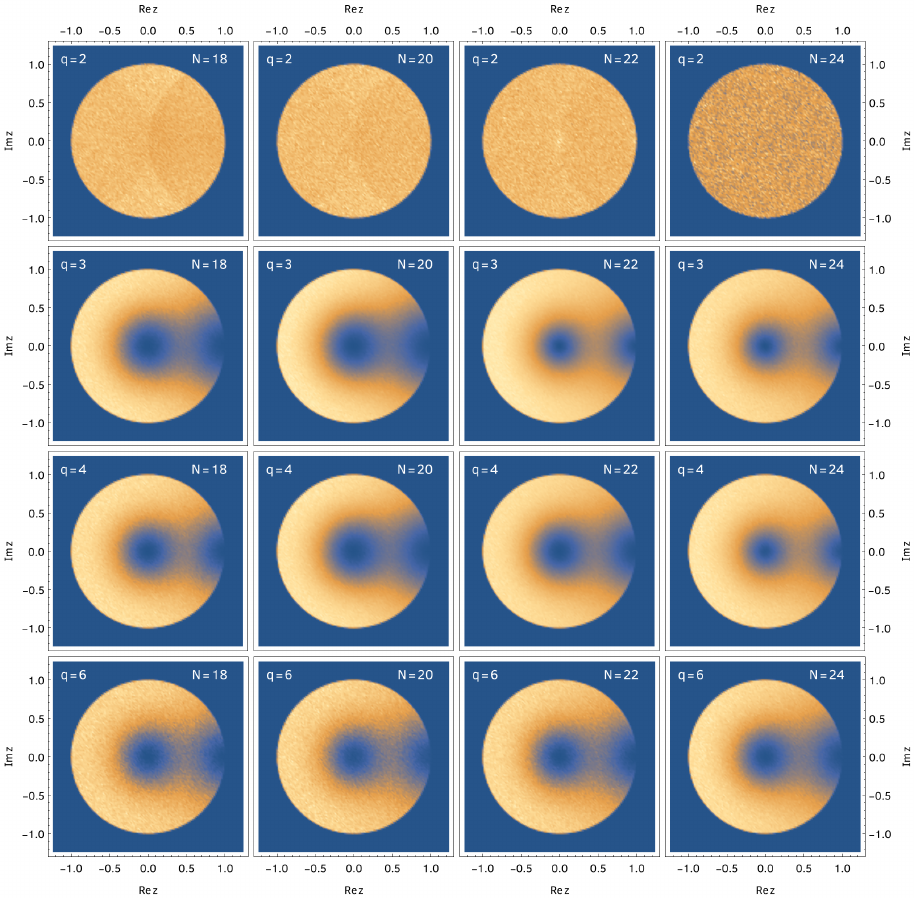}
	\caption{Distribution of CSRs, Eq.~(\ref{eq:cspacing}), in the complex plane for $N=18$, $20$, $22$, and $24$ (left to right) and $q=2$, $3$, $4$, and $6$ (top to bottom). We observe the suppression of the CSR density around the origin and along the real positive semiaxis (signaling level repulsion) for $q>2$. We can also see a variation of the distribution as a function of $N$ for $q=3$ and $4$ (Bott periodicity).}
	\label{fig:Thesis_SYKCSR}
\end{figure}

As before, we define the CSR as
\begin{equation}\label{eq:cspacing}
z_k=\frac{E_k^\mathrm{NN}-E_k}{E_k^\mathrm{NNN}-E_k}.
\end{equation}
where $E_k$ with $k = 1,2,\ldots, 2^{N/2}$ 
is the complex spectrum for a given disorder realization. We perform ensemble averaging, such that we have a minimum of $10^6$ eigenvalues for each set of parameters $N$ and $q$. For even $q$, we consider the two parity sectors separately. The distribution of the resulting averaged CSR $z_k$ is depicted in Figs.~\ref{fig:Thesis_SYKCSR}. We observe qualitative differences between $q=2$ and $q>2$. 
For the former, it is rather unstructured (i.e., flat) with no clear signature of level repulsion for small spacing, which is a signature of the absence of quantum chaos. By contrast, for $q>2$, the CSR is heavily suppressed for small spacings, especially at small angles, a pattern very similar to that observed for non-Hermitian random matrices. We now show that the three universality classes of local bulk correlations---AI$^\dagger$,
A, and AII$^\dagger$~\cite{hamazaki2020PRR}---can be clearly distinguished in the nHSYK model by the CSR distribution.

\begin{figure}[tbp]
	\centering
	\includegraphics[width=0.95\textwidth]{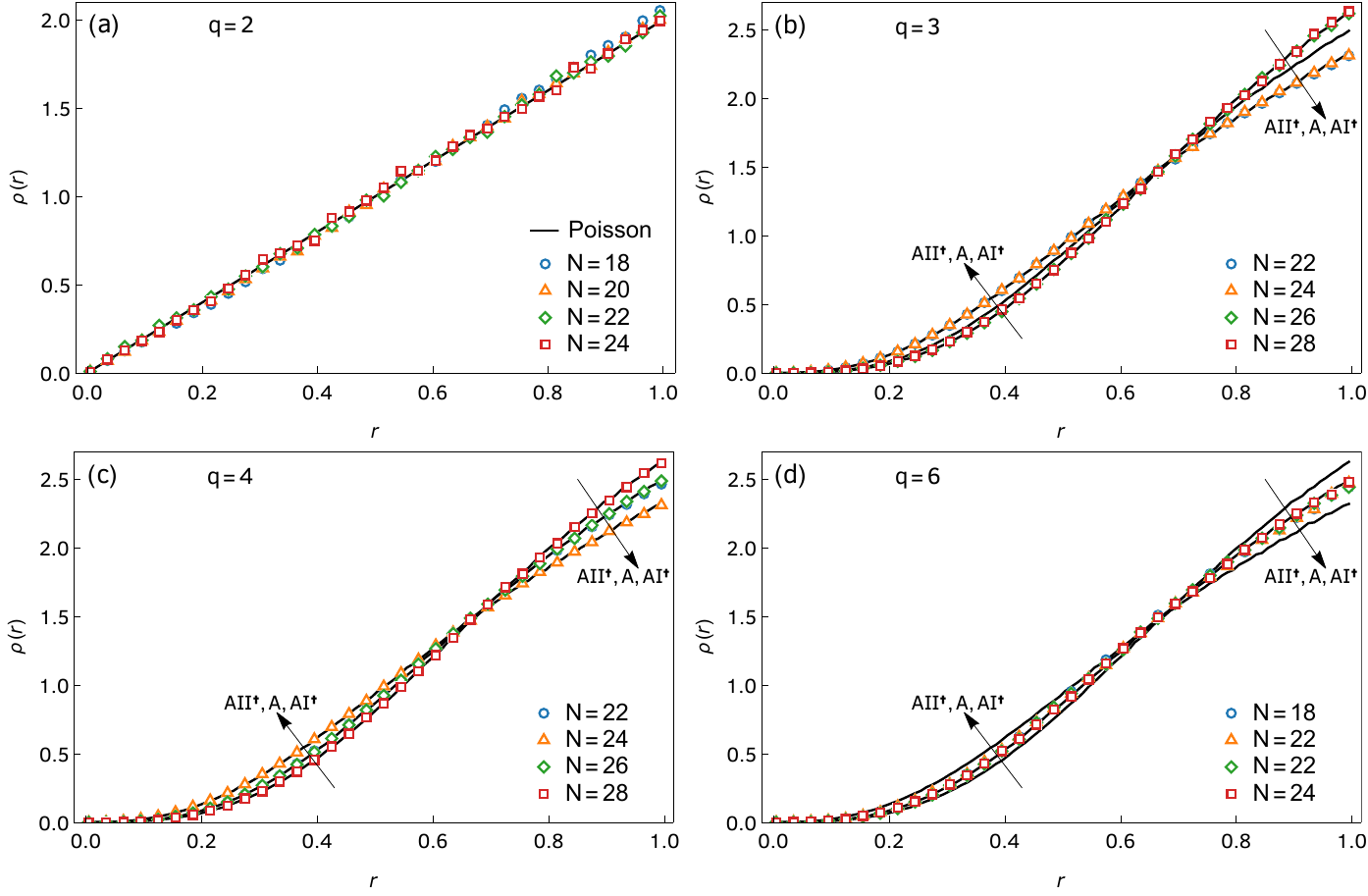}
	\caption{Radial marginal distribution of CSRs, for different values of $q$ and $N$. For $q>2$, we find agreement with the predictions of Table~\ref{tab:nHSYK_class} in all cases. For $q=2$, we see agreement with Poisson statistics.}
	\label{fig:Thesis_nHSYKCSRrad}
\end{figure}

\begin{figure}[tbp]
	\centering
	\includegraphics[width=0.95\textwidth]{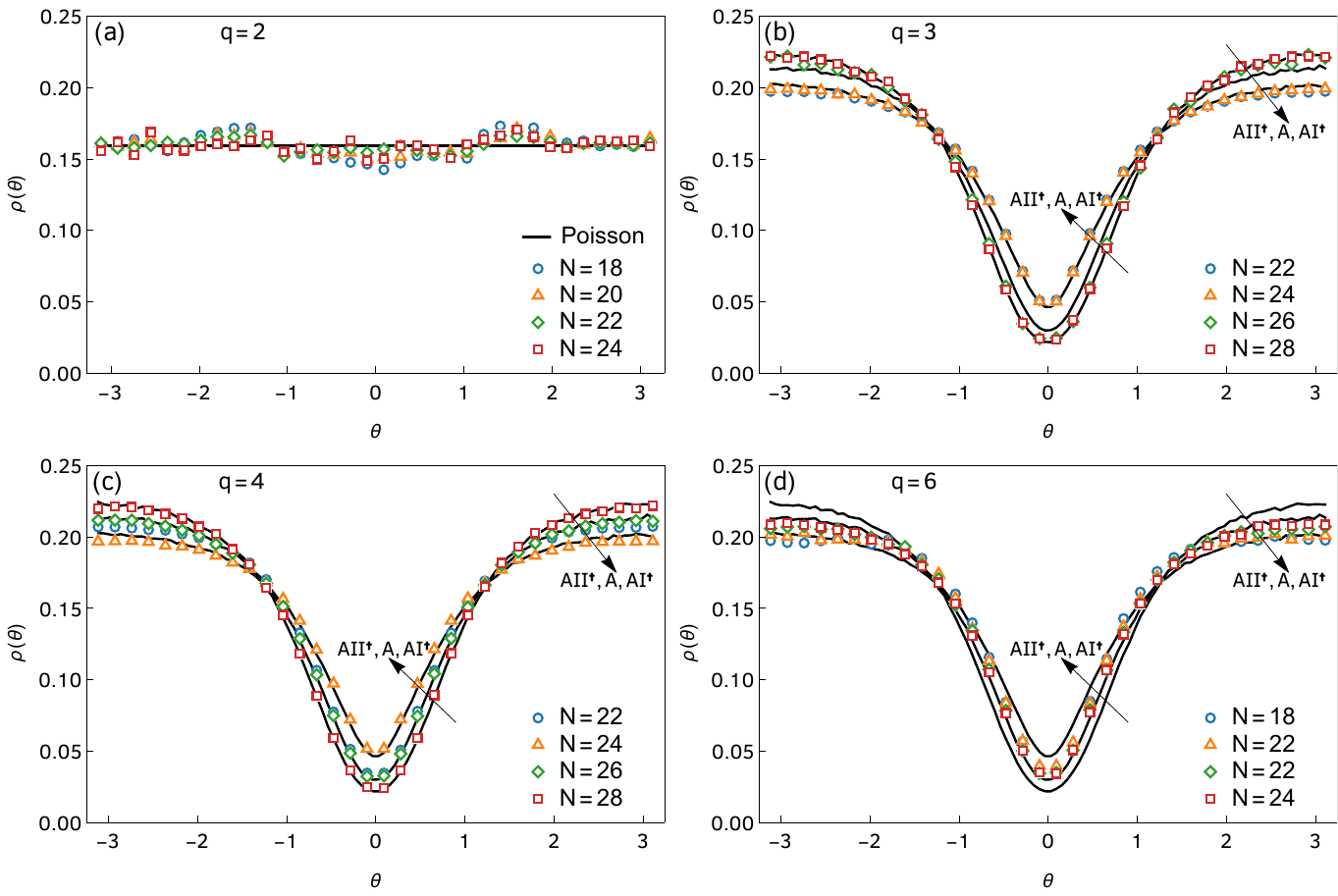}
	\caption{Radial marginal distribution of CSRs, for different values of $q$ and $N$. For $q>2$, we find agreement with the predictions of Table~\ref{tab:nHSYK_class} in all cases, although the converge to the universal large-$N$ results is slower than for the radial distribution. For $q=2$, we see agreement with Poisson statistics.}
	\label{fig:Thesis_nHSYKCSRang}
\end{figure}

In order to gain a more quantitative understanding of the spectral correlations, we compute the marginal angular, $\rho(\theta)$, and radial, $\rho(r)$, complex spacing ratio distributions, where $z_k=r_k e^{\i \theta_k}$. The results, presented in Figs.~\ref{fig:Thesis_nHSYKCSRrad} and \ref{fig:Thesis_nHSYKCSRang}, confirm the existence, depending on $N$ and $q$, of the three universality classes mentioned above. However, only for $q >2$, do we observe agreement with the random matrix prediction, which indicates that, as in the real case, this is a requirement for many-body dissipative quantum chaos.

\begin{figure}[tbp]
	\centering
	\includegraphics[width=0.49\textwidth]{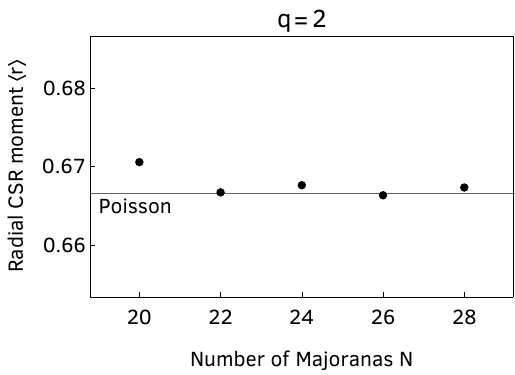}
	\includegraphics[width=0.49\textwidth]{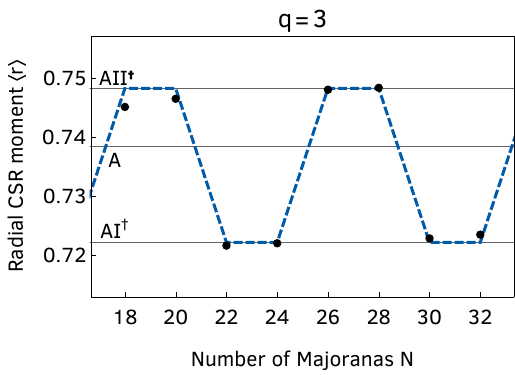}
	\includegraphics[width=0.49\textwidth]{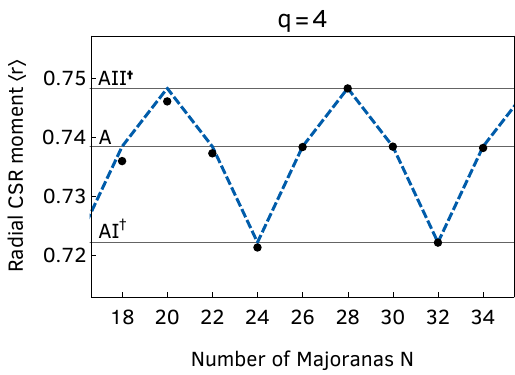}
	\includegraphics[width=0.49\textwidth]{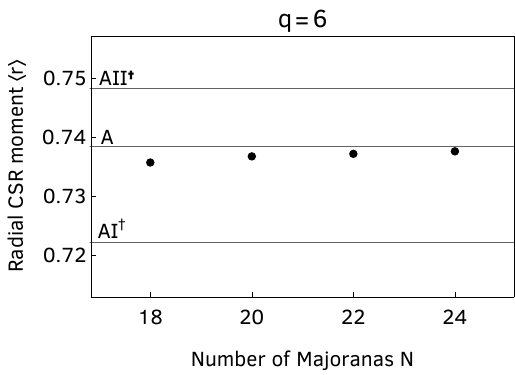}
	\caption{First radial moment $\av{r}$ of the CSR distribution as a function of the number of Majoranas, $N$, for the $q=2,3,4,6$ nHSYK model. The dots are the results of exact numerical diagonalization, the horizontal solid lines are the values of $\av{r}$ for the three universal bulk statistics (A, AI$^\dagger$, and AII$^\dagger$), and the dashed curve follows the classification scheme of Table~\ref{tab:nHSYK_class}. For $q=2$, $\av{r}$ goes to the Poisson value $2/3$ as $N$ increases, showing that no RMT correlations exist around the Heisenberg time. For $q=6$, the three classes realized for different values of $N$ all have the same bulk correlations (those of class A). We see excellent agreement between the nHSYK results and the RMT predictions, except for the smaller values of $N$, where finite-size effects are more pronounced.}
	\label{fig:CSRavr}
\end{figure}

Finally, for a more visual confirmation of the symmetry classification, we compute the first moment of the marginal radial distribution, $\av{r}=\int \d r\, r \rho(r)$, as a function of $N$ and $q$. The values of $\av{r}$ for the three universality classes (A, AI$^\dagger$, and AII$^\dagger$) are reproduced in Table~\ref{tab:numerics}. The results presented in Fig.~\ref{fig:CSRavr} show that the $\av{r}$ of the nHSYK model closely follows the predicted RMT pattern for $q=3$, $4$, and $6$, while it goes to the Poisson value for $q=2$.

\subsection{Hard-edge correlations}
\label{sec:1SYK_hard-edge}

\begin{table}[tbp]
	\caption{Universal single-number signatures of the non-Hermitian universality classes without reality conditions studied in Ch.~\ref{chapter:correlations}. For convenience, this table reproduces the results of Table~\ref{tab:corr_numerics}.}
	\label{tab:numerics}
	\begin{tabular}{@{}l ccc@{}}
		\toprule
		Class    & A      & AI$^\dagger$  & AII$^\dagger$ \\ \midrule
		$\av{r}$ & 0.7384  & 0.7222         & 0.7486         \\ \bottomrule
	\end{tabular}
	\quad
	\begin{tabular}{@{}l ccccccc@{}}
		\toprule
		Class & AIII$^\dagger$ & D      & C      & AI$^\dagger_+$ & AII$^\dagger_+$ & AI$^\dagger_-$ & AII$^\dagger_-$ \\ \midrule
		$R_1$ & 1.129          & 1.228  & 1.102  & 1.222          & 1.096          & 1.123          & 1.138            \\ \bottomrule
	\end{tabular}
\end{table}

The CSRs of the nHSYK Hamiltonian~(\ref{eq:1SYK_hami}) distinguish universality classes A, AI$^\dagger$, and AII$^\dagger$ and confirm the predictions of Table~\ref{tab:nHSYK_class} for the local bulk correlations. Because the full spectrum was employed in the evaluation of the CSRs, $\rho(z)$ cannot distinguish between, for instance, class A and classes C and D, as the last two only differ from class A in the region $|E| \sim 0$ of small eigenvalues. In this section, we study the distribution of the eigenvalue with the lowest absolute value. The shape of this observable is expected to have a universal form for each universality class that should agree with the random matrix prediction provided the spectrum has one of the inversion symmetries studied previously ($\scC_-$ or $\Pi$). This enables us to identify additional universality classes depending on the type of inversion symmetry of the nHSYK Hamiltonian. 

\begin{figure}[tbp]
	\centering
	\includegraphics[width=0.9\textwidth]{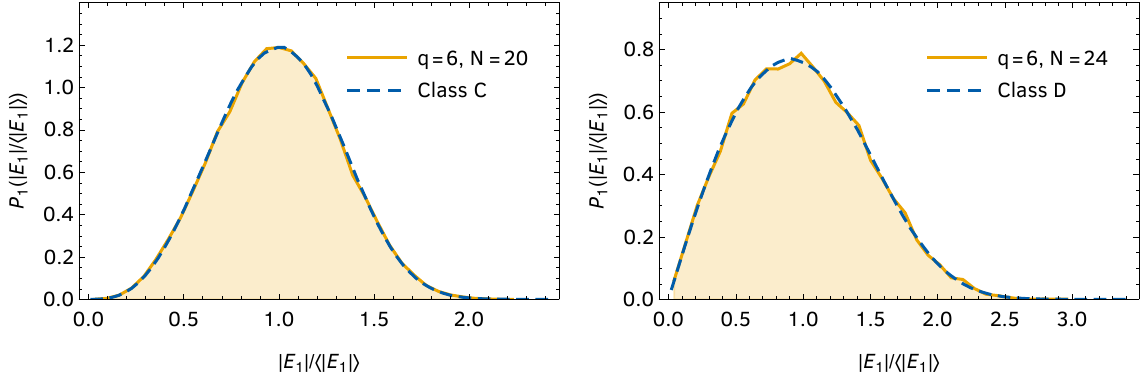}
	\caption{Distribution of the eigenvalues of $H$ with the smallest absolute value for the $q=6$ nHSYK model with $N=20$ and $N=24$ (filled histogram), compared with the random matrix theory predictions for the non-Hermitian classes C and D, respectively (dashed curves). The RMT predictions are obtained by exactly diagonalizing $10^7$ random matrices of dimension $100$ structured according to the fourth column of Table~\ref{tab:correlations_sym_class_noreal}. The nHSYK results (solid curves) are obtained from an ensemble of approximately $4.5\times10^{4}$ and $3\times10^{4}$ realizations for $N=20$ and $N=24$, respectively, resulting in much larger finite-size effects. The comparison is parameter-free and does not involve any fitting.}
	\label{fig:EMINdistq6}
\end{figure}

The distribution of the eigenvalue with the smallest modulus, $P_1(\abs{E_1})$, gives, when combined with the bulk CSR distribution, a measure to uniquely distinguish the ten non-Hermitian symmetry classes without reality conditions. As an example, in Fig.~\ref{fig:EMINdistq6}, we show the distribution of $\abs{E_1}$ for the nHSYK model with $q=6$ and $N=20$ and $24$, and compare it with the prediction of non-Hermitian random matrix theory for classes C and D, respectively.
We thus see that, while the $q = 6$ nHSYK Hamiltonian has the same bulk statistics for all $N$, we can still resolve the Bott periodicity, which enables us to distinguish universality classes, through the statistics of $|E_1|$.

\begin{figure}[tbp]
	\centering
	\includegraphics[width=0.45\textwidth]{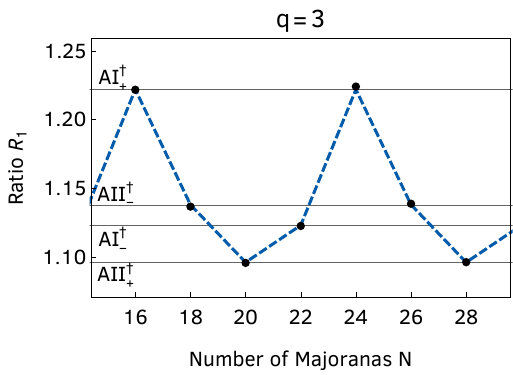}
	\includegraphics[width=0.45\textwidth]{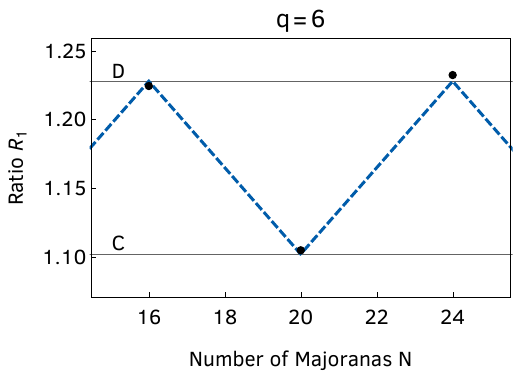}
	\caption{Ratio $R_1$ in Eq.~(\ref{eq:r1}) as a function of the number of Majoranas, $N$, for the $q = 3$ and the $q = 6$ nHSYK model. The dots are the results of exact numerical diagonalization of the corresponding nHSYK Hamiltonian, the horizontal solid lines are the values of the ratio for the six classes with spectral inversion symmetry realized in the nHSYK model, and the dashed curves follow the classification scheme of Table~\ref{tab:nHSYK_class}. For $q = 3$, the Hamiltonian has a chiral symmetry for all even $N$, while for $q = 6$, there is a particle-hole symmetry only if $N$ is a multiple of $4$. We see excellent agreement between the nHSYK results and the RMT predictions for all available system sizes.}
	\label{fig:EMINrat}
\end{figure}

To further confirm our symmetry classification, in Fig.~\ref{fig:EMINrat}, we show the value of $R_1$, defined in Eq.~(\ref{eq:r1}) as a function of $N$ for the $q=3$ and $q=6$ nHSYK model. We again see excellent agreement with the random matrix predictions, thus fully confirming the symmetry classification of Sec.~\ref{sec:1SYK_symm_class}.

\section{Limits to universality in many-body quantum chaos}
\label{sec:1SYK_long-range}

We now probe the dynamics of the nHSYK model for timescales shorter than the Heisenberg time through a detailed comparison of the unfolded dissipative DSFF with the random matrix predictions worked out in Ch.~\ref{chapter:correlations}.

\subsection{Spectral density and unfolding}

\begin{figure}[tbp]
	\centering
	\includegraphics[width=\textwidth]{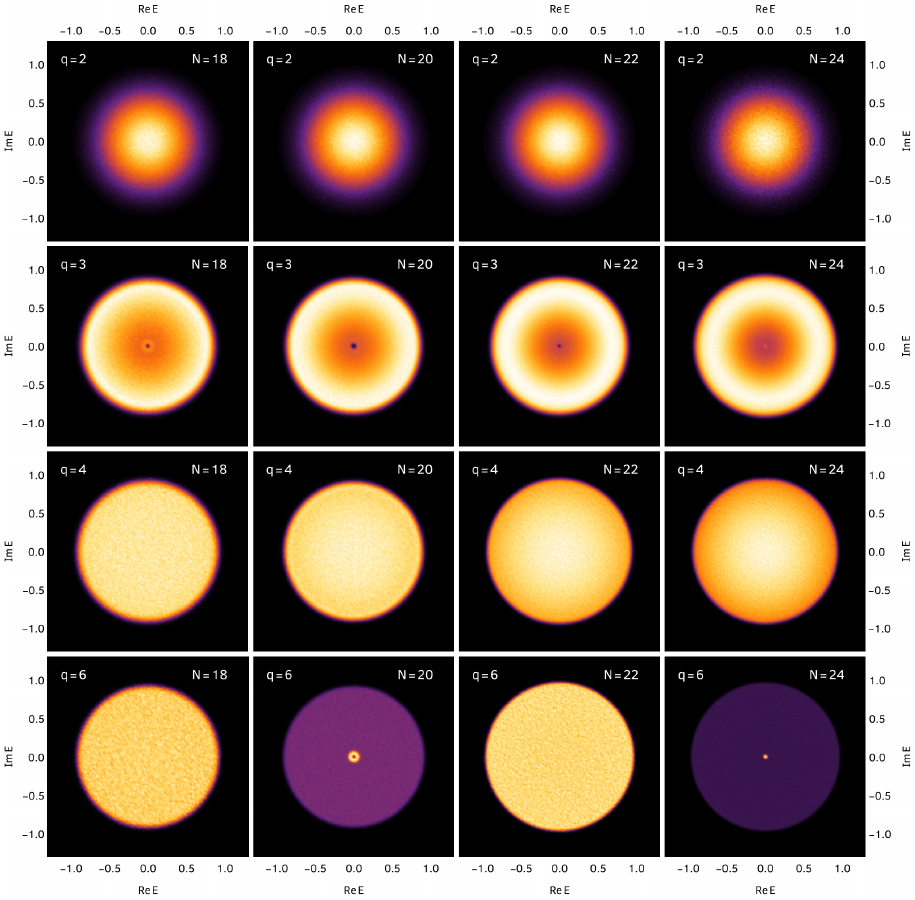}
	\caption{Spectral density of the complex eigenvalues $E_k$ of the nHSYK Hamiltonian of Eq.~(\ref{eq:1SYK_hami}) normalized to the unit disk, for $q=2$, $3$, $4$, and $6$ (top to bottom) and $N=18$, $20$, $22$, and $24$ (left to right).
	}
	\label{fig:1SYK_den}
\end{figure}

As usual, the spectral density of the nHSYK model is nonuniversal as it is determined by the details of the Hilbert space and the dynamics. For small $q<6$, is also far from constant, although it is always isotropic (radially symmetric), see Fig.~\ref{fig:1SYK_den}. 
For $q=2$, see the top row of Fig.~\ref{fig:1SYK_den}, the spectral density has a maximum in the $|E| \sim 0$ region and then decreases monotonically with no sign of a sharp spectral edge. There is no qualitative dependence on $N$. 	
The $q=3$ spectral density, depicted in the second row of Fig.~\ref{fig:1SYK_den}, is qualitatively different from the $q=2$ one. For all $N$, the spectrum has a sharp edge that becomes discontinuous in the thermodynamic limit, and the maximum is not at the center but rather in a ring not far from the edge. The chiral symmetry of the spectrum, $E \to -E$, has a rather profound effect on the density in the region $|E|\sim0$. We can observe characteristic oscillations near $|E|=0$ for $N=18$. The spectral density is, in general, suppressed in this region, although the suppression strength and pattern depend on $N$. These features are related to the different non-Hermitian universality classes.
For $q=4$, unlike the $q=3$ case, the average spectral density, shown in the third row of Fig.~\ref{fig:1SYK_den}, is rather unstructured, with a broad maximum located in the central part, followed by a slow decay for larger energies. Finally, as is the case for $q=3$, a rather sharp edge is observed where the density vanishes abruptly. Because of the absence of inversion symmetry, the spectral density is not suppressed or enhanced in the $|E|\sim0 $ region.
For $q=6$, see bottom row of Fig.~\ref{fig:1SYK_den}, the spectrum has inversion symmetry for $N=20$ and $24$, but, unlike for $q=3$ and to a lesser extent for $q=4$, the density is rather unstructured except in the $|E| \to 0$ region where, depending on $N$, we observe either a sharp suppression ($N = 20$) or a strong enhancement ($N = 24$). We find a sharp edge also in this case.

Because the nHSYK spectrum is not constant in the unit disk, we must radially unfold it before computing the DSFF. 
For a radially symmetric spectrum, $\{E_k\}_{k=1,\dots,D}$, the spectral density $\bar\varrho(E,E^*)$ satisfies
\be
\bar \varrho(E,E^*) \d^2 E = \bar \varrho(|E|)\,|E|\, \d|E|\, \d ({\rm arg}(E)).
\ee
The unfolding is performed using the average radial spectral density $\bar \varrho(|E|)$, which is a smooth function and, therefore, does not affect local statistics. The unfolded eigenvalues, $E_k^{\rm unf}$, are given by
\be
E_k^{\rm unf} = \left(2\pi \int_0^{|E_k|}  \d r\, r\bar \varrho(r) \right )^{1/2}\frac{E_k}{|E_k|}.
\ee
As a check of this transformation, we can take the flat density $\bar \varrho(|E|) =1/\pi$ which results in $E_k^{\rm unf} = E_k$.
When the spectral density is almost constant, as is the case of the nHSYK model for $q\ge 6$, unfolding can be approximated by  just rescaling the eigenvalues to the unit disk by
\be
E_k \to  E_k\sqrt{\frac{ \pi \bar \varrho{(0)}}D}.
\ee
An analytic expression for $\bar \varrho(|E|)$ is not presently known and we unfold the spectral density using eight-order polynomial fitting.

In what follows, we will compare the DSFF of the nHSYK model,
\be
K_c(\tau, \theta) 
&=& \frac1D
\left \langle
\left| \sum_{k=1}^D e^{\i\tau {\rm Re}(e^{\i\theta} E_k)} \right|^{2}
\right \rangle_{c},
\label{form2}
\ee
with the random matrix prediction in the corresponding universality class (Table~\ref{tab:nHSYK_class}). $D=2^{N/2-1}$ for even $q$ (owing to the existence of two fermion parity blocks) in cases without Kramers degeneracy, $D=2^{N/2-2}$ for $q\mod4=0$ and $N\mod8=4$ (due to fermion parity and Kramers degeneracy), and $D=2^{N/2}$ for odd $q$.
We obtain the spectrum by exact diagonalization techniques. We carry out an ensemble average to suppress statistical fluctuations, reaching at least $2\times 10^6$ eigenvalues for a given $q$ and $N$. Since the spectrum is rotationally invariant, we additionally average the DSFF over 10 values of $\theta=\pi j/5$, $j=1,\dots,10$, and denote the resulting averaged DSFF as $K_c(\tau)$.

\subsection{Long-time universality for \texorpdfstring{$q=4$}{q=4}}

\begin{figure}[tbp]
	\centering \includegraphics[width=0.88\columnwidth]{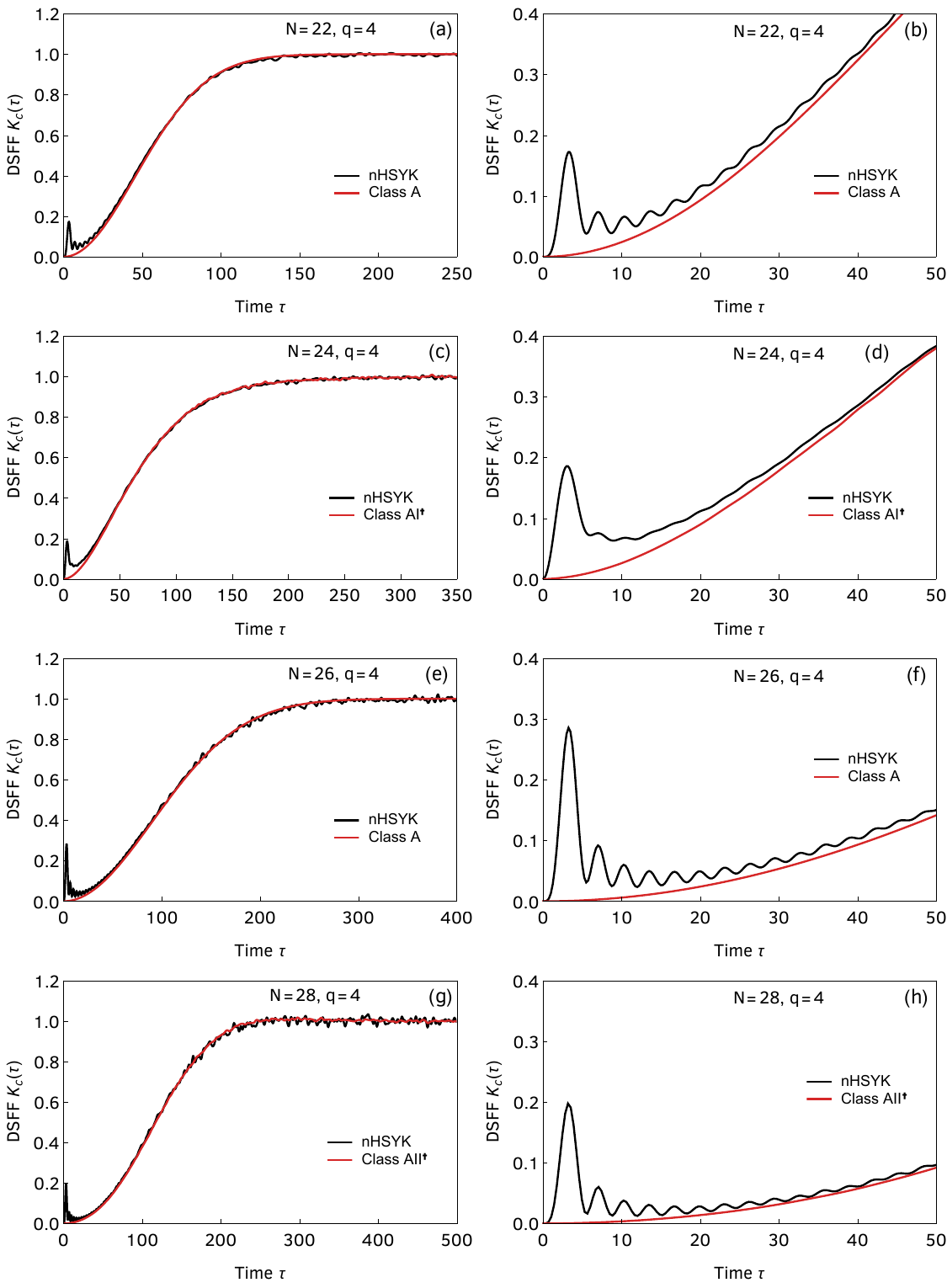}
	\caption{Connected DSFF of the unfolded eigenvalues, Eq.~(\ref{form2}), normalized by the number of eigenvalues $D$, of the $q = 4$ nHSYK model (black curves) for $N =22$, $24$, $26$, and $28$ (from top to bottom). The results are compared to the predictions of RMT (red curves) for the respective symmetry class (see Table~\ref{tab:nHSYK_class}), obtained analytically for class A [$N=22$ and $26$, Eq.~(\ref{form-an})] and numerically for class AI$^\dagger$ ($N=24$) and AII$^\dagger$ ($N=28$), from random matrices of size $2048$ through Eq.~(\ref{eq:1SYK_rescaling_Kc}). The right plots are a magnification of the left ones for small times. We find excellent quantitative agreement without any fitting up to relatively small $\tau$ of the order of the correlation hole, which signals the timescale from which the quantum chaotic dynamics is universal.}
	\label{fig:Thesis_DSFFq4}
\end{figure}

In Fig.~\ref{fig:Thesis_DSFFq4}, we depict the results for the $q=4$ nHSYK Hamiltonian for $N=20$--$28$ and compare them to RMT predictions in the respective universality class. For $N=22$ and $26$, we used the analytical result, Eq.~(\ref{form-an}). For $N=24$ and $28$, which belong to the universality class AI$^\dagger$ and AII$^\dagger$, respectively, no analytical formula is available and the random matrix result was obtained numerically for $D =2048$. The DSFF for other values of $D$ can be obtained by using the scaling relation
\be
\label{eq:1SYK_rescaling_Kc}
K_c^{(D_2)} (\tau) = K_c^{(D_1)} \left(  \sqrt{\frac{D_1}{D_2}} \tau\right ).
\ee
We find excellent agreement with the random matrix predictions for $\tau > \sqrt D$ but the two results differ for smaller times.
This is fully consistent with the results of short-range correlations which are insensitive to these deviations.

\subsection{The limits of universality: Collective scale fluctuations}

To better understand the short-time behavior of the DSFF we have enlarged the region close to the origin in the plots of the right column of Fig.~\ref{fig:Thesis_DSFFq4}.
The local minimum of $K_c(\tau)$
for $\tau > 0$, usually termed correlation hole~\cite{leviandier1986,wilkie1991,alhassid1992,torres2018PRB,schiulaz2019PRB}, defines, for a real spectrum, the maximum timescale for which the dynamics does not fully relax to the universal prediction of RMT.
In the Hermitian SYK model~\cite{garcia-garcia2018PRD}, it is determined by the collective fluctuations
of the spectrum that arise because the number of independent matrix elements ($\sim N^q$) is much smaller than the number of matrix elements of the Hamiltonian ($2^{N/2}$). The same mechanism is at work in the non-Hermitian case, where we have the same mismatch in the number of matrix elements.

The oscillations for small times are mostly due to collective \emph{scale} fluctuations~\cite{jia2020JHEP}. They correspond to fluctuations in the overall scale of eigenvalues from one realization to the next: $x_n\to x_n(1+\xi)$, for all $n$, where $x_n$ are the real parts of the eigenvalues.\footnote{Since our spectrum is isotropic, for the analytical considerations to follow, it is convenient to set $\theta=0$ and regard the DSFF as the spectral form factor (SFF) of the real parts of the eigenvalues.} $\xi$ is a random variable with zero mean and gives rise to the scale fluctuation of the spectral density:
\be
\varrho^\mathrm{scale}_\xi(x) =\frac 1{1+\xi}\,\bar \varrho\left( \frac x{1+\xi} \right ),
\ee
where $\bar\varrho$ is the ensemble-averaged spectral density. 
The connected two-point correlator for these scale fluctuations is given by
\begin{equation}
\begin{split}
\left\langle \varrho^\mathrm{scale}_\xi(x)\varrho^\mathrm{scale}_\xi(y)\right \rangle_c
&=
\left\langle\left[\bar\varrho(x)(1-\xi)-\xi x \bar\varrho'(x)\right]\left[
\bar\varrho(y)(1-\xi)-\xi y\bar\varrho'(y)\right]\right\rangle -\bar\varrho(x) \bar \varrho(y)
\\
&=
\left[\bar\varrho(x)+x\bar\varrho'(x)\right]
\left[\bar\varrho(y)+y\bar\varrho'(y)\right]\langle \xi^2\rangle,
\end{split}
\end{equation}
where the prime denotes the derivative, we have used $\langle\xi\rangle=0$, and we have dropped all terms of order $\langle\xi^4\rangle$ and above.
This contributes to the DSFF as~\cite{berkooz2021JHEPb}
\begin{equation}
\begin{split}
\delta K_c(\tau) 
&=\frac{1}{D}\int \d x \d y \left\langle \varrho^\mathrm{scale}_\xi(x)\varrho^\mathrm{scale}_\xi(y)\right \rangle_c e^{\i \tau (x-y)}
\\
&= \frac {\langle \xi^2\rangle}D
\left |\int \d x\, \frac \d{\d x}[x\bar\varrho(x)] e^{\i \tau x}\right |^2
\\
&= \frac {\langle \xi^2\rangle}D \tau^2
\left |\int \d x\, x\,\bar \varrho(x) e^{\i \tau x}\right |^2,
\label{kt-sc}
\end{split}
\end{equation}
where, in the last equality, we integrated by parts.
The spectral density of the real parts of the eigenvalues is given by the semicircle distribution, normalized as $\int\bar\varrho(x) dx =D$,
\be
\bar \varrho(x) = \frac {2D}{\pi} \sqrt{1-x^2},
\ee
and its Fourier integral is given by
\be
\frac 2\pi D \int_{-1}^{1} dx\, x \, e^{\i \tau x} \sqrt{1-x^2}
= 2 i  D  \frac{J_2(\tau)}\tau,
\ee
where $J_2$ is a Bessel function. The contribution of the scale fluctuations to the DSFF is, thus,
\be
\label{delK}
\delta K_c(\tau) &=& 4 D\langle \xi^2\rangle [J_2(\tau)]^2,
\ee
which decreases as $1/\tau$ for large $\tau$. The analytical result for the total DSFF in class A, including the scale fluctuations factor, is then given by
\be
\label{eq:total_SFF_with_scale}
K_c(\tau) &=& 1-e^{-\tau^2/4D} + 4D\langle \xi^2\rangle [J_2(\tau)]^2.
\ee

\begin{figure}[tbp]
	\centering
	\includegraphics[width=0.65\textwidth]{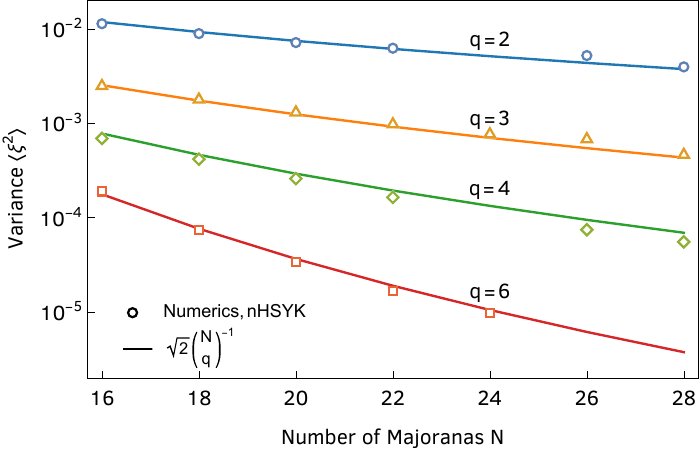}
	\caption{Variance $\langle\xi^2\rangle$ computed numerically for the nHSYK as a function of $N$ for different $q$. We see good agreement with $\sqrt{2}\binom{N}{q}^{-1}$.}
	\label{fig:Thesis_variancexi2}
\end{figure}

The variance $\langle\xi^2\rangle$ can be computed as~\cite{jia2020JHEP}
\be
\langle \xi^2 \rangle =\frac 14 \left(\frac {M_{2,2}}{M_2^2} -1\right ),
\label{xi}
\ee
with the moments defined as
\begin{align}
\label{eq:1SYK_moment2}
M_2 &=  \left \langle \frac 1D\sum_k x_k^2 \right\rangle,
\\
\label{eq:1SYK_moment22}
M_{2,2} &= \left\langle \frac 1D \sum_k x_k^2\frac 1D \sum_k x_k^2 \right \rangle.
\end{align}
For the Hermitian SYK model, $x_k$ are the eigenvalues of the Hamiltonian, and the moments $M_2\equiv M_{2,0}$ and $M_{2,2}$ can be identified as double trace moments of the Hamiltonian, $M_{m,n}=\av{\Tr H^m \Tr H^n}/D^{m+n}$, and can be evaluated exactly~\cite{jia2020JHEP}:
\be
\langle\xi^2\rangle=\frac 12 {\binom{N}{q}}^{-1}.
\label{xi2}
\ee
For $x_k$ the real parts of the eigenvalues of a non-Hermitian matrix with spectral density unfolded to constant density
inside the unit disk, $M_{2,0}$ and $M_{2,2}$ cannot be written as traces of moments of the Hamiltonian, but $\langle \xi^2\rangle$ can be computed numerically from Eqs.~(\ref{eq:1SYK_moment2}) and (\ref{eq:1SYK_moment22}). For our nHSYK model, we find that it is approximately equal to
\be
\langle \xi^2\rangle \approx \sqrt{2}{\binom{N}{q}}^{-1},
\ee
see Fig.~\ref{fig:Thesis_variancexi2}.

\begin{figure}[tbp]
	\centering
	\includegraphics[width=\columnwidth]{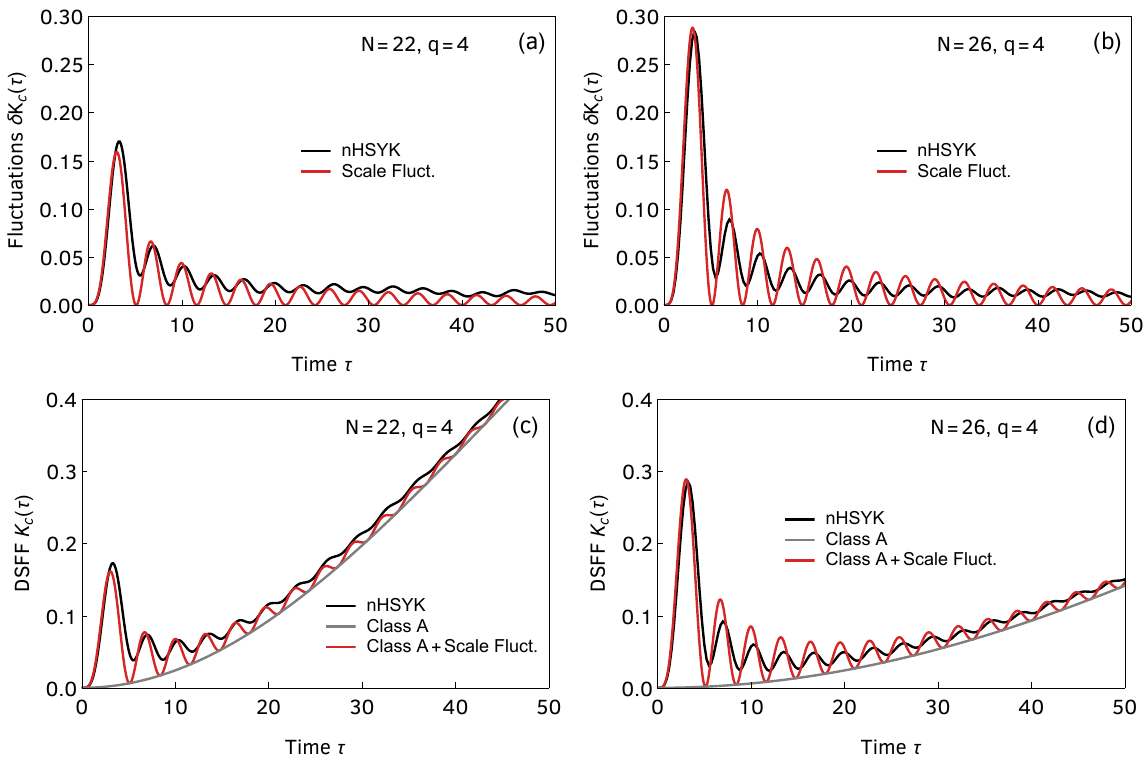}
	\caption{Deviations of the nHSYK DSFF due to collective scale fluctuations as a functions of time, for $q=4$ and $N=22$ (left) and $26$ (right). Top row: Difference $\delta K_c(\tau)$ between the connected DSFF of nHSYK model and the universal class A result, Eq.~(\ref{form-an}), (black curves), compared to the analytical result due to scale fluctuations, Eq.~(\ref{delK}), (red curves).
	Bottom row: Connected DSFF $K_c(\tau)$ obtained numerically for the nHSYK model [same as Figs.~\ref{fig:Thesis_DSFFq4}~(b) and (f)] (black curves), compared to the class A predictions (grey curves), and the sum of the class A and collective fluctuations contribution (red curves). We see that the collective scale fluctuations explain the oscillations in the nHSYK DSFF and account for most of the deviations from RMT universality.}
	\label{fig:Thesis_DSFFq4Fluct}
\end{figure}

In the top row of Fig.~\ref{fig:Thesis_DSFFq4Fluct}, we
show the difference $\delta K_c$ between the DSFF of the nHSYK model for $q=4$ and the RMT result for class A. The numerical results (black curves) for $N=22$ (left) and $N=26$ 
(right) are compared to the analytical result~\eref{delK} due to scale fluctuations (red curves). 
In the bottom row of Fig.~\ref{fig:Thesis_DSFFq4Fluct}, we compare the full analytical DSFF, Eq.~(\ref{eq:total_SFF_with_scale}), with the numerical results. 
Given that higher multipole collective fluctuations also contribute to the difference, the agreement
with the analytical result is better than expected, in particular for small times. We thus conclude that most of the oscillatory behavior is due to the lowest-order multipole, i.e., the scale fluctuations. We emphasize that the results for $\delta K_c(\tau)$ are obtained without
using fitting parameters [we use the numerical value of $\av{\xi^2}$ computed from Eq.~(\ref{xi})].
Note that the period of the oscillations does not depend
on $N$ and is close to the period of the oscillations of $J_2^2(\tau)$. The amplitude
increases with $D$ and also varies as the amplitude of $J_2^2(\tau)$.

The location of the correlation hole can be obtained by equating the two contributions to the DSFF: the universal RMT contribution, Eq.~(\ref{form-an}), and the collective fluctuations contribution, Eq.~(\ref{delK}). By replacing the oscillatory part of $J_2^2$
by its asymptotic envelope, the location of the correlation hole is thus given by the minimum of 
\be
4\langle \xi^2\rangle\frac 1\pi  \frac { D }{\tau} + 1-e^{- \frac {\tau^2}{4D}}.
\label{benv}
\ee
This condition cannot be solved analytically, but it gives the rough position of the correlation hole and can
be studied numerically for small values of $N$. From inspection of the plots of the DSFF in class A, we find good agreement with this estimate.
The condition~(\ref{benv}) can be recast as
\be
\frac{1}{\pi}\langle\xi^2\rangle \sim \frac{\tau^3}{8D^2}e^{-\tau^2/4D}.
\label{scale_nH}
\ee
When $\langle \xi^2\rangle \gtrsim 1/\sqrt D $, this condition no longer has a solution for a real time $\tau$ and there is no correlation hole. In the case $q=4$, this occurs for $N\gtrsim 80$. Since the contribution of the scale fluctuation dominates the $\tau$-dependence of the DSFF
all the way up to the Heisenberg time, the DSFF is no longer a useful measure for spectral fluctuations
due to quantum chaos. 

This result is to be contrasted with the Hermitian case, for which the location of the correlation hole is roughly determined by the condition,
\be
\langle \xi^2 \rangle \sim \frac{\tau^3}{D^2}.
\ee
Although $\langle \xi^2 \rangle$ is approximately the same as before, the Heisenberg time is now of order $D$ (instead of $\sqrt{D}$). As a consequence, there are real solutions $\tau$ for all values of $\langle\xi^2\rangle$. Furthermore, the correlation hole would
only be larger than the Heisenberg time if $\langle \xi^2\rangle\gtrsim D$, a condition that is never satisfied. We conclude that for the Hermitian SYK model, there is always a parametrically large separation between the timescale where collective fluctuations are relevant for the SFF and the Heisenberg time, contrary to the DSFF of the nHSYK model.

Finally, we note that it is possible to eliminate the collective spectral fluctuations by unfolding
the spectrum realization by realization \cite{gharibyan2018JHEP,jia2020JHEP}. Then
these oscillations do not show up in the DSFF.

\subsection{Dependence of nonuniversal features on \texorpdfstring{$q$}{q}}

\begin{figure}[tbp]
	\centering
	\includegraphics[width=\textwidth]{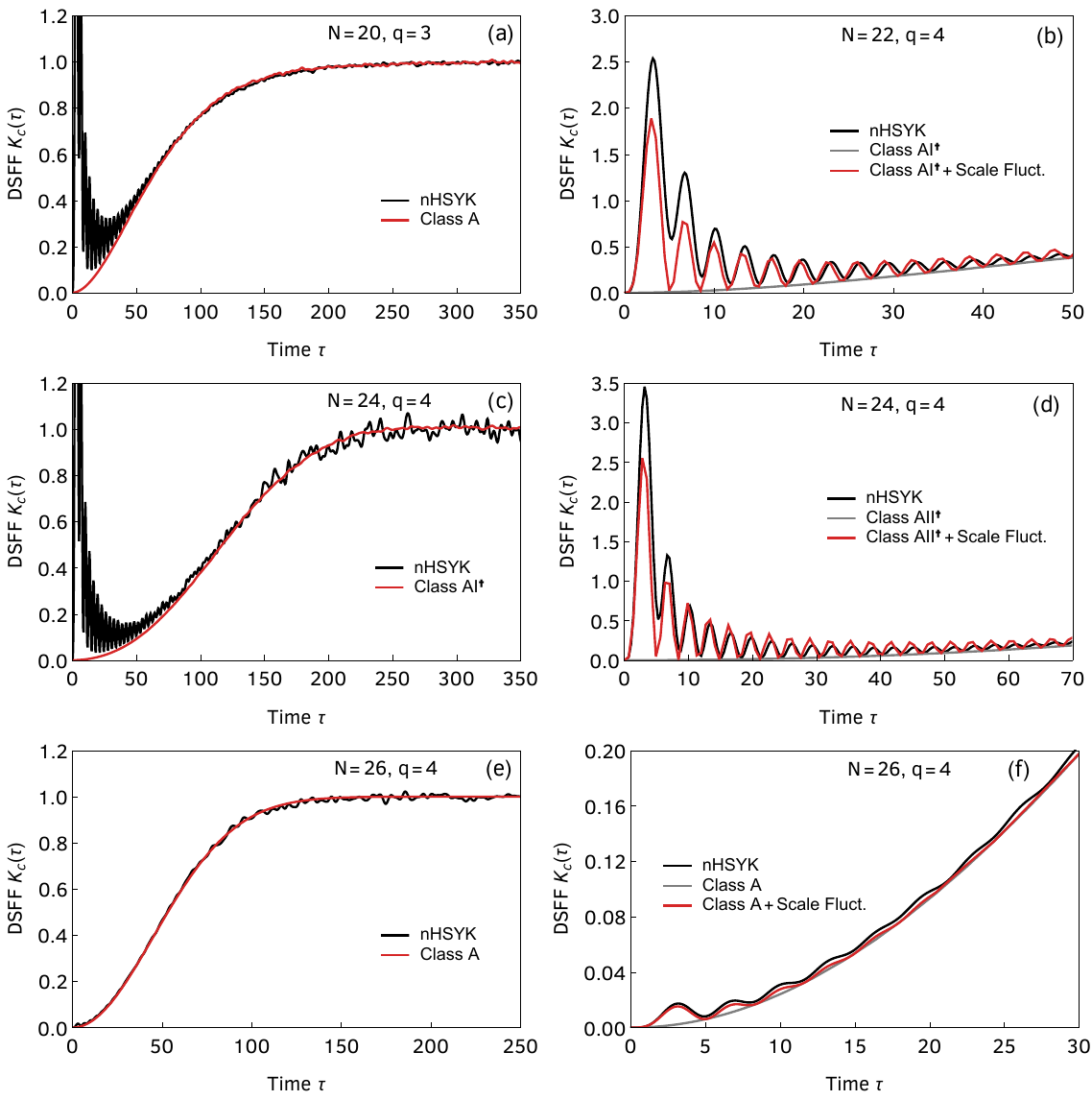}
	\caption{Connected DSFF of the unfolded eigenvalues, Eq.~(\ref{form2}), normalized by the number of eigenvalues $D$, of the nHSYK model for $q=3$ and $q=6$ (black curves). The results are compared to the RMT predictions in the respective class (grey curves) and the sum of the RMT and collective fluctuations contribution (red curves). The right plots are a magnification of the left ones for small times. As for $q=4$, the collective scale fluctuations explain the oscillations in nHSYK DSFF and account for most of the deviations from RMT universality.
	\label{fig:Thesis_DSFFq36}
	}
\end{figure}

The results for the $q = 3$ and $q=6$ nHSYK Hamiltonian, see Fig.~\ref{fig:Thesis_DSFFq36}, confirm the picture obtained for $q=4$. Agreement with the random matrix predictions in the expected universality class is observed for large times.
The area below the small-time peak decreases markedly for increasing values of $q$. This is expected since a larger $q > 2$ brings the nHSYK Hamiltonian closer to a random matrix, as more entries of the Hamiltonian are nonzero. From our previous discussion, the area below the peak is proportional to $2^{N/2}/\binom{N}{q}$. For $N=24$, this is given by $14.84$, $2.02$, $0.39$, and $0.03$ for $q=2$, $3$, $4$, and $6$, respectively.

The oscillatory behavior in the small-$\tau$ region for $q=3$, although not qualitatively different from $q=4$, has a much larger amplitude than in the $q=4$ case (see the right panels of
Fig.~\ref{fig:Thesis_DSFFq36}). This results in
a correlation hole that is shifted to a larger value of $\tau$.
On the other hand, for $q=6$, the amplitude of the oscillations is very small, and we barely observe
any deviation from the random matrix predictions.
For the SYK model with real couplings, it can be shown~\cite{erdos2014MPAG} that for $q \gg \sqrt{N}$ the SYK Hamiltonian resembles a random matrix with a semicircular spectral density.
For the nHSYK model, this corresponds to a constant level density inside the eigenvalue disk
so that the real parts of the eigenvalues are distributed according to a semicircle.
For $q = 6$, we are likely in this asymptotic region.

\subsection{Integrable behavior for \texorpdfstring{$q=2$}{q=2}}

\begin{figure}[tbp]
	\centering
	\includegraphics[width=\textwidth]{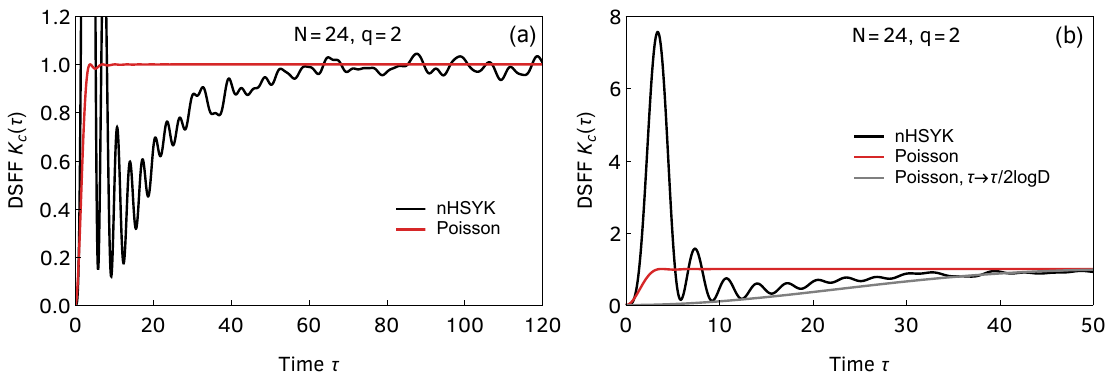}
	\caption{Connected DSFF of the unfolded eigenvalues, Eq.~(\ref{form2}), normalized by the number of eigenvalues $D=2^{N/2-1}$, of the nHSYK model for $q=2$ and $N=24$ (black curves). The results are compared to the Poisson prediction, Eq.~(\ref{eq:DSFF_Poisson}) (red curves).
	The right plot is a magnification of the left one for small times, where we also plot the Poisson prediction with a shift of the horizontal axis $\tau\to\tau/2\log D$, seeing a reasonable fit with the numerics. As was expected, the $q = 2$ nHSYK model shows correlations that are in between Poisson statistics and RMT statistics. The reason is that the integrable many-body spectrum is determined by a small number ($N/2$) of chaotic single-particle energies. The oscillations observed for small, but nonzero, $\tau$ are due to collective fluctuations of the spectral density.}
	\label{fig:Thesis_DSFFq2}
\end{figure}

The $q=2$ SYK and nHSYK models are both integrable with all energy levels determined by $N/2$ single-particle energies. In this case, we expect Poisson level statistics for sufficiently long times, but deviations may be observed for shorter times. Indeed, the DSFF has more structure than plain Poisson statistics of completely uncorrelated random variables.
As illustrated in Fig.~\ref{fig:Thesis_DSFFq2}, the DSFF
saturates to the Poisson limit, $K_c(\tau)= 1$, at a scale of order $\log D$, which is much shorter than for $q>2$,
where the scale is determined by $\sqrt D$ (for $N=24$ the two scales are of the same order of magnitude and our data cannot really distinguish between the two). 
The analytical result for uncorrelated eigenvalues unfolded to
constant density inside the complex unit disk, given by Eq.~(\ref{eq:DSFF_Poisson}),
saturates to Poisson statistics at $\tau = \mathcal{O}(1)$ and
does not match the numerical result (see the red curves in Fig.~\ref{fig:Thesis_DSFFq2}).
A reasonable fit is obtained by replacing
$ \tau \to \tau / 2\log D $ (grey curve in the right panel of Fig.~\ref{fig:Thesis_DSFFq2}), but we have no rigorous argument for this substitution.

Physically, the saturation scale of the $q=2$ DSFF---shorter than RMT, longer than Poisson---is related to the fact that the model can be mapped onto free fermions with single-particle energies correlated according to RMT~\cite{cotler2017JHEP}.
The short-time dynamics, controlled by the single-particle excitations, will be very different from that expected for a generic integrable system. However, for longer times
of the order $\log D$, multiparticle excitations
will reveal the generic integrable nature of the quantum dynamics.

As is the case for $q=3$ and $q=4$, we find oscillations for
small values of $\tau $ with the same period but with a larger amplitude. These oscillations, which dominate the quadratic $\tau$-dependence, are due to scale fluctuations of the average spectral density.

\section{Summary and outlook}

We have proposed a single-site SYK model with complex couplings as a toy model for many-body non-Hermitian quantum chaos. We have shown that 9 out of the 10 non-Hermitian symmetry classes without reality conditions occur naturally in this SYK model. 
It describes both generic features of the universal quantum ergodic state reached around the Heisenberg time and nonuniversal, but still rather generic, properties of quantum interacting systems in its approach to ergodicity.

A detailed spectral analysis of the nHSYK model, involving short-range correlators such as the CSRs in the bulk and the distribution of the smallest eigenvalue near the hard edge, has revealed an excellent agreement with the RMT predictions for nine universality classes occurring for different choices of $N$ and $q$ in the Hamiltonian. 

We have also studied long-range correlations of the non-Hermitian SYK model by means of the DSFF (the SFF of projected eigenvalues) with results for the Ginibre or Ginibre-like ensembles as a benchmark.
For small times, the DSFF deviates from the RMT result and shows an oscillatory behavior with a period conjugate to the overall width of the spectrum and an amplitude decreasing with time that is very sensitive to the number $q$ of interacting Majoranas, a parameter that controls the fraction of
independent matrix elements of the Hamiltonian in Fock space. The area below the peak is proportional to $2^{N/2} / \binom{N}{q}$ and decreases rapidly going from $q=2$ to $q=6$. Averaging over the oscillations, the small-time behavior of the DSFF is similar to that of Hermitian systems with a correlation hole. Although this is a typical feature of strongly interacting quantum chaotic systems, in this case, it is caused by collective ensemble fluctuations rather than by the nonuniversal dynamics at that timescale. 

A feature of spectral correlations of the projection of complex eigenvalues is that eigenvalues that are many level spacings apart, and are essentially uncorrelated, can have projections that are close. This results in Poisson statistics already after a timescale of $\sim \sqrt D$.
The early onset of Poisson statistics has the consequence that the collective spectral fluctuations can no longer be separated from universal eigenvalue fluctuations. 

Natural extensions of this work include the analytical calculation of the spectral density of the nHSYK model by combinatorial techniques and a short-time characterization of many-body dissipative quantum chaos in this nHSYK model by the evaluation of the Lyapunov exponent resulting from an out-of-time-order correlation function~\cite{larkin1969,kitaev2015TALK1,kitaev2015TALK2,kitaev2015TALK3,maldacena2016PRD} for times of the order of the Ehrenfest time. The latter could help dynamically characterize non-Hermitian quantum chaos. For real spectra, we have the Bohigas-Giannoni-Schmit conjecture that relates dynamical and spectral correlations. However, despite the heavy use of terminology borrowed from the real-spectrum case, it is still unclear to what extent agreement with non-Hermitian random matrix predictions is related to quantum chaos in the original sense of quantum dynamics of classically chaotic systems.

Additionally, we mention the possibility of realizing further symmetry classes in two-site nHSYK models. These models, relevant for wormhole physics~\cite{maldacena2018ARXIV,garcia-garcia2021PRD,garcia-garcia2019PRD} and also in the exploration of dominant off-diagonal replica configurations~\cite{garcia-garcia2022PRL}, provide an interesting playground to implement not only the universality classes already realized in the nHSYK but also going beyond by tuning the additional parameters appropriately. This will be the subject of a future publication~\cite{garcia2023ARXIVb}.

%% file: Thesis_ClassificationLindblad.tex

\chapter{Symmetry classification of many-body Lindbladians: Tenfold way and beyond}
\label{chapter:classificationLindblad}

In this chapter, we consider the symmetry classification of the dynamical generators of physically-consistent dynamics.
Non-Hermitian Hamiltonians provide an effective description of open quantum dynamics only when quantum jumps can be neglected, for instance, for short times or postselecting jump-free quantum trajectories. A complete description of an open quantum system coupled to a Markovian (i.e., memoryless) environment must go beyond the non-Hermitian Hamiltonian description, and one should consider systems evolving under the action of Liouvillian superoperators of Lindblad form. It is a question of fundamental interest to find out how many symmetry classes can be realized by many-body Lindbladians, which are far more constrained than arbitrary non-Hermitian Hamiltonians, specifically by the conservation of trace, Hermiticity, and (complete) positivity. In other words, we ask to which subset of the 54-fold classification do physical open quantum systems belong. Lieu, McGinley, and Cooper~\cite{lieu2020PRL} used causality arguments to argue that there are also ten classes of single-particle spectra of noninteracting (quadratic) Lindbladians. However, they did not consider shifting the spectral origin, which avoids the causality restrictions, as pointed out by Kawasaki, Mochizuki, and Obuse~\cite{kawasaki2022PRB}. Once this possibility is accounted for, more than ten classes of non-Hermitian Hamiltonians can be implemented at the level of single-particle spectra. The importance of the shift of the spectral origin, and the associated spectral dihedral symmetry, was already noted for many-body Lindbladians in Refs.~\cite{prosen2012PRL,prosen2012PRA}, but a symmetry classification was not put forward. 

Here, we take this fundamental step and show that many-body Lindbladians possess a rich symmetry classification.
Our classification is based on the behavior of the many-body matrix representation of the Lindbladian under antiunitary symmetries and unitary involutions. We find that Hermiticity preservation reduces the number of symmetry classes, while trace preservation and complete positivity do not, and that the set of admissible classes depends on the presence of additional unitary symmetries: in their absence or in symmetry sectors containing steady states, many-body Lindbladians belong to one of ten non-Hermitian symmetry classes; if however, there are additional symmetries and we consider non-steady-state sectors, they belong to a different set of 19 classes. In both cases, it does not include classes with Kramers degeneracy.
Moreover, the number of distinct symmetry classes of Lindbladian dynamics in symmetry sectors that contain the steady state(s) is exactly the same (ten) as the number of distinct Altland-Zirnbauer symmetry classes of Hermitian steady-state density operators, although the precise correspondence remains to be understood.
Remarkably, our classification admits a straightforward generalization to the case of non-Markovian, and even non-trace-preserving, open quantum dynamics.

Our work is qualitatively different from the previous attempt at a symmetry classification of fermionic open quantum matter by Altland, Diehl, and Fleischhauer~\cite{altland2021PRX}. 
Specifically, Ref.~\cite{altland2021PRX} considers the invariance of the dynamics under linear or antilinear and canonical or anticanonical transformations of fermionic creation and annihilation operators, while our transformations apply to any kind of Hilbert space (including second-quantized Lindbladians in Fock space) and are defined by general transformation properties of the \emph{matrix representation} of the Lindbladian. As such, our classification scheme accurately captures many-body spectral and eigenvector properties, as relevant, e.g., for quantum chaos.

The rest of this chapter is organized as follows. First, in Sec.~\ref{sec:classification}, we establish the general symmetry classification of many-body Lindbladians and determine conditions that the Hamiltonian and jump operators must satisfy in a given class.
While the abstract classification is completely general, we then apply it to general (long-range, interacting, spatially inhomogeneous) spin-$1/2$ chains. Specifically, in Sec.~\ref{sec:examples}, we explicitly build examples in all ten classes of Lindbladians in steady-state sectors, describing standard physical processes such as dephasing, spin injection and absorption, and incoherent hopping, thus illustrating the relevance of our classification for practical physics applications. 
Finally, in Sec.~\ref{sec:rmt}, we show that the examples in each class display unique random-matrix correlations. To fully resolve all symmetries, we employ the combined analysis of bulk complex spacing ratios and the overlap of eigenvector pairs related by symmetry operations. We further find that statistics of levels constrained onto the real and imaginary axes or close to the origin are not universal due to the spontaneous breaking of Liouvillian PT symmetry.

\section{Lindbladian symmetry classification}
\label{sec:classification}

\subsection{Matrix representation of the Lindbladian}

We consider the quantum master equation for the system's density matrix, $\pd_t\rho=\scL \rho$, where the Liouvillian superoperator is of the Lindblad form,
\begin{equation}
\scL \rho= -\i \comm{H}{\rho}+\sum_{m=1}^{M}\(2L_m\rho L_m^\dagger-\acomm{L_m^\dagger L_m}{\rho}\),
\label{lindf}
\end{equation}
with Hamiltonian $H$ and $M$ traceless jump operators $L_m$, $m=1,\dots,M$ acting over a Hilbert space $\mathcal{H}$. The Lindbladian admits a matrix representation (vectorization) over a doubled Hilbert space $\mathcal{H}\otimes\mathcal{H}$ (the so-called Liouville space), $\scL=\scLH+\scLD+\scLJ$, where the Hamiltonian, dissipative, and jump contributions are, respectively, given by
\begin{align}
	\label{eq:scLH}
	\scLH &= -\i \(H\otimes \id -\id \otimes H^*\),
	\\
	\label{eq:scLD}
	\scLD&=-\(
	\sum_{m=1}^M L_m^\dagger L_m \otimes \id 
	+ \id \otimes \sum_{m=1}^M \(L_m^\dagger L_m\)^*
	\),
	\\
	\label{eq:scLJ}
	\scLJ&=2\sum_{m=1}^M  L_m\otimes L_m^*.
\end{align}
$()^*$ denotes complex conjugation in a matrix representation with respect to a fixed basis of $\mathcal H$ (or 
$\mathcal H\otimes \mathcal H$).
We see below that the three contributions have different transformation properties. We further define the traceless shifted Lindbladian~\cite{prosen2012PRL},
\begin{equation}\label{eq:scL_prime}
\scL'=\scL-\alpha\,\mathcal{I},
\qquad
\alpha=\frac{\Tr \scL}{\Tr \scI}=
-2\,\frac{\sum_m \Tr L_m^\dagger L_m}{\Tr \id},
\end{equation} 
where $\scI=\id\otimes \id$ is the identity operator over the Liouville space. As we show below, the symmetry classification of Lindbladians is necessarily formulated in terms of $\scL'$.

\subsection{Superoperator symmetries}

Just as for the Hamiltonian case, the symmetry classification of the Lindbladian follows from the behavior of its irreducible blocks under involutive antiunitary (superoperator) symmetries. More precisely, if there is a unitary superoperator $\scU$ that commutes with the Lindbladian $\scL$,
\begin{equation}
\scU\scL\scU^{-1}=\scL,
\end{equation}
we can block diagonalize (reduce) $\scL$ into sectors of fixed eigenvalues of $\scU$. For the moment, let us assume no such unitary symmetries exist and the Lindbladian is irreducible; we consider unitary symmetries in Sec.~\ref{subsec:unitary_syms}. As in Ch.~\ref{chapter:correlations}, we look for the existence of antiunitary superoperators $\scT_\pm$ and $\scC_\pm$ and unitary superoperators $\scP$ and $\scQ_\pm$, such that $\scL$ satisfies
\begin{alignat}{99}
	\label{eq:nHsym_Tp}
	&\scT_+ \scL \scT_+^{-1} = +\scL,\qquad 
	&&\scT_+^2=\pm 1,
	\\
	\label{eq:nHsym_Tm}
	&\scT_- \scL \scT_-^{-1} = -\scL,\qquad 
	&&\scT_-^2=\pm 1,
	\\
	\label{eq:nHsym_Cp}
	&\scC_+ \scL^\dagger \scC_+^{-1} = +\scL,\qquad 
	&&\scC_+^2=\pm 1,
	\\
	\label{eq:nHsym_Cm}
	&\scC_- \scL^\dagger\scC_-^{-1} = -\scL,\qquad 
	&&\scC_-^2=\pm 1,
	\\
	\label{eq:nHsym_P}
	&\scP \scL \scP^{-1} = -\scL,\qquad 
	&&\scP^2=1,
	\\
	\label{eq:nHsym_Qp}
	&\scQ_+ \scL^\dagger \scQ_+^{-1} =  +\scL,\qquad 
	&&\scQ_+^2=1,
	\\
	\label{eq:nHsym_Qm}
	&\scQ_- \scL^\dagger \scQ_-^{-1} = - \scL,\qquad 
	&&\scQ_-^2=1.
\end{alignat}
As before, we do not consider the existence of more than one antiunitary of a given kind, since we assumed $\scL$ to be irreducible. The unitary involutions, which are nontrivial only in the absence of antiunitary symmetries, can either commute or anticommute with each other and with the antiunitary symmetries; that is,
\begin{align}
	\label{eq:eps_1}
	&\scP \scT =\epsilon_{\scP\!\scT}\ \scT \scP,
	\quad
	&\scP \scC =\epsilon_{\scP\scC}\ \scC \scP,
	\\
	&\scQ \scT =\epsilon_{\scQ\!\scT}\ \scT \scQ,
	\quad
	&\scQ \scC =\epsilon_{\scQ\scC}\ \scC \scQ,
	\\
	\label{eq:eps_5}
	&\scQ \scP =\epsilon_{\scP\!\scQ}\ \scP \scQ,
\end{align}
where all $\epsilon=\pm1$ and $\scT$, $\scC$, and $\scQ$ can be one of $\scT_\pm$, $\scC_\pm$, or $\scQ_\pm$, respectively. Only three of the $\epsilon$ are independent, say, $\epsilon_{\scP\!\scT}$, $\epsilon_{\scQ\!\scT}$, and $\epsilon_{\scP\!\scQ}$. The remaining two are determined by $\epsilon_{\scP\scC}=\epsilon_{\scP\!\scQ}\epsilon_{\scP\!\scT}$ and $\epsilon_{\scQ\scC}=\epsilon_{\scQ\!\scT}$.

\subsection{Lindbladians without unitary symmetries}

The spectrum of the Lindbladian cannot be freely rotated since there is a preferred axis of symmetry (the negative real axis), and hence Lindbladians belong to one of the 54 line-gap spectra classes. However, not all these symmetry classes can be realized in Lindbladian dynamics because of the special structure of the Lindblad superoperator. 

First, we notice that because $\scL$ preserves the Hermiticity of the density matrix, $\(\scL \rho\)^\dagger=\scL \rho^\dagger$, the eigenvalues of $\scL$ come in complex-conjugate pairs and we always have a $\scT_+$ symmetry squaring to $+1$, given by Eq.~(\ref{eq:nHsym_Tp}) with $\scT_+=\scK\swap$, where $\scK$ is the complex-conjugation superoperator defined by $\scK\rho=\rho^*$ and $\scK \scL \scK^{-1}=\scL^*$, and the \textsc{swap} operator $\swap$ exchanges the two copies of the doubled Hilbert space, $\swap \(A\otimes B\) \swap=B\otimes A$ for any operators $A,B$, and satisfies $\swap^2=+1$. Obviously, the same conclusion holds for the shifted Lindbladian $\scL'$. There are 15 symmetry classes out of the 54 that satisfy $\scT_+^2=+1$ (dubbed AI, AI$_+$, AI$_-$, BDI$^\dagger$, DIII$^\dagger$, BDI, CI, BDI$_{++}$, BDI$_{+-}$, BDI$_{-+}$, BDI$_{--}$, CI$_{+-}$, CI$_{++}$, CI$_{--}$, and CI$_{-+}$). Second, a transposition symmetry $\scC_+$ is also allowed and determines the bulk level repulsion~\cite{hamazaki2020PRR}. 

By considering the bare Lindbladian $\scL$, it would seem we have exhausted the possible symmetries. Indeed, because $\scL$ is trace preserving and completely positive, its spectrum always has a zero eigenvalue (corresponding to the steady state) and the remaining eigenvalues have nonpositive real parts, which forbids any possible symmetries that reflect the spectrum across either the origin or the imaginary axis~\cite{lieu2020PRL}, i.e., $\scT_-$ and $\scC_-$. On the other hand, the spectrum of the shifted Lindbladian $\scL'$ is centered at the origin and there are eigenvalues with both positive and negative real parts. Hence, both $\scT_-$ and $\scC_-$ are allowed symmetries of $\scL'$. This is an immediate consequence of the well-known fact that while the involutive symmetries are usually stated as in Eqs.~(\ref{eq:nHsym_Tp})--(\ref{eq:nHsym_Qm}), they need only hold up to addition of multiples of the identity. For instance, we can modify Eq.~(\ref{eq:nHsym_Tm}) to
\begin{equation}
\label{eq:nHsym_Tm_scL'}
\scT_- \scL \scT_-^{-1} = -\scL+2\alpha \scI,
\end{equation}
for some real constant $\alpha$. If we take $\alpha$ to be as defined in Eq.~(\ref{eq:scL_prime}), the previous equation can be rewritten as a standard symmetry condition for $\scL'$:
\begin{equation}
\scT_- \scL' \scT_-^{-1} = -\scL'.
\end{equation}
The $\scC_-$, $\scP$, and $\scQ_-$ symmetry transformations in Eqs.~(\ref{eq:nHsym_Cm}), (\ref{eq:nHsym_P}), and (\ref{eq:nHsym_Qm}), have to be redefined in the same way. On the other hand, no redefinition of $\scT_+$, $\scC_+$, and $\scQ_+$ symmetries is necessary, as we can trivially add $-\alpha\scI$ to both sides of Eqs.~(\ref{eq:nHsym_Tp}), (\ref{eq:nHsym_Cp}), and (\ref{eq:nHsym_Qp}) to rewrite them in terms of $\scL'$.
The possibility of shifting the spectrum is usually ignored because shifts in energy are irrelevant; i.e., we can always choose Hamiltonians to be traceless. However, the trace of the Lindbladian is not arbitrary and generalized transformations in terms of $\scL'$ have to be considered. 

Before proceeding, we note that instead of organizing the 15 classes in terms of the antiunitary symmetries present besides the $\scT_+$ symmetry, it will also prove convenient to alternatively label a class by its unitary involutions $\scP$ and $\scQ_\pm$. This also offers a check on our counting of the classes: there is one class with no unitary involutions; if there is one additional unitary involution, it can be either $\scP$, $\scQ_+$, or $\scQ_-$, and in each case it can either commute or anticommute with $\scT_+$, i.e., $3\times2=6$ classes; if two additional unitary involutions are present we can, without loss of generality, consider them to be $\scP$ and $\scQ_+$ (the other two combinations are obtained by taking one of $\scP$ or $\scQ_+$ and their product as the two independent involutions, since the product $\scP \scQ_+$ is a $\scQ_-$ symmetry), which either commute or anticommute with each other and with $\scT_+$, i.e., $2\times2\times2=8$ classes; there is no class with the three involutions since the product $\scP\scQ_+\scQ_-$ is a unitary symmetry commuting with the Lindbladian, which we assume not to exist; in total, we thus have $1+6+8+0=15$ classes. The $\scP$ and $\scQ_\pm$ symmetries of the Lindbladian then induce antiunitary symmetries through the relations
\begin{equation}\label{eq:antiunitary_from_unitary}
\scT_-=\scP \scT_+,
\quad
\scC_-=\scQ_- \scT_+,
\quad \text{and}\quad
\scC_+=\scQ_+ \scT_+.
\end{equation}
Furthermore, the square of the antiunitary symmetries and the commutation relations of the unitary involutions are related by
\begin{align}
	\label{eq:sq_commT}
	&\scT_-^2=\epsilon_{\scP\!\scT_+}\scT_+^2,
	\\
	&\scC_\pm^2=\epsilon_{\scQ_\pm\!\scT_+}\scT_+^2
	=
	\label{eq:sq_commC}\epsilon_{\scP\!\scT_+}\epsilon_{\scQ_\mp\!\scT_+}\epsilon_{\scP\!\scQ_\mp}\scT_+^2.
\end{align}
These two ways of labeling symmetry classes are equivalent and are used interchangeably in what follows. In the remainder of this section and in Sec.~\ref{sec:examples}, we use the unitary involutions, while the discussion of random-matrix universality in Sec.~\ref{sec:rmt} is based on antiunitary symmetries.

One might be tempted to conclude that no further restrictions on the symmetries of $\scL'$ exist and, thus, that there are 15 symmetry classes of many-body Lindbladians. However, the Lindbladian is not an arbitrary superoperator with $\scT_+^2=1$ symmetry, and has an additional structure in terms of the Hamiltonian and jump operators. In Sec.~\ref{subsec:conditions}, we derive the conditions these operators must satisfy in order to implement a superoperator symmetry of the Lindbladian. Based on these conditions, in Sec.~\ref{subsec:noCp2-1} we argue that, remarkably, a $\scC_-^2=-1$ symmetry is not allowed. Since there are five classes out of the 15 (DIII$^\dagger$, BDI$_{+-}$, BDI$_{--}$, CI$_{++}$, and CI$_{-+}$) with $\scC_-^2=-1$, a Liouvillian without unitary symmetries belongs to one of ten non-Hermitian symmetry classes, which are listed in Table~\ref{tab:Lindbladian classes} together with their defining relations and matrix realizations. 

{
\setlength{\tabcolsep}{1.4pt}
\begin{table*}[t!]
	\caption{Non-Hermitian symmetry classes with $\scT_+^2=+1$ and $\scC_+^2\neq-1$, which can realized by Lindbladians with unbroken $\scT_+$ symmetry. For each class, we list its Bernard-LeClaire (BL) symmetries, the nomenclature following Ref.~\cite{kawabata2019PRX}, the squares of its antiunitary symmetries, its unitary involutions and their commutation relations [as defined in Eqs.~(\ref{eq:eps_1})--(\ref{eq:eps_5})], and an explicit matrix realization. In the second column, we have adopted a shorthand notation, where the commutation relations of $\scP$ symmetry are indicated with a subscript in the class name (class AI$_+$, say, is denoted AI + $\mathcal{S}_+$ in Ref.~\cite{kawabata2019PRX}). Moreover, these class names are not unique (for instance, class AI is also known as D$^\dagger$, and class BDI$_{-+}$ as CI$^\dagger_{+-}$, DIII$_{-+}$, or BDI$^\dagger_{+-}$~\cite{kawabata2019PRX}). In the matrix realizations of the last column, $A$, $B$, $C$, and $D$ are arbitrary non-Hermitian matrices unless specified otherwise and empty entries correspond to zeros.}
	\label{tab:Lindbladian classes}
	\small
	\begin{tabular}{@{}Sl Sl Sc Sc Sc Sc Sc Sc Sc Sc Sc Sl@{}}
		\toprule
		BL symm. & Class & $\scT_+^2$ & $\scC_-^2$ & $\scC_+^2$ & $\scT_-^2$ & $\epsilon_{\scP\!\scT_+}$ & $\epsilon_{\scQ_+\!\scT_+}$ & $\epsilon_{\scQ_-\!\scT_+}$ & $\epsilon_{\scP\!\scQ_+}$ & $\epsilon_{\scP\!\scQ_-}$ & Matrix realization  \\ \midrule
		1, K    & AI             & $+1$ & ---  & ---  & ---  & ---  & ---  & ---  & ---  & ---  & $A=A^*$       \\ \midrule
		2, PK   & AI$_+$         & $+1$ & ---  & ---  & $+1$ & $+1$ & ---  & ---  & ---  & ---  & $\matAIp$     \\ \midrule
		3, PK   & AI$_-$         & $+1$ & ---  & ---  & $-1$ & $-1$ & ---  & ---  & ---  & ---  & $\matAIm$     \\ \midrule
		4, QC   & BDI$^\dagger$  & $+1$ & ---  & $+1$ & ---  & ---  & $+1$ & ---  & ---  & ---  & $\matBDId$    \\ \midrule
		5, QC   & BDI            & $+1$ & $+1$ & ---  & ---  & ---  & ---  & $+1$ & ---  & ---  & $\matBDIalt$  \\ \midrule
		6, QC   & CI             & $+1$ & $-1$ & ---  & ---  & ---  & ---  & $-1$ & ---  & ---  & $\matCIalt$   \\ \midrule
		7, PQC  & BDI$_{++}$     & $+1$ & $+1$ & $+1$ & $+1$ & $+1$ & $+1$ & ---  & $+1$ & ---  & $\matBDIdpp$  \\ \midrule
		8, PQC & BDI$_{-+}$     & $+1$ & $+1$ & $+1$ & $-1$ & $-1$ & $+1$ & ---  & $-1$ & ---  & $\matBDIdpm$  \\ \midrule
		9, PQC & CI$_{+-}$      & $+1$ & $-1$ & $+1$ & $+1$ & $+1$ & $+1$ & ---  & $-1$ & ---  & $\matBDIdmp$  \\ \midrule
		10, PQC & CI$_{--}$      & $+1$ & $-1$ & $+1$ & $-1$ & $-1$ & $+1$ & ---  & $+1$ & ---  & $\matBDIdmm$ \\
		\bottomrule
	\end{tabular}
\end{table*}
}

\subsection{Lindbladians with unitary symmetries}
\label{subsec:unitary_syms}

Let us now consider the consequences of a unitary symmetry $\scU$ commuting with the Lindbladian. These symmetries come in two types~\cite{buca2012} (strong and weak). If the Hamiltonian and jump operators jointly satisfy the symmetry relations
\begin{align}
	\label{eq:strong_symm}
	\comm{u}{H}=\comm{u}{L_m}=0, \quad m=1,\dots, M,
\end{align}
then both unitary superoperators
\begin{align}
	\scU_\mathrm{L}=u\otimes \id 
	\qquad \text{and} \qquad
	\scU_\mathrm{R}=\id \otimes u^*
\end{align}
commute with the Lindbladian and we refer to them as a Liouvillian strong symmetry~\cite{buca2012}. There are $n$ quantum numbers in \emph{each copy} in the doubled Liouville space $\mathcal{H}\otimes\mathcal{H}$ (where $n$ denotes the number of distinct eigenvalues of $u$), which are conserved independently. The Liouville space thus splits into $n^2$ invariant subspaces (symmetry sectors) and there are generically $n$ distinct steady states, one in each diagonal sector with equal quantum numbers in the two copies. 

If the relations of Eq.~(\ref{eq:strong_symm}) are not all satisfied simultaneously, but
\begin{equation}
\scU=\scU_\mathrm{L}\scU_\mathrm{R}=u\otimes u^*
\end{equation} 
still commutes with $\scL'$, we call it a weak Liouvillian symmetry~\cite{buca2012}. There are between $n$ and $n(n-1)$ invariant subspaces (depending on the precise form of the symmetry $u$) and generically a single steady state (in the symmetry sector with eigenvalue $1$). For additional details, see Ref.~\cite{buca2012}.

We block diagonalize $\scL'$, such that each block has a well-defined eigenvalue of $\scU$ (weak symmetry) or $\scU_\mathrm{L}$, $\scU_\mathrm{R}$ (strong symmetry). For a given block to belong to a certain symmetry class, the antiunitary symmetries $\scT_\pm$ and $\scC_\pm$ and the unitary involutions $\scP$ and $\scQ_\pm$ defining that class must act within the block, i.e., they must commute with the projector onto that block. If they mix different blocks (because they connect eigenstates in different symmetry sectors), the superoperator symmetry is broken in those blocks, although the full Lindbladian possesses the symmetry.

Following the previous considerations, we immediately conclude that the presence of commuting unitary symmetries enriches the symmetry classification of the Lindbladian and allows us to go beyond the tenfold classification: if the $\scT_+=\scK\swap$ symmetry is broken by $\scU$ and no other independent $\scT_+$ symmetry is realized, the irreducible block of the Lindbladian belongs to one of 19 symmetry classes with no $\scT_+$ and $\scC_+^2\neq-1$. Note that the same arguments put forward in Sec.~\ref{subsec:noCp2-1} that prohibit a $\scC_+^2=-1$ symmetry also preclude a $\scT_+^2=-1$ symmetry. Moreover, in these classes, because of the absence of $\scT_+$ symmetry, the existence of a $\scP$ or $\scQ_\pm$ symmetry is independent of the existence of a $\scT_-$ or $\scC_\pm$ symmetry, respectively, and, therefore, a careful counting leads to 19 independent symmetry classes. Remarkably, the projection of the Lindbladian into a symmetry sector that contains a steady state (eigenvalue zero) always preserves the $\scT_+$ symmetry, as we discuss in detail below for the two cases of strong and weak symmetries. $\scT_+^2=+1$ symmetry is only broken in the blocks without the steady states (i.e., without the eigenvalue zero), which correspond to short-lived transient dynamics. We thus reach the conclusion that many-body Lindbladians admit a ($10+19$)-fold symmetry classification: in the absence of unitary symmetries \emph{or} in the presence of unitary symmetries in all symmetry sectors containing the steady state(s), the Lindbladian belongs to one of ten non-Hermitian symmetry classes with $\scT_+^2=+1$; if however, there are additional unitary symmetries and we consider non-steady-state sectors, the Lindbladian may belong to a different set of 19 classes with broken $\scT_+$ symmetry.

The simplest way to break $\scT_+$ symmetry occurs when $\scU$ and $\scT_+\scK$ do not commute, and hence do not share a common eigenbasis. As an example, we mention the case of a Liouvillian strong symmetry: since the transformation acts as a symmetry of each copy of Hilbert space individually, by definition it does not commute with the \textsc{swap} operator implementing the $\scT_+$ symmetry. Then, the $\scT_+$ is unbroken in the blocks with the same quantum number in both copies (the blocks containing the $n$ steady states) and is broken in the remaining.

However, even if $\scT_+\scK$ and $\scU$ commute, the $\scT_+$ symmetry can be broken if $\scT_+$ does not commute with the projector onto a specific subspace. Since $\scU$ has complex unimodular eigenvalues and the $\scT_+$ symmetry involves complex conjugation, states in the sector with quantum number $e^{i\theta}$ are transformed into states in the sector with quantum number $e^{-i\theta}$ under the action of $\scT_+$. The $\scT_+$ symmetry is preserved in the sectors with quantum number $\pm 1$ and broken in all others. The sector with quantum number $+1$ always exists (and contains the steady state)~\cite{buca2012}, while the additional $\scT_+$-unbroken sector with eigenvalue $-1$ might or might not exist (it does for a $\mathbb{Z}_2$ symmetry, which will be relevant below).

Before concluding this section, let us briefly comment on the impossibility of implementing a Lindbladian class with $\scC_+^2=-1$. Indeed, as we show in Sec.~\ref{subsec:noCp2-1}, the existence of a $\scC_+^2=-1$ symmetry always requires the presence of a strong symmetry. Moreover, we also show that, under quite general conditions, the $\scC_+^2=-1$ symmetry, when it exists, always breaks the strong symmetry it induces. Therefore, even when a $\scC_+^2=-1$ symmetry exists (with physical consequences such as Liouvillian or open system version of Kramers degeneracy~\cite{lieu2022PRB}), it does not define a symmetry class with $\scC_+^2=-1$ (which has consequences, for instance, for level statistics). 
The same argument also implies that a $\scT_+^2=-1$ symmetry cannot exist. Hence, if a unitary symmetry breaks the $\scT_+=\scK\swap$ symmetry, an alternative $\scT_+^2=-1$ symmetry cannot be implemented, supporting our counting of 19 classes above.

In summary, many-body Lindbladians have a tenfold classification in the absence of unitary symmetries. The presence of the latter allows for 19 additional classes beyond the tenfold way. Since the Lindbladian is specified in terms of its Hamiltonian and jump operators, it is natural to ask what conditions these operators must satisfy for the Lindbladian superoperator to belong to one of the classes discussed above. We address this question in the next section.

\subsection{Conditions on the Hamiltonian and jump operators}
\label{subsec:conditions}

In this section, we derive sufficient operator conditions for inducing superoperator symmetries of the Lindbladian. We see that these conditions are different for the three contributions to the Lindbladian, $\scLH$, $\scLD$, and $\scLJ$. We state the conditions in terms of the unitary involutions $\scP$ and $\scQ_\pm$. As mentioned above, they could be alternatively expressed in terms of the antiunitary symmetries $\scT_-$ and $\scC_\pm$.

\subsubsection{Jump term}
\label{subsec:sym_scLJ}

To impose superoperator $\scP$ and $\scQ_\pm$ symmetries on $\scLJ$, we impose operator $\scP$ and $\scQ_\pm$ symmetries on the jump operators. Since we always work with traceless jump operators, we do not need to consider the symmetries of the shifted jump operators. Furthermore, we do not require that each $L_m$ transforms to itself under the symmetries, only that the complete set of jump operators is closed under it. More precisely, we consider $L_m$ that satisfy
\begin{alignat}{99}
	\label{eq:symL_p}
	&p_a L_m p_a^{-1} &&= \epsilon^L_{pa}\sum_{n=1}^M P_{mn} L_n,\qquad 
	&&p_{a}^2=+1,
	\\
	\label{eq:symL_q}
	&q_a L_m^\dagger q_a^{-1} &&= \epsilon^L_{qa} \sum_{n=1}^M Q_{mn} L_n,\qquad 
	&&q_{a}^2=+1,
\end{alignat}
where $a=1,2$, $\epsilon^L_{pa},\epsilon^L_{qa}=\pm1$, $p_{a}$ and $q_{a}$ are unitary and Hermitian, and $P$ and $Q$ are $M\times M$ unitary Hermitian and unitary symmetric matrices, respectively.
Note that the index $a$ allows for more than one symmetry of each type, but that it is also possible that only one exists, in which case $p_1=p_2$ or $q_1=q_2$.
Next, we define the unitary superoperators:
\begin{alignat}{99}
	\label{eq:sym_scL_Pab}
	&\scP_{ab}=p_a\otimes p_b^*,\qquad 
	&&\scP_{ab}^2=+1,
	\\
	\label{eq:sym_scL_Qab}
	&\scQ_{ab}=q_a\otimes q_b^*,\qquad 
	&&\scQ_{ab}^2=+1.
\end{alignat}
It is straightforward to check that $\scP_{ab}$ acts as either a commuting, unitary, or an anticommuting, chiral, symmetry of $\scLJ$ [defined in Eq.~(\ref{eq:scLJ})]:
\begin{equation}
\label{eq:P_LJ_rel}
\begin{split}
\scP_{ab}\scLJ\scP_{ab}^{-1}
&=2\sum_{m}\(p_a\otimes p_{b}^*\)\(L_m\otimes L_m^*\) \(p_a \otimes p_{b}^*\)
\\
&=2\epsilon^L_{pa}\epsilon^L_{pb}\sum_{mnp} P_{mn} P_{mp}^*\, L_n \otimes L_p^*
\\
&=\epsilon^L_{pa}\epsilon^L_{pb}\,\scLJ,
\end{split}
\end{equation} 
where we use Eq.~(\ref{eq:symL_p}) and the unitarity of $P$. If $\epsilon^L_{pa}=\epsilon^L_{pb}$, $\scP_{ab}$ acts as a commuting unitary symmetry of $\scLJ$; if $\epsilon^L_{pa}=-\epsilon^L_{pb}$, it acts as a chiral symmetry.  

Similarly, we can show that $\scQ_{ab}$ acts as a $\scQ_\pm$ symmetry of $\scLJ$, depending on the signs $\epsilon^L_{qa}$, $\epsilon^L_{qb}$: if $\epsilon^L_{qa}=\epsilon^L_{qb}$, $\scQ_{ab}$ acts as a $\scQ_+$ symmetry; if $\epsilon^L_{qa}=-\epsilon^L_{qb}$, as a $\scQ_-$ symmetry.

\subsubsection{Dissipative term}
\label{subsec:sym_scLD}

The conditions of Eq.~(\ref{eq:symL_p}) and (\ref{eq:symL_q}) are not enough to generate symmetries of $\scLD$. For instance, a chiral symmetry $p_a$, Eq.~(\ref{eq:symL_p}), does not modify the term $\sum_m L_m^\dagger L_m$:
\begin{equation}
\begin{split}
p_a \(\sum_{m} L_m^\dagger L_m\) p_a^{-1}
&=\sum_m p_a L_m^\dagger p_a^{-1}p_aL_m p_a^{-1}
\\
&=+\sum_m L_m^\dagger L_m,
\end{split}
\end{equation}
and, hence,
\begin{equation}
\label{eq:transf_scP_scLD}
\scP_{ab}\scLD \scP_{ab}^{-1}=+\scLD;
\end{equation}
i.e., the condition of Eq.~(\ref{eq:symL_p}) only leads to a commuting unitary symmetry of $\scLD$, not to a chiral symmetry. To generate a $\scP$ symmetry, we have to require that the jump operators additionally satisfy
\begin{equation}\label{eq:sym_scLJ_P}
\sum_m L_m^\dagger L_m = -\frac{\alpha}{2} \id,
\end{equation}
where $\alpha$ is defined in Eq.~(\ref{eq:scL_prime}). In that case, Eq.~(\ref{eq:transf_scP_scLD}) reads as
\begin{equation}\label{eq:scTab_scLD}
\scP_{ab}\scLD \scP_{ab}^{-1}
=-\scLD+2\alpha \scI
\iff
\scP_{ab}\scLD'\scP_{ab}^{-1}=-\scLD',
\end{equation}
in accordance with Eq.~(\ref{eq:nHsym_Tm_scL'}). 
Note that, for this particular symmetry only, we actually have $\scLD'=0$.
We thus see that it is the dissipative contribution that forces us to consider the symmetries of the shifted Lindbladian. One particular way of satisfying Eq.~(\ref{eq:sym_scLJ_P}), which we encounter in the examples below, is to have each jump operator individually satisfy
\begin{equation}\label{eq:sym_scLJ_P_m}
L_m^\dagger L_m = -\frac{\alpha_{m}}{2} \id,
\end{equation}
for some $\alpha_m\in\mathbb{R}$.

Similarly, we can see that a pseudo-Hermiticity transformation, Eq.~(\ref{eq:symL_q}), transforms the term $\sum_m L_m^\dagger L_m$ as
\begin{equation}
q_a\(\sum_m L_m^\dagger L_m\)^\dagger q_a^{-1}=\sum_{m}L_mL_m^\dagger,
\end{equation}
and hence we must impose a condition on the commutator or anticommutator of $L_m$. If we impose that
\begin{equation}\label{eq:sym_scLJ_Q-}
\sum_m\acomm{L_m^\dagger}{L_m}=-\alpha \id,
\end{equation}
then $\scQ_{ab}$ acts as a $\scQ_-$ symmetry:
\begin{equation}
\scQ_{ab}\scLD^\dagger \scQ_{ab}^{-1}
=-\scLD+2\alpha \scI
\iff
\scQ_{ab}\scLD'^\dagger\scQ_{ab}^{-1}=-\scLD'.
\end{equation}
Again, it will often prove convenient for each jump operator to satisfy this condition individually; i.e.,
\begin{equation}\label{eq:sym_scLJ_Q-_m}
\acomm{L_m^\dagger}{L_m}=-\alpha'_m \id,
\end{equation}
for some $\alpha'_m\in\mathbb{R}$. 
If we instead impose that
\begin{equation}\label{eq:sym_scLJ_Q+}
\sum_m\comm{L_m^\dagger}{L_m}=0
\end{equation}
(or $\comm{L_m}{L_m^\dagger}=0$ for each jump operator), then $\scQ_{ab}$ acts as a $\scQ_+$ symmetry, $\scQ_{ab}\scLD'^\dagger \scQ_{ab}^{-1}=\scLD'$. 

\subsubsection{Hamiltonian term}
\label{subsec:sym_scLH}

Finally, we address the conditions one has to impose on the Hamiltonian such that $\scLH$ possesses $\scP$ and $\scQ_\pm$ symmetries. We start from the symmetries of the Hamiltonian:
\begin{align}
	\label{eq:symH}
	&p_a H p_a^{-1} = \epsilon^H_{pa} H,
	\\
	&q_a H q_a^{-1} = \epsilon^H_{qa} H.
\end{align}
Note that for the full Lindbladian to satisfy a superoperator symmetry, the matrices $p_a$ and $q_a$ have to be the same as those in Eqs.~(\ref{eq:symL_p}) and (\ref{eq:symL_q}). Under the action of the superoperator $\scP_{ab}$, $\scLH$ transforms as
\begin{equation}
\begin{split}
\scP_{ab} \scLH \scP_{ab}^{-1} 
&=-\i \(p_a H p_a^{-1}\otimes \id - \id \otimes \(p_b H p_b^{-1}\)^*\)
\\
&=-\i \(\epsilon^H_{pa}H\otimes\id-\epsilon^{H}_{pb}\id\otimes H^*\).
\end{split}
\end{equation}
For $\scP_{ab}$ to be a symmetry of $\scLH$, we must have $\epsilon^H_{pa}=\epsilon^H_{pb}$. Then, if $\epsilon^H_{pa}=\epsilon^H_{pb}=-1$, we have
\begin{equation}
\scP_{ab} \scLH \scP_{ab}^{-1} 
=\i \(H\otimes\id-\id\otimes H^*\)
=-\scLH,
\end{equation}
i.e., $\scP_{ab}$ acts as a $\scP$ symmetry, while if $\epsilon^H_{pa}=\epsilon^H_{pb}=+1$, it acts as a commuting unitary symmetry.

Proceeding analogously for the pseudo-Hermiticity transformations, we find $\epsilon^H_{qa}=\epsilon^H_{qb}$. If $\epsilon^H_{qa}=\epsilon^H_{qb}=-1$, then $\scQ_{ab}$ acts as $\scQ_+$ symmetry and if $\epsilon^H_{qa}=\epsilon^H_{qb}=+1$, it acts as a $\scQ_-$ symmetry. 

Finally, we note that in the case of real Hamiltonian and jump operators, we can define a modified superoperator $\widetilde{\scQ}_{ab}=\scQ_{ab}\swap$ whose action on $\scLH$ is reversed: if $\epsilon^H_{qa}=\epsilon^H_{qb}=+1$, $\widetilde{\scQ}_{ab}$ is a $\scQ_+$ symmetry, while it is a $\scQ_-$ symmetry if $\epsilon^H_{qa}=\epsilon^H_{qb}=-1$. Similarly, if the Hamiltonian is real and the jump operators are symmetric, we can define $\widetilde{\scP}_{ab}=\scP_{ab}\swap$, such that if $\epsilon^H_{pa}=\epsilon^H_{pb}=+1$, $\widetilde{\scP}_{ab}$ is a $\scP$ symmetry of $\scLH$, while it is a commuting unitary symmetry when $\epsilon^H_{pa}=\epsilon^H_{pb}=-1$. In either case, the action on $\scLJ$ and $\scLD$ is not modified.

\subsection{Absence of \texorpdfstring{$\scC_+^2=-1$}{C+2=1} symmetry and Kramers degeneracy}
\label{subsec:noCp2-1}

We now show that, under fairly general conditions, classes with $\scC_+^2=-1$ do not exist in the Lindbladian classification. The proof proceeds in two steps. First, we show that, because of the two-copy tensor-product structure of the Lindbladian, a $\scC_+^2=-1$ symmetry always implies the existence of a Liouvillian strong symmetry. Then, we show that, by construction, the $\scC_+^2=-1$ is always broken by the strong symmetry it induces. As a consequence, if degenerate Kramers pairs exist, they do not occur in the same symmetry sector, and none of the blocks of the Lindbladian displays, by itself, Kramers degeneracy. 
Importantly, the absence of Kramers pairs inside individual symmetry sectors is a rather universal result of systems with a two-copy structure and symmetric intercopy coupling, as it is also observed for fermionic Lindbladians~\cite{kawabata2023PRXQ,garcia2023ARXIVb} and for a Hermitian two-site fermionic Sachdev-Ye-Kitaev Hamiltonian~\cite{garcia2023ARXIV}, where an identical argument holds.
The same mechanism also prevents the existence of a $\scT_+^2=-1$ symmetry, which does not affect the ten classes with unbroken $\scT_+$ swap symmetry, but it is fundamental in the counting of the 19 classes with broken $\scT_+$ symmetry (as it precludes any additional classes with $\scT_+^2=-1$).

The first part of the proof is completely general. Let us assume that a superoperator symmetry $\scQ_+=q_a\otimes q_b^*$ of the Lindbladian exists. From Eq.~(\ref{eq:sq_commC}), we have that $\scC_+^2=\epsilon_{\scQ_+\!\scT_+}$. The commutation relation of $\scQ_+$ with $\scT_+$ is given by:
\begin{equation}
\begin{split}
\scQ_+\scT_+&=(q_a\otimes q_b^*)\scK\swap 
\\&=\scK \swap (q_b\otimes q_a^*)
\\&=\scT_+ \scQ_+ \left[q_a^{-1}q_b\otimes (q_b^{-1}q_a)^* \right],
\end{split}
\end{equation}
We want to impose that $\scC_+^2=-1\Leftrightarrow\scQ_+\scT_+=-\scT_+\scQ_+$. Clearly, that is not possible if $q_a=q_b$, i.e., if the Hamiltonian and jump operators have a single $\scQ_+$ operator symmetry. Consequently, we must consider a $\scQ_+$ symmetry of the form
\begin{equation}
\scQ_+=q_1\otimes q_2^*,
\end{equation}
with $q_1\neq q_2$. Furthermore, to implement the unitary involution $\scQ_+$, the Hamiltonian and jump operators must satisfy $\epsilon_{q1}^H=\epsilon_{q2}^H=-1$ and $\epsilon_{q1}^L=\epsilon_{q2}^L$, according to the previous section. The former condition further precludes that one of the $q_{1,2}$ is the identity operator. It then immediately follows that the product $q_1q_2$ commutes with the Hamiltonian and all jump operators and thus implements a Liouvillian strong symmetry.

To conclude the proof, we must show that if the $\scC_+^2=-1$ symmetry exists, it is always broken by the strong symmetry it induces. Let us define matrices $\varepsilon_{12}$, $\eta_1$, and $\eta_{2}$ through the relations
\begin{equation}
\label{eq:Kramers_assump}
q_1 q_2=\varepsilon_{12}q_2 q_1
\qquad \text{and} \qquad
q_{1,2}=\eta_{1,2} q_{1,2}^*.
\end{equation}
Because $q_{1,2}$ are unitary, it immediately follows that $\varepsilon_{12}$ and $\eta_{1,2}$ must also be unitary. To proceed, we make the mild assumption that $\varepsilon_{12}$ and $\eta_{1,2}$ are unimodular complex numbers (i.e., proportional to the identity).
This assumption holds for any $q_1$ and $q_2$ that can be expressed as a string of Pauli operators (which is true for the spin-chain examples of Sec.~\ref{sec:examples} and for fermionic models discussed in Refs.~\cite{kawabata2023PRXQ,garcia2023ARXIVb}. While the proof we present in the following strictly holds only in this case, we believe the argument extends to general $q_a$ written as sums of such Pauli strings (for which $\varepsilon_{12}$ and $\eta_{1,2}$ are more general unitary matrices) and, consequently, that sectors with $\scC_+^2=-1$ do not exist in general.

Proceeding under the assumption that $\varepsilon_{12}$ is a complex unimodular number, we take the strong symmetry to be implemented by the unitary
\begin{equation}
u=q_1 q_2,
\end{equation}
which satisfies $u^2=\varepsilon_{12}$ and $u^*=\eta_1\eta_2u$.
Since $u$ defines a strong symmetry, both 
\begin{equation}
\scU_\mathrm{L} = u\otimes \id 
\quad \text{and} \quad
\scU_\mathrm{R} = \id \otimes u^* = \eta_1 \eta_2 \(\id\otimes u\)
\end{equation}
are independently conserved, with eigenvalues $p_\mathrm{L,R}=\varepsilon_{12}^{1/2}$. The projectors onto the conserved sectors are 
\begin{equation}
\mathbb{P}_{\mathrm{L},\mathrm{R}}^\pm 
=\frac{1}{2}\(\scI \pm \scU_{\mathrm{L},\mathrm{R}}/p_{\mathrm{L},\mathrm{R}}\)
=\frac{1}{2}\(\scI \pm \varepsilon_{12}^{-1/2}\scU_{\mathrm{L},\mathrm{R}}\).
\end{equation}
We can now start imposing conditions on the choice of operators. The $\scC_+^2$ symmetry squares to
\begin{equation}
\begin{split}
\scC_+^2=\(\scQ_+\scT_+\)^2
&= (q_1\otimes q_2^*)(q_2\otimes q_1^*)
\\
&= \varepsilon_{12} \(q_1 q_2 \otimes q_1^* q_2^*\)
\\
&= \varepsilon_{12}\scU_\mathrm{L}\scU_\mathrm{R}.
\end{split}
\end{equation}
Since we want $\scC_+^2=-1$, we must be in a sector of $\scU_{\mathrm{L},\mathrm{R}}$ with quantum numbers $p_\mathrm{L}p_\mathrm{R}\varepsilon_{12}=-1$. On the other hand, for $\scC_+$ to act inside a given symmetry sector of $u$, it must commute with the projector. The commutation relation is given by
\begin{equation}
\begin{split}
\scC_+ \mathbb{P}_{\mathrm{L},\mathrm{R}}^\pm
&= \scQ_+ \scT_+\mathbb{P}_{\mathrm{L},\mathrm{R}}^\pm
\\
&= (q_1\otimes q_2^*) \mathbb{P}_{\mathrm{R},\mathrm{L}}^{\pm\varepsilon_{12}} \scT_+
\\
&= \frac{1}{2}\(\scI\pm \varepsilon_{12}^{1/2}\scU_{\mathrm{L},\mathrm{R}}\)\scQ_+\scT_+
\\
&= \mathbb{P}_{\mathrm{L},\mathrm{R}}^{\pm p_\mathrm{L}p_\mathrm{R}\varepsilon_{12}}\scC_+.
\end{split}
\end{equation}
From this, it follows that $p_\mathrm{L}p_\mathrm{R}\varepsilon_{12}=+1$, in contradiction with the condition we found above. We conclude that either $\scC_+$ acts inside a sector but squares to $+1$, or it squares to $-1$ but connects different sectors. In either case, a definite symmetry sector does not belong to a class with $\scC_+^2=-1$.

As noted in the previous section, one might define an alternative symmetry operator $\scQ_+=(q_1\otimes q_2^*)\swap$. The calculation proceeds in the same way as above, and we find again two contradicting conditions: to have $\scC_+^2=-1$ we require $\eta_1\eta_2=-1$, while for $\scC_+$ to act inside a single symmetry sector, we must have $\eta_1\eta_2=+1$.

To conclude this section, note that the existence of $\scT_-^2=-1$ or $\scC_-^2=-1$ symmetries does not imply a strong symmetry because the jump operators must satisfy $\epsilon_{q1}^L\neq \epsilon_{q2}^L$ and, consequently, the product $q_1q_2$ anticommutes with the jump operators and can lead, at most, to a weak symmetry. On the other hand, if the $\scT_+=\scK\swap$ symmetry is broken, the same argument prevents the implementation of an alternative $\scT_+^2=-1$ symmetry. Hence, Lindbladian symmetry classes have either $\scT_+^2=+1$ or no $\scT_+$.

\subsection{Generalization to non-Markovian and non-trace-preserving open quantum dynamics}

After the developments of the previous sections, we are now in the position to make the remarkable observation that the classification we have developed is not restricted to Lindbladian dynamics, but to all Hermiticity-preserving dynamics, including non-Markovian and even non-trace-preserving dynamics. The same considerations hold also for generalized PT-symmetric Hamiltonians~\cite{garcia2023ARXIVb}.

To see this, we start from the fact that the Liouvillian generator $\Lambda$ of any Hermiticity-preserving quantum master equation, $\partial_t\rho=\Lambda \rho$, can be written in the form~\cite{hall2014PRA}
\begin{equation}
\label{eq:general_generator}
\Lambda\rho 
=
-\i \comm{H}{\rho}+\acomm{\Gamma}{\rho}
+2\sum_{m=1}^M \gamma_m L_m\rho L_m^\dagger,
\end{equation}
where, in addition to the Hamiltonian and jump operators, we have a second independent ``Hamiltonian'' $\Gamma=\Gamma^\dagger$, and the real rates $\gamma_m$ can be negative in general. Furthermore, we could also assume all of $H$, $L_m$, $\gamma_m$, and $\Gamma$ to be time dependent. 
In the most general case, the master equation~(\ref{eq:general_generator}), while Hermiticity preserving, is not necessarily positivity preserving~\cite{hall2014PRA}.
Trace preservation, $\partial_t\Tr\rho=\Tr\partial_t\rho=0$, is enforced by the restriction
\begin{equation}
\Gamma=\sum_{m=1}^M \gamma_m L_m^\dagger L_m.
\end{equation}
Additionally, Markovianity is implemented by considering only positive rates $\gamma_m>0$. Then (and only then) they can be absorbed into the jump operators, $L_m\to \sqrt{\gamma_m}L_m$, and we recover the Liouvillian of Lindblad 
form (\ref{lindf}), $\Lambda=\scL$.

As before, the Liouvillian $\Lambda$ can be vectorized as $\Lambda=\Lambda_\mathrm{H}+\Lambda_\mathrm{D}+\Lambda_\mathrm{J}$, where the Hamiltonian, $\Lambda_\mathrm{H}$, and jump, $\Lambda_\mathrm{J}$, contributions are still given by Eqs.~(\ref{eq:scLH}) and (\ref{eq:scLJ}), respectively (apart from the real scalar rates $\gamma_m$ that do not change the classification), while the dissipative contribution $\Lambda_\mathrm{D}$ is now given by
\begin{equation}
\Lambda_\mathrm{D}=\Gamma\otimes \id +\id \otimes \Gamma^*.
\end{equation}

It is now immediately clear that the classification (or, more precisely, the set of admissible classes) is not changed in this more general case. There is always a $\scT_+^2=1$ symmetry implemented by the \textsc{swap} operator (which can be broken by a Liouvillian strong symmetry), while the impossibility of $\scC_+^2=-1$ and $\scT_+^2=-1$ symmetries is imposed by the jump contribution and is, hence, unchanged. We thus have ten classes with unbroken $\scT_+$ symmetry and 19 additional ones with broken $\scT_+$ symmetry. 

What does change is the class to which a particular physical example is assigned. On the one hand, the jump operators no longer need to satisfy the strict conditions of Sec.~\ref{subsec:sym_scLD} [Eqs.~(\ref{eq:sym_scLJ_P}), (\ref{eq:sym_scLJ_Q-}), and (\ref{eq:sym_scLJ_Q+})], which facilitates finding examples in the classes with more symmetries. On the other hand, we have to impose constraints on the matrix $\Gamma$. More specifically, we consider that it admits the following symmetries:
\begin{align}
	p_a \Gamma p_a^{-1}=\epsilon_{pa}^\Gamma \Gamma,
	\\
	q_a \Gamma q_a^{-1}=\epsilon_{qa}^\Gamma \Gamma,
\end{align}
where the unitary operators $p_a$ and $q_a$ are the same as those in Secs.~\ref{subsec:sym_scLJ} and \ref{subsec:sym_scLH}, but the signs $\epsilon_{pa}^\Gamma$ and $\epsilon_{qa}^\Gamma$ are independent from the ones in the Hamiltonian and the jump contribution. Now, following exactly the same steps as in Sec.~\ref{subsec:sym_scLH} but noting that there is a factor-of-$\i$ difference between how $H$ and $\Gamma$ appear in the Liouvillian, we conclude that (i) $\scP_{ab}$ [defined in Eq.~(\ref{eq:sym_scL_Pab})] acts as a $\scP$ symmetry if $\epsilon_{pa}^\Gamma=\epsilon_{pb}^\Gamma=-1$ and as a commuting unitary symmetry if $\epsilon_{pa}^\Gamma=\epsilon_{pb}^\Gamma=+1$; (ii) $\scQ_{ab}$ [defined in Eq.~(\ref{eq:sym_scL_Qab})] acts as a $\scQ_+$ symmetry if $\epsilon_{qa}^\Gamma=\epsilon_{qb}^\Gamma=+1$ and as a $\scQ_-$ symmetry if $\epsilon_{qa}^\Gamma=\epsilon_{qb}^\Gamma=-1$; (iii) if $\Gamma$ is real symmetric, then we can define the alternative symmetry superoperators $\widetilde{\scP}_{ab}=\scP_{ab}\swap$ and $\widetilde{\scQ}_{ab}=\scP_{ab}\swap$, as discussed in Sec.~\ref{subsec:sym_scLH}, with the conditions on $\epsilon_{qa}^\Gamma, \epsilon_{qb}^\Gamma$ unchanged.

Since Hermiticity preservation is a physical constraint that one can hardly imagine to be relaxed, we conclude that our framework provides the most general symmetry classification of the dynamical generators of open quantum matter.

\subsection{Physical consequences for correlation functions}
\label{subsec:physical_consequences}

Before proceeding with specific examples of the classification developed so far, we derive general statements about the dynamics of open quantum systems described by any Lindbladian (or more general Liouvillian) with involutive global symmetries. Most importantly, we show that when the involutive symmetry involves a minus sign ($\scP$ or $\scQ_-$) we can derive a time-reversal-like invariance property for an observable in a time-dependent state, or a related correlation function.

We start with the case of a $\scP$ (or $\scT_-$) symmetry. For any fixed observable $O$ and state $\rho$ that are invariant under the $\scP$ operation, i.e., that satisfy the properties $\scP O=O$ and $\scP \rho=\rho$, we define the nonequilibrium correlation function
\begin{equation}
F(t)=\Tr[O \rho(t)],
\end{equation}
where $\rho(t)=\exp{\scL t}\rho$ is the state evolved under the Lindbladian $\scL$ for time $t$. If $\scL$ satisfies Eq.~(\ref{eq:nHsym_P}), it follows that
\begin{equation}
F(t)
=e^{-2\alpha t}\Tr\left[O \scP e^{-\scL t} \scP \rho\right]
=e^{-2\alpha t}\Tr\left[O \rho(-t)\right],
\end{equation}
or, equivalently,
\begin{equation}
\label{eq:correlation_Tm}
F(-t)=e^{2\alpha t}F(t).
\end{equation}
For general open quantum systems, the quantity $F(-t)$ is not well defined: $-\scL$ does not generate a completely positive semigroup and, given a state at time $t$, we can only propagate it forward in time, not backward. The remarkable relation~(\ref{eq:correlation_Tm}) tells us, however, that in systems with a $\scP$ symmetry, $F(-t)$ is written in terms of two well-defined quantities [$F(t)$ and $\exp{2\alpha t}$] and is thus itself well defined. This opens the possibility of knowing the past of a dissipative system solely from the knowledge of its future. In particular, this feature could improve error-canceling schemes on noisy intermediate-scale quantum devices in combination with the recent proposal of Ref.~\cite{minev2022NatPhys}.

Next we consider $\scQ_\pm$ (equivalently, $\scC_\pm$) symmetries. Because these symmetries relate $\scL$ to its adjoint $\scL^\dagger$, we must consider correlation functions of two observables, or fidelity-like correlation functions of two states, $\rho$ and $\sigma$. Focusing on the latter case, we define
\begin{equation}
G_{\rho\sigma}(t)=\Tr[\sigma \rho(t)],
\end{equation}
and consider states that are themselves invariant under the symmetry transformation, $\scQ_\pm \rho=\rho$ and $\scQ_\pm \sigma=\sigma$. If $\scL$ has a $\scQ_-$ symmetry, Eq.~(\ref{eq:nHsym_Qm}), we find, proceeding as before, that
\begin{equation}
G_{\rho\sigma}(-t)=e^{2\alpha t} G_{\sigma \rho }(t),
\end{equation}
which, besides reversing time also swaps the two states. If, instead, the Lindbladian has a $\scQ_+$ symmetry, Eq.~(\ref{eq:nHsym_Qp}), no time reversal takes place and
\begin{equation}
G_{\rho\sigma}(t)=G_{\sigma\rho}(t);
\end{equation}
i.e., $G$ is symmetric under the exchange of the two states $\sigma$ and $\rho$.

\section{Physical examples: Tenfold way in dissipative spin chains}
\label{sec:examples}

In the following sections, we realize the tenfold way of many-body Lindbladians with unbroken $\scT_+$ symmetry in spatially inhomogeneous spin chains. In Sec.~\ref{subsec:examples_IncHop}, we also present an example with a strong symmetry and, hence, sectors with broken $\scT_+$ symmetry. Throughout, we consider chains of $L$ spins $1/2$, represented by local Pauli operators $\sigma^{\alpha}_j=\id_{2\times 2}^{\otimes (j-1)}\otimes\sigma^\alpha\otimes \id_{2\times 2}^{\otimes(L-j)}$, $\alpha=x,y,z$, $j=1,2,\dots,L$, with periodic boundary conditions $\sigma_{L+1}^\alpha\equiv \sigma^\alpha_1$. We realize all ten symmetry classes by considering simple jump operators routinely used in the literature (dephasing, incoherent hopping, and spin injection or removal) and choosing an appropriate Hamiltonian. We thus conclude that the symmetry classes discuss in this work are not an exotic theoretical artifact, but are ubiquitous and implementable in current experimental setups.

\subsection{Dephasing. Classes \texorpdfstring{BDI$_{++}$}{BDI++}, \texorpdfstring{CI$_{+-}$}{CI+-}, \texorpdfstring{BDI$_{-+}$}{BDI-+}, \texorpdfstring{CI$_{--}$}{CI--}, \texorpdfstring{BDI$^\dagger$}{BDIdg}, and AI}
\label{subsec:examples_Deph}

As a first example, we consider local dephasing jump operators,
\begin{equation}
L_j=\sqrt{\gamma_j}\sigma_j^z,
\end{equation}
where $\gamma_j$ are arbitrary positive dephasing rates. The trace of the Lindbladian is $\alpha=-2\sum_{j}\gamma_j$ and the shifted Lindbladian reads as
\begin{equation}
\scL'=-\i H\otimes \id +\id \otimes \i H+2\sum_{j=1}^L \gamma_j \sigma_j^z\otimes \sigma_j^z.
\end{equation}
Introducing the global spin operators,
\begin{equation}
\Sigma^\alpha=\prod_{j=1}^L \sigma^\alpha_j = (\sigma^\alpha)^{\otimes L},
\qquad
(\Sigma^\alpha)^2=+\id,
\end{equation} 
for $\alpha=x,y,z$, we can immediately check that the jump operators satisfy
\begin{align}
	\label{eq:ex_deph_L_Sigma_comm}
	\acomm{L_j}{\Sigma^x}=\acomm{L_j}{\Sigma^y}=\comm{L_j}{\Sigma^z}=0.
\end{align}
The dephasing Lindbladian is extremely rich, as the jump operators are real, Hermitian, and unitary. They thus satisfy all the conditions for symmetries of $\scLD$, Eqs.~(\ref{eq:sym_scLJ_P}), (\ref{eq:sym_scLJ_Q-}), and (\ref{eq:sym_scLJ_Q+}), allowing for the implementation of all three types of symmetries $\scP$ and $\scQ_\pm$ and realizing many different symmetry classes, depending on the choice of Hamiltonian. 

First, we consider a transverse-field Hamiltonian with a time-reversal-breaking interaction (not restricted to nearest neighbors):
\begin{equation}\label{eq:H_deph_noTRS}
H=\sum_{j=1}^L g^x_j \sigma^x_j+\sum_{j<k}K_{jk}\sigma^y_j\sigma^z_k,
\end{equation}
with $g_j^x$ and $K_{jk}$ arbitrary real coupling constants.
The Hamiltonian satisfies
\begin{align}
	\label{eq:ex_deph_H_Sigma_comm}
	\comm{H}{\Sigma^x}=\acomm{H}{\Sigma^y}=\acomm{H}{\Sigma^z}=0.
\end{align}
From Eqs.~(\ref{eq:ex_deph_L_Sigma_comm}) and (\ref{eq:ex_deph_H_Sigma_comm}), it follows that the Lindbladian admits the commuting unitary symmetry (weak Liouvillian symmetry):
\begin{equation}\label{eq:ex_deph_scU}
\scU^x=\Sigma^x\otimes \Sigma^x,
\end{equation}
with eigenvalues $\pm1$. Accordingly, the Liouville space $(\mathbb C^2)^{\otimes L} \otimes (\mathbb C^2)^{\otimes L}$ splits into two sectors of positive or negative transverse parity ($\scU^x=\pm\scI$).
Moreover, Eqs.~(\ref{eq:ex_deph_L_Sigma_comm}) and (\ref{eq:ex_deph_H_Sigma_comm}) imply that 
\begin{equation}\label{eq:anti_ops_deph}
\scP=\Sigma^z\otimes\Sigma^y 
\quad\text{and}\quad
\scQ_+=\Sigma^z\otimes \Sigma^z
\end{equation}
act as chiral symmetry and pseudo-Hermiticity of both the jump and Hamiltonian contributions. Both these symmetries and $\scT_+=\scK\swap$ commute with $\scU^x$ and, hence, act within the irreducible blocks of the Lindbladian. To identify the symmetry class of the (shifted) Lindbladian, we check the commutation relations of the $\scP$ and $\scQ_+$ operators:
\begin{align}
	&\scP \scT_+ =(-1)^L\scU^x\, \scT_+ \scP,
	\\
	&\scQ_+ \scT_+ = \scT_+ \scQ_+,
	\\
	&\scQ_+ \scP =(-1)^L\, \scP \scQ_+.
\end{align}
Depending on the chain length and the parity sector, the Lindbladian belongs to different classes: for even $L$ and even parity ($\scU^x=+\scI$), it belongs to class BDI$_{++}$ (recall Table~\ref{tab:Lindbladian classes}); for even $L$ and odd parity ($\scU^x=-\scI$), to class CI$_{--}$; for odd $L$ and even parity, to class BDI$_{-+}$; and for odd $L$ and odd parity, to class CI$_{+-}$.
We note that the same symmetry classification holds if we add a second set of ``dephasing'' operators $\tilde{L}_j=\sqrt{\tilde{\gamma}_j}\sigma^y_j$.

As a second example, we choose a generic, time-reversal invariant XYZ Hamiltonian in a transverse field,
\begin{equation}
\label{eq:H_XYZ_X}
H=H_\mathrm{XYZ}+\sum_{j=1}^L g^x_j \sigma^x_j,
\end{equation}
with
\begin{equation}\label{eq:H_YXZ}
H_\mathrm{XYZ}=\sum_{j<k}
J_{jk}^x \sigma_j^x \sigma_k^x
+J_{jk}^y \sigma_j^y \sigma_k^y
+J_{jk}^z \sigma_j^z \sigma_k^z,
\end{equation}
and $J_{jk}^\alpha$ arbitrary real coupling constants. The Hamiltonian again commutes with $\Sigma^x$, but the anticommutation relations with $\Sigma^y$ and $\Sigma^z$ are broken. As before, the Lindbladian admits $\scU^x$ [Eq.(\ref{eq:ex_deph_scU})] as a weak Liouvillian symmetry. Because the Hamiltonian is real and the jump operators are real and symmetric, 
\begin{equation}
\scP=\sqrt{p_x}(\Sigma^x\otimes \id)\swap
\quad\text{and}\quad
\scQ_+=\swap
\end{equation}
act as chiral symmetry and pseudo-Hermiticity of the jump and Hamiltonian contributions. Here, $p_x$ denotes the transverse parity, i.e., the eigenvalue of $\scU^x$, and is introduced in the definition of $\scP$ to ensure that $\scP^2=+1$ in both symmetry sectors. Both these symmetries and $\scT_+$ commute with $\scU^x$ and satisfy
\begin{align}
	&\scP \scT_+ =\scT_+ \scP,
	\\
	\label{eq:ex_deph_comm_TQ}
	&\scQ_+ \scT_+ =\scT_+ \scQ_+,
	\\
	&\scQ_+ \scP =\scU^x\, \scP \scQ_+.
\end{align}
The different parity sectors of the Lindbladian belong to different symmetry classes, this time irrespective of the chain length: in the sector of even parity ($\scU^x=+\scI$), the Lindbladian belongs to class BDI$_{++}$, whereas in the sector of odd parity ($\scU^x=-\scI$), it belongs to class CI$_{+-}$.

If we assume a more general Hamiltonian, we reduce the set of symmetries of the Lindbladian. If we add a second transverse component of the magnetic field say,
\begin{equation}
H=H_\mathrm{XYZ}+\sum_{j=1}^L \left(g_j^x\sigma_j^x+h_j \sigma_j^z\right),
\end{equation}
we break the transverse parity conservation and all the commutation relations of the Hamiltonian. Consequently, the Lindbladian is irreducible and chiral symmetry $\scP$ is broken. $\scQ_+=\swap$ still implements a pseudo-Hermiticity transformation commuting with $\scT_+$ and, hence, the Lindbladian belongs to class BDI$^\dagger$.

Adding a third component to the magnetic field, i.e., setting
\begin{equation}
H=H_\mathrm{XYZ}+\sum_{j=1}^L \left(g_j^x\sigma_j^x+g_j^y\sigma_j^y + h_j \sigma_j^z\right),
\end{equation}
implies there is no longer a nontrivial basis in which the Hamiltonian is real and prevents the choice of the \textsc{swap} operator $\swap$ as a pseudo-Hermiticity operator. Since there are no symmetries of the Lindbladian besides $\scT_+$, this case belongs to class AI.

\subsection{Spin injection or removal. Classes BDI and CI}
\label{subsec:examples_InjRem}

We now consider a set of jump operators describing spin injection into the chain (which can occur in the bulk or at the boundaries),
\begin{equation}
L_j=a_j \sigma_j^+,
\end{equation}
where $a_j$ are arbitrary real coefficients. The same considerations apply to the jump operators describing spin removal, $L_j=b_j\sigma_j^-$. In addition, we take the XYZ Hamiltonian of Eq.~(\ref{eq:H_YXZ}), which commutes with all three $\Sigma^{x,y,z}$. Since the jump operators satisfy
\begin{equation}\label{eq:ex_inj_L_Sigma_comm}
\Sigma^x L_j^\dagger \Sigma^x=L_j,
\quad
\Sigma^y L_j^\dagger \Sigma^y=-L_j,
\quad \text{and} \quad
\acomm{L_j}{\Sigma^z}=0,
\end{equation}
we see that the longitudinal parity,
\begin{equation}
\scU^z=\Sigma^z\otimes\Sigma^z,
\end{equation}
is conserved as a Liouvillian weak symmetry, but the transverse parity $\scU^x$ is not. Furthermore, the jump operators satisfy $\acomm{L_j^\dagger}{L_j}=a_j^2 \id_j$, but are neither normal, $\comm{L_j^\dagger}{L_j}=-a_j^2 \sigma^z_j$, nor unitary, $L_j^\dagger L_j=a_j^2(\id_j-\sigma_j^z)/2$. $\scLD'$ can, therefore, only satisfy a $\scQ_-$ symmetry [according to Eqs.~(\ref{eq:sym_scLJ_P}), (\ref{eq:sym_scLJ_Q-}), and (\ref{eq:sym_scLJ_Q+})]. We take 
\begin{equation}
\scQ_-=\Sigma^x\otimes \Sigma^y
\end{equation}
as the pseudo-Hermiticity superoperator, which satisfies the commutation relation:
\begin{equation}
\scQ_- \scT_+=(-1)^L\scU^z\, \scT_+ \scQ_-.
\end{equation}
For even $L$, the spin-injection Lindbladian belongs to class BDI in the even parity sector $\scU^z=\scI$ and to class CI in the odd parity sector ($\scU^z=-\scI$). For odd $L$, the result is reversed.

\subsection{Incoherent hopping. Class \texorpdfstring{BDI$^\dagger$}{BDIdg} and beyond the tenfold way}
\label{subsec:examples_IncHop}

Next, we consider jump operators describing a two-site XY interaction,
\begin{equation}
\label{eq:jumpops_inchop}
L_{jk}=M_{jk}^x \sigma^x_j\sigma^x_k
+ M_{jk}^y \sigma^y_j\sigma^y_k,
\end{equation}
with arbitrary complex couplings $M_{jk}^\alpha$. In the case $M^x_{jk}=M^y_{jk}$, they describe incoherent hopping. The jump operators satisfy
\begin{equation}
\comm{L_{jk}}{\Sigma^x}=\comm{L_{jk}}{\Sigma^y}=\comm{L_{jk}}{\Sigma^z}=0.
\end{equation}
Choosing the XYZ Hamiltonian in a longitudinal field,
\begin{equation}
H=H_\mathrm{XYZ}+\sum_{j=1}^L h_j^z \sigma^z_j,
\end{equation}
that satisfies
\begin{equation}
\comm{H}{\Sigma^z}=0,
\end{equation}
the longitudinal parity $\Sigma^z$ is a Liouvillian strong symmetry, i.e., the Lindbladian conserves independently left and right longitudinal parity, $\scU^z_\mathrm{L}\scL \scU^z_\mathrm{L}=\scL$ and $\scU^z_\mathrm{R}\scL \scU^z_\mathrm{R}=\scL$, with
\begin{equation}
\scU^z_\mathrm{L}=\Sigma^z\otimes \id
\quad \text{and} \quad
\scU^z_\mathrm{R}=\id\otimes \Sigma^z.
\end{equation}
As discussed in Sec.~\ref{subsec:unitary_syms}, a Lindbladian with a strong symmetry preserves the $\scT_+$ symmetry in steady-state sectors and breaks it in all others. In this case, there is thus a $\scT_+^2=+1$ symmetry in the sectors with even total longitudinal parity, $\scU^z=\scU^z_\mathrm{L}\scU^z_\mathrm{R}=+\scI$, while it is broken for odd total parity, $\scU^z=-\scI$.

Because the jump operators are normal but not unitary [i.e., satisfy Eq.~(\ref{eq:sym_scLJ_Q+}), but not Eq.~(\ref{eq:sym_scLJ_P}) or (\ref{eq:sym_scLJ_Q-})], $\scLD$ only admits a $\scQ_+$ symmetry. Because, additionally, the Hamiltonian and jump operators are symmetric, the pseudo-Hermiticity superoperator is given by $\scQ_+=\swap$ and it commutes with the $\scT_+$ operator as stated in Eq.~(\ref{eq:ex_deph_comm_TQ}). Then, it follows that for even parity, $\scU^z=+\scI$, the Lindbladian belongs to class BDI$^\dagger$, while for odd parity, $\scU^z=-\scI$, it belongs to class AI$^\dagger$, which is outside the tenfold classification of Table~\ref{tab:Lindbladian classes}.

\subsection{Simultaneous dephasing and incoherent hopping. Classes \texorpdfstring{AI$_+$}{AI+} and \texorpdfstring{AI$_-$}{AI-}}
\label{subsec:examples_Chiral}

We now consider dephasing and incoherent hopping to occur simultaneously and choose jump operators
\begin{align}
	\label{eq:jumpops_chiral}
	L_{jk\ell}=\sqrt{\gamma_{jk\ell}}\(\sigma^z_{j}+\eta_{jk\ell}\sigma^x_k\sigma^y_\ell\),
\end{align}
with real $\gamma_{jk\ell}$ and complex $\eta_{jk\ell}$, which satisfy
\begin{equation}
\acomm{L_{jk\ell}}{\Sigma^x}=\acomm{L_{jk\ell}}{\Sigma^y}=\comm{L_{jk\ell}}{\Sigma^z}=0.
\end{equation}

We take the same Hamiltonian as in Eq.~(\ref{eq:H_deph_noTRS}), which satisfies the commutation relations of Eq.~(\ref{eq:ex_deph_H_Sigma_comm}). This Lindbladian again conserves transverse parity $\scU^x$ as a weak Liouvillian symmetry and has two symmetry sectors.

When either $j$ equals one of $k$, $\ell$, or $\Re\eta_{jk\ell}=0$, the jump operators are, up to a numerical prefactor, unitary (and, by consequence, normal), and hence, according to Eqs.~(\ref{eq:sym_scLJ_P}), (\ref{eq:sym_scLJ_Q-}), and (\ref{eq:sym_scLJ_Q+}), $\scLD$ admits all of $\scP$ and $\scQ_\pm$ symmetries. However, since the Hamiltonian is not time-reversal symmetric (i.e., is not real in any basis that is trivially related to the representation basis), a $\scQ_-$ symmetry of $\scLH$ requires that both operators $q_a$ and $q_b$ in $\scQ_{ab}=q_a\otimes q_b$ (recall Sec.~\ref{subsec:sym_scLH}) commute with $H$, which would imply $q_a=q_b=\Sigma^x$. This is, however incompatible with a $\scQ_-$ symmetry of $\scLJ$, since it would require that one of $q_a$, $q_b$ commutes with the jump operators and the other one anticommutes (recall Sec.~\ref{subsec:sym_scLJ}). Noting that the jump operators are non-Hermitian, we can also exclude a $\scQ_+$ symmetry. We thus conclude that the Lindbladian only possesses $\scP$ symmetry, which is implemented by the unitary operator:
\begin{equation}
\scP=\Sigma^z\otimes\Sigma^y.
\end{equation}
Because of the commutation relation with the $\scT_+$ symmetry, 
\begin{equation}
\scP \scT_+ = (-1)^L \scU^x \, \scT_+ \scP,
\end{equation}
we find that the Liouvillian belongs to class AI$_+$ if $L$ and the parity $\scU^x$ are both even or both odd, and to class AI$_-$ if one of $L$ or $\scU^x$ is even and the other odd.

\section{Random-matrix correlations and universality}
\label{sec:rmt}

Having established the tenfold classification of irreducible Lindbladians and presented physical examples of all classes, we now look for signatures of random-matrix universality in each of those classes.
We always consider spin chains of $L$ sites with periodic boundary conditions and considered nearest-neighbor and next-to-nearest-neighbor interactions. More specifically, we restrict the couplings $J^{x,y,z}_{jk}$ in Eqs.~(\ref{eq:H_YXZ}) and (\ref{eq:jumpops_inchop}) to $J^{x,y,z}_{j,j+1}$, the couplings $K_{jk}$ in Eq.~(\ref{eq:H_deph_noTRS}) to $K_{j,j+1}$ and $K_{j,j+2}$, and the couplings $\gamma_{jk\ell}$ and $\eta_{jk\ell}$ in Eq.~(\ref{eq:jumpops_chiral}) to $\gamma_{j,j+1,j+2}$ and $\eta_{j,j+1,j+2}$, respectively.

To perform the statistical analysis of random-matrix correlations, we consider random 
(i.e., quenched disordered) spin chains. For a given coupling $g$, we either choose a fixed value $g=g_0$ or sample it from a box distribution in $[g_0-\d g, g_0+\d g]$, in which case we denote it as $g=g_0\pm\d g$. The values of the couplings in the different examples (in the order discussed in Sec.~\ref{sec:examples}) are as follows (we suppress the site indices, which were already discussed above):
\begin{enumerate}
	\item \textit{Dephasing, $H_\mathrm{X}$:}
	$\gamma=1.1\pm0.9$, $K=1.0$, and $g^x=0\pm2.1$.
	\item \textit{Dephasing, $H_\mathrm{XYZ}+H_\mathrm{X}$:}
	$\gamma=1.1\pm0.9$, $J^x=1.0$, $J^y=0.8$, $J^z=0.55$, and $g^x=0\pm0.7$.
	\item \textit{Dephasing, $H_\mathrm{XYZ}+H_\mathrm{X}+H_\mathrm{Y}$:}
	$\gamma=1.1\pm0.9$, $J^x=1.0$, $J^y=0.8$, $J^z=0.55$, $g^x=0\pm0.7$, and $g^y=-0.1\pm0.9$.
	\item \textit{Dephasing, $H_\mathrm{XYZ}+H_\mathrm{X}+H_\mathrm{Y}+H_\mathrm{Z}$:}
	$\gamma=1.1\pm0.9$, $J^x=1.0$, $J^y=0.8$, $J^z=0.55$, $g^x=0\pm0.7$, $g^y=-0.1\pm0.9$, and $h=0.2\pm0.3$.
	\item \textit{Spin injection and removal:}
	$a=0.8\pm0.4$, $b=0.7\pm0.5$, $J^x=1.0$, $J^y=0.8$, and $J^z=0.55$.
	\item \textit{Incoherent hopping:}
	$M^x=(0.3+0.2\i)\pm(0.2+0.5\i)$, $M^y=(0.5-0.4\i)\pm(0.4+0.1\i)$, $J^x=1.0$, $J^y=0.8$, $J^z=0.55$, and $h=3\pm2$.
	\item \textit{Incoherent hopping + dephasing:}
	$\gamma=1.1\pm0.9$, $\eta=0.4$, $K=0.8$, $h=0\pm0.7$.
\end{enumerate}

The eigenvalues and eigenvectors are obtained by numerical exact diagonalization. Since we are interested in the bulk correlators, we select only the eigenvalues with both real and imaginary parts larger than $10^{-6}$ (in absolute value) and their corresponding eigenvectors.

\subsection{Spectral consequences of antiunitary symmetries}
\label{subsec:spectrum}

As we have seen above, a Lindbladian class can be labeled by its antiunitary symmetries $\scT_-$ and $\scC_\pm$ or, equivalently, the closely related unitary involutions $\scP$ and $\scQ_\pm$, see Eq.~(\ref{eq:antiunitary_from_unitary}). In this section, we will use $\scT_-$ and $\scC_\pm$.
We discussed the global consequences of antiunitary symmetries in Ch.~\ref{chapter:correlations} (see, in particular, Fig.~\ref{fig:spectral_consequences}). We recall the important results for our current problem. We denote the eigenvalues of the vectorized shifted Lindbladian by $\lambda_\alpha$ and the, in general distinct, right and left eigenvectors by $\ket{\phi_\alpha}$ and $\ket{\tphi_\alpha}$, respectively, i.e., 
\begin{align}
	\label{eq:evL}
	&\scL' \ket{\phi_\alpha}=\lambda_\alpha \ket{\phi_\alpha},
	\\
	\label{eq:evLd}
	&\scL'^\dagger \ket{\tphi_\alpha}=\lambda_\alpha^* \ket{\tphi_\alpha}.
\end{align}

\begin{figure*}[t]
	\centering
	\includegraphics[width=\textwidth]{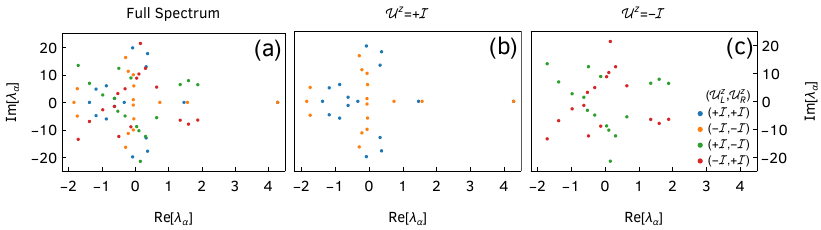}
	\caption{Spectrum of $\scL'$ in the complex plane for the incoherent hopping chain of length $L=3$ discussed in Sec.~\ref{subsec:examples_IncHop}. The remaining parameters are given in the text. The Lindbladian has four strong-symmetry sectors labeled by the pair $(\scU^z_\mathrm{L},\scU^z_\mathrm{R})$ and represented in different colors. For visual clarity, we present the full spectrum in (a), the two sectors with even parity---for which $\scT_+$ is unbroken, contain a steady-state each, and eigenvalues come in complex-conjugated pairs in each---in (b), and the two sectors with odd parity---for which $\scT_+$ is broken---in (c).}
	\label{fig:Strong_Symm}
\end{figure*}

Let us first consider the presence of a $\scT_+$ symmetry. If $\ket{\phi_\alpha}$ is a right eigenvector with eigenvalue $\lambda_\alpha$, $\scT_+\ket{\phi_\alpha}$ is also a right eigenvector of $\scL'$ with complex-conjugated eigenvalue. When $\scT_+$ is unbroken, the spectrum of $\scL'$ is symmetric about the real axis. This is illustrated in Fig.~\ref{fig:Strong_Symm}, where we show the spectrum of $\scL'$ in the complex plane for the incoherent hopping example of Sec.~\ref{sec:examples}. Recall that this example has four strong-symmetry sectors, labeled by the pair of longitudinal parities $(\scU^z_\mathrm{L},\scU^z_\mathrm{R})$. As is clearly visible, and in agreement with our predictions, the full spectrum; see Fig.~\ref{fig:Strong_Symm}(a), and the spectra of the two sectors with total parity $\scU^z=\scU^z_\mathrm{L}\scU^z_\mathrm{R}=+\scI$, see Fig.~\ref{fig:Strong_Symm}(b), are symmetric about the real axis since $\scT_+$ is unbroken; while $\scT_+$ connects the two sectors with $\scU^z=-\scI$ and does not act inside of each (i.e., is broken), leading to each pair of complex-conjugated eigenvalues to be split between the two sectors, see Fig.~\ref{fig:Strong_Symm}(c).

Proceeding similarly, we find that if there is a $\scT_-$ symmetry, $\scT_-\ket{\phi_\alpha}$ is a right eigenvector with eigenvalue $-\lambda_\alpha^*$; i.e., the spectrum is symmetric about the imaginary axis. We conclude that the presence of $\scT_-$ leads to a dihedral symmetry of the Lindbladian spectrum, a phenomenon first pointed out in Ref.~\cite{prosen2012PRL}. 

If there is an antiunitary symmetry $\scC_\pm$, then the operator implementing it connects left and right eigenvectors. Indeed, for $\scC_+$ we have that $\scC_+\ket{\phi_\alpha}$ is a left eigenvector of $\scL'$ with the same eigenvalue as $\ket{\phi_\alpha}$. Accordingly, this symmetry does not affect the global shape of the spectrum. Finally, if there is a $\scC_-$ symmetry, $\scC_-\ket{\phi_\alpha}$ is a left eigenvector of $\scL'$ with eigenvalue $-\lambda_\alpha$; i.e., $\scC_-$ reflects the spectrum across the origin. Therefore, the presence of this symmetry also implies the dihedral symmetry of the spectrum.

\begin{figure}[t]
	\centering
	\includegraphics[width=0.8\columnwidth]{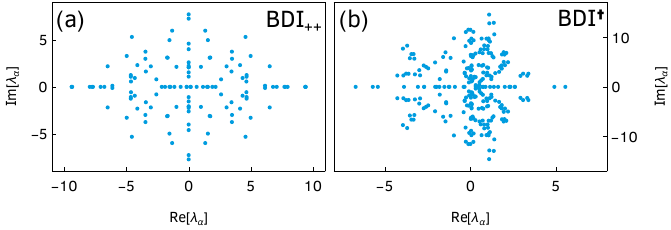}
	\caption{Spectrum of $\scL'$ in the complex plane for (a) the dephasing spin chain with Hamiltonian $H_\mathrm{XYZ}+H_\mathrm{X}$ (Sec.~\ref{subsec:examples_Deph}) and (b) the incoherent hopping chain (Sec.~\ref{subsec:examples_IncHop}). In both cases, we consider $L=4$ and the remaining parameters are given in the text. The dephasing spin chain belongs to class BDI$_{++}$ and the spectrum of the shifted Lindbladian has dihedral symmetry, while the incoherent hopping chain belongs to class BDI$^\dagger$ and its spectrum does not display dihedral symmetry.}
	\label{fig:spectra}
\end{figure}

From the data in Table~\ref{tab:Lindbladian classes} and the preceding discussion, we conclude there are eight classes with dihedral symmetry and two without (AI and BDI$^\dagger$). We illustrate this in Fig.~\ref{fig:spectra}, where we show the spectrum of $\scL'$ in the complex plane for two examples of Sec.~\ref{sec:examples}: the dephasing spin chain belonging to class BDI$_{++}$, which exhibits dihedral symmetry, and the incoherent hopping chain in class BDI$^\dagger$, which does not.

\subsection{Complex spacing ratios}

We now move to the random matrix signatures of the different antiunitary symmetries. First, we consider (bulk) local level statistics, which are sensitive to the value of $\scC_+^2$ (as usual, we denote the absence of the symmetry as $\scC_+^2=0$), captured by the distribution of complex spacing ratios (CSRs). As before, we define the CSR as
\begin{equation}
z_\alpha=\frac{
	\lambda_\alpha^{\mathrm{NN}}-\lambda_\alpha}{
	\lambda_\alpha^{\mathrm{NNN}}-\lambda_\alpha
},
\end{equation}
where $\lambda_\alpha^{\mathrm{NN}}$ and $\lambda_\alpha^{\mathrm{NNN}}$ are the nearest and next-to-nearest neighbors of $\lambda_\alpha$ in the complex plane.
The three types of level repulsion, A ($\scC_+^2=0$), AI$^\dagger$ ($\scC_+^2=+1$), and AII$^\dagger$ ($\scC_+^2=-1$)~\cite{hamazaki2020PRR} were depicted in Fig.~\ref{fig:CSR_RMT}. Level repulsion in class AII$^\dagger$ does not occur in Lindbladian symmetry classes.

\begin{figure*}[t]
	\centering
	\includegraphics[width=\textwidth]{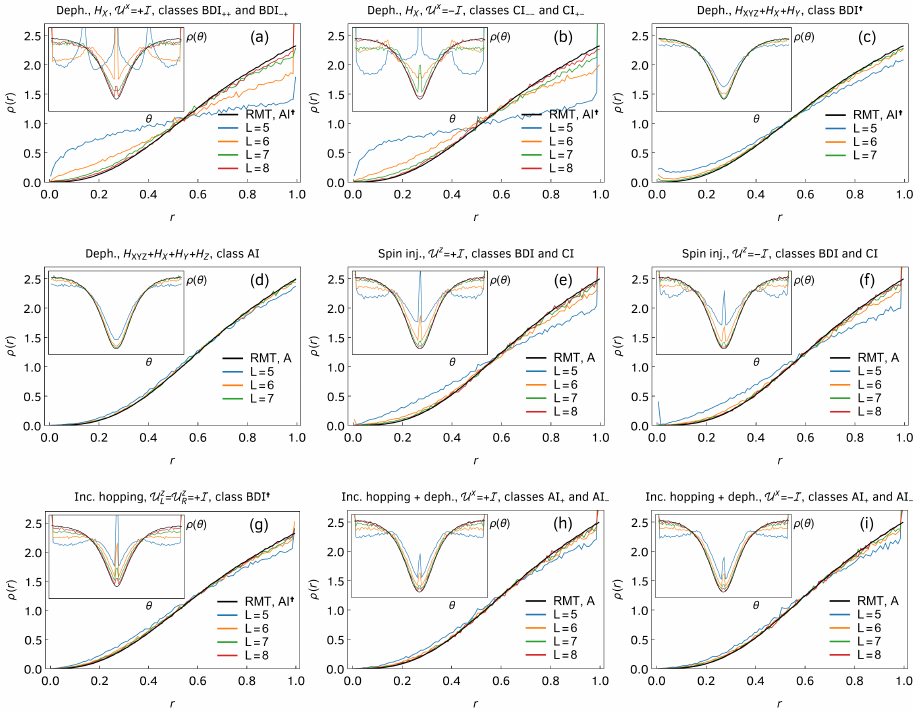}
	\caption{Complex spacing ratio distribution of all the spin-chain examples discussed in Sec.~\ref{sec:examples}. In each panel, we show the marginal radial distribution for different chain lengths (colored lines) and compare it with the random matrix prediction of Fig.~\ref{fig:CSR_RMT} (black line). In the insets, we show the marginal angular distribution. In all cases, we observe excellent agreement with the universal RMT result as $L$ increases, while there are also very strong finite-size effects that difficult a comparison for $L=5$ and $6$.}
	\label{fig:CSR_SpinChains}
\end{figure*}

To identify random-matrix universality in the examples of Sec.~\ref{sec:examples}, we randomly sample disordered spin chains and compute the CSR distribution, as described above.
For the examples with a Liouvillian weak symmetry, we consider chains of length $L=5$, $6$, $7$, and $8$, corresponding to symmetry sectors of size $2^{2L-1}=512$, $2048$, $8192$, and $32768$, respectively. For the example with a Liouvillian strong symmetry, we also consider $L=5$, $6$, $7$, and $8$, which, in this case, correspond to sector dimensions of $2^{2L-2}=256$, $1024$, $4096$ and $16384$, respectively. For the examples without unitary symmetries, we study chains of length $L=5$, $6$, and $7$, corresponding to irreducible Liouvillians of dimension $L=2^{2L}=1024$, $4096$, and $16384$, respectively. At least $10^6$ eigenvalues were considered.
In Fig.~\ref{fig:CSR_SpinChains}, we compare the marginal radial and angular CSR distributions for all the physical spin-chain examples discussed in Sec.~\ref{sec:examples} with the random matrix theory (RMT) predictions [given in Figs.~\ref{fig:CSR_RMT}(d) and (e) for the three bulk RMT ensembles], finding excellent agreement when the length $L$ of the chain becomes large. Our results illustrate RMT universality in the full tenfold classification of Lindbladians with unbroken $\scT_+$ symmetry. 

Through the use of bulk CSR, we can only resolve the value of $\scC_+^2$ (which is manifest in some panels of Fig.~\ref{fig:CSR_SpinChains}, where we group together results for different symmetry classes that share the same level repulsion). We now discuss numerical signatures that can also distinguish the values of $\scC_-^2$ and $\scT_-^2$.

\subsection{Statistics of real and imaginary eigenvalues and eigenvalues close to the origin}

\begin{figure*}[t]
	\centering
	\includegraphics[width=\textwidth]{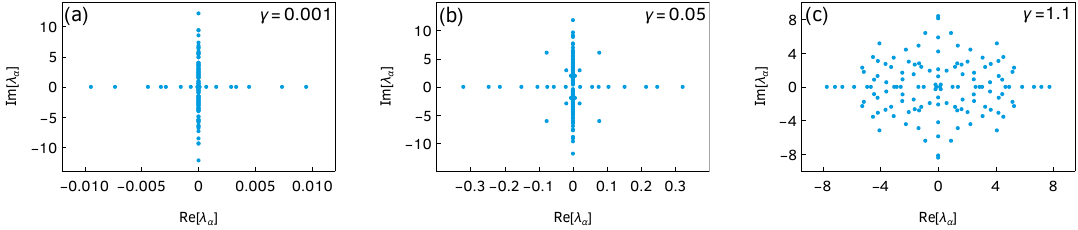}
	\caption{Spectrum of $\scL'$ in the complex plane for the dephasing spin chain with Hamiltonian $H_\mathrm{XYZ}+H_\mathrm{X}$ (Sec.~\ref{subsec:examples_Deph}) for $L=4$ and different dephasing strengths $\gamma$. The remaining parameters are given in the text. For small $\gamma$ (a), PT symmetry is unbroken and the whole spectrum lives on the real and imaginary axes. For large enough $\gamma$ (b), a series of exceptional points occurs, with a pair of eigenvalues on one of the symmetry axes colliding and shooting off into the complex plane, spontaneously breaking PT symmetry. For very large $\gamma$ (c), most of the spectrum lives in the complex plane. This phenomenon of spontaneous PT symmetry breaking renders the number of real and purely imaginary eigenvalues, and their statistics, nonuniversal, and we do not use them to characterize the symmetries and correlations Lindbladians.}
	\label{fig:PT_symmetry_breaking}
\end{figure*}

In Hermitian systems, particle-hole and chiral symmetries manifest themselves in the eigenvalues near the origin, because these symmetries reflect the spectrum across it. Not only are universal local correlations in this region distinct from the bulk, even the spectral density is universal in a certain microscopic limit and determined only by the symmetry~\cite{verbaarschot1993PRL,verbaarschot1994PRL,akemann1997NuclPhysB}. The same reasoning leads to the speculation that in order to obtain local information on $\scC_-$ and $\scT_\pm$ symmetries, we need to restrict our attention to the vicinity of the axis of symmetry of the spectrum. The effects of these symmetries on the correlations near the origin have been addressed in the Ginibre~\cite{akemann2009JPhysA,akemann2009PRE} and non-Ginibre classes~\cite{garcia2022PRX}, while universal statistics of real eigenvalues were studied for the real Ginibre~\cite{lehmann1991PRL,kanzieper2005PRL,forrester2007PRL} and non-Ginibre classes~\cite{xiao2022PRR}.

First, we point out that the statistics of the eigenvalues closest to the origin, employed in Ref.~\cite{garcia2022PRX} as a signature of different symmetry classes, are affected by the existence of exact zero modes of $\scL'$ in some of our families of spin-1/2 chains. The existence of these zero modes for all realizations of disorder induces additional level repulsion from the origin, altering the distribution of nonzero eigenvalues closest to it. As an example of this phenomenon, we mention the chain with spin injection, belonging to class BDI, which supports two exact zero modes. In principle, if the number of zero modes does not scale with system size, one could study the distribution of the eigenvalues closest to the origin for each number of zero modes.

More critically, we also find that for our examples in disordered spin chains, the statistics of eigenvalues on or near the axes of symmetry are nonuniversal because of the spontaneous breaking of PT symmetry~\cite{prosen2012PRL}. In Ref.~\cite{xiao2022PRR} it was found that, in the ergodic regime (in which RMT behavior is expected), the number of real eigenvalues in the spectra of some physical non-Hermitian Hamiltonians is universal and equal to the random matrix value $~\propto\sqrt{D}$, with $D$ the Hilbert space dimension. In contrast, for Lindbladians with dihedral symmetry, the fraction of real and imaginary eigenvalues and its statistics depend on the relative strength $g$ of the non-Hamiltonian part of the Lindbladian ($g$ can be, for instance, the dephasing strength $\gamma$ or the spin-injection rate $a$). For $g<g_{\mathrm{PT}}$, with finite size-dependent critical $g_\mathrm{PT}$, all eigenvalues of the shifted Lindbladian $\mathcal L'$ reside on the cross formed by the real and imaginary axes (PT-unbroken phase)~\cite{prosen2012PRL}. At $g=g_\mathrm{PT}$ (the first exceptional point), a pair of eigenvalues on the cross collides and shoots off into the complex plane, spontaneously breaking PT symmetry, and, consequently, reducing the number of real or imaginary eigenvalues. As $g$ increases further, more collisions of eigenvalues occur. The change in the number of real eigenvalues for the dephasing spin chain in class BDI$_{++}$ as a function of $\gamma$ is illustrated in Fig.~\ref{fig:PT_symmetry_breaking}. 
Concomitantly with the nonuniversality of the number of eigenvalues on the axes of symmetry, we also find their statistics to be nonuniversal and depend sensitively on the coupling $g$. As a consequence, we are not able to employ the statistics of, say, purely imaginary eigenvalues as a diagnostic of the symmetries and RMT universality in Lindbladian classes.
We expect that for $g\gg g_\mathrm{PT}$, i.e., deep in the symmetry-broken phase, the number of real and imaginary eigenvalues becomes universal and their statistics obey RMT. In particular, in the thermodynamic limit, we expect $g_\mathrm{PT}\to0$~\cite{prosen2012PRL}, and, hence RMT statistics for all nonzero dissipation. However, since we have only access to relatively small system sizes and the disorder changes the precise value of $g_\mathrm{PT}$ from realization to realization, we do not pursue this question further in this work, and instead turn to an alternative signature of the different symmetries that works at any coupling $g$.

\subsection{Eigenvector overlaps}

Having ruled out the prospect of inferring antiunitary symmetries of Lindbladians through local spectral information near the symmetry axis, we turn to the possibility of using nonlocal bulk information. To that end, we consider the Chalker-Mehlig eigenvector overlap matrix~\cite{chalker1998PRL,mehlig2000JMP}, discussed in Ch.~\ref{chapter:correlations}:
\begin{equation}
O_{\alpha\beta}=\braket{\tphi_\alpha}{\tphi_\beta}\braket{\phi_\beta}{\phi_\alpha}.
\end{equation}

\begin{figure*}[t]
	\centering
	\includegraphics[width=\textwidth]{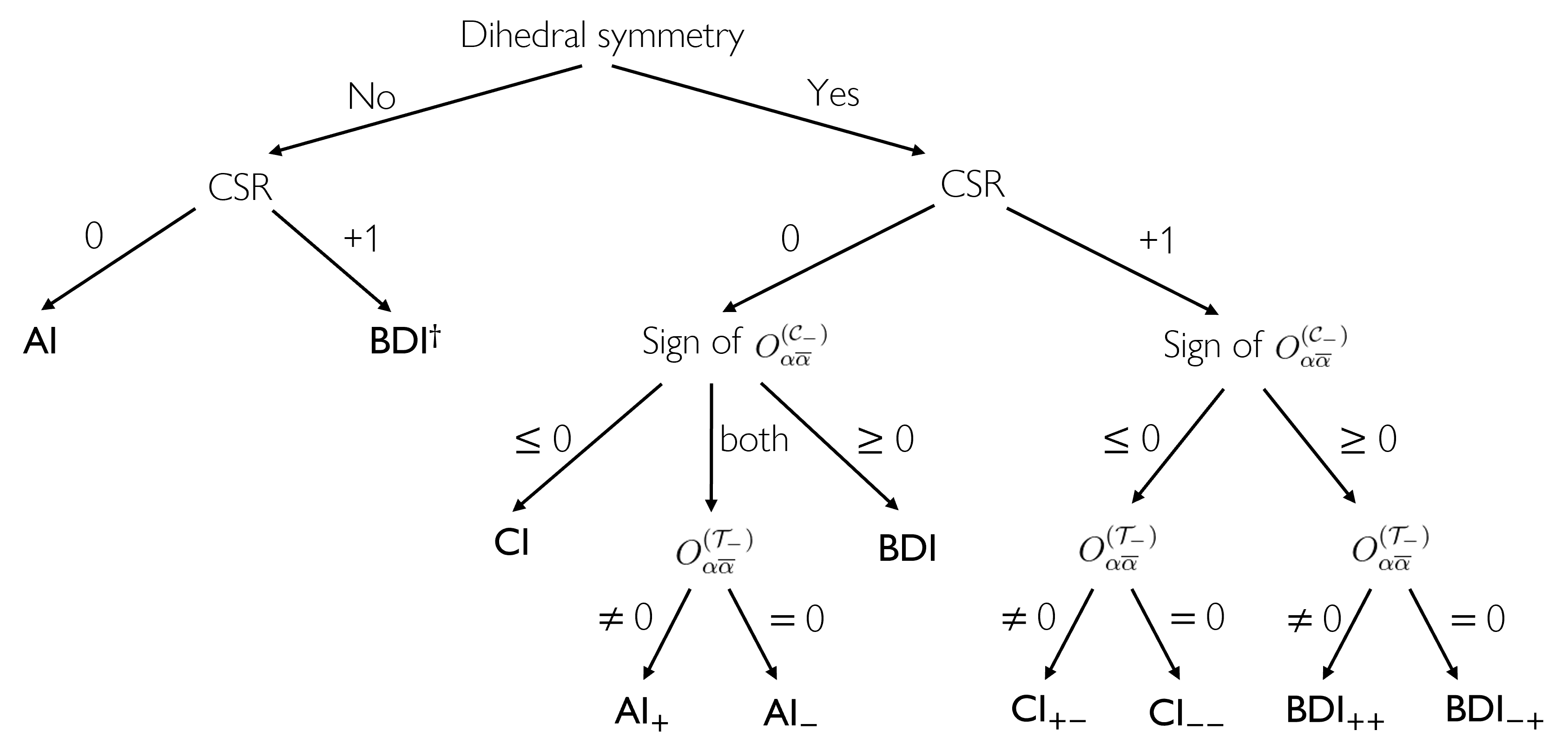}
	\caption{Decision tree illustrating the possibility of distinguishing the full Lindbladian tenfold classification by jointly employing the dihedral symmetry of the spectrum, the complex spacing ratio distribution (denoted as $0$ for bulk level repulsion of class A and $+1$ for class AI$^\dagger$), the sign of the off-diagonal eigenvector overlap $O^{(\scC_-)}_{\alpha\balpha}$, and whether or not the overlaps $O^{(\scT_-)}_{\alpha\balpha}$ are identically zero.}
	\label{fig:decision_tree_classes}
\end{figure*}

In Ch.~\ref{chapter:correlations} we proved the following two results:
\begin{enumerate}
	\item If $\ket{\tphi_\balpha}\propto \scC_-\ket{\phi_\alpha}$, the overlaps $O_{\alpha\balpha}$ (denoted $O_{\alpha\balpha}^{(\scC_-)}$ for clarity) are all non-negative if $\scC_-^2=+1$, and all nonpositive if $\scC_-^2=-1$. If $\scC_-^2=0$ (i.e., if the symmetry is absent and the eigenvectors are independent), the overlaps are still real for spectra with dihedral symmetry, and the fraction of positive and negative matrix elements is $1/2$ each. For classes with no $\scC_-$ symmetry and no dihedral symmetry, the overlaps are complex.
	\item If $\ket{\phi_\balpha}\propto \scT_-\ket{\phi_\alpha}$ and $\scT_-^2=-1$, the overlaps $O_{\alpha\balpha}$ (denoted $O_{\alpha\balpha}^{(\scT_-)}$) all vanish identically. If $\scT_-^2=+1$ or $0$ they assume arbitrary complex values.
\end{enumerate}
Recall also that explicit knowledge of the operator $\scC_\pm$ or $\scT_\pm$ is not required to compute the respective eigenvector overlaps.

{
	\setlength{\tabcolsep}{8pt}
	\begin{table*}[t]
		\caption{Signatures of $\scC_-$ and $\scT_-$ symmetries in eigenvector overlaps for all the spin-chain examples discussed in Sec.~\ref{sec:examples}. Each set of examples (dephasing, spin injection or removal, incoherent hopping, and simultaneous incoherent hopping and dephasing) realizes different classes depending on the choice of Hamiltonian, conserved parity sectors, and the parity of the chain length $L$. The examples are listed in the same order as discussed in Sec.~\ref{sec:examples}. For each, we give the corresponding symmetry class, the values of the square of the two antiunitary symmetries $\scC_-$ and $\scT_-$, whether the sign of the off-diagonal overlap $O_{\alpha\balpha}^{(\scC_-)}$ is mostly non-negative or nonpositive, together with the fraction of times this happens, and whether the off-diagonal overlap $O_{\alpha\balpha}^{(\scT_-)}$ is mostly zero or nonzero, together with the fraction of times this happens. We see that the criteria we put forward for the values of $\scC_-^2$ and $\scT_-^2$ are always satisfied at least $99.9\%$ of the time and all the predictions of Sec.~\ref{sec:examples} are verified.}
		\label{tab:overlap_results}
		\begin{tabular}{@{}llrrll@{}}
			\toprule
			Example                                                        & Class         & $\scC_-^2$ & $\scT_-^2$ & $\mathrm{Re}\,O_{\alpha\balpha}^{(\scC_-)}$ & $O_{\alpha\balpha}^{(\scT_-)}$ \\ \midrule
			Deph., $H_\mathrm{X}$, $\scU^x=+\scI$, $L$ even                 & BDI$_{++}$    & $+1$       & $+1$       & $\geq0$, $100\%$               & $\neq0$, $99.98\%$             \\
			Deph., $H_\mathrm{X}$, $\scU^x=+\scI$, $L$ odd                  & BDI$_{-+}$    & $+1$       & $-1$       & $\geq0$, $100\%$               & $=0$, $99.98\%$                \\
			Deph., $H_\mathrm{X}$, $\scU^x=-\scI$, $L$ even                 & CI$_{--}$     & $-1$       & $-1$       & $\leq0$, $100\%$               & $=0$, $99.92\%$                \\
			Deph., $H_\mathrm{X}$, $\scU^x=-\scI$, $L$ odd                  & CI$_{+-}$     & $-1$       & $+1$       & $\leq0$, $100\%$               & $\neq0$, $99.99\%$             \\
			Deph., $H_\mathrm{XYZ}+H_\mathrm{X}$, $\scU^x=+\scI$            & BDI$_{++}$    & $+1$       & $+1$       & $\geq0$, $100\%$               & $\neq0$, $100\%$               \\
			Deph., $H_\mathrm{XYZ}+H_\mathrm{X}$, $\scU^x=-\scI$            & CI$_{+-}$     & $-1$       & $+1$       & $\leq0$, $100\%$               & $\neq0$, $100\%$               \\
			Deph., $H_\mathrm{XYZ}+H_\mathrm{X}+H_\mathrm{Y}$              & BDI$^\dagger$ & $0$        & $0$        & $\leq0$, $50.12\%$             & $\neq0$, $100\%$               \\
			Deph., $H_\mathrm{XYZ}+H_\mathrm{X}+H_\mathrm{Y}+H_\mathrm{Z}$ & AI            & $0$        & $0$        & $\leq0$, $50.07\%$             & $\neq0$, $99.999\%$            \\
			Spin inj., $\scU^z=+\scI$, $L$ even                            & BDI           & $+1$       & $0$        & $\geq0$, $100\%$               & $\neq0$, $100\%$               \\
			Spin inj., $\scU^z=+\scI$, $L$ odd                             & CI            & $-1$       & $0$        & $\leq0$, $100\%$               & $\neq0$, $100\%$               \\
			Spin inj., $\scU^z=-\scI$, $L$ even                            & CI            & $-1$       & $0$        & $\leq0$, $100\%$               & $\neq0$, $100\%$               \\
			Spin inj., $\scU^z=-\scI$, $L$ odd                             & BDI           & $+1$       & $0$        & $\geq0$, $100\%$               & $\neq0$, $100\%$               \\
			Inc. hopping, $\scU^z_\mathrm{L}=\scU^z_\mathrm{R}=+\scI$      & BDI$^\dagger$ & $0$        & $0$        & $\leq0$, $51.01\%$             & $\neq0$, $100\%$             \\
			Inc. hopping + deph., $\scU^x=+\scI$, $L$ even                 & AI$_+$        & $0$        & $+1$       & $\geq0$, $50.50\%$             & $\neq0$, $100\%$               \\
			Inc. hopping + deph., $\scU^x=+\scI$, $L$ odd                  & AI$_-$        & $0$        & $-1$       & $\geq0$, $50.44\%$             & $=0$, $99.98\%$                \\
			Inc. hopping + deph., $\scU^x=-\scI$, $L$ even                 & AI$_-$        & $0$        & $-1$       & $\geq0$, $50.06\%$             & $=0$, $99.94\%$                \\
			Inc. hopping + deph., $\scU^x=-\scI$, $L$ odd                  & AI$_+$        & $0$        & $+1$       & $\geq0$, $51.20\%$             & $\neq0$, $100\%$               \\ \bottomrule
		\end{tabular}
	\end{table*}
}

The overlaps $O^{(\scC_-)}_{\alpha\balpha}$ and $O^{(\scT_-)}_{\alpha\balpha}$, together with the CSR distribution, are enough to distinguish the ten Lindbladian classes with unbroken $\scT_+$ symmetry, as illustrated in Fig.~\ref{fig:decision_tree_classes}.
We compute $O_{\alpha\balpha}^{(\scA)}$, $\scA=\scC_-,\scT_-$, in the bulk for the randomly sampled disordered spin chains of each example. At least $2\times 10^6$ eigenvectors were considered and we restrict ourselves to sizes $L=5$ and $L=6$. The fraction of positive, negative, and zero $O_{\alpha\balpha}^{(\scC_-)}$ and of zero and nonzero $O_{\alpha\balpha}^{(\scT_-)}$ in each class are listed in Table~\ref{tab:overlap_results}. We conclude that in all of them, the criteria for $\scC_-^2$ and $\scT_-^2$ are satisfied for at least $99.9\%$ of realizations and our examples conform spectacularly to random-matrix universality, confirming the tenfold classification of many-body Lindbladians with unbroken $\scT_+$ symmetry put forward in previous sections.

\section{Summary and outlook}

In this chapter, we put forward a symmetry classification of many-body Lindbladian superoperators and confirmed it through a study of random-matrix correlators in experimentally implementable dissipative spin chains. We found that Lindbladians without unitary symmetries and Lindbladians with symmetries in steady-state symmetry sectors belong to one of ten non-Hermitian symmetry classes. These classes are characterized by the existence of a $\scT_+^2=+1$ symmetry implemented by the \textsc{swap} operator. Going beyond sectors with steady states breaks the $\scT_+$ swap symmetry between the two copies (bra and ket) of the system and, consequently, enriches the symmetry classification.

Interestingly, we found compelling evidence that $\scC_+^2=-1$ and $\scT_+^2=-1$ symmetries cannot be implemented inside individual symmetry sectors, reducing the allowed number of classes of many-body Lindbladian from 54 to 29. This conclusion does not exclude the possibility of a $\scC_+^2=-1$ connecting different sectors. Indeed, such a symmetry can be implemented between two sectors of odd fermionic parity~\cite{lieu2022PRB}. As a consequence, all eigenvalues are doubly degenerate, but the two eigenvalues of a given pair belong to different sectors, thus not defining a symmetry class with Kramers degeneracy.

The $\scC_+$ symmetry of a given class can be detected through the use of bulk complex spacing ratios, while $\scC_-$ and $\scT_-$ symmetries require the study of correlations on or near the axes of symmetry of the spectrum. Because of the spontaneous breaking of PT symmetry, we found eigenvalue correlations on these axes not to be useful in practice.
Intead, we employed the overlap of symmetry connected eigenstates as an empirical detector of antiunitary symmetry. Together with the bulk complex spacing ratios and dihedral symmetry, they allow us to fully resolve the Lindbladian tenfold way without unitary symmetries.

Our work complements ongoing effort to characterize PT-symmetric Lindbladian dynamics~\cite{prosen2012PRL,prosen2012PRA,vancaspel2018PRA,huybrechts2020PRB,huber2020SciPost,nakanishi2022PRA,starchl2022PRL} (not to be confused with conventional Hamiltonian PT symmetry). Note that Liouvillian PT symmetry is nothing but pseudo-Hermiticity of the dynamical generator. The definition of Liouvillian PT symmetry put forward in Ref.~\cite{huber2020SciPost}, which we would propose to call a \emph{strong Liouvillian PT symmetry} (or strong pseudo-Hermiticity), clearly implies the existence of a pseudo-Hermiticity transformation $\scQ_\pm$ of the Lindbladian but, by allowing for shifts of the Lindbladian spectrum, our classification goes beyond that definition and includes Lindbladians with weak pseudo-Hermiticity (\emph{weak Liouvillian PT symmetry}).
Remarkably, pseudo-Hermiticity has observable consequences in the transient quantum dynamics. 
More concretely, the dihedral symmetry of the spectrum implies the existence of a time-reversal-like property of certain correlation functions, despite the dynamics being dissipative.
Furthermore, if the pseudo-Hermiticity is not spontaneously broken, then there is collective decay of the eigenmodes, as all eigenvalues of the shifted Lindbladian are either purely real or purely imaginary.

Finally, our work also does not address the relation between the non-Hermitian classification of dynamical generators and the Hermitian classification of steady states. The two classifications are decoupled for quadratic open quantum systems~\cite{lieu2020PRL}, but it is unclear, at this point, if there exists any correspondence between the Altland-Zirnbauer class~\cite{altland1997} of the steady state and the corresponding dynamical Bernard-LeClair class of the generator in the many-body case. A simple one-to-one correspondence cannot exist because Lindbladians in any of the five classes with $\scC_+^2=+1$ lead to a featureless steady state proportional to the identity [as follows from Eq.~(\ref{eq:sym_scLJ_Q+})], but there could still exist a more limited correspondence between the remaining five classes and a subset of the Altland-Zirnbauer classes. We leave a matching of symmetries on both sides (if any exists) for future work.

%% file: Thesis_Circuits.tex

\chapter{Integrable nonunitary open quantum circuits}
\label{chapter:circuits}

Integrability is a fascinating field of mathematical physics. It provides exact solutions to dynamics and equilibrium in very diverse contexts, ranging from deterministic (i) classical~\cite{faddeevtakhtajan,bernard} and (ii) quantum~\cite{faddeev,korepin} many-body Hamiltonian dynamics, to classical stochastic systems, (iii) {\em in}~\cite{baxter1982}, and (iv) {\em out}~\cite{derrida} of equilibrium.
Although the Liouville-Arnold~\cite{arnold} (i), Bethe-ansatz~\cite{bethe,gaudin} (ii), and 
Onsager~\cite{onsager} (iii) threads of integrability were initially developed independently, they were beautifully united within the techniques of (quantum) inverse scattering~\cite{gardner,korepin,faddeev} and the celebrated Yang-Baxter equation~\cite{baxter1982,cnyang}.
Later, quantum inverse scattering methods (a.k.a.\ algebraic Bethe ansatz) found their way to the exact solution (diagonalization) of classical stochastic systems---many-body Markov chains, such as simple exclusion processes~\cite{kirone}. More recently, related new techniques have been developed for the exact solution of open integrable quantum many-body systems, specifically, by extending the algebraic Bethe ansatz to noncompact (nonunitary) auxiliary spaces~\cite{prosen2015REVIEW} and by providing an exact mapping between Liouvillians of open many-body systems and Bethe-ansatz integrable systems on (thermofield) doubled Hilbert spaces~\cite{medvedyeva2016,ziolkowska2020SciPost}.

Very recently, (local) quantum circuits have become an important paradigm of nonequilibrium many-body physics, in particular, due to the ability to simulate them in emerging quantum computing facilities, where they provide a natural platform for the demonstration of quantum supremacy~\cite{google2019}. Moreover, (open) quantum circuits with local projective measurements have been shown to host an exciting new physics paradigm of measurement-induced phase transitions~\cite{chan2019,skinner2019,li2019,altman,ludwig}.
A natural and significant question arises, namely, if integrability methods can be extended to such the quantum-circuit paradigm. The results on integrable trotterizations of integrable quantum spin chains~\cite{vanicat2018} and classical stochastic parallel update exclusion processes~\cite{vanicat2018b} give very encouraging hints.

In this chapter, we take a key step in this direction by constructing an integrable open (nonunitary) local quantum circuit. We show that Shastry's $\check{R}$-matrix~\cite{shastry1986a,shastry1986b,shastry1988,maassarani1998,essler2005}, the essential integrability concept of the one-dimensional Fermi-Hubbard model, can be interpreted as a completely positive (CP) trace-preserving (TP) map over a pair of qubits (spins 1/2) after a suitable analytic continuation of the interaction and spectral parameters. Our CPTP map represents a convex combination of two coherent (unitary) symmetric nearest-neighbor-hopping (XX) processes, one of them composed with local dephasing. By virtue of the Yang-Baxter equation, we then show the existence of a commuting transfer matrix for the brickwork quantum circuit built from such CPTP maps, generating a family of local superoperators commuting with the dynamical map. The integrability of the map is broken by adding interactions to the local coherent dynamics or by removing the dephasing. In particular, even circuits built from convex combinations of local free-fermion unitaries are nonintegrable. Moreover, the construction allows us to explicitly build circuits belonging to different non-Hermitian symmetry classes, which are characterized by the behavior under transposition instead of complex conjugation. The integrability of the Floquet dynamics is also demonstrated empirically by studying spectral statistics (complex spacing ratios~\cite{sa2020PRX}).

The rest of the chapter is organized as follows. In Sec.~\ref{sec:Hubbard_circuit} we define the dissipative Hubbard circuit and describe in detail its elementary local gates. Then, we prove that the circuit is indeed integrable and CPTP in Secs.~\ref{sec:proof_integrability} and \ref{sec:proof_CPTP}, respectively. The subsequent three sections focus on the physical content of our circuit: We address its Kraus representation in Sec.~\ref{sec:Kraus_representation}, integrability-breaking regimes in Sec.~\ref{sec:integrability_breaking}, and its symmetries in Sec.~\ref{sec:symmetries}. Finally, we present numerical evidence corroborating all our results in Sec.~\ref{sec:numerics}.

This chapter is based on Ref.~\cite{sa2021PRB}.

\section{The dissipative Hubbard circuit}
\label{sec:Hubbard_circuit}
We again consider a spin-$1/2$ chain of even size $L$ with periodic boundary conditions. The density matrix $\rho$ of the system evolves under the action of the discrete-time quantum channel $\Psi$, $\rho(t+1)=\Psi[\rho(t)]$---a linear map over the $4^L$-dimensional state vector $\rho$---that we choose to be of the brickwork circuit form:
\begin{equation}\label{eq:Psi_def}
\begin{split}
\Psi&=
	\left(
	\prod_{j=1}^{L/2}
	\RH_{2j,2j+1}
	\right)
	\left(
	\prod_{j=1}^{L/2}\RH_{2j-1,2j}
	\right)
	\\
	&=\includegraphics[width=0.7\columnwidth,valign=c]{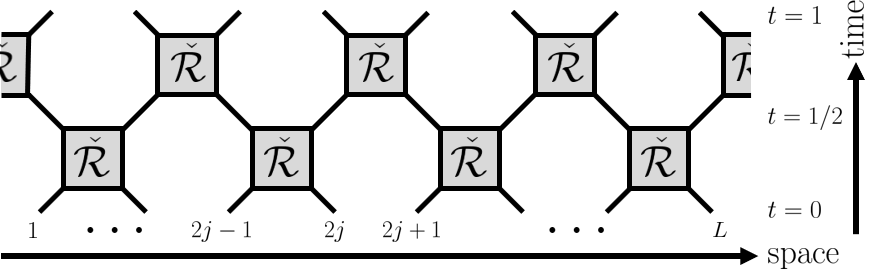}
	\end{split}
	\end{equation}
	Here, $\RH_{kl}$ is the Hubbard $\check{R}$-matrix nontrivially acting on sites $k$ and $l$ [defined explicitly in Eqs.~(\ref{eq:Rch_general6v})--(\ref{eq:RcheckHubb})]. Each wire in Eq.~(\ref{eq:Psi_def}) carries a four-dimensional operator Hilbert space and $\RH$ acts as a two-site ($16\times16$) elementary gate (grey box). One time-step consists of two rows of the circuit---in the second of which the elementary gates are shifted by one site. Accordingly, Eq.~(\ref{eq:Psi_def}) can also be written as
	\begin{equation}
	\Psi=\mathbb{T}^\dagger\Phi\mathbb{T}\Phi,
	\end{equation} 
	where $\Phi=\RH^{\otimes L/2}$ corresponds to a single row of the circuit. We introduced the one-site translation operator $\mathbb{T}$, defined by its action on the computational operator-basis, $\mathbb{T}\ket{e_1,e_2,\dots,e_L}=\ket{e_L,e_1,\dots,e_{L-1}}$, where 
	indices $e_j\in\{0,1,2,3\}$ label four possible spin-1/2 operators at site $j$.
	While each row $\Phi$ of the circuit is factorizable into two-site elementary gates, the checkerboard pattern renders the full circuit $\Psi$ interacting. Because the same gate $\RH$ is applied throughout space and time, the repeated action of $\Psi$ leads to, in general, nonunitary translationally-invariant Floquet dynamics.
	
	After a Jordan-Wigner transformation, the  Hubbard model can be understood as a spin ladder formed of a pair of XX models (corresponding to up- and down-spin fermions or to the \emph{bra} and the \emph{ket} of the density matrix~\cite{medvedyeva2016} in our nonunitary formulation) coupled by the Hubbard interaction along the rungs. Thus, we start with the (two-site) spin-$1/2$ XX $\check{R}$-matrix,
	\begin{equation}\label{eq:Rch_general6v}
	\check{R}
	=\frac{1}{a}
	\begin{pmatrix}
	a & 0 & 0 & 0 \\
	0 & c & -\i b & 0 \\
	0 & \i b & c & 0 \\
	0 & 0 & 0 & a
	\end{pmatrix},
	\end{equation}
	which admits a simple trigonometric parametrization:
	\begin{equation}\label{eq:RchXX}
	a=\cos\lambda,\quad
	b=\sin\lambda,\quad
	c=1.
	\end{equation}
	$\check{R}=\check{R}(\lambda)$ is real orthogonal for imaginary spectral parameter $\lambda\in \i\mathbb R$.\footnote{We have introduced the factors of $\pm \i$ multiplying $b$ (corresponding to the choice $x=-\i$ in Eq.~(12.93) of Ref.~\cite{essler2005}) to compensate for the imaginary spectral parameter (since $\sin \i\lambda=\i\sinh\lambda$). While these factors could be removed by a trivial similarity transformation, this choice will prove convenient below.}
	We introduce a basis $\{e_{\alpha}^{\beta}\}$ of $2\times2$ matrices such that the only nonzero entry (equal to 1) of $e^{\beta}_{\alpha}$ is in row $\alpha$ and column $\beta$. We then consider the action of $\check{R}$ on two copies of the system (corresponding to ket ($\uparrow$) and bra ($\downarrow$) of the vectorized density matrix $\rho=\sum_{mn}\rho_{mn} \ket{m}\bra{n}\mapsto \ket{\rho}=
	\sum_{mn}\rho_{mn}\ket{m}\otimes \ket{n}^*$),
	\begin{equation}  \label{eq:rupdown}
	\begin{split}
	\check{r}_{\uparrow}(\lambda)=\check{R}^{\alpha\gamma}_{\beta\delta}(\lambda)\,e^{\beta}_{\alpha}\otimes \id_2\otimes e^{\delta}_{\gamma}\otimes \id_2,\\
	\check{r}_{\downarrow}(\lambda)=\check{R}^{\alpha\gamma}_{\beta\delta}(\lambda)\,\id_2\otimes e^{\beta}_{\alpha}\otimes \id_2\otimes e^{\delta}_{\gamma},
	\end{split}
	\end{equation}
	where $\id_d$ is the $d\times d$ identity matrix. Summation over repeated indices is assumed throughout. 
	
	In terms of the XX $\check{R}$-matrices, the Hubbard $\check{R}$-matrix reads (choosing the appropriate gauge)~\cite{essler2005}
	\begin{equation}\label{eq:RcheckHubb}
	\RH(\lambda,\mu)
	=\beta\, \check{r}(\lambda-\mu)
	+\alpha\, \check{r}(\lambda+\mu)
	\left(\sigma^{z}\otimes\sigma^{z}\otimes 
	\id_4\right),
	\end{equation}  
	where $\check{r}(\lambda)=\check{r}_{\uparrow}(\lambda)\check{r}_{\downarrow}(\lambda)$ and $\sigma^{x,y,z}$ denote the standard Pauli matrices. The two prefactors $\alpha\equiv\alpha(\lambda,\mu,u)$ and $\beta\equiv\beta(\lambda,\mu,u)$ depend on two independent spectral parameters $\lambda$ and $\mu$ and on the Hubbard interaction strength $u$. The Hubbard $\check{R}$-matrix---which is not of difference form---satisfies the Yang-Baxter equation,
	\begin{equation}\label{eq:YBE}
	\begin{split}
	&\left(\id_4\otimes\RH(\lambda,\mu)\right)
	\left(\RH(\lambda,\nu)\otimes\id_4\right)
	\left(\id_4\otimes\RH(\mu,\nu)\right)=\\
	=&\left(\RH(\mu,\nu)\otimes\id_4\right)
	\left(\id_4\otimes\RH(\lambda,\nu)\right)
	\left(\RH(\lambda,\mu)\otimes \id_4\right)
	,
	\end{split}
	\end{equation}
	if the ratio $\alpha/\beta$ is fixed as
	\begin{equation}
	\frac{\alpha}{\beta}=\frac{\cos(\lambda+\mu)\sinh(h-\ell)}{\cos(\lambda-\mu)\cosh(h-\ell)},
	\end{equation}
	where $h$ and $\ell$ are implicitly defined in terms of $\lambda$, $\mu$, and $u$ through $\sinh(2h)/\sin(2\lambda)=\sinh(2\ell)/\sin(2\mu)=u$. Finally, by choosing
	\begin{equation}
	\beta=
	\frac{\cos(\lambda-\mu)\cosh(h-\ell)}{\cos(\lambda-\mu)\cosh(h-\ell)+\cos(\lambda+\mu)\sinh(h-\ell)},
	\end{equation}
	we have $\alpha+\beta=1$. Furthermore, with this choice of $\beta$, $\RH$ satisfies the unitarity condition:
	\begin{equation}\label{eq:unitarity_condition}
	\RH(\lambda,\mu)\RH(\mu,\lambda)=\id_{16}.
	\end{equation}
	
	\section{Proof of the integrability of the Hubbard circuit}
	\label{sec:proof_integrability}
	By construction, the Hubbard circuit $\Psi$~(\ref{eq:Psi_def}) is integrable. Indeed, since $\RH$ satisfies the (braid) Yang-Baxter equation~(\ref{eq:YBE}), there exists a one-parameter family $\tt(\omega)$ of transfer matrices in involution, i.e.,  $\comm{\tt(\omega_1)}{\tt(\omega_2)}=0$ for all $\omega_1,\omega_2$. After introducing an auxiliary space, labeled $a$, identical to the (local) four-dimensional physical Hilbert space, the transfer matrix is expressed as the partial trace of the monodromy matrix, $\tt(\omega)=\Tr_a\sT_a(\omega)$, with
	\begin{equation}\label{eq:monodromy_matrix}
	\sT_a(\omega)=\prod_{1\leq j\leq L}^{\leftarrow}\RHp_{aj}\left(
	\omega,\frac{\lambda+\mu}{2}-(-1)^j\frac{\lambda-\mu}{2}
	\right),
	\end{equation}
	where $\RHp=\sP \RH$, $\sP$ is a $16\times 16$ permutation matrix defined by $\sP \left( \ket{\rho_1} \otimes \ket{\rho_2} \right) = \ket{\rho_2}\otimes \ket{\rho_1}$, and the symbol {\scriptsize $\displaystyle{\prod^{\leftarrow}}_j$} indicates an ordered product with decreasing index~$j$. The monodromy matrix $\sT_a$ is inhomogeneous (staggered) to account for the checkerboard pattern of the quantum circuit. Evaluating the monodromy matrix~(\ref{eq:monodromy_matrix}) at the two special (a.k.a.\ shift) points $\omega=\lambda$ and $\omega=\mu$, the Floquet propagator $\Psi$, defined by Eq.~(\ref{eq:Psi_def}), can be written as
	\begin{equation}\label{eq:Psi_TransferMatrix}
	\Psi=\tt(\mu)^{-1}\tt(\lambda).
	\end{equation} 
	To verify this claim, we start by computing the monodromy matrix at $\omega=\lambda$. It reads as
	\begin{equation}
	\begin{split}
	\sT_a(\lambda)\,
	&=\prod_{1\leq j\leq L/2}^{\leftarrow}
	\RHp_{a,2j}(\lambda,\mu)\sP_{a,2j-1}\\
	&=\prod_{1\leq j\leq L/2}^{\leftarrow}
	\sP_{a,2j}\RH_{a,2j}(\lambda,\mu)\sP_{a,2j-1}\\
	&=\prod_{1\leq j\leq L/2}^{\leftarrow}
	\sP_{a,2j}\sP_{a,2j-1}
	\RH_{2j-1,2j}(\lambda,\mu),
	\end{split}
	\end{equation}
	where we have used the identities $\RHp(\lambda,\lambda)=\RHp(\mu,\mu)=\sP$ and $\RH_{al}\sP_{ak}=\sP_{ak}\RH_{kl}$. Because all $\sP$ and $\RH$ operators commute with each other when acting on different Hilbert spaces (i.e., when they have no subscript indices in common), taking the trace over the auxiliary space yields
	\begin{equation}
	\tt(\lambda)=\Tr_a\left[
	\prod_{1\leq j\leq L}^{\leftarrow}\sP_{aj}
	\right]\prod_{j=1}^{L/2}\RH_{2j-1,2j}(\lambda,\mu).
	\end{equation} 
	To evaluate the remaining trace, we use $\sP_{al}\sP_{ak}=\sP_{ak}\sP_{kl}$ to permute $\sP_{a1}$ over all the other $\sP_{aj}$ and then use $\Tr_a \sP_{a1}=\id_4$. The transfer matrix finally reads as
	\begin{equation}
	\tt(\lambda)=\left(
	\prod_{2\leq j\leq L}^{\leftarrow}\sP_{1j}
	\right)
	\prod_{j=1}^{L/2}\RH_{2j-1,2j}(\lambda,\mu).
	\end{equation}
	The computation for $\omega=\mu$ proceeds similarly and results in the expression
	\begin{equation}
	\sT_a(\mu)
	=\left(
	\prod_{1\leq j\leq L}^{\leftarrow} \sP_{aj}
	\right)
	\RH_{a1}(\mu,\lambda)
	\prod_{j=1}^{L/2-1}\RH_{2j,2j+1}(\mu,\lambda).
	\end{equation}
	To evaluate the trace over the auxiliary space, we first cycle $\RH_{a1}$ to the left of the product of permutations, permute it over $\sP_{aL}$ to obtain $\RH_{L1}$, take it out of the trace, and, at last, evaluate the resulting trace of permutations as above. Finally, imposing periodic boundary conditions (i.e., identifying $j=L+1$ with $j=1$) and using the unitarity condition (\ref{eq:unitarity_condition}), the transfer matrix at $\omega=\mu$ is given by 
	\begin{equation}
	\tt(\mu)=\left(
	\prod_{2\leq j\leq L}^{\leftarrow}\sP_{1j}
	\right)
	\prod_{j=1}^{L/2}\left(\RH_{2j,2j+1}(\lambda,\mu)\right)^{-1}.
	\end{equation}
	It is now evident that the dynamical generator (\ref{eq:Psi_def}) can be written as in Eq.~(\ref{eq:Psi_TransferMatrix}):
	\begin{equation}
	\begin{split}
	\Psi&=\tt(\mu)^{-1}\tt(\lambda)\\
	&=
	\left(
	\prod_{j=1}^{L/2}\RH_{2j,2j+1}(\lambda,\mu)
	\right)
	\left(
	\prod_{j=1}^{L/2}\RH_{2j-1,2j}(\lambda,\mu)
	\right).
	\end{split}
	\end{equation}
	
	The involution property of the transfer matrix implies the integrability of the circuit since $\Psi$ commutes with $\tt(\omega)$ for all $\omega$ and, in particular, with the two infinite sets of local superoperator charges generated from $\tt(\omega)$ by logarithmic differentiation:
	\begin{equation}
	\sQ_n^{(1)}=\frac{\d^n}{\d\omega^n}\log \tt(\omega)\bigg|_{\omega=\lambda},\ 
	\sQ_n^{(2)}=\frac{\d^n}{\d\omega^n}\log \tt(\omega)\bigg|_{\omega=\mu}.
	\end{equation}
	
	\section{The Hubbard \texorpdfstring{$\check{R}$}{R}-matrix as a local CPTP map}
	\label{sec:proof_CPTP}
	Having proved that the circuit is integrable, it remains to be shown that it describes proper open quantum dynamics, i.e., that it is a CPTP map. It suffices to show this for the elementary gates $\RH$. Indeed, choosing $\lambda,\mu,u \in i\mathbb{R}$ (purely imaginary interaction), then $\alpha,\beta\in\mathbb{R}$ and $\RH$ becomes a bistochastic quantum map~\cite{bengtsson2017,bruzda2009} (i.e., a unital CPTP map). To check this result, we first reshuffle the indices of $\RH$ to obtain the dynamical Choi matrix~$D$~\cite{bengtsson2017}, such that 
	$D^{\alpha\gamma\varepsilon\eta}_{\beta\delta\zeta\theta}=\RH^{\alpha\beta\varepsilon\zeta}_{\gamma\delta\eta\theta}$.
	Due to the channel-state duality~\cite{jamiolkowski1972,choi1975}, the map $\RH$ is CP if $D$ is non-negative; it is TP if the partial trace of $D$ over the first copy of the system is the identity; and it is unital if the partial trace over the second copy of the system is the identity. The TP and unitary conditions can be written as
	\begin{subequations}
		\begin{align}\label{eq:condition_TP}
		&D^{\alpha\gamma\varepsilon\eta}_{\alpha\delta\varepsilon\theta}=
		\RH^{\alpha\alpha\varepsilon\varepsilon}_{\gamma\delta\eta\theta}
		=\delta^{\gamma}_{\delta}\delta^{\eta}_{\theta},
		\\\label{eq:condition_unital}
		&D^{\alpha\gamma\varepsilon\eta}_{\beta\gamma\zeta\eta}=
		\RH^{\alpha\beta\varepsilon\zeta}_{\gamma\gamma\eta\eta}
		=\delta^{\alpha}_{\beta}\delta^{\varepsilon}_{\zeta},
		\end{align}
	\end{subequations}
	respectively. To see that Eq.~(\ref{eq:condition_TP}) holds, we write out the components of the Choi matrix using Eq.~(\ref{eq:RcheckHubb}),
	\begin{equation}
	\begin{split}
	D^{\alpha\gamma\varepsilon\eta}_{\beta\delta\zeta\theta}(\lambda,&\mu)
	=\beta\, \check{R}^{\alpha\varepsilon}_{\gamma\eta}(\lambda-\mu)\check{R}^{\beta\zeta}_{\delta\theta}(\lambda-\mu)\\
	&+\alpha\, \check{R}^{\alpha\varepsilon}_{\iota\eta}(\lambda+\mu)\check{R}^{\beta\zeta}_{\kappa\theta}(\lambda+\mu)\left(\sigma^z\right)^{\iota}_{\gamma}\left(\sigma^z\right)^{\kappa}_{\delta},
	\end{split}
	\end{equation}
	compute the trace of Eq.~(\ref{eq:condition_TP}),
	\begin{equation}\label{eq:proof_TP}
	\begin{split}
	D^{\alpha\gamma\varepsilon\eta}_{\alpha\delta\varepsilon\theta}
	&=\left(\check{R}^\dagger \check{R}\right)^{\gamma\eta}_{\delta\theta}
	\times \begin{cases}
	\beta+\alpha\quad\text{if }\gamma=\delta\\
	\beta-\alpha\quad\text{if }\gamma\neq\delta
	\end{cases}\\
	&=(\beta+\alpha)\delta_\delta^\gamma\delta_\theta^\eta
	=\delta_\delta^\gamma\delta_\theta^\eta,
	\end{split}
	\end{equation}
	and find that the map is indeed TP. In Eq.~(\ref{eq:proof_TP}), the three equalities hold because (i) $\check{R}$ admits a real representation, (ii) it is unitary, and (iii) we have fixed $\alpha+\beta=1$, respectively. Similarly, a computation starting from Eq.~(\ref{eq:condition_unital}) leads to a term proportional to $\check{R}\check{R}^\dagger$, which again is nonvanishing only when the prefactor is $\beta+\alpha=1$, and the map is therefore unital. Finally, since we have a rank-two map, of the sixteen eigenvalues of $D$ fourteen are zero and the remaining two are explicitly found to be $4\alpha>0$ and $4\beta>0$.\footnote{ 
		The analytic continuation of $\lambda$, $\mu$, and $u$ to the imaginary axes can always be chosen to render $\alpha$ and $\beta$ positive.}
	The Choi matrix is therefore non-negative and the map is CP. We have thus shown that a suitable analytic continuation of $\RH$ is a unital CPTP map.
	
	\section{Kraus representation and physical interpretation}
	\label{sec:Kraus_representation}
	The previous result implies that $\RH$ can be written in the Kraus form~\cite{kraus1983,nielsen2002,bengtsson2017}. 
	Abandoning the formal identification with the Fermi-Hubbard model, we identify, by reordering tensor factors, 
	\begin{equation}
	\phi^{\alpha\gamma\varepsilon\eta}_{\beta\delta\zeta\theta}(q_+,q_-,p)\equiv\RH^{\alpha\varepsilon\gamma\eta}_{\beta\zeta\delta\theta}(\lambda,\mu,u)
	\end{equation}
	with a vectorized quantum map parametrized by three independent real parameters: the coherent hopping strengths $q_\pm\equiv-\i(\lambda\pm\mu)\in\mathbb{R}$ and the relative weight of the channels $p\equiv\alpha\in[0,1]$. 
	Swapping the second and third tensor-product factors in Eq.~(\ref{eq:RcheckHubb}), $\phi$ can be written in the vectorized Kraus representation,
	\begin{equation}\label{eq:Kraus_rep}
	\phi(q_+,q_-,p)=K_-\otimes \conj{K}_- + K_+\otimes\conj{K}_+, 
	\end{equation} 
	acting on (local two-site) states as $\phi[\rho]=K_-\rho K_-^\dagger+K_+\rho K_+^\dagger$,
	with Kraus operators
	\begin{equation}\label{eq:Kraus_2channel}
	K_-=\sqrt{1-p}\,\check{R}(i q_-),\;\;
	K_+=\sqrt{p}\,\check{R}(i q_+)
	\left(\sigma^z\otimes \id_2\right).
	\end{equation}
	
	We see that the Kraus map of Eqs.~(\ref{eq:Kraus_rep}) and (\ref{eq:Kraus_2channel}), which we dub the Hubbard-Kraus map, describes the discrete-time dynamics of free fermions (after undoing the Jordan-Wigner transformation) subjected to local dephasing.\footnote{
		Note that, the circuit $\Psi$ does \emph{not} describe a dissipative Hubbard model. $\RH$ is used as a mathematical device to build an integrable circuit which, \textit{a priori}, is unrelated to the original model. For a recent study of an exactly-solvable dissipative Hubbard model, see Ref.~\cite{nakagawa2020PRL}.
	} 
	Indeed, after a suitable change of basis, the $\check{R}$-matrix~(\ref{eq:Rch_general6v}) with parametrization~(\ref{eq:RchXX}) can be written as $\check{R}(iq_\pm)=\exp\{\i\,\mathrm{gd}(q_\pm)H_{\mathrm{XX}}\}$, where $ H_\mathrm{XX}=\left(\sigma^x\otimes\sigma^x+\sigma^y\otimes\sigma^y\right)/2$ is the XX-chain Hamiltonian and $\mathrm{gd}(q)=\int_0^{q}\d x/\cosh{x}$ is the Gudermannian function. In the Trotter limit, $q_\pm\to0$, $\mathrm{gd}(q_\pm)\to q_\pm$, and the quantum map~(\ref{eq:Kraus_rep}) describes the quantum stochastic process in which, at each (discrete) half-time-step, with probability $1-p$, a fermion hops from the first to the second site (or vice-versa) with amplitude $q_-$; or, with probability $p$, it hops with amplitude $q_+$; in the latter case, it is also subject to dephasing when at the first site. We again emphasize that only the local Kraus maps describe free dynamics (with dephasing), as the checkerboard pattern of the circuit makes the full circuit strongly interacting.
	
	To build the extensive quantum circuit of length $L$ out of the elementary two-site building blocks in the Kraus representation, we define a row Kraus operator $F_{\underline{\nu}}$ by tensoring $L/2$ copies of the elementary Kraus operators $K_\pm$, $F_{\underline{\nu}}=\bigotimes_{j=1}^{L/2}K_{\nu_{2j}}$.
	Here, $\underline{\nu}=(\nu_2,\nu_4,\dots,\nu_L)$ is a multi-index with all two-site indices, $\nu_{2j}=\pm$, and $K_{\nu_{2j}}$ is a Kraus operator coupling sites $2j-1$ and $2j$. The quantum map corresponding to the entire row is then $\Phi=\check{\mathcal R}^{\otimes L/2}=\sum_{\underline{\nu}}F_{\underline{\nu}}\otimes F_{\underline{\nu}}^*$ (where tensor-product factors are reordered such that all second-copy degrees of freedom come after the first copy). 
	The second row of the circuit is again obtained by translation by one site, $\mathbb{T}^\dagger\Phi\mathbb{T}$. Then, one complete time step is given by
	\begin{align}
	\label{eq:Psi_Kraus}
	\Psi
	&=\mathbb{T}^\dagger\Phi\mathbb{T} \Phi\\
	\label{eq:circuit_Kraus}
	&=\sum_{\underline{\nu}}\ 
	\includegraphics[width=0.7\columnwidth,valign=c]{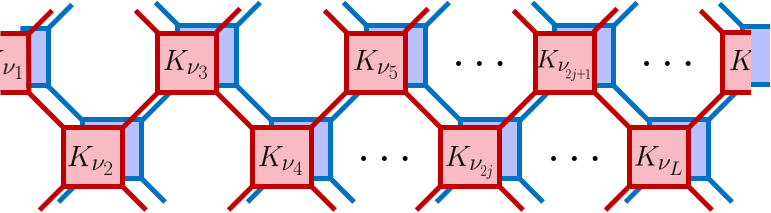}
	\end{align}
	where the superimposed layers represent the two copies of the system---red gates act on the ket of the density matrix while complex conjugate blue gates act on the bra.
	
	\section{Integrability-breaking deformations}
	\label{sec:integrability_breaking}
	A natural way of breaking the integrability of the circuit is by adding interactions to the \emph{local} coherent processes. This can be achieved by replacing, in Eq.~(\ref{eq:Kraus_2channel}), the XX $\check{R}$-matrices by more general XXZ (six-vertex) $\check{R}$-matrices~\cite{baxter1982,samaj2013}, which have the same form of Eq.~(\ref{eq:Rch_general6v}) but admit a two-parameter trigonometric parametrization,
	\begin{equation}\label{eq:RchXXZ}
	a=\sin(\lambda+\gamma),\quad
	b=\sin\lambda,\quad
	c=\sin\gamma,
	\end{equation}
	where $\gamma\in(-\pi,\pi]$ is related to the anisotropy parameter of the XXZ chain and, as before, $\lambda\in i\mathbb{R}$. The XX $\check{R}$-matrix~(\ref{eq:RchXX}) follows from Eq.~(\ref{eq:RchXXZ}) upon setting $\gamma=\pi/2$. The resulting Kraus operators~(\ref{eq:Kraus_2channel}) have five independent real parameters $p$, $q_\pm$, $\gamma_\pm$ and the extensive dephasing-XXZ circuit is built from them exactly as before. Note that integrability is broken because the $\RH$-matrix obtained this way from Eqs.~(\ref{eq:rupdown})~and~(\ref{eq:RcheckHubb}) no longer satisfies the Yang-Baxter equation~(\ref{eq:YBE}).
	
	Furthermore, while one might be tempted to conjecture the integrability of the quantum map~(\ref{eq:Kraus_2channel}) for general $\check{R}$-matrices at a free-fermion point~\cite{fan1970} (i.e., satisfying $a^2=c^2-b^2$), this turns out not to be correct. One such model is obtained from the Hubbard-Kraus map by removing dephasing from the second channel. The resulting quantum map, still of the form~(\ref{eq:Kraus_rep}), is a convex combination of two unitary free Kraus channels,
	\begin{equation}\label{eq:Kraus_2unitary}
	K_-=\sqrt{1-p}\,\check{R}(\i q_-),\quad
	K_+=\sqrt{p}\,\check{R}(\i q_+),
	\end{equation}
	which no longer satisfies the Yang-Baxter equation~(\ref{eq:YBE}). We thus arrive at the strong conclusion that even the simplest local dynamics (i.e., the convex combination of free unitaries) can lead to nonintegrable quantum circuits. This result highlights the special, and rather nontrivial, nature of the construction of the Hubbard-Kraus circuit above.
Below, we will give numerical evidence for the breaking of integrability in the preceding two examples (dubbed XXZ circuit and two-free-channel circuit).

\section{Symmetries and spectral statistics}
\label{sec:symmetries}
According to the quantum chaos conjectures~\cite{berry1977,bohigas1984} of dissipative systems~\cite{grobe1988,sa2020PRX}, the statistics of the complex eigenvalues of an integrable circuit are the same as those of uncorrelated random variables (henceforth, Poisson statistics), while nonintegrable models follow the predictions of random matrix theory (RMT), in the corresponding symmetry class. 
The comparison can only be done once all the (unitary and mutually commuting) symmetries of the model have been resolved, i.e., separating sectors with a fixed set of eigenvalues of the unitary symmetries.\footnote{Otherwise, levels from different symmetry sectors overlap without interacting and one obtains apparent Poisson statistics regardless of the actual statistics.}

Let us describe the unitary symmetries of our circuits. Because of the structure of the quantum circuit~(\ref{eq:Psi_def}), all the considered models are invariant under translation by two sites ($\comm{\Psi}{\mathbb{T}^2}=0$) and, therefore, have conserved quasi-momentum. Since $\mathbb{T}^L=\id$, the eigenvalues of $\mathbb{T}$ are $\exp{2\pi \i(k/L)}$, with $k=0,1,\dots,L-1$. Then, the translational-invariant Kraus circuit with $L$ sites ($L$ even) has $L/2$ sectors of conserved quasi-momentum $k\in\{0,1,\dots,L/2-1\}$. Note that each row Kraus operator $F_{\underline{\nu}}$ is not translationally invariant, only their sum is, and hence there is no conservation of momentum in each bra/ket copy individually.

Furthermore, the circuits are invariant under simultaneous space translation by one site (half a unit cell) and temporal translation by one circuit layer (half a time step), which can be encoded in the commutation relation $\comm{\mathbb{T}\Phi}{\mathbb{T}^\dagger \Phi}=0$.
For a sector of fixed quasi-momentum $k$, $(\Psi)_k=e^{-4\pi \i k/L}(\mathbb{T}\Phi)_k^2$, where $(A)_k\equiv\mathbb{P}_k A \mathbb{P}_k$ and $\mathbb{P}_k$ are orthogonal momentum-projection operators:
\begin{equation}\label{eq:momentum_projector}
\mathbb{P}_k=\frac{2}{L}\sum_{n=0}^{L/2-1}\mathbb{T}^{2n} \exp{-2\pi \i\frac{kn}{L/2}}\,.
\end{equation}
Therefore, resolving the space-time symmetry of $\Psi$ amounts to examining the spectral statistics of $(\mathbb{T}\Phi)_k$.

Besides the kinematic symmetries of the circuit, the XX and XXZ $\check{R}$-matrices display conservation of (total) magnetization (or particle-number in a fermion picture) in each (bra and ket) copy of the system independently.\footnote{This also restricts the allowed incoherent processes to dephasing, which is the case in all our models.}
In these conditions, for a given copy of the system, each row Kraus operator $F_{\underline{\nu}}$ splits into $L+1$ sectors of total magnetization $S_z=M$, where $S_z$ acts on the computational-basis states as $S_z\ket{s_1,\dots,s_L}=\sum_{j=1}^Ls_j\ket{s_1,\dots,s_L}$. Each sector $M$ has dimension $\binom{L}{M}$. Accordingly, the quantum map $\Phi$ splits into $(L+1)^2$ sectors, each of dimension $N=\binom{L}{M_\uparrow}\binom{L}{M_\downarrow}$ (here, $M_\uparrow$ and $M_\downarrow$ denote the magnetization in the two copies).
We restrict ourselves to sectors with $M_{\uparrow,\downarrow}\neq L/2$ to avoid an additional $\mathbb{Z}_2$ spin-flip (particle-hole) symmetry. Finally, there is another $\mathbb{Z}_2$ symmetry connecting the two copies of the system; we also avoid this symmetry by considering only sectors with $M_\uparrow\neq M_\downarrow$.

Once inside a fixed sector of the unitary symmetries, the symmetry class to which each circuit belongs is determined by its behavior under transposition instead of complex conjugation~\cite{hamazaki2020PRR}, as discussed in Ch.~\ref{chapter:correlations}. Transposition symmetry imposes local correlations and completely determines the short-distance spectral statistics~\cite{grobe1989,hamazaki2020PRR}. We argue in Appendix~\ref{app:KrausTransposition} that both the Hubbard-Kraus and two-free-channel circuits admit a transposition symmetry; the latter---being nonintegrable---has, therefore, the same spectral statistics as matrices from class AI$^\dagger$~\cite{kawabata2019PRX,hamazaki2020PRR,sa2020PRX} (complex symmetric random matrices with Gaussian entries). In contrast, the XXZ circuit breaks transposition symmetry and has, therefore, the same spectral statistics as matrices from the Ginibre Orthogonal Ensemble (GinOE, real asymmetric random matrices with Gaussian entries).\footnote{
Because matrices from the Ginibre Orthogonal Ensemble (GinOE) and Ginibre Unitary Ensemble (GinUE) differ by their behavior under complex conjugation and not transposition, they share the same spectral correlations. For this reason, the large-$N$ results presented for the GinUE in Ref.~\cite{sa2020PRX} carry over to the GinOE.}

\begin{figure}[tbp]
	\centering
	\includegraphics[width=\columnwidth]{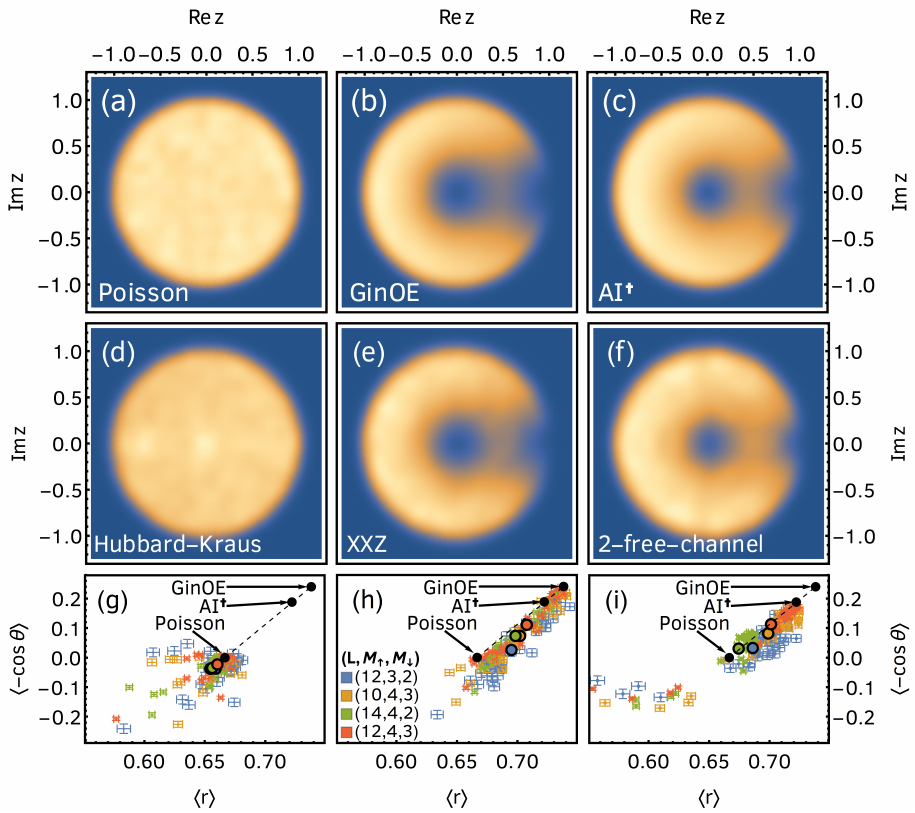}
	\caption{(a)--(c): CSR distributions obtained from sampling $10^5$ uncorrelated random variables (Poisson spectrum) (a) or exact diagonalization of $10^4\times 10^4$ random matrices from the GinOE (b) or class AI$^\dagger$ (c) [$104$ realizations superimposed in (b) and (c)]. (d)--(f): CSR distributions of (a single realization of) the operator $(\mathbb{T}\Phi)_k$ for the Hubbard-Kraus (d), XXZ (e), and two-free-channel (f) circuits. The eigenvalues were obtained by exact diagonalization in sectors of fixed $L=12$, $M_\uparrow=4$, $M_\downarrow=3$, and $k=0,1,2$---leading to three sectors of fixed quasi-momentum $k$ with $18150$ eigenvalues each, the ratios of which were then superimposed. The numerical parameters were chosen as follows: $q_+=-0.2$, $q_-=0.6$, $p=0.55$ (for all three circuits) and $\gamma_+=\gamma_-=1$ (for the XXZ circuit). (g)--(i): $\av{r}$ versus $\av{-\cos\theta}$ plots for the (g) Hubbard-Kraus, (h) XXZ, and (i) two-free-channel circuits, obtained by randomly sampling (40 samples) $p$, $q_\pm$, and $\gamma_\pm$, for different $L$, $M_\uparrow$, and $M_\downarrow$ and fixed $k=1$. Larger, black-rimmed dots mark the average (center of mass) for each symmetry sector. The points accumulate around the Poisson~$(2/3,0)$, AI$^\dagger$~$(0.722,0.188)$, and GinOE~$(0.738,0.241)$ points or spread over the line between them.}
	\label{fig:ratios}
\end{figure}

\section{Numerical results}
\label{sec:numerics}

To probe the statistics of the dissipative quantum circuits, we consider the complex spacing ratios (CSRs) of the eigenvalues of the operator $(\mathbb{T}\Phi)_k$. We denote the set of eigenvalues by $\{\Lambda_j\}$ and, for each $\Lambda_j$, we find its nearest neighbor, $\Lambda_j^\mathrm{NN}$, and its next-to-nearest neighbor, $\Lambda_j^\mathrm{NNN}$. The CSRs are defined by $z_j=(\Lambda_j^\mathrm{NN}-\Lambda_j)/(\Lambda_j^\mathrm{NNN}-\Lambda_j)$. In the thermodynamic limit, the probability distribution of $z_j$ is flat on the unit disk for Poisson statistics; for non-Hermitian random matrices, it has a characteristic C-shape (an analytic surmise is given in Ref.~\cite{sa2020PRX}), see Figs.~\ref{fig:ratios}(a)--\ref{fig:ratios}(c). In Figs.~\ref{fig:ratios}(d)--\ref{fig:ratios}(f) we plot the CSR distribution of $(\mathbb{T}\Phi)_k$ for, respectively, the Hubbard-Kraus, XXZ, and two-free-channel circuits (eigenvalues obtained from exact diagonalization). The flat distribution of the integrable Hubbard-Kraus circuit (d) and the C-shaped distribution of the chaotic XXZ (e) and two-unitary-channel (f) circuits are clearly visible. In the latter two cases, the CSR distribution also allows us to distinguish the different symmetry classes to which the circuits belong.

To provide a more quantitative measure of spectral chaoticity, we express the CSR in polar coordinates, $z=r\exp{i\theta}$, and characterize its distribution by two numbers, namely, the mean ratio $\av{r}$, measuring the degree of radial level repulsion, and the angular correlation $\av{-\cos\theta}$. For each model, we randomly sample the two independent hopping parameters $q_\pm$ from a standard normal distribution (i.e., zero mean, unit variance), the channels' relative weight $p$ from a uniform distribution on $[0,1]$, and, in the case of the XXZ circuit, the anisotropy parameters $\gamma_\pm$ from a uniform distribution on $(-\pi,\pi]$. For a fixed random realization of the parameters, we exactly diagonalize the quantum circuit for four different system sizes and conserved magnetization sectors---$(L,M_\uparrow,M_\downarrow)$=$(12,3,2)$, $(10,4,3)$, $(14,4,2)$, and $(12,4,2)$, corresponding to sector sizes $N$=$2420$, $5040$, $13013$ and $18150$, respectively---and compute the pairs ($\av{r},\av{-\cos\theta}$). A Monte Carlo sampling of the quantum circuits then yields a $\av{r}$ versus $\av{-\cos\theta}$ scatter plot, which can again be compared with the results for Poisson random variables and GinOE and AI$^\dagger$ matrices.

There are three special points: For a set of uncorrelated (Poisson) random variables we have exactly $(\av{r},\av{-\cos\theta})=(2/3,0)$, while for random matrices from the GinOE and class AI$^\dagger$ we have $(\av{r},\av{-\cos\theta})\approx(0.738,0.241)$ and $(\av{r},\av{-\cos\theta})\approx(0.722,0.188)$, respectively~\cite{sa2020PRX}. When approaching the thermodynamic limit, we expect that the points obtained by sampling an integrable circuit concentrate around the Poisson point, while they accumulate around the RMT point of the respective symmetry class when sampling from a nonintegrable model. For finite system sizes $L$, the points spread over the line connecting the two fixed points and one tries to determine to which one of them the scatter points are flowing as $L$ increases. This last step may be difficult to perform if only small system sizes are available and/or finite-size effects are pronounced. Moreover, the flow may be nonmonotonic; for instance, it may depend more strongly on the system length $L$ than on the magnetization sectors $M$, or sectors with different parity of $M$ may flow differently.
Moreover, there are deviations from the pattern described above if the spectrum is close to one-dimensional, for instance, if the quantum map is very close to being unitary (which happens when $p$ is very close to either $0$ or $1$). For this reason, we have only considered quantum circuits for which $\abs{z_\mathrm{max}}-\abs{z_\mathrm{min}}>0.2$, where $z_{\mathrm{\max}(\mathrm{min})}$ are the eigenvalues of $\mathbb{T}\Phi$---in the the appropriate symmetry sector---with largest (smallest) absolute value.

Figures~\ref{fig:ratios}(g)--\ref{fig:ratios}(i) show the $\av{r}$ versus $\av{-\cos\theta}$ plots for the Hubbard-Kraus, XXZ, and two-free-channel circuits, respectively. For the Hubbard-Kraus circuit (g), which is integrable by construction, we obtain a high concentration of points around the Poisson point, even for modest system sizes. For the chaotic XXZ circuit (h) of the same system sizes, while points spread over the line connecting the Poisson and GinOE points, there is now a high accumulation of data around the GinOE point, signaling integrability breaking. Note that the center-of-mass values are flowing to the GinOE point as sector dimension increases. Finally, the two-free-channel circuit (i) displays the same qualitative integrability-breaking behavior as the XXZ circuit and, for the largest system sizes, has reached the AI$^\dagger$ point.

\section{Summary and outlook}%
\label{sec:circuits_conclusions}

Let us summarize the two key findings of this chapter, addressing the critical question we posed at the start, namely, if integrability methods can be extended to the realm of dissipative quantum circuits. First, we answered it in the positive, by showing that Shastry's celebrated $\check{R}$-matrix of the Fermi-Hubbard model can be interpreted as a unital CPTP map of a pair of qubits (spins $1/2$), for imaginary values of interaction and spectral parameters. By consequence of the Yang-Baxter equation, this implies integrability of the brickwork open circuit built from such nonunitary two-qubit maps.
This result opens a new avenue for studying general integrable open (driven/dissipative) quantum Floquet circuits. For example, our result straightforwardly generalizes to $\mathrm{SU}(d)$ open qudit circuits using Maassarani's $\check{R}$-matrix~\cite{maassarani1998}. 
Second, our construction shows that building integrable dissipative circuits is highly nontrivial, in the sense that deformations of the Hubbard-Kraus circuit (even convex combinations of local free-fermion unitaries) are nonintegrable.

Finally, we note that for a real interaction parameter $u\in \mathbb R$ the staggered transfer matrix~(\ref{eq:monodromy_matrix}) generates a \emph{unitary} Floquet circuit (\ref{eq:Psi_TransferMatrix}) for a $2\times L$ spin ladder which represents an integrable trotterization of the Hermitian Fermi-Hubbard model by taking
$\lambda=i\tau$, $\mu=0$, where $\tau\in\mathbb R$ is the time step.
This remarkable side result parallels the result~\cite{vanicat2018} for the Heisenberg chain.

%% file: Thesis_QLaguerre.tex

\chapter{\texorpdfstring{$Q$}{Q}-Laguerre spectral density and quantum chaos in the Wishart-Sachdev-Ye-Kitaev model}
\label{chapter:QLaguerre}

{
\allowdisplaybreaks

In this chapter, we study the Wishart-Sachdev-Ye-Kitaev (WSYK) model consisting of two $\hat{q}$-body Sachdev-Ye-Kitaev (SYK) models with general complex couplings, one the Hermitian conjugate of the other, living in off-diagonal blocks of a larger WSYK Hamiltonian. The spectrum is positive with a hard edge at zero energy.
We employ diagrammatic and combinatorial techniques to compute analytically the low-order moments of the Hamiltonian. In the limit of large number $N$ of Majoranas, we have found striking similarities with the moments of the weight function of the Al-Salam-Chihara $Q$-Laguerre polynomials.
For $\hat{q} = 3, 4$, the $Q$-Laguerre prediction, with $Q=Q(\hat{q},N)$ also computed analytically, agrees well with exact diagonalization results for $28 \leq N \leq 34$ while we observe some deviations for $\hat q =  2$. The most salient feature of the spectral density is that, for odd $\hat{q}$, low-energy excitations grow as a stretched exponential, with a functional form different from that of the supersymmetric SYK model. 
For $\hat q = 4$, a detailed analysis of level statistics reveals quantum chaotic dynamics. More specifically, the spacing ratios in the bulk of the spectrum and the distribution of the smallest eigenvalue (repulsion form the hard edge) are very well approximated by that of an ensemble of random matrices that, depending on $N$, belong to the chiral or superconducting universality classes.  In particular, we report the first realization of level statistics belonging to the chGUE universality class, which completes the tenfold-way classification in the SYK model.

This chapter is based on Ref.~\cite{sa2022PRD}.

\section{The Wishart-Sachdev-Ye-Kitaev model}
\label{sec:QLaguerre_defs}

We start with the definition of the SYK Hamiltonian that describes $q$-body random interactions among $N$ Majorana fermions,
\begin{equation}\label{eq:def_SYK}
H=\i^{q(q-1)/2}\sum_{i_1<\dots<i_q=1}^N 
J_{i_1\cdots i_q}
\gamma_{i_1}\cdots\gamma_{i_q}
\equiv \i^{q(q-1)/2}\sum_{\va} J_\va \Gamma_\va,
\end{equation}
where $N$ and $q$ are even integers, $\va$ is a multi-index accounting for all $q$ indices $i_1,\dots,i_q=1,\dots,N$, $J_{i_1\cdots i_q}\equiv J_\va$ is a totally antisymmetric tensor with independent real Gaussian entries with zero mean, i.e.,
\begin{equation}
\av{J_\va}=0
\quad \text{and} \quad
\av{J_\va J_\vb}=\av{J^2}\delta_{\va,\vb},
\end{equation}
$\gamma_i$ are Majorana fermions satisfying the Clifford algebra $\acomm{\gamma_i}{\gamma_j}=2\delta_{ij}$ and, therefore, represented by $2^{N/2}$-dimensional (Hermitian) Dirac $\gamma$-matrices, and $\Gamma_\va=\gamma_{i_1}\cdots\gamma_{i_q}$ is a product of $q$ $\gamma$-matrices with all indices $i_1,\dots,i_q$ mutually different, satisfying
\begin{equation}\label{eq:relations_Gamma}
\Gamma_\va^2=\i^{q(q-1)}\id
\quad \text{and} \quad
\Gamma_\va\Gamma_\vb=(-1)^{q+\abs{\va\cap\vb}}
\Gamma_\vb\Gamma_\va,
\end{equation}
with $\id$ the $2^{N/2}$-dimensional identity and $\abs{\va\cap\vb}$ the number of $\gamma$-matrices that $\Gamma_\va$ and $\Gamma_\vb$ have in common.

Several restrictions in $H$ can be relaxed. Let us consider a set of $M$ independent \emph{non-Hermitian} $\qb$-body charges $L_\mu$ (with $\qb$ allowed to be odd):
\begin{equation}
L_\mu=\sum_{i_1,\dots,i_\qb=1}^N 
\ell_{\mu,i_1\cdots i_\qb}
\gamma_{i_1}\cdots\gamma_{i_\qb}
\equiv\sum_{\va} \ell_{\mu,\va} \Gamma_\va,
\end{equation}
where the $\ell_{\mu,i_1\cdots i_\qb}\equiv \ell_{\mu,\va}$ are antisymmetric (in $\va$) independent \emph{complex} Gaussian random variables (the exact variances are specified below). From the charges $L_\mu$, we can define a Hermitian and positive-definite Hamiltonian, which we dub the Wishart-Sachdev-Ye-Kitaev (WSYK) Hamiltonian:
\begin{equation}\label{eq:def_WSYK}
W=\sum_{\mu=1}^{M}L_\mu^\dagger L_\mu
=i^{\qb(\qb-1)}\sum_{\va,\vast}
\(\sum_{\mu=1}^{M} \ell_{\mu,\vast}^*\ell_{\mu,\va}\)
\Gamma_\vast \Gamma_\va,
\end{equation}
where the factor $i^{\qb(\qb-1)}$ arises from bringing $\Gamma_\vast^\dagger$ back to $\Gamma_\vast$ by commuting all the (Hermitian) $\gamma$-matrices in $\Gamma_\vast^\dagger$.
Note that while all $\gamma$-matrices in a single $\Gamma_\va$ are different (owing to the antisymmetry of $\ell_{\mu,\va}$), there could be overlaps between the $\gamma$-matrices in $\Gamma_\va$ and those in $\Gamma_\vast$ (which follows from the fact that the tensor $K_{\vast\va}=\sum_{\mu}\ell_{\mu,\vast}^{*}\ell_{\mu,\va}$
is \emph{not} totally antisymmetric). Setting $\qb=q/2$, the positive-definite Hamiltonian~$W$ describes $q$-body interactions like the original Hamiltonian $H$. Note that, for $M=1$, the eigenvalues of $W$ are exactly the squares of the eigenvalues of the off-diagonal block Hamiltonian 
\begin{equation}
\label{eq:chiral_form}
\mathcal{H}=\begin{pmatrix}
0 & L \\
L^\dagger & 0
\end{pmatrix},
\end{equation}
which has a natural chiral structure.

We now address the exact distribution of the random couplings $\ell_{\mu,\va}$. We decompose $\ell_{\mu,\va}=\ell^{(1)}_{\mu,\va}+\i k \ell^{(2)}_{\mu,\va}$, $0\leq k\leq 1$. The real, $\ell^{(1)}_{\mu,\va}$, and imaginary, $\ell^{(2)}_{\mu,\va}$, parts of $\ell_{\mu,\va}$ are independent and identically distributed Gaussian random variables with
\begin{equation}
\av{\ell^{(1)}_{\mu,\va}}
=\av{\ell^{(2)}_{\mu,\va}}=0
\quad \text{and} \quad
\av{\ell^{(1)}_{\mu,\va} \ell^{(1)}_{\mu',\va'}}
=\av{\ell^{(2)}_{\mu,\va} \ell^{(2)}_{\mu',\va'}}
=\frac{1}{2}\av{\ell^2}\delta_{\mu,\mu'}\delta_{\va,\va'}.
\end{equation}
It then follows that $\av{\ell_{\mu,\va}}=\av{\ell_{\mu,\va}^*}=0$,
\begin{equation}\label{eq:variances_ell}
\av{\ell_{\must,\vast}^*\ell_{\mu,\va}}=
\frac{1+k^2}{2}\av{\ell^2}\delta_{\must\!,\mu}\delta_{\vast\!,\va},
\quad \text{and} \quad
\av{\ell_{\must,\vast}\ell_{\mu,\va}}=
\av{\ell_{\must,\vast}^*\ell_{\mu,\va}^*}=
\frac{1-k^2}{2}\av{\ell^2}\delta_{\must\!,\mu}\delta_{\vast\!,\va}.
\end{equation}
We name the three special cases $k=0$, $k=1$, and $0<k<1$ the linear, circular, and elliptic WSYK models, respectively [borrowing the nomenclature of standard random matrix theory (RMT), after the shapes of the support of the random variables $\ell_{\mu,\va}$ seen as random matrices].

Note that only for the $M=1$ linear WSYK model are the eigenvalues of $W$ the squares of the eigenvalues of $L$. When the charge is non-Hermitian, there is no relation between the eigenvalues of $L$ and its singular values (the eigenvalues of $W=L^\dagger L$), apart from some general inequalities. The same is true if there are $M>1$ independent (noncommuting) charges, irrespective of their Hermiticity. As a result, to compute the correct spectral density in those cases, one cannot rely on the known combinatorial expansion of the standard SYK moments (see Appendix~\ref{app:review_standard_SYK} for a review). The WSYK moments have to be computed afresh and a new combinatorial interpretation has to be given.

In the rest of this chapter, we focus on the circular WSYK model. We compute the spectral density of the $M=1$ circular WSYK model in Secs.~\ref{sec:moments_WSYK}--\ref{sec:spectral_density_numerics} (with simple asymptotic formulas for different regimes derived in Appendix~\ref{app:asymptotics}). Then, we propose an ansatz for the spectral density of the $M>1$ circular WSYK model in Sec.~\ref{app:M>1}. We study the quantum chaos properties of the $M=1$ circular model through its level statistics in Sec.~\ref{sec:qlaguerre_qchaos}. Finally, in Sec.~\ref{sec:1SYK_nHWSYK} we consider the non-Hermitian WSYK model and provide its symmetry classification.

\section{Moments of the circular WSYK model}
\label{sec:moments_WSYK}

Our aim is to derive the spectral density of the circular WSYK model using the method of moments. Let us briefly mention how this computation is carried out for the standard SYK model~\cite{garcia-garcia2016PRD,garcia-garcia2017PRD}, referring the reader to Appendix~\ref{app:review_standard_SYK} for a detailed review. One starts by an exact computation of the first few moments, which can be given a combinatorial interpretation in terms of perfect matchings. Arbitrarily high moments cannot be computed exactly in general, but one can find approximations to varying degrees of accuracy and exact results when $q\propto \sqrt{N}$. In particular, with this scaling, the moments can be expressed as a sum over the number of crossings of perfect matchings and, hence, identified with the moments of the weight function of the $Q$-Hermite polynomials~\cite{ismail1987EJC}.

Let us apply this program to the circular WSYK model. Because the random variables $\ell_{\mu,\va}$ are Gaussian, the moments $\av{\Tr W^{p}}$ are evaluated by Wick contraction, i.e., by summing over all possible pair contractions of the indices $\va$, $\vast$, $\vb$, $\vbst$, etc. In contrast to the standard SYK model, the odd moments are nonvanishing and, more importantly, the non-Hermitian couplings suppress certain Wick contractions. For example, by performing all allowed contractions, the second moment of $W$ is explicitly found to be
\begin{equation}\label{eq:W_moment2_example}
\begin{split}
\av{\Tr W^2}&
=\sum_{\mu,\nu}\sum_{\va,\vb,\vc,\vd}
\av{\ell_{\mu,\va}^*\ell_{\mu,\vb}
	\ell_{\nu,\vc}^*\ell_{\nu,\vd}}
\Tr\(\Gamma_\va\Gamma_\vb\Gamma_\vc\Gamma_\vd\)\\
&=\av{\ell^2}^2\sum_{\mu,\nu}\sum_{\va,\vb}
\left[
\delta_{\mu\mu}\delta_{\nu\nu}
\Tr\(\Gamma_{\va}\Gamma_{\va}\Gamma_{\vb}\Gamma_{\vb}\)
+0\cdot
\Tr\(\Gamma_{\va}\Gamma_{\vb}\Gamma_{\va}\Gamma_{\vb}\)
+\delta_{\mu\nu}\delta_{\nu\mu}
\Tr\(\Gamma_{\va}\Gamma_{\vb}\Gamma_{\vb}\Gamma_{\va}\)
\right]\\
&=2^{N/2}\(\av{\ell^2}\binom{N}{\qb}\)^2\(M^2+M\),
\end{split}
\end{equation}
where the second term in the Wick expansion is identically zero because of Eq.~(\ref{eq:variances_ell}) with $k=1$ and we used Eq.~(\ref{eq:relations_Gamma}) to evaluate the traces.

As for the standard SYK model, we can introduce a combinatorial diagrammatic notation to simplify the representation and evaluation of the moments. First, we note that to avoid dealing with allowed and forbidden perfect-matching diagrams we can switch to an expansion in terms of permutation diagrams~\cite{corteel2007AAM,kasraoui2011AAM,josuat-verges2011DM,corteel2016BOOKCH,zeng2020BOOKCH}\footnote{It was noted in Ref.~\cite{berkooz2020JHEP} that the moments $\mathcal{N}=2$ supersymmetric SYK model also admit an expansion in terms of permutations, but the connection to the $Q$-Laguerre polynomials was not made.}. Indeed, at $p$th order, the set of $\Gamma$-matrices in the trace $\Tr\(\Gamma_{\va_1}\Gamma_{\va_2}\cdots \Gamma_{\va_{2p}}\)$ is naturally bipartite: half of the $\Gamma$-matrices (more precisely, the even-numbered ones, $\{\Gamma_{\va_{2j}}\}$, $j=1,\dots,p$) come from insertions of matrices $L_\mu$, the other half ($\{\Gamma_{\va_{2j-1}}\}$) from insertions of $L_\mu^\dagger$, two odd-numbered (or two even-numbered) $\Gamma$-matrices cannot be coupled (according to Eq.~(\ref{eq:variances_ell}) with $k=1$), and each odd-numbered $\Gamma$-matrix is coupled to one and only one even-numbered one. We can, therefore, identify each Wick contraction with a permutation $\sigma\in\mathcal S_{p}$ such that if $\sigma(j)=k$, with $j,k=1,\dots,p$, then $\va_{2j-1}$ is contracted with $\va_{2k}$, leading to a factor $\langle\ell_{\mu,\va_{2j-1}}^*\ell_{\mu,\va_{2k}}\rangle$. The permutation diagram corresponding to $\sigma\in\mathcal{S}_p$ is then given by a set of $p$ dots, labeled $j=1,\dots,p$, with edges connecting all pairs of dots $(j,\sigma(j))$; if $j\leq\sigma(j)$ the edge is drawn above the dots, if $j>\sigma(j)$ it is drawn below. For example, the two allowed contractions in Eq.~(\ref{eq:W_moment2_example}) are represented diagrammatically as follows (we omit the labeling of the dots):
\begingroup
\allowdisplaybreaks
\begin{align}
	\label{eq:W_moment2_diagram1}
	\begin{tikzpicture}[baseline=(b)]
	\begin{feynman}[inline=(a)]
	\vertex[dot] (a) {};
	\vertex[dot, right=0.6cm of a] (b) {};
	\vertex[right=0.3cm of a] (x) {};
	\vertex[below=0.55cm of x] {};
	\end{feynman}
	\draw [/tikzfeynman] (a) to[out=135, in=45, loop, min distance=0.9cm] (a);
	\draw [/tikzfeynman] (b) to[out=135, in=45, loop, min distance=0.9cm] (b);
	\end{tikzpicture}
	\hspace{-0.4cm}
	&=2^{-N/2}\binom{N}{\qb}^{-1}
	\sum_{\mu,\nu}\sum_{\va,\vb}
	\delta_{\mu\mu}\delta_{\nu\nu}
	\Tr\(\Gamma_{\va}\Gamma_{\va}\Gamma_{\vb}\Gamma_{\vb}\)
	\intertext{and}
	\label{eq:W_moment2_diagram2}
	\begin{tikzpicture}[inner sep=2pt,baseline=(b)]
	\begin{feynman}[inline=(a)]
	\vertex[dot] (a) {};
	\vertex[dot, right=0.6cm of a] (b) {};
	\diagram*{(a) --[half left] (b)};
	\diagram*{(b) --[half left] (a)};
	\vertex[right=0.3cm of a] (x) {};
	\vertex[below=0.6cm of x] {};
	\end{feynman}
	\end{tikzpicture}
	\hspace{+0.15cm}
	&=2^{-N/2}\binom{N}{\qb}^{-1}
	\sum_{\mu,\nu}\sum_{\va,\vb}
	\delta_{\mu\nu}\delta_{\nu\mu}
	\Tr\(\Gamma_{\va}\Gamma_{\vb}\Gamma_{\vb}\Gamma_{\va}\).
\end{align}
\endgroup
Note that each dot comes also equipped with an index $\mu_j$ related to which charge $L_\mu$ it belongs. Each closed loop in the diagram then contributes with a factor $M$. Equations~(\ref{eq:W_moment2_diagram1}) and (\ref{eq:W_moment2_diagram2}) thus give a factor of $M^2$ and $M$, respectively, in agreement with Eq.~(\ref{eq:W_moment2_example}).

In full generality, the (normalized) moments of the circular WSYK model can be written as a sum over permutations $\sigma\in\mathcal{S}_p$,
\begin{equation}\label{eq:moments_W_sum_permutations}
\frac{1}{\sigmaL^p}\frac{\av{\Tr W^p}}{\Tr \id}
=\sum_{\sigma\in\mathcal{S}_p}
t(\sigma)M^{\mathrm{cyc}(\sigma)},
\end{equation}
where $\sigmaL=\av{\ell^2}\binom{N}{\qb}$ is the energy scale of the WSYK model, $\Tr\id=2^{N/2}$, the weight~$t(\sigma)$ gives the normalized trace of each diagram~$\sigma$, and $\mathrm{cyc}(\sigma)$ is the number of cycles in the permutation $\sigma$ (number of closed loops in the respective diagram).

For example, the first four moments of $W$ are easily written down and organized in terms of permutation diagrams:
\begingroup
\allowdisplaybreaks
\begin{align}
	\av{\Tr W^0}&=2^{N/2},
	\\
	\av{\Tr W^1}&=2^{N/2}\av{\ell^2} \binom{N}{\qb}
	\begin{tikzpicture}[baseline=(a)]
	\begin{feynman}[inline=(a)]
	\vertex[dot] (a) {};
	\vertex[below=0.35cm of a] {\footnotesize{$1\times M$}};
	\end{feynman}
	\draw [/tikzfeynman] (a) to[out=135, in=45, loop, min distance=0.9cm] (a);
	\end{tikzpicture}
	=2^{N/2}\av{\ell^2} \binom{N}{\qb}\, M,
	\\
	\label{eq:W_moment2}
	\av{\Tr W^2}&=2^{N/2}\(\av{\ell^2} \binom{N}{\qb}\)^2
	\(
	\hspace{-0.3cm}
	\begin{tikzpicture}[baseline=(b)]
	\begin{feynman}[inline=(a)]
	\vertex[dot] (a) {};
	\vertex[dot, right=0.6cm of a] (b) {};
	\vertex[right=0.3cm of a] (x) {};
	\vertex[below=0.55cm of x] {\footnotesize{$1\times M^2$}};
	\end{feynman}
	\draw [/tikzfeynman] (a) to[out=135, in=45, loop, min distance=0.9cm] (a);
	\draw [/tikzfeynman] (b) to[out=135, in=45, loop, min distance=0.9cm] (b);
	\end{tikzpicture}
	\hspace{-0.3cm}
	+
	\hspace{+0.15cm}
	\begin{tikzpicture}[inner sep=2pt,baseline=(b)]
	\begin{feynman}[inline=(a)]
	\vertex[dot] (a) {};
	\vertex[dot, right=0.6cm of a] (b) {};
	\diagram*{(a) --[half left] (b)};
	\diagram*{(b) --[half left] (a)};
	\vertex[right=0.3cm of a] (x) {};
	\vertex[below=0.6cm of x] {\footnotesize{$1\times M$}};
	\end{feynman}
	\end{tikzpicture}
	\hspace{+0.15cm}
	\)
	=2^{N/2}\(\av{\ell^2} \binom{N}{\qb}\)^2
	\(M^2+M\),
	\\
	\begin{split}
		\label{eq:W_moment3}
		\av{\Tr W^3}&=2^{N/2}\(\av{\ell^2} \binom{N}{\qb}\)^3\\
		\times&\left(
		\hspace{-0.3cm}
		\begin{tikzpicture}[baseline=(b)]
		\begin{feynman}[inline=(a)]
		\vertex[dot] (a) {};
		\vertex[dot, right=0.6cm of a] (b) {};
		\vertex[dot, right=0.6cm of b] (c) {};
		\vertex[below=0.635cm of b] {\footnotesize{$1\times M^3$}};
		\end{feynman}
		\draw [/tikzfeynman] (a) to[out=135, in=45, loop, min distance=0.9cm] (a);
		\draw [/tikzfeynman] (b) to[out=135, in=45, loop, min distance=0.9cm] (b);
		\draw [/tikzfeynman] (c) to[out=135, in=45, loop, min distance=0.9cm] (c);
		\end{tikzpicture}
		\hspace{-0.3cm}
		+
		\hspace{-0.3cm}
		\begin{tikzpicture}[inner sep=2pt,baseline=(b)]
		\begin{feynman}[inline=(a)]
		\vertex[dot] (a) {};
		\vertex[dot, right=0.6cm of a] (b) {};
		\vertex[dot, right=0.6cm of b] (c) {};
		\diagram*{(b) --[half left] (c)};
		\diagram*{(c) --[half left] (b)};
		\vertex[below=0.65cm of b] {\footnotesize{$1\times M^2$}};
		\end{feynman}
		\draw [/tikzfeynman] (a) to[out=135, in=45, loop, min distance=0.9cm] (a);
		\end{tikzpicture}
		\hspace{+0.15cm}
		+
		\hspace{+0.15cm}
		\begin{tikzpicture}[inner sep=2pt,baseline=(b)]
		\begin{feynman}[inline=(a)]
		\vertex[dot] (a) {};
		\vertex[dot, right=0.6cm of a] (b) {};
		\vertex[dot, right=0.6cm of b] (c) {};
		\diagram*{(a) --[half left] (b)};
		\diagram*{(b) --[half left] (a)};
		\vertex[below=0.65cm of b] {\footnotesize{$1\times M^2$}};
		\end{feynman}
		\draw [/tikzfeynman] (c) to[out=135, in=45, loop, min distance=0.9cm] (c);
		\end{tikzpicture}
		\hspace{-0.3cm}
		+
		\hspace{+0.15cm}
		\begin{tikzpicture}[inner sep=2pt,baseline=(b)]
		\begin{feynman}[inline=(a)]
		\vertex[dot] (a) {};
		\vertex[dot, right=0.6cm of a] (b) {};
		\vertex[dot, right=0.6cm of b] (c) {};
		\diagram*{(a) --[half left] (c)};
		\diagram*{(c) --[out=260, in=280] (a)};
		\vertex[below=0.65cm of b] {\footnotesize{$1\times M^2$}};
		\end{feynman}
		\draw [/tikzfeynman] (b) to[out=135, in=45, loop, min distance=0.8cm] (b);
		\end{tikzpicture}
		\hspace{+0.15cm}
		+
		\hspace{+0.15cm}
		\begin{tikzpicture}[inner sep=2pt,baseline=(b)]
		\begin{feynman}[inline=(a)]
		\vertex[dot] (a) {};
		\vertex[dot, right=0.6cm of a] (b) {};
		\vertex[dot, right=0.6cm of b] (c) {};
		\diagram*{(a) --[half left] (c)};
		\diagram*{(c) --[half left] (b)};
		\diagram*{(b) --[half left] (a)};
		\vertex[below=0.71cm of b] {\footnotesize{$1\times M$}};
		\end{feynman}
		\end{tikzpicture}
		\hspace{+0.15cm}
		+
		\hspace{+0.15cm}
		\begin{tikzpicture}[inner sep=2pt,baseline=(b)]
		\begin{feynman}[inline=(a)]
		\vertex[dot] (a) {};
		\vertex[dot, right=0.6cm of a] (b) {};
		\vertex[dot, right=0.6cm of b] (c) {};
		\diagram*{(a) --[half left] (b)};
		\diagram*{(b) --[half left] (c)};
		\diagram*{(c) --[out=260, in=280] (a)};
		\vertex[below=0.73cm of b] {\footnotesize{$t_3\times M$}};
		\end{feynman}
		\end{tikzpicture}
		\hspace{+0.15cm}
		\right)
		\\
		&=2^{N/2}\(\av{\ell^2} \binom{N}{\qb}\)^3
		\(M^3+3M^2+(t_3+1)M\),
	\end{split}
	\\
	\begin{split}\nonumber
		\label{eq:W_moment4}
		\av{\Tr W^4}&=2^{N/2}\(\av{\ell^2} \binom{N}{\qb}\)^4
	\end{split}
	\\
	\begin{split}\nonumber
		\times&\left(
		\hspace{-0.3cm}
		\begin{tikzpicture}[baseline=(b)]
		\begin{feynman}[inline=(a)]
		\vertex[dot] (a) {};
		\vertex[dot, right=0.6cm of a] (b) {};
		\vertex[dot, right=0.6cm of b] (c) {};
		\vertex[dot, right=0.6cm of c] (d) {};
		\vertex[right=0.3cm of b] (x) {};
		\vertex[below=0.62cm of x] {\footnotesize{$1\times M^4$}};
		\end{feynman}
		\draw [/tikzfeynman] (a) to[out=135, in=45, loop, min distance=0.9cm] (a);
		\draw [/tikzfeynman] (b) to[out=135, in=45, loop, min distance=0.9cm] (b);
		\draw [/tikzfeynman] (c) to[out=135, in=45, loop, min distance=0.9cm] (c);
		\draw [/tikzfeynman] (d) to[out=135, in=45, loop, min distance=0.9cm] (d);
		\end{tikzpicture}
		\hspace{-0.3cm}
		+
		\hspace{-0.3cm}
		\begin{tikzpicture}[inner sep=2pt,baseline=(b)]
		\begin{feynman}[inline=(a)]
		\vertex[dot] (a) {};
		\vertex[dot, right=0.6cm of a] (b) {};
		\vertex[dot, right=0.6cm of b] (c) {};
		\vertex[dot, right=0.6cm of c] (d) {};
		\vertex[right=0.3cm of b] (x) {};
		\diagram*{(c) --[half left] (d)};
		\diagram*{(d) --[half left] (c)};
		\vertex[below=0.635cm of x] {\footnotesize{$1\times M^3$}};
		\end{feynman}
		\draw [/tikzfeynman] (a) to[out=135, in=45, loop, min distance=0.9cm] (a);
		\draw [/tikzfeynman] (b) to[out=135, in=45, loop, min distance=0.9cm] (b);
		\end{tikzpicture}
		\hspace{+0.15cm}
		+
		\hspace{-0.30cm}
		\begin{tikzpicture}[inner sep=2pt,baseline=(b)]
		\begin{feynman}[inline=(a)]
		\vertex[dot] (a) {};
		\vertex[dot, right=0.6cm of a] (b) {};
		\vertex[dot, right=0.6cm of b] (c) {};
		\vertex[dot, right=0.6cm of c] (d) {};
		\vertex[right=0.3cm of b] (x) {};
		\diagram*{(b) --[half left] (c)};
		\diagram*{(c) --[half left] (b)};
		\vertex[below=0.635cm of x] {\footnotesize{$1\times M^3$}};
		\end{feynman}
		\draw [/tikzfeynman] (a) to[out=135, in=45, loop, min distance=0.9cm] (a);
		\draw [/tikzfeynman] (d) to[out=135, in=45, loop, min distance=0.9cm] (d);
		\end{tikzpicture}
		\hspace{-0.3cm}
		+
		\hspace{+0.15cm}
		\begin{tikzpicture}[inner sep=2pt,baseline=(b)]
		\begin{feynman}[inline=(a)]
		\vertex[dot] (a) {};
		\vertex[dot, right=0.6cm of a] (b) {};
		\vertex[dot, right=0.6cm of b] (c) {};
		\vertex[dot, right=0.6cm of c] (d) {};
		\vertex[right=0.3cm of b] (x) {};
		\diagram*{(a) --[half left] (b)};
		\diagram*{(b) --[half left] (a)};
		\vertex[below=0.635cm of x] {\footnotesize{$1\times M^3$}};
		\end{feynman}
		\draw [/tikzfeynman] (c) to[out=135, in=45, loop, min distance=0.8cm] (c);
		\draw [/tikzfeynman] (d) to[out=135, in=45, loop, min distance=0.8cm] (d);
		\end{tikzpicture}
		\hspace{-0.3cm}
		+
		\hspace{-0.3cm}
		\begin{tikzpicture}[inner sep=2pt,baseline=(b)]
		\begin{feynman}[inline=(a)]
		\vertex[dot] (a) {};
		\vertex[dot, right=0.6cm of a] (b) {};
		\vertex[dot, right=0.6cm of b] (c) {};
		\vertex[dot, right=0.6cm of c] (d) {};
		\vertex[right=0.3cm of b] (x) {};
		\diagram*{(b) --[half left] (d)};
		\diagram*{(d) --[out=260, in=280] (b)};
		\vertex[below=0.635cm of x] {\footnotesize{$1\times M^3$}};
		\end{feynman}
		\draw [/tikzfeynman] (a) to[out=135, in=45, loop, min distance=0.8cm] (a);
		\draw [/tikzfeynman] (c) to[out=135, in=45, loop, min distance=0.8cm] (c);
		\end{tikzpicture}
		\right.
	\end{split}
	\\
	\begin{split}\nonumber
		&+
		\hspace{+0.15cm}
		\begin{tikzpicture}[inner sep=2pt,baseline=(b)]
		\begin{feynman}[inline=(a)]
		\vertex[dot] (a) {};
		\vertex[dot, right=0.6cm of a] (b) {};
		\vertex[dot, right=0.6cm of b] (c) {};
		\vertex[dot, right=0.6cm of c] (d) {};
		\vertex[right=0.3cm of b] (x) {};
		\diagram*{(a) --[half left] (c)};
		\diagram*{(c) --[out=260, in=280] (a)};
		\vertex[below=0.8cm of x] {\footnotesize{$1\times M^3$}};
		\end{feynman}
		\draw [/tikzfeynman] (b) to[out=135, in=45, loop, min distance=0.8cm] (b);
		\draw [/tikzfeynman] (d) to[out=135, in=45, loop, min distance=0.8cm] (d);
		\end{tikzpicture}
		\hspace{-0.3cm}
		+
		\hspace{+0.15cm}
		\begin{tikzpicture}[inner sep=2pt,baseline=(b)]
		\begin{feynman}[inline=(a)]
		\vertex[dot] (a) {};
		\vertex[dot, right=0.6cm of a] (b) {};
		\vertex[dot, right=0.6cm of b] (c) {};
		\vertex[dot, right=0.6cm of c] (d) {};
		\vertex[right=0.3cm of b] (x) {};
		\diagram*{(a) --[half left] (d)};
		\diagram*{(d) --[out=240, in=300] (a)};
		\vertex[below=0.8cm of x] {\footnotesize{$1\times M^3$}};
		\end{feynman}
		\draw [/tikzfeynman] (b) to[out=135, in=45, loop, min distance=0.8cm] (b);
		\draw [/tikzfeynman] (c) to[out=135, in=45, loop, min distance=0.8cm] (c);
		\end{tikzpicture}
		\hspace{+0.15cm}
		+
		\hspace{-0.3cm}
		\begin{tikzpicture}[inner sep=2pt,baseline=(b)]
		\begin{feynman}[inline=(a)]
		\vertex[dot] (a) {};
		\vertex[dot, right=0.6cm of a] (b) {};
		\vertex[dot, right=0.6cm of b] (c) {};
		\vertex[dot, right=0.6cm of c] (d) {};
		\vertex[right=0.3cm of b] (x) {};
		\diagram*{(b) --[half left] (d)};
		\diagram*{(d) --[half left] (c)};
		\diagram*{(c) --[half left] (b)};
		\vertex[below=0.785cm of x] {\footnotesize{$1\times M^2$}};
		\end{feynman}
		\draw [/tikzfeynman] (a) to[out=135, in=45, loop, min distance=0.8cm] (a);
		\end{tikzpicture}
		\hspace{+0.15cm}
		+
		\hspace{+0.15cm}
		\begin{tikzpicture}[inner sep=2pt,baseline=(b)]
		\begin{feynman}[inline=(a)]
		\vertex[dot] (a) {};
		\vertex[dot, right=0.6cm of a] (b) {};
		\vertex[dot, right=0.6cm of b] (c) {};
		\vertex[dot, right=0.6cm of c] (d) {};
		\vertex[right=0.3cm of b] (x) {};
		\diagram*{(a) --[half left] (c)};
		\diagram*{(c) --[half left] (b)};
		\diagram*{(b) --[half left] (a)};
		\vertex[below=0.785cm of x] {\footnotesize{$1\times M^2$}};
		\end{feynman}
		\draw [/tikzfeynman] (d) to[out=135, in=45, loop, min distance=0.8cm] (d);
		\end{tikzpicture}
		\hspace{-0.30cm}
		+
		\hspace{+0.15cm}
		\begin{tikzpicture}[inner sep=2pt,baseline=(b)]
		\begin{feynman}[inline=(a)]
		\vertex[dot] (a) {};
		\vertex[dot, right=0.6cm of a] (b) {};
		\vertex[dot, right=0.6cm of b] (c) {};
		\vertex[dot, right=0.6cm of c] (d) {};
		\vertex[right=0.3cm of b] (x) {};
		\diagram*{(a) --[half left] (d)};
		\diagram*{(d) --[half left] (c)};
		\diagram*{(c) --[out=260, in=280] (a)};
		\vertex[below=0.785cm of x] {\footnotesize{$1\times M^2$}};
		\end{feynman}
		\draw [/tikzfeynman] (b) to[out=135, in=45, loop, min distance=0.8cm] (b);
		\end{tikzpicture}
	\end{split}
	\\
	\begin{split}
		&+
		\hspace{+0.15cm}
		\begin{tikzpicture}[inner sep=2pt,baseline=(b)]
		\begin{feynman}[inline=(a)]
		\vertex[dot] (a) {};
		\vertex[dot, right=0.6cm of a] (b) {};
		\vertex[dot, right=0.6cm of b] (c) {};
		\vertex[dot, right=0.6cm of c] (d) {};
		\vertex[right=0.3cm of b] (x) {};
		\diagram*{(a) --[half left] (d)};
		\diagram*{(d) --[out=260, in=280] (b)};
		\diagram*{(b) --[half left] (a)};
		\vertex[below=0.785cm of x] {\footnotesize{$1\times M^2$}};
		\end{feynman}
		\draw [/tikzfeynman] (c) to[out=135, in=45, loop, min distance=0.8cm] (c);
		\end{tikzpicture}
		\hspace{+0.15cm}
		+
		\hspace{+0.15cm}
		\begin{tikzpicture}[inner sep=2pt,baseline=(b)]
		\begin{feynman}[inline=(a)]
		\vertex[dot] (a) {};
		\vertex[dot, right=0.6cm of a] (b) {};
		\vertex[dot, right=0.6cm of b] (c) {};
		\vertex[dot, right=0.6cm of c] (d) {};
		\vertex[right=0.3cm of b] (x) {};
		\diagram*{(a) --[half left] (b)};
		\diagram*{(b) --[half left] (a)};
		\diagram*{(c) --[half left] (d)};
		\diagram*{(d) --[half left] (c)};
		\vertex[below=0.785cm of x] {\footnotesize{$1\times M^2$}};
		\end{feynman}
		\end{tikzpicture}
		\hspace{+0.15cm}
		+
		\hspace{+0.15cm}
		\begin{tikzpicture}[inner sep=2pt,baseline=(b)]
		\begin{feynman}[inline=(a)]
		\vertex[dot] (a) {};
		\vertex[dot, right=0.6cm of a] (b) {};
		\vertex[dot, right=0.6cm of b] (c) {};
		\vertex[dot, right=0.6cm of c] (d) {};
		\vertex[right=0.3cm of b] (x) {};
		\diagram*{(a) --[out=80, in=100] (d)};
		\diagram*{(d) --[out=260, in=280] (a)};
		\diagram*{(b) --[half left] (c)};
		\diagram*{(c) --[half left] (b)};
		\vertex[below=0.785cm of x] {\footnotesize{$1\times M^2$}};
		\end{feynman}
		\end{tikzpicture}
		\hspace{+0.15cm}
		+
		\hspace{-0.30cm}
		\begin{tikzpicture}[inner sep=2pt,baseline=(b)]
		\begin{feynman}[inline=(a)]
		\vertex[dot] (a) {};
		\vertex[dot, right=0.6cm of a] (b) {};
		\vertex[dot, right=0.6cm of b] (c) {};
		\vertex[dot, right=0.6cm of c] (d) {};
		\vertex[right=0.3cm of b] (x) {};
		\diagram*{(b) --[half left] (c)};
		\diagram*{(c) --[half left] (d)};
		\diagram*{(d) --[out=260, in=280] (b)};
		\vertex[below=0.785cm of x] {\footnotesize{$t_3\times M^2$}};
		\end{feynman}
		\draw [/tikzfeynman] (a) to[out=135, in=45, loop, min distance=0.8cm] (a);
		\end{tikzpicture}
		\hspace{+0.15cm}
		+
		\hspace{+0.15cm}
		\begin{tikzpicture}[inner sep=2pt,baseline=(b)]
		\begin{feynman}[inline=(a)]
		\vertex[dot] (a) {};
		\vertex[dot, right=0.6cm of a] (b) {};
		\vertex[dot, right=0.6cm of b] (c) {};
		\vertex[dot, right=0.6cm of c] (d) {};
		\vertex[right=0.3cm of b] (x) {};
		\diagram*{(a) --[half left] (b)};
		\diagram*{(b) --[half left] (c)};
		\diagram*{(c) --[out=260, in=280] (a)};
		\vertex[below=0.785cm of x] {\footnotesize{$t_3\times M^2$}};
		\end{feynman}
		\draw [/tikzfeynman] (d) to[out=135, in=45, loop, min distance=0.8cm] (d);
		\end{tikzpicture}
		\hspace{-0.3cm}
	\end{split}
	\\
	\begin{split}\nonumber
		&+
		\hspace{+0.15cm}
		\begin{tikzpicture}[inner sep=2pt,baseline=(b)]
		\begin{feynman}[inline=(a)]
		\vertex[dot] (a) {};
		\vertex[dot, right=0.6cm of a] (b) {};
		\vertex[dot, right=0.6cm of b] (c) {};
		\vertex[dot, right=0.6cm of c] (d) {};
		\vertex[right=0.3cm of b] (x) {};
		\diagram*{(a) --[half left] (c)};
		\diagram*{(c) --[half left] (d)};
		\diagram*{(d) --[out=240, in=300] (a)};
		\vertex[below=0.785cm of x] {\footnotesize{$t_3\times M^2$}};
		\end{feynman}
		\draw [/tikzfeynman] (b) to[out=135, in=45, loop, min distance=0.8cm] (b);
		\end{tikzpicture}
		\hspace{+0.15cm}
		+
		\hspace{+0.15cm}
		\begin{tikzpicture}[inner sep=2pt,baseline=(b)]
		\begin{feynman}[inline=(a)]
		\vertex[dot] (a) {};
		\vertex[dot, right=0.6cm of a] (b) {};
		\vertex[dot, right=0.6cm of b] (c) {};
		\vertex[dot, right=0.6cm of c] (d) {};
		\vertex[right=0.3cm of b] (x) {};
		\diagram*{(a) --[half left] (b)};
		\diagram*{(b) --[half left] (d)};
		\diagram*{(d) --[out=240, in=300] (a)};
		\vertex[below=0.785cm of x] {\footnotesize{$t_3\times M^2$}};
		\end{feynman}
		\draw [/tikzfeynman] (c) to[out=135, in=45, loop, min distance=0.8cm] (c);
		\end{tikzpicture}
		\hspace{+0.15cm}
		+
		\hspace{+0.15cm}
		\begin{tikzpicture}[inner sep=2pt,baseline=(b)]
		\begin{feynman}[inline=(a)]
		\vertex[dot] (a) {};
		\vertex[dot, right=0.6cm of a] (b) {};
		\vertex[dot, right=0.6cm of b] (c) {};
		\vertex[dot, right=0.6cm of c] (d) {};
		\vertex[right=0.3cm of b] (x) {};
		\diagram*{(a) --[out=80, in=100] (c)};
		\diagram*{(c) --[out=260, in=280] (a)};
		\diagram*{(b) --[out=80, in=100] (d)};
		\diagram*{(d) --[out=260, in=280] (b)};
		\vertex[below=0.785cm of x] {\footnotesize{$t_4\times M^2$}};
		\end{feynman}
		\end{tikzpicture}
		\hspace{+0.15cm}
		+
		\hspace{+0.15cm}
		\begin{tikzpicture}[inner sep=2pt,baseline=(b)]
		\begin{feynman}[inline=(a)]
		\vertex[dot] (a) {};
		\vertex[dot, right=0.6cm of a] (b) {};
		\vertex[dot, right=0.6cm of b] (c) {};
		\vertex[dot, right=0.6cm of c] (d) {};
		\vertex[right=0.3cm of b] (x) {};
		\diagram*{(a) --[out=80,in=100] (d)};
		\diagram*{(d) --[half left] (c)};
		\diagram*{(c) --[half left] (b)};
		\diagram*{(b) --[half left] (a)};
		\vertex[below=0.785cm of x] {\footnotesize{$1\times M$}};
		\end{feynman}
		\end{tikzpicture}
		\hspace{+0.15cm}
		+
		\hspace{+0.15cm}
		\begin{tikzpicture}[inner sep=2pt,baseline=(b)]
		\begin{feynman}[inline=(a)]
		\vertex[dot] (a) {};
		\vertex[dot, right=0.6cm of a] (b) {};
		\vertex[dot, right=0.6cm of b] (c) {};
		\vertex[dot, right=0.6cm of c] (d) {};
		\vertex[right=0.3cm of b] (x) {};
		\diagram*{(a) --[half left] (b)};
		\diagram*{(b) --[half left] (d)};
		\diagram*{(d) --[half left] (c)};
		\diagram*{(c) --[out=260, in=280] (a)};
		\vertex[below=0.805cm of x] {\footnotesize{$t_3\times M$}};
		\end{feynman}
		\end{tikzpicture}
		\hspace{+0.15cm}+
	\end{split}
	\\
	\begin{split}\nonumber
		&+\left.
		\hspace{+0.15cm}
		\begin{tikzpicture}[inner sep=2pt,baseline=(b)]
		\begin{feynman}[inline=(a)]
		\vertex[dot] (a) {};
		\vertex[dot, right=0.6cm of a] (b) {};
		\vertex[dot, right=0.6cm of b] (c) {};
		\vertex[dot, right=0.6cm of c] (d) {};
		\vertex[right=0.3cm of b] (x) {};
		\diagram*{(a) --[half left] (c)};
		\diagram*{(c) --[half left] (d)};
		\diagram*{(d) --[out=260, in=280] (b)};
		\diagram*{(b) --[half left] (a)};
		\vertex[below=0.805cm of x] {\footnotesize{$t_3\times M$}};
		\end{feynman}
		\end{tikzpicture}
		\hspace{+0.15cm}
		+
		\hspace{+0.15cm}
		\begin{tikzpicture}[inner sep=2pt,baseline=(b)]
		\begin{feynman}[inline=(a)]
		\vertex[dot] (a) {};
		\vertex[dot, right=0.6cm of a] (b) {};
		\vertex[dot, right=0.6cm of b] (c) {};
		\vertex[dot, right=0.6cm of c] (d) {};
		\vertex[right=0.3cm of b] (x) {};
		\diagram*{(a) --[half left] (c)};
		\diagram*{(c) --[half left] (b)};
		\diagram*{(b) --[half left] (d)};
		\diagram*{(d) --[out=240, in=300] (a)};
		\vertex[below=0.805cm of x] {\footnotesize{$t_3\times M$}};
		\end{feynman}
		\end{tikzpicture}
		\hspace{+0.15cm}
		+
		\hspace{+0.15cm}
		\begin{tikzpicture}[inner sep=2pt,baseline=(b)]
		\begin{feynman}[inline=(a)]
		\vertex[dot] (a) {};
		\vertex[dot, right=0.6cm of a] (b) {};
		\vertex[dot, right=0.6cm of b] (c) {};
		\vertex[dot, right=0.6cm of c] (d) {};
		\vertex[right=0.3cm of b] (x) {};
		\diagram*{(a) --[out=800, in=100] (d)};
		\diagram*{(d) --[out=260, in=280] (b)};
		\diagram*{(b) --[half left] (c)};
		\diagram*{(c) --[out=260, in=280] (a)};
		\vertex[below=0.805cm of x] {\footnotesize{$t_3\times M$}};
		\end{feynman}
		\end{tikzpicture}
		\hspace{+0.15cm}
		+
		\hspace{+0.15cm}
		\begin{tikzpicture}[inner sep=2pt,baseline=(b)]
		\begin{feynman}[inline=(a)]
		\vertex[dot] (a) {};
		\vertex[dot, right=0.6cm of a] (b) {};
		\vertex[dot, right=0.6cm of b] (c) {};
		\vertex[dot, right=0.6cm of c] (d) {};
		\vertex[right=0.3cm of b] (x) {};
		\diagram*{(a) --[half left] (b)};
		\diagram*{(b) --[half left] (c)};
		\diagram*{(c) --[half left] (d)};
		\diagram*{(d) --[out=240, in=300] (a)};
		\vertex[below=0.805cm of x] {\footnotesize{$t_4\times M$}};
		\end{feynman}
		\end{tikzpicture}
		\hspace{+0.15cm}
		\right)
	\end{split}
	\\
	\begin{split}\nonumber
		&=2^{N/2}\(\av{\ell^2} \binom{N}{\qb}\)^4
		\(
		M^4+6M^3+(6+4t_3+t_4)M^2+(1+4t_3+t_4)M
		\).
	\end{split}
\end{align}
\endgroup
Below each diagram we wrote (i) a factor of $M$ for each closed loop and (ii) the weight~$t(\sigma)$ coming from the trace of $\Gamma$-matrices. Different diagrams can correspond to the same trace because of the latter's cyclic property. 

The first nontrivial diagram, $t_3$, arises in the third moment of $W$ (sixth order in $L$ and $L^\dagger$) and it is explicitly given by~\cite{garcia-garcia2016PRD,garcia-garcia2018JHEP}:
\begin{equation}\label{eq:diagram_t3}
\begin{split}
t_3(\qb,N)&\equiv
\begin{tikzpicture}[inner sep=2pt,baseline=(b)]
\begin{feynman}[inline=(a)]
\vertex[dot] (a) {};
\vertex[dot, right=0.6cm of a] (b) {};
\vertex[dot, right=0.6cm of b] (c) {};
\diagram*{(a) --[half left] (b)};
\diagram*{(b) --[half left] (c)};
\diagram*{(c) --[out=260, in=280] (a)};;
\end{feynman}
\end{tikzpicture}
= 2^{-N/2} \binom{N}{\qb}^{-3} \sum_{\va,\vb,\vc} \Tr\(
\Gamma_\va \Gamma_\vb \Gamma_\vc \Gamma_\va \Gamma_\vb \Gamma_\vc \)
\\
&=\binom{N}{\qb}^{-2} 
\sum_{s=0}^\qb \sum_{r=0}^\qb \sum_{m=0}^r 
(-1)^{\qb+s+m}
\binom{\qb}{s}\binom{N-\qb}{\qb-s}\binom{s}{r-m}
\binom{2(\qb-s)}{m}\binom{N-2\qb+s}{\qb-r}.
\end{split}
\end{equation}
To evaluate the trace, we have to first commute $\Gamma_\va$ with the product $\Gamma_\vb\Gamma_\vc$ and then commute $\Gamma_\vb$ with $\Gamma_\vc$, using Eq.~(\ref{eq:relations_Gamma}). Diagram $t_3$ also arises in the standard SYK model at sixth order in an expansion in powers of $H$; for its perfect-matching representation and a detailed account on how to compute it, see Appendix~\ref{app:review_standard_SYK}. However, the nontrivial diagram 
\begin{equation}\label{eq:diagram_t2}
t_2(\qb,N)
=2^{-N/2}\binom{N}{\qb}^{-2}\sum_{\va,\vb}\Tr\(\Gamma_\va \Gamma_\vb \Gamma_\va \Gamma_\vb\)
=\binom{N}{\qb}^{-1}\sum_{s=0}^\qb 
(-1)^{\qb+s}\binom{\qb}{s}\binom{N-\qb}{\qb-s}
\end{equation}
that arises at fourth order in an expansion in powers of $H$ [see Eqs.~(\ref{eq:H_moment4_example}) and (\ref{eq:diagram_t2_perfect_matching})] does not arise at fourth order in an expansion in powers of $L$ and $L^\dagger$. This is also true for higher orders: no diagrams related to $t_2$ (e.g., its powers) arise, as $t_2$ is a forbidden contraction when $k=1$ (i.e., it corresponds to a perfect matching that is not a permutation). In the context of the standard SYK model, the first nontrivial diagram (i.e., the single-crossing diagram, in that case $t_2$) has the combinatorial interpretation of the deformation parameter $Q$. For the circular WSYK model, we thus anticipate that it is diagram $t_3$ that assumes this role. 

Another nontrivial diagram, $t_4$, appears in the fourth moment. It cannot be reduced to a power of the lower-order diagram $t_3$ and it is evaluated analogously to before~\cite{garcia-garcia2018JHEP}:
\begin{equation}\label{eq:diagram_t4}
\begin{split}
t_4(\qb,N)&\equiv
\begin{tikzpicture}[inner sep=2pt,baseline=(b)]
\begin{feynman}[inline=(a)]
\vertex[dot] (a) {};
\vertex[dot, right=0.6cm of a] (b) {};
\vertex[dot, right=0.6cm of b] (c) {};
\vertex[dot, right=0.6cm of c] (d) {};
\diagram*{(a) --[half left] (b)};
\diagram*{(b) --[half left] (c)};
\diagram*{(c) --[half left] (d)};
\diagram*{(d) --[out=240, in=300] (a)};
\end{feynman}
\end{tikzpicture}
=2^{-N/2}\binom{N}{\qb}^{-4}\sum_{\va,\vb,\vc,\vd}
\Tr\(
\Gamma_\va \Gamma_\vb \Gamma_\vc \Gamma_\va
\Gamma_\vd \Gamma_\vc \Gamma_\vb \Gamma_\vd
\)
\\
&=
\binom{N}{\qb}^{-3} 
\sum_{s=0}^\qb \binom{\qb}{s}\binom{N-\qb}{\qb-s}
\left\{\sum_{r=0}^\qb \sum_{m=0}^r 
(-1)^{m} \binom{s}{r-m} \binom{2(\qb-s)}{m}\binom{N-2\qb+s}{\qb-r}\right\}^2.
\end{split}
\end{equation}
To arrive at the explicit expression, we performed independent commutations of $\Gamma_\vb$ and $\Gamma_\vc$ over the product $\Gamma_\va\Gamma_\vd$, using again Eq.~(\ref{eq:relations_Gamma}).

In principle, one can compute $t(\sigma)$ exactly for every diagram and for all $p$ by explicitly computing traces. However, the computations quickly become intractable. Alternatively, when $\qb\propto\sqrt{N}$ and $M=1$, the weights $t(\sigma)$ can be computed exactly.

\section{Analytic spectral density of the circular WSYK model with \texorpdfstring{$M=1$}{M=1}}
\label{sec:spectral_density_WSYK}

When $\qb\propto \sqrt{N}$, the weight $t(\sigma)$ is fully characterized by the numbers of commutations required to bring the $\Gamma$-matrices to a trivial ordering. The number of commutations in a trace corresponds to the number of crossings in the corresponding perfect-matching, and not permutation, diagram. In this scaling limit, it is possible to show~\cite{erdos2014MPAG,feng2019PMJ} that, for a perfect matching $\pi$ with $\mathrm{cross}(\pi)$ crossings, $t(\pi)=Q^{\mathrm{cross}(\pi)}$, where $Q=(-1)^\qb\exp(-2\alpha)$, $\alpha=\qb^2/N$ is $N$ independent and $N\to\infty$. To use this exact scaling-limit result, the sum in Eq.~(\ref{eq:moments_W_sum_permutations}) would have to be performed over a subset of perfect matchings. Unfortunately, such a restricted sum cannot be performed in closed form. Instead, we noticed that if we approximate the number of commutations in the trace by the number of crossings in the permutation diagram, the sum in Eq.~(\ref{eq:moments_W_sum_permutations}) is feasible and the moments correspond to the moments of the weight function of certain $Q$-Laguerre polynomials, where $-1 < Q(\qb,N) < 1$ becomes independent of $N$ in the scaling limit. When $\qb$ is fixed and finite, the same combinatorial arguments still hold to order $1/N$~\cite{garcia-garcia2018JHEP}. Hence, for any $\qb$, we can compute the spectral density to leading and next-to-leading order in $1/N$.

For fixed $\qb$ and in the limit $N\to\infty$, there are, to leading order in $1/N$, no $\gamma$-matrices common to different $\Gamma$-matrices and, thus the commutations of the latter can be ignored, i.e., $t(\sigma)=1$ for all $\sigma\in\mathcal{S}_p$. The leading-order moments simply count the number of allowed diagrams at each order, which for permutations $\sigma\in\mathcal{S}_p$ are
\begin{equation}
\frac{1}{\sigmaL^p}\frac{\av{\Tr W^p}}{\Tr \id}=p!.
\end{equation}
These are the moments of the exponential distribution and, hence, to leading order, the spectral density of the $M=1$ circular WSYK model is $\varrho(E)=\exp(-E)$.

To next-to-leading order, we take the commutations of $\Gamma$-matrices into account but ignore correlations between different commutations. Therefore, we completely characterize a diagram~$\sigma$ simply by its number of crossings, $\mathrm{cross}(\sigma)$. Some additional care is required when counting crossings diagrammatically, compared to the perfect-matching case: $\mathrm{cross}(\sigma)$ is the number of pairs of edges above the line of dots that cross
(~$
\begin{tikzpicture}[inner sep=2pt,baseline=(b)]
\begin{feynman}[inline=(a)]
\vertex[dot] (a) {};
\vertex[dot,right=0.4cm of a] (b) {};
\vertex[dot,right=0.4cm of b] (c) {};
\vertex[dot,right=0.4cm of c] (d) {};
\diagram*{(a) --[half left,min distance=0.4cm] (c)};
\diagram*{(b) --[half left,min distance=0.4cm] (d)};
\end{feynman}
\end{tikzpicture}
$~) \emph{or touch} 
(~$
\begin{tikzpicture}[inner sep=2pt,baseline=(b)]
\begin{feynman}[inline=(a)]
\vertex[dot] (a) {};
\vertex[dot,right=0.4cm of a] (b) {};
\vertex[dot,right=0.4cm of b] (c) {};
\diagram*{(a) --[half left,min distance=0.4cm] (b)};
\diagram*{(b) --[half left,min distance=0.4cm] (c)};
\end{feynman}
\end{tikzpicture}
$~), plus the number of pairs of edges below the line of dots that cross
(~$
\begin{tikzpicture}[inner sep=2pt,baseline=(l)]
\begin{feynman}[inline=(a)]
\vertex[dot] (a) {};
\vertex[dot,right=0.4cm of a] (b) {};
\vertex[dot,right=0.4cm of b] (c) {};
\vertex[dot,right=0.4cm of c] (d) {};
\vertex[right=0.2cm of b, below=0.3cm of b] (l) {};
\diagram*{(a) --[half right,min distance=0.4cm] (c)};
\diagram*{(b) --[half right,min distance=0.4cm] (d)};
\end{feynman}
\end{tikzpicture}
$~)~\cite{corteel2007AAM}\footnote{This is the reason why the permutation $\sigma(1,2,3)=(2,3,1)$ has a crossing despite no edges actually crossing in its diagram, see Eq.~(\ref{eq:diagram_t3}).}.
The elementary single-crossing diagram is $t_3(\qb,N)$ and, as for the standard SYK model, we approximate more complicated diagrams with $\mathrm{cross}(\sigma)$ crossings by ascribing a factor $t_3$ to each crossing, i.e., we approximate $t(\sigma)=t_3^{\mathrm{cross}(\sigma)}$\footnote{The number of permutations in $\mathcal{S}_p$ with $k$ crossings can be explicitly computed for arbitrary $(p,k)$ and is tabulated as sequence A263776 in the Online Encyclopedia of Integer Sequences (OEIS)~\cite{OEIS_A263776}.}. As an example, we have $t_4\approx t_3^2$. As mentioned before, this identification does not become exact when $\qb\propto\sqrt{N}$. 

Summing over all diagrams, we arrive at
\begin{equation}
\frac{1}{\sigmaL^p}\frac{\av{\Tr W^p}}{\Tr \id}
=\sum_{\sigma\in\sS_p} t_3^{\mathrm{cross(\sigma)}},
\end{equation}
which are recognized as the moments of the orthogonality weight function of the Al-Salam-Chihara $Q$-Laguerre polynomials~\cite{alsalam1976,kasraoui2011AAM} with $Q=t_3(\qb,N)$ and $y=1$:
\begin{equation}\label{eq:spectral_density_QLaguerre}
\varrho_{\mathrm{QL}}(E;Q)=
\frac{(Q;Q)_\infty^2(-Q;Q)_\infty^2}{(-Q^2;Q^2)_\infty^2}
\frac{2}{\pi \EL}\sqrt{\frac{1-E/\EL}{E/\EL}}
\prod_{k=1}^\infty 
\frac{
	1-\frac{4\(1-2E/\EL\)^2}{2+Q^{k}+Q^{-k}}}{
	\(1-\frac{2\(1-2E/\EL\)}{Q^k+Q^{-k}}\)^2},
\end{equation}
supported on $0\leq E\leq \EL$, where the (dimensionless) spectral edge of the WSYK model is
\begin{equation}\label{eq:E0_QLaguerre}
\EL=\frac{4}{1-Q},
\end{equation}
and $(a;Q)_\infty=\prod_{k=0}^\infty \(1-aQ^k\)$ is the $Q$-Pochhammer symbol. Note that the spectral density is of the form of a single-channel Marchenko-Pastur distribution~\cite{marchenko1967,forrester2010},
\begin{equation}\label{eq:Marchenko-Pastur}
\varrho_{\mathrm{MP}}(E)=
\frac{2}{\pi \EL}\sqrt{\frac{1-E/\EL}{E/\EL}},
\end{equation}
times a $Q$-dependant multiplicative correction\footnote{The Marchenko-Pastur distribution depends itself on $Q$ through its endpoint. In the random-matrix limit, $Q\to0$, the endpoint becomes $\EL\to4$.}.

Let us briefly comment on the nature of the $Q$-Laguerre approximation. It is exact to linear order in $Q$ or, alternatively, to next-to-leading order in $1/N$. The first correction is diagram $t_4$ that appears in the fourth moment. In the scaling limit $\qb^2=\alpha N$, we have the exact results $Q=t_3=-\exp[-6\alpha]$ and $t_4=\exp[-8\alpha]$~\cite{erdos2014MPAG,feng2019PMJ}; the $Q$-Laguerre approximation is, instead, $t_4\approx Q^2=\exp[-12\alpha]$; higher-order corrections proceed similarly. 
We thus see that, contrary to standard SYK case for which the $Q$-Hermite spectral density becomes exact in the scaling limit, the $Q$-Laguerre density is only an approximation of the true WSYK density both for fixed $\qb$ and in the scaling limit $\qb\propto \sqrt{N}$. Its accuracy has to be checked on a case-by-case basis, which we do for $\qb=2$, $3$, $4$ and $N\leq34$ in Sec.~\ref{sec:spectral_density_numerics}.

Notwithstanding the preceding considerations, it is important to emphasize that the $Q$-Laguerre mapping has two key features necessary to accurately predict the WSYK spectral density, namely, it accounts for the correct number of diagrams and it correctly identifies the deformation parameter as $Q=t_3$ (i.e., the minimal number of commutations in a trace is three). The other natural candidates for describing the spectral density ($Q$-Hermite and RMT) do not possess either of these properties.

It is also important to note that the $Q$-Laguerre density~(\ref{eq:spectral_density_QLaguerre}) is \emph{not equal} to the $Q$-Hermite density after the change of variables $E\to E^2$, as it is the case for the $M=1$ linear WSYK model (equivalently, the $\mathcal{N}=1$ supersymmetric SYK model), defined in Eq.~(\ref{eq:def_WSYK}). 
First, $Q$ depends exclusively on the symmetries of the model (i.e., Hermiticity) and appropriate combinatorics are required to obtain its correct value: $Q=t_2$ [Eq.~(\ref{eq:diagram_t2})] for the linear WSYK model, while $Q=t_3$ [Eq.~(\ref{eq:diagram_t3})] for the circular WSYK model. In particular, the spectral edge, $\EL$, of both models is formally the same [see Eqs.~(\ref{eq:E0_QLaguerre}) and (\ref{eq:E0_linear_WSYK}) below] but it is evaluated at different $Q$. Second, even if we put in the correct value of $Q$ by hand, for $Q>0$ the multiplicative corrections are different for the circular and linear WSYK models. The latter's spectral density can be trivially obtained from the standard SYK model by a change of variables. Indeed, because the couplings are real we have $W=L^2$ and, hence, the eigenvalues of $W$ are the squares of the eigenvalues of $L$. Effecting the change of variables $E\to E^2$ and multiplying by the associated Jacobian $1/\sqrt{E}$, we obtain the following spectral density, which has to be evaluated at $Q=t_2(\qb,N)$ (recall that each $\Gamma$-matrix now only has $\qb=q/2$ $\gamma$-matrices):
\begin{equation}\label{eq:spectral_density_linearWSYK}
\begin{split}
\varrho(E;Q)&=\frac{1}{\sqrt{E}}\varrho_{\mathrm{QH}}(\sqrt{E};Q)\\
&=(Q;Q)_\infty(-Q;Q)_\infty^2\frac{2}{\pi \EL}
\sqrt{\frac{1-E/\EL}{E/\EL}}
\prod_{k=1}^\infty
\(1-\frac{4E/\EL}{2+Q^{k}+Q^{-k}}\),
\end{split}
\end{equation}
supported on $0\leq E\leq \EL$, where the (dimensionless) spectral edge of the WSYK model is
\begin{equation}\label{eq:E0_linear_WSYK}
\EL=\EH^2=\frac{4}{1-Q}.
\end{equation}
As would be expected, the spectral density assumes the form of the single-channel Marchenko-Pastur distribution times the $Q$-Hermite multiplicative correction. Specifically, as $Q\to1$, in the circular model $\varrho(E)\to\exp\(-E\)$, while in the linear model $\varrho(E)\to(1/\sqrt{2\pi E})\exp\(-E/2\)$. On the other hand, as $Q\to0$ (the random-matrix limit) both spectral densities go to the Marchenko-Pastur distribution. Indeed, the relevant diagrams for RMT are noncrossing and because all noncrossing perfect matchings are also noncrossing permutations, the allowed diagrams for the Gaussian and Wishart-Laguerre ensembles coincide.

\subsection{Simple asymptotic formulas for the spectral density in the large-\texorpdfstring{$N$}{N} limit}

To conclude this section, we note that the $Q$-Laguerre density~(\ref{eq:spectral_density_QLaguerre}) has a simple asymptotic form in the bulk (i.e., for $0\ll E\ll \EL$), close to the hard edge $E\approx 0$, and near the soft edge $E\approx \EL$. These asymptotic formulas are obtained in the large-$N$ limit after a Poisson resummation of the $Q$-Laguerre spectral density; see Appendix~\ref{app:asymptotics} for their derivation. The asymptotic densities depend on the sign of $Q$. For a finite but large enough $N$ (the limit we are mostly interested in), $Q=t_3(\qb,N)$ is positive for even $\qb$ and negative for odd $\qb$. 

\paragraph*{Positive $Q$ (even $\qb$).}
The asymptotic bulk density is given by
\begin{equation}\label{eq:asympt_initial}
\varrho_{Q>0}^{(\mathrm{bulk})}(E;Q)=C''_Q 
\exp\left[
\frac{2\arcsin^2\(1-\frac{2E}{\EL}\)-\arccos^2\(\frac{2E}{\EL}-1\)}{\log Q}\right],
\end{equation}
where the constant $C''_Q$ is
\begin{equation}
C''_Q
=\frac{(Q;Q)_\infty^2(-Q;Q)_\infty^2}{(-Q^2;Q^2)_\infty^2}\frac{\exp\left[\pi^2/\(4\log Q\)\right]}{\(1+\exp[\pi^2/\log Q]\)^2}
\frac{2}{\pi \EL}.
\end{equation}

Near the hard edge, $E\approx 0$, the asymptotic density is
\begin{equation}
\varrho_{Q>0}^{(\mathrm{hard})}(E;Q)=
C''_Q 
\exp\left[-\frac{\pi^2}{2\log Q}\right]
\coth\left[-\frac{2\pi}{\log Q}\sqrt{\frac{E}{\EL}} \right],
\end{equation}
which has the expected $1/\sqrt{E}$ divergence. Intriguingly, unlike the standard and supersymmetric SYK models, the spectral density in this low-energy region does not increase exponentially. As was mentioned previously, a density of low-energy excitations proportional to $\exp[\gamma\sqrt{E-E_0}]$, with $\gamma > 0$ and $E_0$ the ground-state energy, is typical of both quantum black holes~\cite{maldacena2016PRD} and nuclear matter~\cite{bethe1936,garcia-garcia2017PRD}. Its absence in the WSYK model, for $Q(\qb,N) > 0$, is a strong indication that no gravity dual interpretation exists in this range of parameters. We shall see shortly that for $Q(\qb,N) < 0$ the situation is different. 

Near the soft edge $E\approx \EL$ corresponding to the high-energy region, the asymptotic density is found to be
\begin{equation}
\varrho_{Q>0}^{(\mathrm{soft})}(E;Q)= 
C''_Q\exp\left[\frac{\pi^2}{2\log Q}\right]
\,2\sinh\left[-\frac{4\pi}{\log Q}\sqrt{1-\frac{E}{\EL}}\right],
\end{equation}
Note that, apart from some multiplicative constants, this is the same density found near the edges of the standard SYK model. This is not surprising as, far away from the origin, the positive-definiteness of the spectrum is irrelevant and one does not expect the density to be very sensitive to the exact distribution of matrix elements of the Hamiltonian. Since $E$ is smaller than, or comparable to, $\EL$, there is no exponential growth of excitations in this region which in any case would not be expected to be related to gravity systems as holographic relations in the context of the SYK model are restricted to the low-energy, strong-coupling region.

\paragraph*{Ngative $Q$ (odd $\qb$).}
The bulk spectral density instead reads as:
\begin{equation}
\begin{split}
\varrho_{Q<0}^{\mathrm{(bulk)}}=
C'_\abs{Q}
&\cosh\left[
\frac{\pi}{\log\abs{Q}}
\abs{\arcsin\(1-\frac{2E}{\EL}\)}
\right]
\\
\times&\exp\left[
\frac{2\arcsin^2\(1-\frac{2E}{\EL}\)-\frac{1}{2}\arccos^2\(1-\frac{2E}{\EL}\)-\frac{1}{2}\arccos^2\(\frac{2E}{\EL}-1\)}{\log \abs{Q}}
\right],
\end{split}
\end{equation}
where the global constant is
\begin{equation}
C'_\abs{Q}
=\frac{(Q;Q)_\infty^2(-Q;Q)_\infty^2}{(-Q^2;Q^2)_\infty^2}
\exp\left[\frac{\pi^2}{4\log\abs{Q}}\right]
\frac{2}{\pi \EL}.
\end{equation}

\begin{figure}[tbp]
	\centering
	\includegraphics[width=\textwidth]{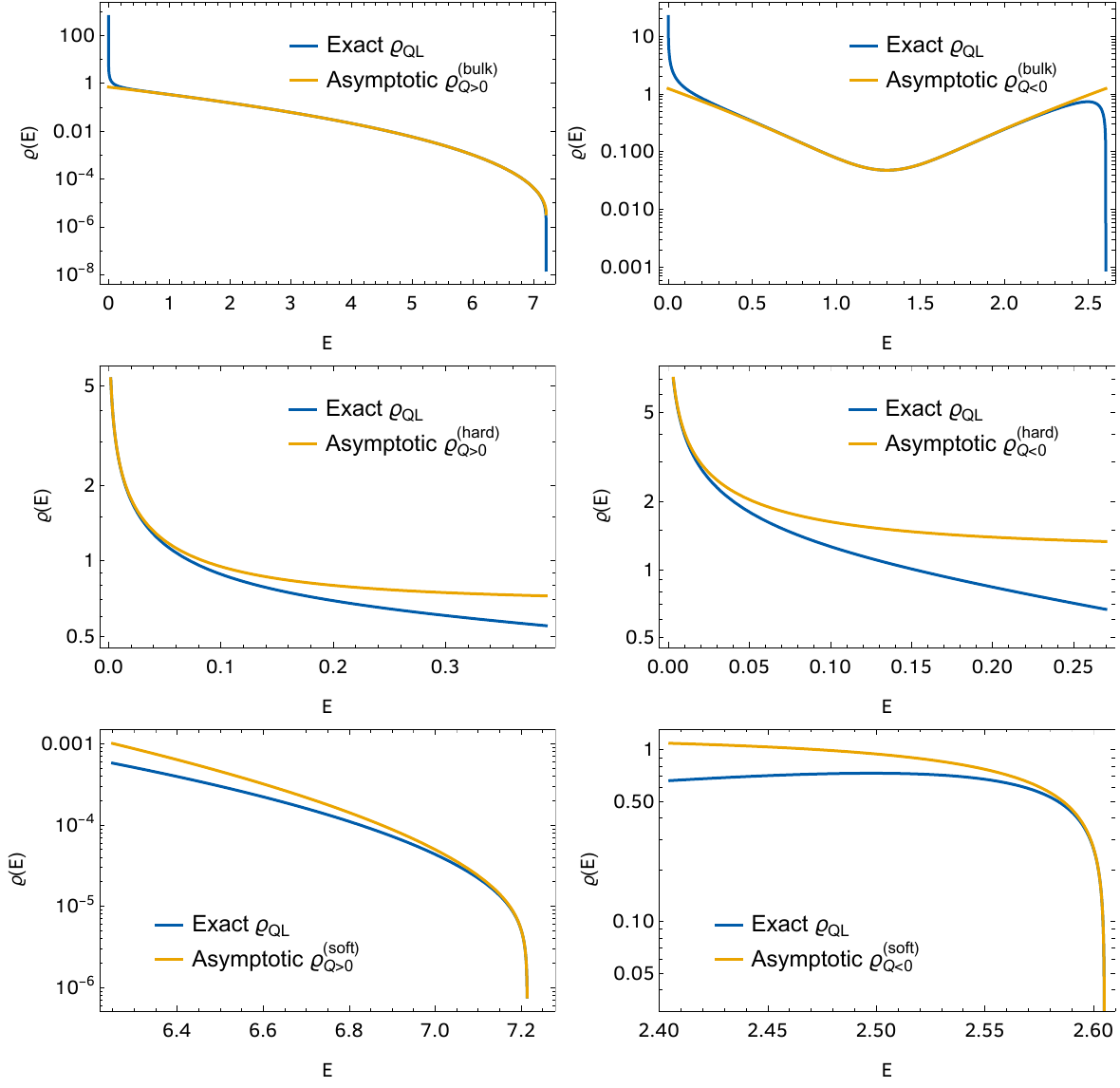}
	\caption{Comparison of the asymptotic spectral densities (orange curves) for the bulk, hard edge, and soft edge with the $Q$-Laguerre density (blue curve), in the respective domains of applicability. (Left) Densities for $Q>0$, namely, $Q=t_3(\qb=2,N=36)$. (Right) Densities for $Q<0$, namely, $Q=t_3(\qb=3,N=96)$.}
	\label{fig:M1AsymptDensity}
\end{figure} 

The spectral density close to the hard edge at $E = 0$ is
\begin{equation}\label{eq:expdenhard}
\varrho_{Q<0}^{\mathrm{(hard)}}=
C'_\abs{Q}
\coth\left[
-\frac{\pi}{\log\abs{Q}}\sqrt{\frac{E}{\EL}}
\right]
\exp\left[
-\frac{2\pi}{\log \abs{Q}}\sqrt{\frac{E}{\EL}}
\right]
\cosh\left[
\frac{\pi^2}{2\log\abs{Q}}\(1-\frac{4}{\pi}\sqrt{\frac{E}{\EL}}\)\right],
\end{equation}
which shows, besides the $1/\sqrt{E}$ divergence at the origin, a stretched exponential growth in the triple scaling limit~\cite{cotler2017JHEP} corresponding to low-energy excitations slightly above the ground-state $E=0$. Interestingly, this is precisely one of the expected features of a field theory with a gravity dual. We note that the functional form is different from the result for the supersymmetric SYK model~\cite{fu2017PRD,stanford2017JHEP,garcia-garcia2018PRD} whose spectrum also has a hard edge at zero energy.

Finally, the negative-$Q$ soft-edge asymptotic density is given by
\begin{equation}\label{eq:asympt_final}
\begin{split}
\varrho_{Q<0}^{\mathrm{(soft)}}=
C'_\abs{Q}
&\tanh\left[
-\frac{\pi}{\log\abs{Q}}\sqrt{1-\frac{E}{\EL}}
\right]
\exp\left[
-\frac{2\pi}{\log \abs{Q}}\sqrt{1-\frac{E}{\EL}}
\right]
\\
\times
&\cosh\left[
\frac{\pi^2}{2\log\abs{Q}}\(1-\frac{4}{\pi}\sqrt{1-\frac{E}{\EL}}\)\right].
\end{split}
\end{equation}
In this limit, the comments made for the positive $Q$ also apply here. 

In Fig.~\ref{fig:M1AsymptDensity}, we compare these simple asymptotic formulas, Eqs.~(\ref{eq:asympt_initial})--(\ref{eq:asympt_final}), with the $Q$-Laguerre weight function, Eq.~(\ref{eq:spectral_density_QLaguerre}), for both positive and negative $Q$. We see excellent agreement in the respective domains of applicability, even for $\qb=2$ and relatively small $N=36$ ($Q\approx 0.45$). As $N$ increases, the bulk asymptotic formula describes the density very well almost all the way down to the soft edge. However, it does not, of course, capture the $1/\sqrt{E}$ divergence close to the origin. For odd $\qb=3$, the asymptotic formulas are also very accurate, but the asymptotic limit is only attained for much larger values of $N$ (e.g., $Q\approx -0.54$ for $N=96$). This confirms the validity of the analytical calculation and, for negative $Q$, the possible existence of a holographic duality. 

\clearpage

\section{Numerical spectral density of the circular WSYK model for fixed \texorpdfstring{$\qb$}{q}}
\label{sec:spectral_density_numerics}

\begin{figure}[tbp]
	\centering
	\includegraphics[width=0.55\textwidth]{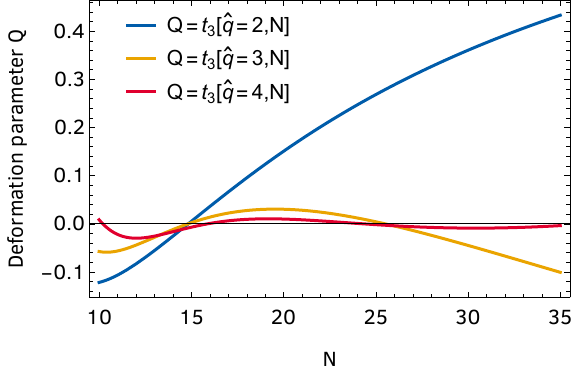}
	\caption{Dependence of the deformation parameter $Q$ on the number of Majoranas $N$ for $\qb = 2,3,4$.}
	\label{fig:QvsN}
\end{figure}

In this section, we study the accuracy of the $Q$-Laguerre approximation for the range of parameters accessible to numerical exact diagonalization, i.e., for a fixed $\qb = 2,3,4\ldots$ that does not scale with $N$. 

For $\qb=3, 4,\dots$ and $N\leq34$, $Q=t_3(\qb,N)$ is very close to $0$ (e.g., we have $Q=t_3(\qb=3,N=28)\approx -0.024$ and $Q=t_3(\qb=4,N=28)\approx -0.008$), see Fig.~\ref{fig:QvsN}. Hence, deviations from plain RMT are small. In Fig.~\ref{fig:M1LinearDensityq}, we compare the $Q$-Laguerre prediction~(\ref{eq:spectral_density_QLaguerre}) against numerical results obtained from exact diagonalization of the circular WSYK Hamiltonian~(\ref{eq:def_WSYK}). As can be seen in the center and right panels of Fig.~\ref{fig:M1LinearDensityq}, for these small values of $Q$, the $Q$-Laguerre density is almost indistinguishable from the random matrix result, the Marchenko-Pastur distribution, Eq.~(\ref{eq:Marchenko-Pastur}). For $\qb=3$ a modest deviation from random matrix behavior can be seen for $N=34$ in the tail of the spectrum, and there is qualitative agreement with the numerical results.

\begin{figure}[tbp]
	\centering
	\includegraphics[width=\textwidth]{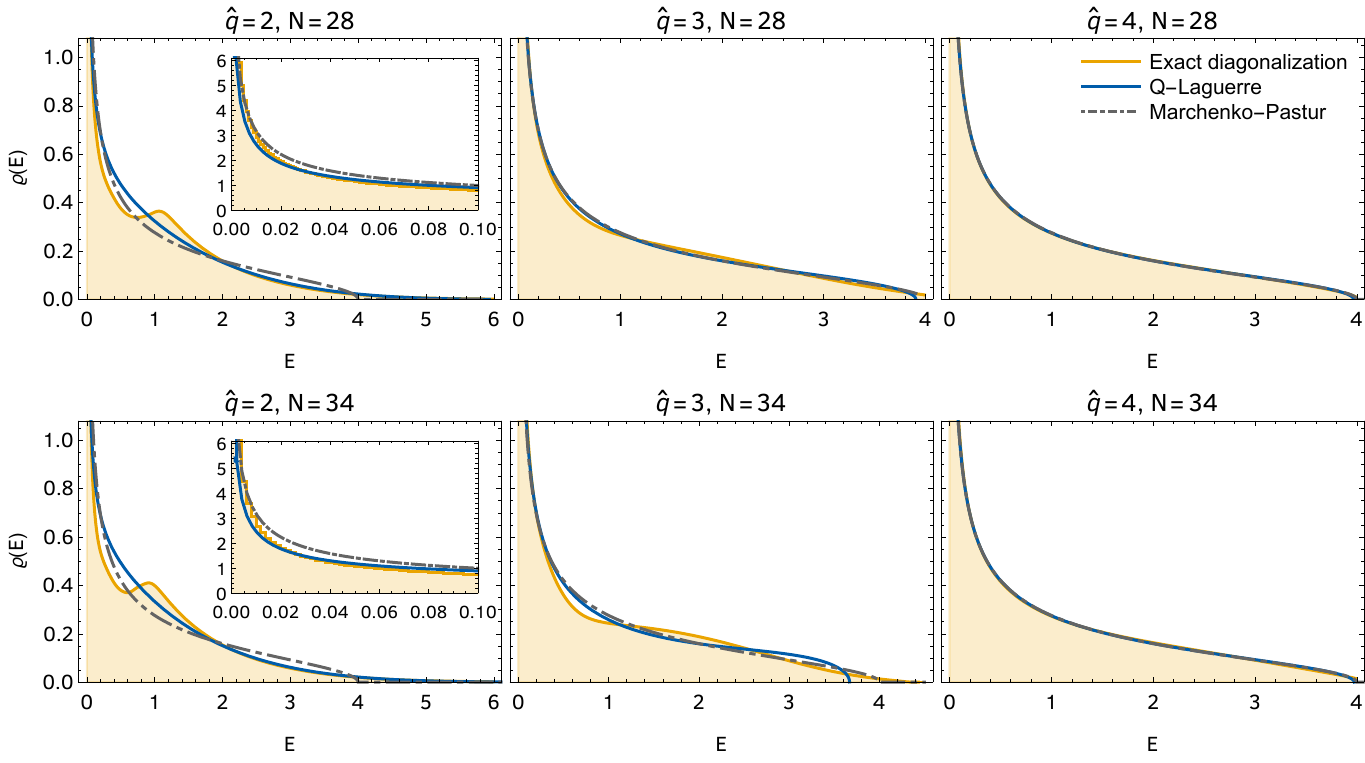}
	\caption{Spectral density of the circular ($k=1$) WSYK model with $M=1$ and $N=28$, $34$ Majorana fermions for $\qb=2$, $3$, and $4$. The orange (shaded) histograms are obtained by exact diagonalization of the WSYK Hamiltonian~(\ref{eq:def_WSYK}) for different disorder realizations totaling at least $2^{19}$ eigenvalues. The blue (full) and gray (dot-dashed) curves correspond to the $Q$-Laguerre [Eq.~(\ref{eq:spectral_density_QLaguerre})] with $Q=t_3(\qb,28)$ and Marchenko-Pastur [Eq.~(\ref{eq:Marchenko-Pastur})] predictions, respectively. We see the spectral density approaching plain random matrix results as $\qb$ increases with $N$ fixed. Insets: zoom on the hard edge for $\qb=2$. We see excellent agreement with both RMT and $Q$-Laguerre near the hard edge.}
	\label{fig:M1LinearDensityq}
\end{figure}

\begin{figure}[tbp]
	\centering
	\includegraphics[width=\textwidth]{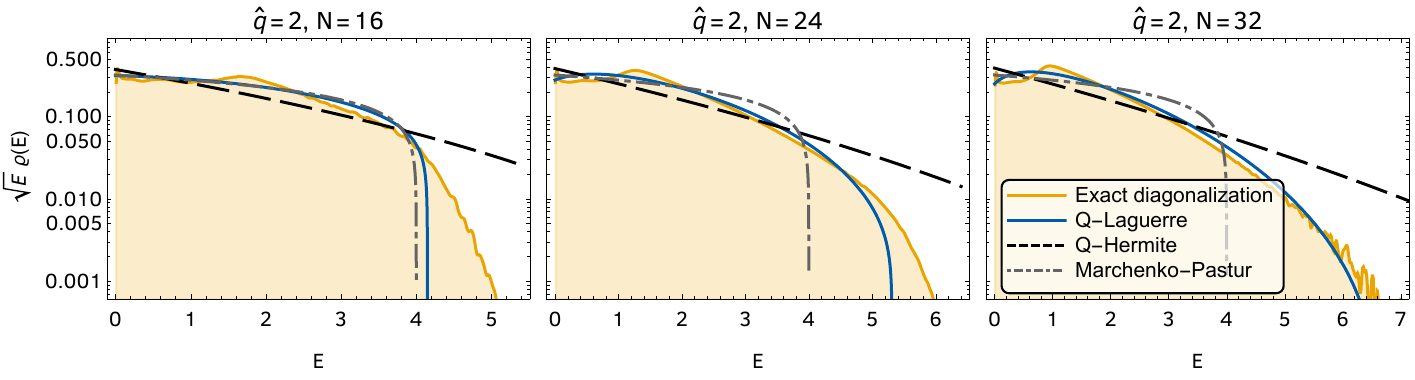}
	\caption{Spectral density (with $1/\sqrt{E}$ divergence factored out) of the circular ($k=1$) WSYK model with $M=1$ and $\qb=2$ ($q=4$) for $N=16$, $24$, and $32$ Majorana fermions. The orange (shaded) histograms are obtained by exact diagonalization of the WSYK Hamiltonian~(\ref{eq:def_WSYK}) for different disorder realizations totaling $2^{18}$ eigenvalues. The blue (full), black (dashed), and gray (dot-dashed) curves correspond to the $Q$-Laguerre [Eq.~(\ref{eq:spectral_density_QLaguerre})] with $Q=t_3(2,N)$, $Q$-Hermite [Eq.~(\ref{eq:spectral_density_linearWSYK})] with $Q=t_2(2,N)$, and Marchenko-Pastur [Eq.~(\ref{eq:Marchenko-Pastur})] predictions, respectively.}
	\label{fig:M1LaguerreDensity}
\end{figure}

For $\qb=2$, we have access to a much larger range of values of $Q=Q(2,N)$ (for instance, $Q=t_3(\qb=2,N=28)\approx 0.326$), see Fig.~\ref{fig:QvsN}. In this case, one can clearly distinguish the $Q$-Laguerre density from the random matrix one across the entire spectrum, see the left panel of Fig.~\ref{fig:M1LinearDensityq}. 

In the right half of the spectrum, $E\gtrsim 2$,
we observe an excellent agreement with the analytical $Q$-Laguerre prediction, which in this case is clearly different from both the $Q$-Hermite and the Marchenko-Pastur distribution, see Figs.~\ref{fig:M1LinearDensityq} and \ref{fig:M1LaguerreDensity}.
In particular, the $Q$-Laguerre density decays as $\exp[-E/\EL]$, while the $Q$-Hermite density as $\exp[-E/2\EL]$ (after the change of variables $E\to E^2$), and the Marchenko-Pastur distribution decays $\sqrt{1-E/\EL}$. Therefore, in this region of the spectrum, all three distributions can be clearly distinguished.
We stress that the comparison is completely determined by the analytical expression with no fitting parameters. In particular, the endpoint $\EL$ is fixed in each case: $\EL=4/(1-t_3(\qb,N))$ for $Q$-Laguerre, $\EL=4/(1-t_2(\qb,N))$ for $Q$-Hermite, and $\EL=4$ for Marchenko-Pastur. The tail of the spectrum is best seen on a logarithmic scale, see Fig.~\ref{fig:M1LaguerreDensity}. As mentioned before, to the right of the mid-spectrum bump, there is excellent agreement between the $Q$-Laguerre prediction and the numerical results. As $N$ grows, this agreement increasingly extends all the way down to the soft edge at $E=\EL$. Moreover, $Q=t_3(\qb,N)$ also increases and we start to see deviations from RMT. It is also clearly visible that the numerical results are \emph{not} well described by the $Q$-Hermite density after the change of variables $E\to E^2$ (conversely, the square roots of the eigenvalues of the circular WSYK do not follow the $Q$-Hermite density of the standard SYK model).

For smaller values of the energy, still for the $\qb=2$ case, we observe a bump around the middle of the spectrum. The bump shifts slightly to lower energies as $N$ increases (not shown). Within the range of $N$ that we can reach numerically, it is unclear whether it is a finite-size effect that will go away in the large-$N$ limit or if it has another origin. In any case, it is not predicted by the $Q$-Laguerre, $Q$-Hermite, or random matrix analytical expressions. The bump is much milder for $\qb=3$ (but visible when $N=34$) and it is not observed for $\qb=4$ in the available range of $N$. A qualitatively similar nonmonotonic behavior was observed~\cite{kanazawa2017JHEP} in a SYK model with Dirac fermions and also two $\hat q = 2$ blocks.  

Near the hard edge ($E=0$), and sufficiently away from the bump, see the insets in the left panels of Fig.~\ref{fig:M1LinearDensityq}, the divergence of the spectral density is well-described by the $Q$-Laguerre density which, in this range of parameters, agrees with the Marchenko-Pastur distribution. The presence of the bump, together with the singularity at $E = 0$ and the limited range of $N$ for which $Q$ is negative, makes it difficult to test the exponential growth, once the divergence is factored out, that characterizes the $Q$-Laguerre spectral density, Eq.~(\ref{eq:expdenhard}), in this region.

In Fig.~\ref{fig:M1LaguerreDensity}, there appears to be a discrepancy between numerics and the $Q$-Laguerre density at the soft (right) edge, which is more pronounced for smaller $N$. However, it is an artifact of the combination of the logarithmic scale (which amplifies small deviations), finite sampling, and relatively small values of $N$. Indeed, we know that the edge (largest eigenvalue), $E_{\mathrm{max}}$, which depends on the disorder realization, is itself a random variable whose distribution only becomes sharply peaked around its mean, $\EL$, for sufficiently large $N$. Therefore, the spectral density beyond the edge is, in reality, the distribution of the largest eigenvalues. Strictly speaking, we note that not only the largest eigenvalue contributes to the tail of the spectral density because, due to ensemble-ensemble fluctuations, it may occur that a few of the largest eigenvalues of one disorder realization are larger than the largest of another disorder realization. In order to illustrate explicitly that the discrepancy in the tail of the distribution is an artifact of the combination of a soft edge and ensemble average, we compare the largest eigenvalue of $W$ averaged over many disorder realizations with the $Q$-Laguerre prediction. Results depicted in Fig.~\ref{fig:M1Edge} show a very good agreement between this quantity and the analytical prediction, Eq.~(\ref{eq:E0_QLaguerre}), for $16\leq N\leq 30$.

\begin{figure}[tbp]
	\centering
	\includegraphics[width=0.5\textwidth]{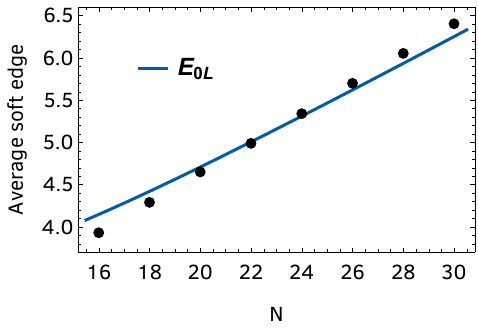}
	\caption{Spectral edge of the circular WSYK model~(\ref{eq:def_WSYK}) with $M=1$ and $\qb=2$ ($q=4$) as a function of the number of Majoranas $N$. The black dots correspond to the ensemble-averaged largest eigenvalue of $W$ obtained from exact diagonalization ($8192$ disorder realizations for $N=16$--$24$, $2048$ realizations for $N=26$, $28$, and $1472$ realizations for $N=30$), while the full line gives the $Q$-Laguerre prediction~(\ref{eq:E0_QLaguerre}) with $Q=t_3(2,N)$.}
	\label{fig:M1Edge}
\end{figure}

\begin{figure}[tbp]
	\centering
	\includegraphics[width=\textwidth]{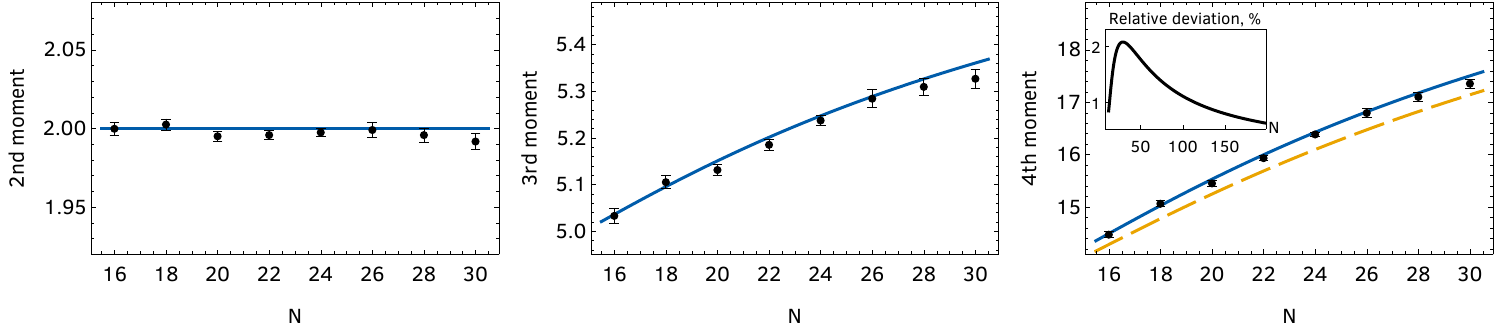}
	\caption{Low-order (normalized) moments of the circular ($k=1)$ WSYK model~(\ref{eq:def_WSYK}) with $M=1$ and $\qb=2$ ($q=4$) as a function of the number of Majoranas $N$. The black dots are obtained from ensemble-averaged exact diagonalization ($8192$ disorder realizations for $N=16$--$24$, $2048$ realizations for $N=26$, $28$, and $1472$ realizations for $N=30$), while the full blue lines gives the exact analytic expressions, Eqs.~(\ref{eq:W_moment2})--(\ref{eq:W_moment4}). For the fourth moment, the $Q$-Laguerre approximation is also plotted (dashed orange line), while the inset shows the relative deviation between the exact analytic prediction and the $Q$-Laguerre approximation. The relative deviation is maximal for the values of $N$ currently accessible, yet it is only around $2\%$.}
	\label{fig:M1Moments}
\end{figure}

We get further numerical evidence corroborating the applicability of our analytic results to systems with fixed $\qb=2$ by evaluating some low-order moments. In Fig.~\ref{fig:M1Moments}, we plot the second, third, and fourth moments as a function of the number of Majoranas $N$. As expected, there is excellent agreement between the moments obtained from exact diagonalization and the exact moments, Eqs.~(\ref{eq:W_moment2})--(\ref{eq:W_moment4}). The fourth moment is the first where there are permutation diagrams with more than one crossing and we can, therefore, compare the $Q$-Laguerre approximation against the exact result. As seen in the inset, the relative deviation between the two expressions is always below $2\%$, a value that is attained for the small values of $N$ considered here, and then decreases as $N$ increases. 

In summary, we observe a good agreement between the $Q$-Laguerre density and moments with numerical results obtained by exact diagonalization of $W$. The approximation of uncorrelated commutations and the replacement of perfect-matching crossings by permutation crossings are, therefore, well justified, at least for $\qb = 3,4$ and the values of $N$ that are numerically accessible. We again emphasize that the good agreement between numerical and analytic predictions relies crucially on the value of $Q=t_3(\qb,N)$, which could not have been obtained starting from the $Q$-Hermite (perfect-matching) combinatorics.

\section{Ansatz for the spectral density of the circular WSYK model with \texorpdfstring{$M>1$}{M>1}}
\label{app:M>1}

We now turn to the the circular WSYK model with $M>1$. Proceeding as before, we will first give the leading-order spectral density in the strict limit $N\to\infty$ with fixed $\qb$. Contrary to the $M=1$ case, we were not able to find a density that is exact to next-to-leading order, as we did not find the $Q$-orthogonal polynomials whose moments exactly reproduce the moments of our model. Nevertheless, we propose an ansatz in terms of generalized Al-Salam-Chihara $Q$-Laguerre polynomials that describes the numerical results for large $M$ (i.e., scaling with $N$) to very high accuracy.

The leading-order moments (ignoring commutations) are obtained by setting, in Eq.~(\ref{eq:moments_W_sum_permutations}), $t(\sigma)=1$ for all $\sigma\in\mathcal{S}_p$,
\begin{equation}\label{eq:M>1_leading_order}
\frac{1}{\sigmaL^p}\frac{\av{\Tr W^p}}{\Tr \id}
=\sum_{\sigma\in\mathcal{S}_p} M^{\mathrm{cyc}(\sigma)}=\sum_{k=1}^p c(p,k)M^k,
\end{equation}
where we recall that $\mathrm{cyc}(\sigma)$ is the number of cycles in the permutation $\sigma$ (number of closed loops in the respective diagram) and we rewrote the sum in terms of the number of permutations of $p$ elements with $k$ cycles, which is known as the unsigned Stirling number of the first kind, $c(p,k)$.\footnote{The unsigned Stirling numbers of the first kind are tabulated as sequence A130534 in the OEIS~\cite{OEIS_A130534}.} 
Now, it is known that $c(p,k)$ are also the coefficients of the polynomial $M(M+1)(M+2)\cdots(M+p-1)=(M+p-1)!/(M-1)!$, expanded in increasing powers of $M$, i.e., $M(M+1)(M+2)\cdots(M+p-1)=\sum_{k=1}^p c(p,k)M^k$. Thus, we conclude that the traces of $W$ are given by
\begin{equation}
\frac{1}{\sigmaL^p}\frac{\av{\Tr W^p}}{\Tr \id}
=\frac{(M+p-1)!}{(M-1)!}
\end{equation}
and are immediately identified as the moments of a $\chi^2$ distribution with $2M$ degrees of freedom. To leading order in the limit $N\to\infty$ we thus find the spectral density of the $M>1$ circular WSYK model to be given by $\varrho(E)=E^{M-1}\exp(-E)/(M-1)!$.

Let us now address the next-to-leading-order, where we consider only uncorrelated commutations.
In Sec.~\ref{sec:spectral_density_WSYK}, we saw that the relevant orthogonal polynomials for the $M=1$ circular WSYK model are the Al-Salam-Chihara $Q$-Laguerre polynomials. Furthermore, in RMT, the relevant (classical) orthogonal polynomials for the multichannel Wishart-Laguerre ensemble are the generalized Laguerre polynomials $L_n^{(\alpha)}(x)$. It is, therefore, natural to conjecture that the relevant orthogonal polynomials for the circular WSYK model with $M>1$ are the generalized Al-Salam-Chihara $(Q,y)$-Laguerre polynomials $L_{n}^{(\alpha)}(x;Q,y)$, recently introduced in Ref.~\cite{pan2020SLC}. These polynomials are orthonormal with respect to the weight function
\begin{equation}\label{eq:spectral_density_gen_QyLaguerre}
\begin{split}
\varrho_{\mathrm{QL}}^{(\alpha)}(E;Q,y)=
&\frac{(Q;Q)_\infty (Q^{\alpha+1};Q)_\infty}{(-Q^2/y;Q^2)_\infty(-Q^{2(\alpha+1)}y;Q^2)_\infty}
\frac{1-Q}{2\pi E}\sqrt{\(E_+-E\)\(E-E_-\)}\\
&\times\prod_{k=1}^\infty 
\frac{
	1-\frac{4v^2(E)}{(1+Q^k)(1+Q^{-k})}}{
	\(1-\frac{2v(E)}{Q^k/\sqrt{y}+Q^{-k}\sqrt{y}}\)
	\(1-\frac{2v(E)}{Q^{k+\alpha}\sqrt{y}+Q^{-k-\alpha}/\sqrt{y}}\)},
\end{split}
\end{equation}
supported on $E_-<E<E_+$, where $-1\leq Q\leq 1$ as before, $y\geq 1$, $\alpha$ is a positive integer, the left and right edges are
\begin{equation}\label{eq:E0_gen_QyLaguerre}
E_\pm=\frac{(\sqrt{y}\pm 1)^2}{1-Q},
\end{equation} 
and $v(E)$ is the recentered and rescaled energy,
\begin{equation}
v(E)=\frac{\bar{E}-E}{\Delta E/2},
\end{equation}
with 
\begin{equation}
\bar{E}=\frac{E_++E_-}{2}=\frac{y+1}{1-Q}
\quad\text{and}\quad
\Delta E=E_+-E_-=\frac{4\sqrt{y}}{1-Q}.
\end{equation}
If we set $y=1$ and $\alpha=0$, we recover the spectral density~(\ref{eq:spectral_density_QLaguerre}) with	 $E_+=\EL$ and $E_-=0$. The spectral density~(\ref{eq:spectral_density_gen_QyLaguerre}) is of the form of the multichannel Marchenko-Pastur distribution times a multiplicative correction. 
The combinatorial interpretation of its $p$th moment, $\mu_p$, was found in Ref.~\cite{pan2020SLC} to be
\begin{equation}
\mu_p=\sum_{\sigma\in\mathcal{S}_p} 
\beta^{\mathrm{rec}(\sigma)}
y^{\mathrm{wex}(\sigma)}
Q^{\mathrm{cross}({\sigma})},
\end{equation}
where $\beta=[\alpha+1]_Q:=(1-Q^{\alpha+1})/(1-Q)$ is the $Q$-analog of the integer $\alpha+1$, $\mathrm{rec}(\sigma)$ is the number of records in the permutation $\sigma$ (in terms of diagrams, the number of edges drawn above the dots with no other edges on top), and $\mathrm{wex}(\sigma)$ is its number of weak excedances (the total number of edges drawn above the dots). The lowest moments explicitly read as
\begin{align}
\label{eq:gen_QyLaguerre_mu1}
\mu_1&=\beta y,
\\
\label{eq:gen_QyLaguerre_mu2}
\mu_2&=\beta^2 y^2+\beta y,
\\
\label{eq:gen_QyLaguerre_mu3}
\mu_3&=\beta^3 y^2 +\beta^2 y^2\(2+Q\)+\beta y\(1+y\),
\\
\label{eq:gen_QyLaguerre_mu4}
\mu_4&=\beta^4 y^4 +\beta^3 y^3 \(3+2Q+Q^2\)+
\beta^2y^2\(\(3+2y\)\(1+Q\)+Q^2\)+
\beta y \(1+\(3+Q\)y+y^2\).
\end{align}

It remains to determine $Q$, $y$, and $\alpha$ in terms of the physical parameters $t_3$ and $M$ (under the approximation of uncorrelated commutations). As mentioned before, we were not able to \emph{derive} the correct parameters, but propose an ansatz that describes the numerical results for large $M$ to very high accuracy. We propose that
\begin{equation}\label{eq:ansatz_ybeta}
\alpha=M-1 \Leftrightarrow \beta=[M]_Q
\quad\text{and}\quad
y=\frac{M}{[M]_Q},
\end{equation}
while $Q$ should be the solution of the equation
\begin{equation}\label{eq:ansatz_Q}
Q=\frac{t_3}{[M]_Q}.
\end{equation}
Conversely, we can use Eqs.~(\ref{eq:ansatz_ybeta}) and (\ref{eq:ansatz_Q}) to specify the physical parameters $M$ and $t_3$ as
\begin{equation}\label{eq:ansatz_Mt3}
M=\beta y\quad\text{and}\quad
t_3=\beta Q.
\end{equation}

\begin{figure}[tbp]
	\centering
	\includegraphics[width=0.45\textwidth]{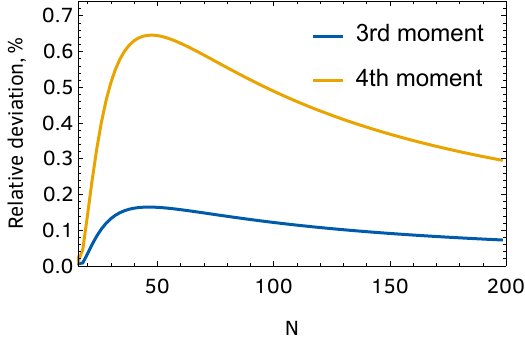}
	\caption{Relative deviation between the exact analytical prediction for the third and fourth moments of $W$, Eqs.~(\ref{eq:W_moment3}) and (\ref{eq:W_moment4}), and the generalized $(Q,y)$-Laguerre ansatz, Eqs.~(\ref{eq:gen_QyLaguerre_mu3}), (\ref{eq:gen_QyLaguerre_mu4}), (\ref{eq:ansatz_ybeta}), and (\ref{eq:ansatz_Q}), as a function of $N$, for $\qb=2$ ($q=4$) and $M=N$. The relative deviation is maximal (yet very small, below $1\%$) for the numerically accessible system sizes and decreases with increasing $N$.}
	\label{fig:MNRelativeDev}
\end{figure}

\begin{figure}[tbp]
	\centering
	\includegraphics[width=\textwidth]{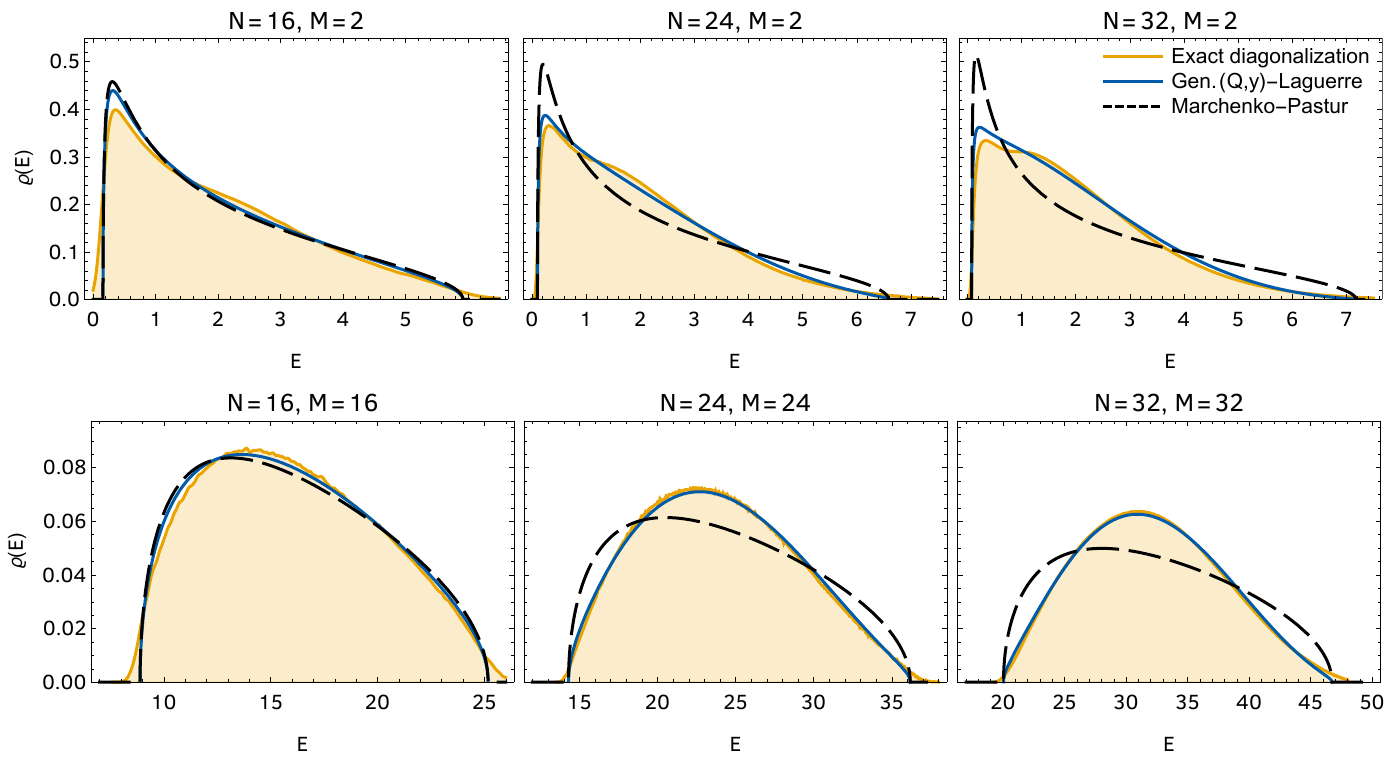}
	\caption{Spectral density of the circular WSYK model with $\qb=2$ ($q=4$) for small, $M=2$, and large, $M=N$, number of charges and three different number of Majoranas, $N=16$, $24$, and $32$. The orange (shaded) histograms are obtained from numerical exact diagonalization of the Hamiltonian~(\ref{eq:def_WSYK}). The blue (full) curves correspond to the generalized Al-Salam-Chihara $(Q,y)$-Laguerre weight [Eq.~(\ref{eq:spectral_density_gen_QyLaguerre})] with parameter ansatz~(\ref{eq:ansatz_ybeta}) and (\ref{eq:ansatz_Q}), while the black (dashed) curve is given by the multichannel Marchenko-Pastur distribution.}
	\label{fig:MNLaguerreDensity}
	\vspace{+2ex}
	\includegraphics[width=\textwidth]{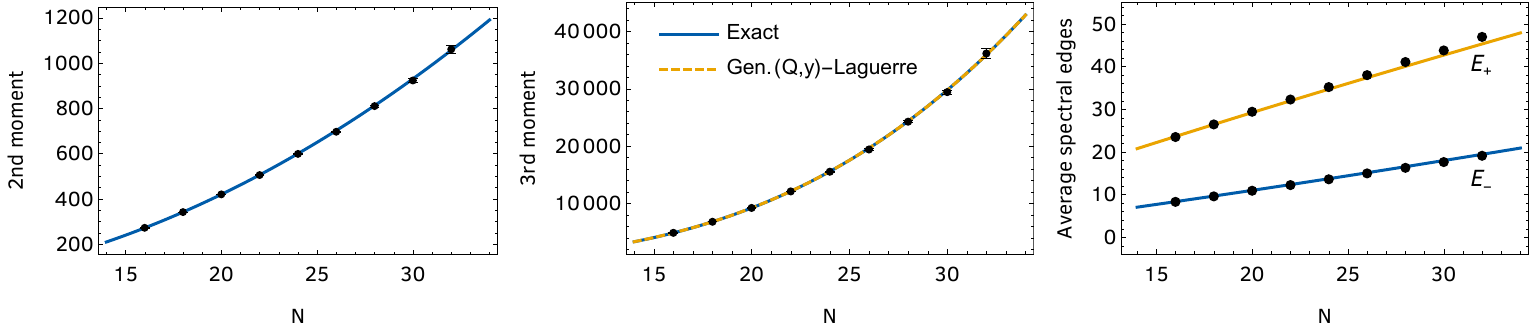}
	\caption{Low-order (normalized) moments and spectral edges of the circular WSYK model with $M=N$ and $\qb=2$ ($q=4$) as a function of the number of Majoranas $N$. The black dots are obtained form ensemble-averaged exact diagonalization ($2^{18}$ eigenvalues). For the moments (left and center panel) the blue line gives the exact analytic expression, Eqs.~(\ref{eq:W_moment2}) and (\ref{eq:W_moment3}), respectively. For the third moment, the generalized Al-Salam-Chihara $(Q,y)$-Laguerre prediction [Eqs.~(\ref{eq:gen_QyLaguerre_mu3}), (\ref{eq:ansatz_ybeta}), and (\ref{eq:ansatz_Q})] is also plotted (dashed orange line) and is seen to perfectly overlap with the exact result (and the numerics). For the spectral edge, the curves are given by the generalized Al-Salam-Chihara $(Q,y)$-Laguerre prediction, Eq.~(\ref{eq:E0_gen_QyLaguerre}).}
	\label{fig:MNMomentsEdge}
\end{figure}

This ansatz for $(Q,y,\alpha)$ has several desirable features: (i) it reproduces exactly the two lowest moments (with no crossings); (ii) in the random matrix limit ($Q\to0$), it recovers the $M$-channel Marchenko-Pastur distribution, since $y\to M$ and $\beta\to1$; (iii) in the limit $Q\to 1$, we have $\alpha\to M-1$ and $y\to 1$ and, because of the the combinatorial interpretation of the moments in this limit~\cite{hwang2020EJC},
\begin{equation}
\mu_p=\sum_{\sigma\in\mathcal{S}_p}y^{\mathrm{wex}(\sigma)}(\alpha+1)^{\mathrm{cyc}(\sigma)},
\end{equation}
$M$ counts the number of cycles in each permutation, in agreement with our leading-order result, Eq.~(\ref{eq:M>1_leading_order}); and (iv) when $M=1$, we find $Q\to t_3$, $y\to1$, and $\alpha\to0$, such that we recover the results of Sec.~\ref{sec:spectral_density_WSYK}. Note that our ansatz is not the only one with these properties. It is also important to note that the third and fourth moments of the weight of generalized $(Q,y)$-Laguerre polynomials, given by Eqs.~(\ref{eq:gen_QyLaguerre_mu3}) and (\ref{eq:gen_QyLaguerre_mu4}), \emph{do not match exactly} the (normalized) moments of $W$ obtained by approximating $t_4\approx t_3^2$ and inserting the ansatz~(\ref{eq:ansatz_Mt3}) into Eqs.~(\ref{eq:W_moment3}) and (\ref{eq:W_moment4}). However, in the case $M=N$, the two expressions are within around $1\%$ of each other and the relative deviation decreases with increasing $N$, see Fig.~\ref{fig:MNRelativeDev}. This result justifies, \emph{a posteriori}, the validity of the approximations and our ansatz, for large $M$. For smaller $M$, e.g., $M=2$, the relative deviation plateaus at a finite value for large $N$ and the value of this plateau in turn decreases as $M$ increases (for instance, for $M=5$ it is already below $2\%$). It is not clear to us, at this point, the reason why our ansatz is effective only for large $M$. Our results also indicate that large $M$ here means \emph{scaling with $N$}. Indeed, we see a decrease of the relative deviation with $N$ not only for $M=N$ but also, e.g., for $M=\sqrt{N}$. On the other hand, even for fixed $M=60$---which is larger than the $M=N$ for the numerically-accessible $N$---we see the plateau of nonvanishing relative deviation (although at a very small value).

Figure~\ref{fig:MNLaguerreDensity} shows the spectral density of the circular WSYK model with $\qb=2$ ($q=4$) and different $N$, for $M=2$ and $M=N$. As expected from the previous discussion, there is excellent quantitative agreement between the numerical exact diagonalization results and the generalized $(Q,y)$-Laguerre ansatz, for large $M=N$. (It also shows that, once again, the Marchenko-Pastur distribution does not give the correct density). For small and fixed $M=2$, while there are noticeable deviations, our ansatz still captures the main qualitative features of the spectral density (much better than standard RMT, in any case). Finally, to further highlight the very accurate description our ansatz gives of the numerical results for $M=N$, we plot, in Fig.~\ref{fig:MNMomentsEdge}, the second and third moments and the left and right (soft) spectral edges of $W$.

\section{Quantum chaos in the Wishart-Sachdev-Ye-Kitaev model}
\label{sec:qlaguerre_qchaos}

In this section, we go beyond the spectral density and consider the quantum chaos properties of the WSYK model. The standard SYK model realizes nine out of the ten Altland-Zirnbauer (AZ) classes, depending on $N$ and $q$. First, we show that the missing universality class, AIII or chGUE, can be realized in the $M=1$ circular WSYK model. Similarly, considering a non-Hermitian WSYK (nHWSYK) model, we can realize Bernard-LeClair (BL) class AIII$^{\dagger}$, the missing one from the non-pseudo-Hermitian tenfold way discussed in Ch.~\ref{chapter:classificationSYK}. Then, we confirm that both models exhibit RMT correlations characteristic of quantum chaotic behavior.

\subsection{Symmetry classification of the WSYK model and its non-Hermitian extension}
\label{sec:1SYK_nHWSYK}

As before, we consider the chiral representation of the $M=1$ WSYK model, in which it is represented by two off-diagonal blocks, where each block is a non-Hermitian SYK Hamiltonian (or, in our previous terminology, charge). By tuning $N$ and $\qb$, we obtain the additional symmetry classes.  
More specifically, we consider the Hamiltonian,
\begin{equation}\label{haminh}
\cH=\adiag(L,L'):=\begin{pmatrix}
& L \\
L' & 
\end{pmatrix},
\end{equation}
equivalent to the two-matrix model $W=L L'$, where $L$ and $L'$ are independent non-Hermitian SYK Hamiltonians (both with the same $N$ and $\qb$). Throughout, $\diag$ and $\adiag$ represent diagonal and antidiagonal block matrices, respectively. The model $\cH$ gives the non-Hermitian generalization of the WSYK defined in Sec.~\ref{sec:QLaguerre_defs}, with the latter satisfying $L'=L^\dagger$ to ensure Hermiticity. As mentioned before, this model has built-in chiral symmetry, which is implemented by the unitary operator $\Pi=\diag(\id,-\id)$ that anticommutes with the Hamiltonian (here, we follow the same nomenclature as in Ch.~\ref{chapter:classificationSYK}). Since the (nH)SYK model with odd $\qb$ already has a chiral symmetry and two symmetries of the same type only generate an additional commuting unitary symmetry, we consider only even $\qb$.

\paragraph*{Hermitian WSYK model.}
For all even $\qb$, if $N\mod8=2,6$, $L$ has no discrete symmetries (it belongs to BL class A), and the chiral symmetry is the only symmetry of $\cH$, which thus belongs to AZ class AIII. We note that, unlike the standard supersymmetric SYK model~\cite{fu2017PRD,kanazawa2017JHEP}, where the supercharge equivalent has an odd number of Majoranas, the WSYK does realize class AIII, thus completing the tenfold way in SYK physics.

If $N\mod8=0,4$, $L$ has a single antiunitary symmetry $\sP$, see Eq.~(\ref{eq:P_transform_H}), that acts on $L$ as $\sP L^\dagger \sP^{-1}=\epsilon L$ and satisfies $\sP^2=\eta$, with $\epsilon,\eta=\pm1$ depending on $N$ and $\qb$. This antiunitary symmetry passes down to $\cH$ as $\scA \cH \scA^{-1}=\epsilon \cH$, where $\scA=\adiag(\sP,\sP)$, satisfying $\scA^2=\eta$. Furthermore, the simultaneous existence of the antiunitary $\scA$ and the chiral symmetry $\Pi$ implies the existence of a second antiunitary symmetry $\scB=\Pi\scA=\adiag(\sP,-\sP)$, which acts on $\cH$ as $\scB\cH \scB^{-1}=-\epsilon \cH$ and squares to $\scB^2=-\eta$. We must consider four possibilities:
\begin{enumerate}
	\item $\qb\mod4=0$ and $N\mod8=0$. $\epsilon=\eta=1$ and $L$ belongs to BL class AI$^\dagger$ (see Ch.~\ref{chapter:classificationSYK}). $\scA$ is a time-reversal symmetry (TRS) of $\cH$ squaring to $+1$, while $\scB$ is a particle-hole symmetry (PHS) of $\cH$ squaring to $-1$. $\cH$ thus belongs to AZ class CI.
	\item $\qb\mod4=0$ and $N\mod8=4$. $\epsilon=1$, $\eta=-1$, and $L$ belongs to BL class AII$^\dagger$. $\scA$ is a TRS of $\cH$ squaring to $-1$, while $\scB$ is a PHS of $\cH$ squaring to $+1$. $\cH$ thus belongs to AZ class DIII.
	\item $\qb\mod4=2$ and $N\mod8=0$. $\epsilon=-1$, $\eta=+1$, and $L$ belongs to BL class D. $\scA$ is a PHS of $\cH$ squaring to $+1$, while $\scB$ is a TRS of $\cH$ squaring to $-1$. $\cH$ thus belongs to AZ class DIII.
	\item $\qb\mod4=2$ and $N\mod8=0$. $\epsilon=\eta=-1$ and $L$ belongs to BL class C. $\scA$ is a PHS of $\cH$ squaring to $-1$, while $\scB$ is a TRS of $\cH$ squaring to $+1$. $\cH$ thus belongs to AZ class CI.
\end{enumerate}
These two classes (CI and DIII) were already realized in the standard SYK model, but only for odd $\qb$.

The symmetry classification of the Hermitian WSYK model $\cH$ with non-Hermitian SYK blocks $L$ and $L'=L^\dagger$ is summarized in Table~\ref{tab:class_WSYK}.

\paragraph*{Non-Hermitian WSYK model.}
Exactly the same procedure can be followed for the non-Hermitian WSYK model, where $L$ and $L'$ are independent. 

For all even $\qb$, if $N\mod8=2,6$, $L$, $L'$ have no discrete symmetries (they belong to BL class A), and the chiral symmetry is the only symmetry of $\cH$, which thus belongs to BL class AIII$^\dagger$. This is the only non-Hermitian class without reality conditions (see Table~\ref{tab:correlations_sym_class_noreal}) not realized by the original nHSYK model~(\ref{eq:1SYK_hami}), thus completing the corresponding tenfold way.

If $N\mod8=0,4$, $L$ and $L'$ satisfy the same symmetries as $L$ did above. $\cH$ is now non-Hermitian, and $\scA$ and $\scB$ act on it as $\scA \cH^\dagger \scA^{-1}=\epsilon \cH$ and $\scB \cH^\dagger \scB^{-1}=-\epsilon \cH$, while still squaring to $\scA^2=\eta$ and $\scB^2=-\eta$. It is clear that $\scA$ and $\scB$ act as $\scC_\pm$ symmetries. The four cases are:
\begin{enumerate}
	\item $\qb\mod4=0$ and $N\mod8=0$. $\epsilon=\eta=1$ and $L$, $L'$ belong to BL class AI$^\dagger$. $\scA$ is a $\scC_+$ symmetry of $\cH$ squaring to $+1$, while $\scB$ is a $\scC_-$ symmetry of $\cH$ squaring to $-1$. $\cH$ thus belongs to BL class AI$^\dagger_-$.
	\item $\qb\mod4=0$ and $N\mod8=4$. $\epsilon=1$, $\eta=-1$, and $L$, $L'$ belong to BL class AII$^\dagger$. $\scA$ is a $\scC_+$ symmetry of $\cH$ squaring to $-1$, while $\scB$ is a $\scC_-$ symmetry of $\cH$ squaring to $+1$. $\cH$ thus belongs to AZ class AII$_-^\dagger$.
	\item $\qb\mod4=2$ and $N\mod8=0$. $\epsilon=-1$, $\eta=+1$, and $L$, $L'$ belong to BL class D. $\scA$ is a $\scC_-$ symmetry of $\cH$ squaring to $+1$, while $\scB$ is a $\scC_+$ symmetry of $\cH$ squaring to $-1$. $\cH$ thus belongs to AZ class AII$_-^\dagger$.
	\item $\qb\mod4=2$ and $N\mod8=0$. $\epsilon=\eta=-1$ and $L$, $L'$ belong to BL class C. $\scA$ is a $\scC_-$ symmetry of $\cH$ squaring to $-1$, while $\scB$ is a $\scC_+$ symmetry of $\cH$ squaring to $+1$. $\cH$ thus belongs to AZ class AI$_-^\dagger$.
\end{enumerate}
These two classes were already realized in the standard nHSYK model, but, again, only for odd $\qb$.

The symmetry classification of the non-Hermitian WSYK model $\cH$ with non-Hermitian SYK blocks $L$ and $L'$ is summarized in Table~\ref{tab:class_nHWSYK}.

\begin{table}[tb]
	\centering
	\caption{Symmetry classes realized in the WSYK model with non-Hermitian SYK blocks $L$ and $L'=L^\dagger$.}
	\label{tab:class_WSYK}
	\begin{tabular}{@{}lcccc@{}}
		\toprule
		$N\,\mathrm{mod}\,8$   & 0    & 2    & 4    & 6    \\ \midrule
		$q\,\mathrm{mod}\,4=0$ & CI   & AIII & DIII & AIII \\
		$q\,\mathrm{mod}\,4=2$ & DIII & AIII & CI   & AIII \\ \bottomrule
	\end{tabular}
	\caption{Symmetry classes realized in the nHWSYK model with non-Hermitian SYK blocks $L$ and $L'$.}
	\label{tab:class_nHWSYK}
	\begin{tabular}{@{}lcccc@{}}
		\toprule
		$N\,\mathrm{mod}\,8$   & 0               & 2              & 4               & 6              \\ \midrule
		$q\,\mathrm{mod}\,4=0$ & AI$^\dagger_-$  & AIII$^\dagger$ & AII$^\dagger_-$ & AIII$^\dagger$ \\
		$q\,\mathrm{mod}\,4=2$ & AII$^\dagger_-$ & AIII$^\dagger$ & AI$^\dagger_-$  & AIII$^\dagger$ \\ \bottomrule
	\end{tabular}
\end{table}

\subsection{Random matrix correlations in the WSYK model}

As just discussed, the Bott periodicity observed in the standard SYK model~\cite{you2017PRB} is also applicable here but for the chiral and superconducting random matrix ensembles~\cite{verbaarschot1994PRL,verbaarschot1993PRL,altland1997}. We now check the local bulk and hard-edge correlations of the WSYK and nHWSYK models, finding excellent agreement with the RMT predictions, thus confirming the symmetry classification of the previous section, as well as the quantum chaotic nature of the models.

To study the correlations of the Hermitian WSYK model, we employ a combination of the distribution of consecutive spacing ratios $r$ (bulk local level statistics), and the distribution of the eigenvalue closest to the origin $E_1$ (hard-edge local level correlations).
Out of the seven AZ classes with spectral mirror symmetry, the three realized in the WSYK model (AIII, CI, and DIII) share the same distribution of $E_1$~\cite{sun2020PRL}, which has been computed exactly for class AIII (chGUE)~\cite{wilke1998PRD,nishigaki1998PRD,damgaard2001PRD}, and was given in Eq.~(\ref{eq:emin_CI}):
\begin{equation}
\label{eq:wishart_P1(E1)}
P_\mathrm{AIII}(x_1)=P_\mathrm{CI}(x_1)=P_\mathrm{DIII}(x_1)=
\frac{\pi}{2}x_1e^{-\pi x_1^2/4},
\end{equation}
where, $x_1=E_1/\av{E_1}$ is normalized to unit mean. 
In Fig.~\ref{fig:Thesis_nHWSYK_qc} (upper row), we compute the distribution $P(x_1)$ for the WSYK model for $N=20$, $22$, and $24$. In all cases, we find a perfect agreement with Eq.~(\ref{eq:wishart_P1(E1)}). 

To distinguish the three symmetry classes, we computed the distribution $P(r)$, for which, as stated in Eq.~(\ref{eq:surmise}), the RMT prediction is 
\begin{equation}
\label{eq:surmise_Wishart}
P(r)=\frac{2}{Z_\beta}\frac{(r+r^2)^\beta}{(1+r+r^2)^{3\beta/2}},
\end{equation}
with $\beta=1$ and $Z_1=8/27$ for the class CI (since it has the same level statistics as the GOE), $\beta=2$ and $Z_2=4\pi/81\sqrt{3}$ for class AIII (same statistics as the GUE), and $\beta=4$ and $Z_4=4\pi/729\sqrt{3}$ for class DIII (same statistics as the GSE). The results of Fig.~\ref{fig:Thesis_nHWSYK_qc} (bottom row) confirm these predictions.

\begin{figure}[tp]
	\centering
	\includegraphics[width=\textwidth]{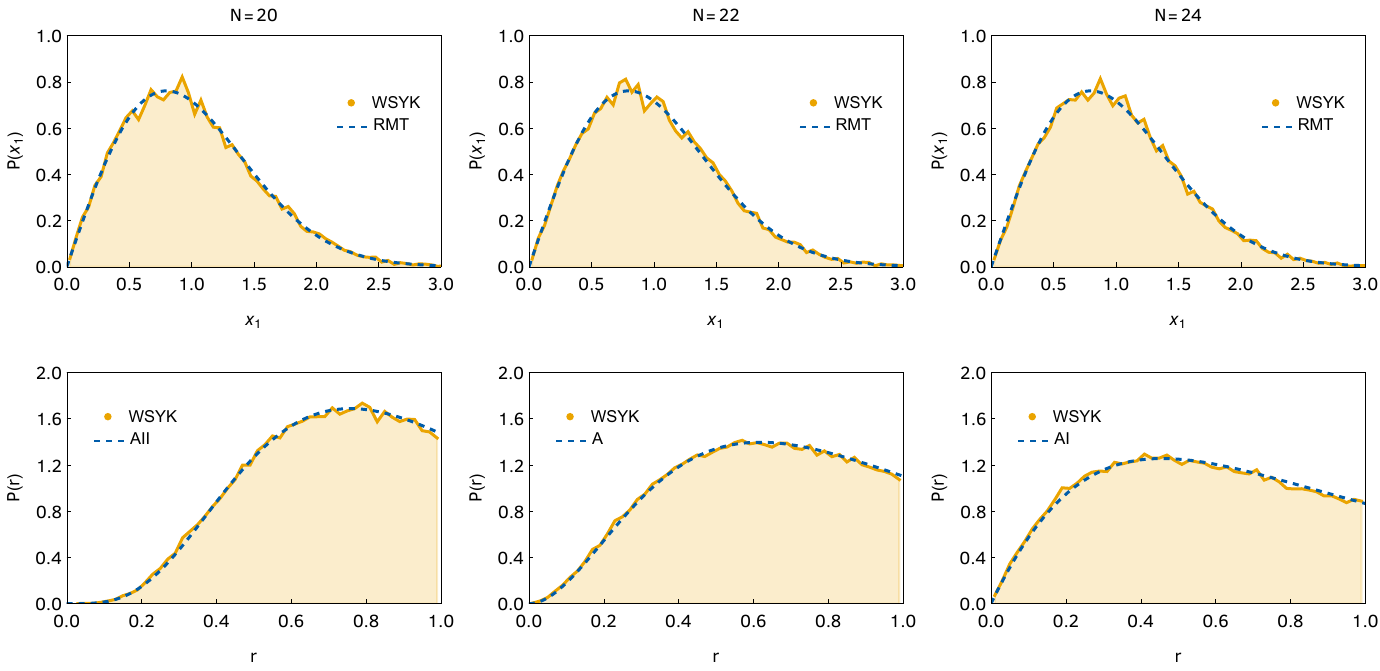}
	\caption{Local spectral correlations of the Hermitian WSYK model~(\ref{haminh}), with $L'=L^\dagger$, for $\qb=4$ and $N=20$, $22$, and $24$ (from left to right), near the hard edge (top) and in the bulk (bottom). Top row: Distribution of the eigenvalues with the smallest absolute value normalized to unit mean (filled histogram), compared with the RMT predictions for the AZ classes DIII ($N=20$), AIII ($N=22$), and CI ($N=24$) (dashed curves). The WSYK results are obtained from ensembles with $2^{14}$ realizations. The RMT predictions coincide for the three classes and are given by Eq.~(\ref{eq:wishart_P1(E1)}). Bottom row: Spacing ratio distribution (filled histogram), compared with the RMT predictions for classes AII ($N=20$), A ($N=22$), and AI ($N=24$) (dashed curves). For the nHWSYK results, we obtained $2^{18}$ eigenvalues. The RMT predictions are given by the Wigner-like surmise of Eq.~(\ref{eq:surmise_Wishart}). We find excellent agreement with the predictions in all cases, confirming the classification of Table~\ref{tab:class_WSYK}.}
	\label{fig:Thesis_WSYK_qc}
\end{figure}

Turning now to the non-Hermitian WSYK model, we consider the analogous quantities for non-Hermitian matrices, namely, the distribution of complex spacing ratios and that of the eigenvalue with the smallest absolute value. In the bulk, class AIII$^\dagger$ has the same statistics than universality class A, class AI$^\dagger_-$ those of class AI$^\dagger$, and class AII$^\dagger_-$ those of class AII$^\dagger$. The distribution of $|E_1|$ is unique to each of the three classes and was computed analytically for class AIII$^\dagger$ in Eq.~(\ref{eq:AIIId_P1_lambda_1}) and numerically for classes AI$^\dagger_-$ and AII$^\dagger_-$ in Ch.~\ref{chapter:correlations}. The results of Fig.~\ref{fig:Thesis_nHWSYK_qc} show very good agreement with the RMT predictions, thus confirming our symmetry classification and the quantum chaotic nature of the nHWSYK model.

\begin{figure}[tp]
	\centering
	\includegraphics[width=\textwidth]{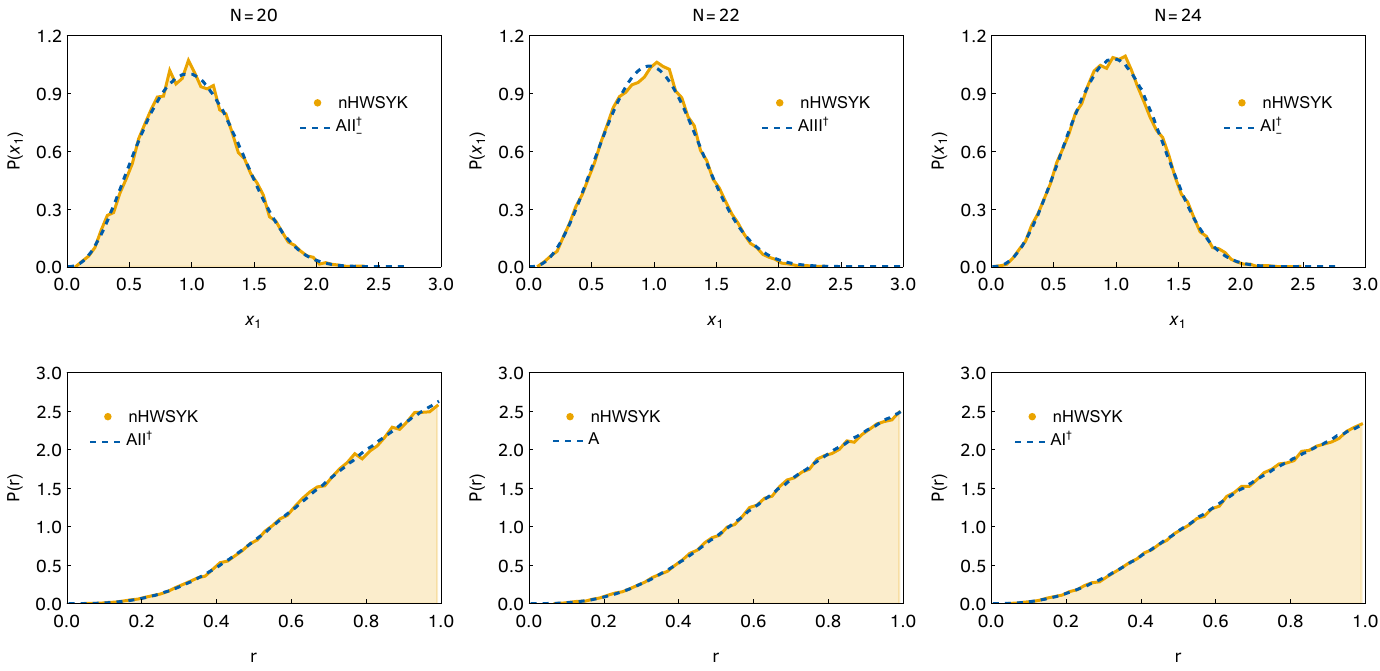}
	\caption{Local spectral correlations of the non-Hermitian WSYK model~(\ref{haminh}), for $\qb=4$ and $N=20$, $22$, and $24$ (from left to right), near the hard edge (top) and in the bulk (bottom). Top row: Distribution of the eigenvalues with the smallest absolute value normalized to unit mean (filled histogram), compared with the RMT predictions for the non-Hermitian classes AII$^\dagger_-$ ($N=20$), AIII$^\dagger$ ($N=22$), and AI$^\dagger_-$ ($N=24$) (dashed curves). The nHWSYK results are obtained from ensembles with $2^{14}$ realizations. The RMT predictions for AI$^\dagger_-$ and AII$\dagger_-$ are obtained by exactly diagonalizing $10^7$ random matrices of dimension $100$ structured according to the fourth column of Table~\ref{tab:correlations_sym_class_noreal}. The RMT prediction of class AIII$^\dagger$ is the analytical surmise of Eq.~(\ref{eq:AIIId_P1_lambda_1}). The comparison is parameter-free and does not involve any fitting. Bottom row: Marginal radial distribution of complex spacing ratios (filled histogram), compared with the RMT predictions for classes AII$^\dagger$ ($N=20$), A ($N=22$), and AI$^\dagger$ ($N=24$) (dashed curves). For the nHWSYK results, we obtained $2^{18}$ eigenvalues. The RMT predictions are obtained by numerical exact diagonalization of $2^{15}\times2^{15}$ random matrices of the corresponding class averaging over an ensemble of $2^8$ realizations.  We find excellent agreement with the predictions in all cases, confirming the classification of Table~\ref{tab:class_nHWSYK}.}
	\label{fig:Thesis_nHWSYK_qc}
\end{figure}

\section{Summary and outlook}

In conclusion, we have investigated a Wishart extension of the SYK model consisting of the product of two Hermitian conjugate SYK matrices, each with complex random couplings. By using combinatorial and diagrammatic techniques, we have computed analytically the low-order moments of the spectral density and have found striking similarities with those of the weight function of Al-Salam-Chihara $Q$-Laguerre polynomials~\cite{alsalam1976,kasraoui2011AAM}, a type of $Q$-Laguerre polynomials~\cite{moak1981}. Based on these similarities, we have carried out a parameter-free comparison between the spectral density of the WSYK model computed numerically and this $Q$-Laguerre weight function where $Q(\hat q, N)$ has been computed analytically. For $\qb= 3,4$, and sufficiently large $N$, we have found good agreement between numerical and analytical results. 
For $\hat q = 4$, we have also reported that the short-range correlations of the model are in good agreement with the random matrix prediction, an indication of quantum chaotic dynamics. Depending on $N$, the universality class corresponds to that of superconducting or chiral random matrix ensembles. In particular, we found chGUE correlations for $\hat{q}=4$ and $N\mod 8=2,6$. This was the only remaining symmetry class of the tenfold-way classification whose level statistics had not been realized in the SYK model.

A distinctive feature of the WSYK is the existence of a microscopic spectral density close to the hard edge of the spectrum whose features are close to the random matrix prediction. However, especially for smaller $\qb$ and not too large $N$, we expect deviations from this universal result. We could not find them analytically but we believe that the combinatorial techniques are versatile enough to provide a viable approach to this problem. It would be necessary to identify a novel scaling limit in the evaluation of the moments that likely leads to some deformation of the universal random matrix results.

Another problem that deserves further attention is the study of elliptic WSYK models where the coupling $k$ of the imaginary part is $0 < k < 1$. Combinatorially, it involves a mix of perfect matchings and permutations that may be related to other flavors of $Q$-polynomials. More generally, it would be interesting to investigate whether there is a general relation between the weight of $Q$-polynomials and combinatorial problems related to the moments of random Hamiltonians of strongly interacting systems. 

This question is not restricted to Hermitian models. For instance, the spectral density of non-Hermitian SYK models is not known. For each class unveiled in Ch.~\ref{chapter:classificationSYK}, it would be interesting to identify the combinatorial problem, the relevant Touchard-Riordan expression, and the explicit form of the spectral density.

}

%% file: Thesis_Conclusions.tex

\chapter{Conclusions and outlook}
\label{chapter:conclusions}

The past five years saw a flurry of activity in dissipative quantum chaos, in which I was fortunate to participate actively. In this chapter, we reflect on the contributions of this thesis to the field and its future prospects.
Despite all the exciting progress, dissipative quantum chaos is still only in its infancy and much remains to be done, be it from the point of view of fundamental physics or technological applications.

\section{Summary of main results}

\subsection{Random matrix theory of Markovian dissipation}

In Part I of this thesis, we addressed the question \emph{How to model a generic dissipative quantum system?} The simplest complete description of such a system relies on the Markovian approximation, which neglects any memory effects of the environment, and leads to the so-called Lindblad equation. Embracing the quantum chaos perspective, we applied non-Hermitian random matrix theory to model these dynamics---both in continuous time, in Ch.~\ref{chapter:randomLindblad}, and in discrete time, in Ch.~\ref{chapter:kraus}. Most importantly, we established the universality of steady states of random dissipative systems.

While the results for fully random dynamics are quite encouraging, physically motivated models have few-body interactions, rendering them very different. The simplest example is that of open free fermions, which we also reviewed in Ch.~\ref{chapter:randomLindblad}. Remarkably, if the number of dissipation channels is large enough, we found the same universal properties as for random Lindbladians, pointing again towards a high degree of universality in dissipative quantum chaos. On the other hand, if the number of dissipation channels is below a critical value, the system can exhibit nonchaotic features and suppress dissipation. 

To model strongly-coupled Lindbladians, we instead initiated the study of an open version of the celebrated Sachdev-Ye-Kitaev (SYK) model. Despite its strong interactions and dissipation, we could still setup an dynamical mean-field theory that becomes exact in the thermodynamic limit. We found that the open SYK model exhibits dissipation-driven relaxation at strong dissipation (and we were able to compute the rate analytically), while at weak dissipation chaos takes over the dynamics, leading to anomalous relaxation with a finite rate even in the absence of an explicit coupling to the bath.
Additionally, the open SYK model establishes a fruitful connection between dissipative quantum matter and gravity~\cite{garcia2023PRD2}, which was not discussed in this thesis.

\subsection{Symmetry classes of dissipative quantum matter}

In Part II of the thesis, we then turned to the question \emph{What do generic quantum systems have in common?} 
The quantum chaos conjecture tells us that, if a system is chaotic, it displays the statistical correlations of a random matrix with the same symmetries. For this reason, symmetry classifications are powerful statistical tools offering universal information not accessible otherwise. 
In Ch.~\ref{chapter:classificationSYK}, we provided a symmetry classification of the non-Hermitian SYK model, which we introduced as paradigmatic model of many-body dissipative quantum chaos. We showed that the nHSYK model describes both generic features of the universal quantum ergodic state reached around the Heisenberg time and nonuniversal, but still rather generic, properties of quantum interacting systems in their approach to ergodicity.
Going beyond effective non-Hermitian Hamiltonians, we provided a general and abstract symmetry classification of the Markovian generators of dissipative quantum matter in Ch.~\ref{chapter:classificationLindblad}.
We found a tenfold classification without conserved quantities, which can be further enriched by unitary symmetries. Interestingly, this is the same number found in the celebrated tenfold classification of isolated systems. Remarkably, the physical constraints prevent the existence of classes with Kramers degeneracy. 

With these rich classifications in hand, we also studied the correlations and universality in dissipative quantum chaos, in Ch.~\ref{chapter:correlations}. We reviewed and extensively employed the increasingly popular complex spacing ratios, which serve as a clean, simple, and quantitative empirical detector of integrability and chaos. We also contributed to the study of correlations over longer energy scales by deepening our understanding of the dissipative spectral form factor. Finally, we proposed, proved, and used off-diagonal eigenvector overlaps as an empirical detector of antiunitary symmetries. Using these tools, we investigated the correlations in spin-chain Lindbladians and the non-Hermitian SYK models and confirmed that they are described by universal random matrix theory---the hallmark of quantum chaotic behavior.

\section{Outlook: the future of dissipative quantum chaos}

As we saw in this thesis, dissipative quantum chaos has seen a tremendous growth in recent years. Yet many directions remain unexplored, e.g.: how to use dissipation and strong interactions to induce transitions from order to chaos and prepare special states of matter? What are the best tools to distinguish dissipative order from chaos? And why does dissipative quantum chaos work in the first place?

First, while we have started to understand how chaotic dissipative systems approach equilibrium steady states, a natural question is how to engineer nonequilibrium steady states stabilized by dissipation in the presence of strong interactions and, possibly, transitions between them.
The open SYK model is the ideal platform for studying both universal quantum chaotic states (described by random matrix theory) and the nonuniversal, but still generic, approach to chaos (for instance, its dynamics feature a system-dependent correlation hole typical of strongly-interacting quantum chaotic systems). Our foregoing work addressed only the simplest case of memoryless evolution towards a featureless (infinite-temperature) stationary state. It would be interesting to investigate a more general setting, where the bath has memory and the system can relax to a nonequilibrium steady state, characteristic of actual physical systems.

Second, we must understand how universal the properties we have found so far are. This will require more tools to detect and precisely quantify chaos in open systems.
A grand goal of dissipative quantum chaos is to develop a more extensive toolbox of signatures that distinguish order from chaos and different universality classes in chaotic phases and to establish the relations between different diagnostics (the web of dissipative quantum chaos).

And third, we have extensively employed a working definition of quantum chaos as the presence of random-matrix signatures but a connection to classical dissipative physics, in the spirit of the Bohigas-Giannoni-Schmit conjecture, is still missing. That is, \textit{why} does the quantum chaos conjecture hold in the presence of dissipation? One possibility for addressing this issue is to consider signatures of chaos that depend directly on the dynamics, for instance, the Loschmidt echo or fidelity (overlap of a perturbed quantum trajectory with an unperturbed one), and apply semiclassical methods.

While theoretical in nature, we expect our work in dissipative quantum chaos to have an impact on experimental physics and technological applications, namely, the design of novel nonequilibrium quantum materials and using dissipation as a resource for quantum information processing. In the latter case, by both fighting and overcoming dissipation, e.g., through quantum error correction, or harnessing it for improved quantum simulations, it will impact the search for applications of current-term noisy intermediate-scale quantum computers and pave the way for improved future ones.

%% file: Thesis_App_KeldyshLindblad.tex

\chapter{Keldysh path integral for the Lindblad equation}
\label{app:KeldyshLindblad}

In this Appendix, we give two derivations of the Keldysh path integral for the Lindblad evolution, a direct formal one (Sec.~\ref{app:Lindbladian_path_integral}) and a microscopic one (Sec.~\ref{app:micro_derivation_Lindblad_Keldysh}). 

\section{Formal derivation of the Lindbladian path integral}
\label{app:Lindbladian_path_integral}

In this section, we briefly review the derivation of the Keldysh path integral for the unitary evolution of bosons, complex fermions, and Majorana fermions. We then follow the same procedure to directly obtain the Keldysh path integral from the Lindblad equation.

\subsection{The Keldysh path integral}

The Keldysh generating function is defined as~\cite{kamenevbook}
\begin{align}
Z_{t_{\mathrm{f}}} & 
=\Tr\left[\rho_{t_\mathrm{f}}\right]
=\Tr\left[U\left(0,t_{\mathrm{f}}\right)U\left(t_{\mathrm{f}},0\right)\rho_{0}\right],
\end{align}
where $\rho_0$ and $\rho_{t_\mathrm{f}}$ are the initial and final density matrices, $U\left(t,t'\right)=T_t\, e^{-\i\int_{t'}^{t}d\tau H(\tau)}$ for $t>t'$ ($T_t$ the time-ordering operator), and $U\left(t,t'\right)=U\left(t',t\right)^{\dagger}$ for $t<t'$. As it stands, $Z_{t_{\mathrm{f}}}=1$, however, we will assume that the forward, $U\left(t_{\mathrm{f}},0\right)$, and backward, $U\left(0,t_{\mathrm{f}}\right)$, propagation can, in principle, be different. In practice, we can consider different source terms in each branch and vary the generating function with respect to the sources to obtain correlation functions of observables. 

Using a set of complex or Grassmann variables, $\xi^\pm(t)$, for bosons or fermions propapagating forward ($+$) or backward ($-$) in time and the associated coherent states, $\ket{\xi^\pm_t}$, the partition function reads as~\cite{kamenevbook}
\begin{align}
Z_{t_{\mathrm{f}}} & =\int \sD\xi\ \tilde{\rho}_{0}
\exp\left\{\i\int_{0}^{t_{\mathrm{f}}}\d t\left[ \bar{\xi}^{+}(t)\,\i\pd_{t}\xi^{+}(t)-\bar{\xi}^{-}(t)\,\i\pd_{t}\xi^{-}(t)-H^{+}(t)+H^{-}(t)\right] \right\},
\end{align}
where 
\begin{align}\label{eq:SM_H_trot}
\tilde{\rho}_{0}=\frac{\bra{\xi_{0}^{+}}\rho_{0}\ket{\zeta\xi_{0}^{-}}}{\sqrt{\braket{\xi_{0}^{-}}{\xi_{0}^{-}}\braket{\xi_{0}^{+}}{\xi_{0}^{+}}}},
\quad 
H^{+}(t)=\frac{\bra{\xi_{t+\d t}^{+}}H\ket{\xi_{t}^{+}}}{\braket{\xi_{t+\d t}^{+}}{\xi_{t}^{+}}},
\quad \text{and} \quad H^{-}(t)=\frac{\bra{\xi_{t}^{-}}H\ket{\xi_{t+\d t}^{-}}}{\braket{\xi_{t}^{-}}{\xi_{t+\d t}^{-}}}.
\end{align}
with $\zeta=\pm1$ for bosons or fermions, and the continuum limit $\d t\to 0$ is understood.

Allowing for more than one bosonic/fermionic field $\xi^\pm_i(t)$ and defining the Keldysh contour of integration, $\sC=\sC^+\cup \sC^- $, with $\d z=\pm \d t$
and $\xi_{i}\left(z\right)=\xi_{i}^{\pm}\left(t\right)$ for $z\in\mathcal{C^{\pm}}$,
we get 
\begin{align}\label{eq:SM_Keldysh_path_integral}
Z_{t_{\mathrm{f}}}= & \int \sD\xi\ \tilde{\rho}_{0}\exp{\i\int_{\mathcal{C}}\d z\left[ \sum_i\bar{\xi}_{i}(z)\,\i\pd_{z}\xi_{i}(z)-H(z)\right] }.
\end{align}

\subsection{Majorana fermions}

Consider $2N$ Majorana operators, defined as
\begin{align}
\begin{split}
\chi_{2i}  &=\frac{1}{\sqrt{2}}\(c_{i}+c_{i}^{\dagger}\),\\
\chi_{2i-1} &=\frac{\i}{\sqrt{2}}\left(c_{i}-c_{i}^{\dagger}\right),
\end{split}
\end{align}
where $c^\dagger_i$ and $c_i$ are canonical creation and annihilation operators, that satisfy the Clifford algebra
\begin{align}
\left\{\chi_{i},\chi_{j}\right\}  & =\delta_{ij}.
\end{align}
We can then define the real Grassmann quantities $a_i$,
\begin{align}
\left(\begin{array}{c}
\xi_{i}\\
\bar{\xi}_{i}
\end{array}\right)=
& \frac{1}{\sqrt{2}}\left(\begin{array}{c}
a_{2i}-\i a_{2i+1}\\
a_{2i}+\i a_{2i+1}
\end{array}\right)=
\frac{1}{\sqrt{2}}\left(
\begin{array}{cc}
1 & -\i\\
1 & \i
\end{array}\right)
\left(\begin{array}{c}
a_{2i}\\
a_{2i+1}
\end{array}\right),
\end{align}
in terms of which the kinetic term of the action reads as
\begin{align}
\int_{\mathcal{C}}\d z\ \bar{\xi}_{i}(z)\,\i\pd_{z}\xi_{i}(z) & =\int_{\mathcal{C}}\d z\ \frac{1}{2}\left(\begin{array}{c}
\xi_{i}(z)\\
\bar{\xi}_{i}(z)
\end{array}\right)^{\dagger}\i\pd_{z}\left(\begin{array}{c}
\xi_{i}(z)\\
\bar{\xi}_{i}(z)
\end{array}\right)=\int_{\mathcal{C}}\d z\frac{1}{2}a_{i}(z)\,\i\pd_{z}a_{i}(z).
\end{align}
The Hamiltonian term is simply defined from the complex case, Eq.~(\ref{eq:SM_H_trot}).

\subsection{Lindbladian time evolution}

We proceed similarly to the standard Hamiltonian case. We start from the Lindblad equation 
\begin{align}
\pd_{t}\rho=\mathcal{\mathcal{L}}\left(\rho\right) 
& = -\i\left[H,\rho\right]
+\sum_{m}\(
2L_{m}\rho L_{m}^{\dagger}
-L_{m}^{\dagger}L_{m}\rho
-\rho L_{m}^{\dagger}L_{m}
\).
\end{align}
From its differential form,
\begin{align}
\rho(t+\d t) & =\rho(t)+\left(
-\i H\rho(t) + \i\rho(t)H
+\sum_{m}2L_{m}\rho(t)L_{m}^{\dagger}
-L_{m}^{\dagger}L_{m}\rho(t)
-\rho(t)L_{m}^{\dagger}L_{m}\right) \d t,
\end{align}
it is possible to write the matrix element $\bra{\xi_{t+\d t}^{+}}\rho(t+\d t)\ket{\xi_{t+\d t}^{-}}$
as a function of $\bra{\xi_{t}^{+}}\rho(t)\ket{\xi_{t}^{-}}$ using
the partition of the identity of coherent states  
\begin{equation}
\begin{split}
\bra{\xi_{t+\d t}^{+}}\rho(t+\d t)\ket{\xi_{t+\d t}^{-}} 
&=\int \sD\xi_{t}\bra{\xi_{t}^{+}}\rho(t)\ket{\xi_{t}^{-}}
\exp\left\{\left[
		\frac{\bar{\xi}_{t+\d t}^{+}-\bar{\xi}_{t}^{+}}{\d t}
	\xi_{t}^{+}
	+\bar{\xi}_{t}^{-}\frac{\xi_{t+\d t}^{-}-\xi_{t}^{-}}{\d t}
	\right.\right.
	\\
&	\left.\left.
	-\i H^{+}(t)+\i H^{-}(t)
	+\sum_{m}\(2L_{m}^{+}(t)\bar{L}_{m}^{-}(t)
	-\Gamma^{+}(t)-\Gamma^{-}(t)\)
	\right] \d t\right\},
\end{split}
\end{equation}
where we defined
\begin{equation}
\begin{split}
&L_{m}^{+}(t)=\frac{\bra{\xi_{t+\d t}^{+}}L_m\ket{\xi_{t}^{+}}}{\braket{\xi_{t+\d t}^{+}}{\xi_{t}^{+}}},
\quad
\bar{L}_{m}^{-}(t)=\frac{\bra{\xi_{t}^{-}}L_m^\dagger\ket{\xi_{t+\d t}^{-}}}{\braket{\xi_{t}^{-}}{\xi_{t+\d t}^{-}}},
\\
&\Gamma^+(t)=\frac{\bra{\xi_{t+\d t}^{+}}\sum_{m}L_{m}^{\dagger}L_{m}\ket{\xi_{t}^{+}}}{\braket{\xi_{t+\d t}^{+}}{\xi_{t}^{+}}},
\quad \text{and} \quad
\Gamma^-(t)=\frac{\bra{\xi_{t}^{-}}\sum_{m}L_{m}^{\dagger}L_{m}\ket{\xi_{t+\d t}^{-}}}{\braket{\xi_{t}^{-}}{\xi_{t+\d t}^{-}}}.
\end{split}
\end{equation}
The generating function is given by iterating the procedure above, expressing $\rho_{t_{\mathrm{f}}}$ in terms of $\rho_{0}$: 
\begin{align}
\begin{split}
Z_{t_{\mathrm{f}}}  =\Tr\left[\rho_{t_{\mathrm{f}}}\right]
&=
\int \sD\xi\ 
\frac{\bra{\xi_{0}^{+}}\rho_{0}\ket{\zeta\xi_{0}^{-}}}{\braket{\xi_{0}^{+}}{\xi_{0}^{-}}}
\exp\left\{
	\sum_{t}\left[ 
	\frac{\bar{\xi}_{t+\d t}^{+}-\bar{\xi}_{t}^{+}}{\d t}\xi_{t}^{+}
	+\bar{\xi}_{t}^{-}\frac{\xi_{t+\d t}^{-}-\xi_{t}^{-}}{\d t}
	\right.\right.
	\\
&	\left.\left.
	-\i H^{+}(t)+\i H^{-}(t)
	+\sum_{m}\(
	2L_{m}^{+}(t)\bar{L}_{m}^{-}(t) -\Gamma^{+}(t)-\Gamma^{-}(t)
	\) \right] \d t\right\}.
\end{split}
\end{align}
In the continuum limit, we obtain
\begin{equation}
\begin{split}
Z_{t_{\mathrm{f}}}= \int \sD\xi\ \tilde{\rho}_0\,
&\exp\left\{\i\int \d t\left[\bar{\xi}^{+}\,\i\pd_{t}\xi^{+}-\bar{\xi}^{-}\,\i\pd_{t}\xi^{-}
	\right.\right.
\\
&	\left.\left.
-\left(H^{+}(t)-\i\Gamma^{+}(t)\right)+\left(H^{-}(t)+\i\Gamma^{-}(t)\right)-2\i \sum_m L_{m}^{+}(t)\bar{L}_{m}^{-}(t)\right]\right\}.
\end{split}
\end{equation}
To recover the expression obtained from the microscopic theory (see Appendix \ref{app:micro_derivation_Lindblad_Keldysh}), we simply make the change of variables $\xi_{t}^{-}\to\zeta\xi_{t}^{-}$
\begin{equation}\label{eq:SM_Keldysh_Lindblad}
\begin{split}
Z_{t_{\mathrm{f}}}= \int \sD\xi\ \tilde{\rho}_0&\exp\left\{\i\int \d t\left[\bar{\xi}^{+}\,\i\pd_{t}\xi^{+}-\bar{\xi}^{-}\,\i\pd_{t}\xi^{-}
		\right.\right.
	\\
	&	\left.\left.
	-\left(H^{+}(t)-\i\Gamma^{+}(t)\right)+\left(H^{-}(t)+\i\Gamma^{-}(t)\right)-2\i\zeta' \sum_m L_{m}^{+}(t)\bar{L}_{m}^{-}(t)\right]\right\},
\end{split}
\end{equation}
where $\zeta'=\pm1$ depending on the bosonic or fermionic nature of the jump operators $L_m$. Note that $\zeta$ and $\zeta'$ may differ, since, e.g., for a system of fermions ($\zeta=-1$), the jump operators may be bosonic ($\zeta'=1$). This is the case for the quadratic jump operators of Ch.~\ref{chapter:SYKLindblad}.

\section{Microscopic derivation of the Lindbladian path integral}
\label{app:micro_derivation_Lindblad_Keldysh}

In this section, we provide an alternative derivation of the Lindbladian path integral, starting from a microscopic unitary theory for the system plus environment. Averaging over the system-reservoir coupling to one-loop allows us to treat both Markovian and non-Markovian time evolution in the Keldysh formalism. We then specialize to the Markovian case to recover the results for the Lindbladian path integral. This procedure also allows us to provide a microscopic derivation of the memory kernel $K(z,z')$ that is consistent with causality.

\subsection{System-reservoir coupling to one loop}
\label{app:pertubation_theory}

Our starting point is the Keldsyh action~(\ref{eq:SM_Keldysh_path_integral}) for the Hamiltonian time evolution. Assuming a system, S, in contact with a macroscopic reservoir,
R, we can partition the degrees of freedom into S and
R. We will further assume the Born approximation, i.e., the S--R coupling can be treated in a one-loop approximation. Assuming a coupling
of the form 
\begin{align}\label{eq:coupling_SR}
H_{\mathrm{SR}} & =\sum_{\mu}\hat{X}_{\mu}\hat{Y}_{\mu}^{\dagger},
\end{align}
where $\hat{X}_{\mu}$ and $\hat{Y}_{\mu}$ are operators in S and R, respectively, that $\av{\hat{Y}_{\mu}}=0$, and that at
the initial time S and R are uncorrelated, we obtain 
\begin{equation}\label{eq:SM_Keldysh_general}
\begin{split}
Z_{t_{\mathrm{f}}} & =\int \sD\xi\ \tilde{\rho}_{0}
\exp{\i\int_{\mathcal{C}}\d z\left\{ \bar{\xi}(z)\,\i\pd_{z}\xi(z)-H(z)\right\} -\int_{\mathcal{C}}\d z\d z'\sum_{\mu\mu'}\bar{X}_{\mu}(z)\av{T_{z}\hat{Y}_{\mu}(z)\hat{Y}_{\mu'}^{\dagger}(z')}X_{\mu'}(z')}\\
& =\int \sD\xi\ \tilde{\rho}_{0}
\exp{\i\int_{\mathcal{C}}\d z\left\{ \bar{\xi}(z)\,\i\pd_{z}\xi(z)-H(z)\right\} -\int_{\mathcal{C}}\d z\d z'\sum_{\mu\mu'}\bar{X}_{\mu}(z)\,\i\Omega_{\mu\mu'}\left(z,z'\right)X_{\mu'}(z')},
\end{split}
\end{equation}
where the integration is now done solely over the system's degrees of freedom,
\begin{align}\label{eq:Omega_def}
\Omega_{\mu\mu'}\left(z,z'\right) & =-\i\av{T_{z}\hat{Y}_{\mu}(z)\hat{Y}_{\mu'}^{\dagger}(z')}
\end{align}
are the contour-ordered correlation functions of the environment, and
\begin{align}
X_{\mu}(z)= & \frac{\bra{\xi_{z+\d z}}\hat{X}_{\mu}\ket{\xi_{z}}}{\braket{\xi_{z+\d z}}{\xi_{z}}},\\
\bar{X}_{\mu}(z)= & \frac{\bra{\xi_{z+\d z}}\hat{X}_{\mu}^{\dagger}\ket{\xi_{z}}}{\braket{\xi_{z+\d z}}{\xi_{z}}}.
\end{align}
Equation~(\ref{eq:SM_Keldysh_general}) is obtained to one-loop order in the system-reservoir coupling. It becomes exact in the case of Hamiltonians quadratic in the creation and annihilation operators of the environment with linear system-environment couplings $\hat{Y}_{\mu}$. 
For the Hamiltonian to be Hermitian it is required that
$\sum_{\mu}\hat{X}_{\mu}\hat{Y}_{\mu}^{\dagger}=\zeta'\sum_{\mu}\hat{X}_{\mu}^{\dagger}\hat{Y}_{\mu}$, where $\zeta'=\pm1$ depending on the bosonic or fermionic nature of the operators $\hat{Y}_{\mu}$ (equivalently $\hat{X}_{\mu}$), i.e., the sum over $\mu$ has to run over all operators $\hat{X}_{\mu}$
and their conjugates $\hat{X}_{\mu}^{\dagger}$. It is convenient to define the notation $\hat{X}_{\mu}^{\dagger}=\hat{X}_{\bar{\mu}}=\hat{X}_{-\mu}$,
such that $\sum_{\mu}\hat{X}_{\mu}\hat{Y}_{\mu}^{\dagger}=\sum_{\mu>0}\left(\hat{X}_{\mu}\hat{Y}_{\mu}^{\dagger}+\zeta'\hat{X}_{\bar{\mu}}^{\dagger}\hat{Y}_{\bar{\mu}}\right)$.
Therefore we can also write
\begin{align}
\label{eq:SM_XOmegaX1}
\nonumber
\sum_{\mu\mu'}&\bar{X}_{\mu}(z)\,\Omega_{\mu\mu'}\left(z,z'\right)X_{\mu'}(z')\\
& =-\i\sum_{\mu\mu'}\bar{X}_{\mu}(z)\av{T_{z}\hat{Y}_{\mu}(z)\hat{Y}_{\mu'}^{\dagger}(z')}X_{\mu'}(z')\\
\label{eq:SM_XOmegaX2}
& =-\i\sum_{\mu\mu'} X_{\mu'}(z')\av{T_{z}\hat{Y}_{\mu'}^{\dagger}(z')\hat{Y}_{\mu}(z)}\bar{X}_{\mu}(z)=\sum_{\mu\mu'}X_{\mu}(z)\,\Omega_{\bar{\mu}\bar{\mu}'}\left(z,z'\right)\bar{X}_{\mu'}(z')\\
\label{eq:SM_XOmegaX3}
& =-\i\zeta' \sum_{\mu\mu'}X_{\mu}(z)\av{T_{z}\hat{Y}_{\mu}^{\dagger}(z)\hat{Y}_{\mu'}^{\dagger}(z')}X_{\mu'}(z')=\zeta'\sum_{\mu\mu'} X_{\mu}(z)\,\Omega_{\bar{\mu}\mu'}\left(z,z'\right)X_{\mu'}(z')\\
\label{eq:SM_XOmegaX4}
& =-\i\sum_{\mu\mu'}\bar{X}_{\mu}(z)\av{T_{z}\hat{Y}_{\mu}(z)\hat{Y}_{\mu'}}\bar{X}_{\mu'}(z')=\zeta'\sum_{\mu\mu'} \bar{X}_{\mu}(z)\,\Omega_{\mu\bar{\mu}'}\left(z,z'\right)\bar{X}_{\mu'}(z'),
\end{align}
where we defined 
\begin{align}
\Omega_{\mu\mu'}\left(z,z'\right) & =-\i\av{T_z\hat{Y}_{\mu}(z)\hat{Y}_{\mu'}^{\dagger}(z')},\\
\Omega_{\bar{\mu}\mu'}\left(z,z'\right) & =-\i\av{T_z\hat{Y}_{\mu}^{\dagger}(z)\hat{Y}_{\mu'}^{\dagger}(z')},\\
\Omega_{\mu\bar{\mu}'}\left(z,z'\right) & =-\i\av{T_z\hat{Y}_{\mu}(z)\hat{Y}_{\mu'}(z')},\\
\Omega_{\bar{\mu}\bar{\mu}'}\left(z,z'\right) & =-\i\av{T_z\hat{Y}_{\mu}^{\dagger}(z)\hat{Y}_{\mu'}(z')}.
\end{align}

In terms of real time, $\d z=\pm \d t$ for the forward and backward contour, respectively, the S-R coupling term reads as
\begin{align}
\begin{split}
-\i\int_{\mathcal{C}}\d z\d z'&\sum_{\mu\mu'} 
\bar{X}_{\mu}(z)\Omega_{\mu\mu'}\left(z,z'\right)X_{\mu'}(z')= 
-\i\int \d t\d t' \sum_{\mu\mu'}
\left[
\bar{X}_{\mu}^{+}(t)\,\Omega_{\mu\mu'}^{\rmT}\left(t,t'\right)X_{\mu'}^{+}(t')\right.\\
&\left.
+\bar{X}_{\mu}^{-}(t)\,\Omega_{\mu\mu'}^{\rmTb}\left(t,t'\right)X_{\mu'}^{-}(t')
-\bar{X}_{\mu}^{+}(t)\,\Omega_{\mu\mu'}^{<}\left(t,t'\right)X_{\mu'}^{-}(t')
-\bar{X}_{\mu}^{-}(t)\,\Omega_{\mu\mu'}^{>}\left(t,t'\right)X_{\mu'}^{+}(t')\right],
\end{split}
\end{align}
where we introduced the greater, lesser, time-ordered, and anti-time-ordered components of $\Omega(z,z')$:
\begin{align}
\label{eq:SM_Omega>}
\Omega_{\mu\mu'}^{>}\left(t,t'\right) & =-\i\av{\hat{Y}_{\mu}(t)\hat{Y}_{\mu'}^{\dagger}(t')},\\
\label{eq:SM_Omega<}
\Omega_{\mu\mu'}^{<}\left(t,t'\right) & =-\i\zeta' \av{\hat{Y}_{\mu'}^{\dagger}(t')\hat{Y}_{\mu}(t)},\\
\label{eq:SM_OmegaT}
\Omega_{\mu\mu'}^{\rmT}\left(t,t'\right) & =\Theta\left(t-t'\right)\Omega_{\mu\mu'}^{>}\left(t,t'\right)+\Theta\left(t'-t\right)\Omega_{\mu\mu'}^{<}\left(t,t'\right),\\
\label{eq:SM_OmegaTb}
\Omega_{\mu\mu'}^{\rmTb}\left(t,t'\right) & =\Theta\left(t'-t\right)\Omega_{\mu\mu'}^{>}\left(t,t'\right)+\Theta\left(t-t'\right)\Omega_{\mu\mu'}^{<}\left(t,t'\right).
\end{align}
We can rewrite the terms of the action that couple different branches of the Keldysh contour as
\begin{equation}\label{eq:SM_action_coupling_branches}
\begin{split}
&-\i\int \d t\d t'\sum_{\mu\mu'}\left[ -\bar{X}_{\mu}^{+}(t)\,\Omega_{\mu\mu'}^{<}\left(t,t'\right)X_{\mu'}^{-}(t')-\bar{X}_{\mu}^{-}(t)\,\Omega_{\mu\mu'}^{>}\left(t,t'\right)X_{\mu'}^{+}(t')\right]
\\
=&-\i\int \d t\d t'\sum_{\mu\mu'}\left[ -\bar{X}_{\mu}^{+}(t)\,\Omega_{\mu\mu'}^{<}\left(t,t'\right)X_{\mu'}^{-}(t')-\zeta'\bar{X}_{\mu}^{+}(t)\,\Omega_{\bar{\mu}'\bar{\mu}}^{>}\left(t',t\right)X_{\mu'}^{-}(t')\right] \\
=&\int \d t\d t'\sum_{\mu\mu'}\bar{X}_{\mu}^{+}(t)\left[2\i\Omega_{\mu\mu'}^{<}\left(t,t'\right)\right]X_{\mu'}^{-}(t'),
\end{split}
\end{equation}
where the last equality follows from the identity
\begin{align}
\Omega_{\bar{\mu}'\bar{\mu}}^{>}\left(t',t\right) & =-\i\av{\hat{Y}_{\mu}^{\dagger}(t)\hat{Y}_{\mu'}(t')}=\zeta'\Omega_{\mu\mu'}^{<}\left(t,t'\right),
\end{align}
while we also have
\begin{align}\label{eq:SM_hermiticity_Omega<}
\Omega_{\mu'\mu}^{<}\left(t',t\right)^{*} & =i\zeta'\av{\hat{Y}_{\mu'}^{\dagger}(t')\hat{Y}_{\mu}(t)}=-\Omega_{\mu\mu'}^{<}\left(t,t'\right).
\end{align}

We conclude that the Keldysh path integral for the (generally non-Markovian) dynamics of the reduced system is given by:
\begin{equation}\label{eq:SM_non_Markovian_action}
\begin{split}
Z_{t_{\mathrm{f}}} & =\int \sD\xi\ \tilde{\rho}_{0}
\exp\left\{\i\int \d t\left[\bar{\xi}^{+}(t)\,\i\pd_{t}\xi^{+}(t)-\bar{\xi}^{-}(t)\,\i\pd_{t}\xi^{-}(t)-H^+(t)+H^-(t)\right]\right.
\\
&-\int \d t\d t'\sum_{\mu\mu' } \( \bar{X}_{\mu}^{+}(t)\,\i\Omega_{\mu\mu'}^{\rmT}\left(t,t'\right)X_{\mu'}^{+}(t')+\bar{X}_{\mu}^{-}(t)\,\i\Omega_{\mu\mu'}^{\rmTb}\left(t,t'\right)X_{\mu'}^{-}(t')\right.
\\&\left.\left.
-\bar{X}_{\mu}^{+}(t)\,2\i\Omega_{\mu\mu'}^{<}\left(t,t'\right)X_{\mu'}^{-}(t')\) \right\}.
\end{split}
\end{equation}

\subsection{Markovian approximation}

Assuming that the time-scales of the system are much smaller than those of the reservoir, we can approximate the correlation functions by an equal-time correlation function:
\begin{align}
\label{eq:SM_Markovian_Omega<}
\i\Omega_{\mu\mu'}^{<}\left(t,t'\right) & \simeq\zeta'\delta\left(t-t'\right)M_{\mu\mu'}\left(t\right),
\\
\label{eq:SM_Markovian_Omega>}
\i\Omega_{\mu\mu'}^{>}\left(t,t'\right) & =\i\zeta'\Omega_{\bar{\mu}'\bar{\mu}}^{<}\left(t',t\right)\simeq\delta\left(t-t'\right)M_{\bar{\mu}'\bar{\mu}}\left(t\right).
\end{align}
By Hermiticity of $\i\Omega_{\mu\mu'}^{<}\left(t,t'\right)$, Eq.~(\ref{eq:SM_hermiticity_Omega<}), we have
that $M^{\dagger}=M$. Moreover, the positive-definiteness of $\sum_m L_m^\dagger L_m$ implies that $M$ is also positive-definite. (This can easily be seen in the time-independent case as follows. By inverting Eq.~(\ref{eq:SM_Markovian_Omega<}), we can write
\begin{align}
M_{\mu\mu'} = \int \d t \d t' \ \i\Omega_{\mu\mu'}^{<}\left(t,t'\right)
= \av{
	\int \d t'\, \hat{Y}_{t'\mu'}^{\dagger}
	\int \d t\, \hat{Y}_{t\mu}
}.
\end{align}
The last term is the average of a positive-definite operator and therefore is non-negative.)
We can then decompose
\begin{equation}\label{eq:SM_M_def} M_{\mu\mu'}\left(t\right)=\sum_{m}v_{\mu}^{m}(t)\lambda_{m}v_{\mu'}^{m*}(t),
\end{equation}
with $\lambda_{m}\ge0$. For the term in the action coupling different branches, Eq.~(\ref{eq:SM_action_coupling_branches}), this implies 
\begin{equation}
\begin{split}
\int \d t \d t' \sum_{\mu\mu' }
\bar{X}_{\mu}^{+}(t)\left[2\i\Omega_{\mu\mu'}^{<}\left(t,t'\right)\right]X_{\mu'}^{-}(t') 
& =\int \d t\sum_{\mu\mu'}\bar{X}_{\mu}^{+}(t)\left[2\zeta' M_{\mu\mu'}\left(t\right)\right]X_{\mu'}^{-}(t)\\
& =\int \d t\sum_{m}2\zeta'\left[\bar{X}_{\mu}^{+}(t)v_{\mu}^{m}(t)\sqrt{\lambda_{m}}\right]\left[\sqrt{\lambda_{m}}v_{\mu'}^{m*}(t)X_{\mu'}^{-}(t)\right]\\
& =\int \d t\ \sum_{m}2\zeta' L_{m}^{+}(t)\bar{L}_{m}^{-}(t),
\end{split}
\end{equation}
where we defined the jump operators
\begin{align}\label{eq:SM_jump_op_def}
L_{m}(t) & =\sum_\mu\hat{X}_{\mu}^{\dagger}v_{\mu}^{m}(t)\sqrt{\lambda_{m}}
\end{align}
and their contour representation
\begin{align}
\label{eq:SM_jump_op_contour+}
L_{m}^{+}(t) & =\sum_\mu \bar{X}_{\mu}^{+}(t)v_{\mu}^{m}(t)\sqrt{\lambda_{m}}=\frac{\bra{\xi_{t+\d t}^{+}}\sum_\mu\hat{X}_{\mu}^{\dagger}v_{\mu}^{m}(t)\sqrt{\lambda_{m}}\ket{\xi_{t}^{+}}}{\braket{\xi_{t+\d t}^{+}}{\xi_{t}^{+}}},\\
\label{eq:SM_jump_op_contour-}
\bar{L}_{m}^{-}(t) & =   \sum_\mu \sqrt{\lambda_{m}}v_{\mu'}^{m*}(t)X_{\mu'}^{-}(t)=\frac{\bra{\xi_{t}^{-}}\sum_\mu\sqrt{\lambda_m}v_{\mu'}^{m*}(t)\hat{X}_{\mu'}\ket{\xi_{t+\d t}^{-}}}{\braket{\xi_{t}^{-}}{\xi_{t+\d t}^{-}}}.
\end{align}

The terms acting within a single branch of the contour are much more delicate to deal with because of time-ordering.
We use the intuition from what we expect for the Lindblad case to postulate the following equal-time limits
\begin{align}
\label{eq:SM_limit1}
\lim_{t\to t'}\bar{X}_{\mu}^{+}(t)X_{\mu'}^{+}(t')\,\Theta\left(t-t'\right) & =\frac{\bra{\xi_{t+\d t}^{+}}\hat{X}_{\mu}^{\dagger}\hat{X}_{\mu'}\ket{\xi_{t}^{+}}}{\braket{\xi_{t+\d t}^{+}}{\xi_{t}^{+}}}\,\Theta\left(t-t'\right),
\\
\label{eq:SM_limit2}
\lim_{t\to t'}\bar{X}_{\mu}^{+}(t)X_{\mu'}^{+}(t')\,\Theta\left(t'-t\right) & =\zeta'\frac{\bra{\xi_{t+\d t}^{+}}\hat{X}_{\mu'}\hat{X}_{\mu}^{\dagger}\ket{\xi_{t}^{+}}}{\braket{\xi_{t+\d t}^{+}}{\xi_{t}^{+}}}\,\Theta\left(t-t'\right),
\\
\label{eq:SM_limit3}
\lim_{t\to t'}\bar{X}_{\mu}^{-}(t)X_{\mu'}^{-}(t')\,\Theta\left(t-t'\right) & =\zeta'\frac{\bra{\xi_{t}^{-}}\hat{X}_{\mu'}\hat{X}_{\mu}^{\dagger}\ket{\xi_{t+\d t}^{-}}}{\braket{\xi_{t}^{-}}{\xi_{t+\d t}^{-}}}\,\Theta\left(t-t'\right),
\\
\label{eq:SM_limit4}
\lim_{t\to t'}\bar{X}_{\mu}^{-}(t)X_{\mu'}^{-}(t')\,\Theta\left(t'-t\right) & =\frac{\bra{\xi_{t}^{-}}\hat{X}_{\mu}^{\dagger}\hat{X}_{\mu'}\ket{\xi_{t+\d t}^{-}}}{\braket{\xi_{t}^{-}}{\xi_{t+\d t}^{-}}}\,\Theta\left(t-t'\right).
\end{align}
For the forward branch, we obtain
\begin{equation}\label{eq:SM_forward_action_1}
\begin{split}
&\int \d t \d t'\sum_{\mu\mu' }
\bar{X}_{\mu}^{+}(t)\left[\i\Omega_{\mu\mu'}^{\rmT}\left(t,t'\right)\right]X_{\mu'}^{+}(t')
\\
=&\int \d t\d t'\sum_{\mu\mu' }\left\{ \bar{X}_{\mu}^{+}(t)\left[\i\Omega_{\mu\mu'}^{>}\left(t,t'\right)\Theta\left(t-t'\right)\right]X_{\mu'}^{+}(t')+\bar{X}_{\mu}^{+}(t)\left[\i\Omega_{\mu\mu'}^{<}\left(t,t'\right)\Theta\left(t'-t\right)\right]X_{\mu'}^{+}(t')\right\}
\\
=&\int \d t\d t'\sum_{\mu\mu' }\left\{ \frac{\bra{\xi_{t+\d t}^{+}}\hat{X}_{\mu}^{\dagger}\hat{X}_{\mu'}\ket{\xi_{t}^{+}}}{\braket{\xi_{t+\d t}^{+}}{\xi_{t}^{+}}}\left[\i\Omega_{\mu\mu'}^{>}\left(t,t'\right)\Theta\left(t-t'\right)\right]\right.
\\&+\left.
\zeta'\frac{\bra{\xi_{t+\d t}^{+}}\hat{X}_{\mu'}\hat{X}_{\mu}^{\dagger}\ket{\xi_{t}^{+}}}{\braket{\xi_{t+\d t}^{+}}{\xi_{t}^{+}}}\left[\i\Omega_{\mu\mu'}^{<}\left(t,t'\right)\Theta\left(t'-t\right)\right]\right\},
\end{split}
\end{equation}
where we used the definition of the time-ordered and anti-time-ordered correlation functions, Eqs.~(\ref{eq:SM_OmegaT}) and (\ref{eq:SM_OmegaTb}), and the limits of Eqs.~(\ref{eq:SM_limit1}) and (\ref{eq:SM_limit2}) to obtain the first and second equalities, respectively.
By renaming indices and recalling that $\hat{X}_{\bar{\mu}}=\hat{X}^\dagger_{\mu}$, the second term in the last line of Eq.~(\ref{eq:SM_forward_action_1}) can be rewritten as
\begin{equation}\label{eq:SM_forward_action_2}
\frac{\bra{\xi_{t+\d t}^{+}}\hat{X}_{\mu'}\hat{X}_{\mu}^{\dagger}\ket{\xi_{t}^{+}}}{\braket{\xi_{t+\d t}^{+}}{\xi_{t}^{+}}}\i\Omega_{\mu\mu'}^{<}\left(t,t'\right)
=
\frac{\bra{\xi_{t+\d t}^{+}}\hat{X}_{\mu}^{\dagger}\hat{X}_{\mu'}\ket{\xi_{t}^{+}}}{\braket{\xi_{t+\d t}^{+}}{\xi_{t}^{+}}}\i\Omega_{\bar{\mu}'\bar{\mu}}^{<}\left(t,t'\right).
\end{equation}
Plugging Eq.~(\ref{eq:SM_forward_action_2}) into Eq.~(\ref{eq:SM_forward_action_1}) and using the Markovian approximation of Eqs.~(\ref{eq:SM_Markovian_Omega<}) and (\ref{eq:SM_Markovian_Omega>}), we obtain
\begin{equation}\label{eq:SM_forward_action_3}
\int \d t \d t' \sum_{\mu\mu' }
\bar{X}_{\mu}^{+}(t)\left[\i\Omega_{\mu\mu'}^{\rmT}\left(t,t'\right)\right]X_{\mu'}^{+}(t')
=\int \d t\d t'\sum_{\mu\mu'}\frac{\bra{\xi_{t+\d t}^{+}}\hat{X}_{\mu}^\dagger\hat{X}_{\mu'}\ket{\xi_{t}^{+}}}{\braket{\xi_{t+\d t}^{+}}{\xi_{t}^{+}}}M_{\bar{\mu}'\bar{\mu}}\left(t\right)\delta\left(t-t'\right).
\end{equation}
We can further use Eq.~(\ref{eq:SM_XOmegaX2}) to rewrite Eq.~(\ref{eq:SM_forward_action_3}) as
\begin{equation}\label{eq:SM_forward_action_4}
\int \d t \d t' \sum_{\mu\mu'}
\bar{X}_{\mu}^{+}(t)\,\i\Omega_{\mu\mu'}^{\rmT}\left(t,t'\right) X_{\mu'}^{+}(t')
=\int \d t\d t'\sum_{\mu\mu'}\frac{\bra{\xi_{t+\d t}^{+}}\hat{X}_{\mu}\hat{X}_{\mu'}^\dagger\ket{\xi_{t}^{+}}}{\braket{\xi_{t+\d t}^{+}}{\xi_{t}^{+}}}M_{\mu'\mu}\left(t\right)\delta\left(t-t'\right),
\end{equation}
and, finally, use Eqs.~(\ref{eq:SM_M_def}) and (\ref{eq:SM_jump_op_def}) to arrive at
\begin{equation}
\begin{split}
\int \d t \d t' \sum_{\mu\mu'}
&\bar{X}_{\mu}^{+}(t)\,\i\Omega_{\mu\mu'}^{\rmT}\left(t,t'\right)X_{\mu'}^{+}(t')
\\
=&\int \d t\ \frac{\bra{\xi_{t+\d t}^{+}}\sum_{m}\left(\sum_\mu\hat{X}_{\mu}v_{\mu}^{m*}(t)\sqrt{\lambda_{m}}\right)\left(\sum_{\mu'}\sqrt{\lambda_{m}}v_{\mu'}^{m}(t)\hat{X}_{\mu'}^{\dagger}\right)\ket{\xi_{t}^{+}}}{\braket{\xi_{t+\d t}^{+}}{\xi_{t}^{+}}}
\\
=&\int \d t\ \frac{\bra{\xi_{t+\d t}^{+}}\sum_{m}L_{m}^{\dagger}(t)L_{m}(t)\ket{\xi_{t}^{+}}}{\braket{\xi_{t+\d t}^{+}}{\xi_{t}^{+}}}
=\int \d t\ \Gamma^{+}(t).
\end{split}
\end{equation}
We proceed similarly for the backward branch:
\begin{align}
\int \d t \d t'\sum_{\mu\mu'}\bar{X}_{\mu}^{-}(t)\,\i\Omega_{\mu\mu'}^{\rmTb}\left(t,t'\right)X_{\mu'}^{-}(t') & =\int \d t\ \frac{\bra{\xi_{t}^{-}}\sum_{m}L_{m}^{\dagger}(t)L_{m}(t)\ket{\xi_{t+\d t}^{-}}}{\braket{\xi_{t}^{-}}{\xi_{t+\d t}^{-}}}=\int \d t\ \Gamma^{-}(t).
\end{align}

Finally, the partition function in the Markovian limit becomes 
\begin{equation}\label{eq:SM_Z_micro_final}
\begin{split}
Z_{t_{\mathrm{f}}} & =\int \sD\xi\ \tilde{\rho}_{0}
\exp\left\{\i\int \d t\left[\bar{\xi}^{+}\,\i\pd_{t}\xi^{+}-\bar{\xi}^{-}\,\i\pd_{t}\xi^{-}
	\right.\right. \\ &\left.\left.
	-\left(H^{+}(t)-\i\Gamma^{+}(t)\right)+\left(H^{-}(t)+\i\Gamma^{-}(t)\right)-2\i\zeta'\sum_m L_{m}^{+}(t)\bar{L}_{m}^{-}(t)\right]\right\},
	\end{split}
	\end{equation}
which coincides with the result obtained directly from the Lindblad equation in Appendix~\ref{app:Lindbladian_path_integral}, Eq.~(\ref{eq:SM_Keldysh_Lindblad}).

\subsection{The memory kernel}
To make contact with the expression for the action written down in Ch.~\ref{chapter:SYKLindblad}, Eq.~(\ref{eq:C_Keldysh_action}), it remains to derive the explicit form of the memory kernel $K(z,z')$ from Eq.~(\ref{eq:SM_Z_micro_final}), namely,
\begin{align}
\label{eq:SM_K_kernel_T}
K^\rmT(t,t')&=K(t^+,t'^+)=\zeta'\dirac{t-t'},
\\
\label{eq:SM_K_kernel_aT}
K^\rmTb(t,t')&=K(t^-,t'^-)=\zeta'\dirac{t-t'},
\\
\label{eq:SM_K_kernel_<}
K^<(t,t')&=K(t^+,t'^-)=2\zeta'\dirac{t-t'},
\\
\label{eq:SM_K_kernel_>}
K^>(t,t')&=K(t^-,t'^+)=0.
\end{align}
Note there is a relative minus sign for the backward branch $\sC^-$, $\d z=\pm\d t$ for $z\in\sC^\pm$, that leads to all components of the kernels having the same sign.

To identify the correlation functions of the environment, $\Omega_{\mu\mu'}\left(z,z'\right)$, we have to relate the Markovian action (\ref{eq:SM_Keldysh_Lindblad}) with the non-Markovian one, Eq.~(\ref{eq:SM_Keldysh_general}), essentially running the procedure of the previous section backwards. However, such a procedure is not unique. The simplest choice is to identify the jump operators with the system operators $\hat{X}_{\mu}$ themselves. This amounts to restricting ourselves to microscopic kernels $\Omega_{\mu\mu'}$ without inter-channel coupling (no anomalous terms in the bath Hamiltonian), i.e., that satisfy 
\begin{equation}\label{eq:Omega_restricted}
\Omega_{\mu\mu'}(z,z')=
\delta_{\mu\mu'}\Omegatilde_\mu(z,z').
\end{equation}
More concretely, if we choose
\begin{align}
v_{\mu}^{m}(t)\sqrt{\lambda_{m}} & =\delta_{\mu m},
\end{align}
then Eq.~(\ref{eq:SM_jump_op_contour+}) gives
\begin{equation}\label{eq:SM_simple_jump_ops}
\begin{cases}
\bar{X}_{\mu=m}^{+}(t) =L_{m}^{+}(t), & \text{for\ }\mu>0,\\
\bar{X}_{\mu=-m}^{+}(t)=X_{\mu=m}^{+}(t)=\bar{L}_{m}^{+}(t), &\text{for\ }\mu<0,
\end{cases}
\end{equation}
and similarly for $L_{m}^-(t)$. In this way, we get
\begin{align}
M_{\mu\mu'} & =\sum_{m}v_{\mu}^{m}(t)\lambda_{m}v_{\mu'}^{m*}(t)=\delta_{\mu\mu'}\Theta_{\mu},
\end{align}
with
\begin{align}
\Theta_{\mu} & =\begin{cases}
1, & \mu>0\\
0, & \mu<0
\end{cases}.
\end{align}
Therefore,
\begin{align}
\label{eq:Omega_tilde_components<}
& \i \Omegatilde_\mu^<(t,t')=\zeta'\, f(t,t')\, \Theta_\mu,
\\
\label{eq:Omega_tilde_components>}
& \i \Omegatilde_\mu^>(t,t')=\zeta'\, \i \Omegatilde^<_{\bar{\mu}}(t',t)=f(t',t)\, \Theta_{\bar{\mu}}.
\end{align}
where for the moment we have allowed for a non-Markovian time kernel $f(t,t')$. (We note that the $f(t,t')$ must be independent of the channel index $\mu$ for us to be able to do the averages over $\Gamma_{ijkl}$ in the Lindbladian SYK model. At most, we can introduce a channel-dependent multiplicative constant that would weight the variance of each channel in $\Gamma_{ijkl}$.)

We now focus on the dissipative contribution to the general Keldysh action, Eq.~(\ref{eq:SM_Keldysh_general}),
\begin{equation}\label{eq:action_micro_X}
-\i \int_\sC \d z \d z'\sum_{\mu\mu'} \conj{X}_\mu(z)\Omega_{\mu\mu'}(z,z')X_{\mu'}(z').
\end{equation}
We expand the sum over (negative and positive) $\mu$ in Eq.~(\ref{eq:action_micro_X}) as sums over (positive only) $m$:
\begin{equation}
\begin{split}
\sum_{\mu\mu'} \conj{X}_\mu(z)\Omega_{\mu\mu'}(z,z')X_{\mu'}(z')
=\sum_{m,m'=1}^M \( \conj{X}_m(z)\Omega_{mm'}(z,z')X_{m'}(z')
+\conj{X}_m(z)\Omega_{m,-m'}(z,z')X_{-m'}(z')
\right.\\
\left.+
\conj{X}_{-m}(z)\Omega_{-mm'}(z,z')X_{m'}(z') +\conj{X}_{-m}(z)\Omega_{-m,-m'}(z,z')X_{-m'}(z')
\).
\end{split}
\end{equation}
Because of the constraint (\ref{eq:Omega_restricted}), we have $\Omega_{m,-m'}=\Omega_{-m,m'}=0$, and using Eq.~(\ref{eq:SM_simple_jump_ops}), the dissipative contribution to Eq.~(\ref{eq:action_micro_X}) reads as
\begin{equation}
\begin{split}
-\i \int_\sC \d z \d z'\sum_{\mu\mu'}& \conj{X}_\mu(z)\Omega_{\mu\mu'}(z,z')X_{\mu'}(z')
\\
&=-\int_\sC \d z \d z'\sum_m\left[
L_m(z)\,\i\Omegatilde_m(z,z')\conj{L}_m(z')
+\conj{L}_m(z)\,\i\Omegatilde_{-m}(z,z')L_m(z')
\right]
\\
&= -\int_\sC \d z \d z'\sum_m \, L_m(z) K(z,z')\conj{L}_m(z'),
\end{split}
\end{equation}
where we defined the memory kernel
\begin{equation}\label{eq:micro_def_K}
K(z,z')=\i\Omegatilde_m(z,z')+\zeta' \i\Omegatilde_{-m}(z',z).
\end{equation}
Using Eqs.~(\ref{eq:Omega_tilde_components<}) and (\ref{eq:Omega_tilde_components>}), we can compute the lesser ($(z,z')\in \sC^+\times\sC^-$),
\begin{equation}
K^<(t,t')
=\i\Omegatilde^<_m(t,t')
+\zeta' \i\Omegatilde^>_{-m}(t',t)
=2\i \Omegatilde_m^<(t,t')=2\zeta' f(t,t'),
\end{equation}
greater ($(z,z')\in \sC^-\times\sC^+$),
\begin{equation}
K^>(t,t')
=\i\Omegatilde^>_m(t,t')
+\zeta' \i\Omegatilde^<_{-m}(t',t)
=0,
\end{equation}
time-ordered ($(z,z')\in \sC^+\times\sC^+$),
\begin{equation}
\begin{split}
K^\rmT(t,t')
&=\i\Omegatilde^\rmT_m(t,t')
+\zeta' \i\Omegatilde^\rmT_{-m}(t',t)
\\
&=\heav{t-t'}\i\Omegatilde_m^>(t,t')
+\heav{t'-t}\i\Omegatilde_m^<(t,t')
\zeta'\heav{t'-t}\i\Omegatilde_{-m}^>(t',t)
+\zeta'\heav{t-t'}\i\Omegatilde_{-m}^<(t',t)
\\
&=2\heav{t'-t}\i\Omegatilde^<_m(t,t')
\\
&=2\zeta'\heav{t'-t}f(t,t'),
\end{split}
\end{equation}
and anti-time-ordered ($(z,z')\in \sC^-\times\sC^-$),
\begin{equation}\label{eq:KrmTB_micro}
K^\rmTb(t,t')=2\zeta'\heav{t-t'}f(t,t'),
\end{equation}
components of the memory kernel $K(z,z')$.

In the Markovian limit, $f(t,t')\to\dirac{t-t'}$ and $\heav{t-t'}f(t-t') \to(1/2)\,\delta(t-t'-0^+)$ and the memory kernel derived from the microscopic theory coincides with the one in Eqs.~(\ref{eq:SM_K_kernel_T})--(\ref{eq:SM_K_kernel_>}).

%% file: Thesis_App_Kraus_GinUECUE.tex

\chapter{G\texorpdfstring{\lowercase{in}}{in}UE-CUE crossover ensemble}
\label{app:kraus_GinUECUE}

In this appendix, we compute the spectral support and eigenvalue density for the general GinUE-CUE crossover ensemble
\begin{equation}\label{eq:RMT_model}
\Phi=a\mathbb{G}+b\mathbb{U},
\end{equation}
where $\mathbb{G}$ is a GinUE matrix, $\mathbb{U}$ a CUE matrix, and $a$, $b$ real constants, which was used in Ch.~\ref{chapter:kraus} to model a random quantum channel. To this end, we employ quaternionic free probability~\cite{feinberg1997a,feinberg1997b,janik1997a,janik1997b,feinberg2001,janik2001,jarosz2004,jarosz2006,feinberg2006,burda2011,burda2015,nowak2017jstat,nowak2017pre,denisov2019PRL}, which we start by briefly reviewing below.

\section{Non-Hermitian free probability review}
\label{sec:RMT_model_analytics}

The quaternionic resolvent (Green's function) of a random matrix $\phi$ is defined by
\begin{equation}
\mathcal{G}(Q)=\av{\frac{1}{N}\mathrm{bTr}(Q-\mathcal{H})^{-1}},
\end{equation}
where $Q$ is a general quaternion parametrized as 
\begin{equation}
Q=\begin{pmatrix}
\alpha & \beta \\
-\conj{\beta} & \conj{\alpha}
\end{pmatrix},
\end{equation}
$\alpha,\beta\in\mathbb{C}$, $\mathcal{H}=\mathrm{diag}(\phi,\phi^\dagger)$, and the block trace $\mathrm{bTr}$ is a partial trace over the Hilbert space variables, returning a $2\times2$ matrix. When inside the block trace, a quaternion Q is to be understood as $Q\otimes \mathbbm{1}$, where $\mathbbm{1}$ is the $N$-dimensional identity matrix. The Green's function is also a quaternion which we parametrize as
\begin{equation}
\mathcal{G}=\begin{pmatrix}
\mathcal{G}_{11} & \mathcal{G}_{12}\\
-\conj{\mathcal{G}}_{12} & \conj{\mathcal{G}}_{11}
\end{pmatrix}.
\end{equation}

Several quantities related to the Green's function prove useful in the following. The functional inverse of the Green's function is the Blue's function $\mathcal{B}(\mathcal{G}(Q))=\mathcal{G}(\mathcal{B}(Q))=Q$, which is related to the $\mathcal{R}$-transform by $\mathcal{R}(Q)=\mathcal{B}-Q^{-1}$. The self-energy $\Sigma(Q)$ is defined as usual by $\mathcal{G}(Q)=(Q-\Sigma(Q))^{-1}$. It then follows that the $\mathcal{R}$-transform evaluated on the Green's function is nothing but the self-energy, $\Sigma(Q)=\mathcal{R}(\mathcal{G}(Q))$.

Next, we recall the scaling properties of the Green's and Blue's functions. If $K$ is some quaternion, we have
\begin{equation}\label{eq:scaling_G}
\mathcal{G}_{K\mathcal{H}}(Q)=\mathcal{G}_\mathcal{H}(K^{-1}Q)K^{-1}.
\end{equation}
The Blue's function and self-energy scale inversely:
\begin{equation}\label{eq:scaling_B}
\mathcal{B}_{K\mathcal{H}}(Q)=K\mathcal{G}_\mathcal{H}(QK),
\end{equation}
and identically for the $\mathcal{R}$-transform.

Turning to the sum of two random matrices $A+B$, one can prove~\cite{nica2006,mingo2017} that the $\mathcal{R}$-transform satisfies the additive property $\mathcal{R}_{A+B}(Q)=\mathcal{R}_A(Q)+\mathcal{R}_B(Q)$. From this property one derives the following non-Hermitian Pastur equation:
\begin{equation}
\mathcal{G}_B\left[Q-\mathcal{R}_A\left(\mathcal{G}_{A+B}(Q)\right)\right]=\mathcal{G}_{A+B}(Q).
\end{equation}
At the end of the calculations, we are interested in returning to the complex plane and hence set $\alpha=z\in\mathbb{C}$, $\beta=0$, and obtain
\begin{equation}\label{eq:nonhermitian_Pastur}
\mathcal{G}_B\left[
\begin{pmatrix} 
z & 0 \\ 0 & \conj{z}
\end{pmatrix}
-\mathcal{R}_A\left[
\begin{pmatrix}
\mathcal{G}_{11}(z,\conj{z}) &
\mathcal{G}_{12}(z,\conj{z})\\
-\conj{\mathcal{G}}_{12}(z,\conj{z}) &
\conj{\mathcal{G}}_{11}(z,\conj{z})\\
\end{pmatrix}
\right]
\right]=
\begin{pmatrix}
\mathcal{G}_{11}(z,\conj{z}) &
\mathcal{G}_{12}(z,\conj{z})\\
-\conj{\mathcal{G}}_{12}(z,\conj{z}) &
\conj{\mathcal{G}}_{11}(z,\conj{z})\\
\end{pmatrix}.
\end{equation}

Finally, to obtain the spectral density, we differentiate the upper-left block of the quaternionic Green's function
\begin{equation}
\varrho(z,\conj{z})=\frac{1}{\pi}\partial_{\conj{z}}\mathcal{G}_{11}(z,\conj{z}).
\end{equation}

In our effective RMT model of the quantum map, the matrix $A$ is a GinUE matrix, while $B$ is drawn form the CUE. Accordingly, we need the $\mathcal{R}$-transform of the GinUE and the quaternionic Green's function of the CUE.

\section{Quaternionic \texorpdfstring{$\mathcal{R}$}{R}-transform for the GinUE}

The $\mathcal{R}$-transform of the GinUE has been obtained in Refs.~\cite{janik1997a,janik1997b,feinberg1997a} by a variety of different methods. For completeness, we briefly lay out the computation starting from the (Hermitian) Gaussian Unitary Ensemble (GUE). A matrix $A$ from the GinUE can be parametrized in terms of two Hermitian matrices $H$, $H'$ from independent Gaussian ensembles, $A=(H+iH')/\sqrt{2}$. The defining feature of the Gaussian ensembles is the additivity of its Green's functions (i.e., the sum of Gaussian matrices is still Gaussian, the result for random matrices analogous to the central limit theorem for classical probability), which implies $\mathcal{G}=\Sigma$ or, equivalently, $\mathcal{R}_H(Q)=Q$. This fact, together with the additivity of the $\mathcal{R}$-transform and the scaling relation~(\ref{eq:scaling_B}), yields
\begin{equation}
\mathcal{R}_A=\mathcal{R}_{H/\sqrt{2}}+\mathcal{R}_{iH'/\sqrt{2}}=\frac{1}{2}(Q+\mathcal{I}Q\mathcal{I}),
\end{equation} 
where $\mathcal{I}=\mathrm{diag}(i,-i)$. We then immediately obtain the $\mathcal{R}$-transform for the GinUE,
\begin{equation}\label{eq:G_GinuE}
\mathcal{R}_\mathrm{GinUE}(Q)=
\begin{pmatrix}
0 & \beta \\
-\conj{\beta} & 0
\end{pmatrix}.
\end{equation}

\section{Quaternionic resolvent for the CUE}

We now consider the quaternionic resolvent for the CUE. Related results were given in Refs.~\cite{jarosz2011,nowak2017pre}, but the end-result of this computation, Eq.~(\ref{eq:G_CUE}), is not explicitly given in the literature, to the best of our knowledge. Because $\mathbb{U}$ is normal (i.e., $\comm{\mathbb{U}}{\mathbb{U}^\dagger}=0$), its left- and right-eigenvectors coincide. We can then write
\begin{equation}
\mathbb{U}=\sum_{n}\ket{n}e^{i\theta_n}\bra{n}
\qquad \text{and}\qquad
\mathbb{U}^\dagger=\sum_{n}\ket{n}e^{-i\theta_n}\bra{n},
\end{equation}
in some eigenbasis $\{\ket{n}\}$. The argument of the Green's function $\mathcal{G}(Q)$ can then be easily inverted,
\begin{equation}
(Q-\mathcal{H})^{-1}=\sum_n\ket{n}\frac{1}{(\alpha-e^{i\theta_n})(\conj{\alpha}-e^{-i\theta_n})+\conj{\beta}\beta}
\begin{pmatrix}
\conj{\alpha}-e^{-i\theta_n} & -\beta \\
\conj{\beta} & \alpha-e^{i\theta_n}
\end{pmatrix}
\bra{n}.
\end{equation}

Taking the large-$N$ limit, we can replace $\av{(1/N)\sum_nF(\theta_n)}$ by $\int \d\theta \varrho(\theta)\av{F(\theta)}$, where $\varrho(\theta)=1/(2\pi)$ is the flat density of the CUE on the unit circle. Performing the block trace, the quaternionic resolvent reads
\begin{equation}
\mathcal{G}(Q)=\int\frac{\d\theta}{2\pi}\frac{1}{(\alpha-e^{i\theta})(\conj{\alpha}-e^{-i\theta})+\conj{\beta}\beta}
\begin{pmatrix}
\conj{\alpha}-e^{-i\theta} & -\beta \\
\conj{\beta} & \alpha-e^{i\theta}
\end{pmatrix}.
\end{equation}

Although the spectrum of $\mathbb{U}$ is one dimensional, and hence depends only on a single real number $\theta$, we next convert the real integral into a contour integral. We define $\zeta=e^{i\theta}$ and, exploiting the fact that $\conj{\zeta}=\zeta^{-1}$ on the unit circle, write
\begin{equation}\label{eq:G_CUE_integral}
\mathcal{G}(Q)=\frac{1}{2\pi i}\oint_{\abs{\zeta}=1}\frac{\d\zeta}{\zeta}\frac{1}{(\alpha-\zeta)(\conj{\alpha}-\zeta^{-1})+\conj{\beta}\beta}
\begin{pmatrix}
\conj{\alpha}-\zeta^{-1} & -\beta \\
\conj{\beta} & \alpha-\zeta
\end{pmatrix}.
\end{equation}
Since the integrand is a holomorphic function of $\zeta$, the Green's function is the sum of the residues of the poles inside the unit circle.
The integrand has three poles,
\[
\zeta_0=0 
\qquad \text{and} \qquad \zeta_\pm=\frac{1+\conj{\alpha}\alpha+\conj{\beta}\beta\pm P}{2\conj{\alpha}},
\]
where $P=\sqrt{(1+\conj{\alpha}\alpha+\conj{\beta}\beta)^2-4\conj{\alpha}\alpha}$. Since $\zeta_-\zeta_+=\alpha/\conj{\alpha}$, i.e., $\abs{\zeta_-\zeta_+}=1$, one of $\zeta_\pm$ is inside the unit circle, the other outside. One can check that $\abs{\zeta_+}>\abs{\zeta_-}$, and hence conclude that $\zeta_0$ and $\zeta_-$ are always inside the unit circle and $\zeta_+$ outside. Computing the residues of the integrand of Eq.~(\ref{eq:G_CUE_integral}), we find
\begin{equation}\label{eq:G_CUE}
\mathcal{G}(Q)=\frac{1}{P}
\begin{pmatrix}
\frac{1}{2\alpha}(\conj{\alpha}\alpha-\conj{\beta}\beta-1+P) & -\beta \\
\conj{\beta} & \frac{1}{2\conj{\alpha}}(\conj{\alpha}\alpha-\conj{\beta}\beta-1+P)
\end{pmatrix}
\end{equation}

\section{Crossover ensemble}

\subsection{Quaternionic resolvent}

We can now put the different pieces together. First, by recalling the scaling properties of $\mathcal{G}$ and $\mathcal{R}$, Eqs.~(\ref{eq:scaling_G})~and~(\ref{eq:scaling_B}), respectively, we have
\begin{equation}
\mathcal{R}_{aA}\left[
\begin{pmatrix}
\alpha & \beta \\
-\conj{\beta} & \conj{\alpha}
\end{pmatrix}
\right]=
a^2
\begin{pmatrix}
0 & \beta \\
-\conj{\beta} & 0
\end{pmatrix}
\end{equation}
and
\begin{equation}
\mathcal{G}_{bB}\left[
\begin{pmatrix}
\alpha & \beta \\
-\conj{\beta} & \conj{\alpha}
\end{pmatrix}
\right]=
\frac{1}{P_b}
\begin{pmatrix}
\frac{1}{2\alpha}(\conj{\alpha}\alpha-\conj{\beta}\beta-b^2+P_b) & -\beta \\
\conj{\beta} & \frac{1}{2\conj{\alpha}}(\conj{\alpha}\alpha-\conj{\beta}\beta-b^2+P_b)
\end{pmatrix},
\end{equation}
where $P_b=\sqrt{(b^2+\conj{\alpha}\alpha+\conj{\beta}\beta)^2-4b^2\conj{\alpha}\alpha}$. The non-Hermitian Pastur equation (\ref{eq:nonhermitian_Pastur}) then reads:
\begin{equation}
\begin{split}
\begin{pmatrix}
\mathcal{G}_{11} & \mathcal{G}_{12}\\
-\conj{\mathcal{G}}_{12} & \conj{\mathcal{G}}_{11}
\end{pmatrix}
&=
\mathcal{G}_{bB}\left[
\begin{pmatrix}
z & -a^2\mathcal{G}_{12} \\
a^2\conj{\mathcal{G}}_{12} & \conj{z}
\end{pmatrix}
\right]
\\
&=\frac{1}{P_b}
\begin{pmatrix}
\frac{1}{2z}(\conj{z}z-a^4\abs{\mathcal{G}_{12}}^2-b^2+P_b) & a^2\mathcal{G}_{12}\\
-a^2\conj{\mathcal{G}}_{12} & \frac{1}{2\conj{z}}(\conj{z}z-a^4\abs{\mathcal{G}_{12}}^2-b^2+P_b)
\end{pmatrix}.
\end{split}
\end{equation}

\subsection{Spectral support}
The off-diagonal terms of the non-Hermitian Pastur equation yield the condition $\mathcal{G}_{12}=\mathcal{G}_{12}a^2/P_b$, which has two solutions. The trivial solution $\mathcal{G}_{12}=0$ is valid outside the eigenvalue support, while the nontrivial solution $P_b=a^2$ is satisfied inside. The value of $\mathcal{G}_{12}$ inside the support then satisfies
\begin{equation}\label{eq:G12_nontrivial}
a^4\abs{\mathcal{G}_{12}}^2=-b^2-\conj{z}z+\sqrt{a^4+4b^2\conj{z}z}.
\end{equation}

The boundaries of the spectral support are found by matching the trivial and nontrivial solutions, i.e., setting $\mathcal{G}_{12}=0$ in Eq.~(\ref{eq:G12_nontrivial}) and solving for $\conj{z}z$. The solutions match for $\abs{z}=\sqrt{b^2\pm a^2}$, and we therefore conclude  that the spectrum is supported on an annulus with inner (outer) radius $R_-(R_+)$, where $R_\pm=\sqrt{b^2\pm a^2}$. When $\abs{a}>\abs{b}$, there is no inner boundary and the spectrum is supported on the disk with radius $R_+$.

\subsection{Eigenvalue density}
The diagonal terms of the non-Hermitian Pastur equation (\ref{eq:nonhermitian_Pastur}) give the condition
\begin{equation}
\mathcal{G}_{11}=\frac{1}{2z}\left(1+\frac{\conj{z}z-a^4\abs{\mathcal{G}_{12}}^2-b^2}{P_b}\right).
\end{equation}
Using $P_b=a^2$ and Eq.~(\ref{eq:G12_nontrivial}), the value of $\mathcal{G}_{11}$ inside the spectral support is 
\begin{equation}
\mathcal{G}_{11}(z,\conj{z})=\frac{a^2+2\conj{z}z-\sqrt{a^4+4b^2\conj{z}z}}{2a^2z}.
\end{equation}

Finally, the eigenvalue distribution can be obtained by differentiating $\mathcal{G}_{11}$:
\begin{equation}
\varrho(z,\conj{z})=\frac{1}{\pi}\partial_{\conj{z}}\mathcal{G}_{11}(z,\conj{z})=\frac{1}{\pi a^2}\left(1-\frac{b^2}{\sqrt{a^4+4b^2\conj{z}z}}\right).
\end{equation}

By setting $a=p/\sqrt{d}$ and $b=(1-p)$, the expressions given in the main text are immediately recovered.

%% file: Thesis_App_KrausTransposition.tex

\chapter{Transposition symmetry of the Kraus circuits}
\label{app:KrausTransposition}%

In this Appendix, we elaborate on the transposition symmetry of Kraus circuits, which determines their spectral statistics. In particular, we argue for the presence of transposition symmetry in the Hubbard-Kraus and two-free-channel circuits, and its absence for the XXZ circuit. 

\begin{figure*}[t]
	\centering
	\includegraphics[width=\textwidth]{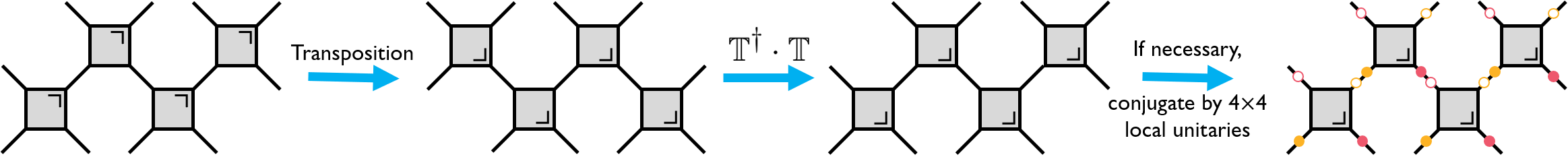}
	\caption{Schematic representation of the steps involved in determining whether there exists a transposition symmetry of the circuit. Gray gates represent local quantum maps $\phi$, while orange and magenta filled circles depict local unitaries $v_-,v_+\in \mathrm{SU}(4)$, respectively, and empty circles their inverses. Transposition is signaled by the flip of the wedge in the corner of the local maps. To respect the kinematical symmetries of the circuit, the allowed unitary transformations are one-site translations and the local unitaries $v_\pm$.}
	\label{fig:SM_transposition_circuit}
\end{figure*}

We start by showing that it suffices to consider the properties of elementary two-site quantum maps. Taking the transpose of the circuit we find
$\Psi^\top=\Phi^\top \mathbb{T}^\top \Phi^\top \mathbb{T}^*=\Phi^\top \mathbb{T}^\dagger \Phi^\top \mathbb{T}$. We want to bring it back to $\Psi$ by a unitary transformation that satisfies the symmetries of the model. These include translations (necessary to bring the circuit back to the correct order of applying first gates as odd-even bonds followed by even-odd bonds) and, possibly, local $4\times 4$ unitaries in each local Hilbert space (local gauge transformations). The procedure is depicted pictorially in Fig.~\ref{fig:SM_transposition_circuit}.

We conclude that the circuit satisfies the transposition symmetry $\Psi=C_+\Psi^\top C_+^\dagger$ if the local quantum maps $\phi$ satisfy
\begin{equation}\label{eq:transposition_condition}
\phi=
\includegraphics[width=0.15\textwidth,valign=c]{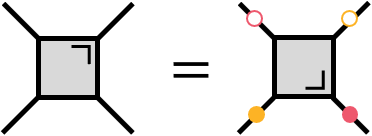}
= (v_-\otimes v_+)\phi^\top (v_+^\dagger \otimes v_-^\dagger),
\end{equation}
for some $v_-,v_+\in \mathrm{SU}(4)$ respecting the dynamical symmetries of the circuits (i.e., magnetization conservation). Moreover, from Eq.~(\ref{eq:Kraus_rep}), we see that the behavior under transposition of $\phi$ is fully determined by the behavior of the Kraus operators $K_\pm$ of each circuit (since tensoring and transposing commute). It thus suffices to analyze the behavior of the Kraus operators under transposition.

\section{Two-free-channel circuit}%
The local quantum map $\phi$ is a convex combination of a pair of unitary channels, whose Kraus operators are given in Eq.~(\ref{eq:Kraus_2unitary}). The transposition symmetry is evident from Eq.~(\ref{eq:Rch_general6v}) after a unitary change of basis  by conjugation with $V=\mathrm{diag}(1,\exp\{i\pi/4\},\exp\{-i\pi/4\},1)$---which corresponds to the choice of gauge $v_-=v_+^*=V$ in Eq.~(\ref{eq:transposition_condition}). Writing out the Kraus operators explicitly (we omit all zero entries) in the new basis,
\begin{equation}
K_\pm(\lambda_\pm)=\begin{pmatrix}
\cos\lambda_\pm & & & \\
& 1 & \sin\lambda_\pm & \\
& \sin\lambda_\pm & 1 & \\
& & & \cos\lambda_\pm
\end{pmatrix},
\end{equation}
we see that they are complex symmetric, where we defined the spectral parameters $\lambda_\pm \equiv\lambda\pm\mu\in i\mathbb{R}$ (related to the hopping amplitudes $q_\pm$ by multiplication by $i$). (The normalization of the $\check{R}$-matrix is irrelevant for the purpose of this appendix and will be dropped throughout.) It follows that, in this basis, $\phi(\lambda_-,\lambda_+)=\phi^\top(\lambda_-,\lambda_+)$, and the circuit enjoys a transposition symmetry.

\section{Hubbard-Kraus circuit}%
Although the Hubbard-Kraus circuit is integrable and, therefore, exhibits Poisson spectral statistics, it is instructive to determine its symmetry class to provide contrast to the XXZ circuit case discussed below. The Kraus operators (\ref{eq:Kraus_2channel}) read as
\begin{equation}
\begin{split}
K_-(\lambda_-)&=\begin{pmatrix}
\cos\lambda_- & & & \\
& 1 & -i\sin\lambda_- & \\
& i\sin\lambda_- & 1 & \\
& & & \cos\lambda_-
\end{pmatrix}
\quad \text{and}\\
K_+(\lambda_+)&=\begin{pmatrix}
\cos\lambda_+ & & & \\
& 1 & i\sin\lambda_+ & \\
& i\sin\lambda_+ & -1 & \\
& & & -\cos\lambda_+
\end{pmatrix}.
\end{split}
\end{equation}
While both Kraus operators cannot be symmetrized simultaneously by a change of basis as before, they still satisfy
\begin{equation}\label{eq:weak_transposition}
K_-^\top(\lambda_-)=K_-(-\lambda_-)=K_-(\lambda_-^*)
\quad \text{and} \quad 
K_+^\top(\lambda_+)=K_+(\lambda_+).
\end{equation}
It follows that $\phi(\lambda_-,\lambda_+)=\phi^\top(\lambda_-^*,\lambda_+)$. For nonintegrable cases, generalizing the local Hubbard-Kraus map to more than two Kraus operators of the form~(\ref{eq:Kraus_2channel}), we have made the empirical observation that the equality of the Kraus operators and their transposes \emph{up to complex conjugation of the imaginary spectral parameters} is enough to guarantee the convergence of their spectral statistics to those of the AI$^\dagger$ class.
This condition is fulfilled by the Hubbard-Kraus map.

\section{XXZ circuit}%
By introducing interactions into the coherent dynamics, i.e., by considering Kraus operators
\begin{equation}
\begin{split}
K_-(\lambda_-)&=\begin{pmatrix}
\sin(\gamma_-+\lambda_-) & & & \\
& \sin\gamma_- & -i\sin\lambda_- & \\
& i\sin\lambda_- & \sin\gamma_- & \\
& & & \sin(\gamma_-+\lambda_-)
\end{pmatrix}
\quad \text{and}\\
K_+(\lambda_+)&=\begin{pmatrix}
\sin(\gamma_++\lambda_+) & & & \\
& \sin\gamma_+ & i\sin\lambda_+ & \\
& i\sin\lambda_+ & -\sin\gamma_+ & \\
& & & -\sin(\gamma_++\lambda_+)
\end{pmatrix},
\end{split}
\end{equation}
it follows that neither can both Kraus operators be simultaneously symmetrized nor do they satisfy Eq.~(\ref{eq:weak_transposition}). Hence, the XXZ circuit does not enjoy a transposition symmetry.

%% file: Thesis_App_SYKCombinatorics.tex

\chapter{Review of the combinatorial approach to the standard SYK model}
\label{app:review_standard_SYK}%

{
	\allowdisplaybreaks
	
In this Appendix, we review the computation of the spectral density of the standard SYK model~(\ref{eq:def_SYK}) using the method of moments, following Refs.~\cite{garcia-garcia2016PRD,garcia-garcia2017PRD,cotler2017JHEP,berkooz2019JHEP}.

Because the random variables $J_\va$ are Gaussian, the odd moments of the SYK Hamiltonian vanish, while the even moments, $\av{\Tr H^{2p}}$, are evaluated by Wick contraction, i.e., by summing over all possible pair contractions of the indices $\va$, $\vb$, $\vc$, etc. For instance, the first nontrivial moment (the fourth) can be explicitly evaluated as:
\begin{equation}\label{eq:H_moment4_example}
\begin{split}
\av{\Tr H^4}
&=\sum_{\va,\vb,\vc,\vd}\av{J_\va J_\vb J_\vc J_\vd}\Tr\(\Gamma_\va\Gamma_\vb\Gamma_\vc\Gamma_\vd\)\\
&=\av{J^2}^2\sum_{\va,\vb}\left[
\Tr\(\Gamma_\va\Gamma_\va\Gamma_\vb\Gamma_\vb\)+
\Tr\(\Gamma_\va\Gamma_\vb\Gamma_\va\Gamma_\vb\)+
\Tr\(\Gamma_\va\Gamma_\vb\Gamma_\vb\Gamma_\va\)
\right]\\
&=2^{N/2}\av{J^2}^2\binom{N}{q}^2\left[
2+2^{-N/2}\binom{N}{q}^{-2}\sum_{\va,\vb}
\Tr\(\Gamma_\va\Gamma_\vb\Gamma_\va\Gamma_\vb\)
\right].
\end{split}
\end{equation}
We can represent graphically the two terms inside square brackets in the last line of Eq.~(\ref{eq:H_moment4_example}) by introducing a diagrammatic notation for the $\Gamma$-matrices and their contractions,
\begin{equation}
\begin{tikzpicture}[baseline=($0.75*(a)+0.25*(x)$)]
\begin{feynman}[inline=(a)]
\vertex[large, dot] (a) {};
\vertex[below=0.3cm of a] (x) {\scriptsize{$\va$}};
\end{feynman}
\end{tikzpicture}
=\Gamma_\va
\qquad \text{and} \qquad
\begin{tikzpicture}[inner sep=2pt,baseline=(a)]
\begin{feynman}[inline=(a)]
\vertex (a) {};
\vertex[right=0.6cm of a] (b) {};
\vertex[below=0.205cm of a] (x) {\scriptsize{$\va$}};
\vertex[below=0.18cm of b] (y) {\scriptsize{$\vb$}};
\diagram*{(a) --[half left,min distance=0.6cm] (b)};
\end{feynman}
\end{tikzpicture}
=\frac{\av{J_\va J_\vb}}{\av{J^2}}=\delta_{\va,\vb}.
\end{equation}
After correctly normalizing $\Tr$ by $2^{-N/2}$ and $\sum_{\va}$ by $\binom{N}{q}^{-1}$, we obviously have (we omit the labels of the dots and edges throughout)
\begin{equation}
t_1(q,N) \equiv 
\begin{tikzpicture}[inner sep=2pt,baseline=(b)]
\begin{feynman}[inline=(a)]
\vertex[dot] (a) {};
\vertex[dot,right=0.4cm of a] (b) {};
\diagram*{(a) --[half left,min distance=0.4cm] (b)};
\end{feynman}
\end{tikzpicture}
= 2^{-N/2} \binom{N}{q}^{-1} \sum_{\va}
\Tr\(\Gamma_\va \Gamma_\va\)
=1,
\end{equation}
while the last term in Eq.~(\ref{eq:H_moment4_example}) can be represented and evaluated as~\cite{garcia-garcia2016PRD}
\begin{equation}\label{eq:diagram_t2_perfect_matching}
t_2(q,N)\equiv
\begin{tikzpicture}[inner sep=2pt,baseline=(b)]
\begin{feynman}[inline=(a)]
\vertex[dot] (a) {};
\vertex[dot,right=0.4cm of a] (b) {};
\vertex[dot,right=0.4cm of b] (c) {};
\vertex[dot,right=0.4cm of c] (d) {};
\diagram*{(a) --[half left,min distance=0.4cm] (c)};
\diagram*{(b) --[half left,min distance=0.4cm] (d)};
\end{feynman}
\end{tikzpicture}
=2^{-N/2}\binom{N}{q}^{-2}\sum_{\va,\vb}\Tr\(\Gamma_\va \Gamma_\vb \Gamma_\va \Gamma_\vb\)
=\binom{N}{q}^{-1}\sum_{s=0}^q 
(-1)^{q+s}\binom{q}{s}\binom{N-q}{q-s}.
\end{equation}
To evaluate $t_2$ we have to commute $\Gamma_\va$ with $\Gamma_\vb$. Let $\Gamma_\va$ and $\Gamma_\vb$ have $s$ $\gamma$-matrices in common, i.e., $s=\abs{\va\cap\vb}$ for fixed $\va$ and $\vb$. We can then express the sum over $\va$ and $\vb$ as a sum over the $q$ indices in $\vb$ and over $s$ by allowing for all possible combinations of indices inside $\va$ and $\vb$: out of the $N$ possible $\gamma$-matrices, we fix the $q$ distinct $\gamma$-matrices in $\Gamma_\vb$ in all $\binom{N}{q}$ possible ways, choose the $s$ indices in $\va$ that coincide with the $q$ indices in $\vb$ in all $\binom{q}{s}$ ways, and allow for the $\binom{N-q}{q-s}$ distinct combinations of the remaining $(q-s)$ $\gamma$-matrices in $\Gamma_\va$ to be any of the $(N-q)$ $\gamma$-matrices still available. The commutation of $\Gamma_\va$ and $\Gamma_\vb$ gives a phase $(-1)^{q+s}$ according to Eq.~(\ref{eq:relations_Gamma}). This procedure yields exactly the factors in Eq.~(\ref{eq:diagram_t2_perfect_matching}).

The diagrammatic representation of higher-order moments can also be immediately written down. The $2p$th moment $\av{\tr H^{2p}}$ will have $2p$ dots ordered on a line corresponding to the $2p$ insertions of $\Gamma$-matrices in the trace. We then connect the $2p$ dots by $p$ edges in all possible ways as required by Wick's Theorem. Each diagram thus obtained corresponds to a \emph{perfect matching} $\pi$. The set of all perfect matchings of $2p$ elements is denoted by $\mathcal{M}_{2p}$ and has $(2p-1)!!$ elements. We can write the moments of the SYK model as a sum over perfect matchings $\pi\in\mathcal{M}_{2p}$,
\begin{equation}
\frac{1}{\sigmaH^{2p}}\frac{\av{\Tr H^{2p}}}{\Tr\id}
=\sum_{\pi\in\mathcal{M}_{2p}}t(\pi),
\end{equation}
where $\sigmaH=\sqrt{\av{J^2}\binom{N}{q}}$ is the energy scale of the SYK model and the weight~$t(\pi)$ is the contribution of the trace associated to the diagram of $\pi$.

With this diagrammatic notation, the first few moments of $H$ are explicitly found to be:
\begingroup
\allowdisplaybreaks
\begin{align}
	\av{\Tr H^0}&=2^{N/2},
	\\
	\av{\Tr H^2}&=
	2^{N/2}\av{J^2}\binom{N}{q}
	\hspace{6pt}
	\begin{tikzpicture}[inner sep=2pt,baseline=(b)]
	\begin{feynman}[inline=(a)]
	\vertex[dot] (a) {};
	\vertex[dot,right=0.4cm of a] (b) {};
	\diagram*{(a) --[half left,min distance=0.4cm] (b)};
	\vertex[right=0.2cm of a] (x) {};
	\vertex[below=0.35cm of x] {$1$};
	\end{feynman}
	\end{tikzpicture}
	=2^{N/2}\av{J}^2\binom{N}{q},
	\\
	\begin{split}
		\av{\Tr H^4}&=
		2^{N/2}\(\av{J^2}\binom{N}{q}\)^2 
		\(
		\begin{tikzpicture}[inner sep=2pt,baseline=(b)]
		\begin{feynman}[inline=(a)]
		\vertex[dot] (a) {};
		\vertex[dot,right=0.4cm of a] (b) {};
		\vertex[dot,right=0.4cm of b] (c) {};
		\vertex[dot,right=0.4cm of c] (d) {};
		\diagram*{(a) --[half left,min distance=0.4cm] (b)};
		\diagram*{(c) --[half left,min distance=0.4cm] (d)};
		\vertex[right=0.2cm of b] (x) {};
		\vertex[below=0.35cm of x] {$1$}; 
		\end{feynman}
		\end{tikzpicture}
		+
		\begin{tikzpicture}[inner sep=2pt,baseline=(b)]
		\begin{feynman}[inline=(a)]
		\vertex[dot] (a) {};
		\vertex[dot,right=0.4cm of a] (b) {};
		\vertex[dot,right=0.4cm of b] (c) {};
		\vertex[dot,right=0.4cm of c] (d) {};
		\diagram*{(a) --[half left,min distance=0.4cm] (c)};
		\diagram*{(b) --[half left,min distance=0.4cm] (d)};
		\vertex[right=0.2cm of b] (x) {};
		\vertex[below=0.35cm of x] {$t_2$};
		\end{feynman}
		\end{tikzpicture}
		+
		\begin{tikzpicture}[inner sep=2pt,baseline=(b)]
		\begin{feynman}[inline=(a)]
		\vertex[dot] (a) {};
		\vertex[dot,right=0.4cm of a] (b) {};
		\vertex[dot,right=0.4cm of b] (c) {};
		\vertex[dot,right=0.4cm of c] (d) {};
		\diagram*{(a) --[half left,min distance=0.4cm] (d)};
		\diagram*{(b) --[half left,min distance=0.4cm] (c)};
		\vertex[right=0.2cm of b] (x) {};
		\vertex[below=0.35cm of x] {$1$}; 
		\end{feynman}
		\end{tikzpicture}
		\)
		\\
		&=2^{N/2}\(\av{J^2}\binom{N}{q}\)^2
		\(2+t_2\),
	\end{split}
	\\
	\begin{split}\nonumber
		\av{\Tr H^6}&=
		2^{N/2}\(\av{J^2}\binom{N}{q}\)^3
	\end{split}
	\\
	\begin{split}\nonumber
		\times&\left(
		\begin{tikzpicture}[inner sep=2pt,baseline=(b)]
		\begin{feynman}[inline=(a)]
		\vertex[dot] (a) {};
		\vertex[dot,right=0.4cm of a] (b) {};
		\vertex[dot,right=0.4cm of b] (c) {};
		\vertex[dot,right=0.4cm of c] (d) {};
		\vertex[dot,right=0.4cm of d] (e) {};
		\vertex[dot,right=0.4cm of e] (f) {};
		\diagram*{(a) --[half left,min distance=0.4cm] (b)};
		\diagram*{(c) --[half left,min distance=0.4cm] (d)}; 
		\diagram*{(e) --[half left,min distance=0.4cm] (f)};
		\vertex[right=0.2cm of c] (x) {};
		\vertex[below=0.35cm of x] {1};
		\end{feynman}
		\end{tikzpicture}
		+
		\begin{tikzpicture}[inner sep=2pt,baseline=(b)]
		\begin{feynman}[inline=(a)]
		\vertex[dot] (a) {};
		\vertex[dot,right=0.4cm of a] (b) {};
		\vertex[dot,right=0.4cm of b] (c) {};
		\vertex[dot,right=0.4cm of c] (d) {};
		\vertex[dot,right=0.4cm of d] (e) {};
		\vertex[dot,right=0.4cm of e] (f) {};
		\diagram*{(a) --[half left,min distance=0.4cm] (b)};
		\diagram*{(c) --[half left,min distance=0.4cm] (e)}; 
		\diagram*{(d) --[half left,min distance=0.4cm] (f)};
		\vertex[right=0.2cm of c] (x) {};
		\vertex[below=0.35cm of x] {$t_2$};
		\end{feynman}
		\end{tikzpicture}
		+
		\begin{tikzpicture}[inner sep=2pt,baseline=(b)]
		\begin{feynman}[inline=(a)]
		\vertex[dot] (a) {};
		\vertex[dot,right=0.4cm of a] (b) {};
		\vertex[dot,right=0.4cm of b] (c) {};
		\vertex[dot,right=0.4cm of c] (d) {};
		\vertex[dot,right=0.4cm of d] (e) {};
		\vertex[dot,right=0.4cm of e] (f) {};
		\diagram*{(a) --[half left,min distance=0.4cm] (b)};
		\diagram*{(c) --[half left,min distance=0.4cm] (f)}; 
		\diagram*{(d) --[half left,min distance=0.4cm] (e)};
		\vertex[right=0.2cm of c] (x) {};
		\vertex[below=0.35cm of x] {1};
		\end{feynman}
		\end{tikzpicture}
		+
		\begin{tikzpicture}[inner sep=2pt,baseline=(b)]
		\begin{feynman}[inline=(a)]
		\vertex[dot] (a) {};
		\vertex[dot,right=0.4cm of a] (b) {};
		\vertex[dot,right=0.4cm of b] (c) {};
		\vertex[dot,right=0.4cm of c] (d) {};
		\vertex[dot,right=0.4cm of d] (e) {};
		\vertex[dot,right=0.4cm of e] (f) {};
		\diagram*{(a) --[half left,min distance=0.4cm] (c)};
		\diagram*{(b) --[half left,min distance=0.4cm] (d)}; 
		\diagram*{(e) --[half left,min distance=0.4cm] (f)};
		\vertex[right=0.2cm of c] (x) {};
		\vertex[below=0.35cm of x] {$t_2$};
		\end{feynman}
		\end{tikzpicture}
		+
		\begin{tikzpicture}[inner sep=2pt,baseline=(b)]
		\begin{feynman}[inline=(a)]
		\vertex[dot] (a) {};
		\vertex[dot,right=0.4cm of a] (b) {};
		\vertex[dot,right=0.4cm of b] (c) {};
		\vertex[dot,right=0.4cm of c] (d) {};
		\vertex[dot,right=0.4cm of d] (e) {};
		\vertex[dot,right=0.4cm of e] (f) {};
		\diagram*{(a) --[half left,min distance=0.4cm] (c)};
		\diagram*{(b) --[half left,min distance=0.4cm] (e)}; 
		\diagram*{(d) --[half left,min distance=0.4cm] (f)};
		\vertex[right=0.2cm of c] (x) {};
		\vertex[below=0.35cm of x] {$t_3'$};
		\end{feynman}
		\end{tikzpicture}
		\right.
	\end{split}
	\\
	\begin{split}
		&+
		\begin{tikzpicture}[inner sep=2pt,baseline=(b)]
		\begin{feynman}[inline=(a)]
		\vertex[dot] (a) {};
		\vertex[dot,right=0.4cm of a] (b) {};
		\vertex[dot,right=0.4cm of b] (c) {};
		\vertex[dot,right=0.4cm of c] (d) {};
		\vertex[dot,right=0.4cm of d] (e) {};
		\vertex[dot,right=0.4cm of e] (f) {};
		\diagram*{(a) --[half left,min distance=0.4cm] (c)};
		\diagram*{(b) --[half left,min distance=0.4cm] (f)}; 
		\diagram*{(d) --[half left,min distance=0.4cm] (e)};
		\vertex[right=0.2cm of c] (x) {};
		\vertex[below=0.35cm of x] {$t_2$};
		\end{feynman}
		\end{tikzpicture}
		+
		\begin{tikzpicture}[inner sep=2pt,baseline=(b)]
		\begin{feynman}[inline=(a)]
		\vertex[dot] (a) {};
		\vertex[dot,right=0.4cm of a] (b) {};
		\vertex[dot,right=0.4cm of b] (c) {};
		\vertex[dot,right=0.4cm of c] (d) {};
		\vertex[dot,right=0.4cm of d] (e) {};
		\vertex[dot,right=0.4cm of e] (f) {};
		\diagram*{(a) --[half left,min distance=0.4cm] (d)};
		\diagram*{(b) --[half left,min distance=0.4cm] (c)}; 
		\diagram*{(e) --[half left,min distance=0.4cm] (f)};
		\vertex[right=0.2cm of c] (x) {};
		\vertex[below=0.35cm of x] {$1$};
		\end{feynman}
		\end{tikzpicture}
		+
		\begin{tikzpicture}[inner sep=2pt,baseline=(b)]
		\begin{feynman}[inline=(a)]
		\vertex[dot] (a) {};
		\vertex[dot,right=0.4cm of a] (b) {};
		\vertex[dot,right=0.4cm of b] (c) {};
		\vertex[dot,right=0.4cm of c] (d) {};
		\vertex[dot,right=0.4cm of d] (e) {};
		\vertex[dot,right=0.4cm of e] (f) {};
		\diagram*{(a) --[half left,min distance=0.4cm] (d)};
		\diagram*{(b) --[half left,min distance=0.4cm] (e)}; 
		\diagram*{(c) --[half left,min distance=0.4cm] (f)};
		\vertex[right=0.2cm of c] (x) {};
		\vertex[below=0.35cm of x] {$t_3$};
		\end{feynman}
		\end{tikzpicture}
		+
		\begin{tikzpicture}[inner sep=2pt,baseline=(b)]
		\begin{feynman}[inline=(a)]
		\vertex[dot] (a) {};
		\vertex[dot,right=0.4cm of a] (b) {};
		\vertex[dot,right=0.4cm of b] (c) {};
		\vertex[dot,right=0.4cm of c] (d) {};
		\vertex[dot,right=0.4cm of d] (e) {};
		\vertex[dot,right=0.4cm of e] (f) {};
		\diagram*{(a) --[half left,min distance=0.4cm] (d)};
		\diagram*{(b) --[half left,min distance=0.4cm] (f)}; 
		\diagram*{(c) --[half left,min distance=0.4cm] (e)};
		\vertex[right=0.2cm of c] (x) {};
		\vertex[below=0.35cm of x] {$t_3'$};
		\end{feynman}
		\end{tikzpicture}
		+
		\begin{tikzpicture}[inner sep=2pt,baseline=(b)]
		\begin{feynman}[inline=(a)]
		\vertex[dot] (a) {};
		\vertex[dot,right=0.4cm of a] (b) {};
		\vertex[dot,right=0.4cm of b] (c) {};
		\vertex[dot,right=0.4cm of c] (d) {};
		\vertex[dot,right=0.4cm of d] (e) {};
		\vertex[dot,right=0.4cm of e] (f) {};
		\diagram*{(a) --[half left,min distance=0.4cm] (e)};
		\diagram*{(b) --[half left,min distance=0.4cm] (c)}; 
		\diagram*{(d) --[half left,min distance=0.4cm] (f)};
		\vertex[right=0.2cm of c] (x) {};
		\vertex[below=0.35cm of x] {$t_2$};
		\end{feynman}
		\end{tikzpicture}
	\end{split}
	\\
	\begin{split}\nonumber
		&+
		\left.
		\begin{tikzpicture}[inner sep=2pt,baseline=(b)]
		\begin{feynman}[inline=(a)]
		\vertex[dot] (a) {};
		\vertex[dot,right=0.4cm of a] (b) {};
		\vertex[dot,right=0.4cm of b] (c) {};
		\vertex[dot,right=0.4cm of c] (d) {};
		\vertex[dot,right=0.4cm of d] (e) {};
		\vertex[dot,right=0.4cm of e] (f) {};
		\diagram*{(a) --[half left,min distance=0.4cm] (e)};
		\diagram*{(b) --[half left,min distance=0.4cm] (d)}; 
		\diagram*{(c) --[half left,min distance=0.4cm] (f)};
		\vertex[right=0.2cm of c] (x) {};
		\vertex[below=0.35cm of x] {$t_3'$};
		\end{feynman}
		\end{tikzpicture}
		+
		\begin{tikzpicture}[inner sep=2pt,baseline=(b)]
		\begin{feynman}[inline=(a)]
		\vertex[dot] (a) {};
		\vertex[dot,right=0.4cm of a] (b) {};
		\vertex[dot,right=0.4cm of b] (c) {};
		\vertex[dot,right=0.4cm of c] (d) {};
		\vertex[dot,right=0.4cm of d] (e) {};
		\vertex[dot,right=0.4cm of e] (f) {};
		\diagram*{(a) --[half left,min distance=0.4cm] (e)};
		\diagram*{(b) --[half left,min distance=0.4cm] (f)}; 
		\diagram*{(c) --[half left,min distance=0.4cm] (d)};
		\vertex[right=0.2cm of c] (x) {};
		\vertex[below=0.35cm of x] {$t_2$};
		\end{feynman}
		\end{tikzpicture}
		+
		\begin{tikzpicture}[inner sep=2pt,baseline=(b)]
		\begin{feynman}[inline=(a)]
		\vertex[dot] (a) {};
		\vertex[dot,right=0.4cm of a] (b) {};
		\vertex[dot,right=0.4cm of b] (c) {};
		\vertex[dot,right=0.4cm of c] (d) {};
		\vertex[dot,right=0.4cm of d] (e) {};
		\vertex[dot,right=0.4cm of e] (f) {};
		\diagram*{(a) --[half left,min distance=0.4cm] (f)};
		\diagram*{(b) --[half left,min distance=0.4cm] (c)}; 
		\diagram*{(d) --[half left,min distance=0.4cm] (e)};
		\vertex[right=0.2cm of c] (x) {};
		\vertex[below=0.35cm of x] {$1$};
		\end{feynman}
		\end{tikzpicture}
		+
		\begin{tikzpicture}[inner sep=2pt,baseline=(b)]
		\begin{feynman}[inline=(a)]
		\vertex[dot] (a) {};
		\vertex[dot,right=0.4cm of a] (b) {};
		\vertex[dot,right=0.4cm of b] (c) {};
		\vertex[dot,right=0.4cm of c] (d) {};
		\vertex[dot,right=0.4cm of d] (e) {};
		\vertex[dot,right=0.4cm of e] (f) {};
		\diagram*{(a) --[half left,min distance=0.4cm] (f)};
		\diagram*{(b) --[half left,min distance=0.4cm] (d)}; 
		\diagram*{(c) --[half left,min distance=0.4cm] (e)};
		\vertex[right=0.2cm of c] (x) {};
		\vertex[below=0.35cm of x] {$t_2$};
		\end{feynman}
		\end{tikzpicture}
		+
		\begin{tikzpicture}[inner sep=2pt,baseline=(b)]
		\begin{feynman}[inline=(a)]
		\vertex[dot] (a) {};
		\vertex[dot,right=0.4cm of a] (b) {};
		\vertex[dot,right=0.4cm of b] (c) {};
		\vertex[dot,right=0.4cm of c] (d) {};
		\vertex[dot,right=0.4cm of d] (e) {};
		\vertex[dot,right=0.4cm of e] (f) {};
		\diagram*{(a) --[half left,min distance=0.4cm] (f)};
		\diagram*{(b) --[half left,min distance=0.4cm] (e)}; 
		\diagram*{(c) --[half left,min distance=0.4cm] (d)};
		\vertex[right=0.2cm of c] (x) {};
		\vertex[below=0.35cm of x] {$1$};
		\end{feynman}
		\end{tikzpicture}
		\right)
	\end{split}
	\\
	\begin{split}\nonumber
		&=
		2^{N/2}\(\av{J^2}\binom{N}{q}\)^3
		\(5+6t_2+3t_3'+t_3\),
	\end{split}
\end{align}
\endgroup
where below each diagram we wrote its value. Note that different diagrams can yield the same value because of the cyclic property of the trace. In the evaluation of the sixth moment, two new diagrams arise, $t_3$ and $t_3'$. We can see that the diagram $t_3'$ corresponds to two copies of $t_2$ glued together---i.e., with fixed $\vb$, one can independently commute $\Gamma_\va$ and $\Gamma_\vc$ with $\Gamma_\vb$---and, therefore, we find
\begin{equation}
t_3'\equiv
\begin{tikzpicture}[inner sep=2pt,baseline=(b)]
\begin{feynman}[inline=(a)]
\vertex[dot] (a) {};
\vertex[dot,right=0.4cm of a] (b) {};
\vertex[dot,right=0.4cm of b] (c) {};
\vertex[dot,right=0.4cm of c] (d) {};
\vertex[dot,right=0.4cm of d] (e) {};
\vertex[dot,right=0.4cm of e] (f) {};
\diagram*{(a) --[half left,min distance=0.4cm] (c)};
\diagram*{(b) --[half left,min distance=0.4cm] (e)}; 
\diagram*{(d) --[half left,min distance=0.4cm] (f)};
\end{feynman}
\end{tikzpicture}
=2^{-N/2}\binom{N}{q}^{-3}\sum_{\va,\vb,\vc} \Tr\(
\Gamma_\va \Gamma_\vb \Gamma_\va \Gamma_\vc \Gamma_\vb \Gamma_\vc
\)
=t_2^2.
\end{equation}
On the other hand, the diagram $t_3$ cannot be reduced to a product of independent lower-order diagrams. It is instead given by~\cite{garcia-garcia2016PRD,garcia-garcia2018PRD}:
\begin{equation}\label{eq:diagram_t3_perfect_matching}
\begin{split}
t_3(q,N)&\equiv
\begin{tikzpicture}[inner sep=2pt,baseline=(b)]
\begin{feynman}[inline=(a)]
\vertex[dot] (a) {};
\vertex[dot,right=0.4cm of a] (b) {};
\vertex[dot,right=0.4cm of b] (c) {};
\vertex[dot,right=0.4cm of c] (d) {};
\vertex[dot,right=0.4cm of d] (e) {};
\vertex[dot,right=0.4cm of e] (f) {};
\diagram*{(a) --[half left,min distance=0.4cm] (d)};
\diagram*{(b) --[half left,min distance=0.4cm] (e)}; 
\diagram*{(c) --[half left,min distance=0.4cm] (f)};
\end{feynman}
\end{tikzpicture}
= 2^{-N/2} \binom{N}{q}^{-3} \sum_{\va,\vb,\vc} \Tr\(
\Gamma_\va \Gamma_\vb \Gamma_\vc \Gamma_\va \Gamma_\vb \Gamma_\vc \)
\\
&=\binom{N}{q}^{-2} 
\sum_{s=0}^q \sum_{r=0}^q \sum_{m=0}^r 
(-1)^{q+s+m}
\binom{q}{s}\binom{N-q}{q-s}\binom{s}{r-m}
\binom{2(q-s)}{m}\binom{N-2q+s}{q-r}.
\end{split}
\end{equation}
The computation of this diagram is as follows. We have to first commute $\Gamma_\va$ with the product $\Gamma_\vb\Gamma_\vc$ and then commute $\Gamma_\vb$ with $\Gamma_\vc$. Let $\Gamma_\vb$ and $\Gamma_\vc$ have $s$ factors in common and $\Gamma_\va$ have $r$ factors in common with the product $\Gamma_\vb\Gamma_\vc$, of which $(r-m)$ are in common with both $\Gamma_\vb$ and $\Gamma_\vc$. The sum over $\va$, $\vb$, and $\vc$ can then be expressed as a sum over the $q$ indices in $\vb$---which yields a factor $\binom{N}{q}$---and over $s$, $r$, and $m$. Now, with all indices in $\vb$ fixed we proceed as before. First, we fix the indices in $\vc$, choosing the $s$ indices in common with $\vb$ out of the $q$ possible values and the remaining $(q-s)$ indices in $\vc$ from the unused $(N-q)$ $\gamma$-matrices. Then, we fix the indices in $\va$, choosing the $(r-m)$ indices in common with both $\vb$ and $\vc$ out of the $s$ indices common to $\vb$ and $\vc$, the $m$ indices in common with only one of $\vb$ or $\vc$ from the $2(q-s)$ indices that $\vb$ and $\vc$ do not share, and the remaining $(q-r)$ indices in $\va$ from the spare $(N-2q+s)$ $\gamma$-matrices.
Finally, the commutation of $\Gamma_\va$ with $\Gamma_\vb\Gamma_\vc$ gives a factor $(-1)^{2q+m}$, while the commutation of $\Gamma_\vb$ with $\Gamma_\vc$ gives a factor $(-1)^{q+s}$.

In principle, one can compute $t(\pi)$ exactly for every diagram and for all $p$ as done for $t_2$ and $t_3$ in Eqs.~(\ref{eq:diagram_t2_perfect_matching}) and (\ref{eq:diagram_t3_perfect_matching}), respectively. However, the computations quickly become intractable (e.g., there are $7!!=105$ diagrams at the next order involving up to six independent commutations of $\Gamma$-matrices). Alternatively, one can, with certain approximations, give a simple combinatorial interpretation to the weight $t(\pi)$ of each diagram at leading and next-to-leading order in $1/N$. The latter becomes exact when $q\propto\sqrt{N}$~\cite{erdos2014MPAG,feng2019PMJ}.

To leading order in $1/N$ and fixed $q$, different $\Gamma$-matrices have no common $\gamma$-matrices. We thus ignore the commutations of the $\Gamma$-matrices altogether and replace all the traces by $1$. We then simply count the number of allowed diagrams (i.e., pair contractions) at each order. Now, these are exactly the moments of the normal distribution,
\begin{equation}
\frac{1}{\sigmaH^{2p}}\frac{\av{\Tr H^{2p}}}{\Tr\id}
=(2p-1)!!,
\end{equation}
and, therefore, to first-order in $1/N$, the spectral density of the SYK model is Gaussian.

To next-to-leading order---or exactly when $q\propto\sqrt{N}$---we take the commutations of $\Gamma$-matrices into account but ignore their correlations. The number of required commutations in a trace equals the number of crossings in the corresponding perfect-matching diagram. Since the trace with a single permutation was evaluated in Eq.~(\ref{eq:diagram_t2_perfect_matching}) and corresponds to diagram $t_2$, a diagram $\pi$ with $\mathrm{cross}(\pi)$ crossings is approximately given by $t_2^{\mathrm{cross}(\pi)}$.\footnote{The number of crossings of a perfect matching corresponds precisely to the number of crossings of lines in its diagram. This is no longer true for permutation diagrams.}%
\footnote{The number of perfect matchings in $\mathcal{M}_{2p}$ with $k$ crossings can be explicitly computed for arbitrary $(p,k)$ and is tabulated as sequence A067311 in the Online Encyclopedia of Integer Sequences (OEIS)~\cite{OEIS_A067311}.}
By summing over all diagrams, the moments are given by
\begin{equation}\label{eq:moments_combinatorics_SYK}
\frac{1}{\sigmaH^{2p}}\frac{\av{\Tr H^{2p}}}{\Tr\id}
=\sum_{\pi\in \sM_{2p}} t_2^{\mathrm{cross}(\pi)}.
\end{equation}
The sum on the right-hand side of Eq.~(\ref{eq:moments_combinatorics_SYK}) can be evaluated by the Touchard-Riordan formula~\cite{riordan1975MC}. Alternatively, it can be recognized as the $2p$th moment of the orthogonality weight of the $Q$-Hermite polynomials with $Q=t_2(q,N)$~\cite{ismail1987EJC}:
\begin{equation}\label{eq:spectral_density_QHermite}
\varrho_\mathrm{QH}(E;Q)
=(Q;Q)_\infty(-Q;Q)_\infty^2\frac{2}{\pi \EH}
\sqrt{1-E^2/\EH^2}
\prod_{k=1}^\infty
\(1-\frac{4E^2/\EH^2}{2+Q^{k}+Q^{-k}}\),
\end{equation}
supported on $-\EH\leq E\leq \EH$, where $\EH$ is the (dimensionless) ground-state energy of the SYK model given by
\begin{equation}\label{eq:E0_QHermite}
\EH=\frac{2}{\sqrt{1-Q}},
\end{equation}
and $(a;Q)_\infty=\prod_{k=0}^\infty \(1-aQ^k\)$ is the $Q$-Pochhammer symbol. Note that the spectral density is of the form of the Wigner semicircle distribution times a $Q$-dependant multiplicative correction.

}

%% file: Thesis_App_Asymptotics_WSYK.tex

\chapter{Simple asymptotic formulas for the WSYK spectral density}
\label{app:asymptotics}

In this Appendix, we derive the simple asymptotic formulas presented in the main text for the $Q$-Laguerre density~(\ref{eq:spectral_density_QLaguerre}) in the different regimes, namely, the bulk, $0\ll E\ll \EL$, the hard edge, $E\to0$, and the soft edge, $E\to\EL$, Eqs.~(\ref{eq:asympt_initial})--(\ref{eq:asympt_final}). The same calculation was done for the standard and supersymmetric SYK models in Refs.~\cite{cotler2017JHEP,garcia-garcia2017PRD} and~\cite{garcia-garcia2018PRD}, respectively. The computation has to be performed separately for positive and negative $Q$. For large enough $N$---the limit we are mostly interested in---$Q=t_3(\qb,N)$ is positive (resp.\ negative) for even (resp.\ odd) $\qb$.

\paragraph*{Positive \texorpdfstring{$Q$}{Q} (even \texorpdfstring{$\qb$}{qhat})}

We start by rewriting Eq.~(\ref{eq:spectral_density_QLaguerre}) as
\begin{equation}
\begin{split}
\log\varrho_{\mathrm{QL}}(E;Q)
&=\log C_Q+\frac{1}{2}\log\(\frac{1-E/\EL}{E/\EL}\)+
\sum_{k=0}^{+\infty}\log\left[
\frac{1-\frac{4\(1-2E/\EL\)^2}{\(Q^{k/2}+Q^{-k/2}\)^2}}{\(1-\frac{2\(1-2E/\EL\)}{Q^k+Q^{-k}}\)^2}
\right]
\\
&=\log C_Q+\frac{1}{2}
\sum_{k=-\infty}^{+\infty}\log\left[
1-\frac{\(1-2E/\EL\)^2}{\cosh^2\(k\log Q/2\)}
\right]
-\sum_{k=-\infty}^{+\infty}\log\left[
{1-\frac{1-2E/\EL}{\cosh\(k\log Q\)}}
\right],
\end{split}
\end{equation}
with
\begin{equation}\label{eq:C_Q} C_Q=\frac{(Q;Q)_\infty^2(-Q;Q)_\infty^2}{(-Q^2;Q^2)_\infty^2}
\frac{2}{\pi \EL}.
\end{equation}

Performing a Poisson resummation, we have
\begin{equation}
\begin{split}
\log\varrho_{\mathrm{QL}}(E;Q)&=
\log C_Q+
\sum_{n=-\infty}^{+\infty}\int_{0}^{\infty}\d x\, \cos\(2\pi n x\)\log\left[
1-\frac{\(1-2E/\EL\)^2}{\cosh^2\(x \log Q /2\)}
\right]\\
&-\sum_{n=-\infty}^{+\infty}\int_{0}^{\infty}\d x\, \cos\(2\pi n x\)\log\left[
1-\frac{1-2E/\EL}{\cosh\(x\log Q\)}
\right].
\end{split}
\end{equation}
Both integrals can be evaluated analytically~\cite{oberhettinger2012}, yielding
\begin{equation}
\begin{split}
&\log\varrho_{\mathrm{QL}}(E;Q)=\log C_Q\\
&-\frac{1}{2}\sum_{n=-\infty}^{+\infty}\frac{
	1-\cosh\left[
	\frac{4\pi n}{\log Q}\arcsin\(1-\frac{2E}{\EL}\)
	\right]-2\cosh\(\frac{\pi^2 n}{\log Q}\)+2\cosh\left[
	\frac{2\pi n}{\log Q}\arccos \(\frac{2E}{\EL}-1\)
	\right]	
}{n\sinh\(\frac{2\pi^2n}{\log Q}\)}.
\end{split}
\end{equation}
We evaluate the $n=0$ term in the sum by taking the limit of the summand as $n\to0$. Then, using the fact that the summand is an even function of $n$, we can rewrite the spectral density as:
\begin{equation}\label{eq:spectral_density_Poisson_resum}
\begin{split}
&\varrho_{\mathrm{QL}}(E;Q)=
C'_Q\exp\left[
\frac{2\arcsin^2\(1-\frac{2E}{\EL}\)-\arccos^2\(\frac{2E}{\EL}-1\)}{\log Q}
\right]\\
&\times\exp\left[-\sum_{n=1}^{\infty}\frac{
	1-\cosh\left[
	\frac{4\pi n}{\log Q}\arcsin\(1-\frac{2E}{E_0}\)
	\right]-2\cosh\(\frac{\pi^2 n}{\log Q}\)+2\cosh\left[
	\frac{2\pi n}{\log Q}\arccos \(\frac{2E}{E_0}-1\)
	\right]	
}{n\sinh\(\frac{2\pi^2n}{\log Q}\)}\right],
\end{split}
\end{equation}
where $C'_Q=C_Q\exp[\pi^2/(4\log Q)]$.

Up to this point, the computation is exact and Eqs.~(\ref{eq:spectral_density_QLaguerre}) and (\ref{eq:spectral_density_Poisson_resum}) are equivalent. For large $N$, we have $Q\to 1^-$ and we can replace the hyperbolic functions by single exponents, i.e., $\cosh(x/\log Q)\approx (1/2)\exp(-\abs{x}/\log Q)$ and $\sinh(x/\log Q)\approx -(1/2)\exp(-\abs{x}/\log Q)$. In this limit, and using the Taylor expansion of $\log(1-x)$, Eq.~(\ref{eq:spectral_density_Poisson_resum}) reads as
\begin{equation}\label{eq:spectral_density_Laguerre_approx}
\begin{split}
&\varrho_{\mathrm{QL}}(E;Q)
\\&\approx C''_Q 
\exp\left[
\frac{2\arcsin^2\(1-\frac{2E}{\EL}\)-\arccos^2\(\frac{2E}{\EL}-1\)}{\log Q}\right]
\frac{
	1-\exp\left[
	\frac{4\pi}{\log Q}
	\arccos\abs{\(1-\frac{2E}{\EL}\)}
	\right]}{\(
	1-\exp\left[
	\frac{2\pi}{\log Q}
	\arccos\(1-\frac{2E}{\EL}\)
	\right]\)^2},
\end{split}
\end{equation}
with
\begin{equation}
C''_Q=
\frac{C'_Q}{\(1+\exp[\pi^2/\log Q]\)^2}
=\frac{C_Q\exp\left[\pi^2/\(4\log Q\)\right]}{\(1+\exp[\pi^2/\log Q]\)^2}.
\end{equation} Equation~(\ref{eq:spectral_density_Laguerre_approx}) approximates the $Q$-Laguerre spectral density~(\ref{eq:spectral_density_QLaguerre}) extremely well, even for relatively small system sizes. For instance, the two expressions are within $0.5\%$ of one another throughout their support for $\qb=2$ and $N=16$, i.e., $Q=t_3(2,16)$, while their relative deviation is below $10^{-10}\%$ for $Q=t_3(2,40)$. 

We can further simplify Eq.~(\ref{eq:spectral_density_Laguerre_approx}), depending on the value of $E/\EL$. For $E$ well inside the bulk, $0\ll E\ll\EL$, we can safely ignore the second term in Eq.~(\ref{eq:spectral_density_Laguerre_approx}) and we obtain the asymptotic formula for the bulk density:
\begin{equation}\label{eq:spectral_density_bulk}
\varrho_{Q>0}^{(\mathrm{bulk})}(E;Q)=C''_Q 
\exp\left[
\frac{2\arcsin^2\(1-\frac{2E}{\EL}\)-\arccos^2\(\frac{2E}{\EL}-1\)}{\log Q}\right].
\end{equation}

Close to the hard edge, $E=0$, we have to expand $\arcsin x\approx \pi/2-\sqrt{2(1-x)}$ around $x=1$ and $\arccos x\approx \pi-\sqrt{2(1+x)}$ around $x=-1$. Inserting these expansions in Eq.~(\ref{eq:spectral_density_Laguerre_approx}), we obtain the asymptotic hard-edge density,
\begin{equation}\label{eq:spectral_density_hard}
\varrho_{Q>0}^{(\mathrm{hard})}(E;Q)=
C''_Q 
\exp\left[-\frac{\pi^2}{2\log Q}\right]
\coth\left[-\frac{2\pi}{\log Q}\sqrt{\frac{E}{\EL}} \right].
\end{equation}

Finally, near the soft edge, $E=\EL$, we have to expand $\arcsin x\approx -\pi/2+\sqrt{2(1+x)}$ around $x=-1$ and $\arccos x\approx \sqrt{2(1-x)}$ around $x=1$. Inserting these expansions in Eq.~(\ref{eq:spectral_density_Laguerre_approx}), we obtain the asymptotic soft-edge density:
\begin{equation}\label{eq:spectral_density_soft}
\varrho_{Q>0}^{(\mathrm{soft})}(E;Q)= 
C''_Q\exp\left[\frac{\pi^2}{2\log Q}\right]
\,2\sinh\left[-\frac{4\pi}{\log Q}\sqrt{1-\frac{E}{\EL}}\right].
\end{equation}

\paragraph*{Negative \texorpdfstring{$Q$}{Q} (odd \texorpdfstring{$\qb$}{qhat})}

We now turn to negative $Q$. The computation proceeds similarly to positive $Q$, but one has to treat separately the product factors with even and odd $k$ in the $Q$-Laguerre spectral density. For negative $Q$, $Q=-\abs{Q}$ and we can rewrite Eq.~(\ref{eq:spectral_density_QLaguerre}) as
\begin{equation}
\begin{split}
\log\varrho_{\mathrm{QL}}(E;&Q)
\\
=\log C_Q
&+\frac{1}{2}\sum_{k=-\infty}^{+\infty}\log\left[
\frac{1-\frac{4\(1-2E/\EL\)^2}{\(\abs{Q}^{k}+\abs{Q}^{-k}\)^2}}{\(1-\frac{2\(1-2E/\EL\)}{\abs{Q}^{2k}+\abs{Q}^{-2k}}\)^2}
\right]
+\frac{1}{2}\sum_{k=-\infty}^{+\infty}\log\left[
\frac{1+\frac{4\(1-2E/\EL\)^2}{\(\abs{Q}^{k-1/2}-\abs{Q}^{-k+1/2}\)^2}}{\(1+\frac{2\(1-2E/\EL\)}{\abs{Q}^{2k-1}+\abs{Q}^{-2k+1}}\)^2}
\right]
\\
=\log C_Q
&+\frac{1}{2}\sum_{k=-\infty}^{+\infty}\log\left[
1-\frac{\(1-2E/\EL\)^2}{\cosh^2\(k\log \abs{Q}\)}
\right]
+\frac{1}{2}\sum_{k=-\infty}^{+\infty}\log\left[
1+\frac{\(1-2E/\EL\)^2}{\sinh^2\((k-1/2)\log \abs{Q}\)}
\right]
\\
&-\sum_{k=-\infty}^{+\infty}\log\left[
{1-\frac{1-2E/\EL}{\cosh\(2k\log \abs{Q}\)}}
\right]
-\sum_{k=-\infty}^{+\infty}\log\left[
{1+\frac{1-2E/\EL}{\cosh\((2k-1)\log \abs{Q}\)}}
\right],
\end{split}
\end{equation}
where $C_Q$ is again given by Eq.~(\ref{eq:C_Q}). As before, we perform a Poisson resummation, obtaining:
\begin{equation}
\begin{split}
\log&\varrho_{\mathrm{QL}}(E;Q)
=\log C_Q
+\sum_{n=-\infty}^{+\infty}\int_0^{\infty}\d x
\cos\(2\pi n x\)
\log\left[
1-\frac{\(1-2E/\EL\)^2}{\cosh^2\(x\log \abs{Q}\)}
\right]
\\
&+\sum_{\substack{n=-\infty\\n\neq 0}}^{+\infty} (-1)^n
\int_0^{\infty}\d x \cos\(2\pi n x\)
\log\left[
1+\frac{\(1-2E/\EL\)^2}{\sinh^2\(x\log \abs{Q}\)}
\right]
+\int_0^{\infty}\d x \log\left[
1+\frac{\(1-2E/\EL\)^2}{\sinh^2\(x\log \abs{Q}\)}
\right]
\\
&-2\sum_{n=-\infty}^{+\infty}\int_0^{\infty}\d x
\cos\(2\pi n x\)\log\left[
{1-\frac{1-2E/\EL}{\cosh\(2x\log \abs{Q}\)}}
\right]
\\
&-2\sum_{n=-\infty}^{+\infty}(-1)^n
\int_0^{\infty}\d x \cos\(2\pi n x\)\log\left[
{1+\frac{1-2E/\EL}{\cosh\(2x\log \abs{Q}\)}}
\right].
\end{split}
\end{equation}
The first, fourth, and fifth integrals were already performed for positive $Q$\footnote{Note that the change of variables $x\to2x$ can be effected by changing $n\to n/2$ in the result of the integration.}, while the second and third can also be evaluated analytically~\cite{garcia-garcia2018JHEP}. We also isolated the $n=0$ term from the second integral, as its values cannot be obtained from the general-$n$ result as the limit $n\to0$. Performing the integrations, the spectral density is exactly rewritten as:
\begin{equation}\label{eq:spectral_density_Poisson_resum_negativeQ}
\begin{split}
&\varrho_{\mathrm{QL}}(E;Q)
\\
&=C'_\abs{Q}
\exp\left[
\frac{2\arcsin^2\(1-\frac{2E}{\EL}\)-\frac{1}{2}\arccos^2\(1-\frac{2E}{\EL}\)-\frac{1}{2}\arccos^2\(\frac{2E}{\EL}-1\)-\pi \abs{\arcsin\(1-\frac{2E}{\EL}\)}}{\log \abs{Q}}
\right]
\\
&\times\exp\left[-\sum_{n=1}^{\infty}\frac{
	1-\cosh\left[
	\frac{2\pi n}{\log \abs{Q}} \arcsin\(1-\frac{2E}{E_0}\)
	\right]-2\cosh\(\frac{\pi^2 n}{2\log \abs{Q}}\)+2\cosh\left[
	\frac{\pi n}{\log \abs{Q}}
	\arccos \(\frac{2E}{E_0}-1\)
	\right]	
}{n\sinh\(\frac{\pi^2n}{\log \abs{Q}}\)}\right]
\\
&\times\exp\left[
2\sum_{n=1}^{\infty}(-1)^n\frac{\cosh\(\frac{\pi^2 n}{2\log \abs{Q}}\)-\cosh\left[
	\frac{\pi n}{\log \abs{Q}}
	\arccos \(1-\frac{2E}{E_0}\)
	\right]}{n\sinh\(\frac{\pi^2n}{\log \abs{Q}}\)}
\right]
\\
&\times\exp\left[
\sum_{n=1}^{\infty}\frac{(-1)^n}{n}\(
1-\exp\left\{\frac{2\pi n}{\log \abs{Q}}\abs{\arcsin\(1-\frac{2E}{\EL}\)}
\right\}\)\right],
\end{split}
\end{equation}
with $C'_\abs{Q}=C_Q\exp[\pi^2/(4\log\abs{Q})]$. Next, in the large-$N$ limit, we again replace the hyperbolic functions by single exponents and use the Taylor expansion of $\log(1-x)$ to approximate the $Q$-Laguerre density by
\begin{equation}\label{eq:spectral_density_Laguerre_approx_Qneg}
\begin{split}
\varrho_{\mathrm{QL}}(E;Q)\approx C'_\abs{Q}
&\cosh\left[
\frac{\pi}{\log\abs{Q}}
\abs{\arcsin\(1-\frac{2E}{\EL}\)}
\right]
\\
\times&\exp\left[
\frac{2\arcsin^2\(1-\frac{2E}{\EL}\)-\frac{1}{2}\arccos^2\(1-\frac{2E}{\EL}\)-\frac{1}{2}\arccos^2\(\frac{2E}{\EL}-1\)}{\log \abs{Q}}
\right]
\\
\times&\frac{
	1-\exp\left[
	\frac{2\pi}{\log\abs{Q}}\arccos\abs{1-\frac{2E}{\EL}}
	\right]}{
	\(
	1-\exp\left[
	\frac{\pi}{\log\abs{Q}}\arccos\(1-\frac{2E}{\EL}\)
	\right]\)^2
	\(
	1-\exp\left[
	\frac{\pi}{\log\abs{Q}}\arccos\(\frac{2E}{\EL}-1\)
	\right]\)^2
}.
\end{split}
\end{equation}
This approximate formula once again describes extremely well the $Q$-Laguerre density. For example, for $\qb=3$ and $N=64$, Eqs.~(\ref{eq:spectral_density_QLaguerre}) and (\ref{eq:spectral_density_Laguerre_approx_Qneg}) are within $0.01\%$ of one another.

Finally, we give the negative-$Q$ asymptotic spectral densities for the bulk, hard edge, and soft edge. As before, the bulk spectral density is well approximated by dropping the last term in Eq.~(\ref{eq:spectral_density_Laguerre_approx_Qneg}):
\begin{equation}
\begin{split}
\varrho_{Q<0}^{\mathrm{(bulk)}}=
C'_\abs{Q}
&\cosh\left[
\frac{\pi}{\log\abs{Q}}
\abs{\arcsin\(1-\frac{2E}{\EL}\)}
\right]
\\
\times&\exp\left[
\frac{2\arcsin^2\(1-\frac{2E}{\EL}\)-\frac{1}{2}\arccos^2\(1-\frac{2E}{\EL}\)-\frac{1}{2}\arccos^2\(\frac{2E}{\EL}-1\)}{\log \abs{Q}}
\right].
\end{split}
\end{equation} 
Expanding the $\arcsin$ and $\arccos$ around $E\approx 0$ and $E\approx \EL$, we obtain the hard- and soft-edge asymptotic densities, respectively:
\begin{align}
\varrho_{Q<0}^{\mathrm{(hard)}}=
C'_\abs{Q}
&\coth\left[
-\frac{\pi}{\log\abs{Q}}\sqrt{\frac{E}{\EL}}
\right]
\exp\left[
-\frac{2\pi}{\log \abs{Q}}\sqrt{\frac{E}{\EL}}
\right]
\cosh\left[
\frac{\pi^2}{2\log\abs{Q}}\(1-\frac{4}{\pi}\sqrt{\frac{E}{\EL}}\)\right],
\\
\begin{split}
\varrho_{Q<0}^{\mathrm{(soft)}}=
C'_\abs{Q}
&\tanh\left[
-\frac{\pi}{\log\abs{Q}}\sqrt{1-\frac{E}{\EL}}
\right]
\exp\left[
-\frac{2\pi}{\log \abs{Q}}\sqrt{1-\frac{E}{\EL}}
\right]
\\
\times
&\cosh\left[
\frac{\pi^2}{2\log\abs{Q}}\(1-\frac{4}{\pi}\sqrt{1-\frac{E}{\EL}}\)\right].
\end{split}
\end{align}